\documentclass[a4paper,12pt,twoside]{report}
\usepackage[left=2cm,right=2cm,top=2cm,bottom=3cm]{geometry}
\usepackage[ruled,vlined]{algorithm2e}
\usepackage{tikz}
\usetikzlibrary{arrows,shapes}
\usetikzlibrary{decorations.markings}
\usetikzlibrary{chains}
\usepackage[bottom]{footmisc}
\usepackage{url}
\usepackage[nottoc]{tocbibind}
\usepackage{enumitem}
\graphicspath{{img/}} 
\usepackage[fleqn]{amsmath}
\usepackage{amssymb}
\usepackage{lineno,hyperref}
\usepackage{graphicx}
\usepackage{adjustbox}
\usepackage[absolute,overlay]{textpos}
\usepackage{multirow, bigstrut}
\usepackage{hhline}
\usepackage{graphicx}
\usepackage{verbatim}
\usepackage{latexsym}
\usepackage{mathchars}
\usepackage{setspace}

\input{blocked.sty}
\input{uhead.sty}
\input{boxit.sty}
\input{icthesis.sty}

\newcommand{\NN}{{\sf I\kern-0.14emN}}   
\newcommand{\ZZ}{{\sf Z\kern-0.45emZ}}   
\newcommand{\QQQ}{{\sf C\kern-0.48emQ}}   
\newcommand{\RR}{{\sf I\kern-0.14emR}}   

\newcommand{\normallinespacing}{\renewcommand{\baselinestretch}{1.5} \normalsize}

\newcommand{\syncc}{~\stackrel{\textstyle \rhd\kern-0.57em\lhd}{\scriptstyle L}~}

\usepackage{color,soul}

\makeatletter
\DeclareRobustCommand{\textsupsub}[2]{{%
  \m@th\ensuremath{%
    ^{\mbox{\fontsize\sf@size\z@#1}}%
    _{\mbox{\fontsize\sf@size\z@#2}}%
  }%
}}
\makeatother

\begin{document}

\title{Connectivity-Driven Parcellation Methods for the Human Cerebral Cortex}

\author{Salim Arslan}
\submitdate{October 2017}

\normallinespacing
\maketitle

\preface


\begin{abstract}

The macro connectome elucidates the pathways through which brain regions are structurally connected or functionally coupled to perform cognitive functions. It embodies the notion of representing, analysing, and understanding all connections within the brain as a network, while the subdivision of the brain into interacting cortical units is inherent in its architecture. As a result, the definition of network nodes is one of the most critical steps in connectivity network analysis. Parcellations derived from anatomical brain atlases or random parcellations are traditionally used for node identification, however these approaches do not always fully reflect the functional/structural organisation of the brain. Connectivity-driven methods have arisen only recently, aiming to delineate parcellations that are more faithful to the underlying connectivity. Such parcellation methods face several challenges, including but not limited to poor signal-to-noise ratio, the curse of dimensionality, and functional/structural variations inherent in individual brains, which are only limitedly addressed by the current state of the art.

In this thesis, we present robust and fully-automated methods for the subdivision of the entire human cerebral cortex based on connectivity information. Our contributions are four-fold: First, we propose a clustering approach to delineate a cortical parcellation that provides a reliable abstraction of the brain's functional organisation. Second, we cast the parcellation problem as a feature reduction problem and make use of manifold learning and image segmentation techniques to identify cortical regions with distinct structural connectivity patterns. Third, we present a multi-layer graphical model that combines within- and between-subject connectivity, which is then decomposed into a cortical parcellation that can represent the whole population, while accounting for the variability across subjects. Finally, we conduct a large-scale, systematic comparison of existing parcellation methods, with a focus on providing some insight into the reliability of brain parcellations in terms of reflecting the underlying connectivity, as well as, revealing their impact on network analysis.

We evaluate the proposed parcellation methods on publicly available data from the Human Connectome Project and a plethora of quantitative and qualitative evaluation techniques investigated in the literature. Experiments across multiple resolutions demonstrate the accuracy of the presented methods at both subject and group levels with regards to reproducibility and fidelity to the data. The neuro-biological interpretation of the proposed parcellations is also investigated by comparing parcel boundaries with well-structured properties of the cerebral cortex. Results show the advantage of connectivity-driven parcellations over traditional approaches in terms of better fitting the underlying connectivity. However, the benefit of using connectivity to parcellate the brain is not always as clear regarding the agreement with other modalities and simple network analysis tasks carried out across healthy subjects. Nonetheless, we believe the proposed methods, along with the systematic evaluation of existing techniques, offer an important contribution to the field of brain parcellation, advancing our understanding of how the human cerebral cortex is organised at the macroscale.

\end{abstract}

\cleardoublepage


\begin{acknowledgements}

I would like to first of all thank my supervisor Daniel Rueckert for giving me the opportunity to undertake my PhD in the BioMedIA group and for his support, guidance, and inspiration throughout this journey. I would like to express my gratitude to Ben Glocker for his comments and feedback during the preparation of this thesis, as well as, Yi-ke Guo and Georg Langs for their time and commitment in the examination process. Many thanks to my colleagues at Imperial College London for providing a great atmosphere and research environment, in particular to Sofia Ira Ktena and Sarah Parisot for their valuable inputs and contributions. Special thanks go to Amani El-Kholy for always being there to sort things out. 

I am also extremely grateful to my family for all the support and encouragement, especially during hard times. Finally, I thank my beloved wife, Dilara, for \textit{everything} - nothing would have been possible without her being by my side. She is the real mvp...



\end{acknowledgements}
\clearpage

\vspace*{\fill}
\begin{originality}
I, Salim Arslan, hereby declare that the work described in this thesis is my own, except where specifically acknowledged.

\end{originality}
\vfill 
\clearpage

\vspace*{\fill}
\begin{copyright}
The copyright of this thesis rests with the author and is made available under a Creative Commons Attribution Non-Commercial No Derivatives license. Researchers are free to copy, distribute or transmit the thesis on the condition that they attribute it, that they do not use it for commercial purposes and that they do not alter, transform or build upon it. For any reuse or redistribution, researchers must make clear to others the license terms of this work.
\end{copyright}
\vfill 
\clearpage
\clearpage
\vspace*{\fill}
\begin{center}
\textit{To Dilara, my amazing wife}
\end{center}
\vfill 
\clearpage
\chapter*{Acronyms}

\textbf{2D} Two-dimensional

\textbf{3D} Three-dimensional

\textbf{dMRI} Diffusion Magnetic Resonance Imaging

\textbf{fMRI} Functional Magnetic Resonance Imaging

\textbf{rs-fMRI} Resting-State Functional Magnetic Resonance Imaging

\textbf{t-fMRI} Task Functional Magnetic Resonance Imaging

\textbf{ACC} Anterior Cingular Cortex

\textbf{ARI} Adjusted Rand Index

\textbf{BA} Brodmann's Area

\textbf{BIC} Bayesian Information Criterion 

\textbf{BOLD} Blood Oxygenation Level Dependent

\textbf{CDP} Connectivity-Driven Parcellation

\textbf{DTI} Diffusion Tensor Imaging 

\textbf{DWI} Diffusion Weighted Imaging 

\textbf{EEG} Electroencephalography 

\textbf{EPI} Echo-planar Imaging

\textbf{GM} Gray Matter

\textbf{HARDI} High Angular Resolution Diffusion Imaging 

\textbf{HCP} Human Connectome Project

\textbf{ICA} Independent Component Analysis

\textbf{MEG} Magnetoencephalography

\textbf{MFC} Medial Frontal Cortex 

\textbf{MNI} Montreal Neurological Institute

\textbf{MRF} Markov Random Field 

\textbf{MRI} Magnetic Resonance Imaging

\textbf{PCA} Principle Component Analysis

\textbf{PET} Positron Emission Tomography

\textbf{ROI} Region of Interest

\textbf{RF} Radio Frequency 

\textbf{RSFC} Resting-State Functional Connectivity

\textbf{RSN} Resting-State Network

\textbf{SAD} Sum of Absolute Differences

\textbf{SC} Silhouette Coefficient

\textbf{SMA} Supplementary Motor Area 

\textbf{SNR} Signal-to-Noise Ratio

\textbf{WM} White Matter


\body

\chapter{Introduction}
\label{chapter:introduction}

\section{Motivation}
Understanding the brain's behaviour and function has been a prominent and ongoing research subject for over a century~\cite{Zilles10,Sporns11}. Neuronal interconnections constitute the primary means of information transmission within the brain and are strongly related to the way the brain functions~\cite{Passingham02,Power11,Smith13}. These connections constitute a complex network that can be estimated at the macroscale via modern neuroimaging techniques such as Magnetic Resonance Imaging (MRI)~\cite{Sporns05,craddock2013imaging,VanEssen13}. While structural connectivity networks are typically inferred from diffusion MRI (dMRI), functional networks can be mapped using resting-state functional MRI (rs-fMRI)~\cite{Honey09,Eickhoff15}. The former allows estimation of the physical (anatomical) connections, while the latter elucidates putative functional connections between spatially remote brain regions. 

Analysing brain connectivity from a network theoretical point of view has shown significant potential for identifying organisational principles in the brain and their connections to cognitive procedures~\cite{Mantini07,Smith09,bullmore2009complex,Sporns11} and brain disorders, such as Alzheimer's disease~\cite{Supekar08,lo2010diffusion}, attention-deficit/hyperactivity disorder~\cite{wang2009altered} and schizophrenia~\cite{Basset08}. This allows to study the brain and its function from a new perspective that accounts for the complexity of its architecture. One of the critical steps in the construction of brain connectivity networks is the definition of network nodes~\cite{Sporns11,Eickhoff15}. Adopting a vertex- or voxel-based representation yields networks that are very noisy and of extremely high dimensionality, making subsequent network analysis steps often intractable~\cite{Thirion14}. An alternative approach to node definition is to parcellate the cerebral cortex into a set of distinct regions of interest (ROI), where each ROI (i.e. parcel) corresponds to a node of the connectivity network. This further allows to reduce the complexity of connectivity, an aspect that is highly critical for the study of brain dynamics with whole-brain models~\cite{craddock2013imaging}.

Traditionally, parcellations derived from neuro-anatomical landmarks~\cite{TzourioMazoyer02,Desikan06,Fischl04} or micro-structural features~\cite{Brodmann09,Vogt19,Nieuwenhuys13} have been used to define ROIs for network analysis~\cite{Sporns11}. Whereas such parcellations are of great importance in order to derive neuro-biologically meaningful brain atlases, they might fail to fully reflect the intrinsic organisation of the brain and capture the functional/structural variability inherent in individual subjects, due to brain maturation or pathology~\cite{elbert2004reorganization,Langs14,Wang15,Gordon16individual}. In addition, they are typically generated on a single or small set of individuals, which can make them biased and unable to accurately represent population variability or adapt to new subjects~\cite{Yeo11,Craddock12}. Alternative approaches include use of random parcellations for node definition~\cite{Sporns11}; however, this kind of approaches could fail to represent the underlying cortical organisation faithfully and may lead to ill-defined nodes in the constructed network~\cite{Smith11}.

\begin{figure}[!t]
\centering
\includegraphics[width=0.8\textwidth]{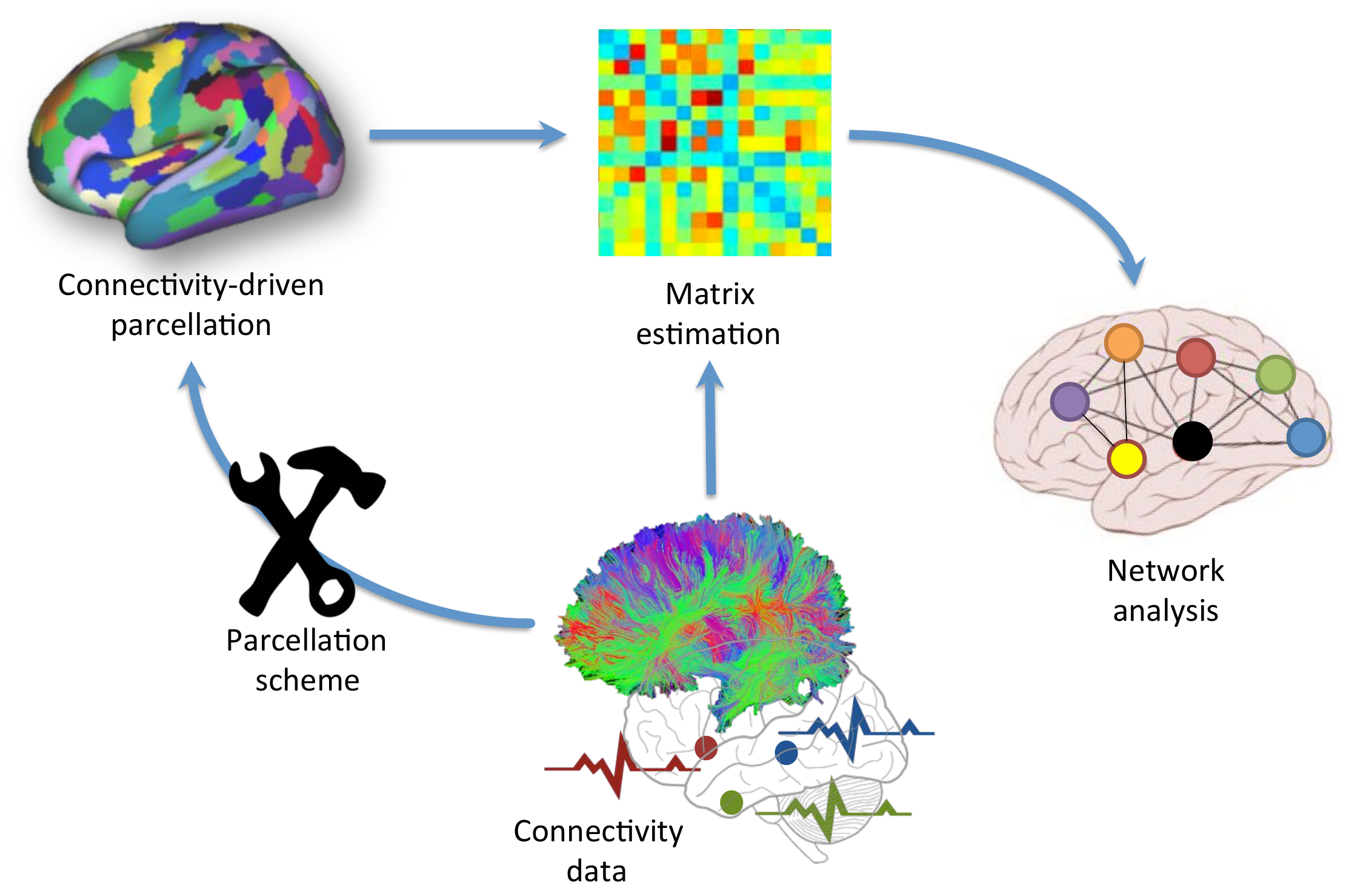} 
\caption{A typical network analysis pipeline driven by a connectivity-based parcellation. }
\label{fig:network_anal_illustration}
\end{figure}

More recent parcellation methods attempt to overcome these problems by using connectivity information, captured from rs-fMRI or dMRI data, to derive a set of network nodes for connectivity analysis~\cite{Thirion14,Eickhoff15} as illustrated in Fig.~\ref{fig:network_anal_illustration}. This type of approaches typically casts the parcellation problem as a clustering problem, in which a connectivity profile is first computed for each individual subunit (e.g. vertices or voxels). These connectivity profiles are then submitted to a parcellation scheme for grouping subunits, such that the connectivity is similar for subunits within the same cluster, but different between clusters~\cite{Eickhoff15}. Since connectivity-based parcellations are directly obtained from the underlying data, such methods can potentially provide highly homogeneous ROIs and separate regions with different patterns of connectivity more accurately. As a result, they are likely to yield a more reliable set of network nodes, as these nodes are typically represented by a single entity (such as the average connectivity profile) in network analysis~\cite{Shen13,Gordon16}.

\begin{figure}[!t]
\centering
\begin{tabular}{cccc}
\includegraphics[width=0.23\textwidth]{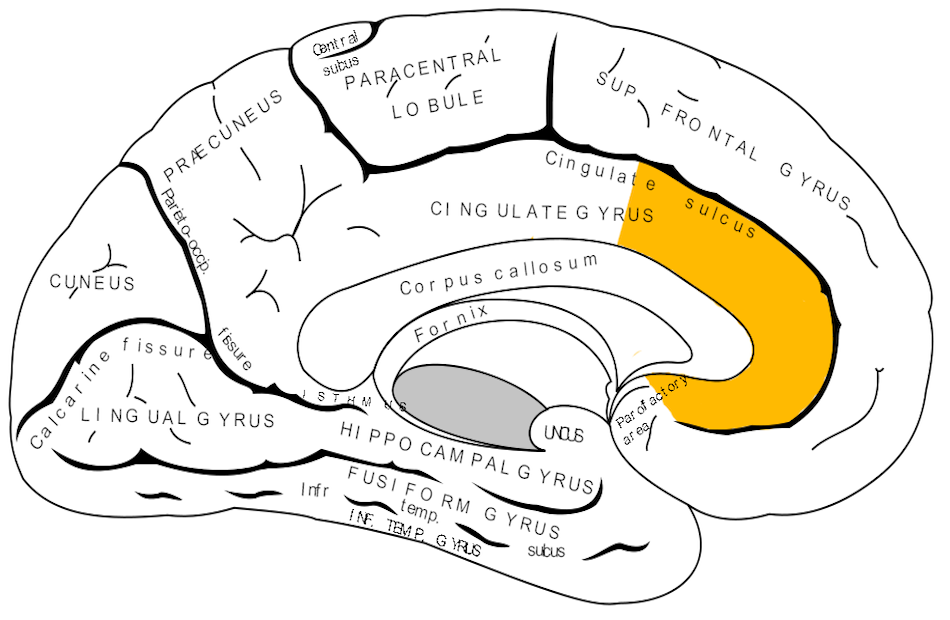} &
\includegraphics[width=0.22\textwidth]{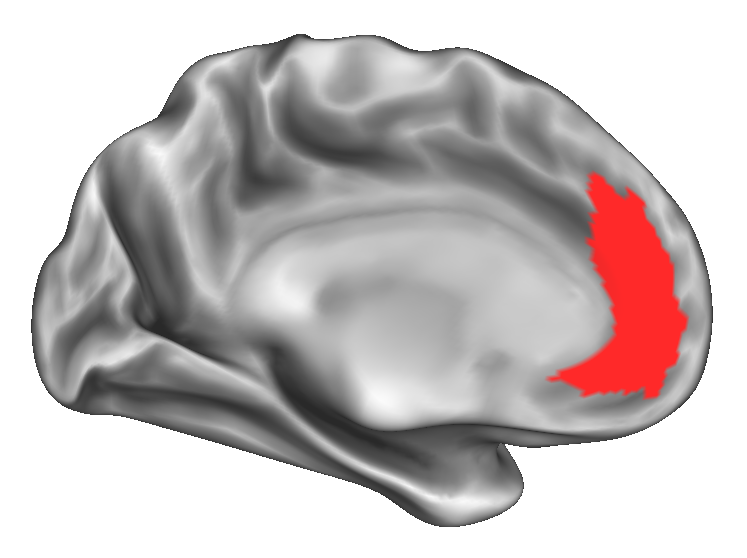} &
\includegraphics[width=0.22\textwidth]{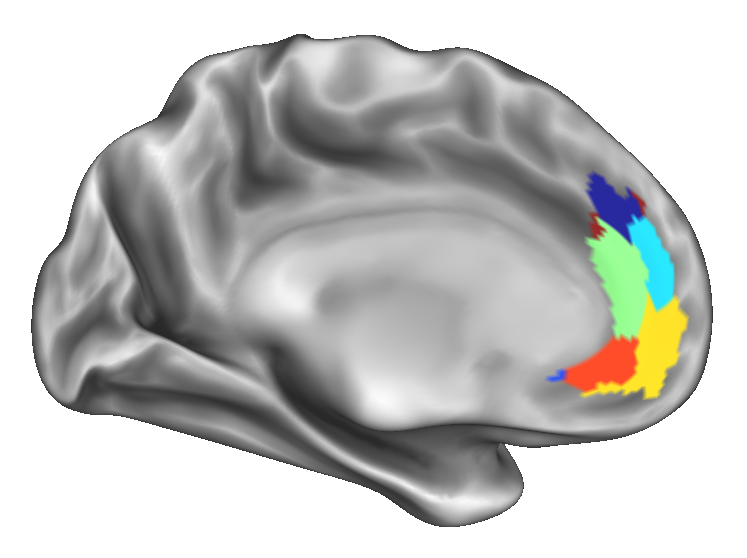} &
\includegraphics[width=0.22\textwidth]{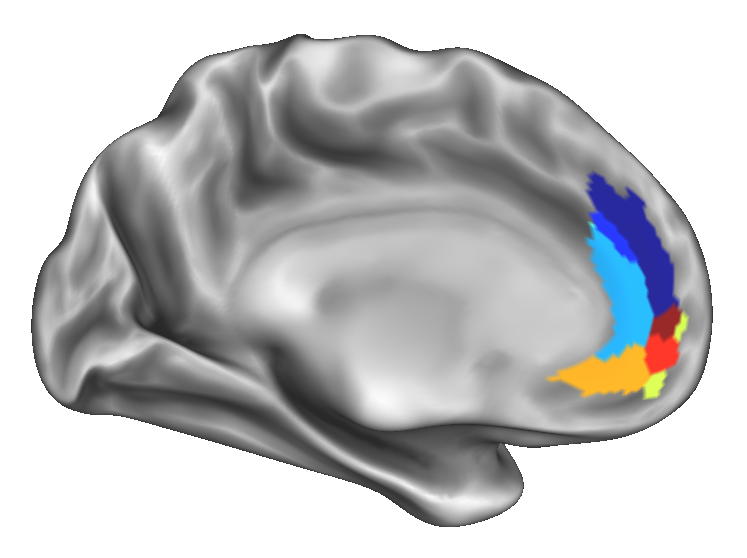} \\
(a) & (b) & (c) & (d) 
\end{tabular}
\caption[Parcellation of anterior cingulate cortex based on different properties.]{(a) Medial surface of left cerebral hemisphere, with anterior cingulate cortex (ACC) is highlighted. (b) ACC as delineated by a cortical-folding based anatomical atlas~\cite{Fischl04}. The cerebral organisation of the same area based on (c) resting-state functional connectivity and (d) dMRI-based structural connectivity, obtained with the connectivity-driven approaches,~\cite{Arslan15b} and~\cite{Arslan16}, respectively. Each colour indicates a distinct region, with homogeneous features (e.g. connectivity).}
\label{fig:acc}
\end{figure}

The role of connectivity in human brain mapping studies is also crucial, as it provides complementary information for subdividing the cortex into anatomically and functionally distinct regions~\cite{Passingham02}. In particular, the functional operations performed by a cortical area are thought to simultaneously depend on its local micro-architecture and connectivity~\cite{Passingham02,Eickhoff15}. As a result, most of the neuro-anatomical parcellations alone cannot precisely match the degree of functional segregation of the cerebral cortex~\cite{Nieuwenhuys13,Glasser16}. For example, the anterior cingulate cortex (ACC), which is involved in certain brain functions such as cognition and emotion~\cite{stevens2011anterior}, is typically represented as a single ROI in anatomical brain atlases~\cite{Talairach88,TzourioMazoyer02,Fischl04} (Fig.~\ref{fig:acc}); however, it exhibits a great amount of heterogeneity in structural~\cite{Beckmann09} and functional connectivity~\cite{Margulies07}. As a result, when ACC is localised from a connectivity-driven parcellation, it typically consists of several subregions with varying shape and size, as shown in Fig.~\ref{fig:acc}(c-d). In general, connectivity-driven parcellations provide a greater flexibility to study the brain function, as they enable the segregation of the cortex at different scales, as opposed to anatomical atlases with fixed resolutions.

Despite many attempts to parcellate the brain with respect to connectivity, the problem is still open to improvements. This is primarily due to the fact that, like all other clustering problems, the parcellation problem is ill-posed, thus, obtaining accurate subdivisions of the cortex depends on the proposed method's fidelity to the underlying data, as well as its capability to encapsulate valuable information from the naturally complex, noisy, and high-dimensional connectivity patterns in the brain~\cite{Parisot15,Eickhoff15}. The heterogeneity of the population under investigation, i.e. inter-subject variability, also possesses an additional challenge, especially regarding the identification of shared patterns of connectivity and the delineation of spatially coherent cortical regions across different subjects~\cite{Blumensath13,Langs15,Wang2014functional}.

\section{Research Contributions and Thesis Outline}
The aim of this thesis is to develop robust and reliable methods for subdividing the cerebral cortex into spatially contiguous, non-overlapping, and distinct regions with respect to underlying connectivity. Parcellations obtained by the proposed methods can provide high-level abstractions of the functional specialisation and segregation in the cerebral cortex. Such parcellations can further be used to define the network nodes in connectivity analysis, thus, help better understand how connectivity changes through development, ageing, and neurological disorders. In addition, our parcellations allow to study the brain's cortical organisation from a multi-scale perspective through the subdivision of the cerebral cortex at varying levels of granularity.  

Chapter~\ref{chapter:background} provides a brief overview of the neuro-biological basis of the brain with a special emphasis on the cerebral cortex, which is responsible for high-level brain functions such as language and memory. It continues with an introduction to imaging of the brain at the macroscale, particularly focusing on rs-fMRI and dMRI, two mostly-used techniques to capture connectivity non-invasively. After summarising the quantitative methods for estimating structural and functional connectivity from the MRI data, we describe the data used to develop and evaluate the proposed parcellation methods throughout the thesis.

Chapter~\ref{chapter:literature} briefly covers the historical foundations of brain parcellation and reviews techniques that allow the segregation of the cerebral cortex based on information obtained from different neuro-anatomical properties, with a particular focus on the connectivity-driven approaches. Both Chapter~\ref{chapter:background} and~\ref{chapter:literature} describe the challenges for obtaining reliable connectivity-driven parcellations, such as low signal-to-noise ratio (SNR) and high dimensionality, as well as, limitations induced by imaging techniques and pre-processing pipelines prior to parcellation. 

Chapter~\ref{chapter:multi-level} presents a novel method for the subject-specific parcellation of the cerebral cortex using resting-state functional connectivity (RSFC). It is based on the idea of combining two different clustering strategies in a parcellation framework that allows subdividing the cortex at varying levels of detail. The chapter also provides an extensive comparison of different subject-level parcellation methods with respect to their fidelity to the underlying connectivity. Most of the statistical techniques used for evaluating parcellations are introduced in this chapter and used throughout the remaining of the thesis. 

Chapter~\ref{chapter:manifold} is motivated by the idea of identifying local connectivity patterns in the brain through dimensionality reduction, which can then be used to compute whole-brain cortical parcellations on a single subject basis. Contrarily to the preceding chapter, it aims to delineate subdivisions of the brain with respect to structural connectivity estimated from dMRI. The proposed method casts the parcellation problem as the localisation of boundaries between cortical regions with distinct connectivity patterns and solves it using image segmentation techniques. 

Both Chapter~\ref{chapter:multi-level} and~\ref{chapter:manifold} discuss the limitations of the presented methods, in particular with respect to different connectivity types used, and investigate the variability across individuals from a cortical parcellation point of view. In addition, multi-modal comparisons are also provided to assess the degree of alignment between connectivity-driven parcellations and well-defined neuro-anatomical properties of the cerebral cortex.  

Chapter~\ref{chapter:joint} proposes a robust group-wise parcellation framework that is simultaneously driven by the within- and inter-subject variability in connectivity. A joint graphical model is formed that can effectively capture the fundamental properties of connectivity within the population, while still preserving individual subject characteristics. The method is presented as an alternative approach to compute a group-wise parcellation, which is typically obtained from either average datasets or \textit{a priori} delineated subject-level parcellations, and shown to produce a more robust and accurate representation of a group of subjects in terms of reflecting the underlying connectivity. 

Chapter~\ref{chapter:survey} provides a large-scale, systematic comparison of existing parcellation methods using publicly available sources. Experiments consist of quantitative assessments of 24 group-level parcellations (including connectivity-driven, random and anatomical parcellations) for different resolutions. Several criteria are simultaneously considered to evaluate parcellations, including (1) reproducibility across different groups, (2) fidelity to the underlying connectivity data estimated from rs-fMRI, (3) agreement with functional activation and well-known properties of the cerebral cortex, and (4) two simple network analysis tasks. This extensive empirical study highlights the strengths and shortcomings of the various methods and aims to provide a guideline for the choice of parcellation technique and resolution according to the task at hand. 

Chapter~\ref{chapter:conclusion} concludes this work with an emphasis on the thesis achievements and possible future directions, including the further use of connectivity-driven parcellations in subject- and group-level connectivity studies.

\chapter{Studying the Human Brain}
\label{chapter:background}

\section*{Abstract}
\textit{This chapter provides an overview of the background required to understand the connectivity-driven parcellation methods presented in this thesis. We first survey the neuro-biological basis of the human brain with an emphasis on the cerebral cortex, the folded surface of the brain, that is responsible for high-level functions. After giving some insight towards the importance of connectivity in the context of brain mapping, we summarise the fundamentals of magnetic resonance imaging (MRI) techniques used for capturing the connectivity that underlie the brain function and anatomy. We then elaborately explain the most commonly used approaches for estimating the functional and structural connectivity from the MRI data. The chapter ends with an overview of the data used to conduct experiments in this thesis. }

\section{Introduction}
The human brain is the `headquarters' of the human nervous system that allows carrying out a variety of operations, ranging from relatively primitive actions such as executing movement, to more complex functions such as thinking and speaking, as well as many different cognitive processes that separate the humans from other animals~\cite{Kendal03,Chen08executive}. Although the neuro-biological foundations of the brain are mostly revealed through advances in neuroscience, the relationship between the brain and cognitive functions that constitute the human behaviour is still not completely uncovered~\cite{Churchland89neurophilosophy,Kendal03}. How is the brain involved in certain cognitive operations that underlie the basis of thought, memory, perception, and act? Going as back as to the ancient Greek times, philosophers and scientists have been endeavouring to answer this question; yet to this time, mapping the brain's function and anatomy still stands as one of the greatest challenges in the field of modern neuroscience~\cite{Churchland89neurophilosophy,Kendal03,bear2007neuroscience}.

In order to understand how the brain functions, one should first know its elemental units and their interactions with each other~\cite{Sporns05}. To this end, the following section explains the fundamentals of the brain anatomy, focusing on the main structures that constitute the brain and their roles in the nervous system. We exclusively cover the cerebral cortex, since many cognitive functions and mental operations take place in this convoluted layer of neural tissue. 

The input and output connectivity of a cortical region is considered to play a critical role to determine its functionality~\cite{Passingham02}. Studying connectivity is therefore important to better understand the link between function and anatomy. Towards this end, Section~\ref{sec:hum_bra_map} gives the rationale behind brain mapping studies that explore connectivity at the micro-, meso-, and macro-scales. The latter is of particular interest, as connectivity analysis at the macroscale allows exploring functional interactions and anatomical pathways between different brain regions. 

In order to provide prior knowledge about connectivity, Section~\ref{sec:imaging_conn} covers the fundamentals of MRI, the most-widely used imaging technology for the \textit{in-vivo} connectivity studies. After briefly explaining the physics behind MRI, the section proceeds with a detailed coverage of diffusion and functional MRI, the main imaging techniques used for capturing the structural and functional connectivity, respectively. In this section, we further provide information about the general drawbacks and problems associated with each technique, as well as the standard preprocessing pipelines necessary to bring the imaging data to an analysable basis.

Given the basics of brain imaging at the macroscale, Section~\ref{sec:struct_func_conn} covers the common approaches for estimating functional and structural connectivity. We first summarise different tractography algorithms used to delineate anatomical pathways with respect to dMRI, and then give a brief overview of the most widely employed statistical techniques for modelling the functional interactions between different cortical regions captured with fMRI.

In Section~\ref{sec:imaging_data}, we wrap up the chapter with an overview of the imaging data used throughout this thesis. We conduct our experiments on the publicly available data collected and distributed by the Human Connectome Project~\cite{VanEssen13}. After a brief introduction to the project, we summarise the image acquisition and preprocessing pipelines, and provide details about the two datasets used in different parts of this thesis.

\section{The Human Brain}
\label{sec:the_brain}
The human brain controls the human nervous system and facilitates mental operations with its highly complex structure~\cite{nieuwenhuys2007human}. It receives input from the environment via sensory organs, processes these signals through a serial -as well as parallel- set of sophisticated pipelines and generates complex responses to coordinate the human body. The integration and interaction of these signals constitute the `mind', a set of operations that leads to observed human behaviour~\cite{Kendal03} and distinguishes humans from each other, despite the fact that the neuro-anatomical structure of the brain is highly similar across individuals. 

\begin{figure}[!bt]
\centering
\begin{tabular}{cc}
\includegraphics[width=0.9\textwidth]{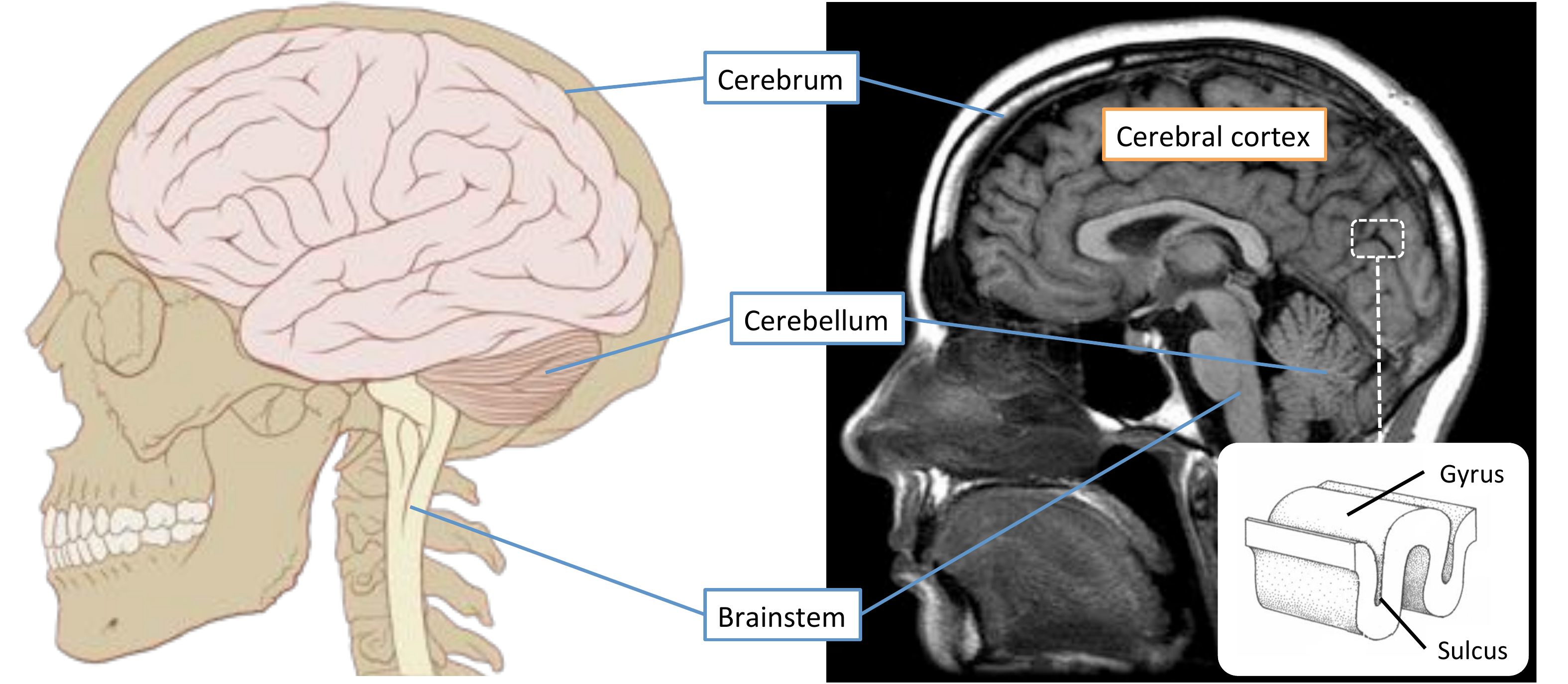} 
\end{tabular}
\caption[An illustration of the human brain and cerebral cortex.]{(\textit{Left}) An illustration of the human brain and skull, showing three distinct structures (image from Wikimedia Commons~\cite{wiki_brain}). (\textit{Right}) Sagittal view of an MR image of the living human brain, in which the cerebral cortex and some other major structures are clearly visible (image is taken from FlickR~\cite{flickr_brain}). Inlay shows a gyrus and sulcus, the characteristic infoldings that give the cerebral cortex its highly convoluted shape.}
\label{fig:brain}
\end{figure}

Like those of all vertebrates (and most of the other animals), the human brain is located in the head and protected by the skull. It consists of three distinct parts: the cerebrum, the cerebellum, and the brain stem, as shown in Fig.~\ref{fig:brain}. The largest of these is the cerebrum, which consists of two approximately symmetric cerebral hemispheres, interconnected by a bundle of nerve fibers. It is covered with a thin layer of neural tissue, the cerebral cortex, and contains several subcortical structures including the basal ganglia, the hippocampus, and the amygdala. The cerebral hemispheres are associated with motor and sensory functions, as well as involved with aspects of different cognitive functions, such as memory, language, and emotion~\cite{Kendal03,bear2007neuroscience}. They are connected to the rest of the brain via the diencephalon, which consists of two substructures, the thalamus and hypothalamus. The former processes motor and sensory information transmitted to the cerebral cortex, while the hypothalamus is involved with the regulation of autonomic and endocrine functions. 

Underneath the cerebrum lies the brain stem that facilitates the brain's integration with the central nervous system by attaching it to the spinal cord. The brain stem is primarily responsible for connecting the motor and sensory systems in the brain to the rest of the body. It is also involved in several vital autonomic functions, including breathing, maintaining consciousness and controlling the heart rate~\cite{nolte2002human}. 

The cerebellum is located at the back of the brain, behind the brain stem. It contributes to motor planning and control, including the regulation of movement, maintenance of balance and learning of motor skills. It is also involved in certain cognitive functions, such as language processing~\cite{booth2007role}. 

\subsection{The Cerebral Cortex}
The cerebral cortex is a thin layer of neural tissue that overlays the cerebral hemispheres (Fig~\ref{fig:brain}). It is primarily associated with most of the mental operations that lead to the observed human behaviour, including but not limited to, complex thought, memory, emotion, learning, planning, acting, control of movement, sensation, vision, and auditory processing~\cite{Kendal03}. As a result, it naturally constitutes the focus of many brain studies and is also targeted in this thesis for parcellation purposes.

The cerebral cortex is mainly composed of two different types of tissue, commonly referred to as `gray matter' (GM) and `white matter' (WM). GM consists of neuronal cell bodies, glial (non-neuronal) cells, and unmyelinated axons. The name `gray' comes from its very light gray appearance that is caused by the blood vessels and neuronal cell bodies~\cite{kolb2009fundamentals}. WM is composed of bundles of long-range myelinated axons that interconnect different areas of GM and transmit nerve impulses between neurons. It is refereed to as WM due to the white fatty substance (myelin) that surrounds the axons, which acts as an electrical insulation and facilitates the transmission of nerve impulses~\cite{klein2006biological}.

Like in all other mammals, the human cerebral cortex has a highly convoluted shape, consisting of several characteristic deep infoldings. The ridges in this convoluted structure are called \textit{gyri}, while the grooves that make these convolutions are called \textit{sulci}. A gyrus and sulcus are shown in Fig.~\ref{fig:brain}. The neuro-developmental process that leads to cortical folding, i.e. gyrification, starts at approximately 15-20 weeks gestational age~\cite{levine1999cortical} and continues even later after birth~\cite{garel2001fetal}. Although the precise reason behind this convoluted shape is not known, it is thought to be an evolutionary strategy to accommodate more GM within the skull's limited volume, and hence, providing a higher brain capacity for processing information and performing cognitive functions~\cite{Kendal03,lui2011development,zilles2013development}.


\section{Mapping the Brain}
\label{sec:hum_bra_map}
Understanding the role of connectivity in brain function is the key to reveal the neural mechanisms facilitating the observed human behaviour~\cite{Sporns05}. The foundations of brain mapping were set in the nineteenth and twentieth centuries, by neuroscientists like Ramon y Cajal and Carl Wernicke, who emphasised the importance of neuronal circuits in understanding the functional organisation of the brain and provided references towards underpinning the anatomical wiring of the brain systems~\cite{Zilles10,kennedy2016micro}. Invasive methods of brain research, such as histological staining and tracking probes, provided avenues for early structural mapping efforts and new insights towards understanding the circuitry of neurons~\cite{kennedy2016micro}. With the development of MRI, it became possible to investigate connectivity within the living brain, as MRI allows to map the anatomical (structural) pathways that underlie human brain function at near millimetre resolution, as well as to capture functional interactions between different cortical areas. 

These advances have collaboratively led to the emerging field of \textit{connectomics}, which can be defined as the scientific efforts for capturing, mapping, and analysing connectivity in the brain. The ultimate aim of connectomics is to produce the `human connectome', a comprehensive map of the brain's neural organisation~\cite{Sporns05} that can be produced and studied at different scales, ranging from single neurons (microscale) to populations of neuronal units (mesoscale) to high-level systems such as distinct brain regions and functional networks (macroscale). 

At the microscale, the connectome refers to a complete map of all neural units and synapses. However, a neuron-by-neuron mapping of the human brain is currently not feasible, due to the amount of neural units comprising the brain. The human cerebral cortex alone contains an estimated number of $20$ billion neurons, each of which is connected to thousands of other cells, yielding an immensely complex network with trillions of synaptic connections~\cite{azevedo2009equal}. Considering current limitations in imaging technologies and analysis tools, the microscale connectome is not likely to be derived in the near future. 

At the mesoscale, the connectome is composed of neuronal populations, the so-called `local processing units', that are formed by linking hundreds or thousands of individual neural cells~\cite{kennedy2016micro}. Several recent animal studies, with a particular focus on mice~\cite{oh2014mesoscale} and rats~\cite{bota2015architecture}, have revealed how the brain is wired at this scale~\cite{Sporns05}. However, such methods currently rely on invasive probing techniques, and hence, cannot be used for mapping the human brain. 

The macroscale connectome refers to functional interactions between brain regions and the anatomical pathways that connect them to each other. Among different modalities used to visualise the macro connectome, MRI is by far the most common technique due to its widespread availability, safety and spatial resolution~\cite{craddock2013imaging}. Diffusion MRI and functional MRI are widely used to capture the brain's structural and functional organisation, respectively. The former allows the reconstruction of anatomical pathways that interconnect different cortical areas in the brain, while the latter enables the delineation of functional connections between spatially remote brain regions (either at rest or while the subject is performing a task). 

Analysis of structural and functional connectivity provides two complementary views of brain mapping and enables the identification of brain's functional and anatomical circuitry~\cite{Sporns11}. Studying their evolution through ageing and their variation across individuals can enable the discovery of biomarkers for various neurological diseases, such as autism and Alzheimer's disease, and may ultimately help develop more effective diagnosis and treatment techniques for brain disorders~\cite{fornito2015connectomics}. Such investigations are made possible through network analysis, in which the connectome is represented as a network (or graph)~\cite{Sporns11}. At the macroscale, the network nodes correspond to the distinct cortical areas (or regions of interest, ROIs) and the edges represent the connections (axonal projections or functional interactions) between them. Definition of the network nodes through parcellation of the cerebral cortex lies at the heart of the macro connectomics, as errors at this stage can propagate into the subsequent stages and consequently reduce the reliability of network analysis. This motivates the generation of connectivity-driven parcellations for node identification, which constitutes the primary focus of this thesis. Fundamentals of brain parcellation and different methods to derive connectivity-based ROIs for network analysis will further be discussed in the following chapter.  

\section{Imaging the Macro Connectome}
\label{sec:imaging_conn}
\subsection{Magnetic Resonance Imaging}
MRI makes use of strong magnetic fields, radio waves, and field gradients to generate detailed images of internal body structures, including but not limited to the heart, the liver, breasts, as well as the brain and other organs. A major advantage of MRI over other imaging techniques, such as Computer Tomography, Positron Emission Tomography (PET) or X-Ray Imaging, is that it does not involve ionising radiation and is therefore relatively safe and allows repeatable scans. On the downside, it may cause discomfort, as MRI scans require patients to stand still inside a narrow tube for a fairly long period of time and to wear headphones for reducing the loud high-pitched noise caused by the scanner.

MRI is based on the physical phenomenon of Nuclear Magnetic Resonance, in which atomic nuclei absorb and emit the energy from radio frequency waves if undergo an external magnetisation. The water and fat molecules in the human body naturally consist of hydrogen atoms, which have their own magnetic fields and inherently spin around their axis at random orientations. An MRI scanner generates a strong magnetic field ${B_0}$ that causes protons in hydrogen nuclei to align with the direction of the ${B_0}$ field, precessing around their axis and generating a longitudinal magnetisation. A pulse of radio frequency (RF) is directed at these precessing protons to temporarily create a second magnetic field perpendicular to ${B_0}$. This process excites the nuclear spin, bringing some of the nuclei into transverse phase coherence with each other. 

Once the RF pulse is removed, the nuclei lose their magnetisation and reach to an equilibrium, that is, recover their initial state with respect to the magnetic field ${B_0}$. The energy transition in the process of realignment (i.e. relaxation) can be measured with a coil and used to generate images. Intensity values in these images result from the different concentration of hydrogen protons in different types of tissue and are characterised by two relaxation factors: T1 (longitudinal) relaxation, which is the realignment of nuclei with the direction of the external magnetic field ${B_0}$, and $T2$ (transverse) relaxation, which can be defined as the lose of coherence among the nuclei, resulting in decay of transverse magnetization. 

\begin{figure}[!ht]
\centering
\begin{tabular}{cccc}
\includegraphics[height=0.25\textwidth]{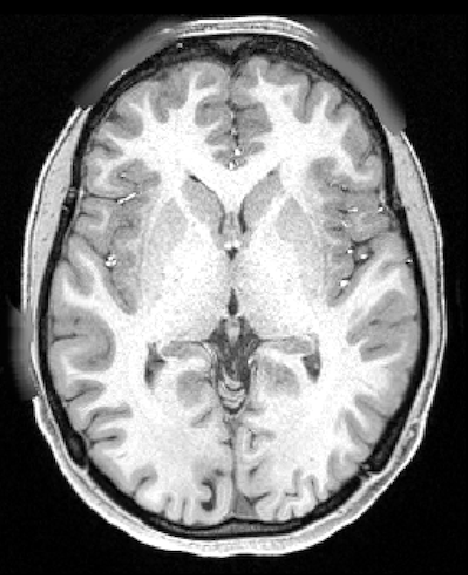} &
\includegraphics[height=0.25\textwidth]{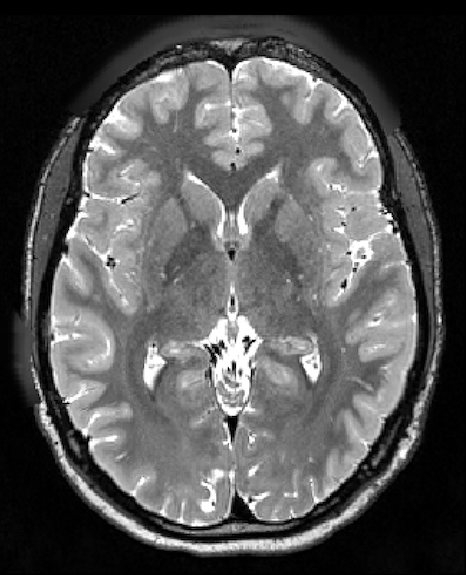} & 
\includegraphics[height=0.25\textwidth]{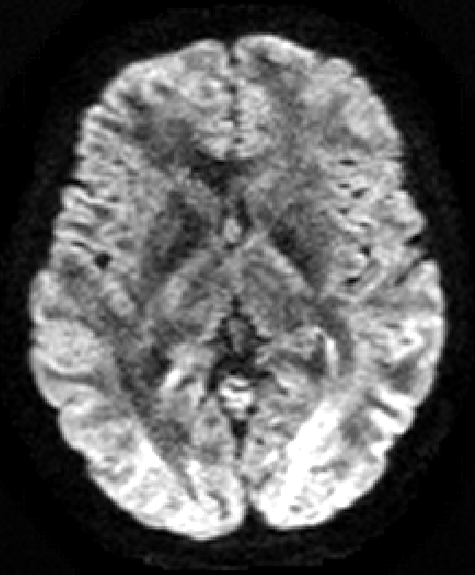} & 
\includegraphics[height=0.25\textwidth]{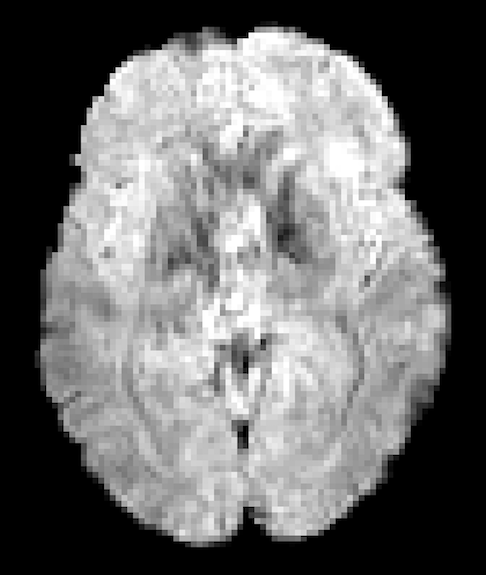} \\
(a) & (b) & (c) & (d) 
\end{tabular}
\caption[Different types of MR images.]{Different types of MR images. (a) T1w image, (b) T2w image, (c) diffusion weighted image, and (d) functional MR image. All images are acquired from a healthy adult brain.}
\label{fig:different_mri}
\end{figure}

By changing the parameters of the MRI pulse sequence, different contrast images can be generated based on the T1 and T2 relaxation factors, e.g. T1-weighted (T1w) and T2-weighted (T2w) images, which can distinguish between different types of tissue. Example T1w and T2w images are presented in Fig.~\ref{fig:different_mri}.

\subsection{Diffusion Magnetic Resonance Imaging}
Diffusion Magnetic Resonance Imaging (dMRI) or Diffusion Weighted Imaging (DWI) for short, is a form of MRI that relies upon the diffusivity of water molecules for generating contrast in MR images. Water molecules tend to move with random replacements (i.e. microscopic movements) in a free (unrestricted) environment. However, molecular diffusion in biological tissues is naturally impeded by physical boundaries, such as cell membranes, fibers, and macromolecules. For example, in the brain, the motion of water molecules is restricted within the neural axons. Water tends to diffuse more rapidly in the direction aligned with the axon's fibrous structure and more slowly in other directions. These anisotropic patterns of diffusion can reveal the microscopic architecture of the underlying tissue, and with the help of computational modelling, can be used to delineate white matter fibers that interconnect different regions in the brain.

In order to capture the displacements of water molecules in the tissue, MRI must be sensitised to diffusion. The principles of diffusion weighting were introduced by Stejskal and Tanner in 1965~\cite{stejskal1965spin}, in which the homogeneity of the magnetic field is linearly altered by using two gradient pulses with the same magnitude but opposite gradient direction, yielding nuclei to precess at different rates. Assuming that the nuclei have not moved in between the pulses, the MR scanner will measure the same signal, as the two pulses will cancel each other. For example this can be observed within the ventricles, where the diffusion is typically unrestricted. On the other hand, if the diffusion is hindered, for example by the myelin sheath, the position of the nuclei will likely be different, consequently leading to a signal loss. This reduced signal is reflected in the DW image as an estimate of the water displacement within each voxel. 

One critical parameter of DWI is the $b$-value, which determines the degree of diffusion weighting on the generated MR image. The choice of $b$-value is usually application-dependent and needs to be tuned based on the task at hand, as it involves a trade-off between sensitivity and signal-to-noise ratio (SNR) (i.e. higher $b$-values yield sharper contrast diffusion images, but also lower SNR)~\cite{jones1999non}. 

The structure of myelinated axonal fibers restricts the motion of water molecules, leading to anisotropic diffusion patterns that are not equal in all directions. Applying diffusion gradients in at least 6 non-collinear directions enables the definition of a diffusion tensor, from which local diffusion patterns can be characterised for each voxel. A diffusion tensor provides quantitative information about the degree of diffusion anisotropy and directions of the fibers through spectral decomposition, and hence, facilitates the delineation of axonal pathways with the help of computational techniques, such as tractography (see Section~\ref{sec:tract}). 

One major drawback of the diffusion tensor imaging (DTI) model is its assumption that the fiber orientation is uniform at each voxel. However, axonal fibers are known to cross each other (i.e. crossing fibers), temporarily merge with one other, or disperse (i.e. fan out) as they approach their destinations, which may lead to heterogeneity in single voxels that can not be accounted for by simple fiber orientation models, such as the diffusion tensor~\cite{craddock2013imaging}. Current estimates show that up to 90\% of the voxels in WM may contain `crossing fibers', suggesting that DTI may not be able to detect axonal pathways in the majority of WM~\cite{jeurissen2013investigating}. More sophisticated diffusion modelling techniques can account for this heterogeneity by using higher order models~\cite{mori2007introduction}, commonly referred to as High Angular Resolution Diffusion Imaging (HARDI)~\cite{tuch2002high}. HARDI approaches are able to resolve multiple orientations~\cite{tuch2002high} by relying on a more complex fiber orientation model, e.g. many more gradient directions, compared to the diffusion tensor. As a result, they provide more precise modelling for the heterogeneous distributions of crossing-fiber orientations~\cite{wedeen2008DSI,craddock2013imaging}. Two well-known HARDI methods are Q-ball imaging~\cite{tuch2004q} and Diffusion Spectrum Imaging~\cite{wedeen2005mapping}, which are currently being used by the major human connnectome projects to generate more accurate mappings of the human brain~\cite{Toga12,VanEssen13}.

\subsubsection{Preprocessing for dMRI}
The most commonly observed artefacts in DW images are caused by spatial distortions related to motion (e.g. head motion, respiration, cardiac pulsation etc.), magnetic field inhomogeneities, and eddy currents~\cite{le2006artifacts}. The latter emerges from rapid switching of gradient pulses during acquisition, yielding to different field gradients than the ones initially programmed. Spatial distortions introduced by eddy currents can be minimised on the acquisition time, for example by using `self-shielded' gradient coils~\cite{le2006artifacts}, or by intentionally altering the shape of the currents sent to the gradient hardware to account for expected eddy-current distortions~\cite{papadakis2000gradient}. 

The local variations in the magnetic field, usually emerging from the interactions between areas with different magnetic susceptibility (e.g. air-tissue interfaces) can also contribute to spatial distortions~\cite{craddock2013imaging}. Impact of such artefacts can be reduced with the use of parallel imaging techniques or corrected via mathematical modelling of the field variations~\cite{andersson2003correct,craddock2013imaging}.

Patient motion typically leads to ghosting effects and large signal variations in the images~\cite{le2006artifacts}. Although various techniques can be used to minimise the impact of motion in dMRI, its complete elimination is not possible without anaesthesia or using an MRI sequence that is less prone to motion artefacts~\cite{le2006artifacts}. This is achieved by echo-planar imaging (EPI)~\cite{poustchi2001principles}, which has now become a golden standard in dMRI due to its ability to effectively `freeze' the motion during acquisition~\cite{le2006artifacts}. However, EPI often provides less spatial resolution than conventional sequences due to hardware limitations and may introduce its own artefacts, e.g. blurry images~\cite{jones2010twenty}. In general, a post-acquisition solution to correct diffusion artefacts involves registering the distorted images to the $b_0$ image, i.e. the image with no diffusion weighting~\cite{soares2013hitchhiker}.  

\subsection{Functional Magnetic Resonance Imaging}
Functional Magnetic Resonance Imaging (fMRI) is a non-invasive, \textit{in-vivo} imaging technique for capturing dynamic changes in neurocognitive activity associated with cerebral blood flow~\cite{huettel2004functional}. The technique is based upon the fact that the amount of blood that is circulated within the cerebral cortex and neural activity regulating mental processes are closely linked. That is, when a brain area is involved in neural activity, blood flow towards the active area increases in response to high energy demand.  

Among others, the most popular form of fMRI uses Blood Oxygenation Level Dependent (BOLD) contrast, which measures the inhomogeneities in MRI signal due to the level of oxygen in blood~\cite{ogawa1990brain}. BOLD reflects the change in heamodynamic response, which facilitates the rapid delivery of blood to the active neuronal tissue to supply their demand for oxygen and other nutrients, such as glucose~\cite{ogawa1992intrinsic,buxton2004modeling}. Haemoglobin is a protein molecule in red blood cells that is responsible for carrying oxygen from the respiratory organs to the tissues and shows different magnetic susceptibility depending on its bond with oxygen. While oxygen-carrying haemoglobin (oxyhaemoglobin) behaves as a diamagnetic substance, its oxygen-free form is typically paramagnetic. When there is a demand for high energy, for example in case of neural firing, oxygen is transferred from oxyhaemoglobin to neurons, making the blood de-oxygenated, and hence, increasing the level of deoxyhaemoglobin. In order to compensate for the greater amount of energy demand, vascular system increases the blood flow, changing the relative level of oxyhaemoglobin and deoxyhaemoglobin in favour of the former. The difference in magnetic susceptibility of haemoglobin molecules facilitates the detection of the increase in blood flow via MRI, as areas with high concentration of oxyhaemoglobin produces a higher signal than areas with low concentration~\cite{amaro2006study}.

Depending on the problem under investigation, two types of fMRI can be employed, each of which provides information about different aspects of brain activation and functional connectivity. The first one is task fMRI (t-fMRI), in which neural activity is recorded while the subject is performing a variety of cognitive tasks designed to activate different brain regions~\cite{Beeck08}. The primary motivation of t-fMRI studies is to examine the brain under a controlled setting, whose limits are delineated by a psychological paradigm. These paradigms may target different aspects of cognition, such as primary sensory processing, information processing, decision making, problem solving, and many more~\cite{Barch13}. Data collection in t-fMRI is typically followed by a statistical analysis stage in which image intensities are compared to task paradigm in order to reveal functional activation~\cite{smith2004overview,amaro2006study}. 

While localisation of function constitutes the primary purpose of t-fMRI studies, more recent approaches also attempt to determine the mappings between certain mental operations and patterns of neural activity~\cite{Norman06}. These so called `mind-reading' methods combine data representation techniques with pattern recognition tools in different applications, including but not limited to lie detection~\cite{Davatzikos05}, object recognition~\cite{Beeck08}, and human behaviour prediction~\cite{Michel12}. 

The second type of fMRI is known as resting-state fMRI (rs-fMRI), as it captures neurocognitive activity while the subject is `at rest', i.e. while the brain is not driven by any external stimuli. In other words, the subject is not requested to perform a neuropsychological task, but told to relax with eyes are shut or fixated to an object. In 1995, Biswal et al. showed that the brain is still active in the absence of an externally prompted stimulus, hence spontaneous fluctuations in BOLD signals can be used to identify functional interactions between different brain regions~\cite{Biswal95}. Since its discovery, rs-fMRI has been the primary tool to explore the brain's functional organisation~\cite{Lowe98,salvador2005neurophysiological,Fransson05,Damoiseaux06,Nir06,DeLuca06,Smith13} as well as to study changes/alterations in functional connectivity due to neurological and psychiatric disorders~\cite{stam2007small,wang2009altered,fornito2012schizophrenia,fornito2015connectomics}. 

It is noteworthy that rs-fMRI allows to explore the brain's functional organisation as a whole, without being biased by a neuropsychological paradigm. On the other hand, t-fMRI only targets certain brain areas in order to investigate their activation in response to external stimuli, but ignores the activation from non-target areas. With regards to this, we focus on rs-fMRI to estimate functional connectivity in the context of brain parcellation, while information derived from t-fMRI is used as a complimentary measure to validate the location of functional areas identified by rs-fMRI based parcellations.

\subsubsection{Preprocessing for fMRI}
\label{sec:fmri_pre}
BOLD fMRI signals are inherently confounded by several potential noise sources, such as head movement, cardiac/respiratory pulsation, or scanner-induced artefacts ~\cite{haller2009pitfalls,Cole10,poldrack2011handbook}. This may lead to a relatively low SNR and pose major challenges for the usability of fMRI data and the interpretability of subsequent analysis. As a result, several preprocessing stages are required to remove noise and correct for artefacts in the acquired BOLD signals prior to any attempt to analyse the data. Noise removal is particularly important for functional connectivity studies, since spurious correlations induced by structured noise may severely increase the amount of falsely identified connections~\cite{Cole10}. Standard preprocessing steps generally include slice-timing correction, head motion correction, distortion correction, temporal filtering, coregistration, spatial normalisation, and spatial smoothing. 

Slice-timing correction is used to realign individual slices in an fMRI brain volume to a reference slice based on their relative timing, as each slice typically records activity at a slightly different time point, leading to a between-slice temporal offset~\cite{sladky2011slice}. 

Head motion is one of the most commonly conferred sources of noise in fMRI. Artefacts induced by motion can lead to mislocalisation of function, activation being detected outside the brain volume, or artefactual fMRI signals~\cite{craddock2013imaging}. It is typically corrected by applying rigid-body registration to a reference volume (such as the first volume) or by identifying structured noise related to motion with the use of statistical analysis techniques, such as independent component analysis~\cite{poldrack2011handbook}. 

Artefacts related to distortion emerges from inhomogeneities in the magnetic field and can be corrected by aligning the functional image with a structural image or by acquiring two fMRI images with different echo time, in which distortion correction is facilitated by mapping the spatial distribution of non-uniform areas in the magnetic field~\cite{hutton2002image}. Alternatively, bias field estimation techniques can be utilised to correct for distortion, in case the magnetic field distribution is not known \textit{a priori}~\cite{guillemaud1997estimating}.

Temporal filtering is the removal of frequencies from the signal that are not of interest, which may be induced by respiratory and cardiac pulsation. BOLD signals related to neuronal activity are typically exist in a particular frequency range, usually located between 0.01 and 0.1 Hz~\cite{toga2015brain}. A band-pass filtering is hence applied to surpass spurious frequencies and only focus on the neurobiologically meaningful range.

Prior to analysis of fMRI data, BOLD signals are registered to a structural MR image of higher resolution of the same subject (i.e. coregistration), so that functionally distinct regions or activated voxels can be mapped into anatomical space. This is generally followed by spatial normalisation to a common space, which could either refer to a volumetric brain template, e.g. Montreal Neurological Institute (MNI) template~\cite{mazziotta1995probabilistic}, or a surface template, e.g. FreeSurfer's fsavereage surface~\cite{Fischl04}. This brings individual subjects to a standard anatomical basis, and hence, facilitates the integration of results across multiple subjects, populations, as well as different analysis pipelines. 

Even by assuming a perfect coregistration between the anatomy and function, a reliable functional alignment across subjects cannot be guaranteed~\cite{Thirion06}. A typical solution to this problem is to sacrifice spatial resolution for group-level analysis, i.e. applying spatial smoothing to BOLD signals, which is generally achieved through convolution with a Gaussian filter. Such a process not only alleviates the impact of misalignment across subjects (e.g. inter-subject variability), but also improves SNR in individual subject data, if the spatial extent of activation matches the width of the filter used~\cite{lindquist2008statistical}. On the downside, smoothing typically reduces the overall resolution in fMRI and may introduce blurring in the group data~\cite{Thirion06,lindquist2008statistical}. As a result, activation in relatively small areas may be mislocalised or may even completely disappear depending on the amount of spatial smoothing~\cite{lindquist2008statistical}.

Regardless of these artefacts, BOLD fMRI constitutes the current state of the art for imaging the brain activation, with its non-invasive nature and relatively high spatial resolution over alternative electro-physiological recordings, such as electroencephalography (EEG) and magnetoencephalography (MEG), which, despite offering a high temporal resolution (typically at the order of milliseconds), suffer from poor spatial resolution and lack of spatial localisation~\cite{logothetis2004interpreting}. Another limitation of BOLD fMRI is the fact that it only provides an indirect measure of the neural activity (i.e. through change in blood flow), while EEG and MEG directly record neural activity in the form of electrical impulses from the brain. In addition, the true biological meaning of BOLD signals is still under investigation~\cite{Eickhoff15} and its lower temporal resolution (typically at the order of seconds) compared to electro-physiological recordings is also a limiting factor for capturing high-frequency patterns.

\section{Structural and Functional Connectivity}
\label{sec:struct_func_conn}
\subsection{Estimating Structural Connectivity}
\label{sec:tract}
Structural connectivity describes the anatomical pathways linking distinct brain areas, which generally refers to WM projections interconnecting cortical and subcortical regions~\cite{sporns2013structure}. These connections form the biological basis for information transfer between remote brain regions and are therefore fundamental for mapping the structural human connectome, and relatedly, a better understanding of brain function~\cite{Sporns05,toga2015brain}. 

\subsubsection{Reconstructing Pathways through Tractography}
As described in the previous section, dMRI facilitates the mapping of the diffusion process of water molecules, and hence, provides information about the location, orientation, and anisotropy of axonal fibers through different diffusion models. With this information available, tractography (fiber tracking) techniques can be utilised to reconstruct trajectories of the major pathways in the brain, from which structural connectivity patterns can be quantified~\cite{mori2002fiber,behrens2009mr}. Tractography estimates the trajectories by tracking the streamlines through a 3D vector field, in which each vector represents the diffusion orientation measured by dMRI~\cite{mori2007introduction}. The majority of the tractography methods can be broadly divided into two categories, depending on how they reconstruct white matter tracts from the underlying diffusion model: i.e. deterministic tractography and probabilistic tractography. 

\textit{Deterministic tractography} relies on the local diffusion information for integrating the streamlines on a step-by-step basis~\cite{behrens2009mr}. Initiated from a seed (a voxel or a region of interest), streamlines are reconstructed with respect to the primary diffusion direction at each voxel. Due to the fact that the directional information is encoded on an imaging grid, each voxel provides only one measurement of orientation. Hence, interpolation methods are essential to transform these discrete measurements into a continuous coordinate system, allowing to estimate the fiber orientation at locations away from the voxel centres~\cite{behrens2009mr}. This could be achieved, for example, by applying each voxel's measurement over the entire voxel~\cite{mori1999three} or using a weighted interpolation of the fiber orientation by incorporating measurements from neighbouring voxels~\cite{basser2000vivo,lazar2003error}. Regardless of the interpolation technique, tracking algorithms continue to delineate streamlines until a termination criterion is reached. Different approaches to terminate this process may include using a white matter mask to constrain the search space of the tracking algorithm or to define heuristic rules to stop the streamline from progressing, for example when the computed anisotropy at a voxel falls beneath a user-defined threshold~\cite{behrens2009mr}. 


The primary intuition behind relying on such a heuristic to stop the tracking algorithm is due to the fact that deterministic tractography does not provide a means to model uncertainties inherently associated with each voxel along a streamline. \textit{Probabilistic tractography}, by contrast, aims to handle the uncertainties so that tractography can keep tracking streamlines through regions of high uncertainty, in which deterministic techniques would be forced to terminate.~\cite{behrens2009mr}. Errors propagating as tractography progress, for example due to imaging noise or crossing fibers, have necessitated the use of probabilistic tractography in order to achieve a more accurate reconstruction of axonal pathways~\cite{behrens2007probabilistic,behrens2009mr}. Probabilistic tractography techniques typically start with characterising the uncertainty of the fiber orientation at each voxel with a probability density function~\cite{behrens2003characterization}. This function allows any streamline to follow an infinite number of orientations, each of which is assigned a different level of probability. Hence, once a voxel is reached, the next orientation is sampled from the probability distribution associated with that voxel. By drawing many such samples (each time providing slightly different orientations), an entire tract connecting two points can be reconstructed~\cite{behrens2009mr}. 

A major benefit of probabilistic tractography, apart from its ability to quantify uncertainty, is its robustness to noise~\cite{behrens2003characterization,behrens2009mr}. Paths progressing towards wrong routes due to noisy voxels are likely to produce low probability values, and hence, are quickly dispersed~\cite{behrens2003characterization,behrens2009mr}. Although probabilistic tractography does not require the definition of a termination criterion, a large curvature threshold is typically employed to prevent streamlines tracing back to where they start, a necessary mechanism to avoid an artificial increase in probability values and generation of implausible pathways~\cite{behrens2009mr}. One major drawback of probabilistic tractography techniques is the computational cost, especially when used to estimate whole-brain connectivity. However, recent advances in GPU-based parallelisation techniques have drastically reduced the computational time required to perform probabilistic tractography for a typical spatial resolution~\cite{hernandez2013accelerating}. 

A major challenge for tractography techniques, either deterministic or probabilistic, emerges from their inability to detect fibers in their entirety~\cite{craddock2013imaging}. Although, tracking algorithms can successfully estimate the location of fibers within the white matter, the diffusion orientation is typically highly ambiguous around WM-GM interface, resulting in the termination of tractography before reaching the GM~\cite{Ng13} and making it impossible to accurately identify the origin or termination of axonal tracts~\cite{Ng13,craddock2013imaging}.

\subsubsection{Quantifying Structural Connectivity}
\begin{figure}[!t]
\centering
\includegraphics[width=\textwidth]{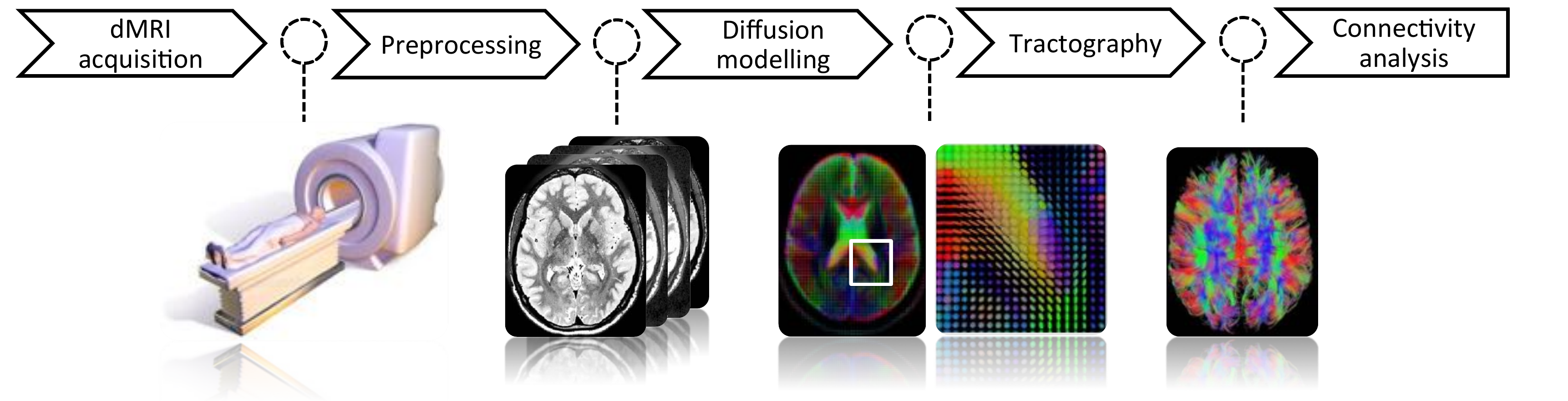} 
\caption[Illustration of a typical dMRI pipeline.]{Illustration of a typical dMRI pipeline. Images of the MRI scanner and diffusion tensor model are from Wikimedia Commons, referenced as~\cite{scanner_image} and~\cite{tensor_image}, respectively.}
\label{fig:dmri_pipelines}
\end{figure}

Tractography techniques facilitate the delineation of trajectories connecting cortical areas through WM and primarily provide information regarding the density and location of connections in the brain~\cite{craddock2013imaging}. However, reconstructed tracts do no inherently allow to quantify the `connection strength' between two cortical regions in the context of brain connectivity~\cite{Jbabdi12}. Rather than, estimates of connectivity can be acquired through some quantitative measurements by utilising different aspects of the diffusion data and reconstructed tracts. While connectivity estimates based on the latter include the number~\cite{Hagmann08} or the total length~\cite{Tomassini07} of reconstructed streamlines between pairs of brain regions, measures driven from diffusion properties of the underlying data usually rely on the computed anisotropy along white matter tracts~\cite{Jbabdi12}. Steps leading to structural connectivity analysis through a typical dMRI pipeline are illustrated in Fig.~\ref{fig:dmri_pipelines}.

Although probabilistic tractography provides a quantitative means for the uncertainty in streamline trajectories, it is still not possible to acquire a true quantitative measurement of the connectivity~\cite{behrens2009mr}. Structural connectivity from probabilistic tractography is typically estimated by counting the number of streamlines passing through a region and dividing it by the total number of streamlines~\cite{behrens2009mr}. However, the reliability of connections obtained in this context is limited by several confounding factors. For example, it is typically assumed that brain regions connected via a major bundle should have a noticeable trace in the diffusion data, and hence, low uncertainty (high confidence) in their trajectories; but, probabilistic tractography is likely to assign a higher confidence to a locally non-dominant fiber bundle than such a major bundle, in case it is crossed by other fibers~\cite{craddock2013imaging}. Similarly, uncertainty along a streamline's path typically increases with the tract length. Other non-interesting factors, such as imaging noise, modelling errors, partial volume effects, and voxel size also affect uncertainty, and hence, may influence the estimation of structural connections~\cite{behrens2009mr,craddock2013imaging}.  

Despite all these limitations and many other aforementioned confounding factors affecting the acquisition and modelling of diffusion data, dMRI and tractography are indispensable for brain mapping, as together they provide the only non-invasive way of measuring (or estimating) the structural connectivity within the brain.


\subsection{Estimating Functional Connectivity}
Functional connectivity is conventionally defined as the temporal dependency of neurophysiological events between spatially remote brain areas~\cite{friston1993functional}. It is used to identify co-activation between different regions that share similar functional characteristics and/or work together to perform cognitive processes. 
Compared to its structural counterpart, functional connectivity is not necessarily supported by physical connections through WM. Similarly, anatomically connected regions (as defined via tractography) also do not need to reflect functional dependency~\cite{Hagmann08}. 


Several mathematical modelling techniques can be used to quantify functional connectivity, each providing a different perspective on the temporal interactions between BOLD signals measured in brain areas~\cite{craddock2013imaging}. In the simplest form, bivariate tests, such as Pearson's correlation~\cite{Biswal95} or coherence~\cite{sun2004measuring}, can be employed between every pair of timeseries to measure their statistical dependence. A limitation of bivariate approaches emerges from the fact that they do no account information from multiple regions~\cite{craddock2013imaging}, and hence, cannot distinguish direct from indirect connections that may be mediated by third-party regions~\cite{Smith11,Eickhoff15}. Whereas such distinction would not be of great importance when correlation is used to measure the distance (similarity) between two BOLD signals, for example in the context of connectivity-driven brain parcellation, it may be a critical consideration point for network analysis~\cite{Smith13}. To this end, partial correlation can correctly estimate the conditional linear dependency between two regions, as it typically accounts for interaction with every single region of interest~\cite{marrelec2006partial}. While the reliability of partial correlations decreases when the number of observations exceeds the sample size (the number of regions)~\cite{peng2009partial}, such as in the case of typically long fMRI acquisitions, regularisation techniques (e.g. graphical lasso~\cite{friedman2008sparse}) can overcome this limitation~\cite{Smith13}. 

\begin{figure}[!t]
\centering
\includegraphics[width=\textwidth]{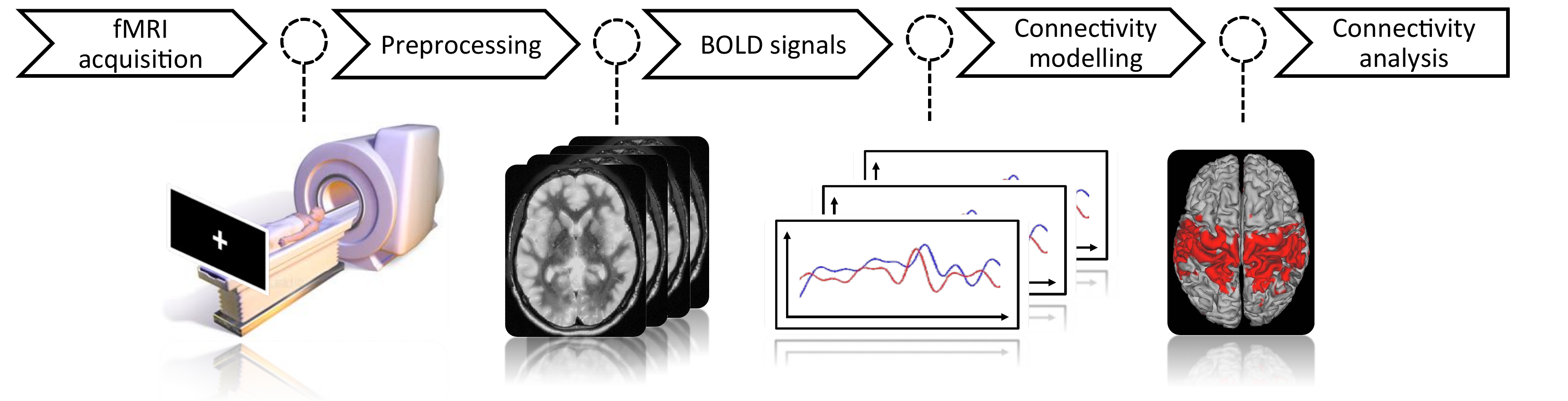} 
\caption[Illustration of a typical fMRI pipeline.]{Illustration of a typical fMRI pipeline. MRI scanner image is from Wikimedia Commons~\cite{scanner_image}.}
\label{fig:fmri_pipelines}
\end{figure}

Although more complex techniques exist to measure functional connectivity, such as methods based on higher-order statistics or lag-based approaches~\cite{Smith13}, results derived from large-scale simulations indicate that correlation-based approaches generally yield more accurate connectivity estimates~\cite{Smith13}. Various data-driven approaches can also be used to study functional connectivity, including but not limited to principle component analysis (PCA), independent component analysis (ICA), and clustering techniques. However such methods in general are more appropriate for identifying spatially distributed networks that are functionally connected during rest (i.e. resting state networks) or deriving nodes in network analysis (for parcellating the brain), and therefore, will be extensively covered in the next chapter. Steps leading to functional connectivity analysis through a typical fMRI pipeline are illustrated in Fig.~\ref{fig:fmri_pipelines}.

While functional connectivity can generally estimate existing connections between brain regions with good accuracy, it typically suffers from false positives (interestingly, structural connectivity is, by contrast, more prone to false negatives)~\cite{Eickhoff15}. False positives usually emerge due to the low SNR inherent in fMRI data, induced by imaging errors, head motion, and physiological noise~\cite{craddock2013imaging}. Special care is taken to reduce the impact of such artefacts via a series of preprocessing steps prior to data analysis as previously discussed (see Section~\ref{sec:fmri_pre}); yet, it is not possible to completely eliminate the inaccurate BOLD signal fluctuations. Some techniques aim to alleviate the impact of false positives by using thresholding mechanisms, with the assumption that negative or weak correlations correspond to spurious activity or noise~\cite{Heuvel08,Yeo11,craddock2013imaging,Langs14}. 


\section{Imaging Data}
\label{sec:imaging_data}
In our experiments throughout this thesis, we rely on the imaging data provided by the Human Connectome Project (HCP)~\cite{VanEssen13}. In the remainder of this section, we first introduce the HCP and then briefly explain the HCP data acquisition and preprocessing pipelines~\cite{Glasser13}. We finally summarise the two datasets used in different chapters of this study.

\subsection{The Human Connectome Project}
HCP (\url{https://www.humanconnectome.org/}) is one of the recent scientific efforts to map the human brain~\cite{VanEssen13}. Led by Washington University, University of Minnesota, and Oxford University (the WU-Minn HCP consortium), it aims to chart the neural circuitry of the human connectome in healthy adults by using non-invasive imaging technologies. HCP provides invaluable information to explore neural mechanisms that underlie the brain function and behaviour, which in turn, contributes to our understanding of the human mind. 

Since its first data release, the HCP datasets of almost 900 healthy adults have been made freely available to the scientific community via the HCP Database, \url{https://db.humanconnectome.org/}. The project focuses on four imaging modalities to acquire data with high spatial and temporal resolution~\cite{Marcus13}. Resting-state fMRI and diffusion MRI respectively provide information about functional and structural brain connectivity and constitute our primary data sources for developing connectivity-driven parcellation algorithms. Task-evoked fMRI reveals much about brain function with respect to various cognitive tasks designed to activate different brain regions and is used as a complementary information source to evaluate the proposed parcellations. Structural MRI captures the shape of the cerebral cortex and subcortical brain areas as well as provides the anatomical (both volumetric and surface-based) templates, allowing the integration of function and anatomy and analysis of results across multiple subjects. 

Prior to HCP, several other projects, including but not limited to the 1000 Functional Connectomes Project~\cite{Biswal10} and the Human Connectome Project led by the UCLA-Harvard consortium~\cite{Toga12} also worked with a similar motivation and enriched the understanding of the human brain with their contributions to the field of macro connectomics. However, HCP has taken the flag one step further by producing MRI data in several different modalities that are preprocessed with novel pipelines~\cite{Glasser13}, and thus, offering ready-to-analyse high-resolution structural, functional, and diffusion datasets for studying the human brain from many different aspects. Besides these attempts to map the adult human connectome, the Developing Human Connectome Project, which aims to create the first connectome of early life (from 20 to 44 weeks post-conceptional age), is still at its preliminary stage and is expected to produce its first data release in 2017~\cite{dHCP}. 

\subsection{HCP Minimal Preprocessing Pipelines} 
HCP brings multiple MRI modalities together in a common framework, allowing to perform cross-subject comparisons and multi-modal analysis of brain architecture, connectivity, and function~\cite{Glasser13}. This is achieved through a series of newly developed preprocessing methods, tailored to many modality-specific challenges faced in both the acquisition and analysis of structural, functional, and diffusion images. Throughout this section, we briefly summarise the key points in the HCP preprocessing pipelines, particularly putting emphasis on the HCP standard cortical space, i.e. the so-called `grayordinate' spatial coordinate system, as well as data acquisition and preprocessing steps for the structural, fMRI and dMRI data. It is worth noting that, this section is compiled from the HCP minimal preprocessing pipelines paper~\cite{Glasser13} and is up-to-date with the reference manual of HCP 900 subjects data release (S900)~\cite{HCP900}. 

\subsubsection{Grayordinate Standard Coordinate System}
Since we attempt to parcellate the cerebral cortex, it is beneficial to use cortical neuroimaging data and surface-constrained methods, rather than relying on volume-based data. Due to the fact that the highly convoluted cortical sheet can be easily analysed as a 2D surface, we can further utilise geodesic distances along the surface, which can be more relevant than 3D Euclidean distances within the volume~\cite{VanEssen12,Smith13,Glasser13,Gordon16}. Relying on the geometry of the cerebral cortex further allows a more effective spatial smoothing and inter-subject registration with greater overlap~\cite{Glasser13}.

\begin{figure}[!t]
\centering
\begin{tabular}{ccc}
\includegraphics[height=0.45\textwidth]{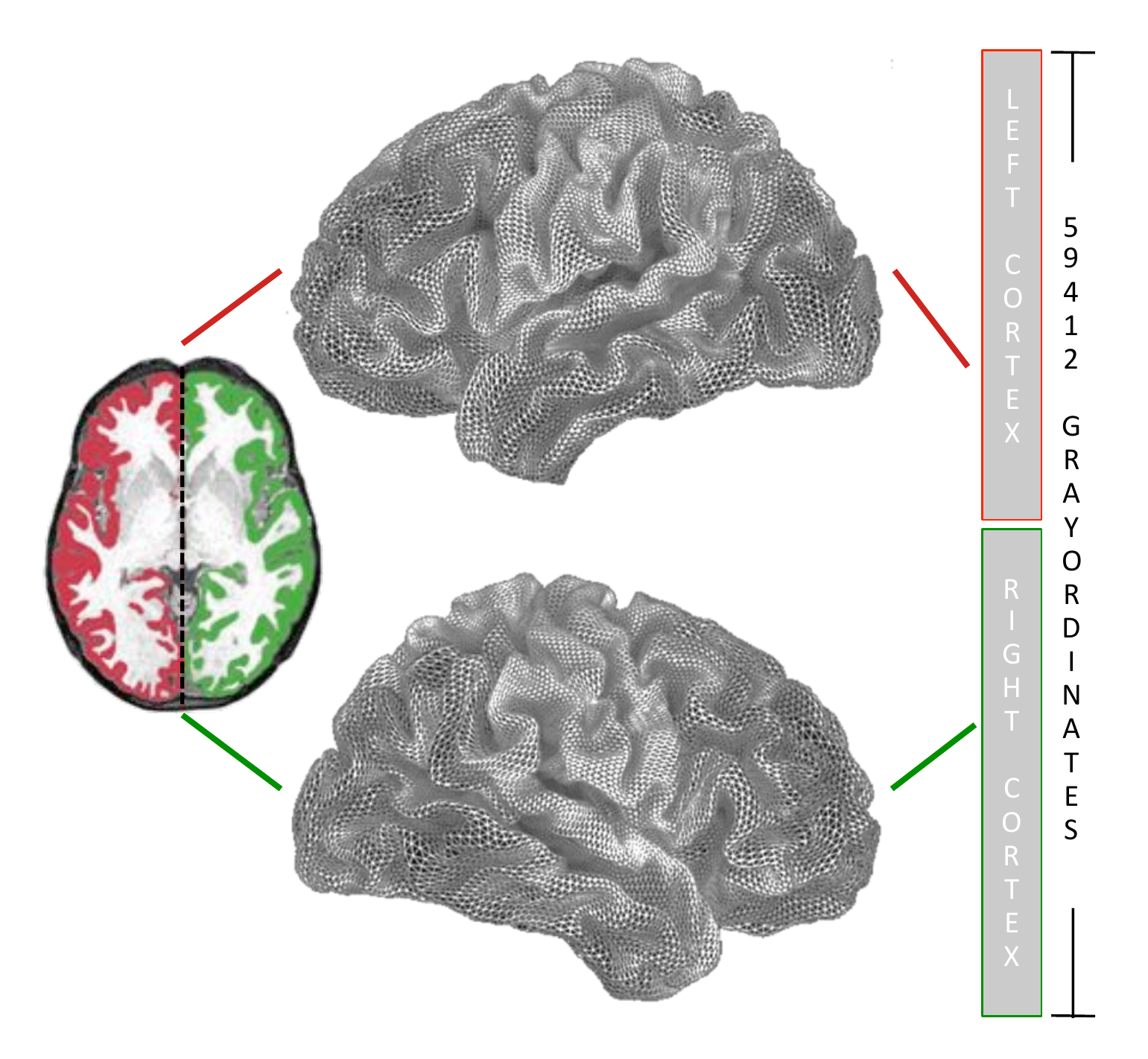} & &
\includegraphics[height=0.45\textwidth]{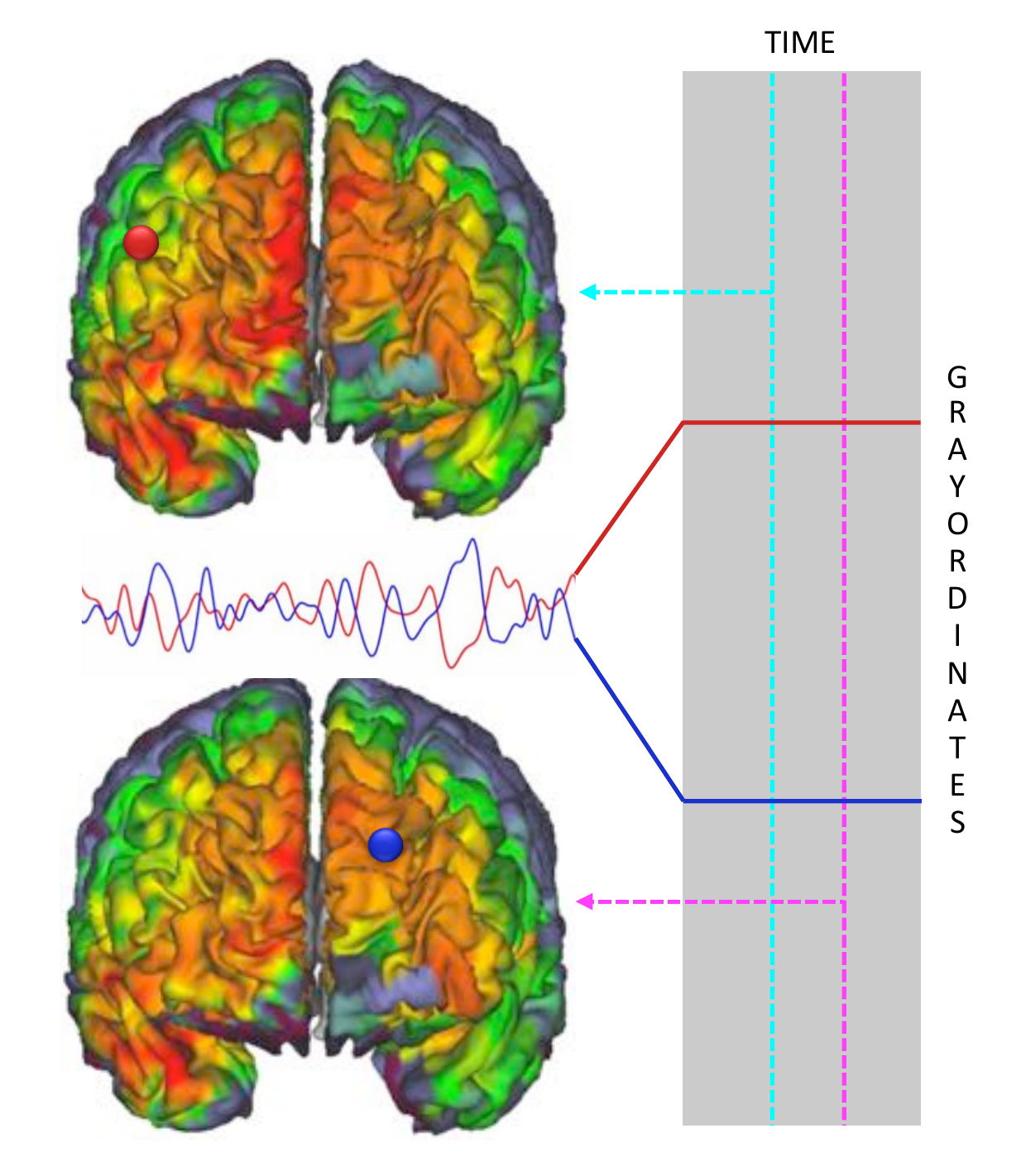} \\
(a) & & (b)
\end{tabular}
\caption[Grayordinate standard coordinate system.]{(a) Grayordinate standard coordinate system. Each cerebral hemisphere roughly contributes about 30k cortical vertices, achieving a total of 59412 grayordinates corresponding to all of the gray matter voxels apart from the non-cortical medial wall. (b) A standard timeseries file in the grayordinate space. Each row represents a grayordinate and the columns record fMRI time points. Each row corresponds to a timeseries mapped to a cortical vertex, as shown in red and blue. Reading along a column produces a spatial pattern, which corresponds to the recorded BOLD signals at a given timepoint across the whole cerebral cortex. Spatial patterns at different timepoints (columns) are shown in cyan and magenta.}
\label{fig:grayordinates}
\end{figure}

HCP datasets are provided in a standard coordinate space obtained by mapping the gray matter voxels onto cortical sheet, that allows combining the left and right cerebral hemispheres into a single file\footnote{It is worth noting that, the original grayordinate space also includes gray matter data from subcortical regions, but since we focus on the cerebral cortex only, we provide information for a simplified version of the grayordinate space with an emphasis on the cerebral hemispheres and omit details from elsewhere in the brain.}. This grayordinate space does not only reduce the computational and storage demands of high resolution MRI data, but also enables a more compact data representation. All gray matter voxels in the cerebral cortex apart from the medial wall are represented at a 2 mm resolution using 59412 grayordinates, including 29696 and 29716 cortical vertices from the left and right hemispheres, respectively (Fig.~\ref{fig:grayordinates}). Data files in the grayordinate system typically contain a single 2D matrix, in which the x axis always represents indices of the standard set of grayordinates, while the y axis may represent, for instance, timepoints as illustrated in Fig.~\ref{fig:grayordinates}. The grayordinate space achieves spatial correspondence across different subjects, and thus, facilitates straightforward cross-subject comparisons.  

\subsubsection{Structural Pipelines}
The HCP structural acquisitions provide T1w and T2w images at 0.7 mm isotropic resolution. This high resolution facilitates the creation of more accurate cortical surfaces and myelin maps compared to conventionally lower resolution MRI data. Please see~\cite{HCP900} for a list of all imaging parameters used in structural MRI acquisitions.

The structural pipelines consist of three stages. The first pipeline produces a `native' structural volume space for each subject, which represents the best approximation of the subject's physical brain. After aligning T1w and T2w images, each native volume is registered to MNI space. The second structural pipeline includes 1) anatomical brain segmentation, 2) reconstruction of white and pial cortical surfaces, and 3) cortical-folding based inter-subject registration to FreeSurfer's standard surface atlas (fsaverage)~\cite{Fischl12} using a novel multimodal surface matching algorithm~\cite{Robinson14}. Finally, the third structural pipeline produces all of the volume and surface files required for visualisation purposes, along with 1) applying the cortical folding-based registration to the Conte69 surface template~\cite{VanEssen12}, 2) downsampling registered surfaces to 32k standard cortical mesh, in which connectivity analysis can be performed, 3) creating the final brain mask, and 4) creating myelin maps, which are computed as the ratio of T1w and T2w intensities~\cite{Glasser11}.

\subsubsection{Functional Pipelines}
Resting-state fMRI data for each subject was acquired in two sessions, divided into four runs of approximately 15 minutes (1200 timepoints) each. The sessions were held on different days and EPI phase encoding was applied in a right-to-left (R-L) direction in one run and in a left-to-right (L-R) direction in the other run for minimising distortion and blurring. During the scans, the subjects were presented a fixation cross-hair, projected against a dark background, which prevented them from falling asleep. The functional data was acquired at a spatial resolution of 2 mm isotropic, allowing an accurate mapping of gray matter BOLD signals onto the cortical sheet~\cite{Glasser13}. Please see~\cite{HCP900} for a list of all imaging parameters used in rs-fMRI acquisitions. 

The pipeline for the rs-fMRI data consists of two stages, which are referred to as `fMRIVolume' and `fMRISurface' in the HCP minimal preprocessing paper~\cite{Glasser13}. fMRIVolume briefly involves the following steps: 1) spatial distortion correction, 2) realignment of the timeseries to correct for subject motion, 3) EPI distortion correction using FSL's `topup' tool~\cite{andersson2003correct} and registration to the T1w image, 4) bias field correction, 5) normalisation of BOLD signals to a global mean of 10000, and 6) masking the data with the final brain mask. The second stage, fMRISurface, is employed to bring the 4D volume data to the grayordinate standard space. This is achieved by first mapping the gray matter voxels onto the native cortical surface and then transforming them onto the 32k standard triangulated mesh (Conte69)~\cite{VanEssen12}, using the cortical folding-driven registration's deformation field. The outcome of this pipeline is a standard set of timeseries for every subject with spatial correspondence, at a spatial resolution of 2 mm average surface vertex spacing. Excluding the non-cortical medial wall vertices, each cortical hemisphere is represented by around 30k vertices (the same number in each subject) in the grayordinate space as illustrated in Fig.~\ref{fig:grayordinates}(b). The fMRI timeseries are then slightly smoothed (2 mm FWHM) on the surface to match the vertex spacing of the standard 32k mesh. The preprocessed data is cleaned of structured noise with ICA-FIX~\cite{Beckmann04,Salimi14}, an ICA-based tool that automatically removes artefactual components from rs-fMRI data. Following these preprocessing and denoising steps, every single timeseries is temporally normalised to zero-mean and unit-variance.

Following completion of rs-fMRI acquisitions in each of the two scanning sessions, subjects were asked to complete a battery of tasks designed to activate different brain regions. Seven major domains were assessed in order to cover a wide range of neural systems, including 1) visual, motion, somato-sensory, and motor systems (MOTOR), 2) category specific representations (GAMBLING), 3) working memory, cognitive control systems (WM), 4) language processing (LANGUAGE), 5) social cognition (SOCIAL), 6) relational processing (RELATIONAL), and 7) emotion processing (EMOTION). These physiological paradigms are described in detail in~\cite{Barch13}. Two fMRI scans were collected for each task. Similar to rs-fMRI acquisition, one scan was acquired with R-L phase encoding direction and the other with the opposite phase encoding direction (L-R).  

Task activation maps for each subject/task were computed using FSL's FEAT, a data-analysis tool based on general linear modelling~\cite{Barch13}. Given the experimental design, FEAT creates a model that best fits the BOLD signals, and provides a statistical map that shows the activated brain areas in response to the stimuli. The analysis is carried out across sessions to obtain activation maps for each subject, which in turn, are used for evaluation purposes throughout this thesis. 

\subsubsection{Diffusion Pipelines and Tractography}
The whole set of DW volumes for each subject was acquired in six separate series (runs) and grouped into three pairs, each representing a different gradient scheme. The paired two runs include the same DW directions, but with reversed phase-encoding (i.e. alternating between R-L and L-R directions in consecutive runs). Each gradient scheme included approximately 90 diffusion weighting directions and six $b_0$ images. Diffusion weighting consisted of three shells of $b=1000$, $2000$, and $3000$ $s/mm^2$. Approximately the same number of images were acquired for each $b-$value within each run. DW images were collected with a spin-echo EPI sequence at a 1.25 mm isotropic resolution, using a Stejskal-Tanner (monopolar) diffusion-encoding scheme. Please see~\cite{HCP900} for a list of all imaging parameters used in dMRI acquisitions. 

The diffusion pipeline briefly consists of the following steps: 1) $b_0$ intensity normalisation, 2) estimation of distortion field by feeding the R-L and L-R $b_0$ images to the FSL's `topup' tool~\cite{andersson2003correct}, 3) estimation of eddy-current distortions and head motion for each image volume using FSL's `eddy' tool~\cite{andersson2012comprehensive}, 4) distortion correction and registration of $b_0$ image to the T1w image, 5) resampling the distortion-corrected diffusion data from eddy into 1.25 mm native structural space, and 6) masking the data with the final brain mask to reduce the file size. The diffusion gradient vectors are also accordingly rotated and registered to the native structural space~\cite{Glasser13}.

Following these preprocessing steps, diffusion data is fed into FSL's multi-shell spherical deconvolution toolbox (bedpostX)~\cite{behrens2003characterization,behrens2007probabilistic,Jbabdi12} for fiber orientation estimation. The output files generated by bedpostX are used by another FSL tool, probtrackX, to perform probabilistic tractography with respect to the estimated fiber orientations. Probabilistic tracking is performed on the native mesh (before registration) representing the grey/white matter interface (specified by the white and pial surfaces). The algorithm generates probability distributions from user-defined seed vertices. In our case, 5000 streamlines are seeded from all cortical vertices (i.e. each vertex is considered as a seed). The output tractography matrix contains a so-called `connectivity' value for each vertex, which corresponds to the number of streamlines that pass through that vertex and reach to the rest of the mesh. This matrix is finally used to estimate structural connectivity in our attempts to generate cortical brain parcellations.

\subsection{Datasets}
Our experiments are based on 200 subjects, acquired from the HCP S900 data release~\cite{HCP900} and grouped into two datasets of 100 subjects each, which will be hereinafter referred to as Dataset 1 and Dataset 2. All subjects had their scans successfully completed for all imaging modalities considered by the HCP.

Dataset 1 consists of 100 subjects (54 female, 46 male adults, age 22-35), whose demographic characteristics are summarised in Table.~\ref{tab:dataset1}. This dataset corresponds to the `Unrelated 100 Subjects' in the HCP database. We use Dataset 1 in particular for the development of connectivity-driven parcellations in Chapters 4, 5, 6, and 7.

\begin{table}[b!]
\centering
\caption{Summary of the demographic information of Dataset 1.}
\begin{tabular}{c}
\includegraphics[width=1\textwidth]{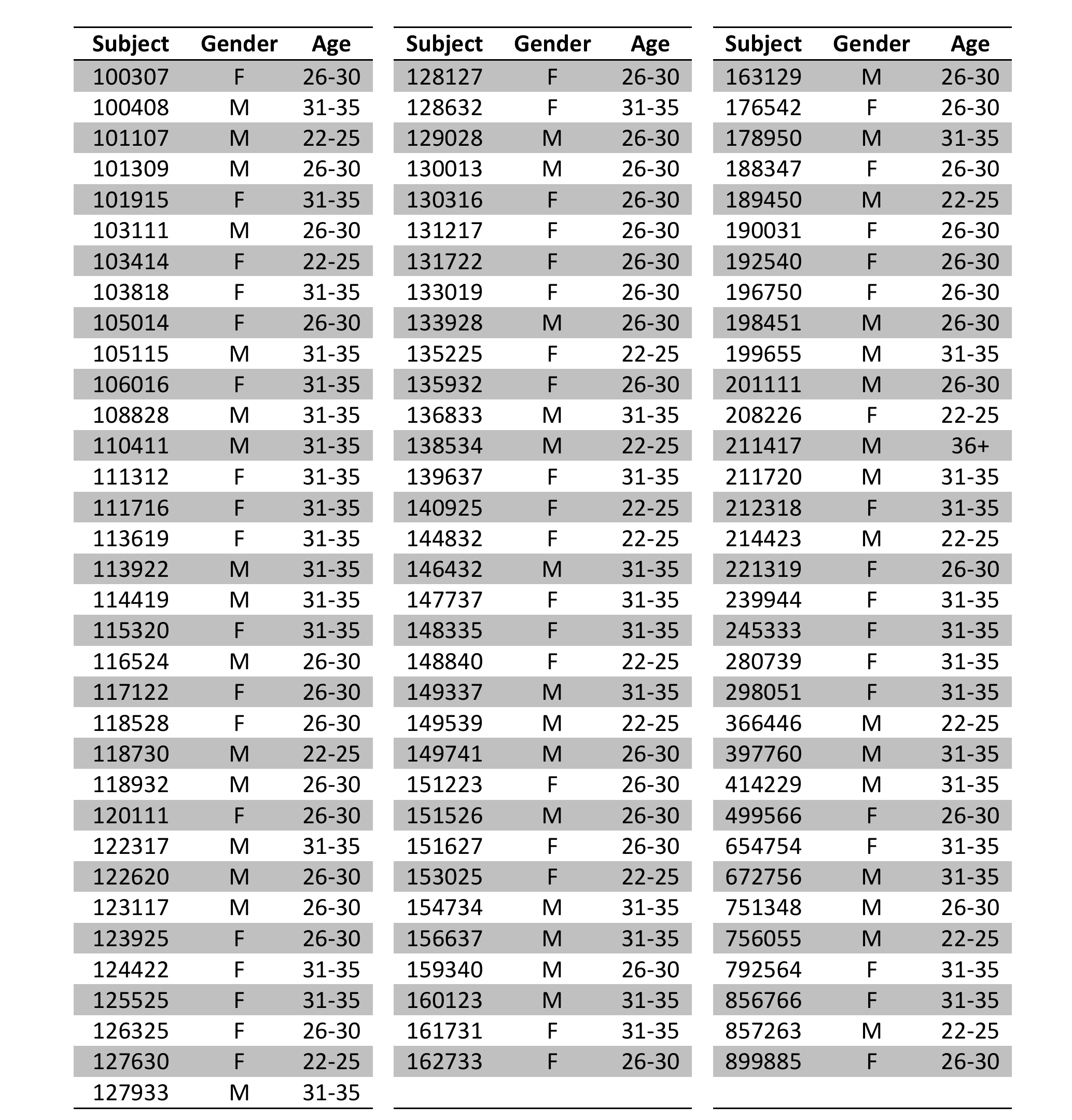}  
\end{tabular}
\label{tab:dataset1}
\end{table}

Dataset 2 consists of 100 randomly selected subjects (50 female, 50 male adults, age 22-35). Summary of the demographic information for all subjects is given in Table.~\ref{tab:dataset2}. We use Dataset 2 primarily for evaluation purposes in Chapter 7; that is, group-level parcellations are computed from Dataset 1, but evaluated on Dataset 2, so as to reduce the possible bias towards the computed parcellations with respect to the publicly available ones.

\begin{table}[ht!]
\centering
\caption{Summary of the demographic information of Dataset 2.}
\begin{tabular}{c}
\includegraphics[width=1\textwidth]{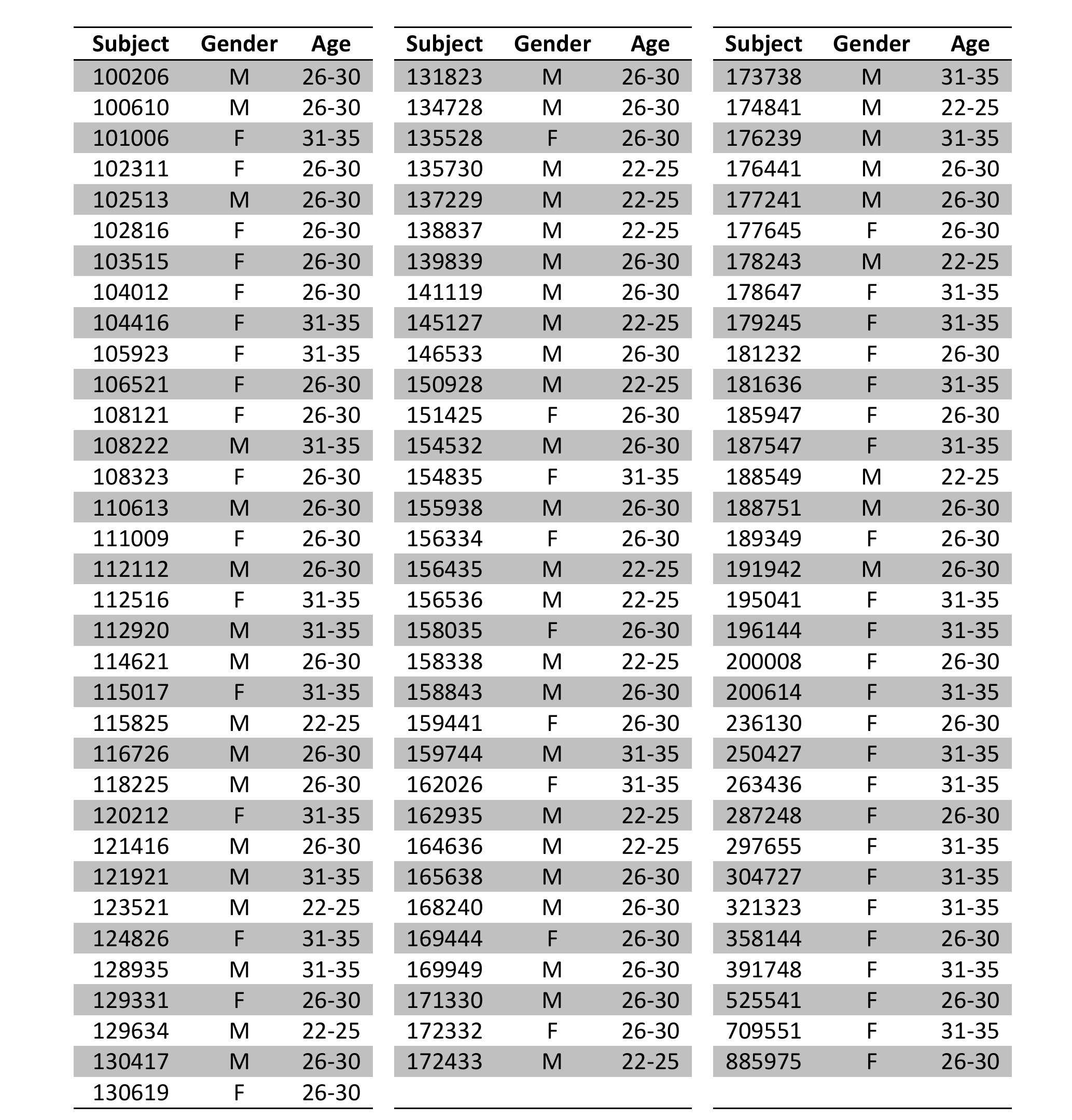}  
\end{tabular}
\label{tab:dataset2}
\end{table}

\chapter{Parcellating the Human Brain}
\label{chapter:literature}

\section*{Abstract}
\textit{This chapter defines the concept of brain parcellation and answers the questions such as `What is a parcellation?' and `In what terms a parcellation is of use?'. In the light of this, we briefly cover the historical foundations of brain parcellation and review widely-used techniques to subdivide the cerebral cortex (and subcortical nuclei) with respect to different neuro-anatomical properties, with a particular focus on functional and structural connectivity. We then extensively review connectivity-driven parcellation approaches by emphasising on their limitations and advantages over each other. We wrap-up the chapter by describing the challenges for obtaining reliable connectivity-driven parcellations and briefly summarise how we aim to overcome each challenge throughout this thesis. }

\section{Introduction}
Subdivision of the brain into anatomically and functionally distinct regions (cortical areas or subcortical nuclei) has been one of the ultimate challenges in the field of neuroscience, as it constitutes a major milestone towards understanding the brain. Cortical areas can be separated from their neighbours based on different properties such as micro-architecture (i.e. cyto-, myle- and receptor-architectonics)~\cite{Zilles02,Eickhoff05}, anatomy (e.g. formations of sulci/gyri)~\cite{TzourioMazoyer02,Desikan06,Fischl12}, function (e.g. involvement in cognitive operations)~\cite{Toga12,Thirion06,Thirion14}, and connectivity (e.g. a set of extrinsic inputs and outputs)~\cite{Passingham02,behrens2003characterization,Kim10,Eickhoff15}. Brain parcellation has many important roles in neuroscience and particularly in macro connectomics, as it provides 1) a reference map of the brain, enabling the comparison of results across subjects, groups, and studies, 2) a common language for researchers to discuss their findings, 3) a snapshot of the functional and structural organisation of the brain at macroscale, and 4) a means to reduce the complexity of connectivity, an aspect that is highly critical for the study of brain dynamics with whole-brain models~\cite{Toga12,craddock2013imaging,Shen13,Glasser16}.

While the first chapter has briefly introduced the rationale behind parcellation, this chapter provides a more detailed explanation of the brain parcellation concept and its implications for neuroscience. Although we refer to different types of neuro-anatomical properties and techniques to segregate the brain, the majority of the chapter focuses on connectivity-driven parcellation. We provide insight towards understanding the role of connectivity in order to define cortical parcellations and identify network nodes for connectome analysis. We further review many different methods developed for parcellating the brain, particularly the cerebral cortex, by using functional and structural connectivity information, with an emphasis on the key points to define a reliable parcellation. Finally, we summarise the challenges that are likely to emerge in the development process of connectivity-driven parcellations.

\section{Brain Parcellation: Aim and Scope}
\subsection{Historical Foundations of Brain Parcellation}
\begin{figure}[!t]
\centering
\includegraphics[width=0.8\textwidth]{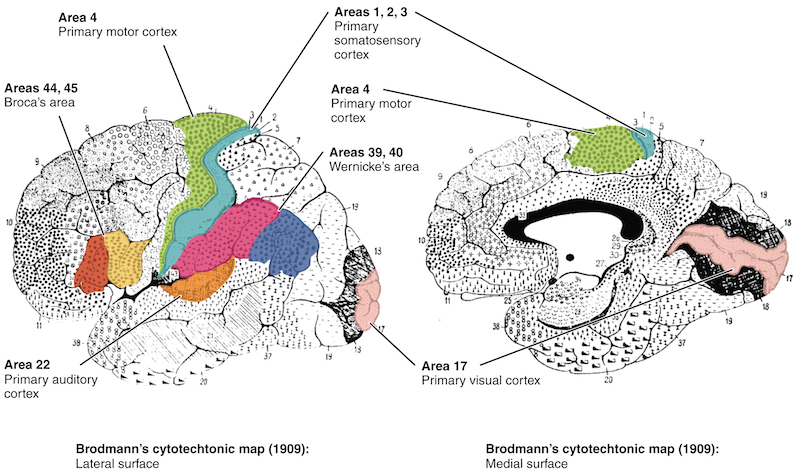} 
\caption[Brodmann's areas of the cerebral cortex.]{Brodmann's areas of the cerebral cortex, with an emphasis on regions specialised in distinct functionality. Image from OpenStax~\cite{brodmann_image}.}
\label{fig:brodmann}
\end{figure}

The concept of parcellation spans more than a century in the field of neuroscience. In 1909, Korbinian Brodmann produced one of the first parcellations of the brain by dividing the cerebral cortex into 52 functionally and anatomically distinct regions (Fig.~\ref{fig:brodmann})~\cite{Brodmann09}. However, prior to his seminal work, the foundations of brain parcellation and functional localisation were already set. As back as 1786, Vicq D'Azry had described the convolution characteristics of the cerebral cortex, emphasizing the differences in its morphology compared to the other animals~\cite{pearce2005brodmann}. In the early nineteenth century, Franz Joseph Gall had published his revolutionary ideas on the functional localisation in the brain, arguing that particular regions of the cerebral cortex were responsible for specific functions~\cite{Kendal03}, while Louis Pierre Gratiolet had divided the cerebral hemispheres into the frontal, temporal, parietal, and occipital lobes, with respect to the overlaying skull bones~\cite{pearce2005brodmann}. In 1861, Pier Paul Broca had discovered that the posterior region of the frontal lobe (now referred to as Broca's area) was involved in speech production. Inspired by Broca's work, in 1876 Karl Wernicke had identified another cortical region (in the posterior part of the temporal lobe, now called Wenicke's area) that was responsible for understanding of speech~\cite{Kendal03}. By the late 1800s, Camillio Golgi and Santiago Ramon y Cajal had revealed the first detailed descriptions of nerve cells by using staining techniques.

Nevertheless, it was Brodmann, who came up with a cortical parcellation by delineating different regions in the cerebral cortex based on the variations in cellular decomposition of gray matter tissues. Brodmann not only influenced the field of neuroscience with a reference map of the cortex, but also provided insight towards the basis of correlation between anatomy and function~\cite{Zilles01}. As a result, Brodmann's areas are still widely used today for the identification of activated foci, attributing its popularity to the modern imaging techniques, which allow \textit{in-vivo} mapping of the architectonic data into a spatial reference system, hence enabling comparison of function with anatomical areas~\cite{Zilles10}.

The clinical relevance of the brain's cellular organisation was already a major topic in the early twentieth century. Following Brodmann's pioneering studies, Cecile Vogt and Oscar Vogt subdivided the cerebral cortex based on the features derived from myeloarchitectonic features. The Vogts focused on the local variations in size and density of fiber bundles in neural tissue and achieved a cortical parcellation with a finer granularity (composed of around 200 distinct regions). Other post-mortem studies on the cellular level, such as by Constantin Freiherr von Economo and Georg Koskinas~\cite{Economo25}, contributed to the field of cytoarchitectonics by defining more cortical areas with homogeneous cellular decomposition. Micro-structural organisation of the cerebral cortex was later illuminated by means of receptor-architectonic mapping, an alternative architectonic technique that relies on the analysis of signals obtained from neurotransmitter receptors after being stimulated by chemical messengers~\cite{Zilles01,Zilles02,Zilles04}.

Until recently, functional localisation and parcellation of the human cortex mostly involved post-mortem examination of the brain, including the aforementioned architectonic techniques as well as observations on lesion-induced brain areas in patients with particular disorders (inability in speech and language processing being the most popular). It was possible to localise function in the living brain, but only via invasive techniques such as using electrodes to stimulate the cortex during surgery~\cite{Kendal03}. Although, all these endeavours extensively contributed to our understanding of the human brain, their extent to reveal the link between anatomy and function was still limited. First of all, the majority of early studies were only based on a single brain or explored very small groups of subjects, hence they were not able to take the inter-subject variability in the organisation of the cerebral cortex into consideration~\cite{Zilles01}. Second, most of the studies were investigator-dependent, as it was not possible to replicate the experiments in a systematic manner~\cite{Zilles10}. In addition, studies targeting the living brain, on top of being invasive, were mostly performed on patients with brain disorders, making the generalisation of observations/findings to healthy people questionable~\cite{Kendal03}.

\subsection{Neuroimaging for Brain Parcellation}
Starting in the early 1980s, structural and functional imaging technologies have revolutionised the study of organisational principles of the brain. MR and PET imaging have opened up new horizons in the field of neuroscience by allowing non-invasive exploration of the living brain from various aspects. Functional MRI has quickly become the most commonly used tool for functional localisation studies with its ability to capture the human brain while the subject is performing a pre-defined task. Parallel advancements in medical image analysis have led to development of various brain atlases and statistical analysis techniques, allowing to map activated foci with anatomically-defined cortical regions in the brain~\cite{Friston96,lindquist2008statistical,smith2004overview,Zilles10,penny2011statistical}, hence, finally linking function with anatomy. 

The Talairach and Tournoux atlas~\cite{Talairach88} has become particularly popular as it is one of the first attempts to provide a 3D spatial coordinate system, where cortical areas are labelled according to surface anatomy (sulci/gyri) and Brodmann's architectonic map; but only based on a single brain. In order to account for anatomical variability across subjects, probabilistic brain atlases (including a reference template and labelled map of distinct brain regions) have been introduced by co-registering and averaging multiple structural and functional images~\cite{collins1994automatic,mazziotta1995digital,mazziotta1995probabilistic}. As the imaging techniques progressed further, these attempts have been followed by many other cortical and subcortical brain parcellations~\cite{Lancaster00,TzourioMazoyer02,Fischl04,Eickhoff05,Desikan06,Zilles09}. Such subdivisions of the brain mainly follow the architectonic boundaries and/or anatomical landmarks in the cerebral cortex and are still being extensively used in neuroimaging studies. 

By the beginning of the new millennium, enough evidence was accumulated to consider the brain as a mosaic of distinct modules that are specialised in certain functions, such that specific facets of cognition, emotion and behaviour can be anatomically localised~\cite{friston2002beyond}. However, it has been also revealed that, no single brain region could perform a particular function by itself, nor would be a one-to-one mapping between brain regions and mental operations~\cite{friston2002beyond,Toga12}. This is further supported by numerous studies targeting the brain at rest (for an extensive review see~\cite{heuvel2010exploring}), which have identified several anatomically separated, but functionally connected cortical regions (commonly referred to as resting-state networks or RSNs) that are associated with a variety of functions, such as sensory/motor, visual processing, auditory processing, memory, and the so-called default mode network (DMN), which gets activated `by default' when subject is not involved in a task~\cite{buckner2007unrest}. 

Most of RSNs are consistently identified across studies, despite differences in the experimental setting, data acquisition and analysis techniques~\cite{Damoiseaux06,Smith09,heuvel2010exploring}. Besides, it has been also reported that functionally linked brain regions composing these large-scale networks continuously exchange information via anatomical connections~\cite{Heuvel09}. As a result, the dynamic interactions and information transmission through functional and structural connections between anatomically distinct regions (such as defined by their local micro-structure) are considered to facilitate performing a particular cognitive, sensory, or motor operation~\cite{Passingham02,Toga12}. 

Put together, the last few decades have seen a tremendous effort to provide a high-level abstraction of the fundamental organisation of the brain at macroscopic scales. This can be easily attributed to the advances in image acquisition techniques, which have facilitated the multi-scale subdivision of the brain using various modalities and methods. This has also brought the scope of parcellation to a much broader extent. As a matter of fact, the term `parcellation' does not refer to a unique map or tool, but rather define a spectrum of subdivisions that encapsulate fundamental neuro-biological information about cortical organisation and allow the mapping of brain function and anatomy with respect to different aspects. In this context, the role of structural and functional connectivity is particularly essential, as they can provide complimentary information towards parcellating the brain into neuro-anatomically distinct areas that are involved in different functions.

\subsection{Parcellation for Connectome Analysis}
In addition to its role in the functional segregation and integration of the cerebral cortex, parcellation constitutes one of the core steps to explore the system organisation of the human connectome at the the macroscale. As previously discussed in Section 2.3 \textit{Mapping the Brain}, the macro connectome is typically modelled as a network (or graph) of interactions between different regions of interest (ROIs) located in the cerebral cortex, where each ROI constitutes a node in the network. Attempts to perform connectivity analysis at the vertex level (i.e. considering each vertex as a distinct region) are generally not feasible due to the high dimensionality of connectomic data, which may yield numerous problems, including computational cost, sensitivity to noise, and lack of interpretability of the resulting networks~\cite{Toga12}. In fact, a higher level of abstraction provides a more neurobiologically meaningful basis for constructing a connectome, considering the aforementioned integration of neural subunits into spatially localised clusters in order to perform a function~\cite{Lu03,Heller06,craddock2013imaging,Thirion14}. As a result, connectivity analysis is generally carried out with respect to a parcellation, which may cover the whole cortex, i.e. whole-brain connectivity analysis, or target a local area of interest, i.e. ROI-based analysis, depending on the study under investigation.

ROI-based analysis is carried out by \textit{a priori} selecting a region or seed, typically based on a traditional task experiment and/or using biological knowledge. The correlations between this predefined ROI and the rest of the cortex (or other ROIs) are then explored via statistical analysis techniques~\cite{nieto2003region}. One major limitation of ROI-based analysis emerges from the fact that it cannot account for the joint interactions among multiple brain regions, as the signal outside the target area is not taken under consideration. The results, therefore, are likely to depend on the choice of the ROI(s), which may be prone to expert's error~\cite{Bellec06}. Besides, such ROIs may not be generalised to new subjects or data~\cite{Thirion14}. 

Whole-brain connectivity analysis targets the entire brain and generally relies on a set of nodes acquired from a cortical parcellation. Parcellations derived from anatomical/cytoarchitectonic brain atlases~\cite{TzourioMazoyer02,Talairach88,Desikan06,Fischl04} or random parcellations~\cite{Hagmann08,Honey09,Sporns11} are traditionally used for node identification, however these approaches might constitute several major limitations~\cite{craddock2013imaging,Thirion14}. Despite many widely-used brain atlases, there exists no consistency across parcellations defined by different atlases~\cite{Bohland09}, which can be attributed to the fact that each atlas represents a labelling of brain according to the accumulated knowledge at the time of atlas creation~\cite{Thirion14}. More importantly, ROIs derived from brain atlases or random parcellations are likely to comprise a mixture of signals, since such parcellations are not defined from the underlying data~\cite{Smith11,Craddock12,Shen13}. One typical example of this is the aforementioned anterior cingulate cortex (ACC), which exhibits a great amount of heterogeneity regarding structural~\cite{Beckmann09} and functional connectivity~\cite{Margulies07}, despite the fact that it is typically represented as a single ROI in traditional brain atlases~\cite{TzourioMazoyer02}. Similarly, many anatomical atlases divide the frontal lobe into relatively large regions, hence a post-parcellation is typically carried out to achieve a finer granularity, which may better fit the data~\cite{Thirion14}. 

Using ill-defined ROIs (ROIs that do not match well the actual structures of interest) may have a great impact on the estimation of network structure. According to large-scale simulations carried out by Smith et al.~\cite{Smith11}, incorrect ROI delineation for especially whole-brain connectivity analysis would severely affect the network modelling, as signals from unrelated parts of the brain are mixed to each other. In addition to network modelling, topological architecture of brain functional networks can also depend on the selection of the underlying parcellation, and thus, network statistics such as small-worldness can vary between networks derived from different atlases~\cite{Wang09}. Parcellation is, therefore, of great importance for the subsequent stages in connectivity analysis, which fundamentally rely on the network components. In this case, connectivity-driven parcellations (or more broadly, data-driven parcellations) provide a more reliable means for the definition of network nodes -hence a better model of the signal-, as they are typically learned from the same connectomic data used to conduct the connectivity analysis~\cite{Thirion06,Flandin02,Lashkari10,craddock2013imaging,Toga12,Thirion14}. Although, connectivity-driven parcellations are not likely to reflect the neuro-biological organisation of the brain, they can be typically associated with known brain structures for a more clinically relevant interpretation of the subsequent analysis~\cite{Thirion14,Finn15,Gordon16}.


\section{Connectivity-Driven Parcellation}
Connectivity-driven parcellation (CDP) aims at dividing the cerebral cortex (or a targeted area) into subregions that differ from their neighbours in connectivity, which can be in the form of long-range anatomical connections or functional interactions with other areas. The idea stems from the fact that each vertex or ROI can be described with a connectivity profile that shows how it is connected with the rest of the cerebral cortex~\cite{Passingham02}. The CDP problem is then cast as a clustering problem, where the objective is to subdivide the cortical surface into a set of clusters in a way that connectivity is similar for vertices in the same cluster but different between clusters~\cite{Eickhoff15}. This consequently allows to represent each parcellated region with a single, consistent connectivity profile in network analysis. CDP can, therefore, provide a well-suited model to the underlying connectivity structure at a lower dimensionality~\cite{Thirion14}.

\subsection{Methods for Connectivity-Driven Parcellation}
Connectivity has been one of the major sources of information for brain parcellation, and consequently, many CDP methods have been proposed, usually in association with a clustering algorithm driven by the similarity between connectivity profiles or fMRI timeseries. Among them \textit{k}-means, spectral clustering, and hierarchical clustering are generally the most commonly employed approaches in CDP studies, while various other techniques, including but not limited to independent component analysis, boundary mapping, non-linear manifold learning, region growing, and mixture models have also been alternatively used, each with its own limitations and advantages. Below we will review the most prominent approaches to obtain connectivity-based parcellations, with an emphasis on the differences between various methods. 

\subsubsection{\textit{K}-means Clustering}
\textit{K}-means clustering constitutes one of the most extensively used methods for CDP studies~\cite{Tomassini07,nanetti2009group,Mezer09,Bellec10}. It typically subdivides a ROI into a predefined number of subregions, which is specified by $k$ as an external parameter. 
After selecting the initial cluster centroids ($k$ seeds), \textit{k}-means clustering algorithm is driven by two alternating steps: (1) associating data samples (e.g. timeseries or connectivity profiles) with clusters (assignment step) and (2) the estimation of new cluster centroids (update step). At each iteration, the cluster assignments take place in a way that minimises the sum of squared differences between the data samples and their associated cluster centroids.

\begin{figure}[thb!]
\centering
\includegraphics[width=\textwidth]{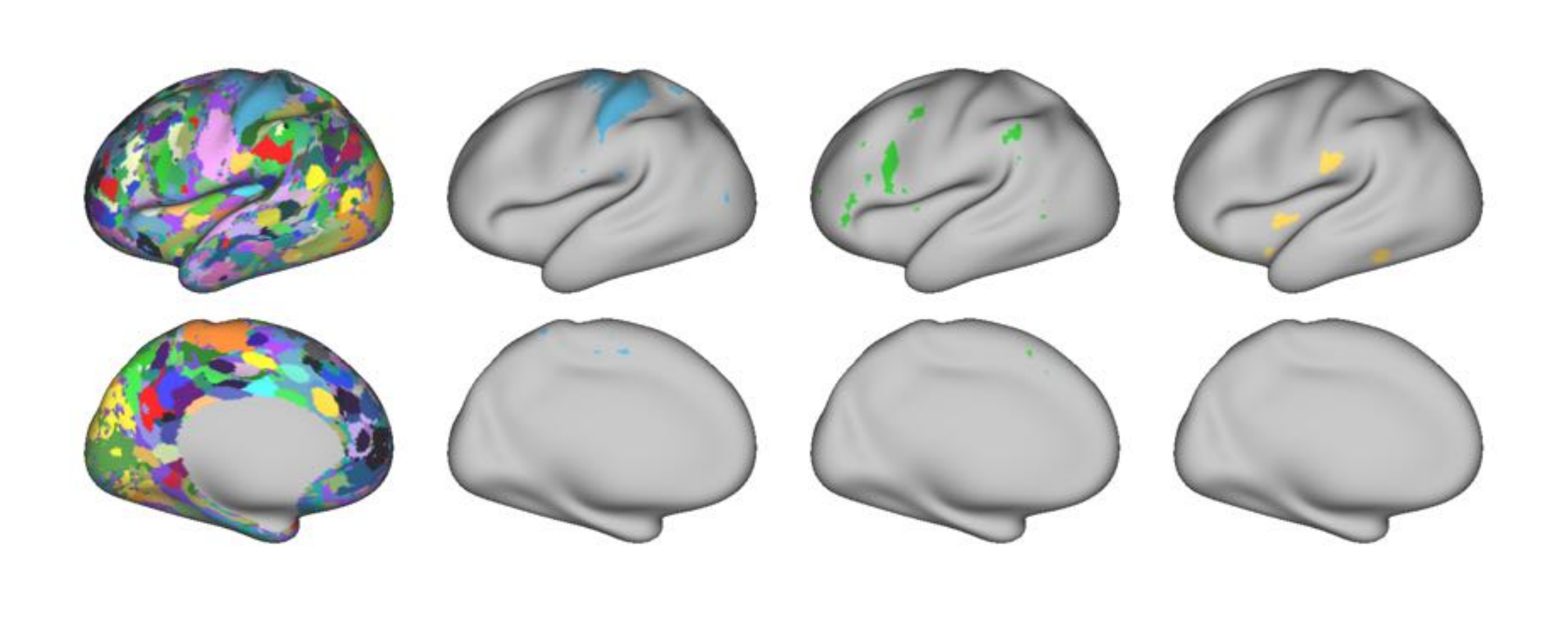} 
\caption[A parcellation of the left hemisphere obtained with \textit{k}-means.]{A parcellation of the left hemisphere obtained with \textit{k}-means using resting-state fMRI connectivity of a single subject ($k=100$). 3 arbitrarily chosen regions are shown individually alongside the parcellation to emphasise the fact that \textit{k}-means does not impose any spatial constraints.}
\label{fig:kmeans}
\end{figure}

The method has been employed in varying aspects of the CDP problem. For example, Mezer et al.~\cite{Mezer09} applied \textit{k}-means to rs-fMRI BOLD timeseries to obtain whole-brain parcellations including subcortical areas and thalamus, while Kahn et al.~\cite{kahnt2012connectivity} relied on a similar approach in order to subdivide the orbitofrontal cortex. Functional connectivity captured at rest is further used to parcellate the medial frontal cortex into supplementary motor areas (SMA) and pre-SMA subregions~\cite{Kim10}. Anwander et al.~\cite{anwander2007connectivity} and Tomassini et al.~\cite{Tomassini07} used \textit{k}-means in conjunction with tractography-driven anatomical connectivity to parcellate Broca's area and lateral premotor cortex, respectively. An analogous approach was followed by Nanetti et al.~\cite{nanetti2009group} to segment the SMA and the insula. \textit{K}-means applied to connectomic data has further revealed the heterogeneous connectivity patterns within the cingulate cortex~\cite{Beckmann09}. A resting-state fMRI-driven parcellation of the left hemisphere obtained with \textit{k}-means is shown in Fig.~\ref{fig:kmeans}.

While \textit{k}-means provides a `hard' clustering (i.e. every vertex can only be assigned to a single cluster), its fuzzy counterpart \textit{c}-means can alternatively be used to obtain a `soft' clustering, where each each vertex is represented by a weighted label vector~\cite{bezdek2013pattern}. For example, by applying \textit{c}-means to the rs-fMRI data, Lee and colleagues~\cite{Lee12} have subdivided the cortical and subcorbital gray matter into several regions, each is associated with a different RSN.

Although being extensively used in the context of CDP, several points should be taken into consideration before adapting a \textit{k}-means based approach. First of all, the algorithm highly depends on the initialisation of cluster centroids, which can be random or based on \textit{a priori} knowledge, such as anatomical landmarks~\cite{Eickhoff15}. Partially due to this randomness, but mostly because of the internal dynamics of the algorithm, it is not likely to obtain the same parcellation every time the algorithm is employed. It is, thus, important to run the algorithm for many times before deciding on the final parcellation~\cite{nanetti2009group}. It is also worth noting that, repeating the clustering process for different values of $k$ yields different solutions and is not expected to obtain a hierarchy of nested regions that are linked to each other in a child-parent relationship~\cite{Eickhoff15}.

\textit{k}-means could become computationally prohibitive especially when the dimensionality of the data is extremely high. This is typically the case when the whole cortex is targeted instead of a relatively small ROI and the parcellation is carried out with a very large $k$. To overcome the impact of computational limitations, Thirion et al.~\cite{Thirion14} suggest to utilise principal component analysis (PCA) as a preprocessing stage to reduce the dimensionality of the data. Another limitation of \textit{k}-means is that it naturally does not provide spatially contiguous parcellations, as the spatial structure of the data is not accounted for. For example, such disconnected parcels can be seen in Fig.~\ref{fig:kmeans}. A typical procedure to acquire spatially more coherent regions is to incorporate external constraints (for example anatomical coordinates of vertices) into the clustering algorithm~\cite{Flandin02,Tomassini07}. A modified \textit{k}-means approach driven by a similar principle will be presented in the next chapter.

\subsubsection{Spectral Clustering}
Another widely-used clustering approach that has been frequently employed to CDP is spectral clustering~\cite{johansen2004changes,klein2007connectivity,Heuvel08,Craddock12,Shen10,Shen13,Parisot16a}, owing its popularity to the fact that the similarity of connectivity profiles between any pair of vertices can be naturally represented as a connectivity matrix~\cite{Eickhoff15}, where nodes correspond to vertices and edges represent the connections between the nodes. Structures of interest in this matrix can be revealed through spectral decomposition and further used to obtain subdivisions of the underlying data. This is typically achieved by computing the first eigenvectors of the normalised graph Laplacian of the similarity matrix~\cite{Shi00}. These eigenvectors provide a feature matrix in a lower dimensional space and can be submitted to a conventional clustering algorithm for parcellation, preferably after being reordered. 

Spectral reordering modifies the structural organisation of the connectivity matrix with respect to the information provided by the eigenvectors, such that vertices with higher similarity are positioned together within the matrix~\cite{barnard1995spectral}. For example, Johansen-Berg et al.~\cite{johansen2004changes} used spectral reordering to parcellate the medial frontal cortex into two subregions (SMA and pre-SMA), in which the number of clusters was not predefined, but estimated by visual inspection~\cite{cloutman2012connectivity}. However, when there is no clear separation between the data points in the reordered connectivity matrix, traditional clustering algorithms, such as \textit{k}-means, are generally employed to obtain the final parcellation borders~\cite{anwander2007connectivity,klein2007connectivity}. This is further beneficial to reduce the inspector bias and obtain parcellations in a more data-driven manner~\cite{cloutman2012connectivity}.

\begin{figure}[thb!]
\centering
\includegraphics[width=\textwidth]{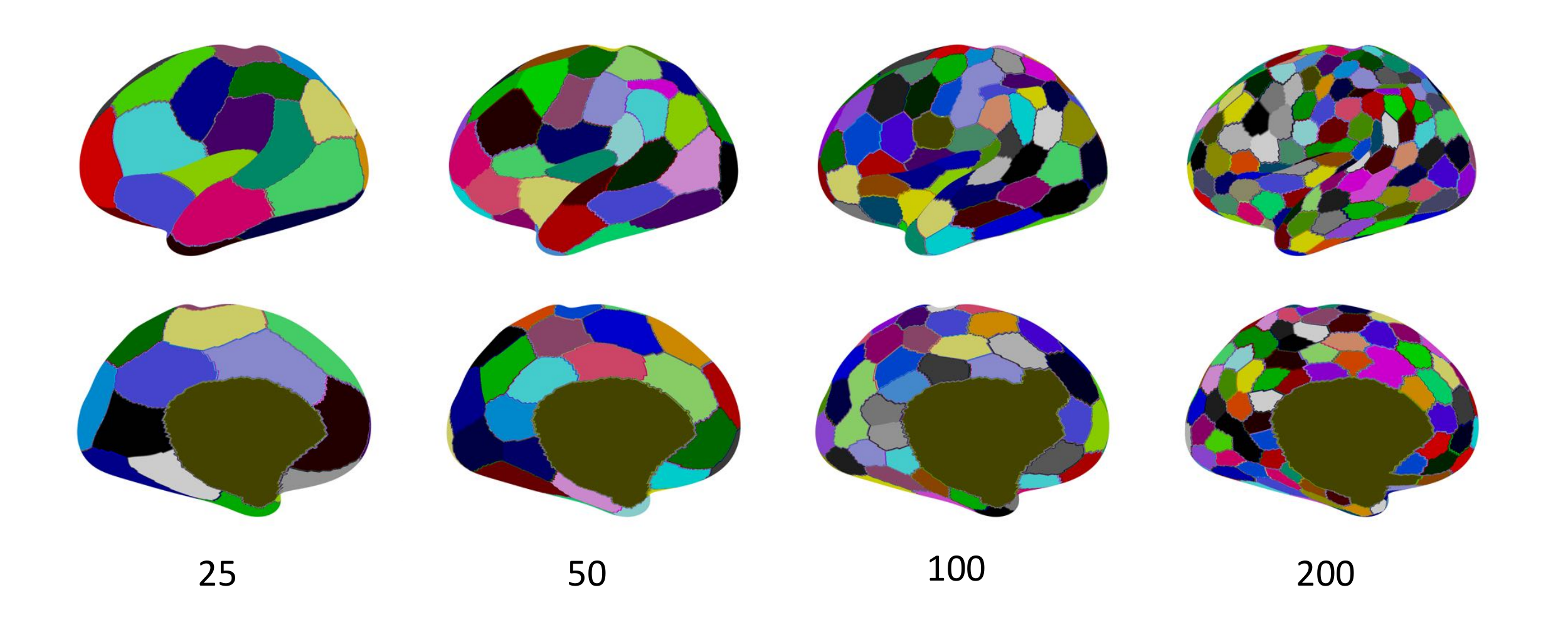} 
\caption[Cortical parcellations obtained using spectral clustering.]{Cortical parcellations of the left hemisphere obtained with the normalised cuts spectral clustering algorithm using resting-state fMRI connectivity of a subject. The numbers indicate the predefined resolution of each parcellation.}
\label{fig:spectral}
\end{figure}

Alternatively to spectral reordering, eigenvectors can be directly fed into a clustering algorithm to compute a parcellation with a predefined number of regions. For example, Thirion et al~\cite{Thirion14} applied spectral clustering to task-based fMRI data in conjunction with \textit{c}-means to obtain ROIs that can be used for functional connectivity analysis. Van den Heuvel et al.~\cite{Heuvel08} modelled the whole-brain CDP problem as a normalised cut (NCut) graph partitioning problem~\cite{Shi00} and used spectral clustering to identify seven RSNs from a combined connectivity graph of several healthy subjects. Craddock et al.~\cite{Craddock12} extended the idea of NCut spectral clustering to obtain spatially contiguous cortical parcellations that can be used as nodes in connectome analysis, by imposing spatial constraints to the input similarity matrix, i.e. only retaining the edges constructed between anatomically adjacent vertices and discarding the others. In such cases, final parcellations can be typically obtained with \textit{k}-means. However, as discussed above, \textit{k}-means heavily depends on initialisation and may yield different parcellations on a run-to-run basis. For a more robust clustering, it can be replaced by an alternative technique proposed in~\cite{Yu03}, that includes learning a rotation to discretise the eigenvector representation, such that the discrete eigenvectors reflect an optimal partitioning of the graph~\cite{Heuvel08,Craddock12,Thirion14}. Several cortical parcellations obtained using spectral clustering with normalised cuts at different levels of detail are shown in Fig.~\ref{fig:spectral}.

There are at least two drawbacks associated with NCut-based spectral clustering, particularly when spatial constraints are in use. NCut graph partitioning tends to yield a `balanced' subdivision of the input graph. However, in cortical areas where a reliable separation does not exist (such as areas with relatively weak connectivity or low SNR), it is likely to generate parcellations with uniformly sized clusters, which can bee seen in Fig.~\ref{fig:spectral}. This relatively simple model may thus fail to reflect the complex structures generally present in the brain~\cite{Eickhoff15}. Due to spatial constraints imposed to the similarity matrix, NCut is driven to a large extent by the underlying structure of the graphical model, which consequently leads to capturing the spatial organisation of the brain rather than the underlying connectivity characteristics~\cite{Craddock12,Thirion14}. Similar to \textit{k}-means and in contrast with hierarchical clustering, spectral clustering does not yield a hierarchy of nested regions when the algorithm is employed for different number of clusters, however there may still be some sort of inherently-captured similarity between parcellations of different granularity~\cite{Eickhoff15}.

An important consideration point in spectral clustering is how to construct a similarity matrix that could summarise the connectivity characteristics shared across individual subjects when a group-wise study is devised. While popular solutions include averaging individual subject connectivity matrices~\cite{Patel08} or computing an intermediate graph representation from subject-level parcellations~\cite{Heuvel08,Craddock12}, more recent techniques construct a joint model of connectivity patterns~\cite{Shen13,Parisot15,Parisot16a}. A novel method to obtain a group-level graphical representation of a population will be provided in Chapter 6.

\subsubsection{Hierarchical Clustering}
\begin{figure}[bht!]
\centering
\includegraphics[width=0.95\textwidth]{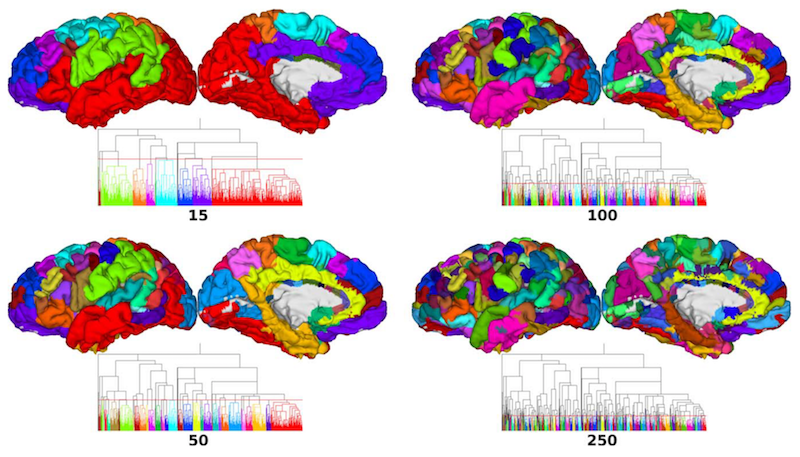} 
\caption[Cortical parcellations with different resolutions obtained via hierarchical clustering.]{Cortical parcellations of the left hemisphere acquired by cutting a hierarchical tree at different levels, denoted by the red horizontal lines. The numbers indicate the predefined resolution of each parcellation. Image is adapted from~\cite{Moreno14}.}
\label{fig:ward}
\end{figure}

Hierarchical clustering is a technique for grouping data points into a set of nested clusters that can be represented as a hierarchical tree, which is often called a dendrogram~\cite{Johnson67}. Cutting this tree at different levels of depth produces parcellations with the desired precision (Fig.~\ref{fig:ward}). Strategies for hierarchical clustering generally fall into two groups, depending on the direction towards which the algorithm is operated, i.e. bottom-up (agglomerative) and top-down (divisive). The former approach starts with each data point as a singleton cluster and the most similar clusters are merged as the algorithm progresses up the hierarchy. The divisive approach initially assigns all data points into a single cluster and progressively divides the least homogeneous cluster at each iteration. However, this strategy is very rarely used for brain parcellation, and hence is not covered here.

Agglomerative hierarchical clustering, on the other hand, is widely adapted as a parcellation technique in numerous CDP studies~\cite{Cordes02,salvador2005neurophysiological,Bellec10,Mumford10,Jenatton11,Eickhoff11,Blumensath13,Moreno14}. Its popularity can be attributed to the fact that, in contrast with the aforementioned clustering approaches, hierarchical clustering provides a spectrum of parcellations that can be explored at different coarseness levels~\cite{Eickhoff15} as shown in Fig.~\ref{fig:ward}. Such a clustering model has a higher tendency to reflect the hierarchical organisation of the human connectome; however, it is worth noting that a hierarchical parcellation does not necessarily provide a neurologically meaningful segregation of the brain~\cite{Eickhoff15}. 

Hierarchical clustering techniques have been adapted to investigate various facets of brain parcellation with regards to both functional and structural connectivity, though the former being more extensively studied. Cordes et al.~\cite{Cordes02} made one of the first attempts to use hierarchical clustering for the identification of functional connectivity patterns in resting state data. Quite a few studies have been carried out to identify consistent RSNs across healthy groups of subjects by using hierarchical clustering at different processing stages~\cite{salvador2005neurophysiological,Bellec10,Mumford10}. For example, Salvador et al.~\cite{salvador2005neurophysiological} revealed six major systems through a hierarchical parcellation of a connectivity network defined from an anatomical template. Bellec et al.~\cite{Bellec10} followed a similar procedure for RSN identification, but instead of relying on an anatomical template for network construction, a data-driven parcellation is acquired through a multi-level approach that combines region growing, \textit{k}-means clustering, and hierarchical clustering. Brain reading studies that aim to predict the subject's behaviour during a t-fMRI session have also utilised hierarchical techniques to obtain structured patterns of activation from task-based data~\cite{Jenatton11,Michel12}. Finally, hierarchical clustering has been applied to both rs-fMRI and dMRI to obtain whole-brain cortical parcellations that can be used to substitute the network nodes in network analysis~\cite{Blumensath13,Moreno14}.

One critical consideration point in these studies is the selection of a `linkage rule' that determines which clusters to merge with respect to a similarity measure. Among many linkage rules, average (mean) link~\cite{salvador2005neurophysiological,Bellec10,Mumford10,Moreno14} and Ward's method~\cite{Smith13,Blumensath13,Eickhoff11,Michel12,Jenatton11} seem to dominate the literature. The former is driven by the similarity measured in terms of the mean connectivity between vertices within each pair~\cite{sokal1958statistical}, whereas Ward's method merges a pair if the resulting cluster leads to minimum increase in the total within-cluster variance~\cite{Ward63}. It is important to note that, hierarchical clustering algorithms can yield very imbalanced parcellations depending on the merging criterion~\cite{Eickhoff15}. As there is no best linkage rule, it needs to be selected experimentally and/or based on the study under investigation. It may also be critical to impose constraints to the merging process, (e.g. two clusters can be merged if and only if they are adjacent to each other), in order to obtain spatially contiguous, and hence, biologically more plausible parcellations~\cite{Blumensath13,Thirion14,Moreno14}.

Although hierarchical approaches generate a nested multi-level parcellation, only few levels of granularity are explored in practice~\cite{Eickhoff15}. While these levels are typically determined arbitrarily, but still accounting for a wide range of resolutions~\cite{Blumensath13,Moreno14}, other methods may rely on more sophisticated techniques to decide where to cut the dendrogram~\cite{Mumford10}. One critical drawback of hierarchical clustering is its sensitivity to noise, which may lead to erroneous (false) boundaries detected at higher levels to progress down the hierarchical tree and subsequently reduce the reliability of parcellations at coarser levels. In other words, once a parcel is formed by merging two clusters, it cannot be undone in later iterations~\cite{Moreno14}. As a result, an optimisation based clustering strategy like \textit{k}-means may be more appropriate for parcellation when dealing with small datasets/ROIs or only a single resolution is desired~\cite{Moreno14}. 

\subsubsection{Independent Component Analysis}
\begin{figure}[t!]
\centering
\includegraphics[width=\textwidth]{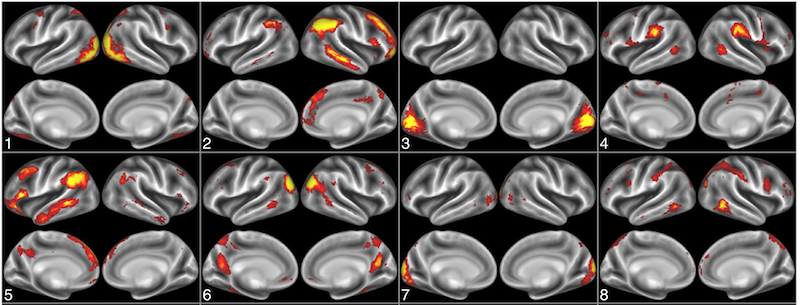} 
\caption[Spatial maps obtained via independent component analysis.]{Spatial maps obtained by applying ICA to resting-state fMRI connectivity data of 20 subjects. Overlays correspond to group-derived $z$-score maps thresholded at $z>5$. Image is adapted from~\cite{Smith13}.}
\label{fig:ica}
\end{figure}

Independent Component analysis (ICA) is currently considered as the reference model for identifying networks that underlie the intrinsic functional organisation of the brain~\cite{Beckmann04}. The key idea behind ICA is to decompose fMRI data into a set of components that are maximally independent from each other. Each component is then interpreted as a spatially distributed connectivity network and represented in the form of a weighted set of vertices (i.e. a spatial map)~\cite{Beckmann04}. Several spatial maps obtained using ICA are shown in Fig.~\ref{fig:ica}. In this figure, each map is thresholded to identify the cortical regions with the strongest connectivity within each component. ICA-based analysis techniques~\cite{kiviniemi2003independent,Beckmann04,Damoiseaux06,Smith09,DeLuca06} have been widely used to subdivide the human connectome into robust functional networks across resting~\cite{Beckmann04,Damoiseaux06} and task-based~\cite{Smith09} experimental settings, as well as across different groups of subjects~\cite{Toga12}. The multivariate signal that forms the basis of ICA can also be formulated in the dictionary learning framework~\cite{Varoquaux11,abraham2013extracting}. Different strategies can be used to combine fMRI data from subjects, such as concatenating individual timeseries~\cite{kiviniemi2003independent,Beckmann04} or using more sophisticated methods that take the inter-subject variability into consideration~\cite{varoquaux2010group,Varoquaux11,abraham2013extracting}. 

ICA may capture different components of the signal, including non-neural components and actual functional networks~\cite{abraham2013extracting}. On one hand, this could allow extraction of artefactual patterns from the data, which is not possible with the aforementioned clustering techniques~\cite{Eickhoff15}. On the other hand, it constitutes a limitation, as manual selection of components may be required~\cite{Shen13}. Spatial maps obtained via ICA may potentially have several separate regions scattered across any given map and overlap with each other, hence, do not provide spatially contiguous and non-overlapping parcellations. As a result, ICA-based techniques may not be preferred by some researchers~\cite{smith2013resting,Eickhoff15}. However, it might be possible to achieve spatial coherence by imposing additional post-processing steps, such as (1) increasing the number of components until spatial contiguity is ensured~\cite{kiviniemi2009functional}, (2) using sparsity-induced regularisation~\cite{abraham2013extracting}, or (3) subdividing components that are not spatially contiguous~\cite{Smith13}. Since ICA inherently assigns each component a distinct timeseries, subsequent network modelling using these components as nodes is guaranteed not to be rank deficient, whereas a hard parcellation is likely to have multiple parcels with very similar timeseries, which may potentially affect the network modelling~\cite{smith2013resting}.

\subsubsection{Boundary Mapping}
\begin{figure}[hbt!]
\centering
\includegraphics[width=0.8\textwidth]{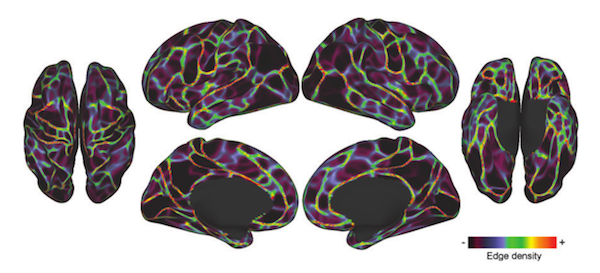} \\
\includegraphics[width=0.8\textwidth]{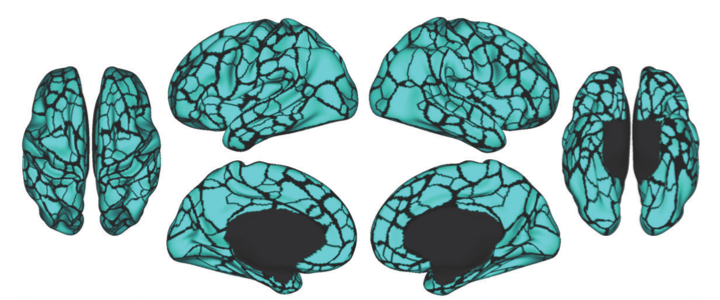} \\
\caption[Whole-brain cortical parcellation with boundary mapping.]{(\textit{Top}) A boundary map obtained from resting-state functional connectivity patterns of 120 healthy young adult subjects. Hot colours indicate the locations where connectivity abruptly changes. (\textit{Bottom}) Cortical parcellations derived from the boundary map. Both images are adapted from~\cite{Gordon16}.}
\label{fig:boundary}
\end{figure}

Boundary mapping is an alternative CDP approach that casts the clustering problem as the identification of abrupt changes between connectivity patterns captured at rest and solves it using image analysis techniques~\cite{Cohen08,Wig13,Wig14,Gordon16}. The key idea behind boundary mapping is to identify locations on the cerebral cortex where the functional connectivity pattern changes rapidly, potentially representing boundaries between functionally localised cortical areas~\cite{Cohen08}. Since the cerebral cortex is treated as a 2D cortical sheet, edge detection algorithms can be used to compute gradient maps from the resting-state correlations. Such maps provide information about the transitional zones in connectivity across the cortex, thus, can be fed into an image segmentation tool (such as watershed algorithm~\cite{Vincent91}) for delineating cortical parcellations. Boundary mapping has been successfully used to obtain areal segregations of the cortical surface from resting-state functional connectivity, targeting both specific cortical regions (such as the frontal cortex~\cite{Cohen08} and the lateral parietal cortex~\cite{nelson2010parcellation}), and the entire cerebral cortex~\cite{Wig13,Wig14,Laumann15,Gordon16,Gordon16individual}. A whole-brain boundary map and the resulting cortical parcellations are provided in Fig.~\ref{fig:boundary}. 

One major drawback of the boundary mapping techniques is that they do not allow model selection, such that the granularity of the parcellation cannot be predefined, which is in contrast with most of the previously mentioned clustering techniques~\cite{Thirion14}. In addition, boundary mapping is very sensitivity to spurious gradients captured from connectivity patterns, especially when the whole cortex is under consideration. As a result, a considerable amount of data is required to reduce the possible impact of falsely identified edges, which makes the technique more suitable for application to averaged-datasets of large populations~\cite{Gordon16}. On the other hand, a subject-level parcellation at the whole-brain scale can still be achieved, for example by scanning a single subject many times~\cite{Laumann15}. However, it should be noted that the high dimensionality may still constrain the detection of robust connectivity patterns at the subject level. 

Considering the fact that detected edges can be represented as a probabilistic map, some researchers position boundary mapping in between hard and soft clustering methods~\cite{Eickhoff15}. While boundary mapping has been a very popular approach for functional CDP, to the best of our knowledge it has not been applied to structural connectivity estimated from dMRI; though, Chapter 5 will cover a newly developed method to fill this gap.

\subsubsection{Non-linear Manifold Learning}
This group of statistical techniques reconceptualises the CDP problem as a non-linear feature reduction problem and proposes using manifold learning to capture patterns of interest within the inherently high dimensional connectivity space~\cite{shen2006nonlinear,thirion2004nonlinear,Langs10,Langs14,Langs15,Langs16}. Contrarily to linear methods such as PCA and ICA, non-linear feature reduction techniques, including but not limited to ISOMAP~\cite{Tenenbaum00}, locally linear embedding (LLE)~\cite{Roweis00lle}, Laplacian eigenmaps (i.e. spectral embedding)~\cite{Belkin03}, and diffusion maps~\cite{Coifman06diffusion}, seek a low-dimensional manifold that is assumed to represent the intrinsic geometry of the underlying data. Such methods encapsulate local proximity information within a nearest-neighbourhood graph, which is then used to compute embedding coordinates in a low dimensional space by applying spectral analysis techniques to certain matrix forms associated with the underlying high dimensional graph~\cite{shen2006nonlinear}

\begin{figure}[t!]
\centering
\includegraphics[width=\textwidth]{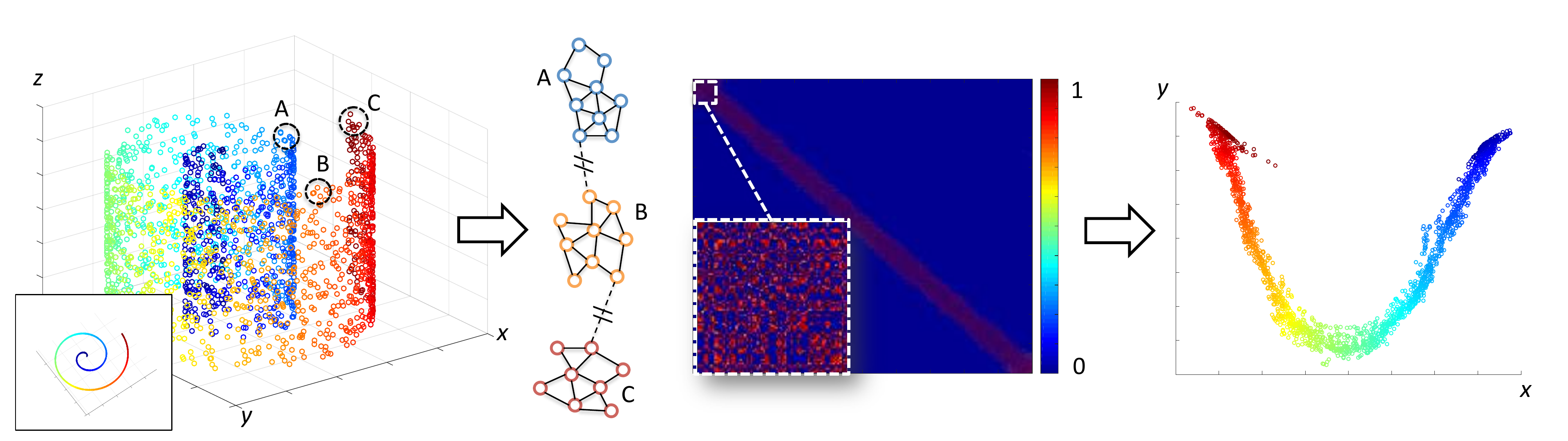} 
\begin{tabular}{ccc}
~~~~(a) ~~~~~~~~~~~~~~~~~~~~~~~~~~~~~~~~~~~~ & (b) ~~~~~~~~~~~~~~~~~~~~~~~~~~~~~ & (c)   
\end{tabular}
\caption[An example of non-linear manifold learning.]{An example of non-linear manifold learning: (a) The well-known Swiss roll dataset generated from 2000 data points, inset showing the bird-view of the space from the \textit{z} axis. (b) The local proximity information in the 3D space is represented with a graphical model (i.e. an adjacency matrix), in which only the closest data points are connected to each other. The graph is partly illustrated for three arbitrary neighbourhoods (A, B, and C). (c) The 2-dimensional embedding obtained using Laplacian eigenmaps, in which locally connected data points in the 3D space remain nearby in the manifold.}
\label{fig:swiss}
\end{figure}

ISOMAP (short for Isometric Mapping) computes a lower-dimensional embedding by first approximating geodesic distances between all points and then applying the classical multi-dimensional scaling (MDS)~\cite{borg2005modern} to the distance matrix. LLE assumes that the manifold in the data is locally linear and finds a non-linear embedding according to the optimal local weights, which are computed for each data point in the form of a linear combination of their nearest neighbours. Laplacian eigenmaps relies on the assumption that the data lies in a low-dimensional manifold in a high-dimensional space~\cite{Belkin03} and uses the graph Laplacian matrix to perform dimensionality reduction. Diffusion maps, while is very similar to Laplacian eigenmaps in terms of underlying graph representation, takes a probabilistic perspective to the dimensionality reduction problem and defines a Markov chain from a dataset, in which the distance between pairs of data points is characterised by their transition probability~\cite{shen2006nonlinear}. An illustrative example featuring the well-known Swiss roll dataset is provided in Fig.~\ref{fig:swiss} to demonstrate typical steps involved in non-linear manifold learning, in which Laplacian eigenmaps is used to compute a 2-dimensional embedding from a three dimensional dataset. 

Various non-linear manifold learning approaches, diffusion maps~\cite{shen2006nonlinear,Langs14,Langs15} and Laplacian eigenmaps~\cite{thirion2004nonlinear} being the most popular, as well as a broader class of spectral decomposition techniques~\cite{shen2006nonlinear,Thirion06,Craddock12,Shen13} are used for learning low-dimensional embeddings from the connectivity data (see~\cite{Langs14} for other related work). The common point in these studies is that the interactions between remote brain regions are represented with a graph in the high dimensional space, which is then decomposed into a feature matrix with lower dimensionality that can be used to reveal, for example, temporal patterns of functional activity~\cite{thirion2004nonlinear,shen2006nonlinear} or further fed into a clustering algorithm to identify cortical parcellations~\cite{Craddock12,Langs14}. Novel CDP approaches that make use of non-linear manifold learning in the form of Laplacian eigenmaps~\cite{Belkin03} and spectral decomposition techniques will be proposed in Chapters 5 and 6, respectively.

\subsubsection{Alternative Methods for CDP}
With the increasing attention in macro connectomics, last decade has seen a tremendous progress towards CDP. While it is not feasible to cover every single CDP approach in as much detail as the aforementioned methods, we provide an overview of alternative techniques that approach the CDP problem from different points of view.

Region growing constitutes a good example to the parcellation techniques that impose a spatial constraint to the clustering process and has been used for many times in CDP studies~\cite{Lu03,Heller06,Bellec06,Blumensath13}. The general idea is to iteratively merge vertices with their most similar neighbour until a stopping criterion is satisfied. Region growing, therefore, inherently maximises the homogeneity within each parcel~\cite{Heller06,Bellec06}. However, this may pose a limitation especially in whole-brain CDP, i.e. imbalanced parcellations with many small but few large parcels. This is typically avoided by capping the parcel size with a predefined threshold~\cite{Bellec06} or driving the algorithm from an initial set of seed vertices that approximately cover the entire cortex~\cite{Blumensath13}. The former leads to parcellations with evenly sized clusters, which may consequently show similar properties as in spectral clustering. Region growing tends to segregate the cortical surface into many, but highly homogeneous clusters, some of which may only consist of a single vertex. As a result, further processing (such as a second level clustering) may be required to obtain more neuro-biologically plausible parcellations~\cite{Bellec06,Bellec10,Blumensath13}. 

A different family of CDP methods consists of probabilistic modelling techniques, which describe the underlying data as a set of probability distributions, such as Gaussian mixture models~\cite{golland07gmm,Golland08,tucholka2008probabilistic,Shen10,Langs14}. Alternatively, von Mises-Fisher distribution is adapted for CDP with regard to its usability for clustering high dimensional data~\cite{Lashkari10,Ryali13}. For example, Yeo et al.~\cite{Yeo11} used such a probabilistic model~\cite{Lashkari10} to subdivide the cerebral cortex into several RSNs based on functional connectivity data collected from 1000 healthy subjects. Langs et al.~\cite{Langs14} combined the functional characteristics of several subjects obtained from a language fMRI study into a joint probabilistic distribution, from which a generative model is learned and used to identify a functional atlas, without relying on the spatial distribution of the contributing functional units (i.e. no explicit spatial constraints imposed to the data)~\cite{Langs14}. A similar stream of processing is later instrumentalised to study the intrinsic functional organisation of the brain at the individual subject level using rs-fMRI~\cite{Langs16}.  


In probabilistic modelling techniques, cluster labels and other parameters are usually inferred via an iterative expectation maximisation procedure~\cite{Ryali13,golland07gmm,Langs14}. While parametric models typically require fixing the number of clusters \textit{a priori}, it may also be uncovered automatically by incorporating a label cost to the model~\cite{Ryali13}. Alternatively, using a mixture model within a non-parametric setting (such as Dirichlet processes) may allow inferring the number of clusters directly from the data~\cite{jbabdi2009multiple}. Non-parametric models are particularly popular tools in the context of whole-brain CDP, as they do not rely on any assumptions on the granularity of parcellation~\cite{Baldassano15,janssen2016let}. 

In general, mixture models may be highly sensitive to initialisation and require multiple runs to achieve a reliable parcellation~\cite{golland07gmm} or to infer the right number of clusters~\cite{jbabdi2009multiple}. Spatial regularisation in this context is generally enforced through Markov Random Field (MRF) priors~\cite{jbabdi2009multiple,Ryali13}, which apply a penalty whenever neighbouring regions are assigned to different labels. However, this may not ensure specially coherent parcellations, as spatial contiguity is a global property that cannot be enforced locally~\cite{Honnorat15}. To overcome this, Honnorat et al.~\cite{Honnorat15} suggest using a convexity prior that guarantees connectedness within parcels; however, such a constraint yields parcels with regular (convex) shapes and comes at the expense of computational inefficiency depending on the spatial dimensionality (i.e. number of nodes). While this method is applied to resting-state fMRI data to compute whole-brain cortical parcellations, another MRF-based approach tackles the same problem, but with application to dMRI-based structural connectivity~\cite{parisot2015continuous}, which is initialised from random parcellations and iteratively updates parcel boundaries in a coarse-to-fine multi-resolution procedure.


\subsection{What Defines A Reliable Connectivity-Driven Parcellation?}
As extensively reviewed in the previous section, there exit many techniques that attack the parcellation problem from a connectivity-driven perspective. Put together, previous work has enhanced our capacity to understand the brain's functional and structural organisation at multiple scales, ranging from high-level spatially distributed structures of connectivity networks to whole-brain segregations of the cerebral cortex. In this thesis, we position ourselves next to the latter, and propose novel solutions to subdivide the cortex into spatially contiguous, non-overlapping, and homogeneous regions, which can subsequently be used to (1) map the cortical organisation of the brain with respect to connectivity and (2) to define the network nodes in connectome analysis. Towards this end, we follow several notable criteria defined by the CDP literature to evaluate the `reliability' of a whole-brain parcellation in the context of cortical segregation and connectome analysis:



\begin{itemize}
\item Subdivided regions should be highly homogeneous, in particular, comprise vertices with similar timeseries or connectivity profiles, since network nodes derived from parcellations are typically represented by a single feature vector (such as the average timeseries)~\cite{Thirion06,Craddock12,Shen13,Gordon16}.

\item Each region should specify a functionally and anatomically distinct area that has a different timeseries or connectivity profile from its neighbours, so as the entire parcellation can provide an abstraction of the functional specialisation and segregation in the cerebral cortex, and can provide a more accurate network modelling for brain mapping~\cite{Smith11,Blumensath13,Eickhoff15,Gordon16}.

\item ROIs should be spatially contiguous and non-overlapping to distinguish homogeneous network nodes (used to construct graphical models for connectivity analysis) from large-scale spatially-distributed networks (heterogeneous regions, usually spanning across the brain)~\cite{Thirion06,Bellec06,Smith09}. Spatial connectedness further yields a physiologically more interpretable connectivity analysis, since enforcing spatial contiguity allows producing anatomically localised regions~\cite{Craddock12,Shen13,janssen2016let}. 


\item Parcellations should be reproducible to a certain extent, meaning that, individual subject parcellations obtained from different scans of the same subject as well as group-wise parcellations computed from different groups of people who share some defining characteristics (e.g. healthy young adults) should exhibit some amount of functional and structural similarity~\cite{Craddock12,Blumensath13,Shen13,Honnorat15,Parisot16a}. 

\item A CDP method should ideally provide cortical parcellations at different levels of detail, such that it would be possible to study the brain's functional/structural organisation in a multi-scale manner~\cite{Craddock12,Blumensath13,Shen13,Honnorat15,Parisot16a}.   

\end{itemize}

In practice, it may not be possible to address all these criteria simultaneously, as there may typically exist trade-offs between different points, e.g. in order to achieve higher fidelity one may need to sacrifice reproducibility. As a result, clustering algorithms may \textit{a priori} make various assumptions or introduce implicit/explicit constraints, depending on the parcellation problem under consideration. As a general note, it is highly important to realise that each assumption and processing decision made by a clustering techniques comes with different consequences and inevitably bias the resulting parcellations in different aspects, including the shape, number, size, and spatial contiguity of the parcels~\cite{Eickhoff15}. 

For example, ICA assumes that the fMRI data consists of a mixture of statistically independent components and that spatially distributed functional networks can be effectively separated from signals of non-neural (e.g. artefactual) origin. With a similar objective, but from a different perspective, nonlinear manifold learning techniques rely on the assumption that structures of interest in the connectivity data live in a low dimensional embedding, which can be captured using spectral decomposition~\cite{thirion2004nonlinear,shen2006nonlinear,Langs14}. Other techniques alter the structure of the underlying data to obtain more robust parcellations, for instance, by applying thresholding to suppress negative and weak correlations, assuming that correlations under a threshold correspond to spurious connections~\cite{Heuvel08,Power11,Craddock12,Arslan15a}. It is also common to rely on spatial constraints~\cite{Craddock12} or regularisation~\cite{abraham2013extracting} for computing what is expected to be physiologically more plausible parcellations.

\subsection{Challenges in Connectivity-Driven Parcellation}

Despite many promising results in the literature, the CDP problem is still open to improvements. Majority of the methodologies we have discussed so far either target relatively small cortical regions (e.g. the insula or SMA) or attempt to reveal the intrinsic connectivity that underlie the brain function (e.g. RSNs). However, the parcellation of the entire cortex is a more challenging task and the aforementioned whole-brain CDP methods only partly satisfy the above criteria. The main challenges that may have a critical impact on the reliability of a parcellation and how we overcome them throughout this thesis are summarised as follows:

\begin{itemize}

\item Connectivity data is typically very noisy because of the previously discussed (Chapter 2, \textit{Imaging the Macro Connectome}) limitations associated with the imaging modalities used to capture connectivity. In particular, fMRI mainly suffers from low SNR, while dMRI and tractography lack sensitivity towards delineating long-range connections. Considering the `garbage-in garbage-out' principle, it is highly critical to clean the data before carrying out any parcellation. We conduct our experiments on the HCP datasets, which have been subject to novel pre-processing pipelines, providing relatively high quality data~\cite{Glasser13,VanEssen13}. 

\item The inherently high dimensionality of the connectivity data from which parcellations are acquired poses another challenge, particularly when subdividing the entire cerebral cortex. Today, MRI captures the brain at millimetre scale and allows longer scanning time than ever. This is obviously beneficial towards obtaining images in greater detail; however, as the dimensionality increases, the computational requirements for analysing the data also grows rapidly. It may as well be more challenging to separate structures of interest (or true signal) from noise. Towards this end, we provide effective solutions to deal with both spatial and temporal dimensionality, where the former is particularly important, as it helps improve the SNR in rs-fMRI timeseries. 

\item Inter-subject variability stands as a challenge especially for group-level studies that aim to identify shared patterns of connectivity across subjects. As connectivity-driven parcellations are directly learnt from the underlying data, they can better capture the variability across subjects compared to traditional brain atlases~\cite{Parisot16a,Abraham16}. However, it is still important to pay particular attention towards incorporating individual functional/structural characteristics into the group-wise CDP computation~\cite{Shen13,Langs15,Nie16}. From a different perspective, inter-subject variability constitutes a critical source of information towards patient-centred medicine and understanding the neural basis of variation in the brain function and human behaviour~\cite{Wang15,Tavor16}. Therefore, it is critical to study the brain's functional and structural organisation at the subject level. Taking both views into consideration, we propose methodologies to compute different cortical parcellations that can be effectively (1) tailored to individual subjects and (2) used to summarise connectional characteristics shared across a group of subjects. 

\item Evaluation in CDP is a challenge in itself, since there exists no widely-accepted parcellation that can be used as the `golden standard'. It is, therefore, highly critical to validate computed parcellations considering different facets of the CDP problem~\cite{Thirion14,Eickhoff15}. Given the aforementioned criteria that define the reliability in the context of CDP, we mainly perform evaluation through measuring reproducibility across subjects/groups, homogeneity within parcels, and the degree of separation between neighbouring regions, as well as carry out multi-modal comparisons with well-known neuro-anatomical patterns (such as Brodmann's cortical regions). More on evaluation will be discussed in the next chapter.
\end{itemize}

\chapter{Single Subject Parcellation of the Cerebral Cortex}
\label{chapter:multi-level}

This chapter is based on:

S. Arslan, S. I. Ktena, A. Makropoulos, E. C. Robinson, D. Rueckert, and S. Parisot, \textit{Human Brain Mapping: A Systematic Comparison of Parcellation Methods for the Human Cerebral Cortex}, NeuroImage, 2017. (\textit{In Press})

S. Arslan and D. Rueckert, \textit{Multi-Level Parcellation of the Cerebral Cortex Using Resting-State fMRI}. International Conference on Medical Image Computing and Computer Assisted Intervention (MICCAI), vol. 9351 of LNCS. Springer, pp. 47-54, 2015.

S. Arslan, S. Parisot, and D. Rueckert, \textit{Supervertex Clustering of the Cerebral Cortex Using Resting-State fMRI}, Organization for Human Brain Mapping (OHBM), Honolulu, 2015.

\section*{Abstract}
\textit{This chapter describes a method for the subject-specific parcellation of the cerebral cortex. We propose a two-level parcellation framework which deploys a different clustering strategy at each level. We adapt a \textit{k}-means clustering approach to group vertices into a relatively large number of highly homogeneous regions with respect to functional and spatial similarities between them. These clusters constitute a robust, high-level representation of the cerebral cortex, which, at the second level, are combined into a spectrum of parcellations with different resolutions using a spatially-constrained hierarchical clustering approach. Using data from 100 healthy subjects, we show that our algorithm segregates the cerebral cortex into distinct parcels at different levels of granularity with higher reproducibility and functional homogeneity, compared to a state-of-the-art two-level approach. We also discuss the pros and cons of our approach over other data-driven clustering techniques, including but not limited to \textit{k}-means, hierarchical clustering, and spectral clustering. Finally, we show that functional connectivity at the individual subject level can be more effectively represented by connectivity-driven parcellation techniques, rather than random or anatomical parcellations. }   

\section{Introduction}
\label{sec:intro}

In this chapter, our aim is to compute robust and homogeneous parcellations for individual subjects, which would provide a reliable abstraction of the brain's functional organisation, and thus, enable generation of graphical models for a more effective analysis of the brain connectivity. Such models are currently derived from anatomical landmarks~\cite{TzourioMazoyer02,Desikan06}, cytoarchitectonic features~\cite{Brodmann09}, or rely on random parcellations~\cite{Sporns11}, thus may fail to fully reflect the underlying function in the brain. We propose a method driven by rs-fMRI, which records neurocognitive activity by measuring the fluctuations in the BOLD signals while the subject is at rest. Since the brain is still active in the absence of external stimuli, resting-state functional connectivity (RSFC) estimated from BOLD timeseries can be used to identify the brain's functional organisation~\cite{Biswal95,Lowe98}.


RSFC-based parcellation techniques generally focus on identifying patterns of brain connectivity that are shared across individuals~\cite{Power11,Yeo11}. Studies replicated under different experimental settings and across many groups of people have discovered that the human brain is organised into several large-scale functional networks that are linked to brain cognitive functions~\cite{Smith09,Van10intrinsic} and human behaviour~\cite{Laird11}. Although these studies are typically conducted on RSFC datasets averaged across many subjects, similar functional systems are also detected at the subject level~\cite{Shehzad2009resting}. However a great amount of individual variability is also observed~\cite{Mueller13,Wang2014functional}, especially in the spatial distribution of these networks~\cite{Wang2014functional}. In addition, evidence suggests that the human connectome possesses connectional traits that are unique to each subject~\cite{Mueller13,Barch13,Gordon16individual}. A recent study~\cite{Finn15} has further shown that RSFC can be used to derive distinct features to successfully distinguish one individual from another. These features, however, may not be observed in group-averaged datasets~\cite{Gordon16individual}. Therefore, parcellating the cerebral cortex on a single subject basis can provide a natural start point for detecting such features, and help better understand how connectivity changes across individual brains. 

With this motivation, we propose a two-level parcellation framework that could be applied to individual subjects for subdividing the cerebral cortex in its entirety. At the first level, we utilise a \textit{k}-means clustering approach to group the cortical vertices into relatively large number of homogeneous regions. This stage reduces the high dimensionality of the data at the vertex level and improves the poor SNR in single subject data. One critical issue that needs to be addressed in order to ensure the success of such a clustering approach is the definition of the distance function that drives the \textit{k}-means clustering. To this end, we propose a hybrid distance function based on rs-fMRI correlations and geodesic distance, which allows to cluster functionally homogeneous vertices, while enforcing spatial contiguity within a cluster. At the second level, we build a hierarchical tree on top of the pre-segmented regions to obtain individual parcellations reflecting the functional organisation of the cortex without losing the spatial integrity within the parcels. Rather than parcellating the cortex into a fixed number of regions, our framework provides parcellations within a range of 50 to 500 parcels, nested within a hierarchical tree, and thus, allowing the analysis of functional connectivity at different levels of detail. 


The most closely related work to our approach is another two-level parcellation method composed of region growing and hierarchical clustering~\cite{Blumensath13}. The major difference between two methods emerges from the first parcellation stage. We show that the proposed framework can provide a more reliable abstraction of the individual brain organisation, achieving higher reproducibility and functional consistency than the other approach. In addition, we expand our experiments by including other RSFC-driven parcellations that are obtained by some widely-used clustering algorithms such as \textit{k}-means, agglomerative hierarchical clustering~\cite{Ward63} and spectral clustering with normalised cuts~\cite{Craddock12} as well as another region growing approach~\cite{Bellec06}. We show that connectivity-driven parcellations have the potential to provide a more reliable representation of the underlying data compared to traditional parcellations obtained from anatomical landmarks~\cite{Desikan06,Fischl04} and random parcellations~\cite{Schirmer2015,Thirion14}.
 
The remainder of this chapter is organised as follows: In Section~\ref{sec:multi-level-methods}, we summarise the proposed parcellation method, with a particular emphasis on the initial parcellation and hierarchical clustering stages. In Section~\ref{sec:multi-level-experiments}, we define several measures for the evaluation of parcellations and summarise other clustering techniques used for comparison. In Section~\ref{sec:multi-level-results} we show the performance of different approaches with respect to visual and quantitative results. Finally, in Section~\ref{sec:multi-level-discussion}, we discuss the pros and cons of the proposed method with respect to the other clustering approaches, and provide some insight towards future research directions.

\section{Methodology}
\label{sec:multi-level-methods}
\subsection{Data}
We conducted our experiments on Dataset 1, which contains rs-fMRI scans from 100 subjects. The details of the dataset are provided in Chapter~\ref{chapter:background} and briefly summarised here. The data for each subject was acquired in two sessions that were held on different days and divided into four runs of approximately 15 minutes each. All data was preprocessed and denoised by the HCP minimal preprocessing pipelines~\cite{Glasser13}. Following the preprocessing steps, all timeseries were normalised to zero-mean and unit-variance. For each subject, we temporally concatenated the 15-minute scans acquired on the same day into two 30-minute rs-fMRI datasets and used them to evaluate our approach.

\subsection{A Two-Level Parcellation Framework}
\begin{figure}[hbt!]
\centering
\includegraphics[width=\linewidth]{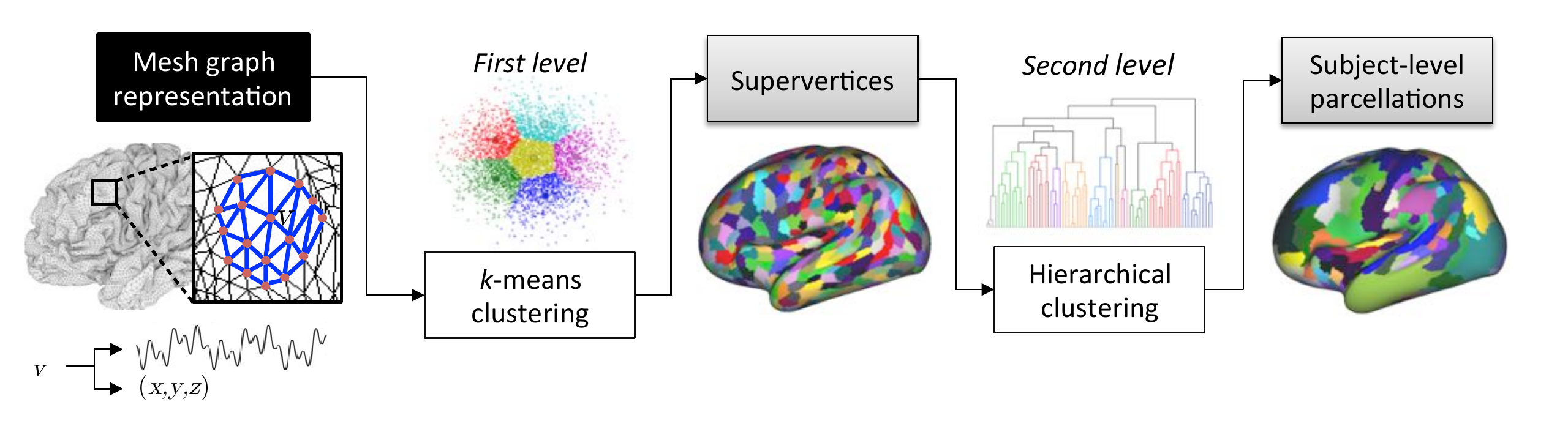}  
\caption[The proposed two-level parcellation framework.]{The two-level parcellation framework. \textit{Initial representation:} The cerebral cortex is represented as a standard triangulated mesh, in which each vertex $v$ is associated with a rs-fMRI timeseries and a set of anatomical coordinates $(x,y,z)$ in the 3D space. \textit{First level:} The cortex is parcellated into a relatively large number of supervertices using a \textit{k}-means clustering variant. \textit{Second level:} Subject-specific parcellations at different resolutions are obtained by merging supervertices with hierarchical clustering.}
\label{fig:2level}
\end{figure}

The proposed method starts with subdividing the cerebral cortex into highly homogeneous and relatively small regions using a modified \textit{k}-means algorithm applied to the BOLD timeseries. This high-resolution parcellation is then submitted to a hierarchical clustering algorithm to obtain a multi-scale, nested set of subject-level parcellations. A visual representation of the parcellation framework is given in Fig.~\ref{fig:2level}.

\subsubsection{Initial Parcellation via Supervertex Clustering}
\label{sec:initial}
The parcellation process begins with clustering the cortical vertices into a set of functionally uniform regions. Inspired by the superpixel approaches used in image segmentation~\cite{Achanta11}, we develop a similar method to construct the initial clusters, i.e. \textit{supervertices}. Our approach is a variation of the \textit{k}-means clustering algorithm, but distinguishes itself from the other variants in two aspects: (1) We limit the search space of each supervertex to the expected
average cluster size, which greatly reduces the number of distance calculations, thus improves the computational performance. (2) We define a hybrid distance function to measure the functional similarity between two vertices. This distance function is capable of assigning highly correlated vertices into same clusters, while enforcing spatial contiguity within a cluster. 

\begin{figure}[b!]
\centering
\includegraphics[width=0.8\linewidth]{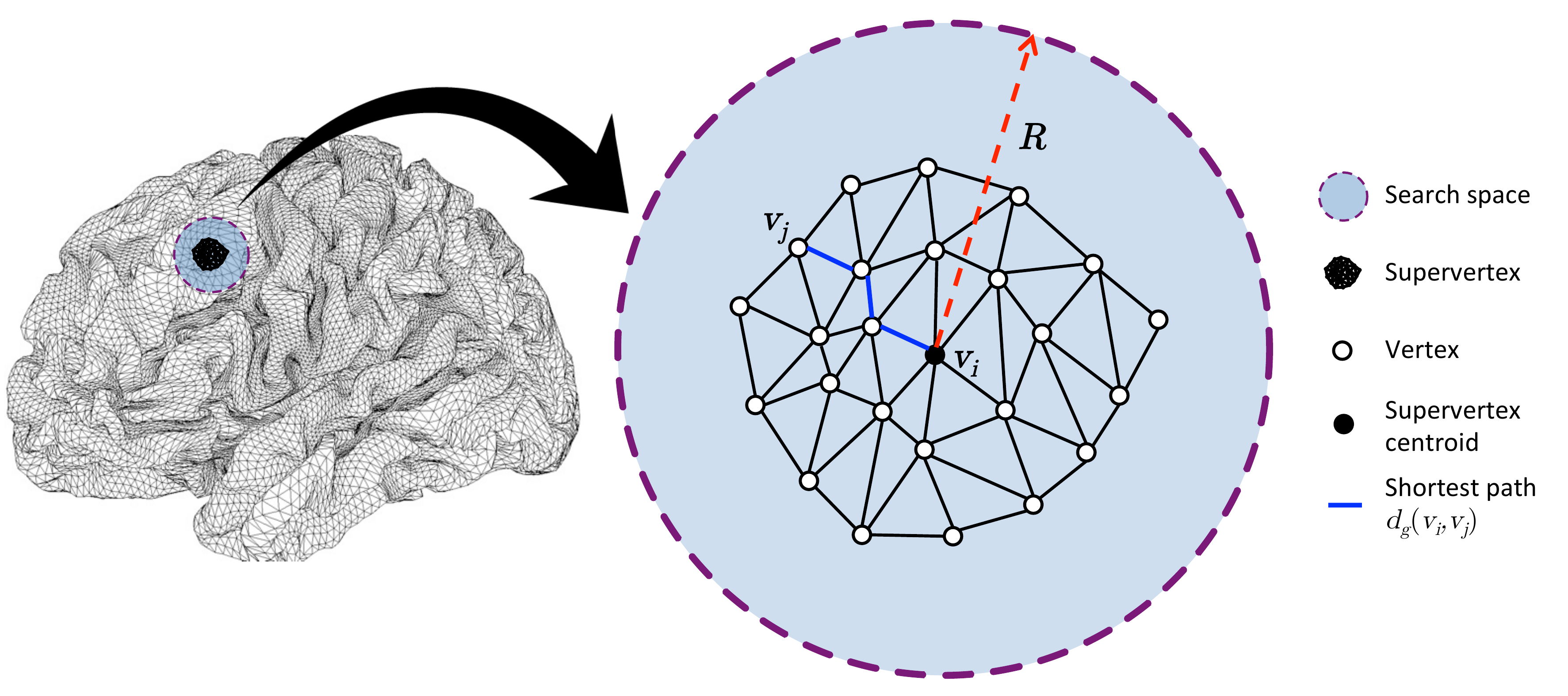}  
\caption[Illustration of the mesh graph that represents the cerebral cortex and the supervertex clustering algorithm.]{Illustration of the mesh graph that represents the cerebral cortex (\textit{left}) and the supervertex clustering algorithm, emphasizing on a single supervertex (\textit{right}). Note that the search space is not necessarily a circle, but displayed here as one for simplicity.}
\label{fig:sv_slic_process}
\end{figure}

The algorithm is designed to operate on the cerebral cortex, where the functional activity can be identified. We represent the cerebral surface as a smooth, triangulated mesh with no topological defects (Fig.~\ref{fig:sv_slic_process}). The mesh vertices and their associations are modelled as a weighted graph $G = (V, E)$, in which $V$ is the set of vertices (nodes) and $E$ is the set of edges connecting neighbour vertices. Each edge $e \in E$ is associated with a weight $w$ that measures the Euclidean distance between two vertices with respect to their anatomical coordinates $(x,y,z)$, i.e. for vertices $v_i$ and $v_j$ connected via $e_{ij}$, $w_{ij} = \|v_i-v_j\|_{2} = \sqrt[]{(x_i-x_j)^2+(y_i-y_j)^2+(z_i-z_j)^2}$.  

Initially, $k$ seed vertices are selected as the supervertex centroids from $V$ by uniformly sub-sampling the mesh graph using the iso2mesh library~\cite{Fang09}. For every supervertex centroid $S_i$ we compute its distance to the vertices which fall within a predefined range $R$ of the supervertex centroid. Limiting the search space significantly increases the efficiency of the algorithm without affecting the final parcellations, since the cortical vertices are likely to be assigned to clusters within their local neighbourhoods due to spatial constraints imposed by the distance function. 

The algorithm iteratively assigns each vertex to a supervertex by computing their dissimilarity with a hybrid distance function. At the end of each iteration, the geometric centres of newly-formed supervertices are chosen as the new centroids. The algorithm converges when all supervertices remain unchanged between two consecutive iterations. These steps are algorithmically summarised in Algorithm~\ref{algo} and an illustration of the approach is given in Fig.~\ref{fig:sv_slic_process}.
  
The supervertex clustering algorithm takes both the functional similarity and spatial proximity of the vertices into account while assigning cluster memberships. In order to avoid inconsistencies in clustering behaviour, the distance function should bring these different measures into a common space. We address this issue by defining separate distance functions for the functional and spatial domains, normalised by their maximal values within a cluster, and then combine them in a 2D Euclidean distance function. 

\begin{algorithm}[!tb]
\DontPrintSemicolon
 \tcc{$k$ initial \textit{supervertex} centroids are selected by uniform sampling.}
 \ForEach{vertex $v$}{
 	$labels(v) \leftarrow 0 $ \;
 	$distances(v) \leftarrow \infty $ \;
 }
 
 \Repeat{$changed$ $\neq true$}{
 $changed \leftarrow false$\;
  \ForEach{supervertex centroid $S_i$}{
  \tcc{Distance calculated only for vertices within a range.}
   \ForEach{vertex $v$ within a range of R around $S_i$}{
      $D=$ distance between $S_i$ and $v$\;
	  \If{$D < distances(v)$}{
		$distances(v) \leftarrow D$\;
		$labels(v) \leftarrow i$\; 
		$changed \leftarrow true$\;
	  }    
    }
  }
  Compute the new supervertex centroids\;
 }
\caption{Supervertex Clustering\label{algo}}
\end{algorithm}

In the functional domain, a vertex $v_i$ is represented by its time series $t_i$. Functional similarity between two vertices $v_i$ and $v_j$ is measured by the Pearson's correlation coefficient $r$ defined as follows: 

\begin{equation}\label{eq:cor}
 	r(t_i, t_j) = \frac{\sum\limits_{l=0}^L (t_{il} - \mu_i)(t_{jl} - \mu_j)}{(L-1)\sigma_i\sigma_j} 
\end{equation}

where $l$ $(0 \leq l \leq L)$ corresponds to a sample of a timeseries of length $L$, $\mu_i$ ($\mu_j$) is the mean and $\sigma_i$ ($\sigma_j$) is the sample standard deviation of $t_i$ ($t_j$). $r$ always produces values within the range [-1, 1]. The extrema at both ends indicate strongly (anti-)correlated time series. In order to convert Pearson's correlation into a distance measure, we define the following transformation:

\begin{equation}\label{eq:pear-dist}
	d_c(v_i,v_j) = 1 - r(t_i, t_j)
\end{equation}

where $d_c$ is referred as the Pearson's distance and yields values $ 0 \leq d_c \leq 2$. This transformation ensures the distance between highly correlated vertices being close to zero, thus increases their likelihood of being assigned to the same cluster.

In the spatial domain, instead of computing the distance directly from the spatial coordinates of two vertices, we propose to use the geodesic distance along the cortical surface, since it more naturally reflects the spatial geometry of the cerebral cortex. Given that each edge is subject to a weight, the geodesic distance $d_g$ between $v_i$ and $v_j$ can be approximated as the sum of edge weights along the shortest path that connect $v_i$ to $v_j$ in the mesh graph, as illustrated in Fig.~\ref{fig:sv_slic_process}.

To combine two measures into a hybrid function, we introduce the functional and spatial normalisation factors, $N_c$ and $N_g$, respectively. These normalisation factors are set to their corresponding maximal distance values in a cluster. We straightforwardly set $N_c$ to 2, since the Pearson's distance values fall within the range $[0, 2]$. Similarly, $N_g$ is set to the predefined local search limit, $R$, since the
maximum distance within a cluster cannot exceed it. Finally, we combine these normalised measures in a Euclidean distance function $D$ as follows:

\begin{equation} \label{eq:final}
D = \sqrt{\alpha{d_c}^2 + (1-\alpha){d_g}^2} 
\end{equation}

Here, $\alpha$ is introduced as a weighting parameter ($0 \leq \alpha \leq 1$) that controls the influence of functional dissimilarity over spatial closeness. We further investigate its effect on the parcellation performance in the following section.

\subsubsection{Subject-Level Parcellation via Hierarchical Clustering}
\label{sec:second-level}
The supervertex clustering stage parcellates the cerebral cortex into a relatively large (1000 to 3000) number of highly homogeneous, non-overlapping regions. Although parcellations at this scale can be used for applications when accuracy is of great importance~\cite{Craddock12,Thirion14}, a second stage clustering must be performed to obtain more neuro-biologically interpretable sub-divisions of the cortex~\cite{Craddock12,Blumensath13}.

\begin{figure}[hb!]
\centering
\includegraphics[width=0.8\linewidth]{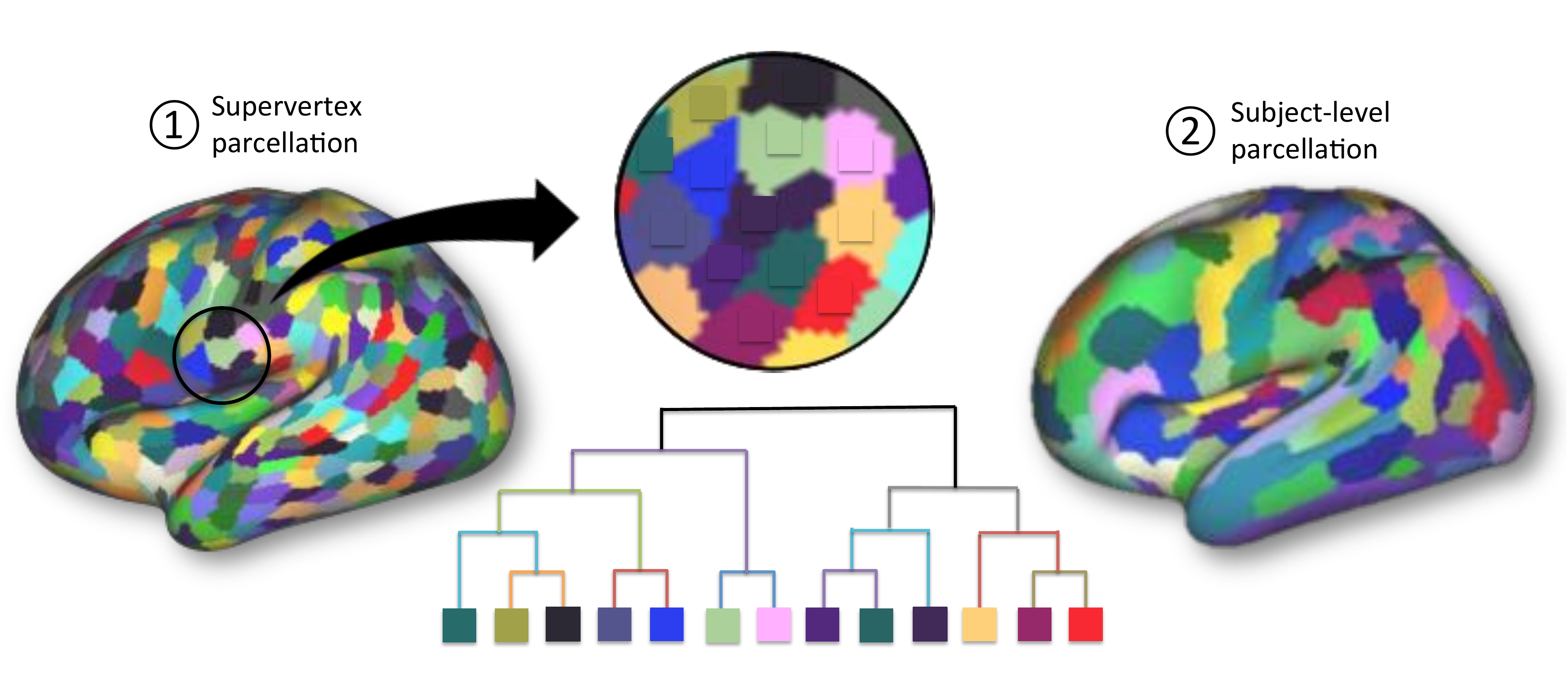}  
\caption{Illustration of the agglomerative hierarchical clustering algorithm.}
\label{fig:sv_hierarchical}
\end{figure}

Towards this end, we adapt an agglomerative hierarchical clustering algorithm~\cite{Johnson67} to join supervertices into non-overlapping parcels, without impairing their functional uniformity. This approach starts with each supervertex as singleton clusters and builds a hierarchy of clusters using a bottom-up strategy in which pairs of supervertices are merged into a single cluster with respect to a linkage rule. An illustration of the clustering algorithm is provided in Fig.~\ref{fig:sv_hierarchical}. 

\begin{figure}[t!]
\centering
\includegraphics[width=0.75\linewidth]{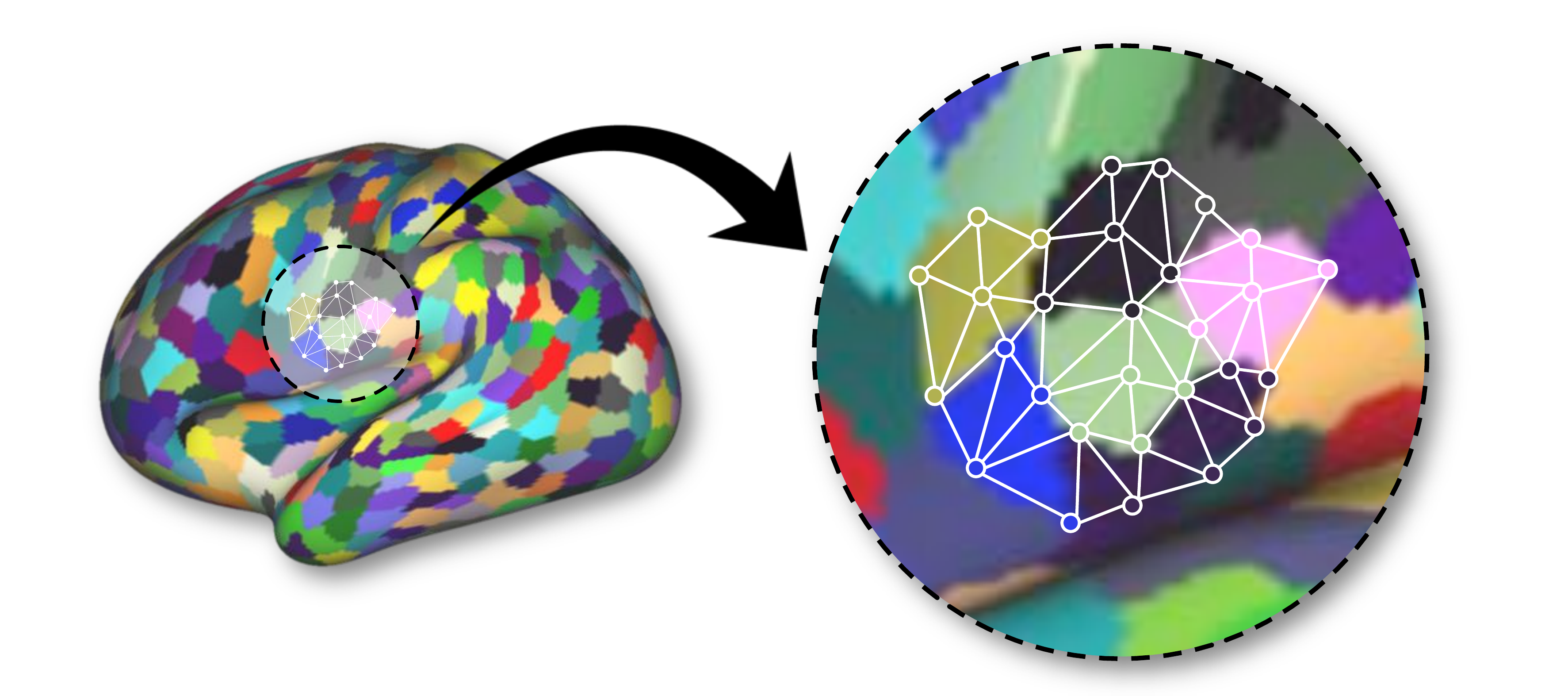}  
\caption[Illustration of the neighbourhood criterion for a set of supervertices.]{Illustration of the neighbourhood criterion for a set of supervertices. Two supervertices $S_a$ and $S_b$ are considered as neighbours, if vertices $v_i \in S_a$ and $v_j \in S_b$ are immediately connected in the cortical mesh.}
\label{fig:sv_adjacency}
\end{figure}

In order to ensure the spatial coherency throughout the parcellation process, a connectivity constraint is imposed, so that only neighbouring clusters are merged into a higher level. Here, two clusters $S_a$ and $S_b$ are considered as neighbours, or adjacent, if vertices $v_i \in S_a$ and $v_j \in S_b$ are connected by a direct edge, $e_{ij}$, in the cortical mesh. An illustration is provided in Fig.~\ref{fig:sv_adjacency} to make the neighbouring criterion more clear. For instance, the yellow, black, and blue supervertices in the figure are all adjacent to each other, since there is at least one edge that directly connects one to another. On the other hand, the pink and blue supervertices are only connected through a chain of neighbouring vertices, hence, are not considered as neighbours.

The algorithm is driven by Ward's minimum variance method~\cite{Ward63} and the similarity between pairing clusters is computed by the Euclidean distance. Among other linkage rules, we use Ward's method as we have observed that it yields the best quantitative results. Similar observations are reported in~\cite{Blumensath13} and a wide range of fMRI-based parcellation methods in the literature are driven by Ward's linkage rule as well~\cite{Goutte99,Jenatton11,Smith13,Thirion14}. The proposed hierarchical clustering technique produces a dendrogram~\cite{Johnson67}, in which the leaves represent the supervertices and the root represents an entire hemisphere. Cutting this tree at different levels of depth produces individual subject parcellations with the desired precision. We investigate the effects of different granularities on the parcel reproducibility and functional reliability in the following section.

\section{Experiments}
\label{sec:multi-level-experiments}
\subsection{Performance Measures}
Evaluating the quality of parcellation methods is a challenging task since there is no ground-truth parcellation of the cerebral cortex. Considering this, we gather the most commonly used evaluation techniques from the literature to evaluate parcellations with respect to varying resolutions and quantitatively compare different clustering techniques to each other. These techniques can be separated into three categories with regards to the parcellation aspects they assess: (1) reproducibility, (2) homogeneity, and (3) Silhouette analysis. These techniques are explained in detail in this chapter and used throughout the thesis for measuring the clustering quality of parcellations.

We also provide a qualitative assessment of the parcellations provided by the proposed two-level framework in terms of its agreement with task activation, the cytoarchitecture of the cerebral cortex, and well-defined structures of myelination. 

\subsubsection{Reproducibility}
\label{sec:multi-repro}
Reproducibility is a widely-accepted technique for evaluating the robustness of a parcellation method with respect to the underlying data/subjects. It measures the extent of alignment in parcellation boundaries between different parcellations. It can be used to evaluate parcellations obtained from a) different subjects (inter-subject reproducibility), b) the same subject but different rs-fMRI acquisitions (within-subject, or scan-to-scan reproducibility), and c) different groups (group-to-group reproducibility). Due to the high inter-subject variability within a population, it is not expected to obtain high reproducibility between different subjects. Nevertheless, a robust parcellation method should yield very similar parcellations for the same subject with different acquisitions. A high reproducibility should be expected of group level parcellations, assuming the group size is large enough. 

We measure the reproducibility of parcellations using two well-known techniques, Dice coefficient~\cite{Dice45} and adjusted Rand index (ARI)~\cite{Hubert85}, each of which provides a means of assessing reproducibility from a different aspect. The former measures the amount of overlap at the parcel level, while ARI provides a more direct evaluation of the similarity between two parcellations by considering each as a whole.  

\paragraph{Dice coefficient.} The Dice coefficient~\cite{Dice45} is a very popular measure of overlap between two labelled areas. It has been extensively used for evaluating brain parcellations~\cite{Yeo11,Craddock12,Honnorat15,Blumensath13,Arslan15b,Parisot16a}. Given two parcels $X$ and $Y$, the Dice coefficient is calculated as:

\begin{equation}
Dice = \frac{2|X \bigcap Y|}{|X|+|Y|}
\end{equation}

\noindent where $|\cdot|$ indicates the number of vertices in a parcel. In order to obtain a global measure of parcellation reproducibility, we follow the approach proposed in~\cite{Blumensath13}. We first compute Dice coefficients for every pair of parcels and match the ones with the highest overlap. The Dice coefficients of matching parcels are then averaged to obtain a global reproducibility score for the whole parcellation. The matching process is performed in an iterative manner, where matching pairs identified in one iteration cannot be matched with other parcels at the next iterations. This process is repeated  until all pairs are identified. A Dice coefficient of 1 implies a perfect match (identical parcellations) and 0 indicates no match.

Low SNR in functional connectivity data or high variability within a group may yield a subdivision of some regions from one parcellation to the next, even when the same algorithm is performed on different acquisitions/subsets. To account for this effect and reduce its impact on reproducibility, we also use a modified version of Dice coefficient that merges the subdivided regions so as to maximise the overlap with the other parcellation as described in~\cite{Blumensath13}. This is done by iteratively matching each parcel in one parcellation to those in the other, if their overlap ratio is $ \geq 0.5$ (i.e. one parcel comprises at least half of the other parcel). After this process, regions in one parcellation that are matched with the same parcel in the other are merged and the average Dice coefficient is computed between the matched pairs as described above. This measure will be referred to as the \textit{joined} Dice coefficient in the remainder of this chapter.

\paragraph{Adjusted Rand index.} ARI~\cite{Hubert85} is another technique for the evaluation of parcellation reproducibility~\cite{Thirion14}. In contrast to Dice coefficient, it measures the agreement of two parcellations without the necessity of initially matching parcels. As a result, it can more effectively measure the agreement between two parcellations with different numbers of clusters~\cite{Milligan86}.

ARI is built upon counting the number of items (in our case, vertices) on which two parcellations agree or disagree~\cite{Vinh09}. It classifies $\bigl(\begin{smallmatrix} M \\ 2\end{smallmatrix}\bigr)$ pairs of vertices into one of the four sets ($M_{11},M_{00},M_{01},M_{10}$), based on their labeling in each parcellation. For parcellations $\mathbf{U}$ and $\mathbf{V}$, $M_{11}$ corresponds to the number of pairs that are assigned to the same parcel in both $\mathbf{U}$ and $\mathbf{V}$, $M_{00}$ corresponds to the number of pairs that are assigned to different clusters in both $\mathbf{U}$ and $\mathbf{V}$, $M_{01}$ corresponds to the number of pairs that are assigned to the same parcel in $\mathbf{U}$, but different parcels in $\mathbf{V}$, and $M_{10}$ corresponds to the number of pairs that are assigned to the same parcel in $\mathbf{V}$, but different parcels in $\mathbf{U}$. Intuitively, $M_{00}$ and $M_{11}$ account for the agreement of parcellations, whereas $M_{01}$ and $M_{10}$ indicate their disagreement~\cite{Vinh09}. After counting the number of pairs, ARI for parcellations $\mathbf{U}$ and $\mathbf{V}$ is computed as follows:

\begin{equation}
ARI(\mathbf{U},\mathbf{V}) = \frac{2(M_{00}M_{11} - M_{01}M_{10})}{(M_{00}+M_{01})(M_{01}+M_{11})+(M_{00}+M_{10})(M_{10}+M_{11})} 
\end{equation}

An ARI of 1 indicates a perfect correspondence between parcellations, whereas a value of 0 implies that the parcellations do not agree on any of the labels.
 
\subsubsection{Homogeneity}
\label{sec:Homogeneity}
Homogeneity is a highly popular parcellation evaluation technique~\cite{Craddock12,Shen13,Gordon16,Arslan15b,Parisot16a,Honnorat15} that aims to measure the similarity of vertices aggregated in the same parcel, since a good parcellation should have the ability to group vertices with highly similar functional connectivity~\cite{Craddock12,Gordon16}. A high homogeneity is particularly important for subsequent network analysis where network nodes are typically represented by the average signal (e.g. BOLD timeseries) within each parcel~\cite{Shen13,Gordon16}. Given a parcellation $\mathbf{U} = \{U_1,U_2, \ldots U_K\}$, the homogeneity $h_k$ of a parcel $U_k$ is measured by calculating the average similarity between every pair of vertices assigned to $U_k$ as
 
\begin{equation}
h_k = \frac{1}{n_k(n_k-1)} \displaystyle\sum_{i,j \in U_{k}, i \neq j }{s(v_i,v_j)}
\end{equation}
 
\noindent where, $n_k$ denotes the number of vertices in parcel $U_k$ and $s(v_i,v_j)$ corresponds to the similarity between vertices $v_i$ and $v_j$. While this similarity can be directly measured by comparing timeseries of two vertices, we use an alternative approach and first define `connectivity fingerprints' for each vertex as suggested in~\cite{Craddock12,Blumensath13}. A connectivity fingerprint (also known as connectivity profile) is a feature vector which indicates the similarity of a vertex with all other vertices, and hence, provides a global measure for how a vertex is connected to the rest of the cerebral cortex. When measuring functional similarity, such maps are computed by correlating timeseries of a vertex with the timeseries of other vertices. Two of these maps can then be compared by using Pearson's correlation, after being subject to Fisher's $r$-to-$z$ transformation~\cite{Cohen08}. 

A global homogeneity value for $\mathbf{U}$ is finally obtained by averaging the homogeneity values across all parcels~\cite{Craddock12}.

\subsubsection{Silhouette analysis} 
Another useful and popular technique to quantify parcellation reliability is Silhouette coefficient (SC)~\cite{Rousseeuw87}, which can be used as an indicator of how well vertices fit in their assigned parcel. For each vertex, it compares the \textit{within-parcel dissimilarity} defined as the average distance to all other vertices in the same parcel, to the \textit{inter-parcel dissimilarity} obtained from those assigned to other parcels~\cite{Yeo11,Craddock12}. SC not only evaluates the compactness of parcels, but also their degree of separation from each other. It is defined as follows:

\begin{equation}
SC_i  = \frac{b_i-a_i}{\mathrm{max}(a_i, b_i)}
\end{equation}

Given a parcellation $\mathbf{U} = \{U_1,U_2, \ldots U_K\}$, $a_i$ and $b_i$ correspond to within-parcel and inter-parcel dissimilarity of vertex $v_i \in U_k$, respectively, and are defined in the following equations. 

\begin{equation}
a_i = \frac{1}{n_k-1} \displaystyle\sum_{j \in U_{k}, i \neq j }{d(v_i,v_j)}
\end{equation}

\begin{equation}
b_i = \frac{1}{M} \displaystyle\sum_{j \in \mathbb{N}(U_k)}{d(v_i,v_j)}
\end{equation}

Here, $n_k$ denotes the number of vertices in $U_k$, $\mathbb{N}(U_k)$ denotes the set of parcels that are neighbours with $U_k$, with $M = \sum_{j \in \mathbb{N}(U_k)}{n_j}$ being the number of vertices within these neighbouring parcels. $d(v_i,v_j)$ stands for the distance measure defined as $1-r$, where $r$ is Pearson's correlation coefficient computed between the connectivity fingerprints of $v_i$ and $v_j$. Instead of computing the inter-parcel dissimilarity with respect to the vertices in all other parcels, we restrict the computations to the neighbouring parcels. This is because (1) it is unlikely for a vertex to be assigned to a remote parcel due to spatial constraints imposed on the parcellations, and (2) computing inter-parcel dissimilarity with respect to all vertices outside a parcel quickly yields a bias towards obtaining high Silhouette coefficients, as the inter-parcel dissimilarity tends to be extremely high due to the many vertices with low similarity contributing to its computation.

Due to the fact that we use correlation distance as the dissimilarity measure, SC ranges within $[-1, +1]$. A negative SC implies misclassification of a vertex, while a value close to 1 indicates that the vertex is clustered with a high degree of confidence. If most vertices possess high Silhouette values, the parcellation is considered to be of high quality. A global score is obtained for each parcellation by averaging the Silhouette coefficients across all vertices.

\subsection{Qualitative Assessment of Parcellations with Other Modalities}
The previously proposed measurements assess the performance of a cortical parcellation from a clustering point of view and show how well the underlying connectivity data is effectively represented at a lower dimensionality. However, when defining regions of interest for neuro-anatomical purposes, the agreement of the parcellation boundaries with other neuro-biological properties also constitutes a critical aspect of parcellation quality. To this end, we provide additional visual results showing the agreement of the parcellation boundaries with several other cortical features, including myelin maps, task fMRI activation maps, and cytoarchitectural areas as defined by Brodmann. 

fMRI data recorded while the subject is performing a behavioural or psychological task is typically used to discover the activated areas in the cerebral cortex~\cite{Beeck08}. Task-evoked fMRI images thus provide, although with limited coverage, a means to delineate the functional organisation of the brain. As a result, local agreements are likely to be observed between RSFC parcellation boundaries and highly activated cortical regions~\cite{Blumensath13}. Similarly, a strong alignment has been observed between myelin maps and resting-state fMRI gradients~\cite{Glasser11}. We should therefore expect the boundaries of RSFC-driven parcellations to align with well-structured myelination patterns. We also assess the agreement of our parcellations with Brodmann's cytoarchitectonic areas~\cite{Brodmann09}. Despite the fact that connectivity obtained from BOLD timeseries does not necessarily need to reflect the cytoarchitecture of the cerebral cortex, several studies report some alignment with certain cytoarchitectonic areas~\cite{Blumensath13,Wig14,Gordon16}, such as the motor and visual cortex.

For our visual assessments, we use the complimentary data provided by the HCP. The Brodmann parcellations contain labels for the primary somato-sensory cortex (BA[3,1,2]), the primary motor cortex (BA4), the pre-motor cortex (BA6), Broca's area (BA[44,45]), the visual cortex (BA17 and MT), and the perirhinal cortex (BA[35,36]).  For the comparisons with the task activation, we select the following subset of contrasts from five different task protocols: the RH-AVG and T-AVG contrasts of the MOTOR protocol; the STORY-MATH contrast of the LANGUGAGE protocol; the RANDOM contrast of the SOCIAL protocol, the PUNISH contrast of the GAMBLING protocol; the MATCH contrast of the RELATIONAL protocol.

\subsection{Inter-Subject Variability Across Proposed Parcellations}
The proposed parcellation scheme provides a subdivision of the cerebral cortex that is unique to each single subject. Due to high functional and structural variability across subjects as well as low SNR in the BOLD timeseries, it is not likely to obtain a high similarity across different subject-level parcellations. However, local agreements between parcellation borders are likely to be expected within some cortical areas, depending on the robustness of the registration algorithm used to align subjects.

In order to measure the extent of this agreement across subjects, we make use of the Dice-based overlapping measure previously introduced in Section~\ref{sec:multi-repro}. To this end, a `consistency map' is computed for each subject by matching its parcellation with the other subject-level parcellations and each cortical vertex is assigned a score of 1 if it is within a matched parcel, and 0, otherwise. This procedure is repeated for all individuals and a consistency map is obtained on a per subject basis. We then compute a global consistency map for the entire cortex by averaging the maps across subjects. Cortical regions with high consistency scores are likely to be parcellated in a similar way (e.g. having parcels of similar size/shape) for most of the subjects . 

\subsection{Comparison Methods}
We compare our method to a set of different parcellation techniques and assess the effectiveness of the two-level parcellation framework separately for each level. Initially, we evaluate the performance of the supervertex clustering scheme by comparing it with two alternative approaches based on region growing~\cite{Bellec06,Blumensath13}. At the subject level, different strategies are applied to the RSFC data to obtain individual subject parcellations, including a state-of-the-art two-level approach that combines region growing with hierarchical clustering~\cite{Blumensath13}, as well as popular clustering algorithms adopted for the RSFC data, namely, \textit{k}-means, Ward's hierarchical clustering~\cite{Ward63}, and spectral clustering with normalised cuts~\cite{Heuvel08,Craddock12}. Finally, we also include parcellations obtained from non-connectivity data to show the potential of the connectivity-driven methods for computing more reliable subject level parcellations. These methods include anatomical parcellations~\cite{Fischl04,Desikan06}, random parcellations~\cite{Schirmer2015}, and geometric parcellations~\cite{Thirion14}. All methods are described below and summarised in Table~\ref{tab:subject-level-methods}. We rely on the following method descriptions and implementations for the comparison methods throughout the thesis, unless otherwise noted.  

\begin{table}[t!]
\centering
\includegraphics[width=\textwidth]{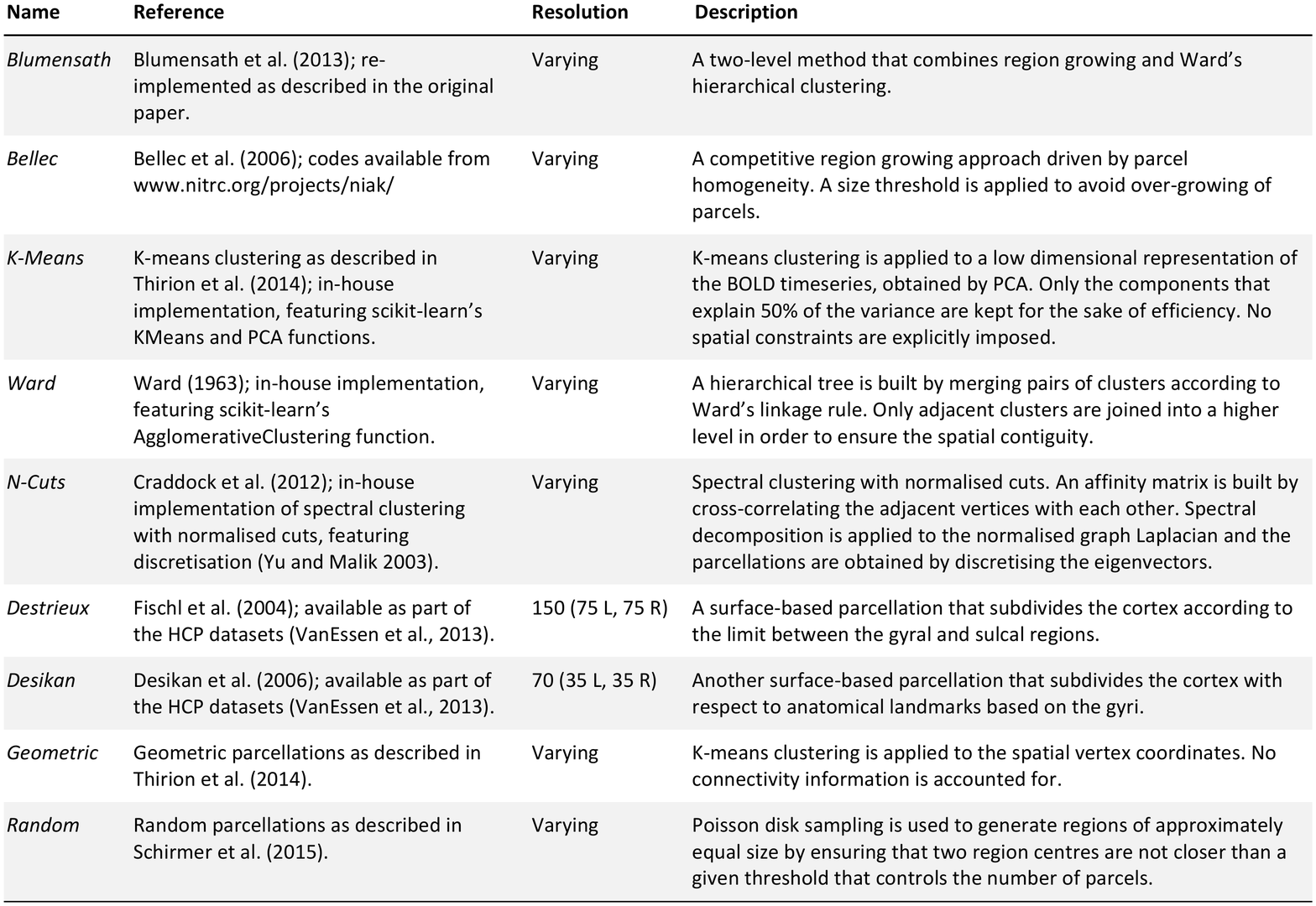}
\caption{Summary of the parcellation methods used for comparison.}
\label{tab:subject-level-methods}
\end{table}


\paragraph{Blumensath's region growing~\cite{Blumensath13}.} This two-level method combines region growing and Ward's hierarchical clustering to derive subject-level parcellations of the cerebral cortex. At the initial level, each seed vertex is grown into a region (cluster) by an iterative process based on the similarity between the cluster centroids and the adjacent vertices to them. In order for a cluster to obtain a vertex, the correlation between the vertex' timeseries and the cluster's timeseries should exceed $p$ times the maximal correlation between all other regions' timeseries and their associated neighbourhood vertex timeseries~\cite{Blumensath13}. 

After all vertices have been assigned to a cluster, subject-level parcellations are obtained by merging adjacent clusters using hierarchical clustering with Ward's linkage rule~\cite{Ward63} as in our proposed two-level framework. However, this approach differs from our method in three ways. (1) Instead of relying on a fixed number of seed vertices to start the parcellation process, it \textit{a priori} locates potentially homogeneous cortical regions that can be used to compute the seeds. (2) These seeds are used as the cluster centroids, but never updated during the region growing process. (3) The representative timeseries of a cluster is computed by averaging the timeseries within the immediate neighbourhood of the seed vertex, instead of typically including all within-cluster vertices.

\paragraph{Bellec's region growing~\cite{Bellec06}.} This region growing approach is originally proposed to reduce the spatial dimensionality of the RSFC data. The objective function is based on maximising the global homogeneity of the parcellation. Parcel homogeneity in this context is defined as the average correlation between the timeseries associated with any pair of vertices within a region. 

Since this method is driven by homogeneity, a size constraint $(t)$ is imposed in order to avoid having non-uniformly evolved parcellations in terms of parcel size (e.g. having many small, highly homogeneous parcels with a few very large parcels across the cortex)~\cite{Bellec06}. During the region growing process, regions that exceed the size parameter $t$ are validated and excluded from further consideration. This ensures that the size of a validated region cannot exceed $(2t - 2)$, as it is possible that two regions of size $(t - 1)$ can still be merged during the growing process. Different from the original paper, we specify the number of regions beforehand in order to be consistent with the other approaches; hence the algorithm keeps merging the smallest adjacent regions until the desired resolution has been reached.

\paragraph{\textit{K}-means algorithm~\cite{Thirion14}.} As an alternative technique, the well-known \textit{k}-means clustering algorithm is used to compute subject-level parcellations for varying resolutions. To overcome the computational limitations inherent to \textit{k-means} due to high dimensionality, we follow the instructions in~\cite{Thirion14} and reduce the dimensionality of the data for each subject via PCA, capturing about $50\%$ of the variance. \textit{K}-means naturally does not compute spatially contiguous parcellations, as the spatial structure of the data is not accounted for. Therefore, resulting parcellations can consist of extremely disjoint regions.  

\paragraph{Ward's hierarchical clustering~\cite{Ward63}.} We cluster the RSFC data using another popular clustering algorithm, Ward's agglomerative hierarchical clustering~\cite{Ward63}. As opposed to \textit{k}-means, spatial coherency within parcels is explicitly ensured by only merging adjacent clusters at each iteration. Different from the proposed approach and~\cite{Blumensath13}, this method is directly applied to the vertex-level RSFC data, instead of relying on an initial fine-resolution parcellation.

\paragraph{Spectral clustering with normalised cuts~\cite{Craddock12}.}

A spectral clustering technique based on normalised cuts is applied to the connectivity data as described in~\cite{Craddock12}. An affinity matrix $W$ is built by cross-correlating vertices with each other. In order to obtain spatially contiguous parcellations, we only retain the edges constructed between adjacent vertices and discard the others. After ensuring that the resulting affinity matrix is fully-connected and positive-semidefinite (i.e. all $W_{ij} \geq 0$), we apply spectral decomposition to the normalised graph Laplacian, defined as $L=D^{-1/2}(D-W)D^{-1/2}$, where $D$ is a diagonal matrix with each entry $D_{ii} = \sum_{j}W_{ij}$ representing the degree of vertex $i$. The eigenvectors corresponding to the $K$ smallest non-zero eigenvalues are clustered via discretisation~\cite{Yu03} to obtain the final parcellations.

\paragraph{Anatomical parcellations.} We also assess the performance of surface-based anatomical parcellations in order to represent the functional organisation of the brain at the subject level. These atlases are distributed as part of the HCP datasets~\cite{Desikan06,Fischl04} and tailored to each individual subject with respect to anatomical features, such as cortical folding. 

The Destrieux atlas~\cite{Fischl04} subdivides the cerebral cortex into a fixed number of 75 parcels per hemisphere according to the limit between the gyral and sulcal regions, determined by the curvature of the surface. The Desikan atlas~\cite{Desikan06} offers a coarser anatomical parcellation (35 parcels per hemisphere) based on gyral landmarks, where a gyrus is defined as the visible part on the pial surface and limited by the adjacent banks of the sulci~\cite{Desikan06}.

\paragraph{Other parcellations.} In addition, we include two other approaches to our experiments that do not account for any connectivity information, with the aim of having a baseline towards data-driven approaches. First, we compute random parcellations using Poisson disk sampling as described in~\cite{Schirmer2015}. Poisson disk sampling is used to generate regions of approximately equal size by ensuring that two region centres are not closer than a given threshold that controls the number of parcels. Second, we obtain geometric parcellations by applying \textit{k}-means algorithm to the anatomical vertex coordinates as described in~\cite{Thirion14}. Since the parcellation is directly driven by spatial proximity between cortical vertices, it is likely to obtain regularly shaped parcels, with roughly similar size. 

\subsection{Experimental Setup}
\begin{figure}[hb!]
\centering
\includegraphics[width=0.5\textwidth]{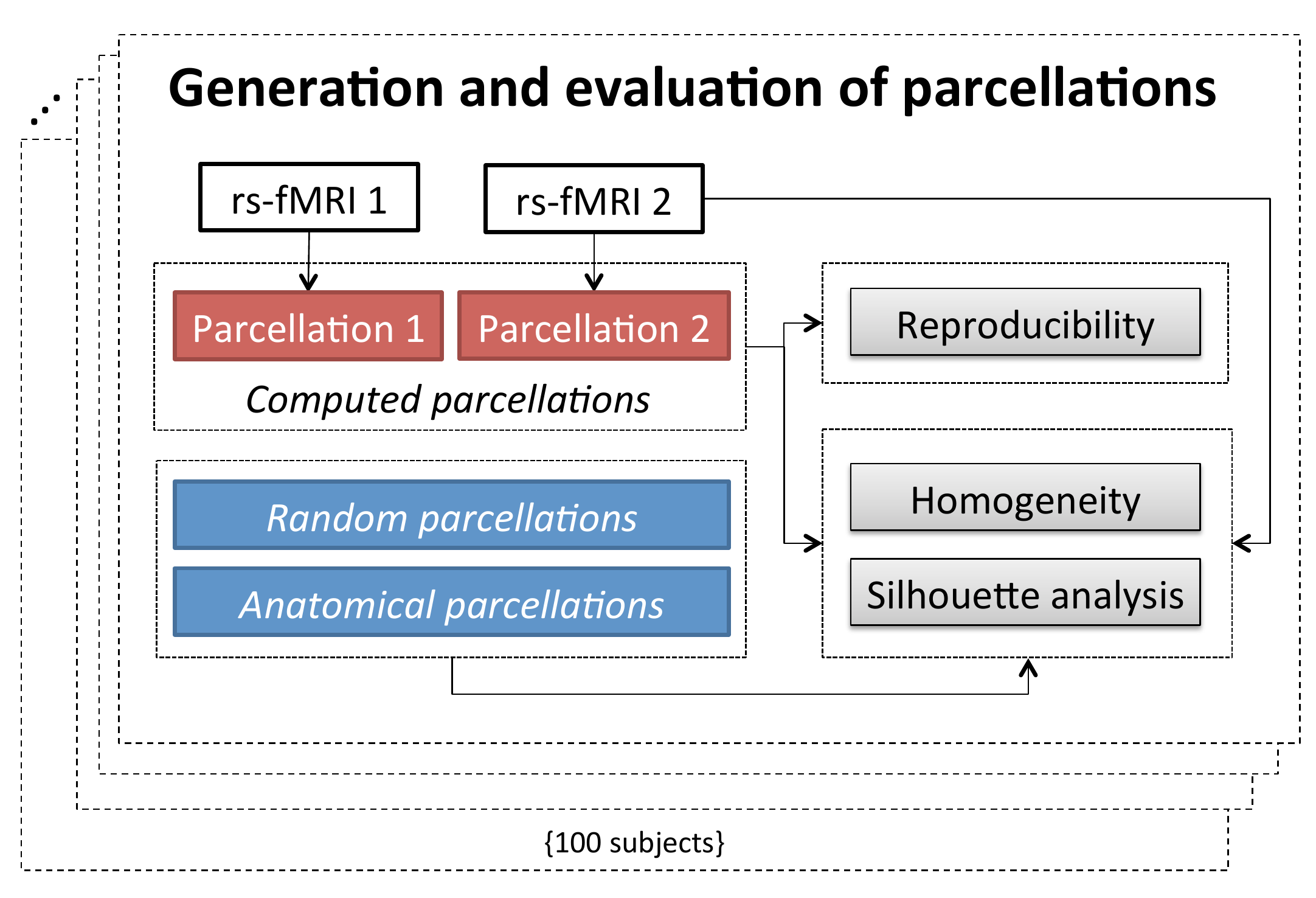}
\caption{Visual outline of the parcellation evaluation pipelines.}
\label{fig:eval-pipelines}
\end{figure}

All subjects in our cohort are represented by two 30-minute rs-fMRI datasets (rs-fMRI 1 and rs-fMRI 2). We perform reproducibility analysis for each subject by comparing their parcellations obtained from rs-fMRI 1 and rs-fMRI 2 (i.e. scan-to-scan reproducibility). Clustering accuracy (homogeneity and Silhouette analysis) is computed based on the first parcellation, but using data from rs-fMRI 2, to avoid possible biases that may have emerged during the computation of the parcellations. There is only one random, \textit{Desikan}, and \textit{Destrieux} parcellation is available for each subject, therefore we skip the reproducibility analysis for these methods. A summary of the evaluation pipelines is given in Fig.~\ref{fig:eval-pipelines}.

\section{Results}
\label{sec:multi-level-results}
\subsection{Parameter Selection}
Other than the number of parcels $K$, the proposed algorithm has two external model parameters: (1) the weighting parameter $\alpha$, and (2) the number of supervertices $N$. 

\begin{figure}[tbh!]
\centering
\includegraphics[width=\linewidth]{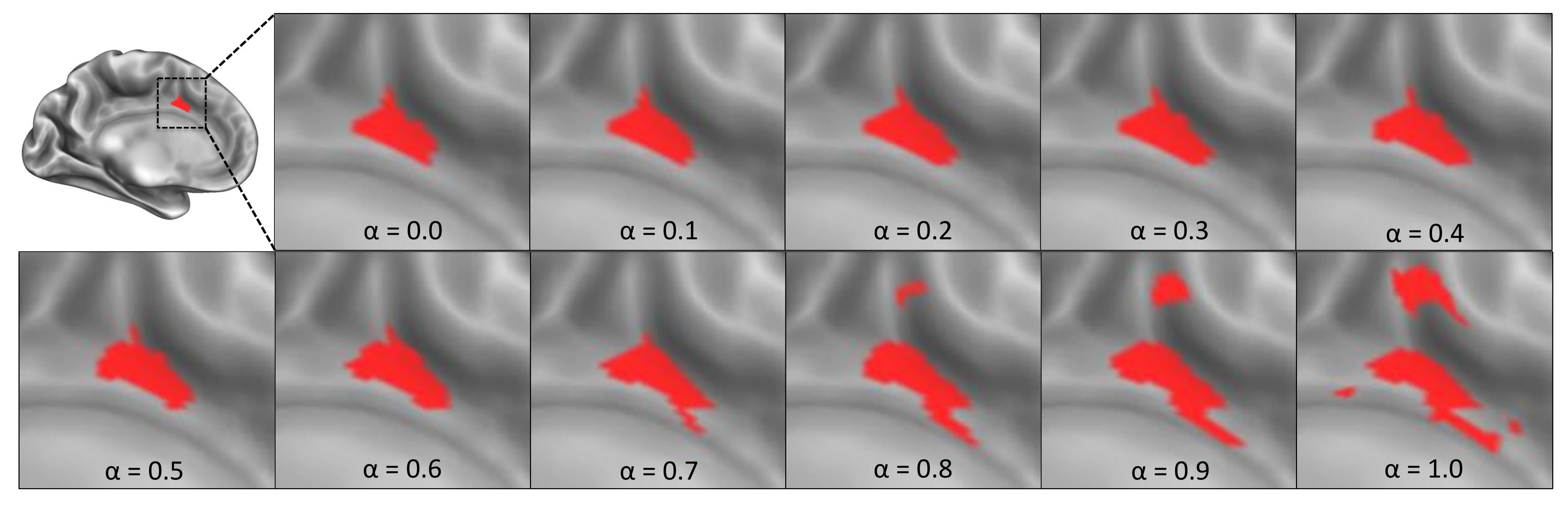}  
\caption{Evolution of the boundaries of a supervertex with different values of $\alpha$.}
\label{fig:sv_evolve}
\end{figure}

The weighting parameter $\alpha$ is used to determine the influence of functional connectivity over spatial proximity, and thus, has a direct impact on the supervertices. Whereas values of $\alpha$ closer to 1 tend to group more functionally correlated vertices together, the spatial contiguity of the clusters can be severely affected, reducing the reproducibility of the final parcels. Decreasing $\alpha$ to an extent would produce more regularly-shaped supervertices, but ultimately may lead to lower homogeneity, as the clustering is only driven by the spatial closeness of vertices to supervertex centroids. Given this trade-off, we set $\alpha$ to $0.7$, which offers the highest homogeneity without violating the spatial integrity of supervertices. The impact of $\alpha$ over the spatial contiguity of a supervertex can be seen in Fig.~\ref{fig:sv_evolve}. 

\begin{figure}[hbt!]
\centering
\includegraphics[width=0.5\textwidth]{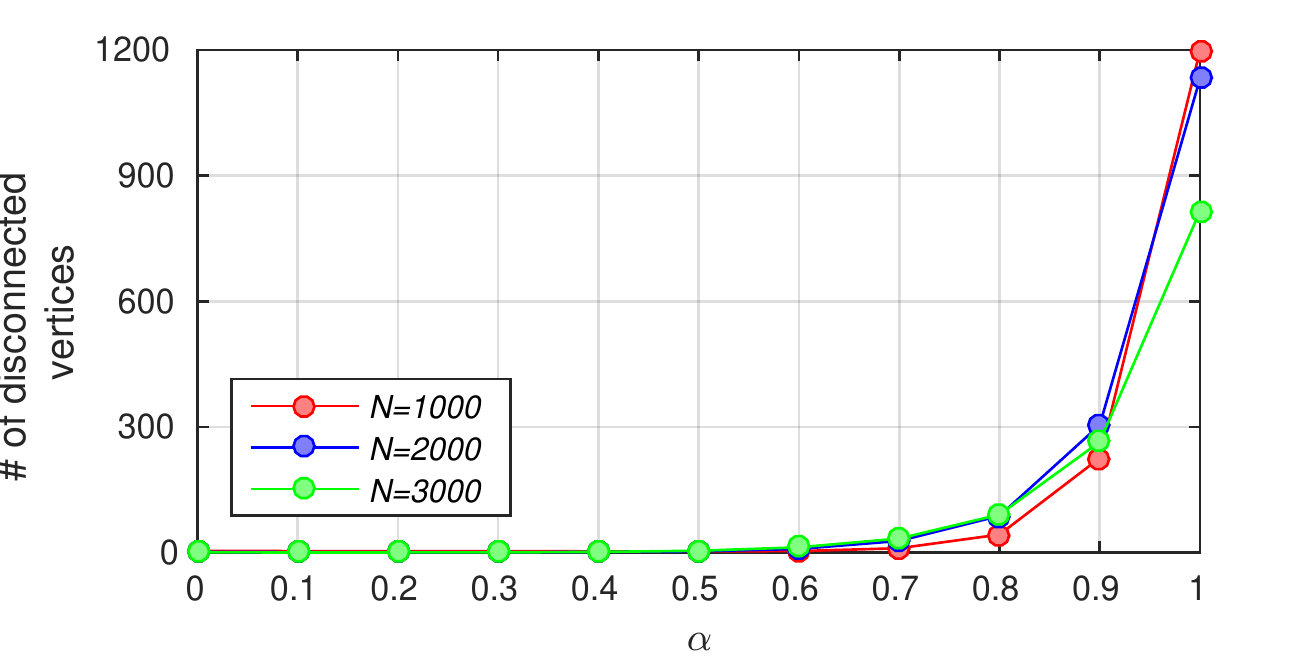} \\
\caption[The number of spatially disconnected vertices with respect to $\alpha$ and $N$.]{The number of vertices that are not spatially linked to the largest connected component in their associated supervertices with respect to different values of $\alpha$ and $N$.}
\label{fig:param-spat}
\end{figure}

In order to show that this tendency holds across all supervertices/subjects, we located the disjoint supervertices and counted the number of vertices that are not spatially connected to their assigned supervertex using the neighbouring criterion introduced in Section~\ref{fig:sv_adjacency}. The results obtained for different numbers of supervertices are given in Fig.~\ref{fig:param-spat}. As can be seen in the figure, $\alpha=0.7$ stands as a cut-off point where supervertices start to become disjoint due to higher influence of functional similarity over spatial constraints.  

\begin{figure}[tbh!]
\centering
\begin{tabular}{cc}
\includegraphics[width=0.5\textwidth]{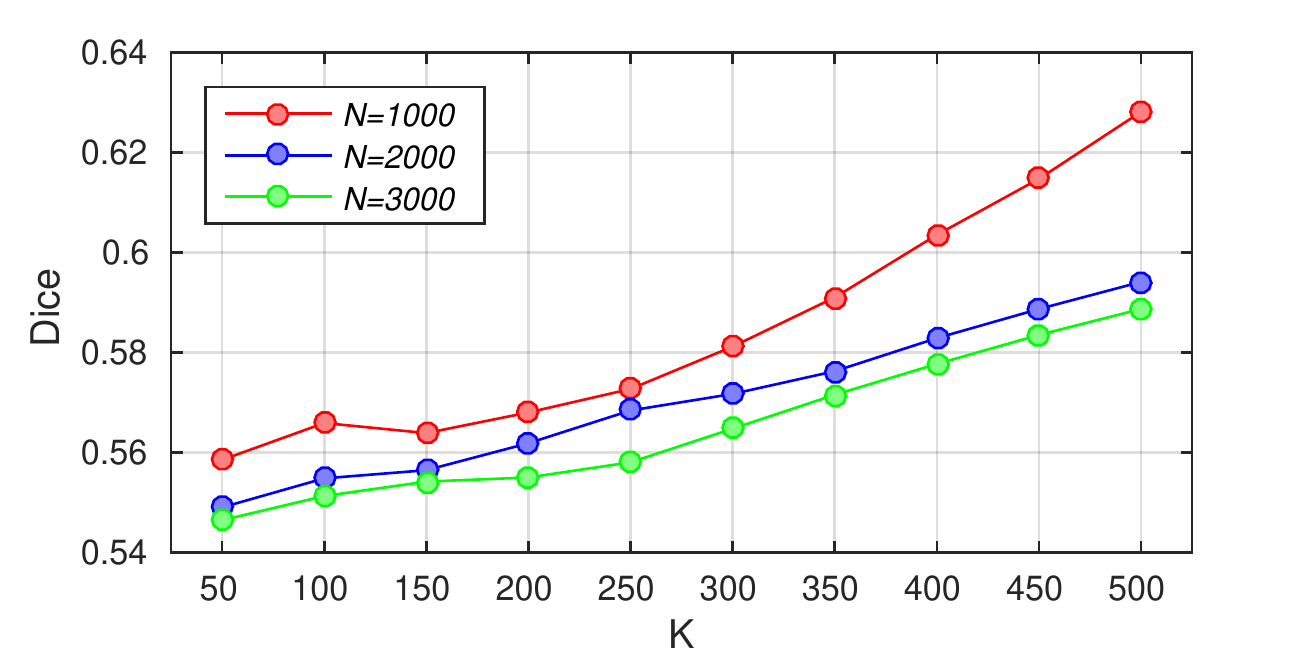}  \kern-1.5em & 
\includegraphics[width=0.5\textwidth]{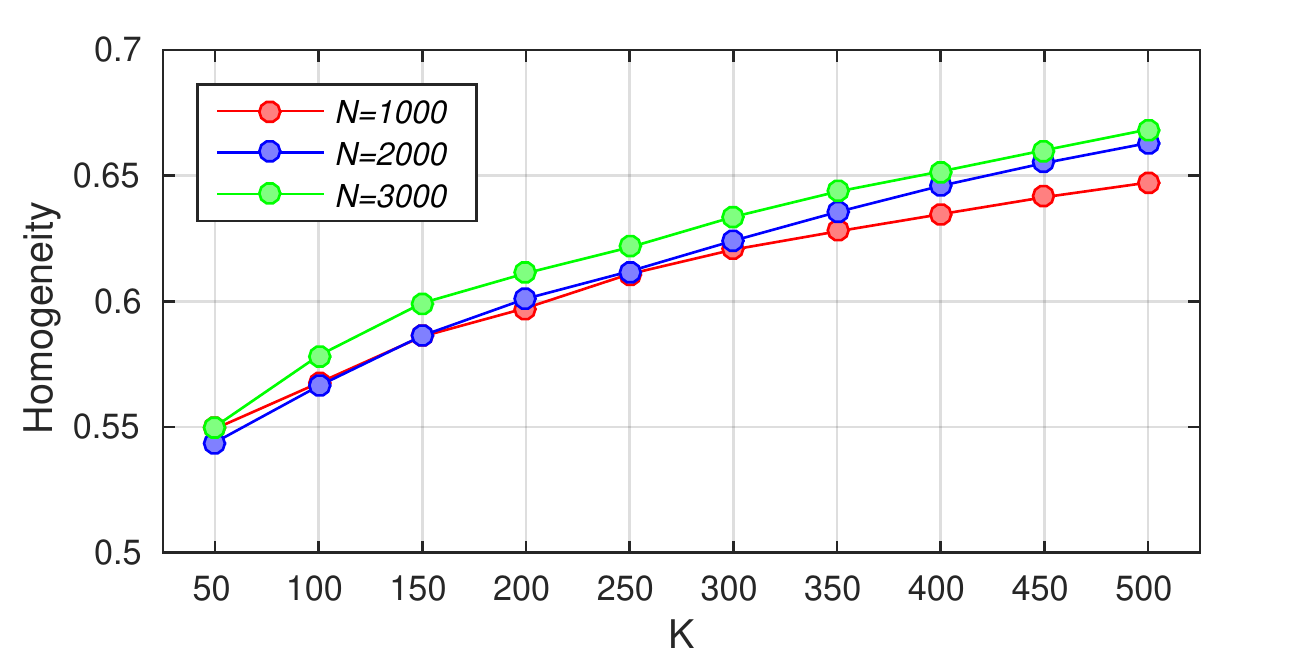}  \\
\end{tabular}
\caption[Reproducibility and homogeneity as a function of the number of parcels.]{Reproducibility (\textit{left}) and homogeneity (\textit{right}) as a function of the number of regions ($K$) for different number of supervertices ($N$).}
\label{fig:param-two-level}
\end{figure}

The number of supervertices, $N$, is the second external parameter imposed to the method. It provides a means of spatial dimensionality reduction, and therefore, can help alleviate the impact of noise at the vertex level. While selecting a very low $N$ may provide more reproducible parcellations, it may lead to over-smoothing of the underlying functional data (i.e. low homogeneity). Increasing $N$ gradually yields more homogeneous supervertices (and hence parcellations), but at the cost of lower reproducibility. The accuracy (by means of reproducibility and homogeneity) of the parcellations with respect to different number of supervertices is shown in Fig.~\ref{fig:param-two-level}. Given this trade-off, we initially parcellate the cortex into approximately $1000$ supervertices per hemisphere (i.e. $N=2000$ initial regions per subject). 

Two other methods, \textit{Blumensath} and \textit{Bellec}, are also driven by external parameters. The $p$ parameter in \textit{Blumensath} determines the degree of similarity between vertices and regions competing for them. Based on our observations and the discussion in the original paper, we set this parameter to 0.9, which significantly increased the computational speed of the approach with only a minimal impact on the final parcellations~\cite{Blumensath13}. The other region growing approach, \textit{Bellec} relies on a size constraint $t$ that avoids over-growing of regions~\cite{Bellec06}. For each supervertex resolution we set $t$ to the expected average parcel size (i.e. number of vertices / number of supervertices).

\subsection{First-Level Clustering Results}
\begin{figure}[!t]
\centering
\begin{tabular}{cccc}
\textit{Arslan} & \textit{Blumensath} & \textit{Bellec} & \textit{Geometric} \\

\includegraphics[width=0.2\textwidth]{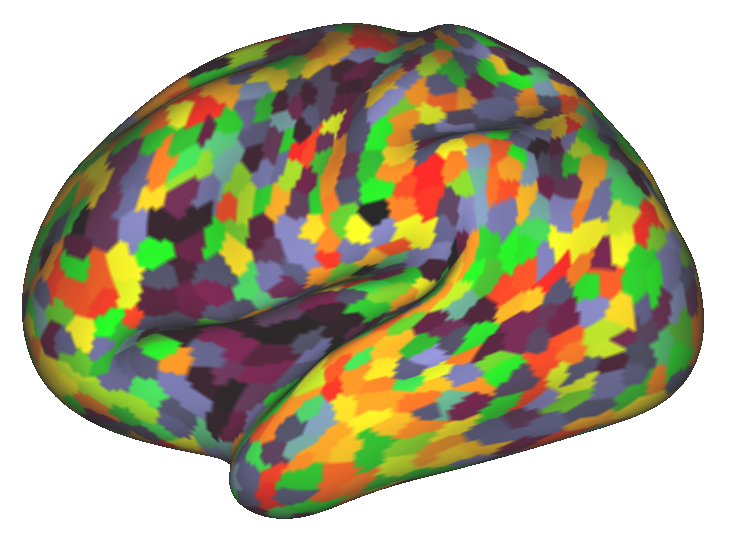} &
\includegraphics[width=0.2\textwidth]{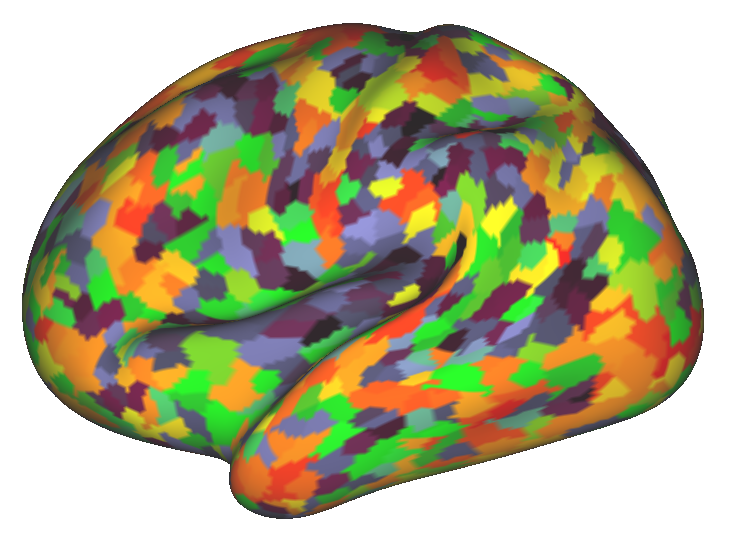} & 
\includegraphics[width=0.2\textwidth]{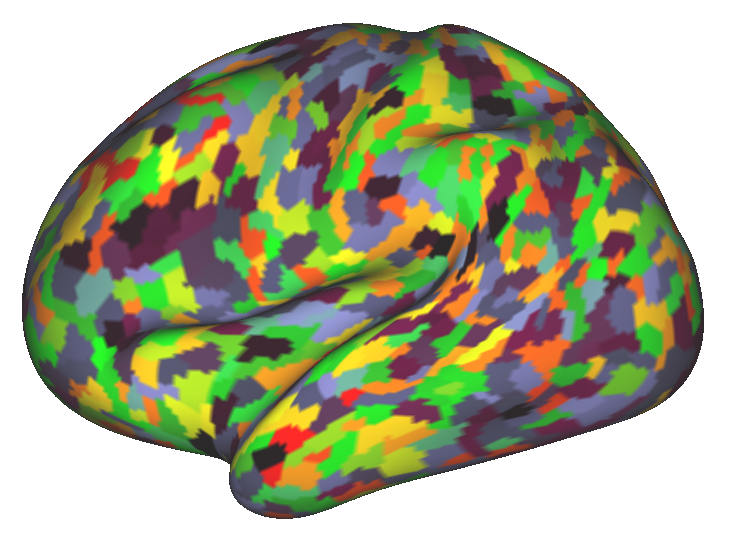} &
\includegraphics[width=0.2\textwidth]{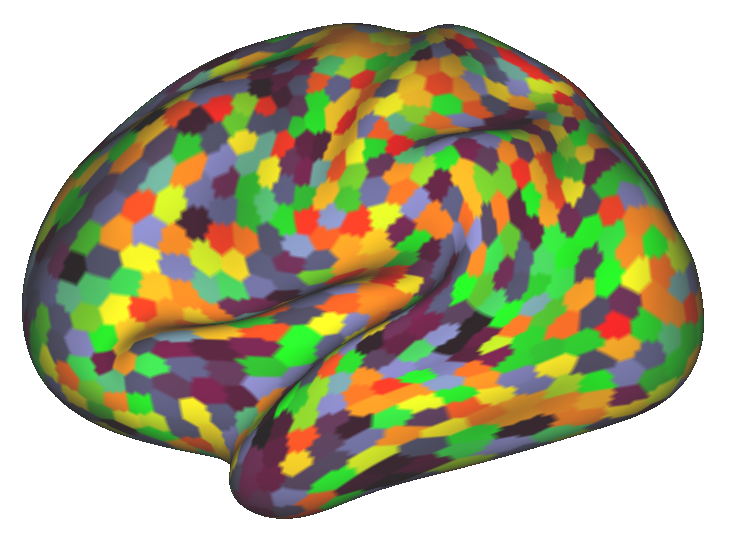} \\

\includegraphics[width=0.2\textwidth]{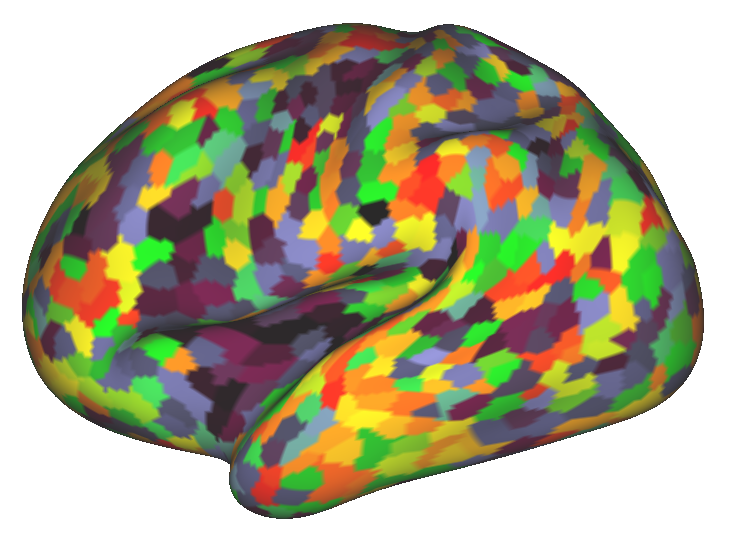} &
\includegraphics[width=0.2\textwidth]{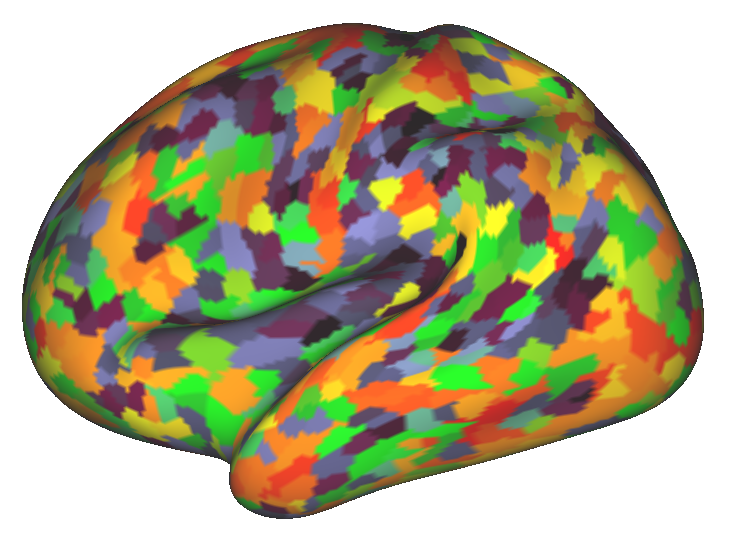} & 
\includegraphics[width=0.2\textwidth]{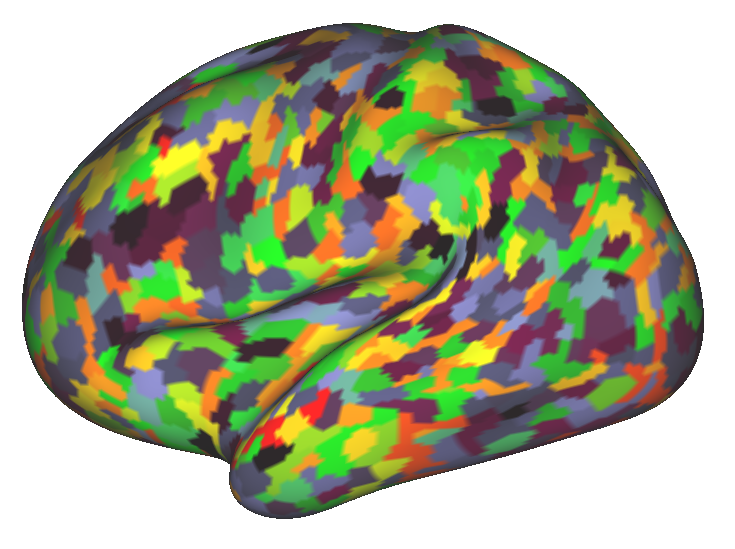} &
\includegraphics[width=0.2\textwidth]{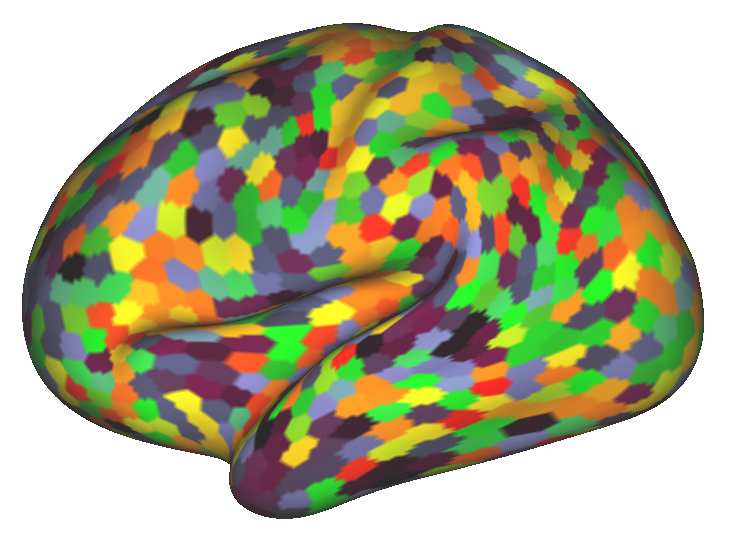} \\

\includegraphics[width=0.2\textwidth]{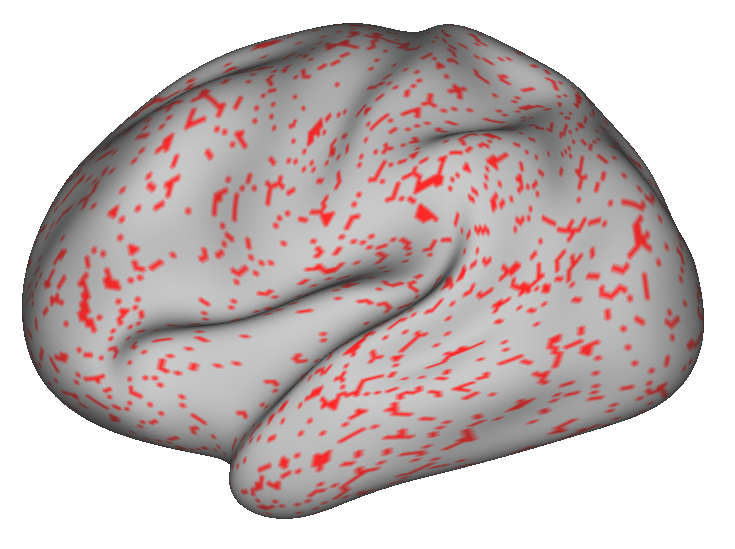} &
\includegraphics[width=0.2\textwidth]{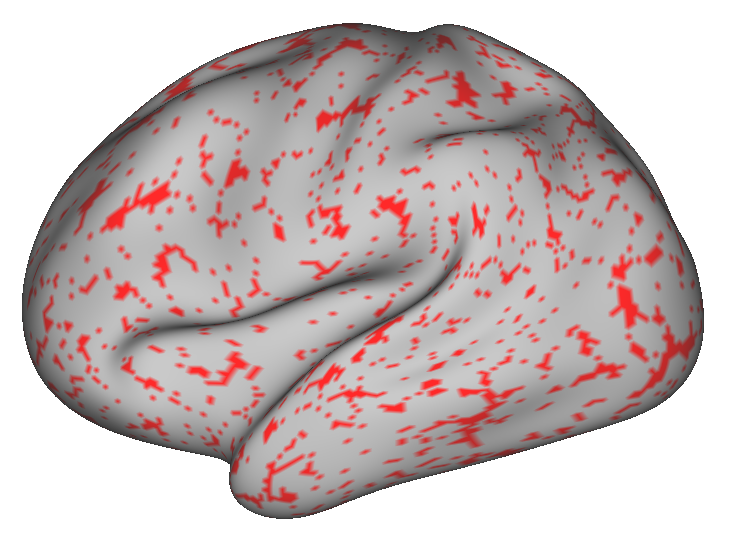} & 
\includegraphics[width=0.2\textwidth]{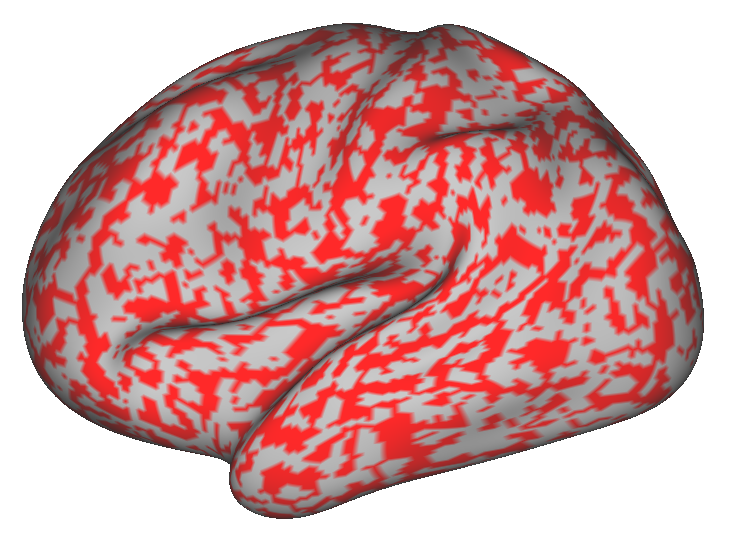} &
\includegraphics[width=0.2\textwidth]{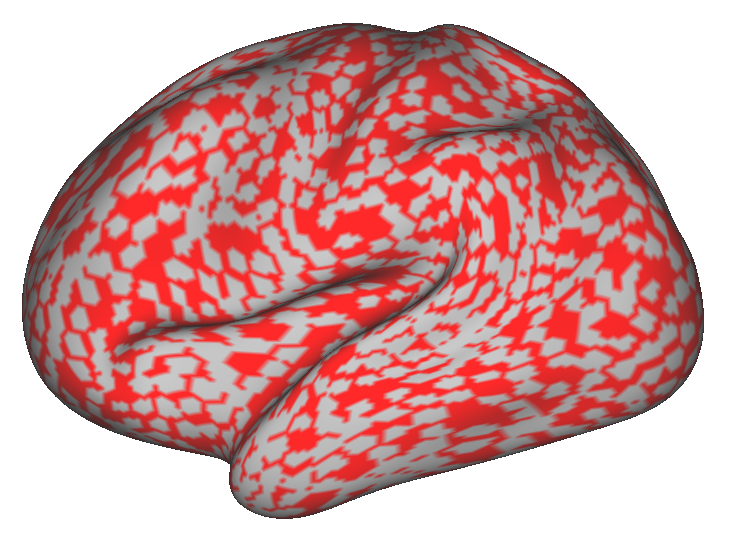} \\

\end{tabular}
\caption[Parcellations of the left lateral cortex derived from one subject for 1000 regions.]{Parcellations of the lateral cortex of the left hemisphere derived from one subject using different approaches for \textit{N} = 1000 regions. The parcellations in the first and second row were obtained from different scans of the same subject in order to evaluate scan-to-scan reproducibility. Parcel colours were matched for better visualisation and easier comparison. The last row shows the non-matching regions. }
\label{fig:first-visual}
\end{figure}

We first provide the evaluation results for the initial parcellation stage (i.e. first level). We assess the performance of the supervertex clustering scheme (denoted as \textit{Arslan}), against two functional connectivity-driven region growing approaches (\textit{Blumensath} and \textit{Bellec}) and geometric (\textit{Geometric}) parcellations, which do not take any connectivity information into account, and hence, provide a baseline for comparison. We test the methods with respect to three different resolutions of approximately $N$= 1000, 2000, and 3000 regions. We determine the exact number of regions for each subject from the \textit{Blumensath}~\cite{Blumensath13} parcellations, as it is the only method that is driven by seed vertices generated from the underlying data, rather than a set of pre-determined centroids. Using the same initial parcellation resolution for each method facilitates their comparison and avoids unfairness with respect to the number of regions which may have emerged otherwise. Parcellations of the lateral cortex of the left hemisphere derived from one subject for a resolution of \textit{N} = 1000 regions are given in Fig.~\ref{fig:first-visual} for visual inspection.

\begin{figure}[!b]
\centering
\begin{tabular}{cc}
\includegraphics[width=0.5\textwidth]{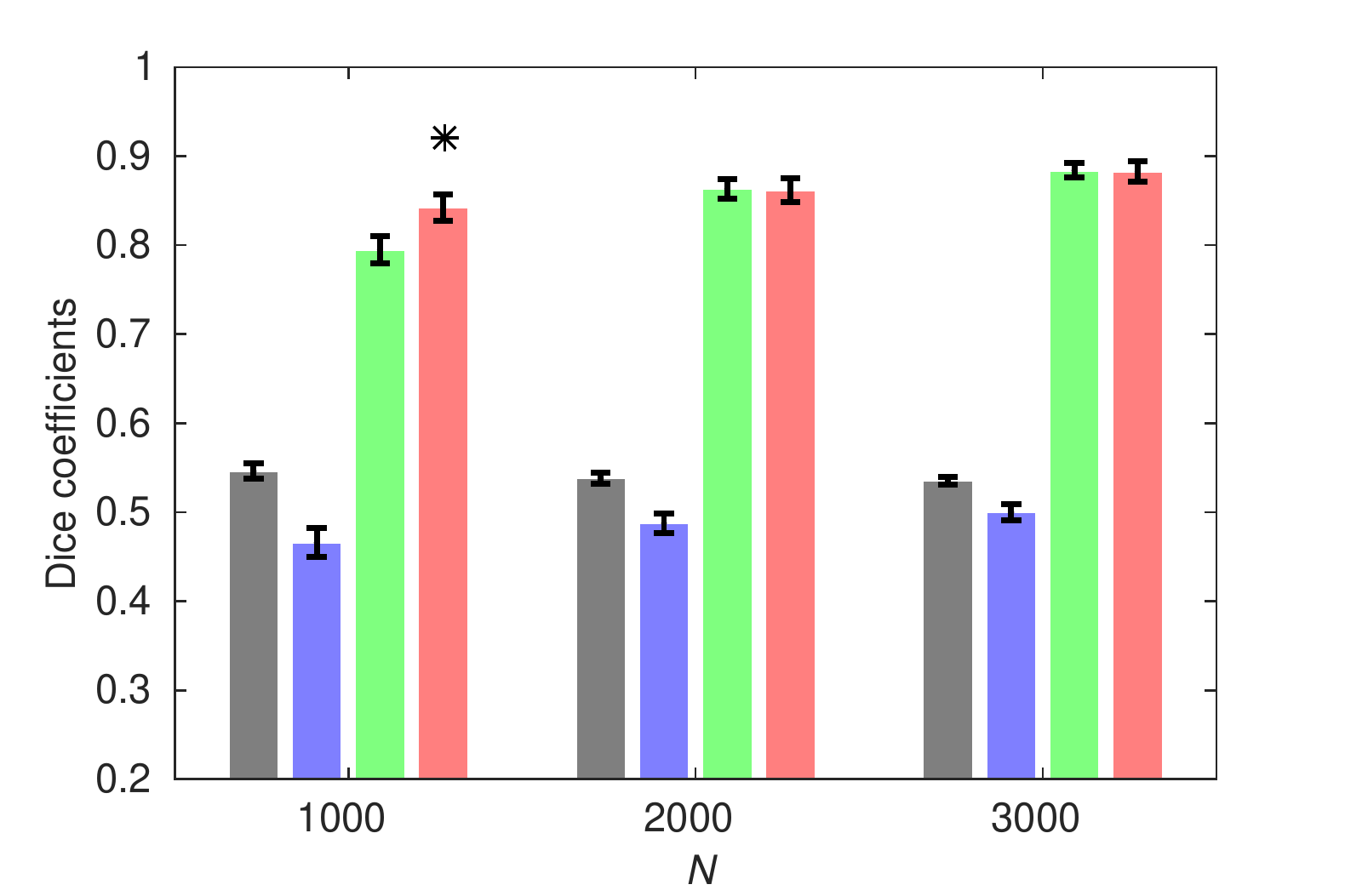} \kern-2em &
\includegraphics[width=0.5\textwidth]{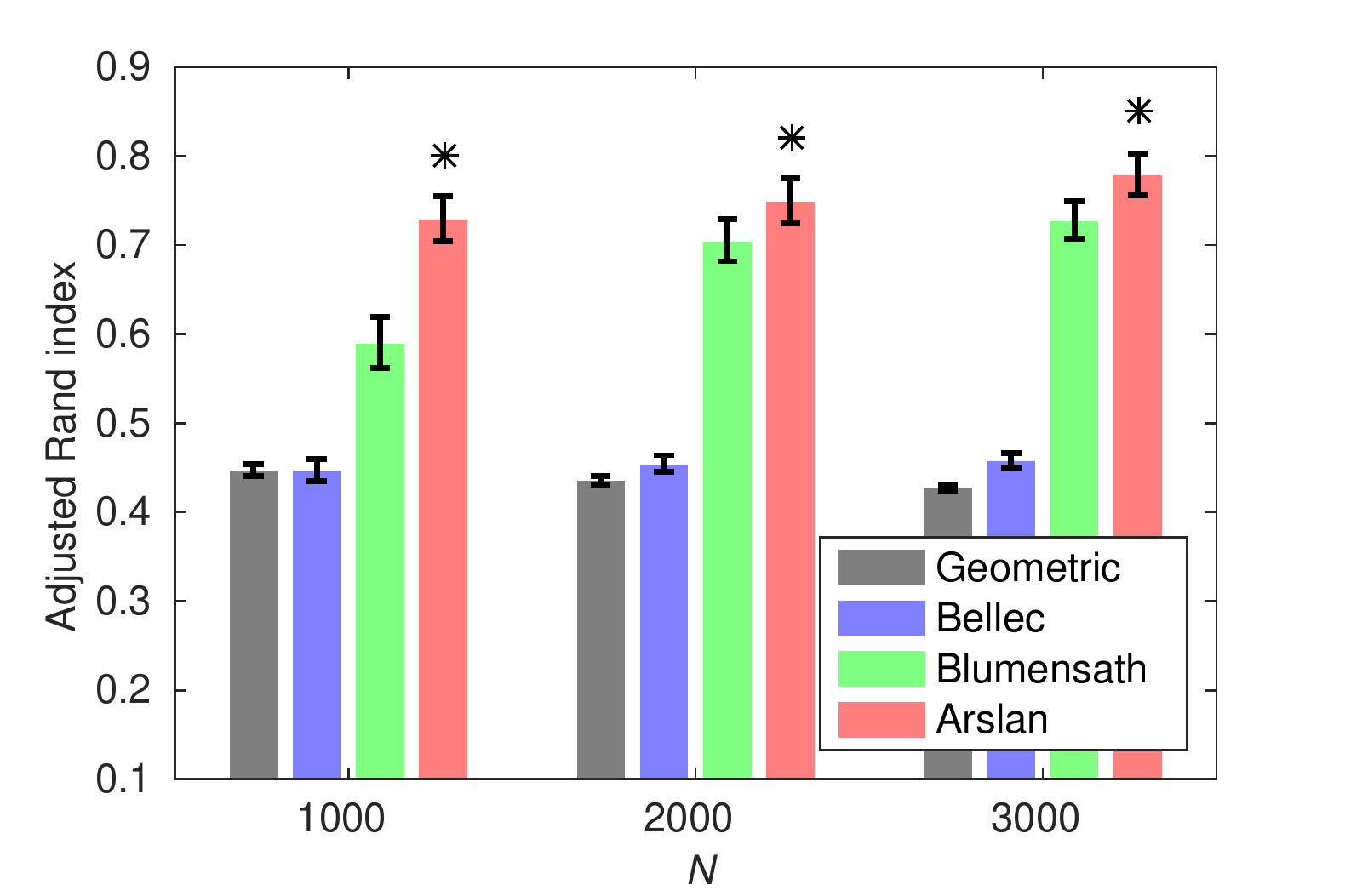}
\end{tabular}
\caption[First-level reproducibility results.]{First-level reproducibility results obtained using Dice coefficient (left) and adjusted Rand index (right). Whiskers in the bars indicate the variability across repetitions as measured by standard deviation. Stars (*) show statistical significance between the winner and the runner-up based on Wilcoxon signed rank test with $p < 0.01$.}
\label{fig:first-repro}
\end{figure}

Scan-to-scan reproducibility results computed by Dice coefficients and ARI are given in Fig.~\ref{fig:first-repro}. It is immediately clear from the figure that, the proposed supervertex clustering approach (\textit{Arslan}) and \textit{Blumensath} yield the most reproducible parcellations across all resolutions. Although similar Dice results are achieved by both methods, \textit{Arslan} appears to produce more robust regions with respect to ARI. \textit{Bellec} can not generate reproducible regions, most likely due to the fact that it does not rely on any spatial constraints in the region growing process. \textit{Geometric} is purely driven by the spatial proximity of the parcels, but due to randomness of the \textit{k}-means algorithm, the resulting parcellations tend to be very different.  

\begin{figure}[!t]
\centering
\begin{tabular}{cc}
\includegraphics[width=0.5\textwidth]{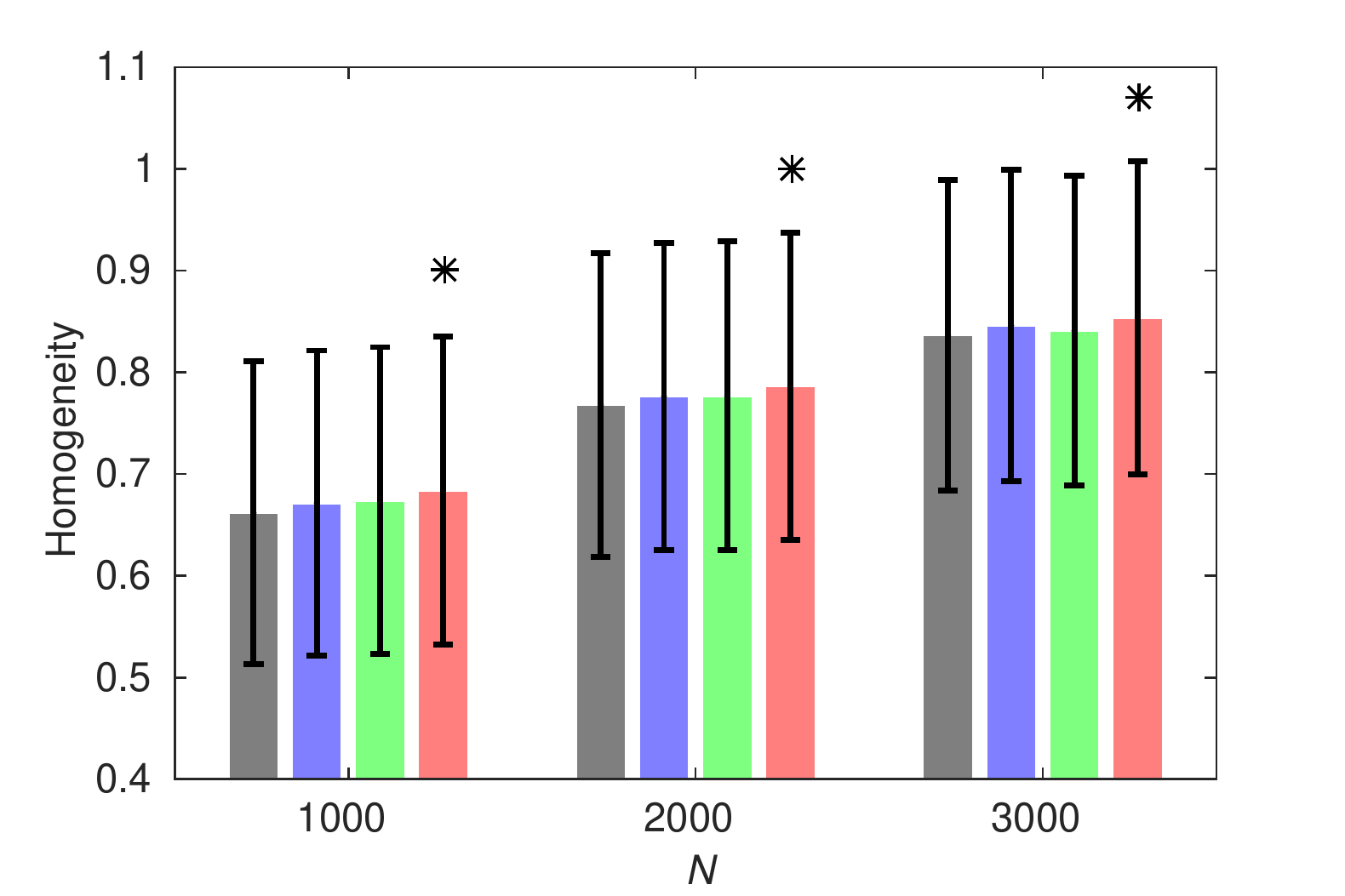} \kern-2em &
\includegraphics[width=0.5\textwidth]{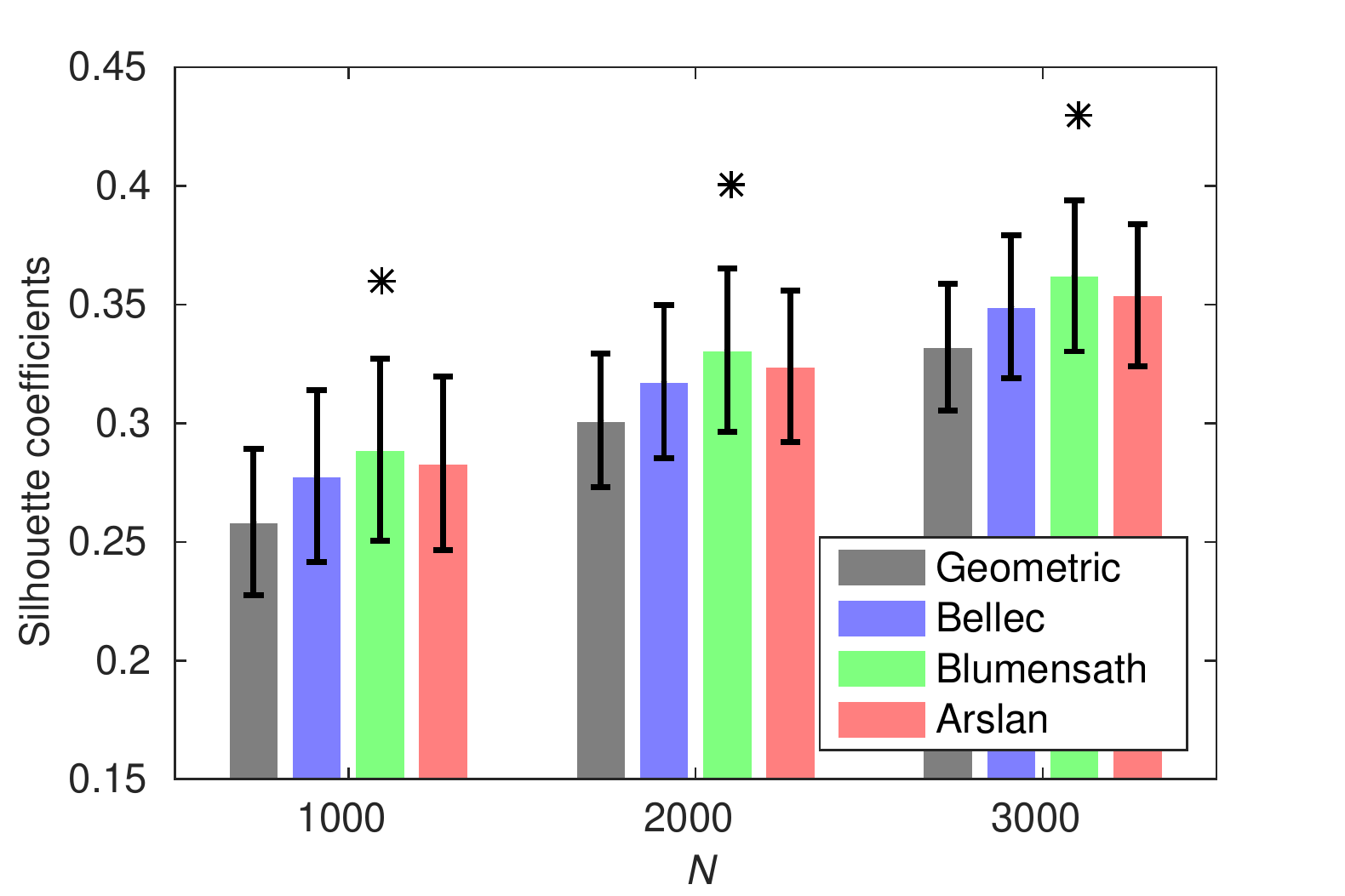}
\end{tabular}
\caption[First-level homogeneity values and Silhouette coefficients.]{First-level homogeneity values (left) and Silhouette coefficients (right). Whiskers in the bars indicate the variability across repetitions as measured by standard deviation. Stars (*) show statistical significance between the winner and the runner-up based on Wilcoxon signed rank test with $p < 0.01$.}
\label{fig:first-homo-silh}
\end{figure}

Homogeneity values and Silhouette coefficients indicating the fidelity of the parcellations to the underlying data are given in Fig.~\ref{fig:first-homo-silh}. Contrarily to reproducibility, stark differences in performance are not observed across methods, which even tend to perform more equivalently with increasing resolution. This is an expected result given the high resolution of the parcellations, which consist of very small regions. Since the BOLD timeseries are smoothed to some extent during preprocessing, this yields an inherent correlation between vertices that are spatially close to each other. 

Nevertheless, \textit{Arslan} and \textit{Blumensath} consistently perform better than the others in terms of homogeneity and Silhouette analysis, respectively. Although driven by a criterion that promotes homogeneity in regions, \textit{Bellec} performs worse than the other two connectivity-driven approaches, most likely due to the size constraint that avoids over-growing. As expected, \textit{Geometric} shows the worst performance, indicating that parcellations driven by the spatial proximity of cortical vertices do not fit the underlying connectivity well, even at very fine resolutions.

\subsection{Subject-Level Parcellation Results}
\begin{figure}[!t]
\centering
\begin{tabular}{lccccl}

\raisebox{1.8em}{\rotatebox{90}{\textit{Arslan}}} &
\includegraphics[width=0.2\textwidth]{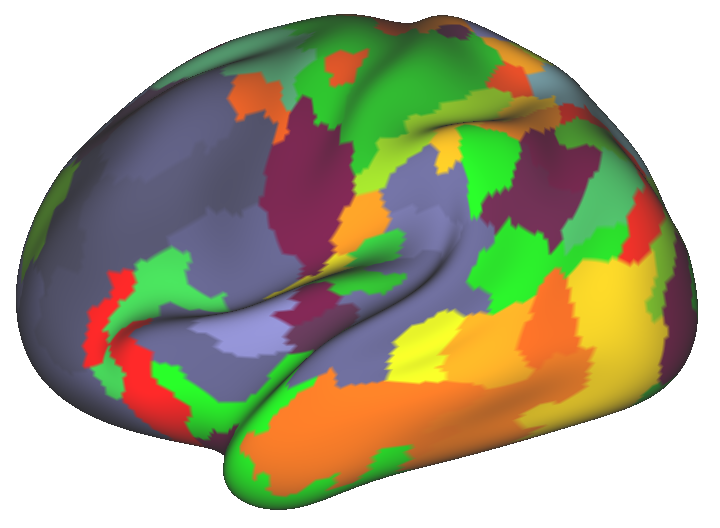} &
\includegraphics[width=0.2\textwidth]{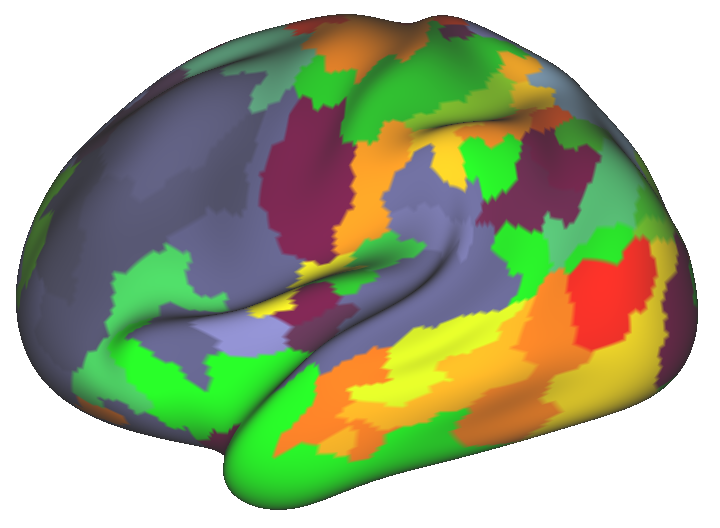} & 
\includegraphics[width=0.2\textwidth]{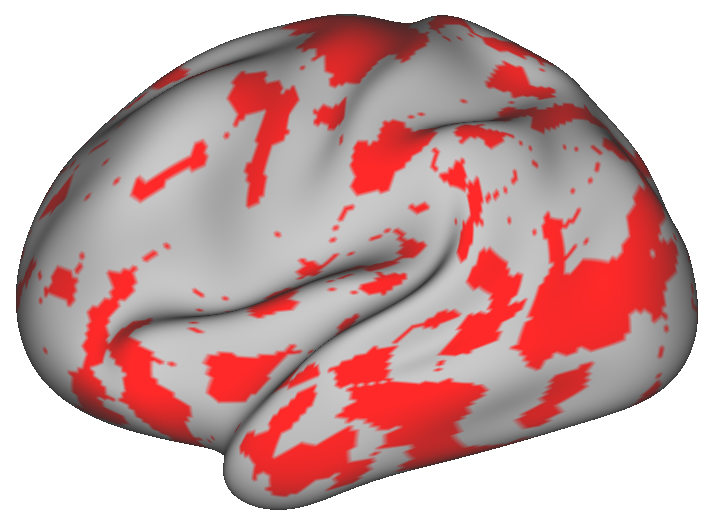} \\

\rotatebox{90}{  \textit{Blumensath}} &
\includegraphics[width=0.2\textwidth]{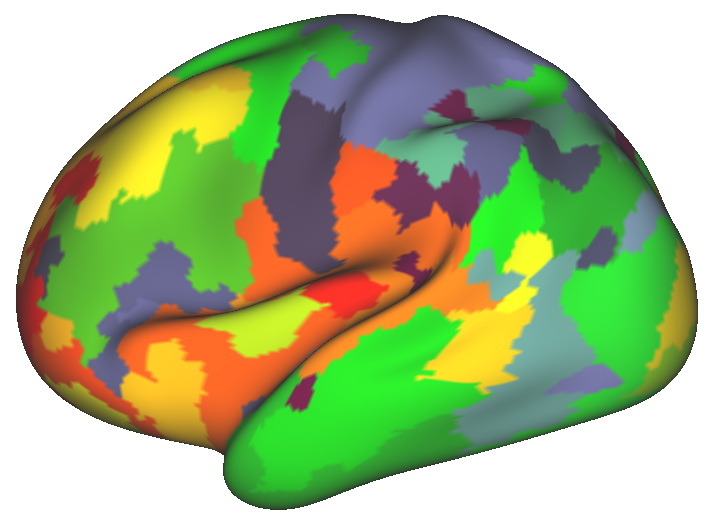} &
\includegraphics[width=0.2\textwidth]{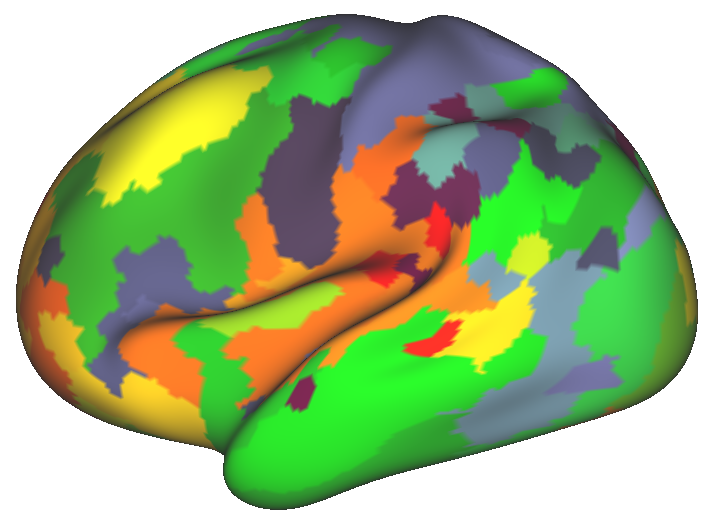} & 
\includegraphics[width=0.2\textwidth]{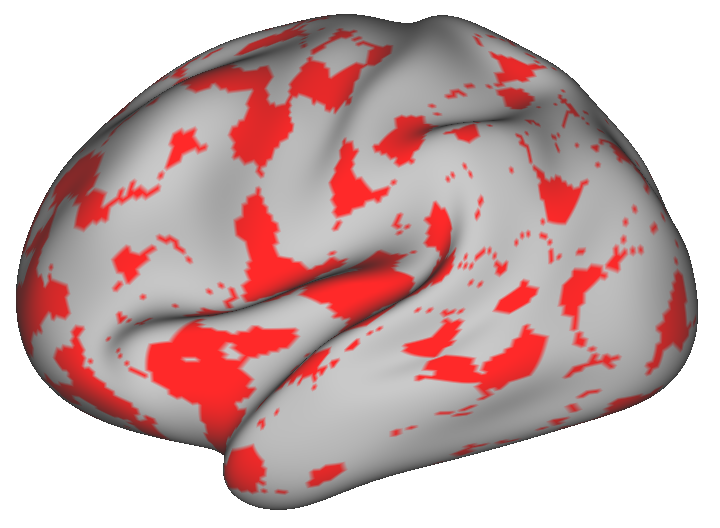} \\

\raisebox{1.8em}{\rotatebox{90}{\textit{Bellec}}} &
\includegraphics[width=0.2\textwidth]{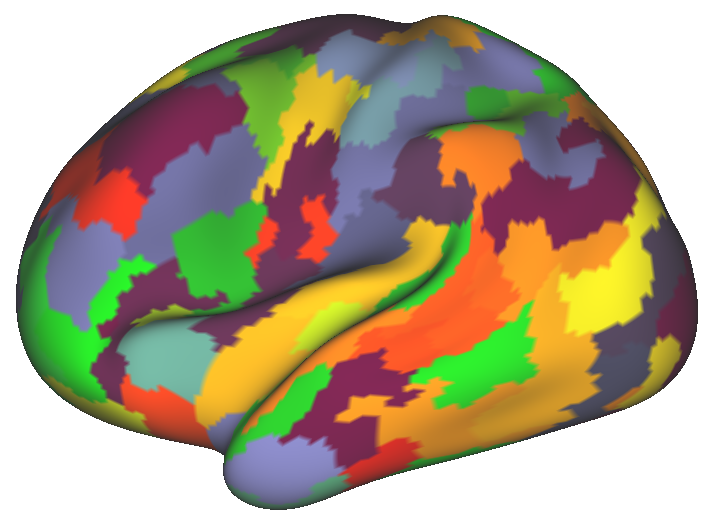} &
\includegraphics[width=0.2\textwidth]{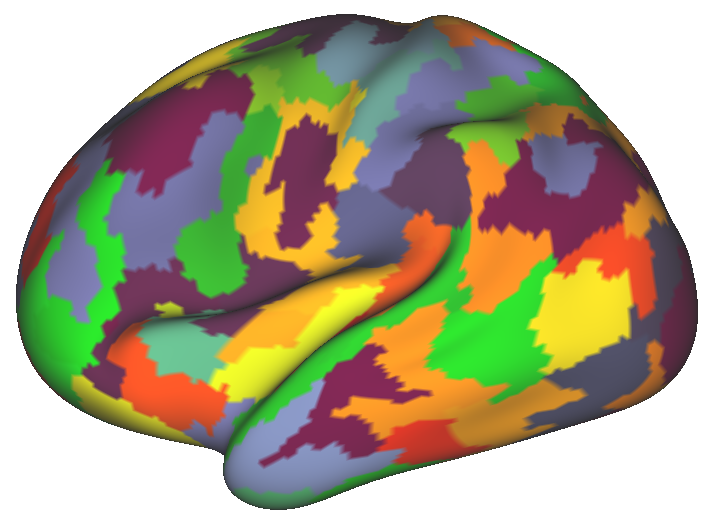} & 
\includegraphics[width=0.2\textwidth]{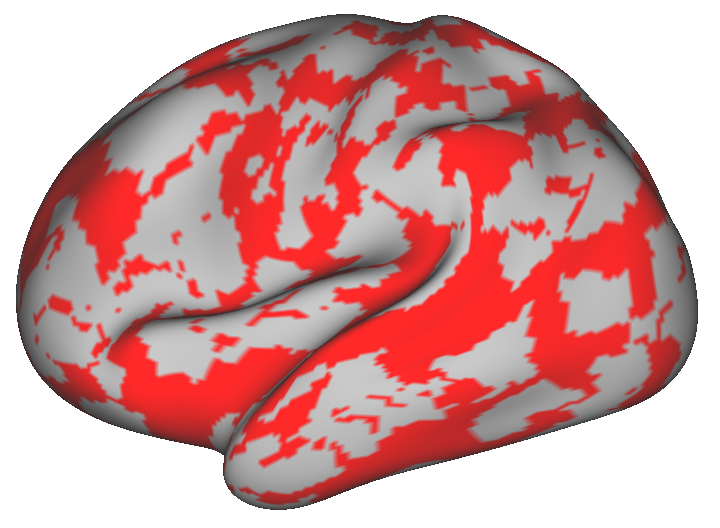} &
\includegraphics[width=0.2\textwidth]{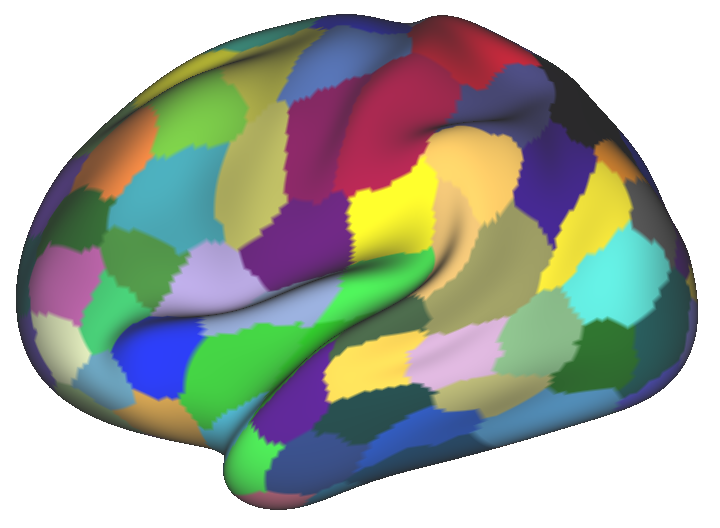} &
\raisebox{4.7em}{\rotatebox{270}{\textit{Random}}} \\

\raisebox{2em}{\rotatebox{90}{\textit{Ward}}} &
\includegraphics[width=0.2\textwidth]{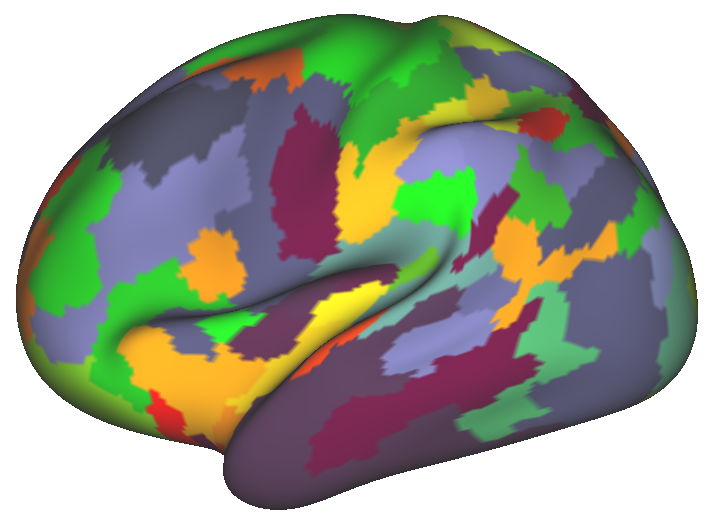} &
\includegraphics[width=0.2\textwidth]{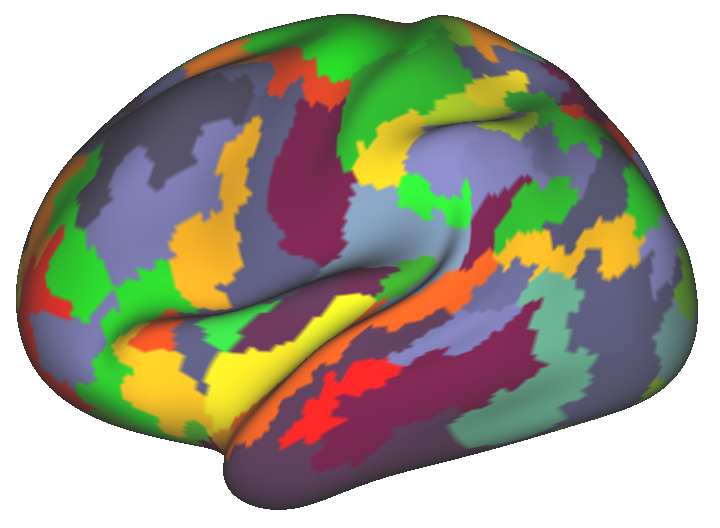} & 
\includegraphics[width=0.2\textwidth]{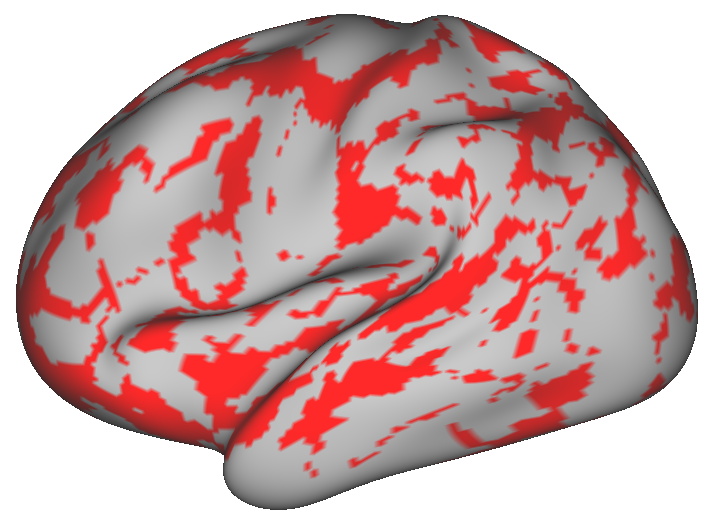} &
\includegraphics[width=0.2\textwidth]{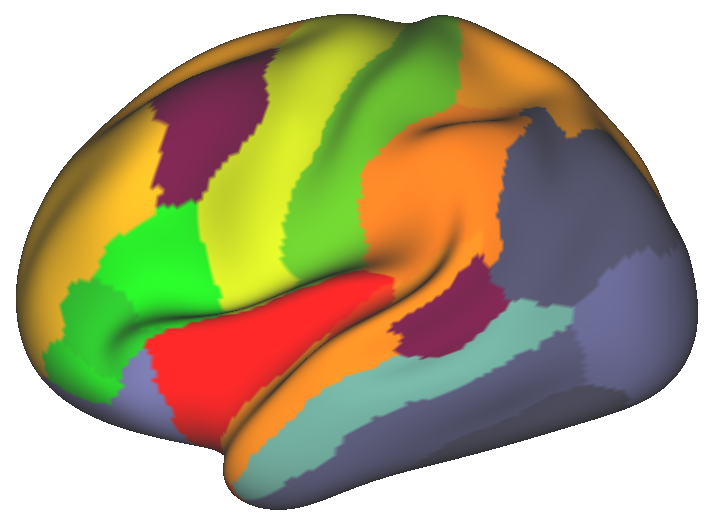}  &
\raisebox{4.8em}{\rotatebox{270}{\textit{Desikan}}} \\

\raisebox{1.2em}{\rotatebox{90}{\textit{K-Means}}} &
\includegraphics[width=0.2\textwidth]{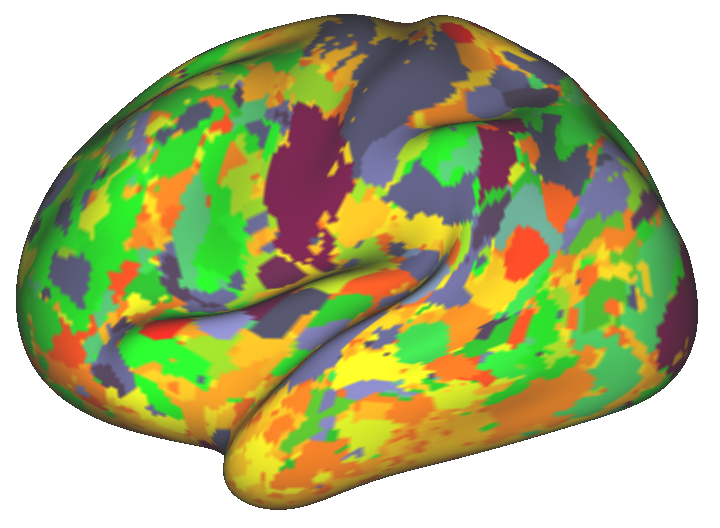} &
\includegraphics[width=0.2\textwidth]{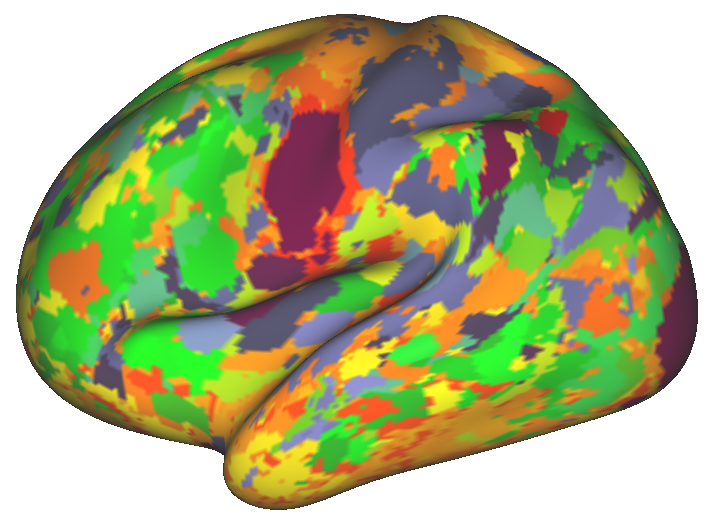} & 
\includegraphics[width=0.2\textwidth]{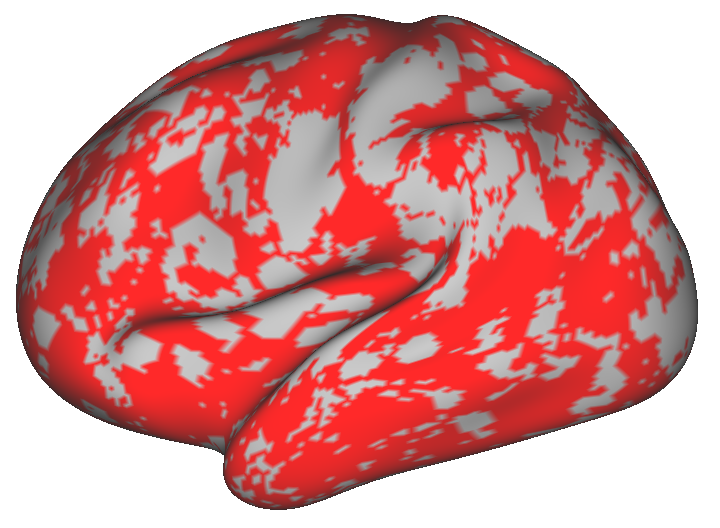} &
\includegraphics[width=0.2\textwidth]{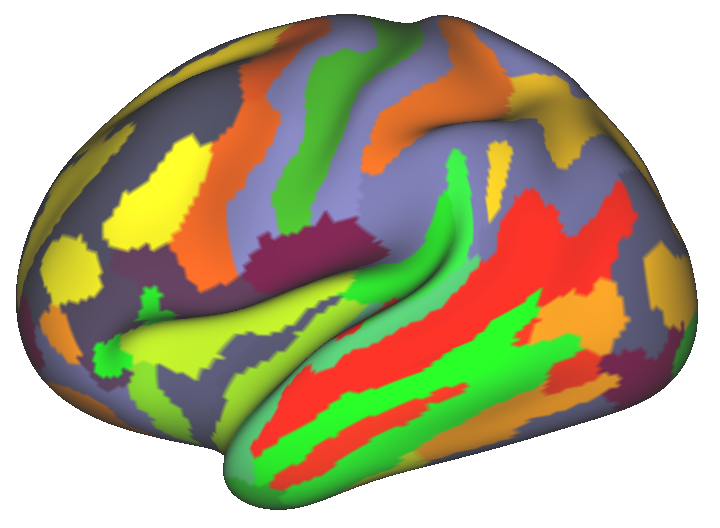}   &
\raisebox{5.0em}{\rotatebox{270}{\textit{Destrieux}}} \\

\raisebox{1.6em}{\rotatebox{90}{\textit{N-Cuts}}} &
\includegraphics[width=0.2\textwidth]{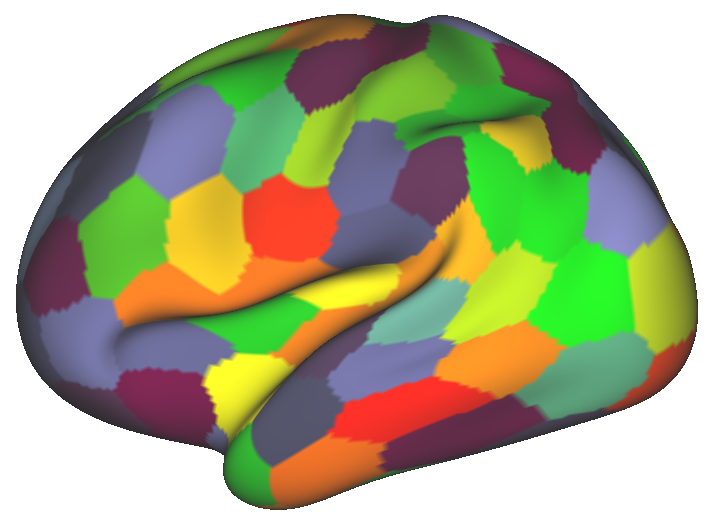} &
\includegraphics[width=0.2\textwidth]{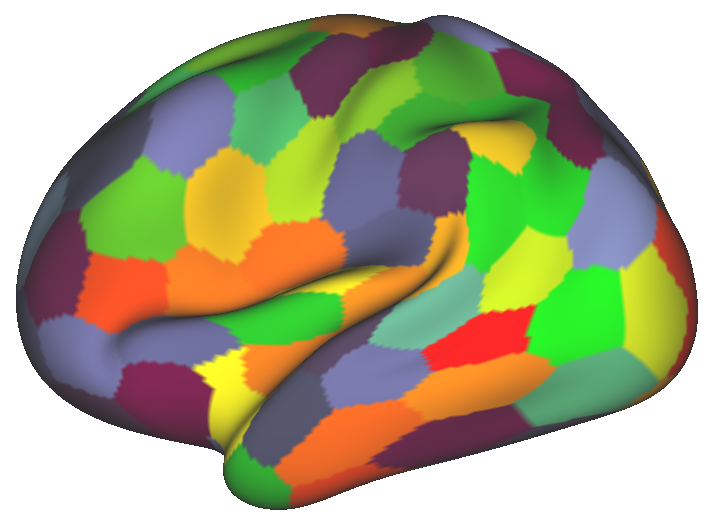} & 
\includegraphics[width=0.2\textwidth]{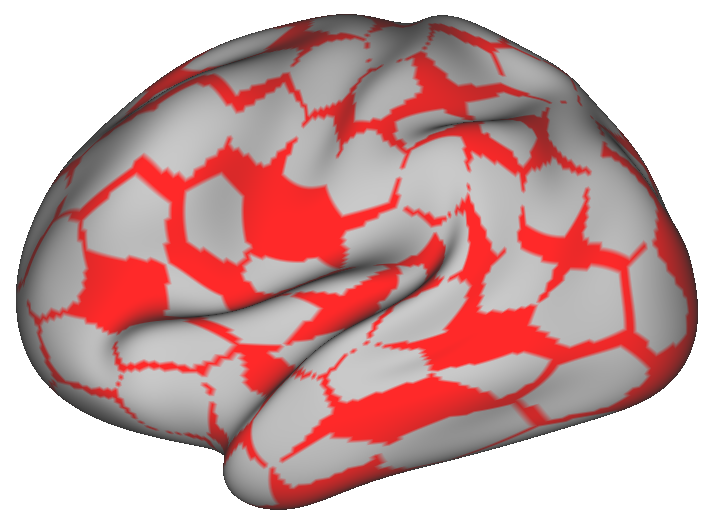} \\

\raisebox{0.8em}{\rotatebox{90}{\textit{Geometric}}} &
\includegraphics[width=0.2\textwidth]{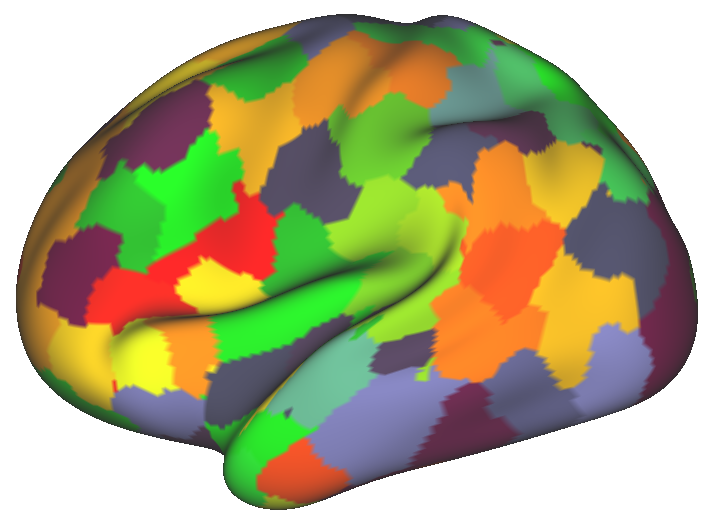} &
\includegraphics[width=0.2\textwidth]{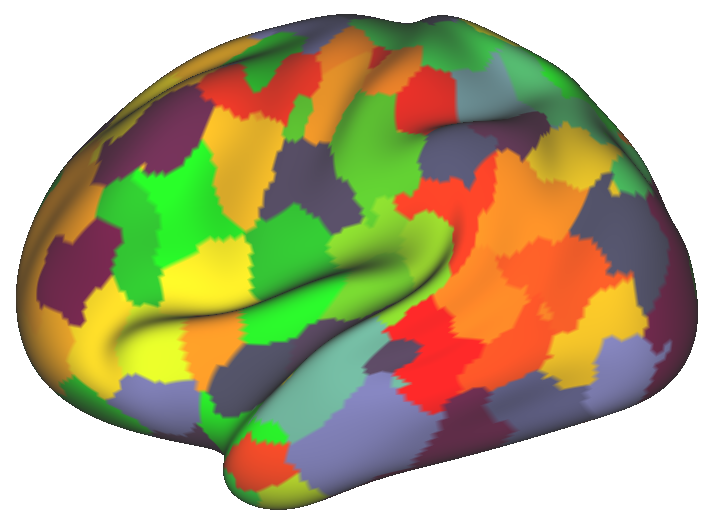} & 
\includegraphics[width=0.2\textwidth]{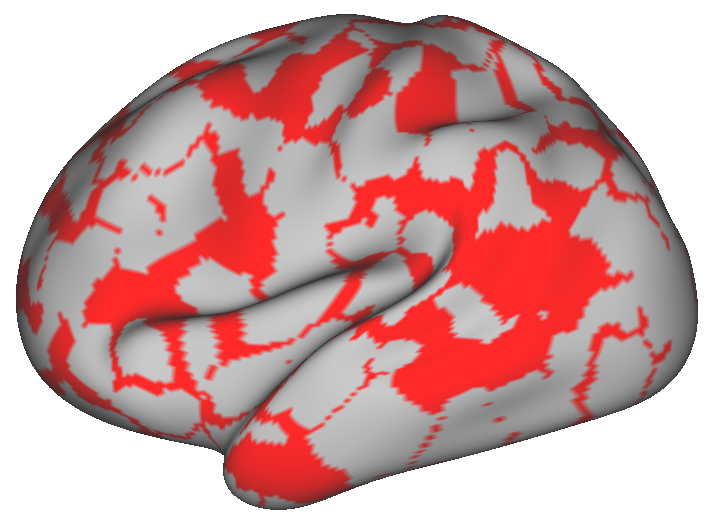} \\

\end{tabular}
\caption[Parcellations of the left lateral cortex derived from one subject for 100 regions.]{Parcellations of the lateral cortex of the left hemisphere derived from one subject for \textit{K} = 100 regions using different approaches. The parcellations in the first and second column are obtained from different scans of the same subject in order to evaluate scan-to-scan reproducibility. Parcel colours are matched for better visualization and easier comparison. The third column shows the non-matching regions. The last column shows the two anatomical parcellations, \textit{Desikan} and \textit{Destrieux}, with 35 and 75 regions, respectively and one random parcellation with the same resolution as the connectivity-driven parcellations.}
\label{fig:subject-visual}
\end{figure}

\begin{figure}[!tbh]  
\centering
\begin{tabular}{ll}
\raisebox{0.1\height}{\includegraphics[width=0.45\textwidth]{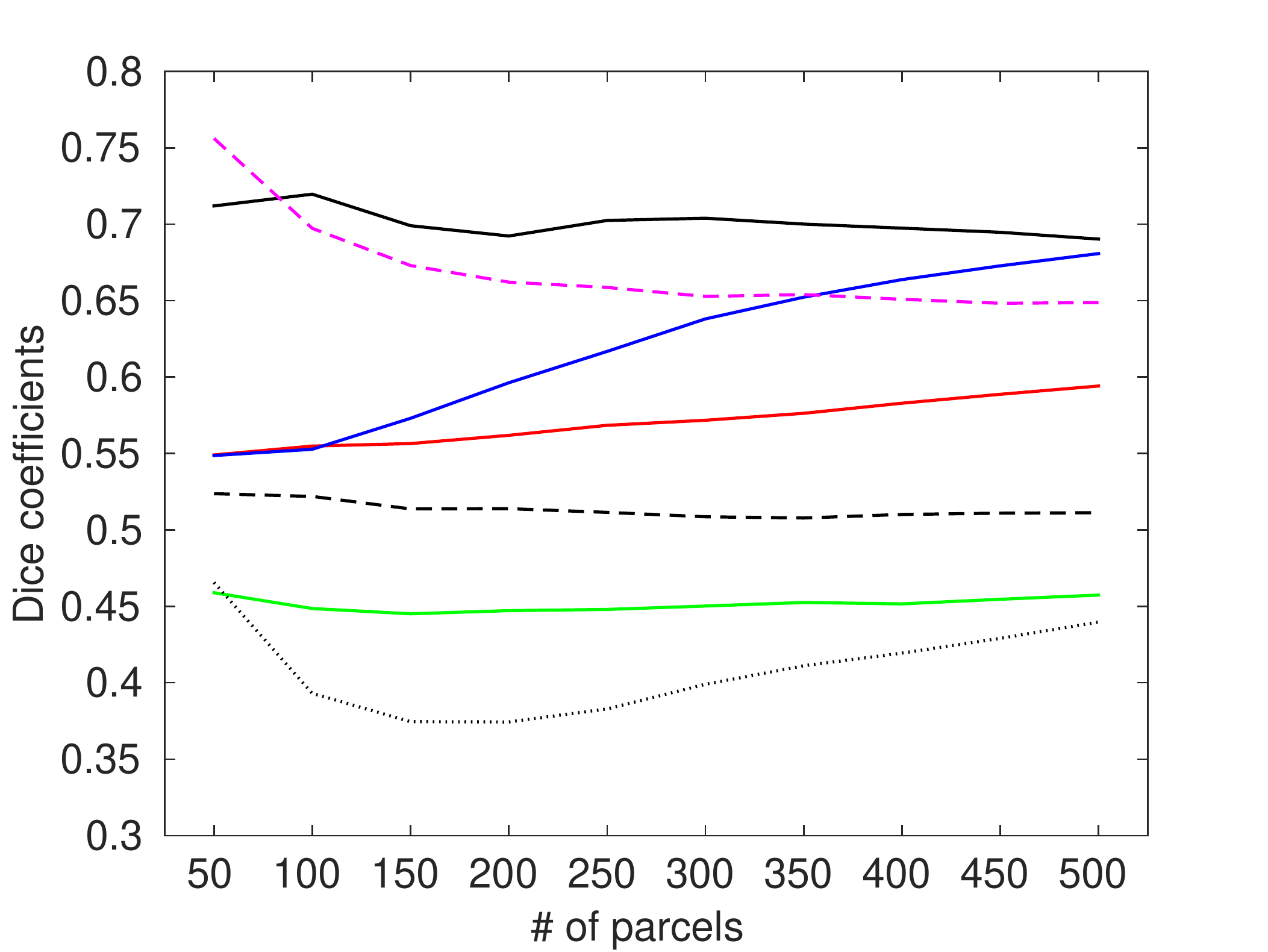}} \kern-2.5em & 
\raisebox{0\height}{\includegraphics[width=0.55\textwidth]{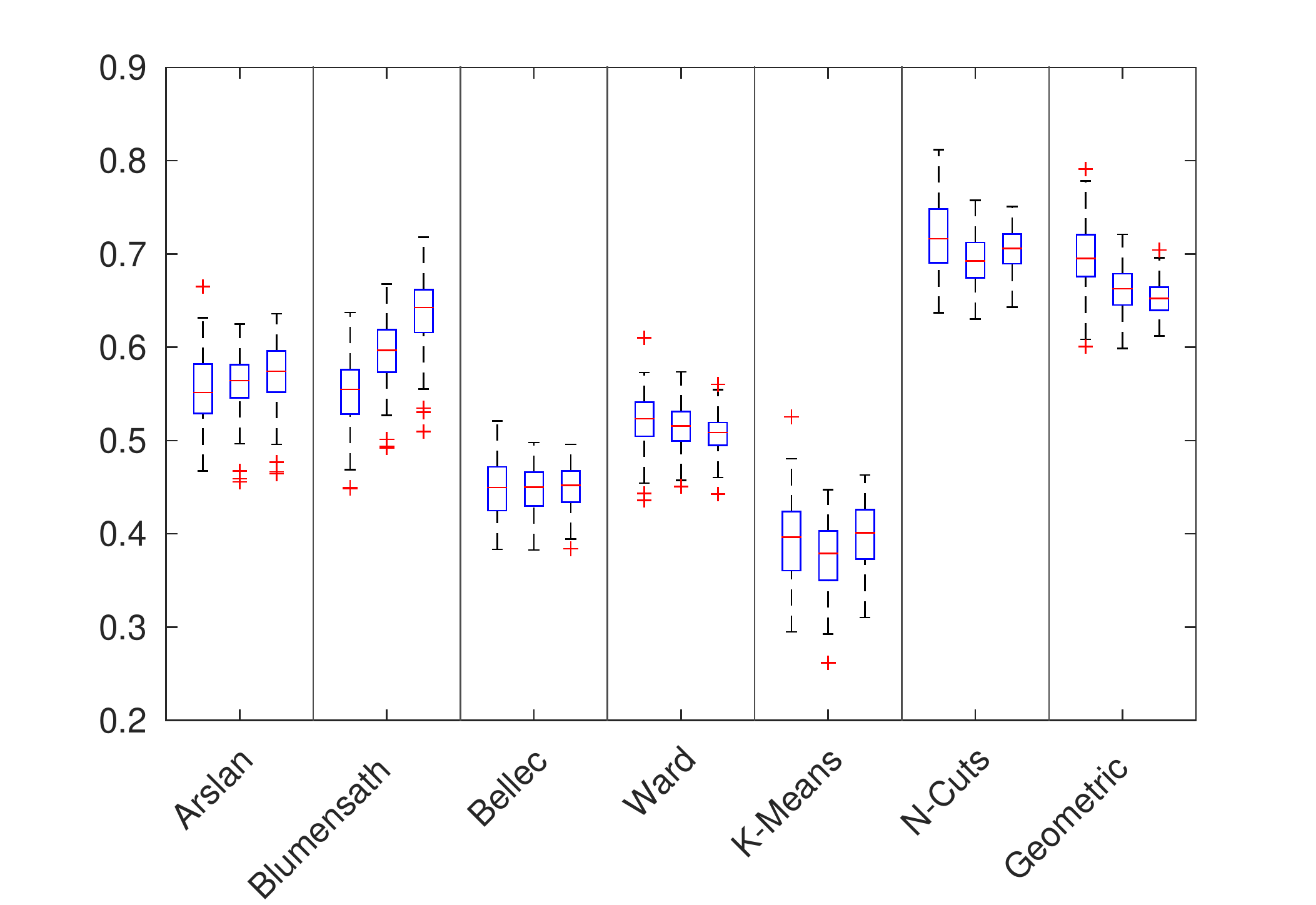}} \\

\raisebox{0.1\height}{\includegraphics[width=0.45\textwidth]{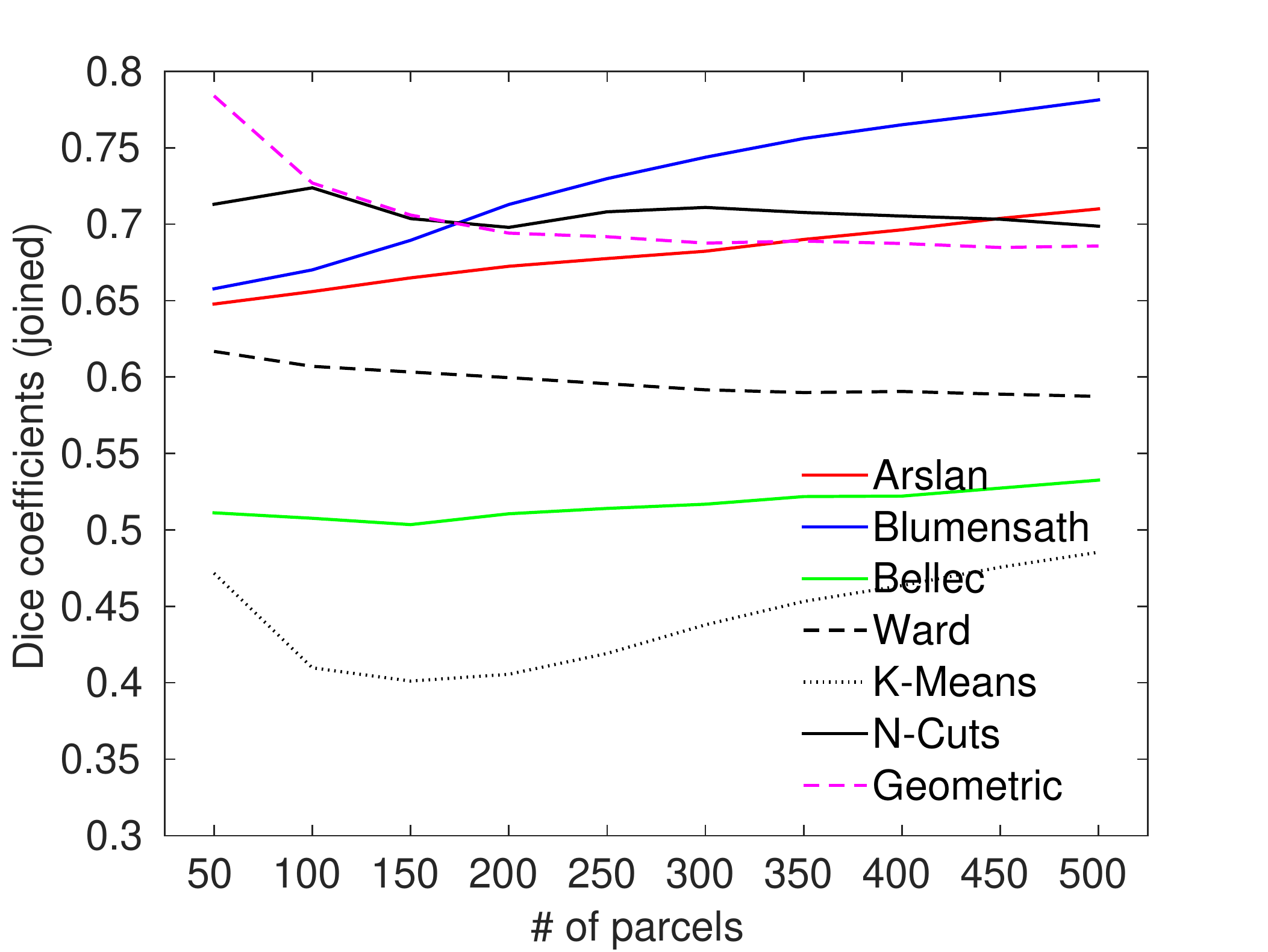}} \kern-2.5em & 
\raisebox{0\height}{\includegraphics[width=0.55\textwidth]{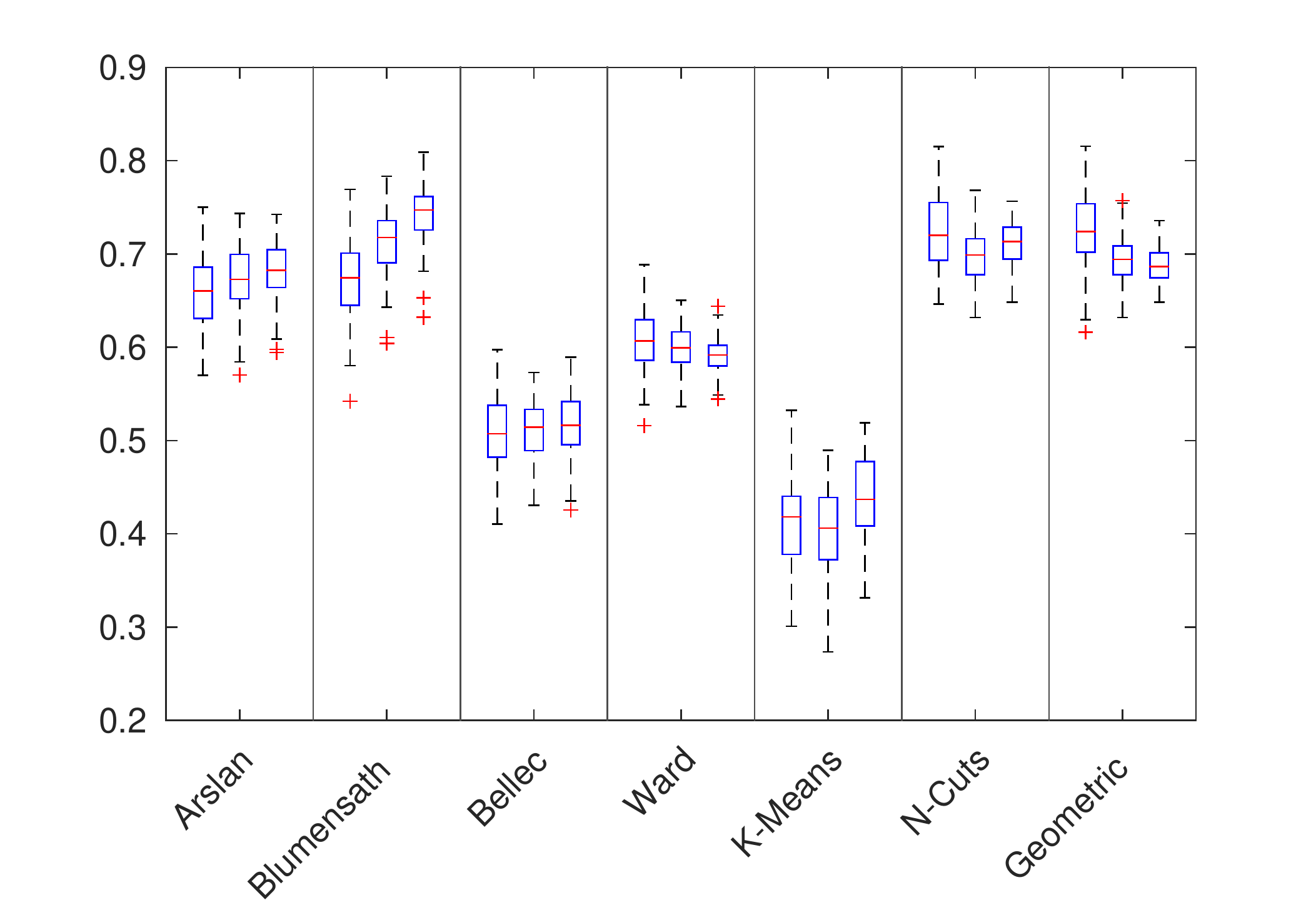}} \\

\raisebox{0.1\height}{\includegraphics[width=0.45\textwidth]{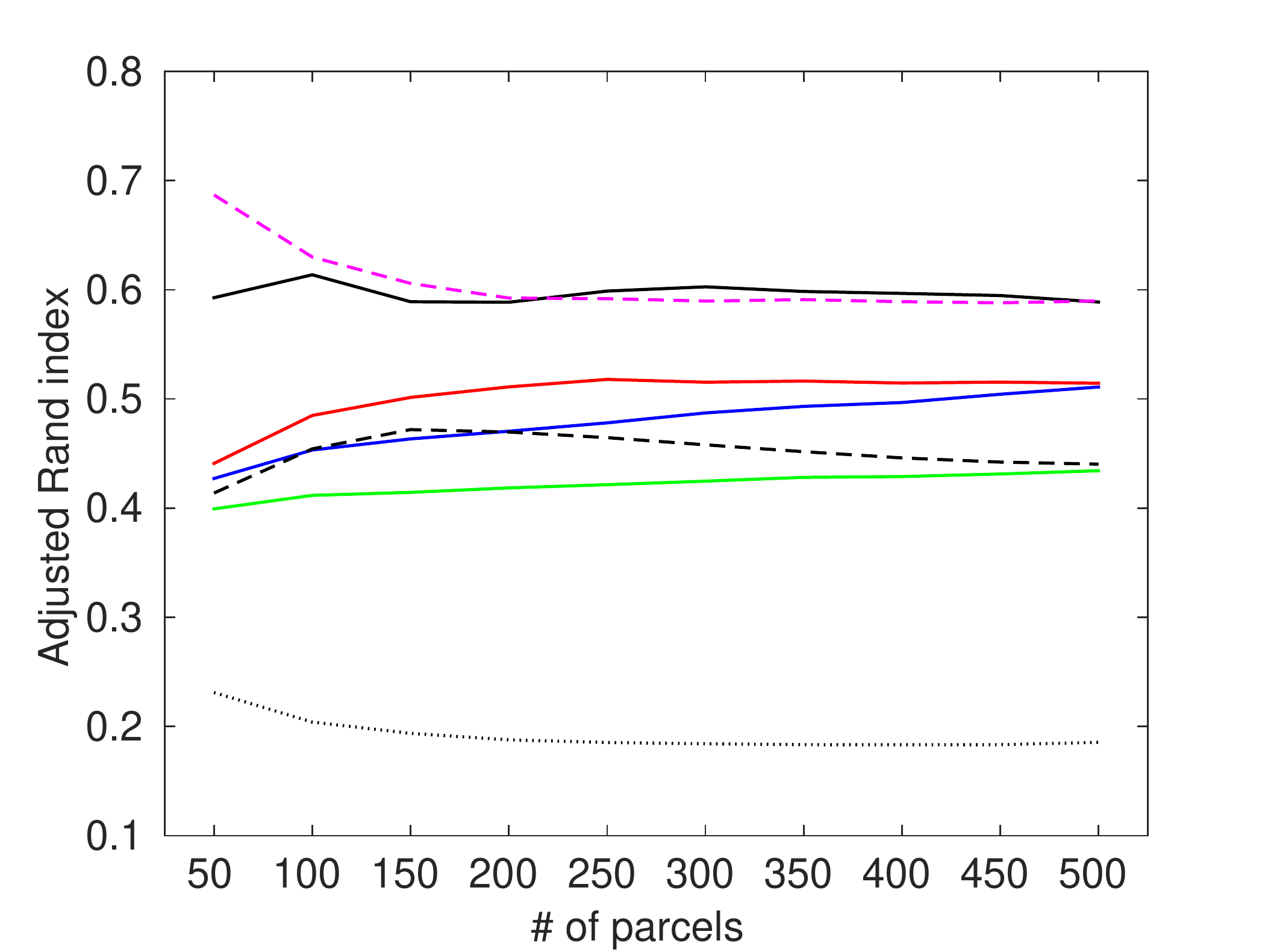}} \kern-2.5em & 
\raisebox{0\height}{\includegraphics[width=0.55\textwidth]{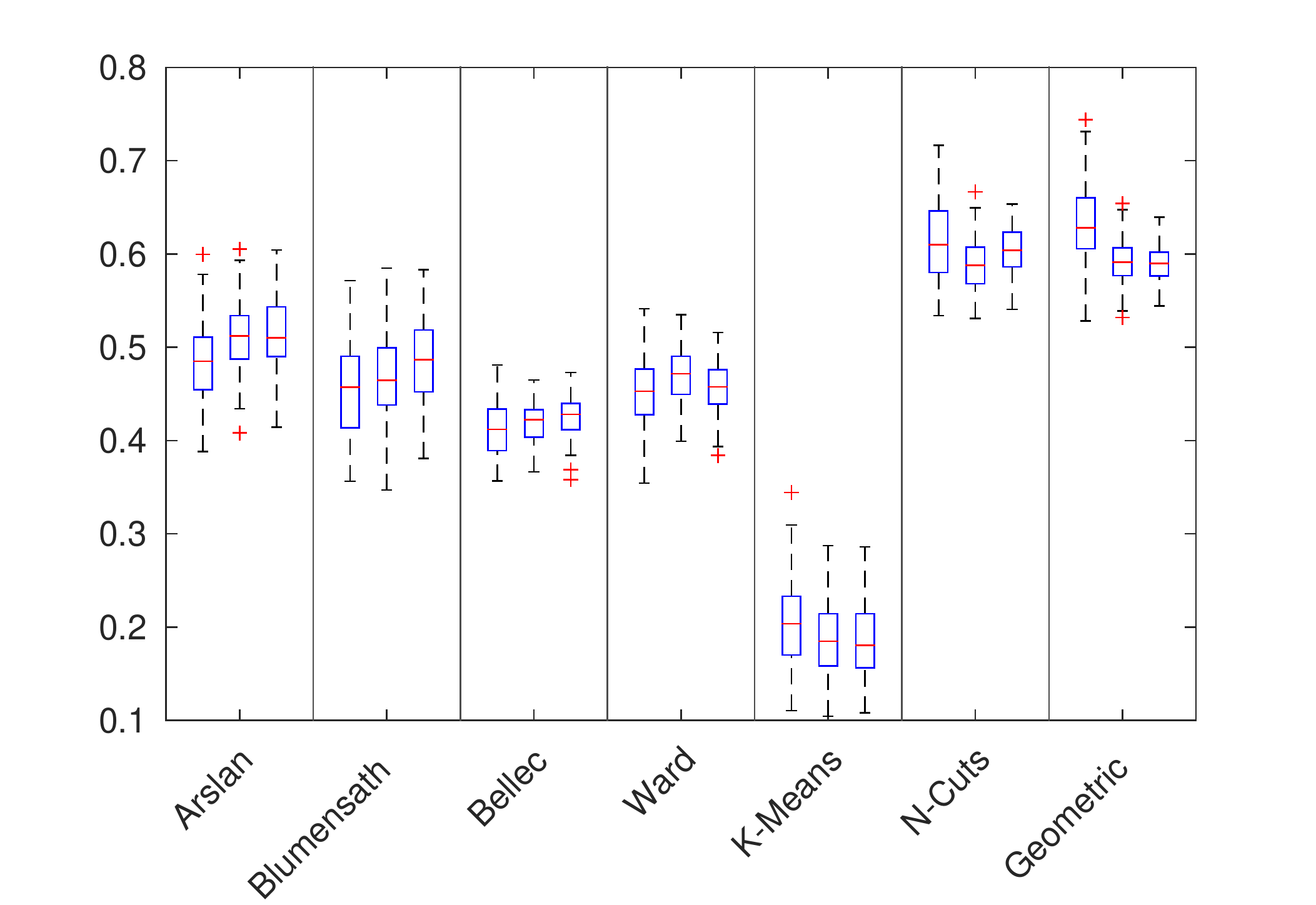}} \\

\end{tabular} 
\caption[Subject-level reproducibility results.]{Subject-level reproducibility results. \textit{Left}: Average reproducibility values obtained using Dice coefficient (top), joined Dice coefficient (middle), and adjusted Rand index (bottom). \textit{Right}: Box plots indicate the reproducibility distribution across subjects for 100, 200, and 300 regions, from left to right, for each method.}
\label{fig:subject-reproducibility}
\end{figure}

\begin{figure}[!t]  
\centering
\begin{tabular}{ll}
\raisebox{0.1\height}{\includegraphics[width=0.45\textwidth]{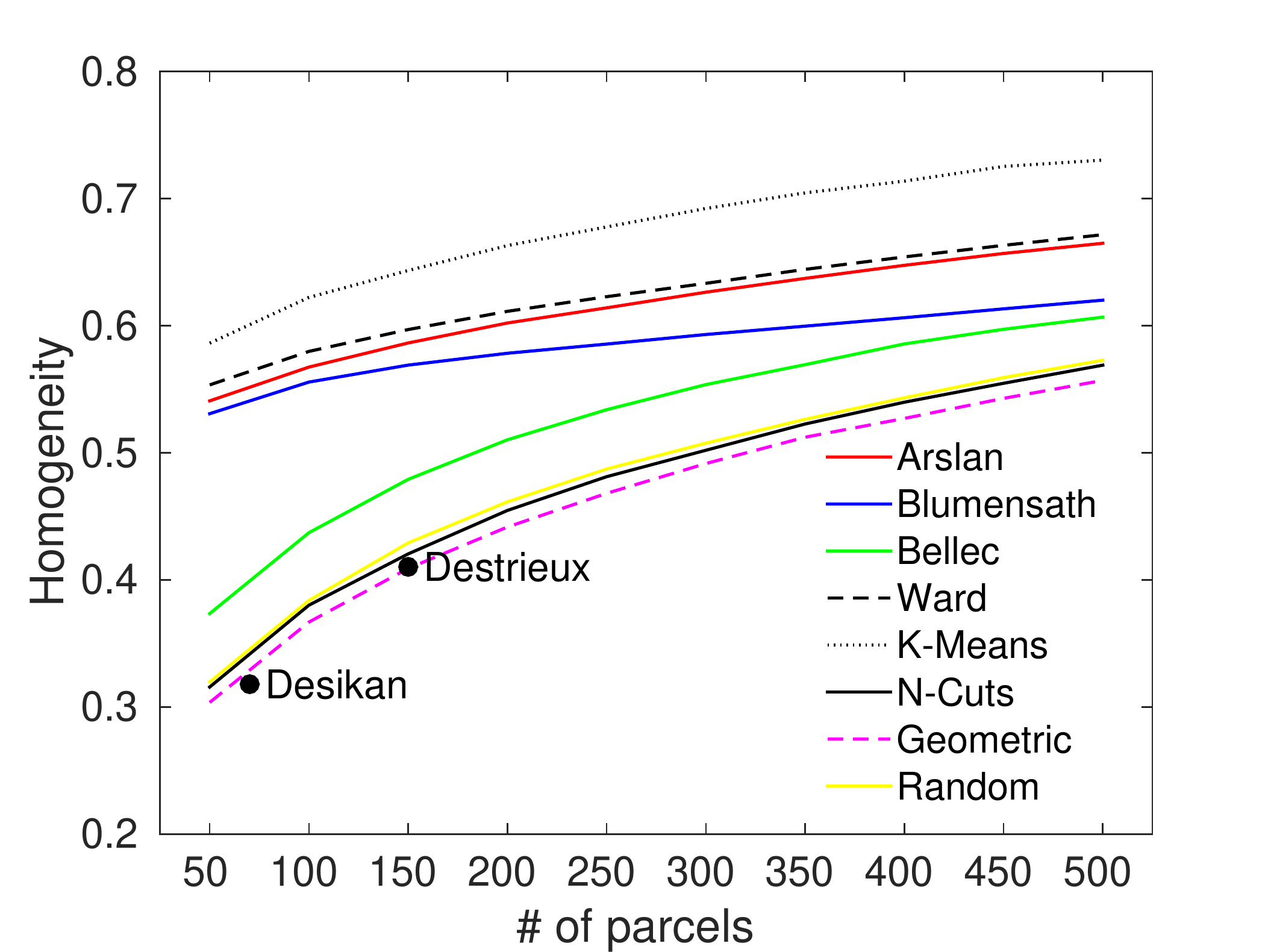}} \kern-2.5em & 
\raisebox{0\height}{\includegraphics[width=0.55\textwidth]{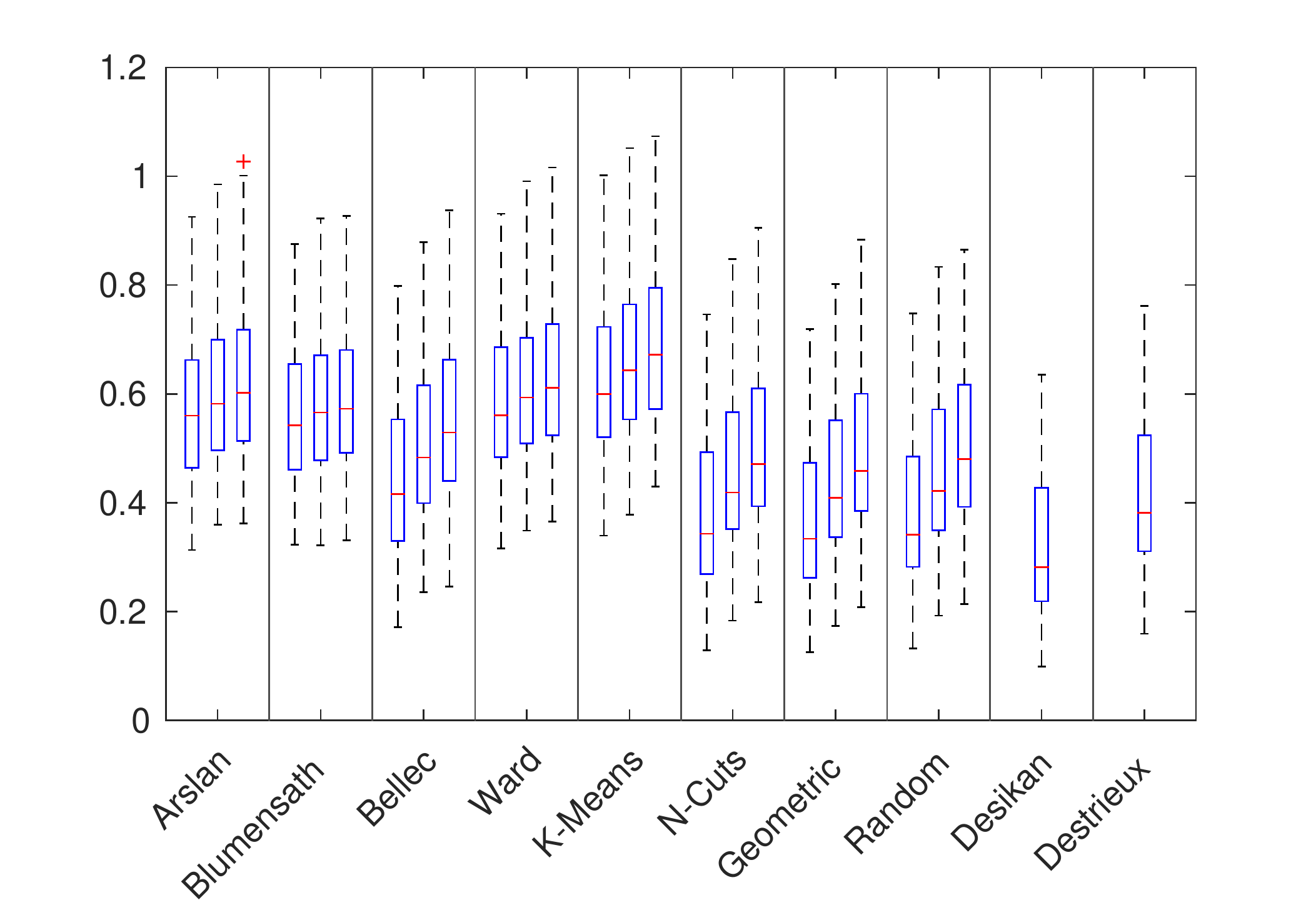}} 
\end{tabular} 
\caption[Subject-level homogeneity results.]{Subject-level homogeneity results. \textit{Left}: Lines show homogeneity values for all resolutions, averaged across subjects, whereas black dots correspond to the average homogeneity obtained from the \textit{Desikan} and \textit{Destrieux} atlases, at a fixed resolution of 70 and 150 parcels, respectively. \textit{Right:} Box plots indicate the homogeneity distribution across subjects for K = 100, 200, and 300 parcels, from left to right for each computed method.}
\label{fig:subject-homogeneity}
\end{figure}

\begin{figure}[!h]  
\centering
\begin{tabular}{ll}
\raisebox{0.1\height}{\includegraphics[width=0.45\textwidth]{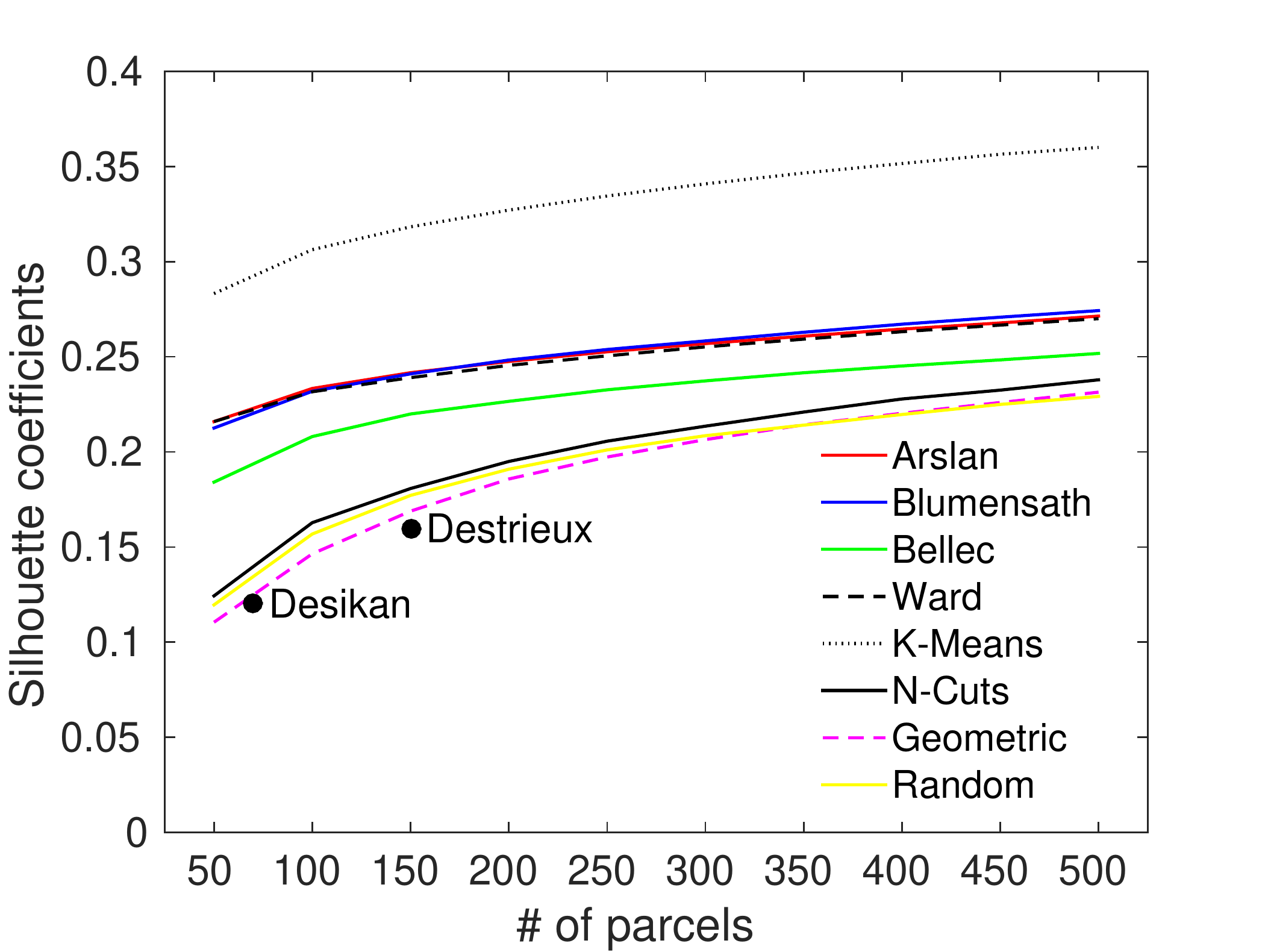}} \kern-2.5em & 
\raisebox{0\height}{\includegraphics[width=0.55\textwidth]{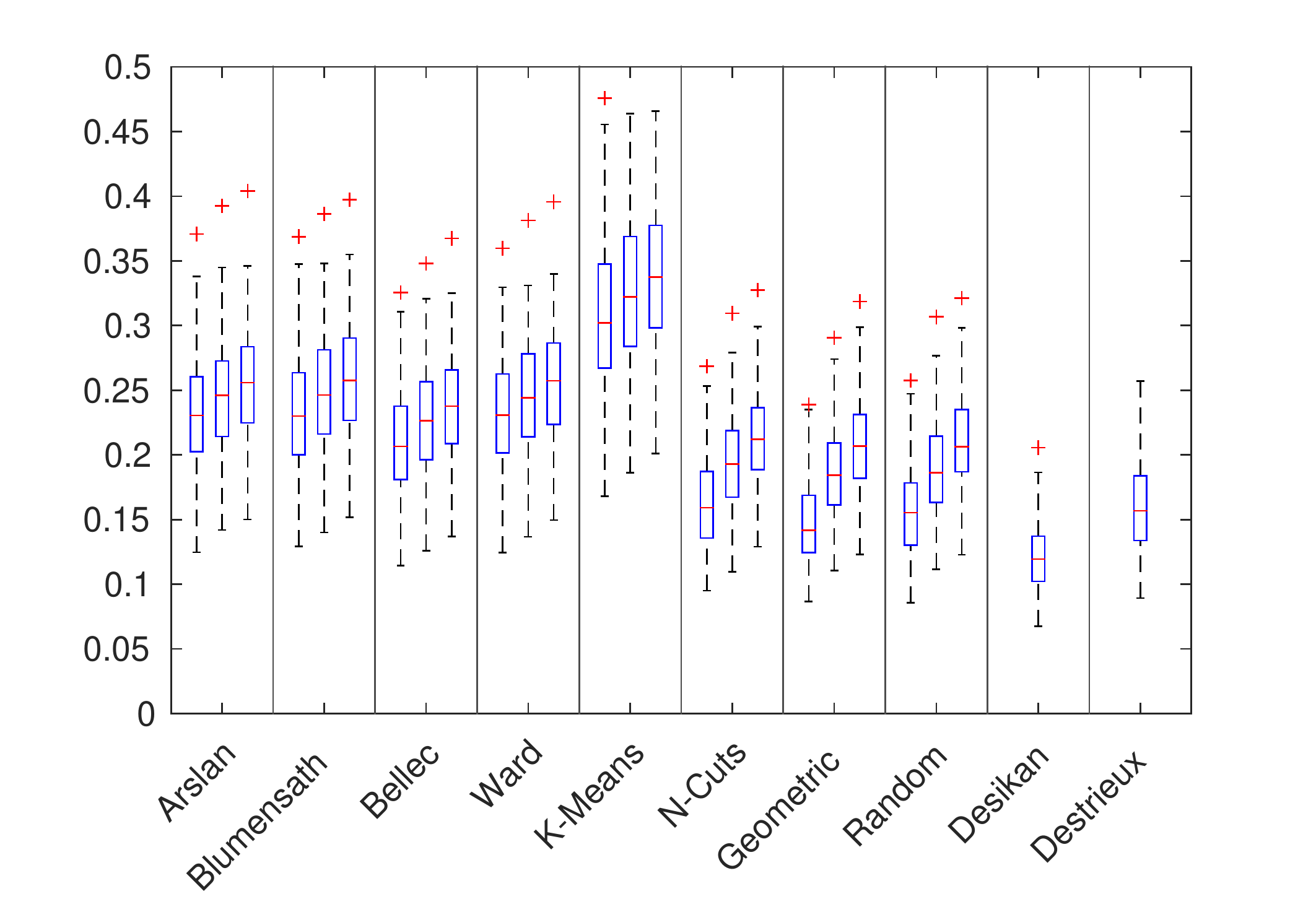}} 
\end{tabular} 
\caption[Subject-level Silhouette analysis results.]{Subject-level Silhouette analysis results. \textit{Left}: Lines show Silhouette coefficients (SC) for all resolutions, averaged across subjects, whereas black dots correspond to the average SC obtained from the Desikan and Destrieux atlases, at a fixed resolution of 70 and 150 parcels, respectively. \textit{Right:} Box plots indicate the SC distribution across subjects for K = 100, 200, and 300 parcels, from left to right for each computed method.}
\label{fig:subject-silhouette}
\end{figure}

We next provide the evaluation results for the second parcellation stage (i.e. subject-level parcellation). For ease of comparison between different methods, we report average evaluation measures in the form of line graphs for all computed resolutions. In order to show the variability across individuals we provide box plots alongside the line graphs, but only for a subset of granularity levels (i.e. for 100, 200, and 300 regions). Parcellations of the lateral cortex of the left hemisphere derived from one subject for a resolution of \textit{K} = 100 regions are given in Fig.~\ref{fig:subject-visual} for visual inspection. Reproducibility results are provided in Fig.~\ref{fig:subject-reproducibility}. Cluster validity results, including homogeneity values and Silhouette coefficients, are presented in Fig.~\ref{fig:subject-homogeneity} and~\ref{fig:subject-silhouette}, respectively. 

Reproducibility results computed by the Dice coefficient and ARI indicate that \textit{Geometric} and \textit{N-Cuts} yield the most reproducible results. Although \textit{Geometric} shows a better performance than \textit{N-Cuts} at relatively low resolutions, this trend shifts towards the favour of \textit{N-Cuts} at higher resolutions. The performance of \textit{N-Cuts} can be explained by the hard spatial constraints imposed to the adjacency matrices that drive the spectral clustering algorithm, which promotes uniformly-sized parcels~\cite{Craddock12,Blumensath13}. Obtaining highly reproducible parcellations for \textit{Geometric} is also expected, as the parcellations of a subject are generated from the same set of spatial coordinates.

Hierarchical clustering applied directly to the underlying BOLD timeseries (\textit{Ward}) shows a poor performance, in particular when compared to the other two hierarchical methods derived from an initial finer parcellation, namely, \textit{Arslan} and \textit{Blumensath}. It is worth noting that Dice overlap measurements indicate a more favourable performance for \textit{Blumensath} with respect to \textit{Arslan}, while a reverse trend is observed in ARI. This can be attributed to the fact that, \textit{Blumensath} parcellations rely on a set of cluster centroids that are learned from the underlying data and are not updated during the region growing process. This inherent bias is propagated to the second level and yields higher Dice scores between pre-matched parcels. On the other hand, higher ARIs indicate a better reproducibility on a cortex-wise scale. In general, results obtained from these two-level approaches suggest that methods initialised with a finer parcellation may be more robust, which can be linked to the fact that the impact of noise is reduced by the initialisation scheme. \textit{Bellec} generally yields low reproducibility scores, but still shows a better performance than \textit{K-Means}. Nevertheless, it should be noted that this method is originally developed to obtain parcellations with much finer resolutions (over 1000 regions per hemisphere)~\cite{Bellec06}, hence, it may not be adapted to this range of resolutions. \textit{K-Means} unsurprisingly yields the least reproducible parcellations, most likely due to the high degree of spatial discontiguity inherent in parcellations obtained via this clustering technique, which is purely driven by the affinity of the underlying BOLD timeseries. Spatially disjoint regions in \textit{K-Means} parcellations can be seen in Fig.~\ref{fig:subject-visual}.

As expected, the Dice coefficient is strongly increased by merging subdivided regions. In particular, this process yields more favourable results for the methods based on hierarchical clustering, i.e. \textit{Ward}, \textit{Arslan} and \textit{Blumensath}, for which the improvement is up to $15\%$. \textit{Blumensath} even surpasses \textit{N-Cuts} and \textit{Geometric} for resolutions with more than 150 parcels, becoming the top performing method regarding reproducibility. Other approaches tend to have a less significant improvement, mostly at a rate of $5-8\%$, while \textit{N-Cuts} and \textit{Geometric} are minimally affected. This trend can be attributed to the fact that hierarchical clustering subdivides the cortex with a bottom-up process, where boundaries derived at lower resolutions are propagated to higher levels. Joining over-parcellated regions may therefore increase the similarity between parcellations that subdivide the same regions at different levels of the hierarchical clustering tree.

Cluster validity measurements show a clear tendency in favour of connectivity-driven approaches. The most prominent trend is that, regardless of the parcellation resolution, \textit{K-Means} outperforms all other methods in terms of both homogeneity (Fig.~\ref{fig:subject-homogeneity}) and Silhouette analysis (Fig.~\ref{fig:subject-silhouette}). This would indicate that \textit{K-Means} generates the best clustering of the underlying data. It is followed by the hierarchical approaches, each of which performs almost equally regarding Silhouette coefficients, while \textit{Ward} is the best with respect to homogeneity. In particular, \textit{Arslan} consistently generates more homogeneous parcellations than \textit{Blumensath}, which might be attributed to the different techniques used by each method for computing an initial parcellation of the cerebral cortex before applying hierarchical clustering. We represent the cortical surface by functionally uniform supervertices and incorporate a flexible spatial constraint into our distance function, enabling any vertex to be assigned to a cluster if they exhibit high correlation and are spatially close (but not necessarily neighbours) as opposed to the \textit{Blumensath}'s region growing, which is based on stable seed points and a more strict definition of spatial proximity. In general, this initial stage helps obtain parcellations with a slightly higher degree of confidence than \textit{Ward}. 

Amongst the connectivity-driven parcellations, \textit{N-Cuts} shows the poorest performance. This can be due to the size bias inherent in this parcellation scheme that could limit the agreement with the underlying data. On the other hand, anatomical parcellations \textit{Desikan} and \textit{Destrieux}, yield the worst measurements and are surpassed by \textit{Geometric} and \textit{Random}. This might suggest that anatomical information alone does not allow to map the brain's functional organisation.

All methods show a performance increasing with the number of parcels computed. This is linked to the fact that evaluation measurements depend on the size of the parcels (e.g. smaller parcels yield better results). It should be noted that this trend may benefit the \textit{K-Means} parcellations, which comprise many small discontinuous parcels. 

Another important observation is the higher inter-subject variability of homogeneity and Silhouette analysis results compared to reproducibility. While one can infer that cluster validity measures are more sensitive than Dice coefficients, this could also be attributed to the fact that reproducibility measures the spatial similarity of parcellations that have been registered onto the same standard cortical surface; as a result, an inherent alignment already exists across subjects. This yields a lower inter-subject variability, especially for the spatially constrained methods and with respect to increasing resolution. On the other hand, functional organisation of the brain as estimated by RSFC can dramatically change from one subject to the next and even between different acquisitions of the same subject. Combining this with the impact of low SNR inherent to rs-fMRI, it may not be possible to parcellate all subjects with high homogeneity and/or confidence. This can be a critical point for consideration, for example, when a group-level study is devised. 

\subsection{Inter-Modality Assessment of the Proposed Parcellations }
While the previously given measurements assess the performance from a clustering point of view, the consistency of the parcellation boundaries with the task-related activated regions as well as other neuro-biological properties, such as cytoarchitecture and myelination, constitute another important aspect of evaluation. It provides a means of assessing the reliability of functional parcellations to represent the cortical organisation of the brain as delineated by other features.

\begin{figure}[!b]
\centering
\begin{tabular}{ccl}
\includegraphics[width=0.2\textwidth]{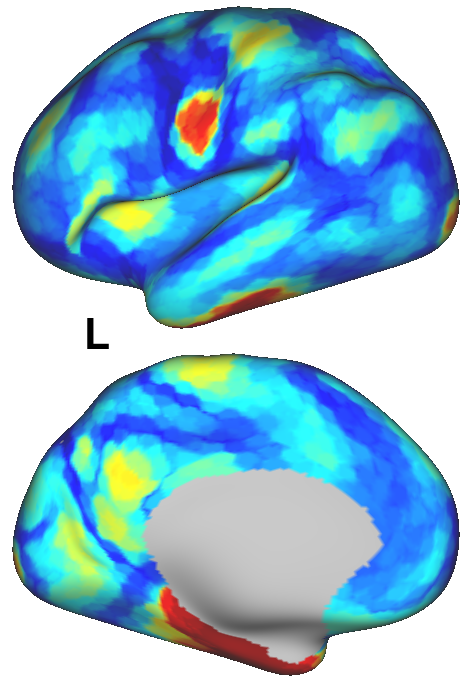}        & 
\includegraphics[width=0.2\textwidth]{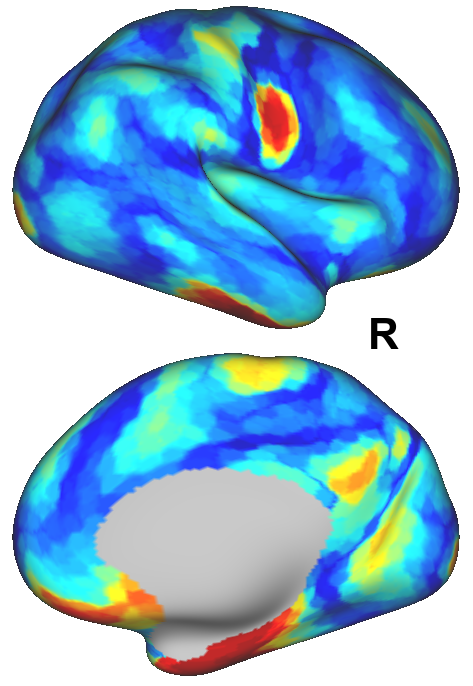}        & 
\multicolumn{1}{c}{\includegraphics[width=0.04\textwidth]{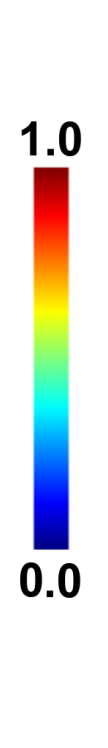}} \\ 
\end{tabular}
\caption[Inter-subject consistency maps obtained from all subjects.]{Inter-subject consistency maps obtained from all subjects in the dataset for a granularity of 100 regions. Colour maps are normalised to $[0,1]$ for better visualisation. Hotter colours indicate higher consistency between parcels across subject-level parcellations.}
\label{fig:ml_inter_subject}
\end{figure}

\begin{figure}[!t]
\centering
\begin{tabular}{lcccc}

 & Parcellation & MOTOR & MOTOR & RELATIONAL \\
 & $K=100$ & RH-AVG & T-AVG & MATCH \\

\\
\raisebox{4.0em}{\rotatebox{90}{\textit{Subject - 1}}} &
\includegraphics[width=0.2\textwidth]{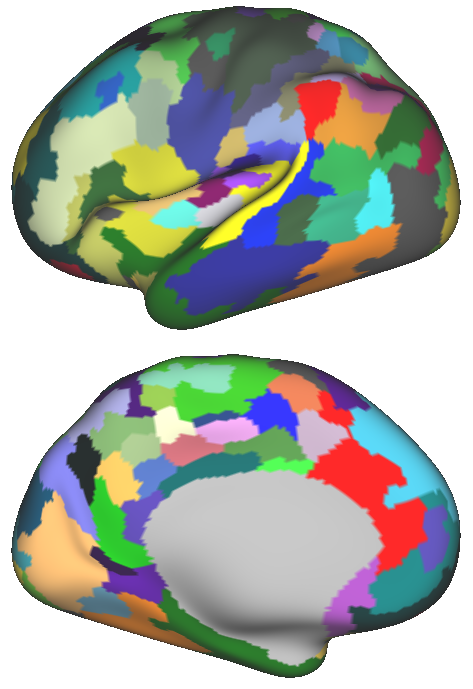} &
\includegraphics[width=0.2\textwidth]{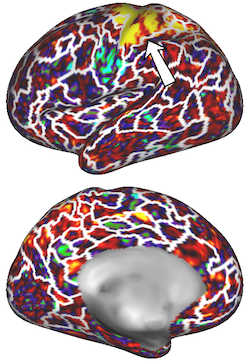} &
\includegraphics[width=0.2\textwidth]{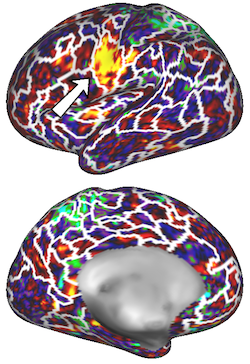} & 
\includegraphics[width=0.2\textwidth]{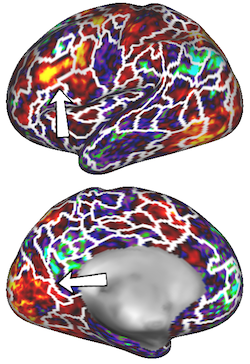} \\

\\
\raisebox{4.0em}{\rotatebox{90}{\textit{Subject - 2}}} &
\includegraphics[width=0.2\textwidth]{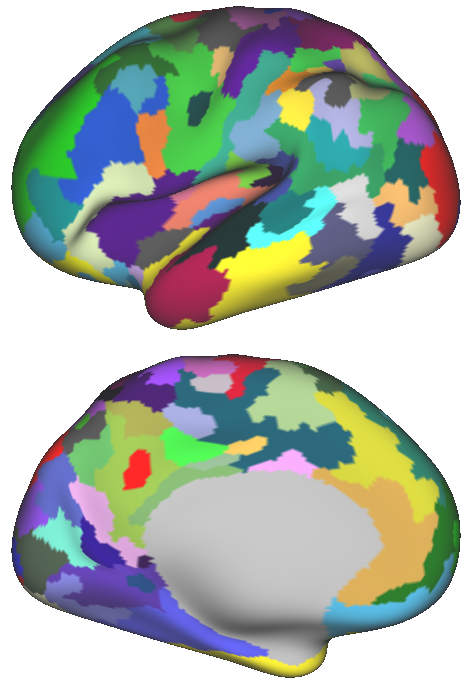} &
\includegraphics[width=0.2\textwidth]{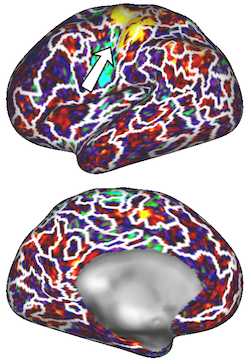} &
\includegraphics[width=0.2\textwidth]{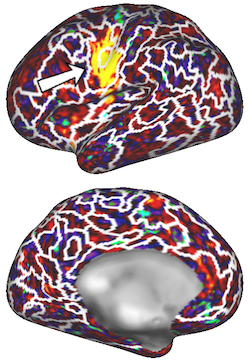} & 
\includegraphics[width=0.2\textwidth]{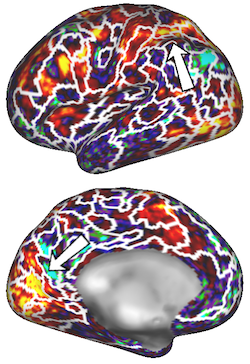} \\

\\
\raisebox{4.0em}{\rotatebox{90}{\textit{Subject - 3}}} &
\includegraphics[width=0.2\textwidth]{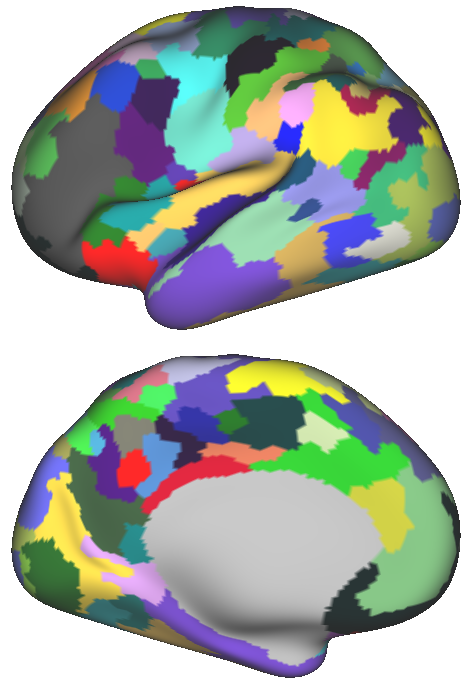} &
\includegraphics[width=0.2\textwidth]{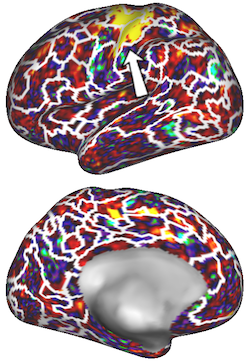} &
\includegraphics[width=0.2\textwidth]{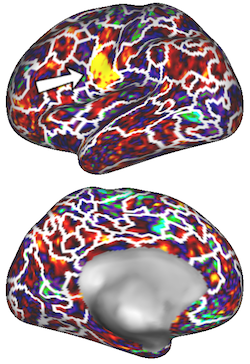} & 
\includegraphics[width=0.2\textwidth]{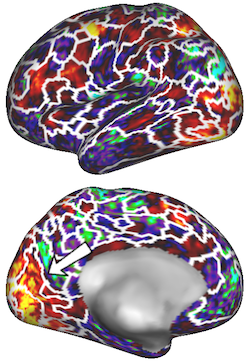} \\

\end{tabular}
\caption[Parcellation boundaries compared to task fMRI activation maps.]{Parcellations of the left hemisphere obtained from three different subjects using the proposed approach (\textit{K} = 100). Parcellation borders are superimposed onto the subject-specific task activation maps represented in terms of $z$ scores. Hot colours indicate activation (yellow being higher than red) and white arrows show aligning borders with the activated areas.}
\label{fig:visual-task-1}
\end{figure}

\begin{figure}[!t]
\centering
\begin{tabular}{lcccc}

 & Parcellation & LANGUGAGE & GAMBLING & SOCIAL \\
 & $K=100$ & STORY-MATH & PUNISH & RANDOM \\

\\
\raisebox{4.0em}{\rotatebox{90}{\textit{Subject-1}}} &
\includegraphics[width=0.2\textwidth]{figs/multi-level-100307_parcels_K100.png} &
\includegraphics[width=0.2\textwidth]{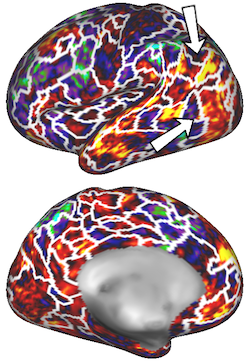} &
\includegraphics[width=0.2\textwidth]{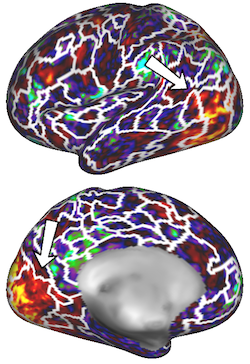} & 
\includegraphics[width=0.2\textwidth]{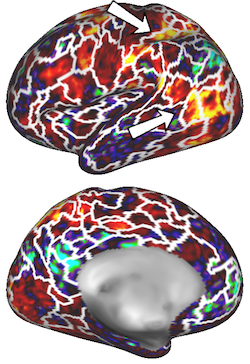} \\

\\
\raisebox{4.0em}{\rotatebox{90}{\textit{Subject-2}}} &
\includegraphics[width=0.2\textwidth]{figs/multi-level-100408_parcels_K100.png} &
\includegraphics[width=0.2\textwidth]{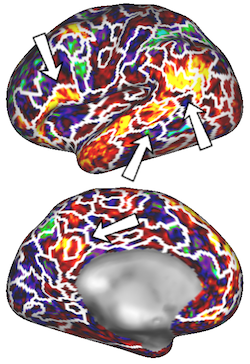} &
\includegraphics[width=0.2\textwidth]{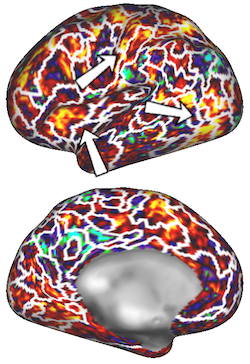} & 
\includegraphics[width=0.2\textwidth]{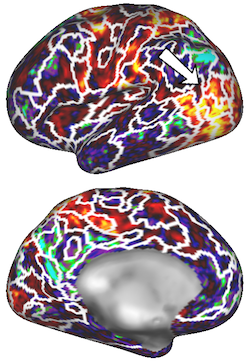} \\

\\
\raisebox{4.0em}{\rotatebox{90}{\textit{Subject-3}}} &
\includegraphics[width=0.2\textwidth]{figs/multi-level-190031_parcels_K100.png} &
\includegraphics[width=0.2\textwidth]{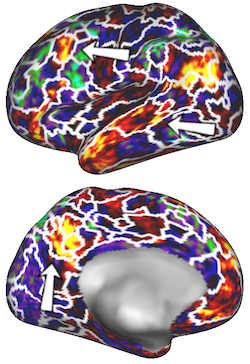} &
\includegraphics[width=0.2\textwidth]{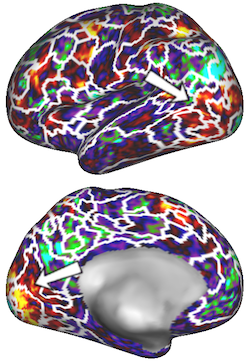} & 
\includegraphics[width=0.2\textwidth]{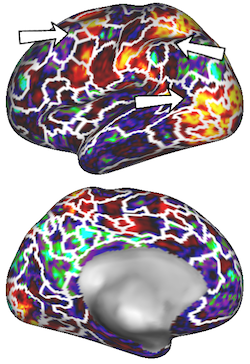} \\

\end{tabular}
\caption[Parcellation boundaries compared to task fMRI activation maps.]{Parcellations of the left hemisphere obtained from three different subjects using the proposed approach (\textit{K} = 100). Parcellation borders are superimposed onto the subject-specific task activation maps represented in terms of $z$ scores. Hot colours indicate activation (yellow being higher than red) and white arrows show aligning borders with the activated areas.}
\label{fig:visual-task-2}
\end{figure}

\begin{figure}[!t]
\centering
\begin{tabular}{lccl}
\multicolumn{1}{c}{\includegraphics[width=0.2\textwidth]{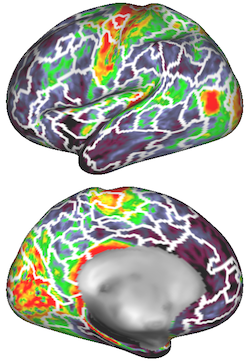}}  &   \includegraphics[width=0.2\textwidth]{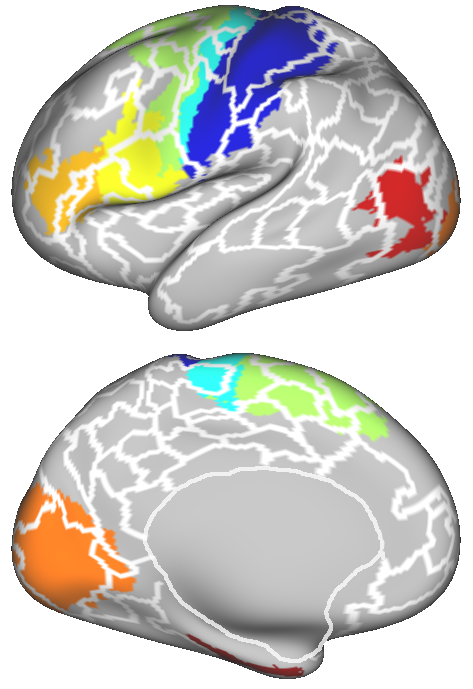}           & ~~ \includegraphics[width=0.2\textwidth]{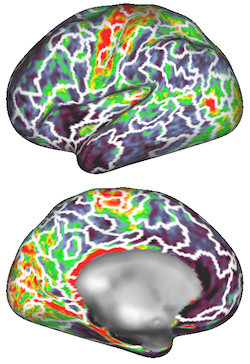}            & 
\multicolumn{1}{c}{\includegraphics[width=0.2\textwidth]{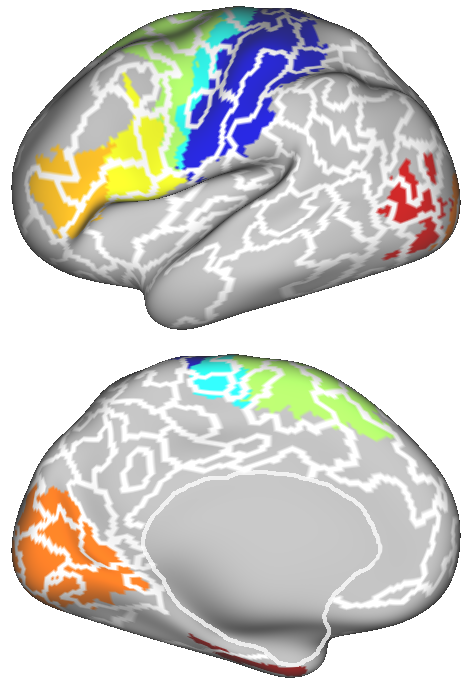}} \\
\multicolumn{2}{c}{\textit{Subject-1}}           & 
    \multicolumn{2}{c}{\textit{  Subject-2}}           \\
    \\
                        & \includegraphics[width=0.2\textwidth]{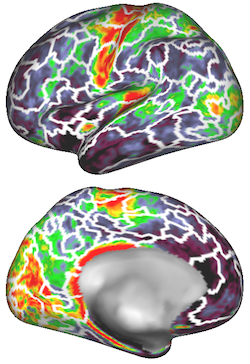}           & \includegraphics[width=0.2\textwidth]{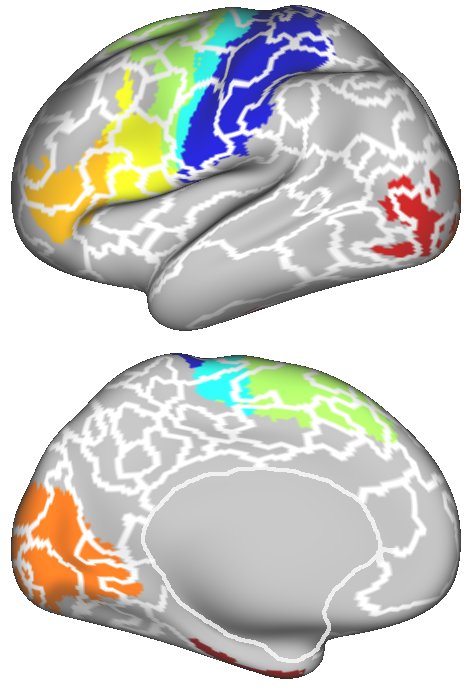}           & \includegraphics[width=0.12\textwidth]{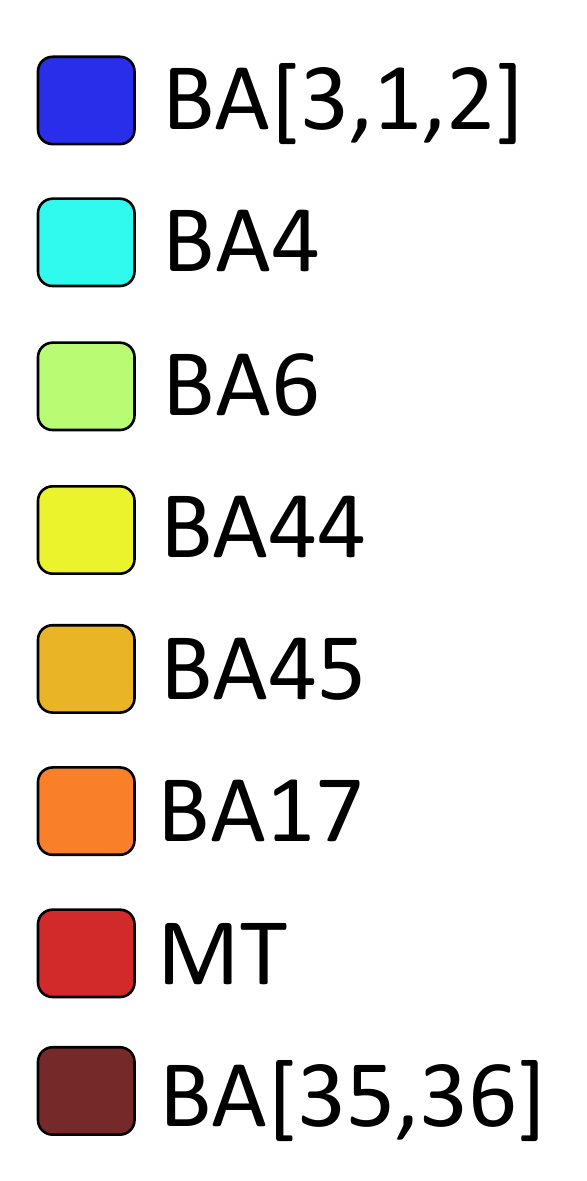}                \\
                        & \multicolumn{2}{c}{\textit{Subject-3}} &                       
\end{tabular}
\caption[Parcellation boundaries compared to myelin maps and Brodmann areas.]{Parcellations of the left hemisphere obtained from three different subjects using the proposed approach at a resolution of \textit{K} = 100 regions are overlaid onto subject-specific myelin maps and Brodmann areas.}
\label{fig:bro_myel}
\end{figure}

To this end, we present example task activation maps of three different subjects in Figs.~\ref{fig:visual-task-1} and~\ref{fig:visual-task-2}. Myelin maps and cytoarchitectonic regions of the same subjects are given in Fig.~\ref{fig:bro_myel}. The white borders in the figures show the parcellation boundaries obtained by the proposed two-level parcellation algorithm at a granularity level of $K = 100$ regions. In order to better interpret the consistency of the agreement between different modalities, we show the consistency maps obtained for the proposed parcellations at the same resolution in Fig.~\ref{fig:ml_inter_subject}. 

It appears that the proposed parcellations show a good degree of alignment with the task-activated areas and well-structured patterns of myelination. This agreement is more prominent within the primary motor, somato-sensory, and visual cortex as indicated by the arrows. In addition, the boundaries around area MT are also found to align well with the changes in the myelo-architecture of the cortex and some activation maps. Although we only present visual results for several subjects, similar results are likely to occur for other subjects, especially within the motor area, which is typically subdivided into alike regions across subjects as indicated in the inter-subject consistency maps in Fig.~\ref{fig:ml_inter_subject}. 


Our findings regarding the similarity between the rs-fMRI parcellations and sharp changes in myelin maps agree with the previous literature, where strong alignment has been observed between the highly myelinated cortical areas and resting-state fMRI gradients~\cite{Glasser11}. These observations might also indicate that the connectivity estimated at rest may truly reflect the functional organisation of the brain, especially within certain areas of the cortex. Although the functional connectivity obtained from BOLD timeseries does not necessarily match the cytoarchitecture of the cerebral cortex, some alignment is observed in Brodmann's areas BA17 (part of visual cortex) and BA[4,6] (motor cortex), for which several other studies have also reported similar observations~\cite{Blumensath13,Wig14,Gordon16}.


\section{Discussion}
\label{sec:multi-level-discussion}
In this chapter, we proposed a new two-level clustering approach to parcellate the cerebral cortex using publicly available data provided by the HCP. The generation and evaluation of the parcellations is based on RSFC, which is thought to represent the functional organisation of the brain on a whole-cortex scale. As there is no gold standard parcellation of the cerebral cortex, we evaluated the reliability of parcellations in terms of two widely-accepted criteria~\cite{Thirion14}: (1) reproducibility (i.e. stability) of a parcellation on a scan-to-scan basis, and (2) fidelity (i.e. accuracy) of a parcellation with respect to the underlying data. We tested the proposed approach with respect to different parcellation schemes and popular clustering methods on a coarse-to-fine multi-resolution basis, considering 1000 to 3000 and 50 to 500 regions per subject. We not only assessed the performance from a clustering point of view, but also showed the extent of agreement between the parcellation boundaries and functionally active regions obtained via task fMRI as well as other cortical features such as the cytoarchitectonic areas and well-structured patterns of myelination. 

We observed that data-driven parcellations in general have a much better agreement with the underlying connectivity compared to anatomical and random parcellations. It is worth noting that, when considering cluster validity measures, there typically exists a trade-off between stability and fidelity to the underlying data~\cite{abraham2013extracting}. For instance, \textit{k}-means clustering largely leads in terms of clustering accuracy due to the fact that it is solely driven by functional connectivity. It, however, shows an extremely poor performance regarding reproducibility. Contrarily, methods that are dominated by spatial constraints, such as spectral clustering, are likely to capture stable features regarding the geometry of the cortical mesh. This appears to confer a strong advantage for reproducibility, but it also constrains the parcellation task and leads to inaccurate alignment with the brain's underlying functional organisation~\cite{Eickhoff15}. 

Hierarchical clustering yields a performance that is likely to reside in-between: it offers the advantage of generating spatially joint parcels, which can help obtain more reproducible parcellations, while still capturing the functional features with high fidelity. It can even compete with spectral clustering in terms of stability when a finer resolution parcellation is initially used to smooth the underlying data. For instance, the proposed parcellation scheme achieves higher reproducibility than the typical hierarchical clustering with only limited impact on the accuracy of the parcellations. This can be linked to the supervertex clustering stage, which relies on a modified \textit{k}-means algorithm that allows injection of flexible spatial constraints into the parcellation process. As a result, it can compute a cortical parcellation that can simultaneously maximise reproducibility and accuracy to a large extent. 

Another advantage of the proposed method in the context of brain mapping emerges from the fact that it provides a hierarchy of nested regions that can be explored at different resolutions~\cite{Blumensath13,Eickhoff15}. It is not necessary to pre-specify a cluster number, as the clustering algorithm inherently provides a multi-scale segregation of the brain. Since not all parts of the cerebral cortex can be subdivided with the same precision, this would allow a fine-to-coarse delineation of cortical areas under investigation without the necessity to use a pre-defined mask. However, this may also confer a disadvantage in terms of propagating errors from one resolution to the next~\cite{Parisot15}. 

Qualitative assessment of RSFC parcellations revealed some degree of alignment with task-activated cortical regions and other neuro-anatomically defined areas. However, it is worth noting that this agreement is highly variable across subjects and RSFC-based parcels do not necessarily match an entire task-activated region or cytoarchitectonic area. Rather than, some of the parcellation borders are likely to align with the gradients in cortical maps or the boundaries of anatomically defined regions. While a direct comparison between RSFC parcellations and other modalities can be difficult to interpret, it may still provide additional confidence for a parcellation method's ability to delineate meaningful cortical segregations and assess its reliability from a neuro-biological point of view~\cite{Blumensath13}. 

Subject-level parcellations are of great importance to study the functional organisation of the brain and its variability within a population. Whereas a high reproducibility can be obtained on a scan-to-scan basis, the functional and structural differences inherent in each individual brain do not allow a subject-to-subject comparison in a similar manner. Instead, we used cross-subject comparisons to evaluate the degree of variability between individuals as shown in Fig.~\ref{fig:ml_inter_subject}. Our experiments revealed that certain areas, such as the motor cortex, can be consistently parcellated into similar regions and/or at a similar granularity; however, typically a great amount of variance exists within the rest of the cortex. 

Understanding the source of variability across subjects is an important research problem and constitutes an obstacle in itself. While a possible alteration in connectivity could be associated with a brain disorder, it could also be attributed to genetic variations~\cite{Dubois16}, topological differences between subjects~\cite{Langs15}, varying strengths of connections between brain areas in individuals~\cite{Gordon16individual}, as well as simply could be due to potential registration errors or low SNR in the underlying data. Interestingly, the fact that we obtain less variance (i.e. more similarity) within the motor and visual cortex can be linked to the HCP's cortical-folding based registration technique~\cite{Robinson14}, which promotes a more consistent registration for these areas~\cite{Parisot16a}. 

The variability across individual parcellations does not allow constructing spatial correspondence across subjects. However, population-level connectome analysis typically rely on the assumption that vertex (or voxel) level correspondence has been ensured \textit{a priori}. This can be achieved by computing a group-wise parcellation that ideally represents the shared characteristics within a population. Such a group representation can be useful in the context of connectome analysis, for instance, to identify how connectivity changes within a population through ageing. Similarly, they can be used for deriving biomarkers in order to better understand disease-related differences in the brain connectivity. In addition, the inter-subject variability problem can be approached from a different perspective: cortical regions that are most consistent and/or least similar across subjects can be located by comparing individual parcellations to a group representation obtained from the same population.

There are two typical extensions that can be attached to our method for computing group-wise parcellations: i) averaging BOLD timeseries across subjects and submitting it to the proposed parcellation framework, or ii) mounting a third-level clustering stage on top of the individual subject parcellations. Given the accuracy of the proposed parcellations at the subject level, we implemented the latter approach and evaluate its performance in Chapter 7 along with a large set of other group-wise parcellation techniques in the literature. While this chapter focused on RSFC captured in the form of BOLD timeseries for a connectivity-based delineation of the brain's cortical organisation, parcellations can also be obtained from other types of connectivity data, such as structural connectivity estimated from dMRI. Next chapter explores this phenomenon within the context of cortical brain parcellation and proposes a novel method driven by dMRI-derived structural connectivity. 
\chapter{Cortical Boundary Mapping through Manifold Learning}
\label{chapter:manifold}

This chapter is based on:

S. Arslan, S. Parisot, and D. Rueckert, \textit{Boundary Mapping through Manifold Learning for Connectivity-Based Cortical Parcellation}. International Conference on Medical Image Computing and Computer Assisted Intervention (MICCAI), vol. 9900 of LNCS. Springer, pp. 115-122, 2016.

\section*{Abstract}
\textit{This chapter proposes a new parcellation method that can be used to model the brain's cortical organisation with respect to structural connectivity estimated from dMRI. To this end, we learn a manifold from the high-dimensional local connectivity patterns of an individual subject and identify parcellation boundaries as points in this low-dimensional embedding where the connectivity patterns change. We compute spatially contiguous and non-overlapping parcels from these boundaries after projecting them back to the cortical surface. This is achieved by boundary mapping, a technique that separates cortical regions using well-known image segmentation algorithms. Our experiments with a set of 100 healthy subjects show that the proposed method can produce spatially contiguous parcels with distinct patterns of connectivity and a higher degree of homogeneity at varying resolutions compared to the state-of-the-art parcellation methods, hence can potentially provide a more reliable set of network nodes for connectome analysis.}

\section{Introduction}
Connectivity-driven parcellation (CDP) techniques have recently gained a lot of attention due to their potential to reveal the functional and structural architecture of the human brain. As these approaches are directly applied to the rs-fMRI and/or dMRI data, they can learn a somewhat optimal representation of connectivity in a lower dimensional space, and therefore, can better reflect the network organisation of the brain compared to traditional parcellations derived from anatomical landmarks or randomly obtained ROIs~\cite{Sporns11}. CDP methods tend to model the connectivity data in association with clustering algorithms~\cite{Eickhoff15}. As we have extensively reviewed in the previous chapters, many clustering approaches such as hierarchical clustering, spectral clustering (normalised-cuts being the most popular), and \textit{k}-means clustering (as well as their variants) have been repeatedly used for the parcellation of the human cerebral cortex~\cite{Thirion14}. However, despite promising results, the CDP problem is still open to improvements. This is primarily due to the facts that (1) the problem is not well-posed and (2) there is not a reference parcellation to supervise the model selection. As a result, obtaining accurate parcellations directly depends on the proposed method's fidelity to the underlying data~\cite{Parisot15} and its capacity to differentiate regions with different connectivity profiles~\cite{Eickhoff15}.

An alternative cortical parcellation approach known as boundary mapping casts the clustering problem as the identification of transitions between connectivity patterns and solves it using image segmentation techniques~\cite{Cohen08,Wig13,Wig14,Gordon16}. This technique has been successfully used to derive areal parcellations of the cortical surface from resting-state functional connectivity (RSFC)~\cite{Gordon16}. The key idea emerges from the fact that RSFC patterns are likely to show rapid changes between different cortical locations in the human brain, similarly to the changes in connectivity that segregate cortical areas in non-human primates~\cite{felleman1991distributed}. The locations where connectivity is in transition are therefore assumed to represent the parcellation boundaries in the human cerebral cortex~\cite{Cohen08,Gordon16}. Although boundary mapping has been extensively used to derive local or whole-brain cortical parcellations from RSFC~\cite{Wig13,Wig14,Gordon16}, to the best of our knowledge, it has not been applied to structural connectivity estimated from dMRI. In contrast to the indirect estimation of connectivity achieved with rs-fMRI, dMRI can estimate the physical white matter connections in the brain. Parcellations derived from dMRI have, therefore, a more intuitive interpretation and tend to be more robust than RSFC parcellations~\cite{Parisot16a}. 

One limitation of boundary mapping is its sensitivity to spurious spatial gradients captured from connectivity patterns. A considerable amount of data is required to obtain a robust parcellation from these gradient maps, and hence, it is typically applied to averaged-datasets of large populations~\cite{Gordon16}. This may obscure patterns of brain organisation specific to individuals~\cite{Laumann15,Gordon16individual}. It also poses an extra challenge for the method's adaptability to the individual subjects: a subject-level parcellation may only be achieved if sufficient data is collected, e.g. the subject is scanned repeatedly over time~\cite{Laumann15}. However, most neuroscience datasets typically provide at most few scans per subject/modality. Another challenge emerges from the correlation/connectivity maps used to obtain the gradients. The high dimensionality of these maps may constrain the detection of robust connectivity patterns at the subject level. To this end, a dimensionality reduction stage may be beneficial for a more robust and interpretable representation of the brain connectivity and may facilitate the use of boundary mapping on a single subject basis. 

With this motivation, we introduce a new parcellation method, in which we learn a non-linear manifold from local connectivity characteristics of an individual subject and develop an effective way of computing parcels from this manifold based on boundary mapping. Our approach rests on the assumption that through dimensionality reduction, we can capture the underlying connectivity structure that may not be visible in high-dimensional space~\cite{Langs15}. We use a low-dimensional embedding to locate transition points where connectivity patterns change and interpret them as an abstract delineation of the parcellation boundaries. After projecting back to the native cortical space, these boundaries are used to compute non-overlapping and spatially contiguous parcels. We achieve this with a watershed segmentation technique, which has been effectively used for whole-cortex boundary mapping before~\cite{Laumann15,Gordon16}. Non-linear manifold learning has been previously used to identify functional networks from fMRI~\cite{Langs14,Langs16} and for surface matching~\cite{Langs15}, as well as within many other fMRI analysis techniques, such as~\cite{Thirion06,shen2006nonlinear}. Here, we propose to use such technique in association with dMRI-based structural connectivity and boundary mapping, in order to compute cortical parcellations for individual subjects, which can be used to investigate the brain's cortical organisation at the subject level and to derive network nodes for a whole-brain connectome analysis. 

Given the lack of a ground truth parcellation of the cortex, we assess the performance of the proposed method from a clustering point of view and evaluate the degree of fidelity to the underlying structural connectivity data using two well-known techniques: (1) parcel homogeneity~\cite{Shen13,Thirion14,Gordon16} and (2) Silhouette analysis~\cite{Yeo11,Craddock12,Eickhoff15,Parisot16b}, as introduced in the previous chapter. Besides structural connectivity that is used for model estimation, we evaluate the parcellations with respect to functional connectivity data obtained from resting-state fMRI as a means of external validation~\cite{Eickhoff15}. Our method is compared to three well-known clustering algorithms applied to the connectivity data, namely, hierarchical clustering, spectral clustering with normalised cuts, and \textit{k}-means clustering, as well as one dMRI-based state-of-the-art connectivity-based parcellation technique~\cite{Parisot15}. In addition, two parcellation schemes which do not take into account any connectivity information~\cite{Schirmer2015,Thirion14} are also incorporated to our experiments to form a baseline for comparisons. Besides the quantitative evaluation measures, we provide visual results of the parcellations across different subjects/resolutions and provide complimentary information to understand the inter-subject variability within the population by means of parcellation consistency across subjects. Finally, we show the extent to which our parcellation boundaries agree with well-established patterns of myelination and cytoarchitecture as defined by the Brodmann atlas~\cite{Brodmann09}. 

The remainder of this chapter is organised as follows: In Section~\ref{sec:manifold-methods}, we summarise the proposed parcellation method, starting with the estimation of structural connectivity and following with a detailed explanation of the manifold learning technique used for dimensionality reduction. The section then covers the steps for boundary map generation and cortical parcellation. In Section~\ref{sec:manifold-experiments}, we summarise the quantitative measures and clustering techniques used for evaluation. In Section~\ref{sec:manifold-results} we provide the performance measurements of the proposed method as well as the other approaches and show the visual results for qualitative and inter-modality assessment of the parcellations. Finally, in Section~\ref{sec:manifold-discussion}, we conclude the chapter with a detailed discussion on the advantages and limitations of the proposed method, and provide some insight towards future research directions.

\section{Methodology}
\label{sec:manifold-methods}
\subsection{Data}
We conduct our experiments on Dataset 1, which comprises 100 subjects (54 female, 46 male healthy adults, ages 22-35). The details of the dataset are provided in Chapter~\ref{chapter:background} and briefly summarised here. Data was acquired and preprocessed following the HCP minimal preprocessing pipelines~\cite{Glasser13}. For each subject, the cortical gray-matter voxels were registered onto the 32k standard triangulated mesh with a 2 mm spacing yielding a standard set of cortical vertices per hemisphere. The dMRI data was acquired using a multi-shell approach, in which diffusion weighting consisted of three shells at $b$-values 1000, 2000, and 3000 $s/mm^2$ and 90 gradient directions obtained per shell~\cite{sotiropoulos2013advances}. 

\subsection{Method} 
\begin{figure}[t!]
\centering
\includegraphics[width=\textwidth]{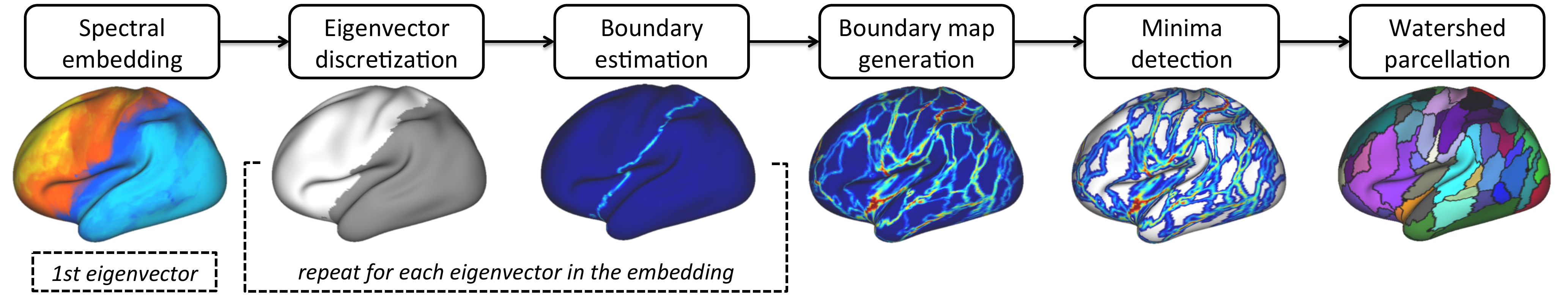}
\caption{Parcellation pipeline summarising all steps after preprocessing.}
\label{fig:pipeline}
\end{figure}

Our method starts with preprocessing the dMRI data using probabilistic tractography
to estimate a structural connectivity network, which is then reduced in dimensionality through non-linear manifold learning with Laplacian eigenmaps~\cite{Belkin03}. Next, we identify boundaries in this low-dimensional embedding as points where connectivity patterns change. Finally, driven by these boundaries, we make use of watershed segmentation to achieve a whole-brain cortical parcellation. All steps are explained thoroughly in the subsequent sections and the parcellation pipeline following preprocessing is summarised in Fig.~\ref{fig:pipeline}.

\subsubsection{Estimating Structural Connectivity}
The tractography matrix has been obtained from the preprocessed dMRI data by performing probabilistic tractography with FSL's bedpostX and probtrackX methods~\cite{behrens2007probabilistic,Jbabdi12}, as described in Chapter~\ref{chapter:background}. bedpostX allows to automatically determine the number of fibers passing through each voxel within the brain and estimate their orientation distribution. probtrackX performs probabilistic tractography with respect to estimated fiber orientations. We seeded 5000 streamlines from each of the $n$ cortical vertices and obtained a tractography matrix that records the number of streamlines originating from each vertex and reaching the rest of the cerebral cortex. 

Long-range connections computed with probabilistic tractography cannot be as effectively estimated as the short-range ones, due to the accumulation of errors along the tracts~\cite{jbabdi2009multiple,Moreno14,Parisot16a}. As a result, the estimated strength of connectivity is likely to be much higher for short-range connections compared to the long-range ones, even though the actual connections may have the same strength. This can potentially influence the connectivity matrix and have an impact on the resulting parcellations. To reduce the bias towards short-range connections, we apply an element-wise log transformation to the tractography matrix as suggested in~\cite{jbabdi2009multiple,Moreno14}. The process reduces the dynamic range of fiber counts, and hence, helps alleviate the bias towards short connections while not losing any information~\cite{Parisot16a}, as opposed to an alternative solution including thresholding short-range connections~\cite{Roca09tractography}. 

We estimate a structural connectivity matrix $C\in{\mathbb R}^{n \times n}$ for each subject in the dataset by cross-correlating the log-transformed tractography matrix using Pearson's correlation coefficient. Each row in $C$ represents how a vertex is connected to the other vertices in the cortical surface and therefore represents a connectivity profile, showing how a vertex is connected to the rest of the cerebral cortex. We excluded the medial wall vertices from further processing as they do not possess reliable information for connectivity analysis.

\subsubsection{Manifold Learning via Laplacian Eigenmaps}
\textit{\textbf{Overview}}
We use \textit{Laplacian eigenmaps}\footnote{It is worth noting that, this section is compiled mainly from the original Laplacian eigenmaps paper by Belkin and Niyogi~\cite{Belkin03}. Readers are referred to the original manuscript for more detailed information.}~\cite{Belkin03} to compute a non-linear embedding from a structural connectivity network. This method can reveal the intrinsic geometry of the underlying connectivity by forming an affinity matrix based on how vertices are connected within their neighbourhoods, hence preserving local proximity information between vertices. Based on the notion of the graph Laplacian obtained from the affinity matrix, a low-dimensional representation of the data can then be computed by using spectral decomposition.


Several key features of the algorithm distinguish it from other non-linear manifold learning techniques and make it a powerful tool to explore structural (or functional) brain connectivity. First of all, the core algorithm is very simple. It only requires to define a nearest-neighbour graph and solves one sparse eigenvector problem. Second, the method exhibits stability with respect to the embedding, that is, regardless of the resolution of the underlying network (i.e. different choices of $n$), the resulting embeddings will recover the same underlying manifold.  Last but not least, Laplacian eigenmaps are closely related to spectral clustering techniques in a sense that the algorithm inherently provides a natural clustering of the data, which can be attributed to the fact that the eigenmaps preserve local proximity information in the embedding. As a result, the resulting low-dimensional embedding can be used for obtaining soft clusters~\cite{Craddock12,Thirion14}. Furthermore, this locality-preserving characteristic makes the algorithm more robust to outliers and noise~\cite{Belkin03}.

\begin{figure}[!t]
\centering
\includegraphics[width=0.75\textwidth]{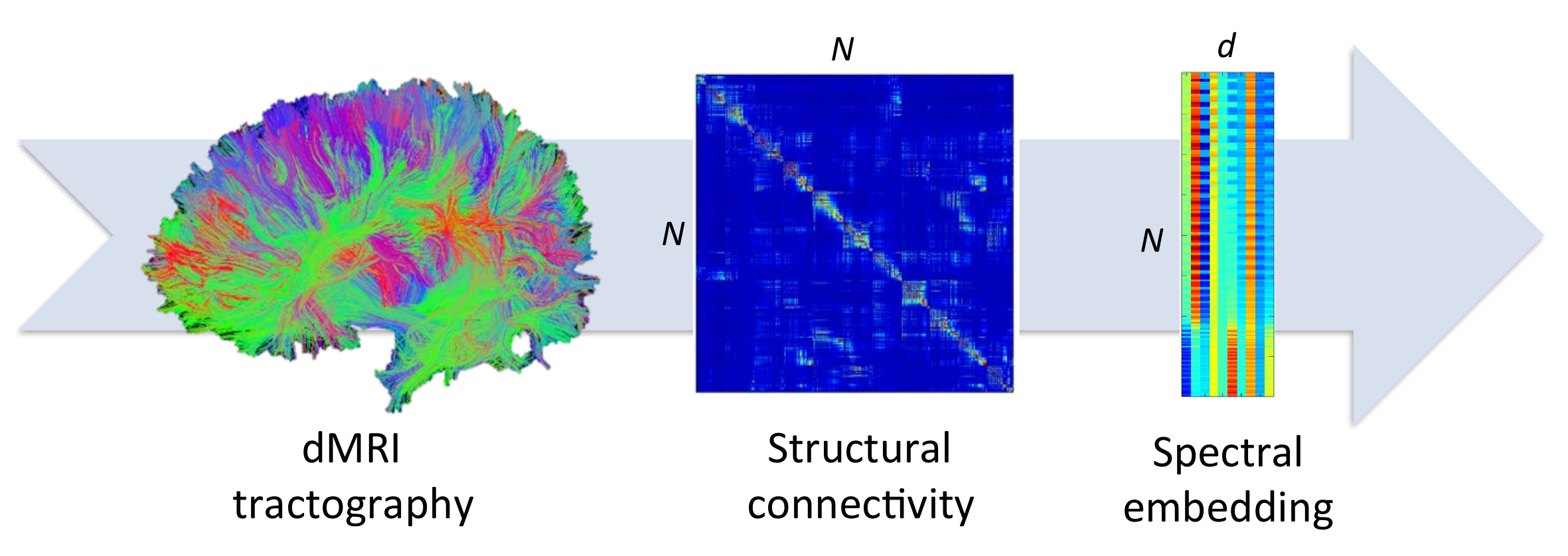}  
\caption{Illustration of non-linear feature reduction using Laplacian eigenmaps.}
\label{fig:feature_reduction}
\end{figure}

\textit{\textbf{The Algorithm}}
The generic dimensionality reduction problem can be defined as the following. Given
a set $x_1, \ldots, x_n$ of $n$ points in ${\mathbb R}^m$, find a new set of $n$ points $y_1,...,y_n$ in ${\mathbb R}^d$ such that $y_i$ represents $x_i$ in a low dimensional space, where $d << m$. Here, we assume that the data points (i.e. vertices) $x_1, \ldots, x_n$ $\in \cal{M}$ and $\cal{M}$ is a manifold embedded in ${\mathbb R}^m$. Specific to our problem, each vertex is represented by an $n$-dimensional connectivity profile, hence, $m = n$. The procedure that leads to eigenmaps can be outlined in three key steps as follows:

\begin{enumerate}
\item \textbf{Constructing an affinity matrix}: An affinity matrix $W$ is formed by constructing an edge between vertices $i$ and $j$ if $x_i$ and $x_j$ are 'close' to each other. Among different methods to define this closeness, we use the $k$ nearest neighbour approach ($k \in \mathbb N$) and connect vertices $i$ and $j$ by an edge if $i$ is among $k$ nearest neighbours of $j$ or $j$ is among $k$ nearest neighbours of $i$, forming a symmetric relationship. \textit{k} is chosen large enough to effectively capture the local connectivity structure and to ensure that the affinity matrix is connected for each subject. 

\item \textbf{Choosing the weights}: The edges are typically weighted by a heat kernel if $i$ and $j$ are connected

\begin{equation}
W_{ij} = \exp\Big(-\frac{\lVert x_i - x_j \lVert^2}{t}\Big),
\end{equation}

otherwise $W_{ij} = 0$. Here, $\lVert \cdot \lVert$ is Euclidean norm and $t$ controls the weight decay. In practice $t$ can be chosen infinite~\cite{thirion2004nonlinear}, leading to $W_{ij} \in \{0,1\}$. Alternatively, the heat kernel can be replaced by an appropriate similarity measure, such as Pearson's correlation, together with a thresholding or scaling scheme to discard negative correlations~\cite{Langs15,Langs14}. In this chapter, we choose this approach and transform the connectivity matrix $C$ into a locality-preserving affinity matrix $W\in{\mathbb R}^{+n \times n}$ by only retaining the correlations of the \textit{k} nearest neighbours of each vertex.


\item \textbf{Spectral decomposition}: Given that $W$ is connected, we compute eigenvalues and eigenvectors by solving the generalized eigen decomposition problem

\begin{equation}
L\mathbf{f} = \lambda D \mathbf{f}, 
\label{eq:eig}
\end{equation} 


where $D$ is a diagonal matrix with each entry $D_{ii}={\sum}_j W_{ij}$ representing the degree of $x_i$ and $L = D-W$ is the graph Laplacian, which is a symmetric and positive semidefinite matrix. The Laplacian matrix can be considered as an operator on functions defined on vertices in $W$~\cite{Belkin03}. Solving Eq.~\ref{eq:eig} reveals the eigenvectors $\mathbf{f}_0, \mathbf{f}_1, \ldots, \mathbf{f}_{n-1}$, ordered according to their eigenvalues $0 = \lambda_0 \leq \lambda_1 \leq \ldots \leq \lambda_{n-1}$. After omitting the eigenvector $\mathbf{f}_0$ corresponding to $\lambda_0$, we can use the next \textit{d} eigenvectors to define an embedding that can approximate a low dimensional manifold~\cite{Belkin03}. Hence, each cortical vertex $i$ in this high-dimensional structural connectivity matrix can now be expressed as a row in the spectral embedding, i.e. $i \mapsto (\mathbf{f}_{1}(i), \ldots, \mathbf{f}_{d}(i))$ as illustrated in Fig.~\ref{fig:feature_reduction}.
\end{enumerate}

\textit{\textbf{An Optimal Embedding}}
The local proximity between pairs of vertices is optimally preserved by the spectral embedding generated using the Laplacian eigenmaps algorithm. The goal of the embedding function $\mathbf{f}$ is to map nearby (or strongly connected) vertices in the high-dimensional space to a $d$-dimensional Euclidean space and ensure that they still remain nearby.

Let us first consider the problem of mapping the vertices to a line, i.e. where $d=1$. We are seeking a mapping $\mathbf{y} = (y_1, y_2, \ldots, y_n)^T$, such that the following criterion is minimised:

\begin{equation}
\sum_{i,j}(y_i-y_j)^2 W_{ij} 
\label{eq:minimise1}
\end{equation} 

Given how the weights $W_{ij}$ are assigned, the objective function incurs a heavy penalty if points that are close in high-dimensional space are mapped far apart in $\mathbf{y}$. For any $\mathbf{y}$, we can rewrite Eq.~\ref{eq:minimise1} as

\begin{equation}
\frac{1}{2} \sum_{i,j}(y_i-y_j)^2 W_{ij} = \mathbf{y}^TL\mathbf{y},
\end{equation} 

where $L$ is the Laplacian matrix. Recall that $W_{ij}$ is symmetric and $D_{ii}={\sum}_j W_{ij}$. Thus,

\begin{equation}
\begin{split}
\sum_{i,j}(y_i-y_j)^2 W_{ij} {}& = \sum_{i,j}(y_i^2 + y_j^2 -2y_iy_j)W_{ij} \\
							   & = \sum_{i}y_i^2D_{ii} + \sum_{j}y_j^2D_{jj} - 2\sum_{i,j}y_iy_jW_{ij} \\
                               & = 2\mathbf{y}^TL\mathbf{y}.
\label{eq:minimise2}
\end{split}
\end{equation}

Since all $W_{ij}$ are non-negative, Eq.~\ref{eq:minimise2} shows that $\mathbf{y}^TL\mathbf{y} \geq 0$, which proves that $L$ is positive semidefinite. The minimisation problem is reduced to finding

\begin{equation}
\underset{\substack{\mathbf{y} \\ \mathbf{y}^TD\mathbf{y}=1}} {\text{argmin}} \; \mathbf{y}^TL\mathbf{y}.
\end{equation} 

Here, the additional constraint $\mathbf{y}^TD\mathbf{y}=1$ removes an arbitrary scaling factor in the embedding~\cite{Belkin03}. The $D$ matrix provides a means to account for the impact of vertices in the graph. A larger $D_{ii}$ implies more edges connected to vertex $i$, hence increases its importance for the minimisation~\cite{Belkin03}. Given that $L$ is a symmetric and positive semidefinite matrix, the Rayleigh-Ritz theorem~\cite{lutkepohl1997handbook} indicates that the solution of the minimisation problem is given by the vector $\mathbf{y}$, which corresponds to the eigenvector associated with the minimum non-zero eigenvalue of $L$~\cite{von2007tutorial}. 

Considering the more general problem of embedding the graph into $d$-dimensional Euclidean space where $d > 1$, we can represent the embedding by the ${n \times d}$ matrix $Y = [\mathbf{y}_1, \ldots, \mathbf{y}_d ]$, in which the $i$th row corresponds to the embedding coordinates of the $i$th vertex. Similarly to Eq.~\ref{eq:minimise1}, we need to minimise

\begin{equation}
\sum_{i,j}{\lVert\mathbf{y}^{(i)}-\mathbf{y}^{(j)}\lVert}^2 W_{ij} = \text{tr}(Y^{T}LY), 
\label{eq:minimise3}  
\end{equation}   

where $\mathbf{y}^{(i)} = [\mathbf{y}_1(i), \ldots, \mathbf{y}_d(i) ]^T$ is the $d$-dimensional representation of the $i$th vertex. By introducing a constraint that prevents an embedding from collapsing onto a subspace of fewer than $d$-1 dimensions, the objective function is reduced to finding

\begin{equation}
\underset{{Y}^TD{Y}=I}{\text{argmin}} \; \text{tr}(Y^{T}LY). 
\end{equation} 

As in the one-dimensional case, standard methods~\cite{lutkepohl1997handbook} show that the solution that minimises the objective function is provided by the matrix of eigenvectors associated with the lowest eigenvalues of the graph Laplacian~\cite{Belkin03}.

\textit{\textbf{Connections to Spectral Clustering}} 
The eigenvectors of the graph Laplacian provide an embedding that reflects the intrinsic geometry of the data and interestingly this solution can also be interpreted in the context of spectral clustering~\cite{Belkin03}, which can be used to sub-partition a graph into $K$ pre-defined clusters. Here, we briefly outline the ideas behind spectral clustering with a particular focus on the normalised-cuts (N-Cuts) technique proposed by Shi and Malik~\cite{Shi00} and show the role of graph Laplacian in providing a set of feature vectors for clustering purposes.

Let $G=(V,E)$ be a weighted graph with $V$ being the set of vertices and $E$ being the set of connections between them. $W$ is the affinity matrix associated with $G$ and defined as before, hence, it is symmetric, connected, and $W_{ij} \geq 0$ for all $i,j$. The ultimate goal of N-Cuts is to find an optimal cut to partition $W$ into $K$ pre-defined clusters, such that vertices within a cluster are more similar to each other than those in different clusters.

For subdividing $W$ into two disjoint clusters $A$ and $B$, where $A \cup B = V$,  the normalised-cut cost function is defined as

\begin{equation}
\label{eq:ncut}
Ncut(A,B)= cut(A,B) \Big( \frac{1}{vol(A)}+\frac{1}{vol(B)}\Big),
\end{equation}

where $cut(A,B)= {\sum}_{i \in A, j \in B} W_{ij}$ represents the sum of the weights on edges that connect vertices in $A$ to vertices in $B$~\cite{Wu1993optimal} and $vol(\cdot)$ represents the total connections from within-cluster vertices to all vertices in the graph. \textit{Ncut} normalises the cost as a fraction of the sum of the edge weights connecting vertices in a cluster to every other vertex in the graph. It simultaneously minimises the inter-cluster similarity and maximises the within-cluster similarity, and hence, achieves a balanced partitioning.

While the minimisation problem is NP-hard~\cite{Shi00}, a near global-optimal solution can be approximated by allowing relaxation of the indicator functions to real values and then solving the relaxed problem~\cite{Belkin03}.

Recall that

\begin{equation}
\mathbf{x}^TL\mathbf{x} = \sum_{i,j}(x_i-x_j)^2 w_{ij}.
\end{equation} 

Let $a = vol(A)$ and $b=vol(B)$. Put

\begin{equation}
  x_i =
  \begin{cases}
    \; \frac{1}{vol(A)}, & \text{if $V_i \in A$} \\ 
    -\frac{1}{vol(B)}, & \text{if $V_i \in B$}
  \end{cases}.
\end{equation}

We now have

\begin{equation}
\mathbf{x}^TL\mathbf{x} = \sum_{i,j}(x_i-x_j)^2 w_{ij} = \sum_{V_i \in A, V_j \in B} \Big( \frac{1}{a} + \frac{1}{b}\Big)^2cut(A,B)
\end{equation} 

and 

\begin{equation}
\begin{split}
\mathbf{x}^TD\mathbf{x} = \sum_{i}x_i^2d_{ii} &= \sum_{V_i \in A} \frac{1}{a^2}d_{ii} + \sum_{V_i \in B} \frac{1}{b^2}d_{ii} \\
&= \frac{1}{a^2}vol(A) + \frac{1}{b^2}vol(B) = \frac{1}{a} + \frac{1}{b},
\end{split}
\end{equation} 

where $D$ is the degree matrix as defined before. Thus,

\begin{equation}
\frac{\mathbf{x}^TL\mathbf{x}}{\mathbf{x}^TD\mathbf{x}} = cut(A,B) \Big(\frac{1}{a} + \frac{1}{b}\Big)=Ncut(A,B).
\label{eq:relaxed2}
\end{equation} 

The relaxed problem is minimising Eq.~\ref{eq:relaxed2}, subject to $\mathbf{x}^TD\mathbf{1}=\mathbf{0}$ where $\mathbf{1}$ is a column vector of ones. Given that $D$ is invertible, $G$ has no isolated vertices (which is ensured while constructing the adjacency matrix) and substituting $\mathbf{y} = D^{1/2}\mathbf{x}$ we can rewrite Eq.~\ref{eq:relaxed2} as

\begin{equation}
\frac{\mathbf{x}^TL\mathbf{x}}{\mathbf{x}^TD\mathbf{x}} = \frac{\mathbf{y}^TD^{-1/2}LD^{-1/2}\mathbf{y}}{\mathbf{y}^T\mathbf{y}}=\frac{\mathbf{y}^T\mathcal{L}\mathbf{y}}{\mathbf{y}^TD\mathbf{y}},
\end{equation} 

where $\mathbf{x}$ and $\mathbf{y}$ are perpendicular to $D^{1/2}\mathbf{1}$. Here the matrix $\mathcal{L}=D^{-1/2}LD^{-1/2}$ is the \textit{normalised} graph Laplacian, a symmetric and positive semidefinite matrix, which serves in the approximation of the minimisation of $Ncut$. That is, the solution to the minimisation problem is given by the eigenvector associated with the smallest non-zero eigenvalue of $\mathcal{L}$.

Similarly to the multi-dimensional embedding problem discussed above, this solution can be easily generalised to obtain a $K$-way partitioning of the graph where $K>2$. That is, by using the set of eigenvectors corresponding to the $K$ smallest non-zero eigenvalues of $\mathcal{L}$, we can obtain a near global-optimal solution to the multi-way normalised-cut problem and feed these eigenvectors to a general purpose clustering algorithm, e.g. $k$-means~\cite{von2007tutorial} or use discretisation to transform the real-valued representation to hard clusters~\cite{Shi00}.

\subsubsection{Eigenvector Discretisation}
As discussed in the preceding section, the locality-preserving process of dimensionality reduction  not only provides an optimal embedding for non-linear manifold learning, but also imposes a soft clustering of the data~\cite{Belkin03}. Therefore, the parcellation problem can be cast as a graph partitioning problem and one would attempt to subdivide the connectivity graph with spectral clustering, e.g. using the normalised cuts criterion~\cite{Yu03}. In particular, each of the smallest eigenvectors corresponds to a real-valued solution that optimally sub-partitions the graph. These partitions can be approximated by transforming the real valued eigenvectors into discrete forms (i.e. discretisation), ideally by dividing them into two parts with respect to a splitting point~\cite{Shi00}. This can further be generalised towards a multi-way partitioning with a recursive or simultaneous discretisation of the smallest eigenvectors~\cite{Shi00}, and thus, can be used to obtain a parcellation~\cite{Craddock12}.

However, by definition, our affinity matrix does not impose any spatial constraints, hence such spectral methods cannot guarantee spatial contiguity within the parcels. Instead, we propose a more effective way of deriving parcellations from discrete eigenvectors, i.e. we use them to identify cortical boundaries where connectivity patterns show distinct changes, and later show that this method can produce more reliable parcellations compared to spatially constrained spectral clustering.

\begin{figure}[!t]
\centering
\includegraphics[width=0.8\textwidth]{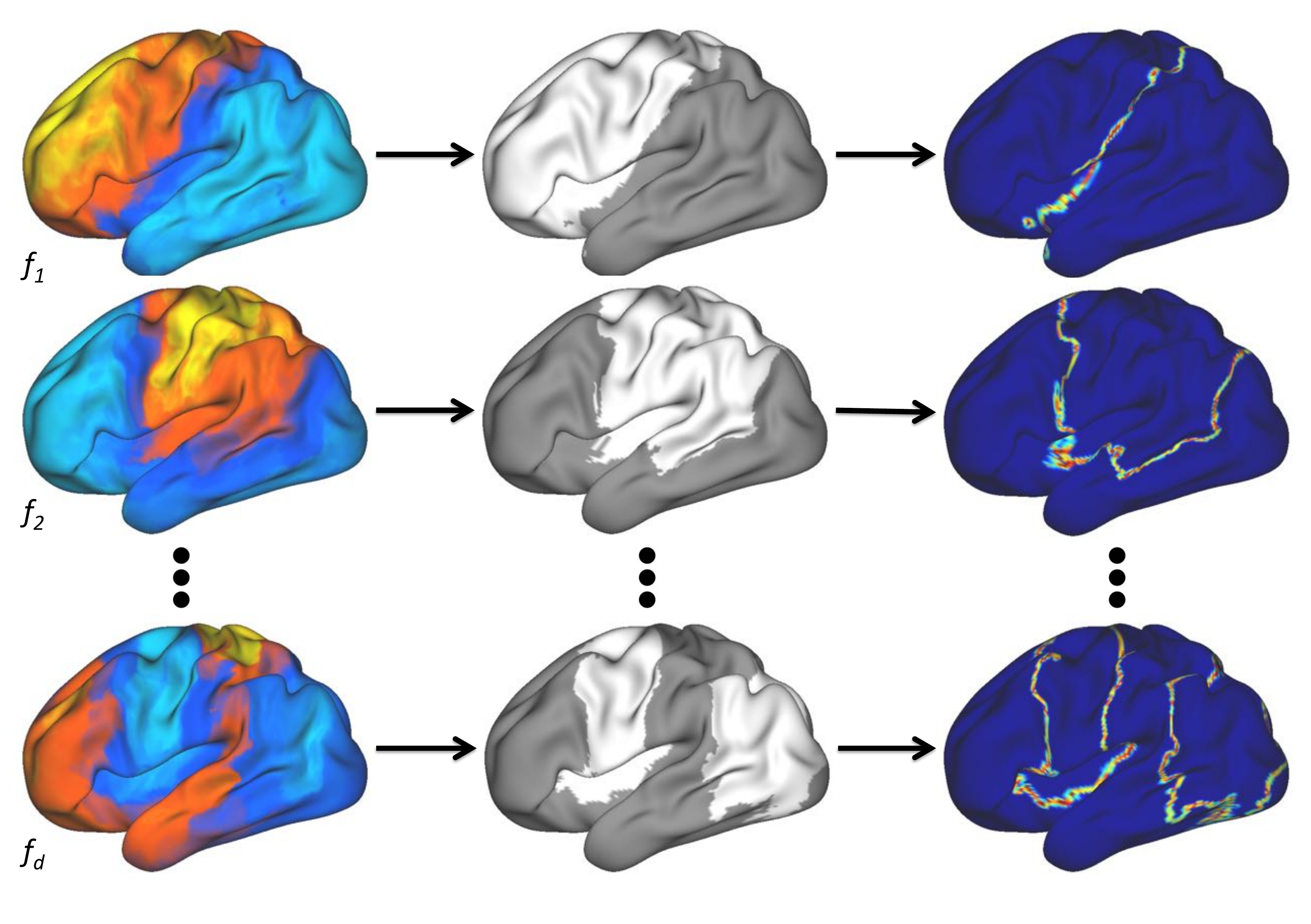}  
\begin{tabular}{ccc}
(a) ~~~~~~~~~~~~~~~~~~~~~~~~~~~~ & (b) ~~~~~~~~~~~~~~~~~~~~~~~~~~~~~ & (c) 
\end{tabular}
\caption[Illustration of eigenvector discretisation and boundary estimation.]{(a) Eigenvectors overlaid onto the lateral left hemisphere. (b) Discretised eigenvectors obtained via \textit{k}-means. (c) Boundary estimation from discrete eigenvectors.}
\label{fig:boun_mapping}
\end{figure}

\begin{figure}[!t]
\centering
\includegraphics[width=\textwidth]{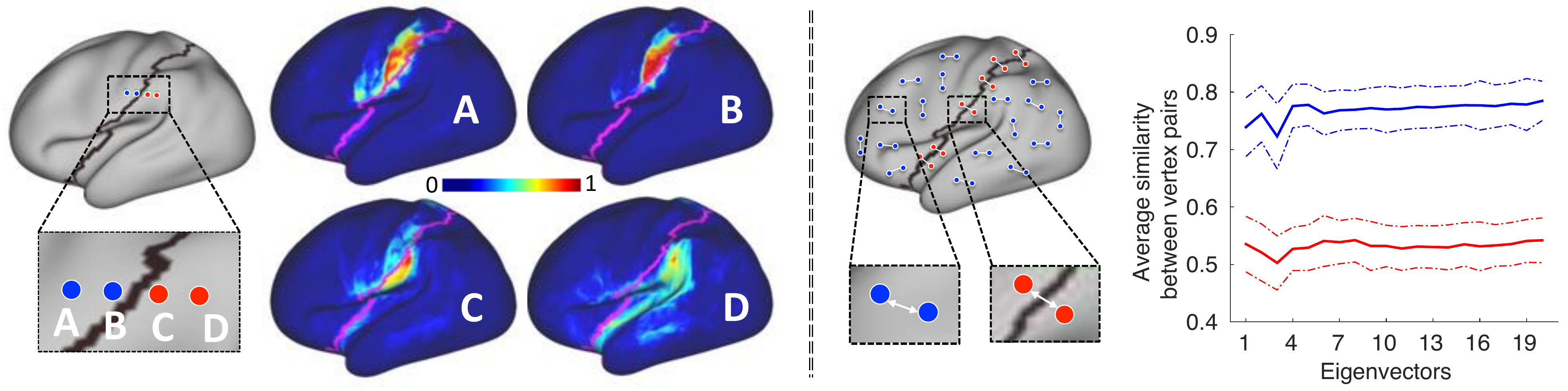}  
\begin{tabular}{cc}
(a) ~~~~~~~~~~~~~~~~~~~~~~~~~~~~~~~~~~~~~~~~~~~~~~~~ & (b)   
\end{tabular}
\caption[Connectivity profiles of different vertices.]{(a) Connectivity profiles of vertices from different sides of a boundary. (b) \textit{Left:} Illustration of the vertex selection procedure. \textit{Right:} Average similarity (correlation) between paired vertices for each eigenvector. Dotted lines show the variability across subjects as means of standard deviations.}
\label{fig:transition}
\end{figure}

To this end, we first subdivide each real-valued eigenvector into two regions using \textit{k}-means as illustrated in Fig.~\ref{fig:boun_mapping}. The edges between these subregions potentially provide good separation points towards obtaining a parcellation, as the vertices within the same subregions tend to have similar connectivity properties, whilst the points closer to the boundary attribute to the cortical areas where the connectivity is in transition. For example, Fig.~\ref{fig:transition}(a) shows that connectivity profiles of different vertices may exhibit similar or varying patterns, depending on their relative location to an edge. In order to show that this tendency holds across the whole cortex, we randomly selected vertices from one subregion adjacent to the edge and paired them with their closest neighbours residing in the other subregion. Keeping the distance between the vertices in pairs approximately the same, we selected new pairs of vertices, but this time from within the same subregions. We then measured the average correlation between the paired vertices' connectivity profiles in each set and repeated this for all eigenvectors and subjects. Fig.~\ref{fig:transition}(b) shows that, the similarity between the connectivity profiles of vertices drops by at least 20$\%$ if they reside on different sides of a boundary. 

\subsubsection{Boundary Map Generation and Cortical Parcellation}
\begin{figure}[!t]
\centering
\includegraphics[width=0.8\textwidth]{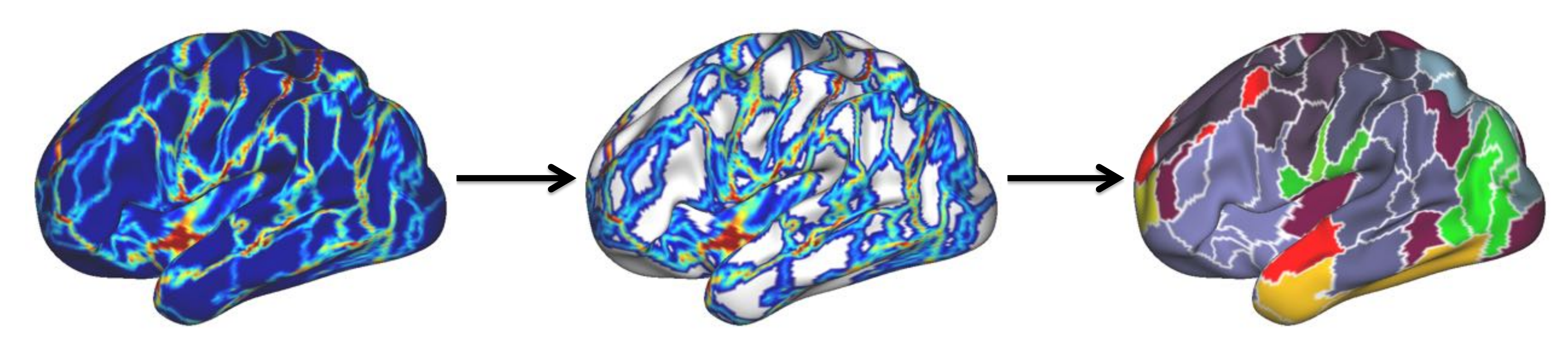}  
\begin{tabular}{ccc}
~~ (a) ~~~~~~~~~~~~~~~~~~~~~~~~~~~ & (b) ~~~~~~~~~~~~~~~~~~~~~~~~~~~~ & (c) 
\end{tabular}
\caption[Boundary map generation and cortical parcellation.]{(a) A boundary map is obtained by combining gradients of the discrete eigenvectors. (b) Local minima detection: White regions correspond to markers that initiate the watershed process. (c) Output of watershed segmentation is a set of spatially contiguous, non-overlapping cortical parcels subdividing the whole cerebral cortex. }
\label{fig:watershed}
\end{figure}

To locate the connectivity transition points and construct a boundary map, we first transfer the discrete eigenvectors back to the native high-dimensional space (project them onto the cortical surface) as shown in Fig.~\ref{fig:boun_mapping}(b). We then compute the first spatial derivative of each eigenvector across the cortical surface [Fig.~\ref{fig:boun_mapping}(c)] and combine them into a boundary map as illustrated in Fig.~\ref{fig:watershed}(a). This map constitutes a more robust substitution for the boundary maps based on gradients directly calculated from the spatial correlations~\cite{Gordon16}, since it can adjust for possible spurious gradients. 

In order to obtain the final parcellations from the boundary map, we use a marker-controlled watershed algorithm~\cite{Vincent91}, a powerful image processing technique used for the segmentation of salient objects in images~\cite{arslan2014color,Gordon16}. The marker-controlled watershed technique typically consists of a marker detection stage and a marking function, where the former identifies markers (or seeds) that correspond to estimated locations of structures of interest in the image. The marking function governs the watershed transformation in which markers grow until they touch each other or reach watershed ridge lines. 

In our case, we use a watershed implementation adapted for cortical meshes~\cite{Gordon16}. To this end, we define a set of markers on the boundary map where each marker corresponds to an estimated parcel position as shown in Fig.~\ref{fig:watershed}(b). The marker definition is typically performed by choosing a threshold on the boundary map. We set this threshold to the 25th percentile of the boundary map intensities, since in many empirically tested cases, this effectively revealed approximate parcel locations to be used as ideal markers for a watershed transformation. Driven by the boundary map, the marking function grows these markers until high-intensity cortical vertices are reached or two regions touch each other in the flooding process of the watershed. As a post-processing step, very small regions (comprising 25 or fewer vertices) are merged with their largest neighbours. The final cortical parcellation is shown in Fig.~\ref{fig:watershed}(c).

\section{Experiments}
\label{sec:manifold-experiments}
\subsection{Quantitative and Qualitative Evaluation Measures}
In the absence of a gold standard parcellation for comparison, we adapted two well-known cluster validity measures, homogeneity~\cite{Arslan15b,Gordon16} and Silhouette analysis~\cite{Craddock12,Eickhoff15}, to assess how well a parcellation reflects the underlying connectivity. Both methods are previously explained in Chapter 3 and summarised here. Homogeneity measures a parcellation's ability to cluster vertices with similar connectivity by calculating average cross-correlations within each parcel after applying Fisher's $r$-to-$z$ transformation to resulting correlation coefficients. Silhouette analysis combines within-parcel similarity with inter-parcel separation and measures how vertices within a parcel are similar to each other, compared to the vertices within the nearest parcels~\cite{Yeo11,Eickhoff15}. 

The goodness-of-fit is estimated with respect to the structural connectivity data from which the parcellations have been derived. However, performing the evaluation based on the same connectivity data initially used to obtain the parcellations may lead to a circular analysis~\cite{Eickhoff15}. To reduce the impact of any bias towards model selection, we also evaluate parcellations by measuring their extent to reflect the underlying connectivity estimated from resting-state fMRI. Although dMRI-based cortical parcellations do not necessarily reflect the functional organisation of the cortex derived from rs-fMRI, this type of evaluation provides an external data source for validation~\cite{Eickhoff15}. 

Standard silhouette coefficient represents how well a vertex lies in its cluster~\cite{Rousseeuw87} and is typically computed for each vertex individually. However, when the model generation and validation are based on different data sources, both within-cluster and out-of-cluster similarity of a vertex are likely to be very low. As a result, Silhouette coefficients tend to be close to zero, indicating neither bad nor good clustering regardless of the clustering configuration. In such a case, considering all vertices within a parcel collectively and computing a parcel-wise Silhouette coefficient may provide a more robust means to assess the quality of clustering~\cite{Craddock12}. To this end, we make use of the modified Silhouette analysis proposed in~\cite{Craddock12}, which normalises the average similarity within a parcel by the average out-of-parcel similarity obtained from the parcels in its neighbourhood. 

We not only evaluate the parcellations from a clustering perceptive, but also assess the agreement of the connectivity-driven parcellations with the cyto- and neuro-anatomical architecture of the cortex, identified with Brodmann's areas and myelin maps, respectively. Visual examination of parcellation boundaries with Brodmann's areas is accompanied by Dice-based overlapping maps to provide a quantitative basis for evaluation. We achieve this by considering the Brodmann atlas as an alternative parcellation of the cerebral cortex and measuring the amount of overlap between the proposed parcellations and Brodmann's areas using Dice coefficients. To this end, we first match each parcel with a Brodmann area, if their overlap ratio is $ \geq 0.5$ (i.e. the Brodmann area comprises at least half of the other parcel) and then calculate the Dice coefficient between the matching pairs. It is worth noting that several parcels can be matched to the same cytoarchitectonic region and therefore merged into a larger parcel, which could potentially yield a bias towards small, evenly sized regions having high Dice scores after merging.

In addition, we assess the inter-subject variability/consistency across parcellations by quantitatively identifying the cortical areas that are repeatedly assigned to the same/different parcels across subjects. This is achieved by computing a consistency map for all subjects/resolutions using the Dice-based matching algorithm, as defined in Chapter 3. This method calculates a consistency score for each vertex within a range of $[0,1]$. Cortical regions with a high inter-subject consistency (values close to 1) are assumed to be parcellated in a similar way (e.g. having parcels of similar size/shape) for most of the subjects. On the other hand, consistency scores close to 0 indicate a high variability across subjects in terms of subdividing a particular cortical area (i.e. vertices being assigned to parcels with highly different size/shape). 

It is worth noting that, one should interpret such maps with great care and always account for possible biases towards the underlying modality and/or processing techniques before drawing any conclusions. For example, one known issue associated with tractography algorithms is the propensity of streamlines terminating within gyri instead of sulci~\cite{VanEssen2013mapping}. This gyral bias may influence the parcellation boundaries obtained from dMRI and consequently yield an inherent alignment with cortical folding~\cite{Parisot16a}. To understand the impact of the gyral bias on the proposed parcellations, we make use of the average `sulc' map, computed across all subjects in the dataset. A sulc map provides information about the depth of vertices within sulci and allows to visualise the macro-anatomy of the folded cortex for inflated meshes~\cite{fischl1999cortical}. In other words, it indicates how deep and how high the brain folds are with respect to the distance of cortical vertices to the gyri and sulci. In our experiments, the average sulc map is obtained from the sulcal depth surface files, which are generated by the HCP structural preprocessing pipelines and made publicly available for each subject.

\subsection{Comparison Methods}
We compare our parcellations to the ones obtained by several other clustering methods, including spectral clustering with normalised cuts~\cite{Craddock12} (\textit{N-Cuts}) applied to a spatially constrained version of the affinity matrix computed for the proposed method; \textit{k}-means clustering (\textit{K-Means}) applied to the low-dimensional embedding; Ward's hierarchical clustering applied to the low-dimensional embedding (\textit{HC-Low}) and driven by the connectivity profiles in the high-dimensional space (\textit{HC-High}); a connectivity-driven, multi-scale spectral clustering technique based on dMRI~\cite{Parisot16a} (\textit{M-Scale}); random parcellations by Poisson disk sampling (\textit{Random}); geometric parcellations, i.e. \textit{k}-means clustering of the vertex coordinates~\cite{Thirion14} (\textit{Geometric}). 

\textit{M-Scale} parcellations are obtained from the spatially-constrained affinity matrices using the publicly available code at \url{https://github.com/parisots/SpectralParcellation}. All other methods are adapted as described in Chapter 3, unless otherwise noted. Apart from \textit{K-Means}, which is exclusively driven by the connectivity information, all methods are spatially constrained and do not yield disjoint parcels. \textit{M-Scale} and \textit{HC-High} are based on an initial connectivity-based over-parcellation of the cortex to compensate for the noise, and thus, to obtain higher accuracy (1000, 2000 and 3000 regions for \textit{M-Scale}; 3000 regions for \textit{HC-High}). \textit{Random} and \textit{Geometric} parcellations do not account for any connectivity information, therefore provide a baseline for the assessment~\cite{Thirion14}. 

\section{Results}
\label{sec:manifold-results}
\subsection{Parameter Selection}
As there is no known optimal number of parcels, we evaluate the proposed method at different scales, determined by the number of eigenvectors incorporated into the boundary map. We present results for $d$ $=$ $10$, $15$, and $20$ eigenvectors per hemisphere, which on average, yield parcellations with around 180, 230, and 280 regions for each subject, respectively. For a fair comparison, other methods are tuned to use the same number of parcels as inferred by our models.

\begin{figure}[!t]  
\centering
\includegraphics[width=0.7\textwidth]{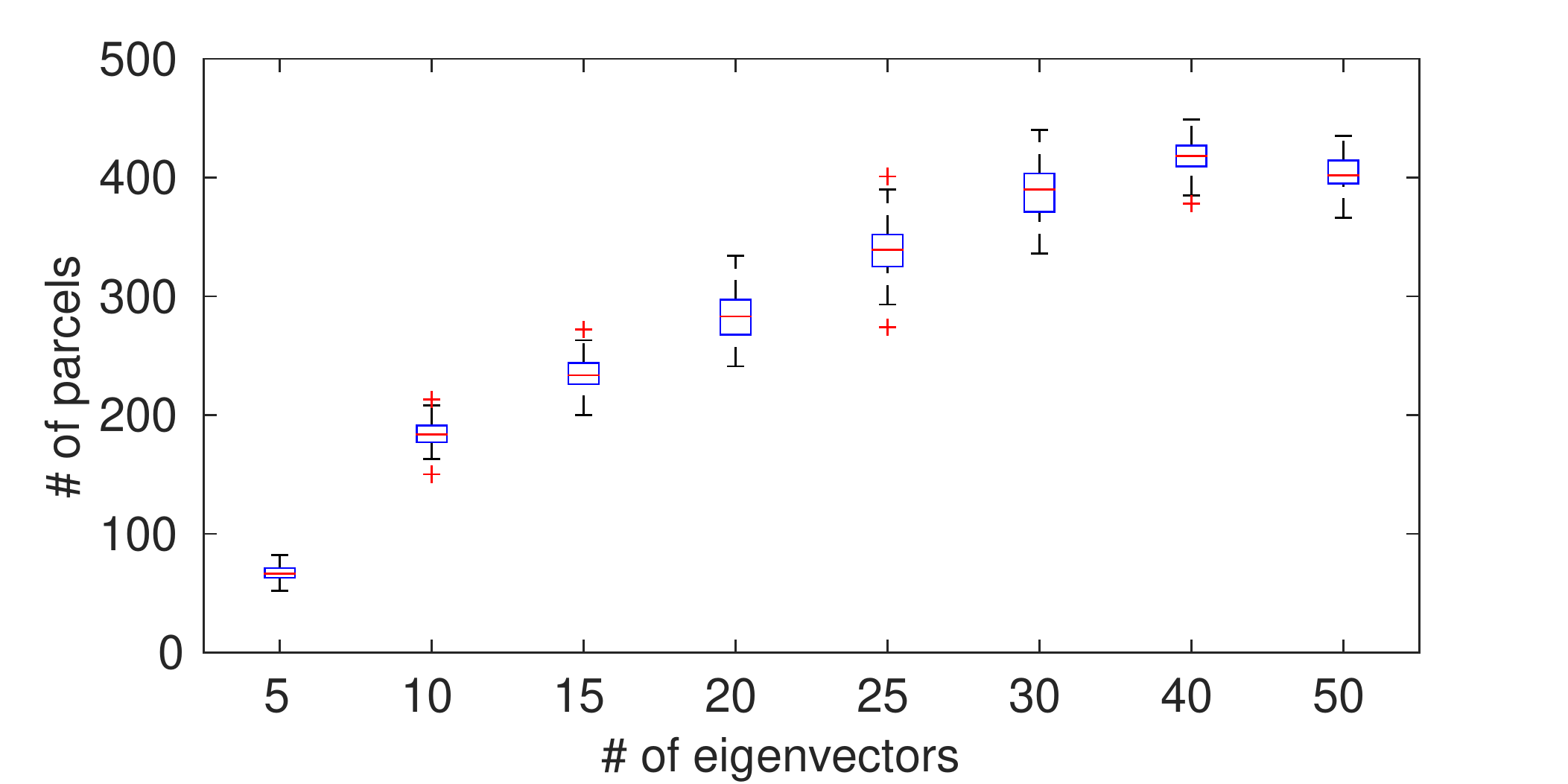} 
\caption[The number of parcels obtained by the proposed method as a function of the number of eigenvectors.]{The number of parcels generated with respect to different number of eigenvectors used in the spectral embedding. Box plots show the variability across subjects. }
\label{fig:parcels_vs_d}
\end{figure}

The number of parcels obtained by the proposed method as a function of the number of eigenvectors is given in Fig.~\ref{fig:parcels_vs_d}. As shown in the figure, fewer eigenvectors lead to very coarse parcellations, which may potentially under-represent the underlying connectivity. Increasing $d$ beyond 20 eigenvectors, although at first introduce more parcels, gradually leads to lower-resolution parcellations with unusual borders. This is due to the fact that incorporating too many eigenvectors to the embedding yield very dense boundary maps and consequently reduce the number of markers that drive the watershed transform.

Another external parameter is the number of nearest neighbours $k$ that is used to determine the connectedness of the affinity matrix. Intuitively, $k$ should be large enough to effectively capture the local connectivity patterns across the cerebral cortex. Our experiments with values of $k$ within a range of $[50, 200]$ revealed that $k=100$ tends to provide locality-preserving affinity matrices for all subjects with most connections having high correlation coefficients ($r>0.5$). If $k$ is set to a very low value, the affinity matrix is likely to capture stable features regarding the spatial geometry of the cortical mesh. This may consequently result in regularly shaped parcels with similar size, and hence, may not be able to accurately represent the underlying structural connectivity. Picking a very large $k$ leads to edges with negative correlation coefficients, indicating that connections tend to become weaker with increasing $k$. It also comes with an additional computational bottleneck, as the size of the matrix dramatically increases when more edges are formed between vertices. 

\subsection{Quantitative Assessment of the Parcellation Quality}
\begin{figure}[!t]  
\centering
\begin{tabular}{cl}
\includegraphics[width=0.7\textwidth]{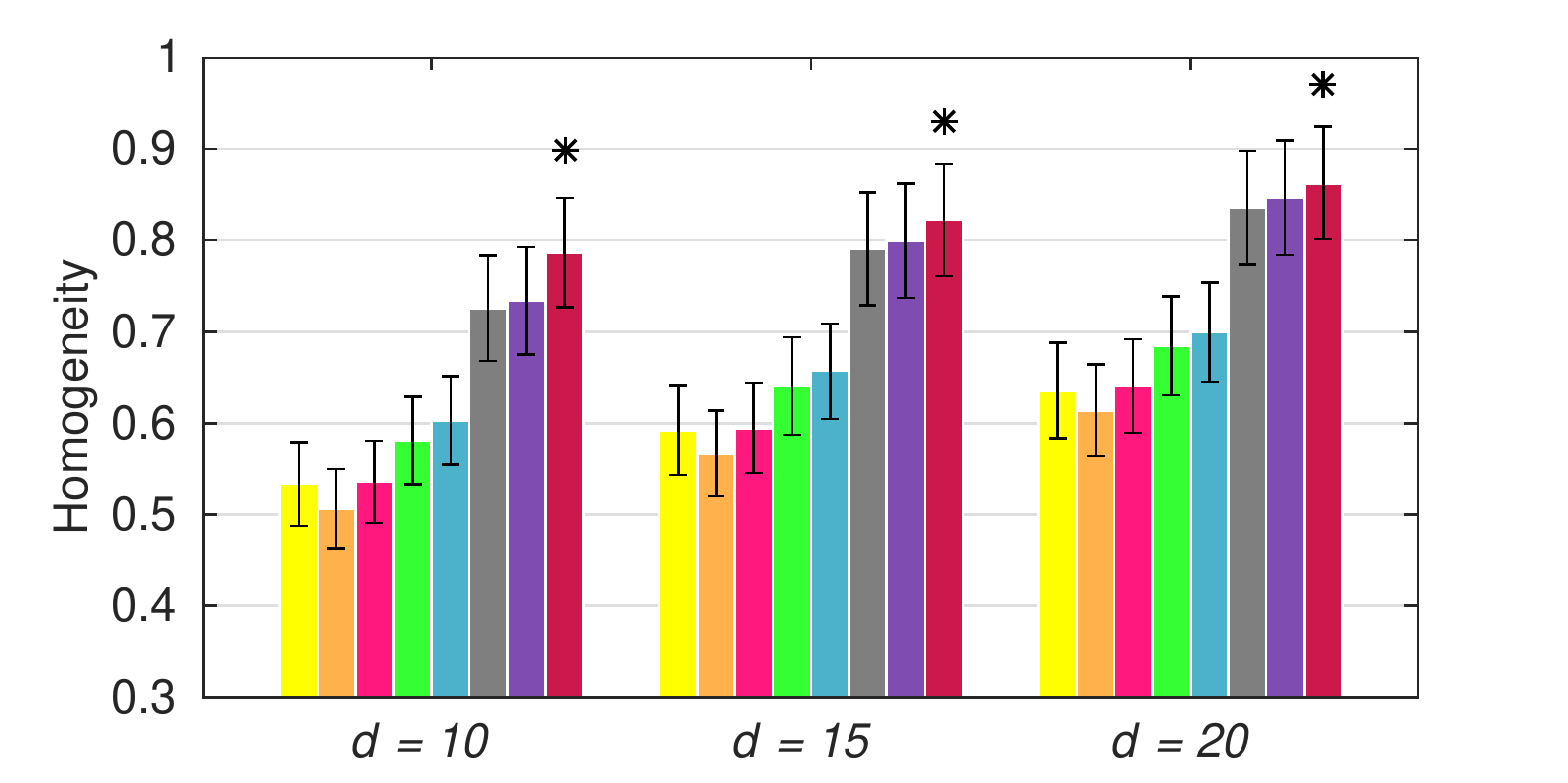} \kern-2.5em & \\ 
\includegraphics[width=0.7\textwidth]{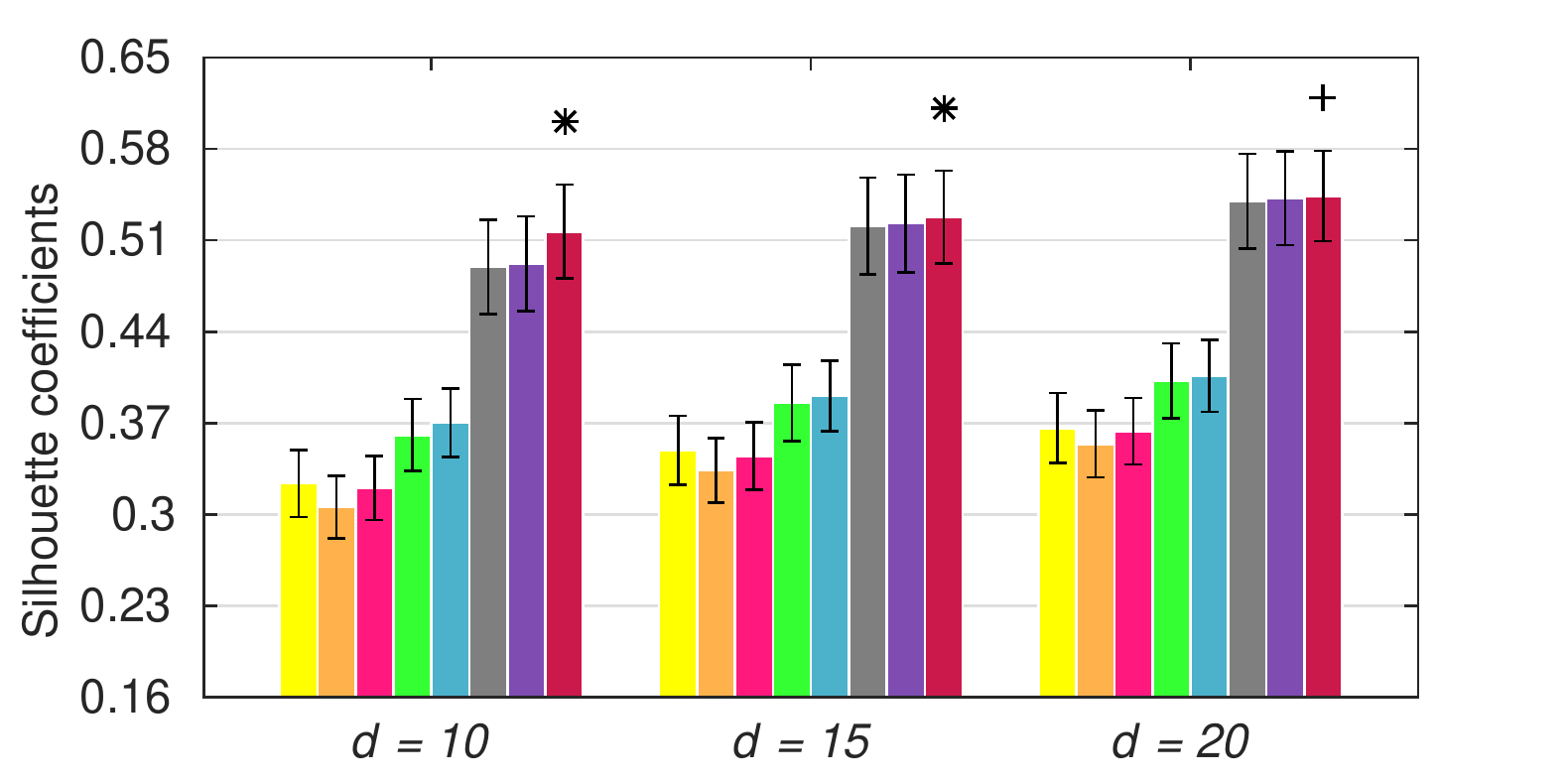} \kern-2.5em &  
\raisebox{.13\height}{\includegraphics[width=0.18\textwidth]{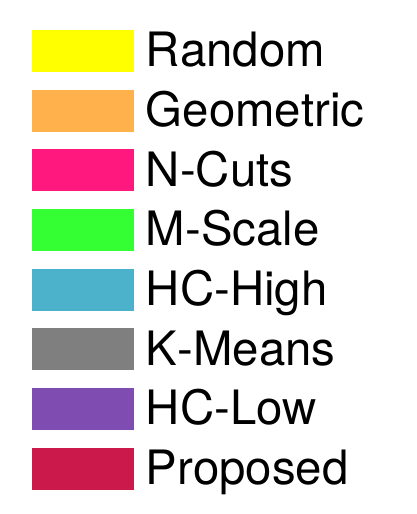}} \\
\end{tabular}
\caption[Homogeneity values and Silhouette coefficients based on structural connectivity.]{Homogeneity values and Silhouette coefficients based on structural connectivity estimated from dMRI. Error bars show the variability across subjects. Stars (*) and cross (+) indicate statistical significance between the winner and the runner-up with $p < 0.01$ and $p < 0.05$, respectively.}
\label{fig:str_quan}
\end{figure}

\begin{figure}[!tb]  
\centering
\begin{tabular}{cl}
\includegraphics[width=0.7\textwidth]{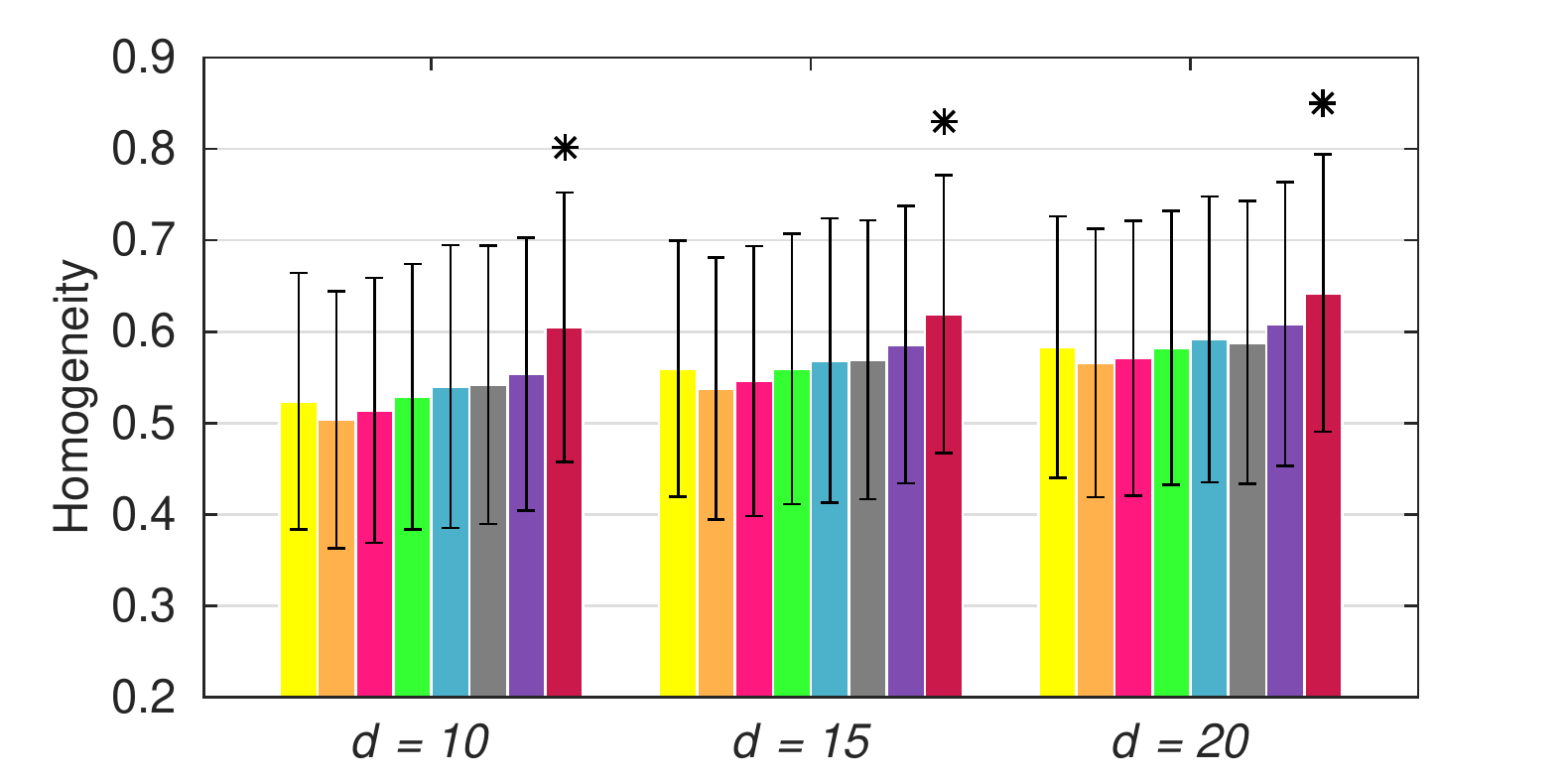} \kern-2.5em & \\ 
\includegraphics[width=0.7\textwidth]{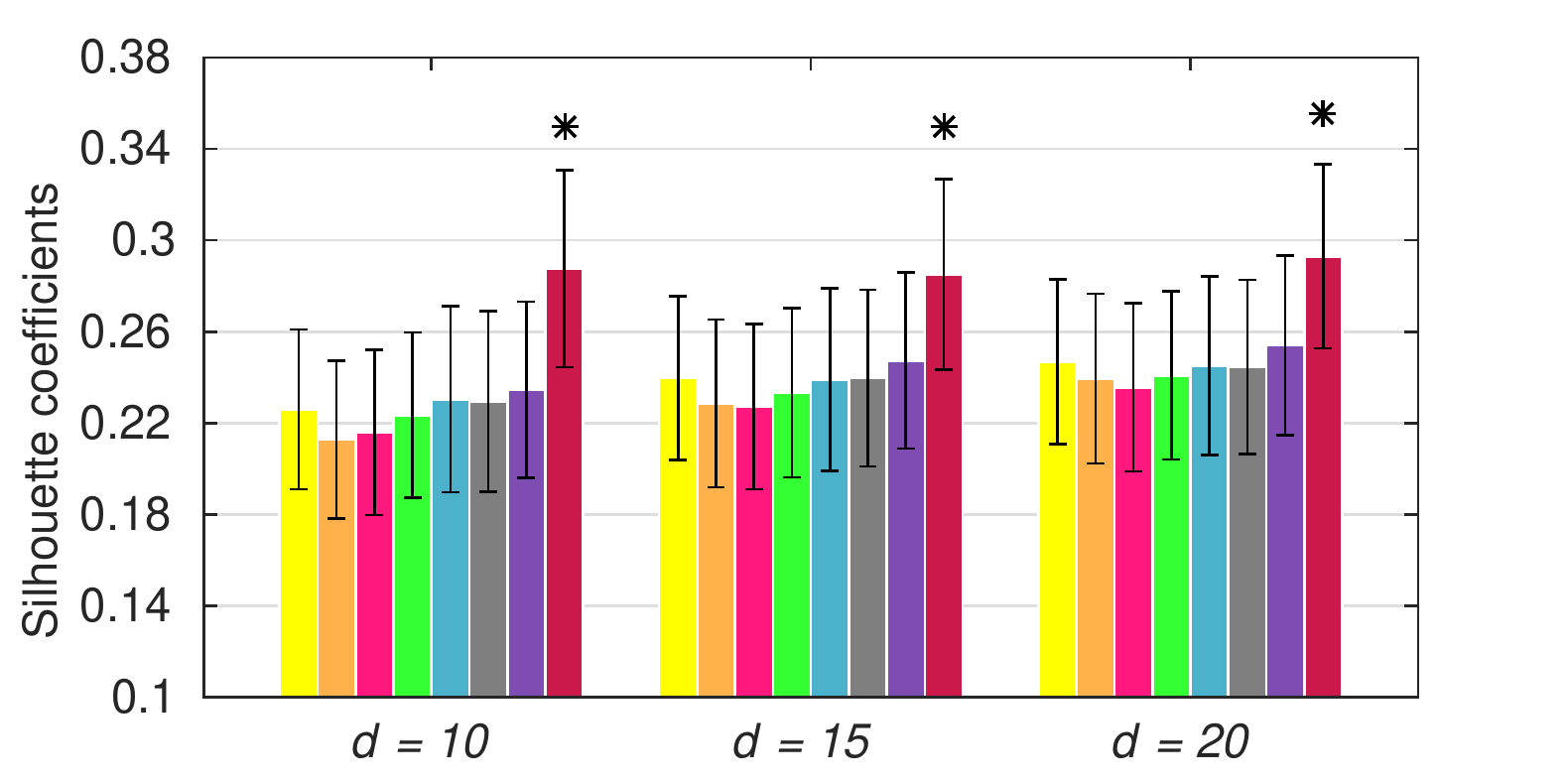} \kern-2.5em &  
\raisebox{.13\height}{\includegraphics[width=0.18\textwidth]{figs/manifold-legend}} \\
\end{tabular}
\caption[Homogeneity values and Silhouette coefficients based on functional connectivity.]{Homogeneity values and Silhouette coefficients based on functional connectivity estimated from resting-state fMRI. Error bars show the variability across subjects. Stars (*) indicate statistical significance between the winner and the runner-up with $p < 0.01$.}
\label{fig:func_quan}
\end{figure}

We present the cluster validity results, including homogeneity values and Silhouette coefficients, based on structural and functional connectivity data in Fig.~\ref{fig:str_quan} and Fig.~\ref{fig:func_quan}, respectively. Statistical significance of the results between the winner and the runner-up method is tested using a Wilcoxon sign-rank test. 

Fig.~\ref{fig:str_quan} shows that our method surpasses other approaches in terms of homogeneity and Silhouette coefficients. The results are statistically significant across all resolutions at $p<0.05$. It is followed by two clustering techniques, \textit{K-Means} and \textit{HC-Low}, both are driven by the low-dimensional embedding. This may indicate that, Laplacian eigenmaps can successfully reveal the intrinsic geometry of the underlying connectivity, and hence, provides a more robust set of features towards parcellating the cortical surface. In addition, the way we utilize discrete eigenvectors
for boundary mapping allows the delineation of more distinct parcels compared to the
traditional clustering approaches directly applied to the spectral coordinates. Considering the results obtained by \textit{HC-High}, we can infer that non-linear dimensionality reduction can identify local connectivity patterns which may not be directly detected in the high dimensional space. 

On the other hand, \textit{M-Scale} and \textit{N-Cuts} can provide accurate parcellations only to some extent and are even surpassed by \textit{Random} for most of the resolutions. This may be linked to the fact that spatially-constrained spectral techniques are more likely to capture the structure of the cortical mesh rather than the connectivity patterns~\cite{Thirion14}, since only the spatially-adjacent vertices are taken into consideration while constructing the connectivity matrix. \textit{Geometric} unsurprisingly yields the least favourite performance regardless of the resolution, potentially indicating that spatial information alone may not be sufficient to reveal the connectivity-based organisation of the cerebral cortex.

The difference in performance between our approach and the others becomes more prominent with the resting-state functional connectivity results as indicated in Fig.~\ref{fig:func_quan}. Both homogeneity and Silhouette analysis show a tendency in favour of the proposed method. Despite the fact that parcellations derived from dMRI do not necessarily represent the functional connectivity estimated with rs-fMRI, these trends indicate that the proposed method better reflect the underlying function with respect to the other approaches. \textit{HC-Low} is the only other method that consistently shows a higher performance than \textit{Random}. Other parcellations, although yield homogeneous regions to some degree, generally fail to separate vertices with different signals from each other, as indicated by Silhouette coefficients.  

\begin{figure}[!thb]
\centering
\begin{tabular}{lccc}

& $K=103$ & $K=124$ & $K=142$ \\

\raisebox{1.0em}{\rotatebox{90}{\textit{Proposed}}} &
\includegraphics[width=0.2\textwidth]{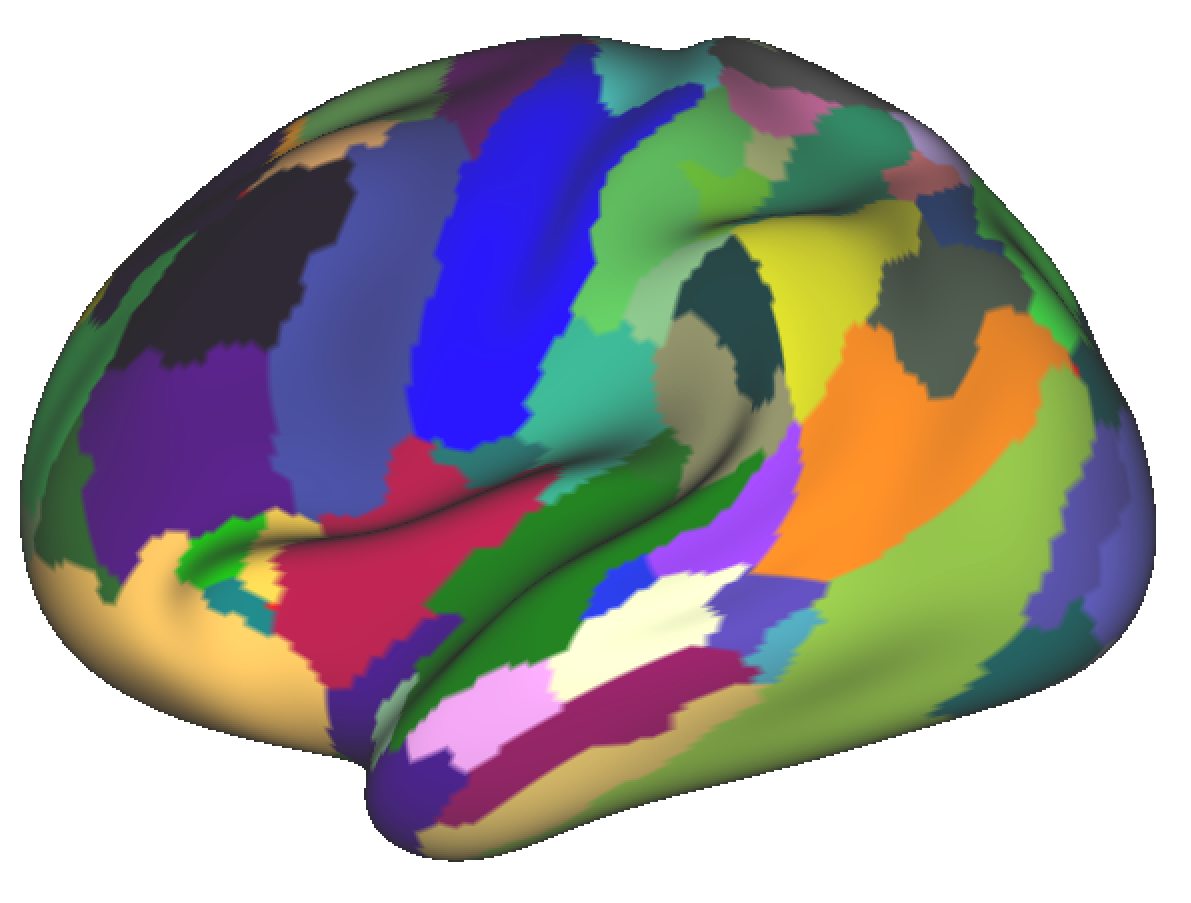} &
\includegraphics[width=0.2\textwidth]{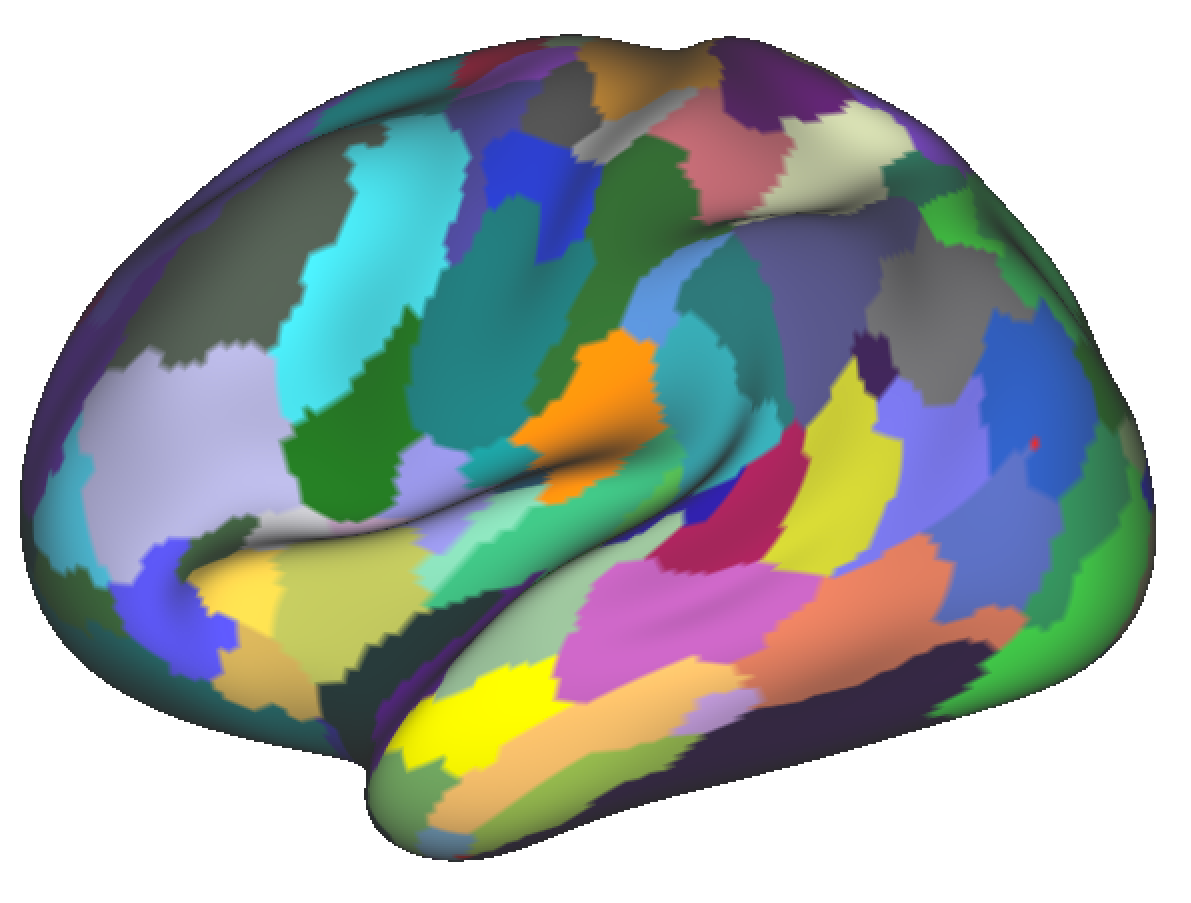} & 
\includegraphics[width=0.2\textwidth]{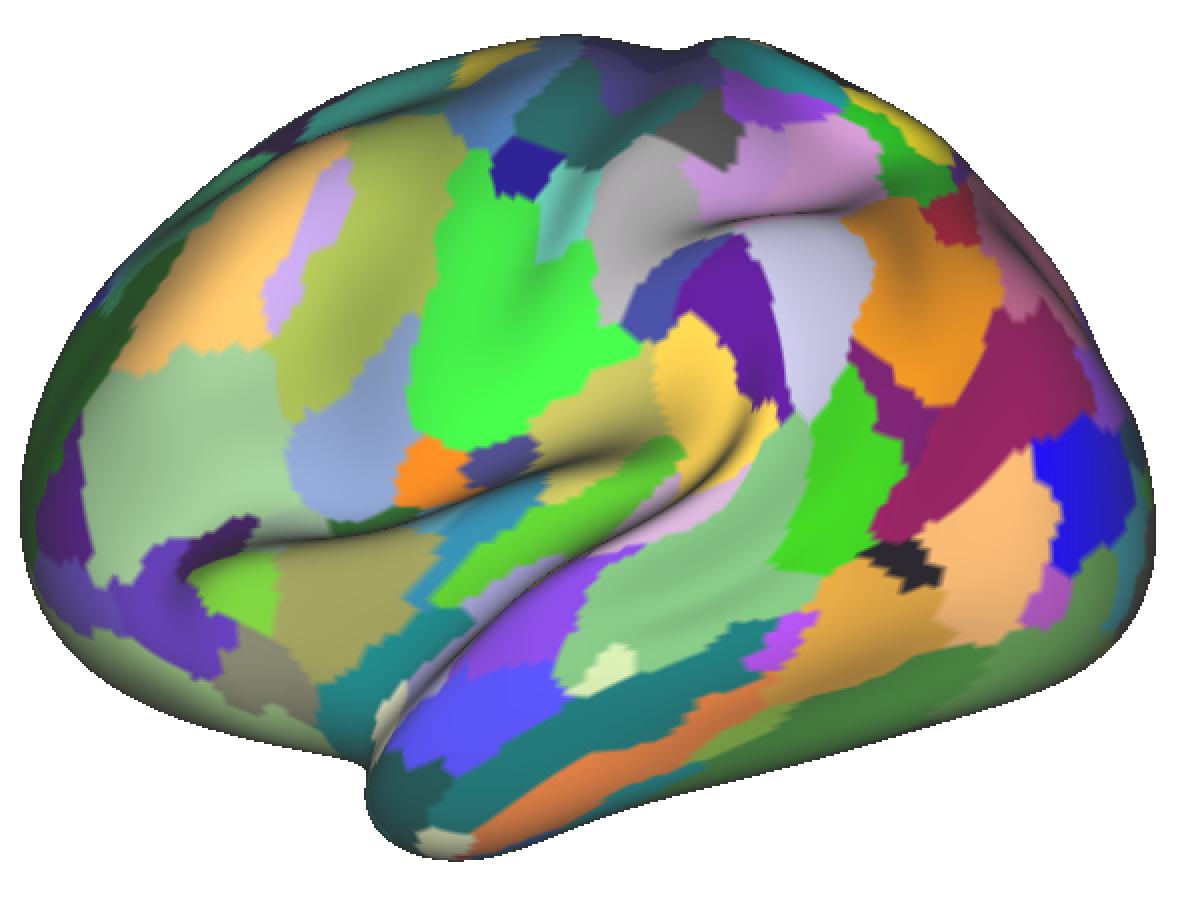} \\

\raisebox{1.2em}{\rotatebox{90}{\textit{HC-Low}}} &
\includegraphics[width=0.2\textwidth]{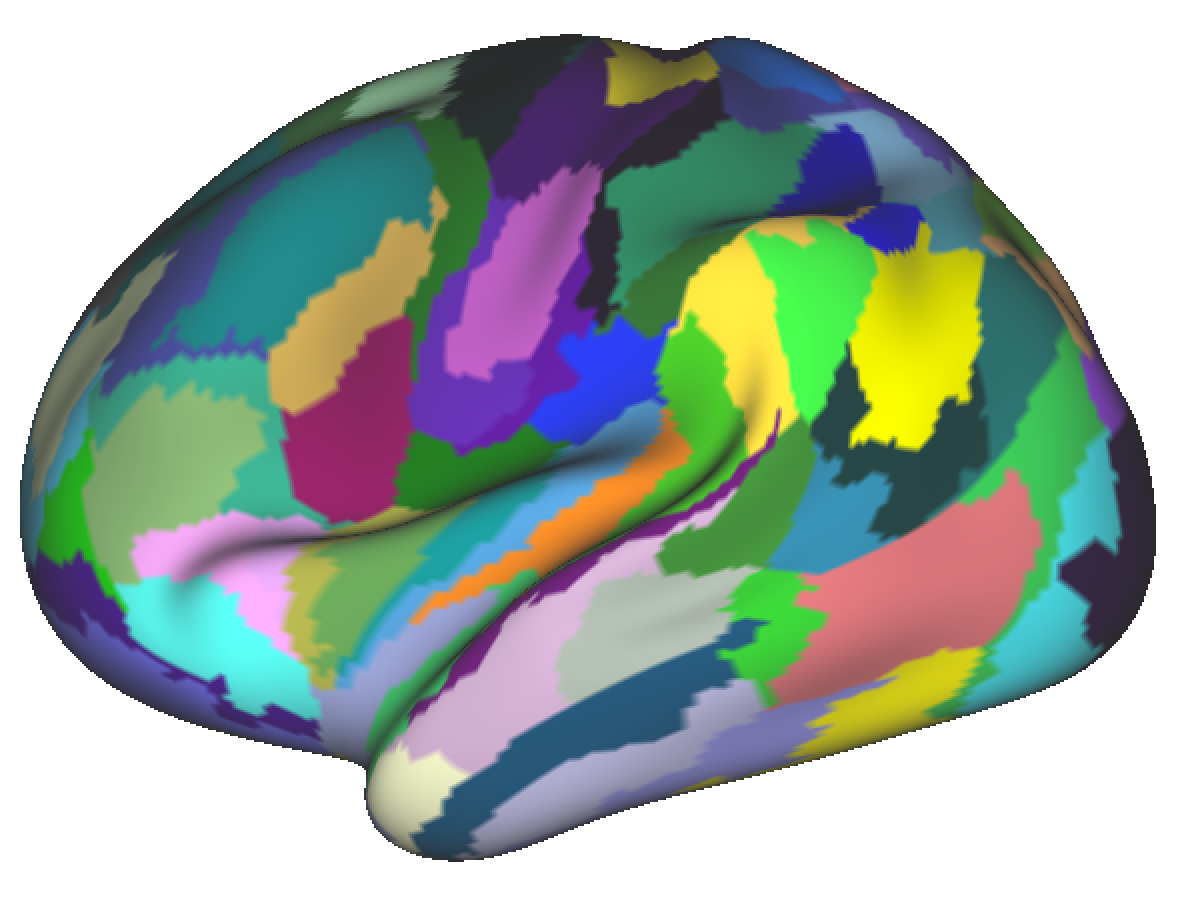} &
\includegraphics[width=0.2\textwidth]{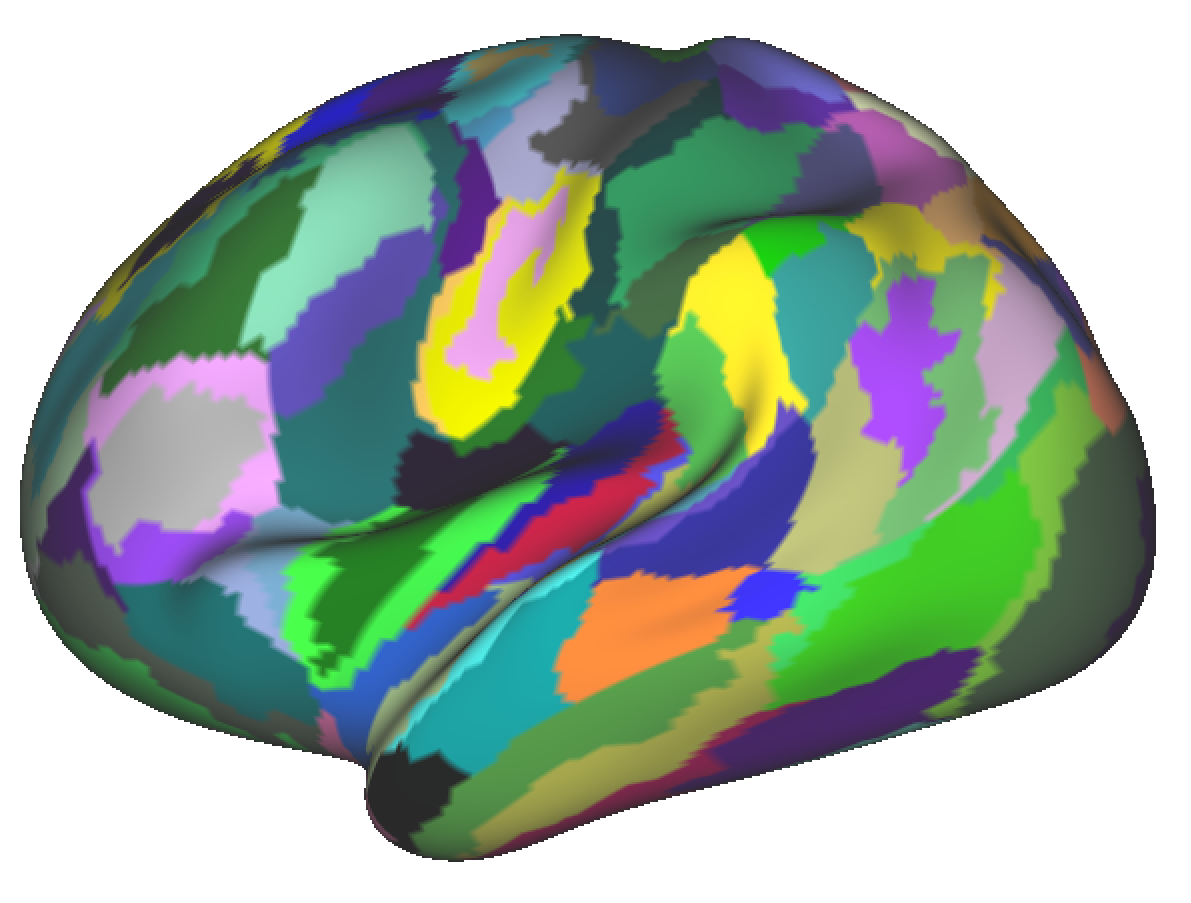} & 
\includegraphics[width=0.2\textwidth]{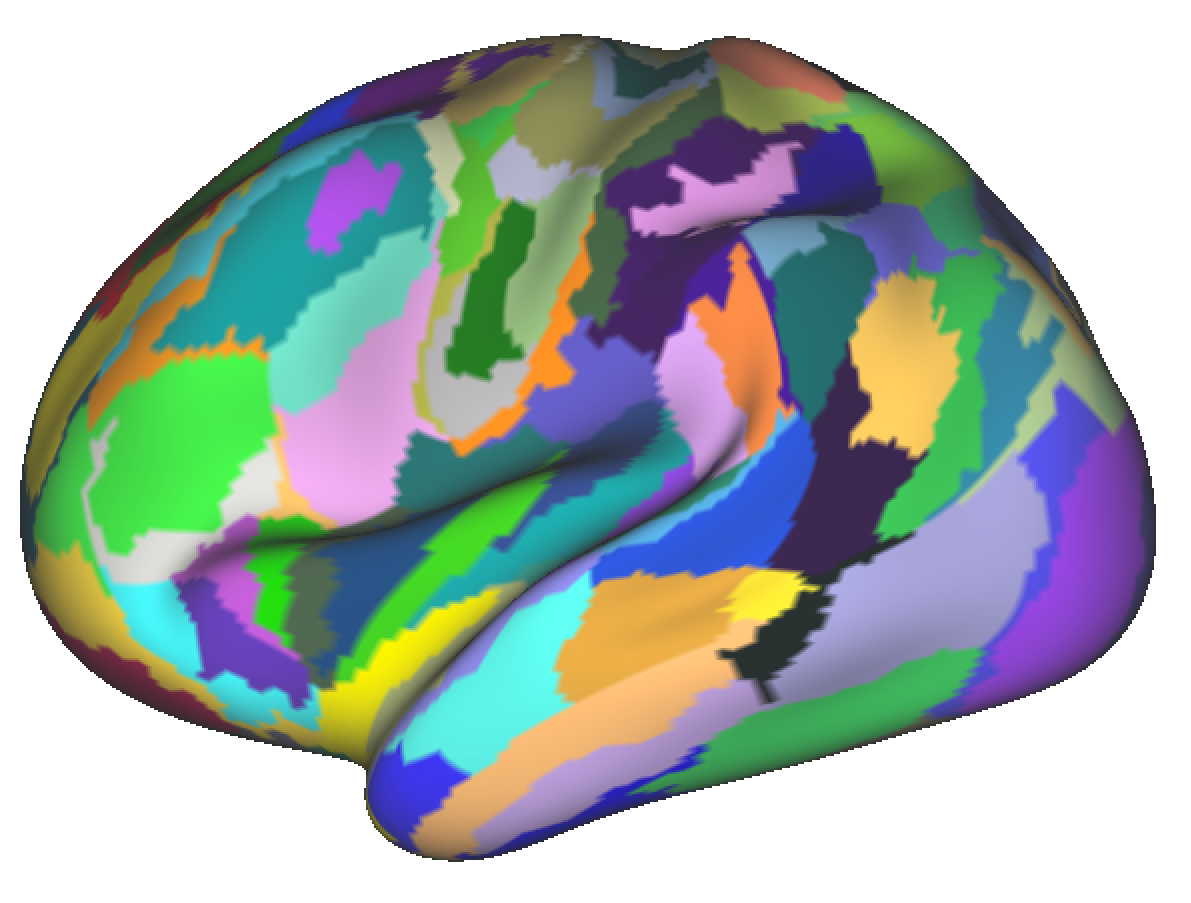} \\

\raisebox{1.0em}{\rotatebox{90}{\textit{K-Means}}} &
\includegraphics[width=0.2\textwidth]{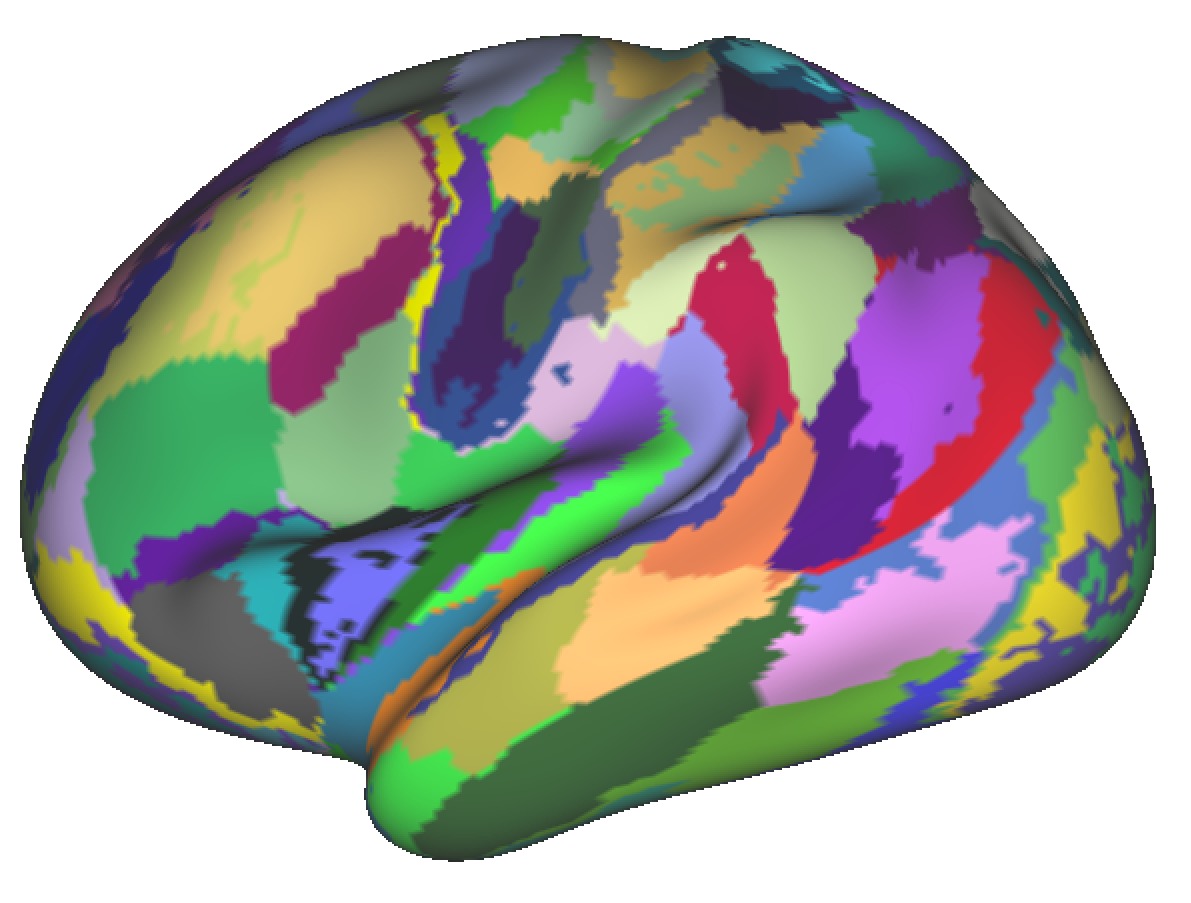} &
\includegraphics[width=0.2\textwidth]{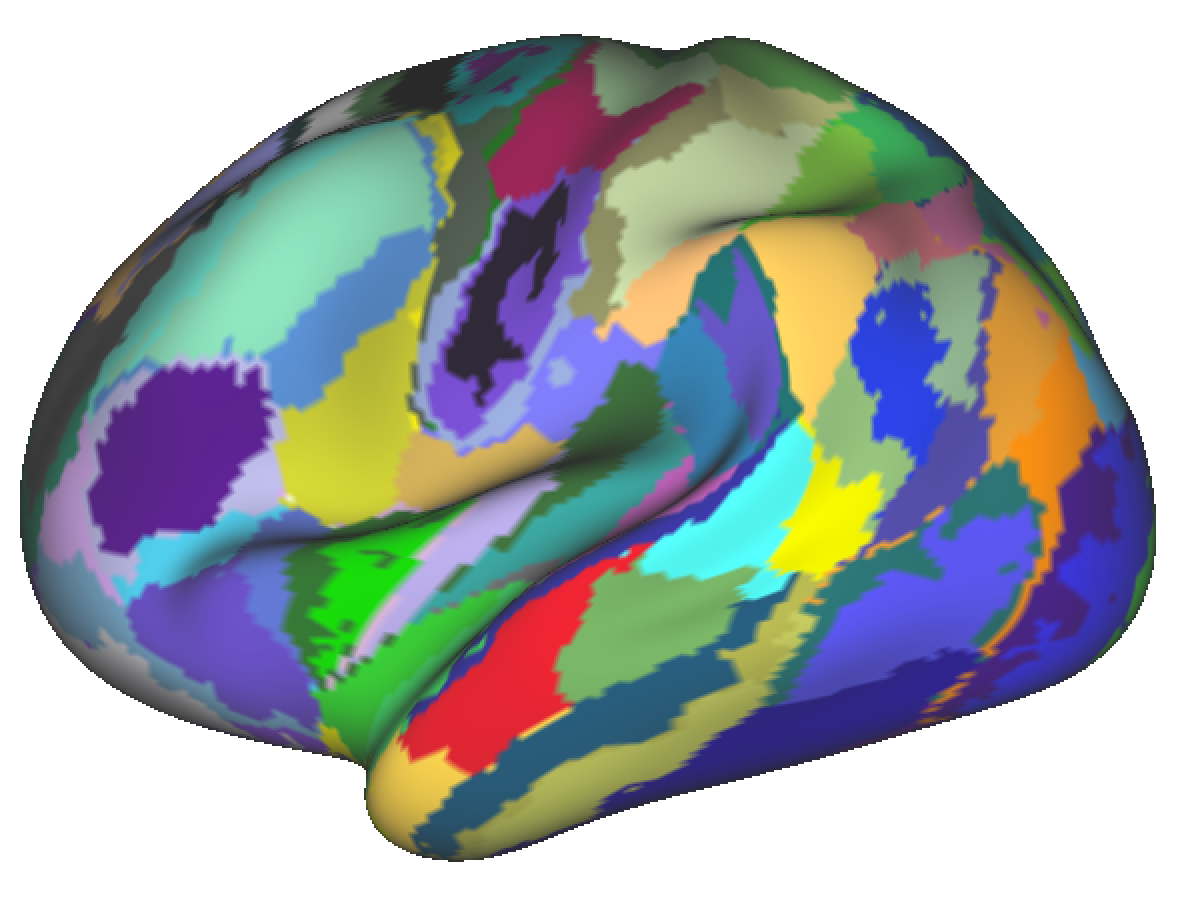} & 
\includegraphics[width=0.2\textwidth]{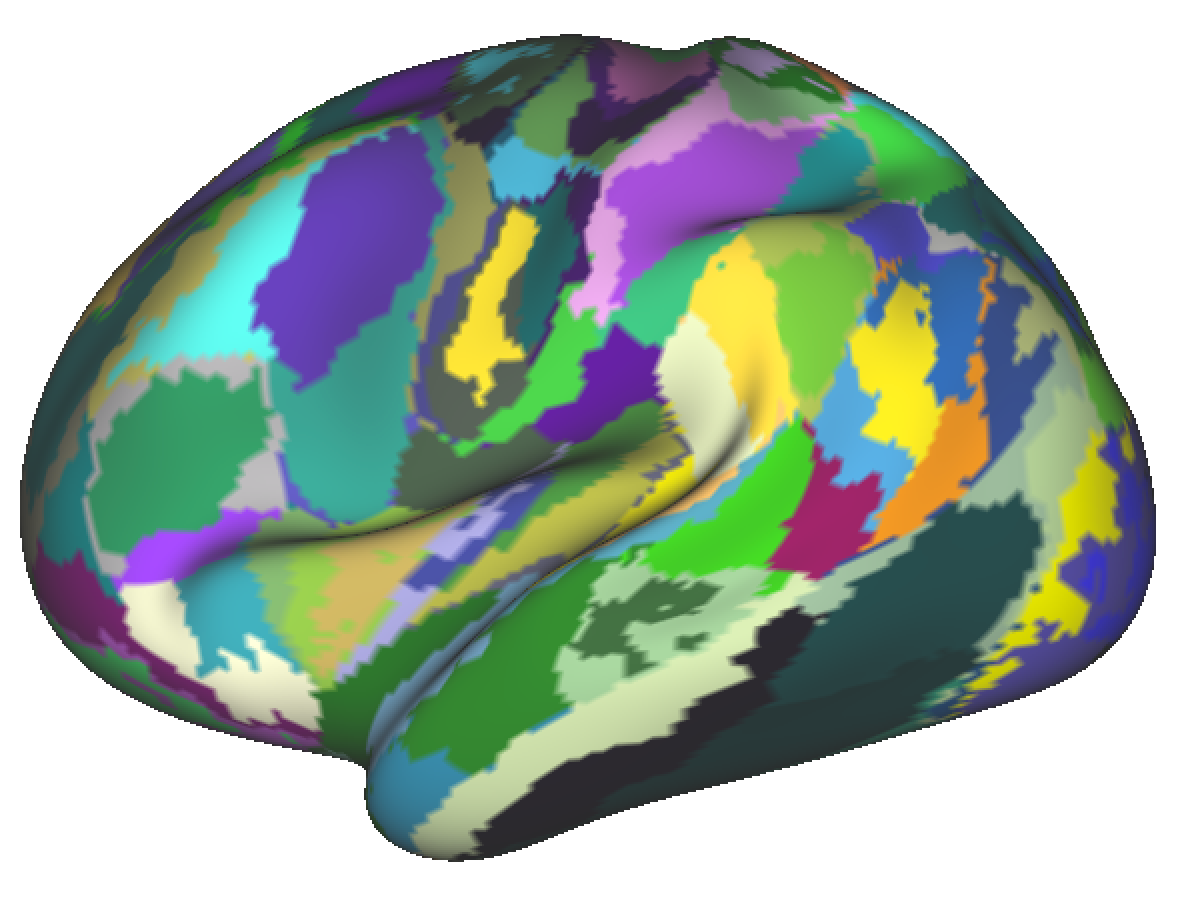} \\

\raisebox{1.0em}{\rotatebox{90}{\textit{HC-High}}} &
\includegraphics[width=0.2\textwidth]{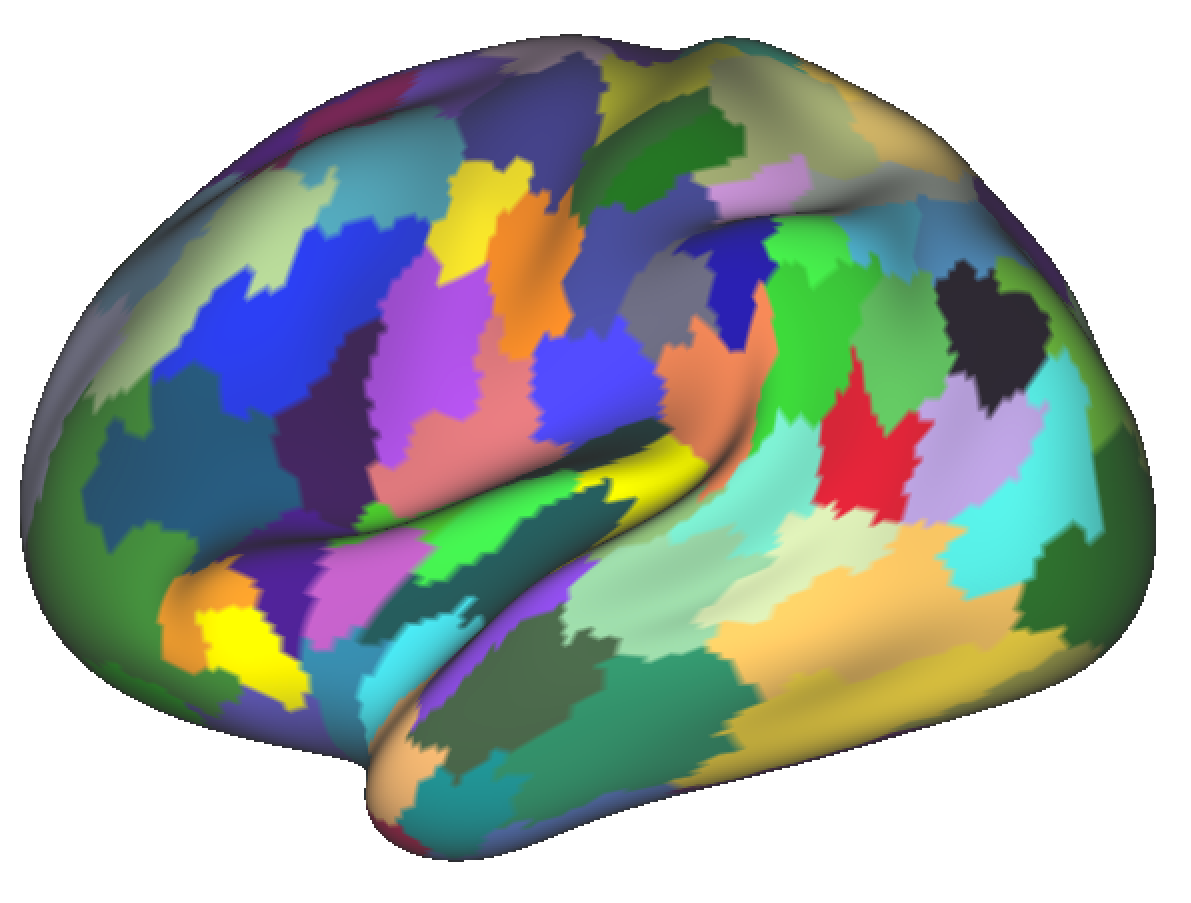} &
\includegraphics[width=0.2\textwidth]{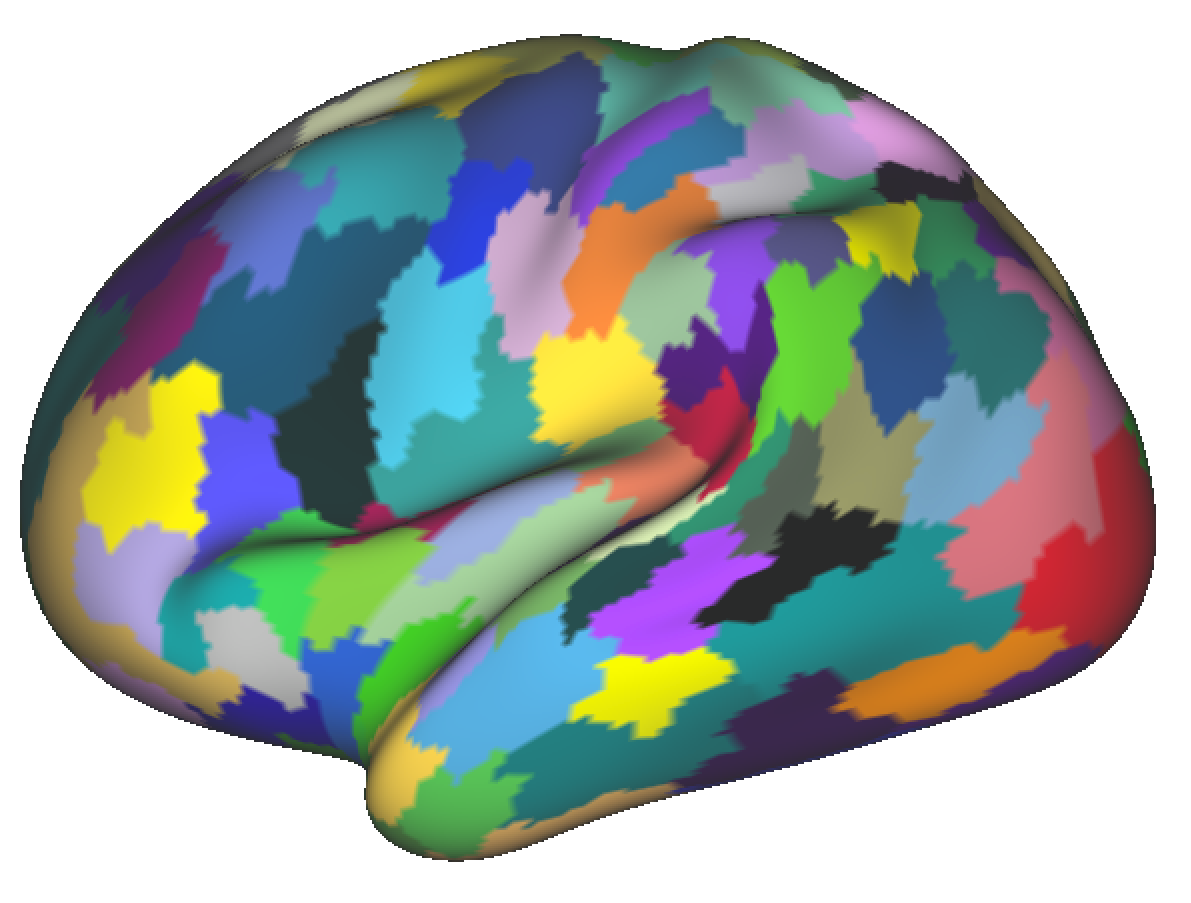} & 
\includegraphics[width=0.2\textwidth]{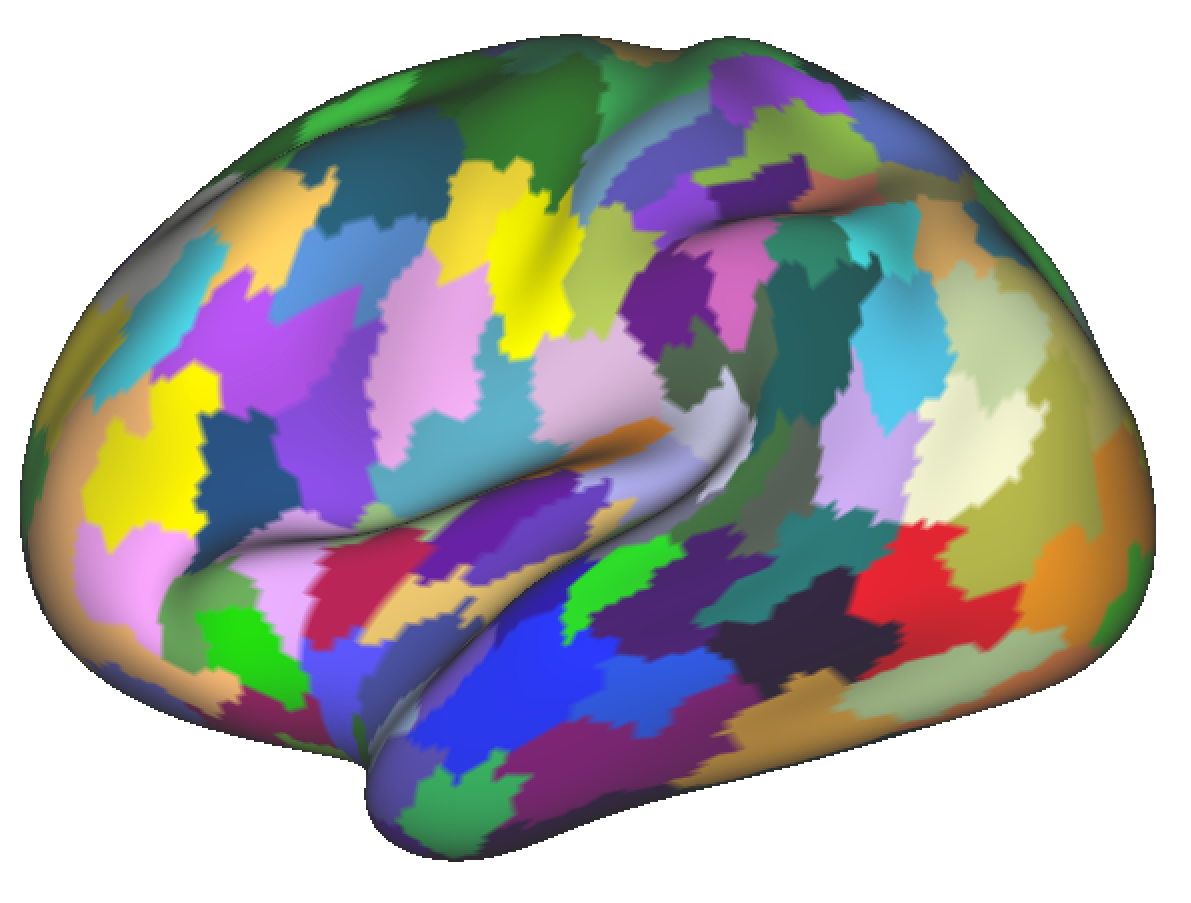} \\

\raisebox{1.4em}{\rotatebox{90}{\textit{M-Scale}}} &
\includegraphics[width=0.2\textwidth]{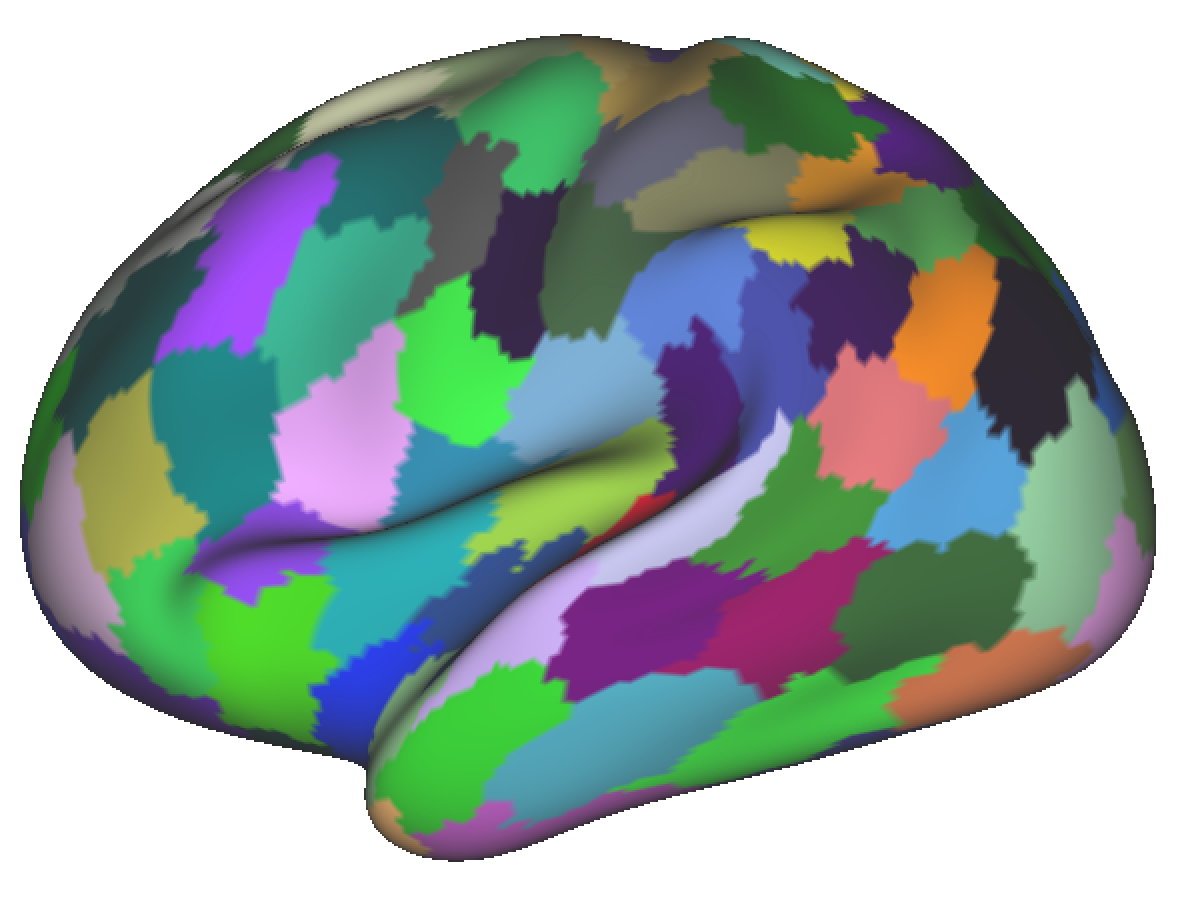} &
\includegraphics[width=0.2\textwidth]{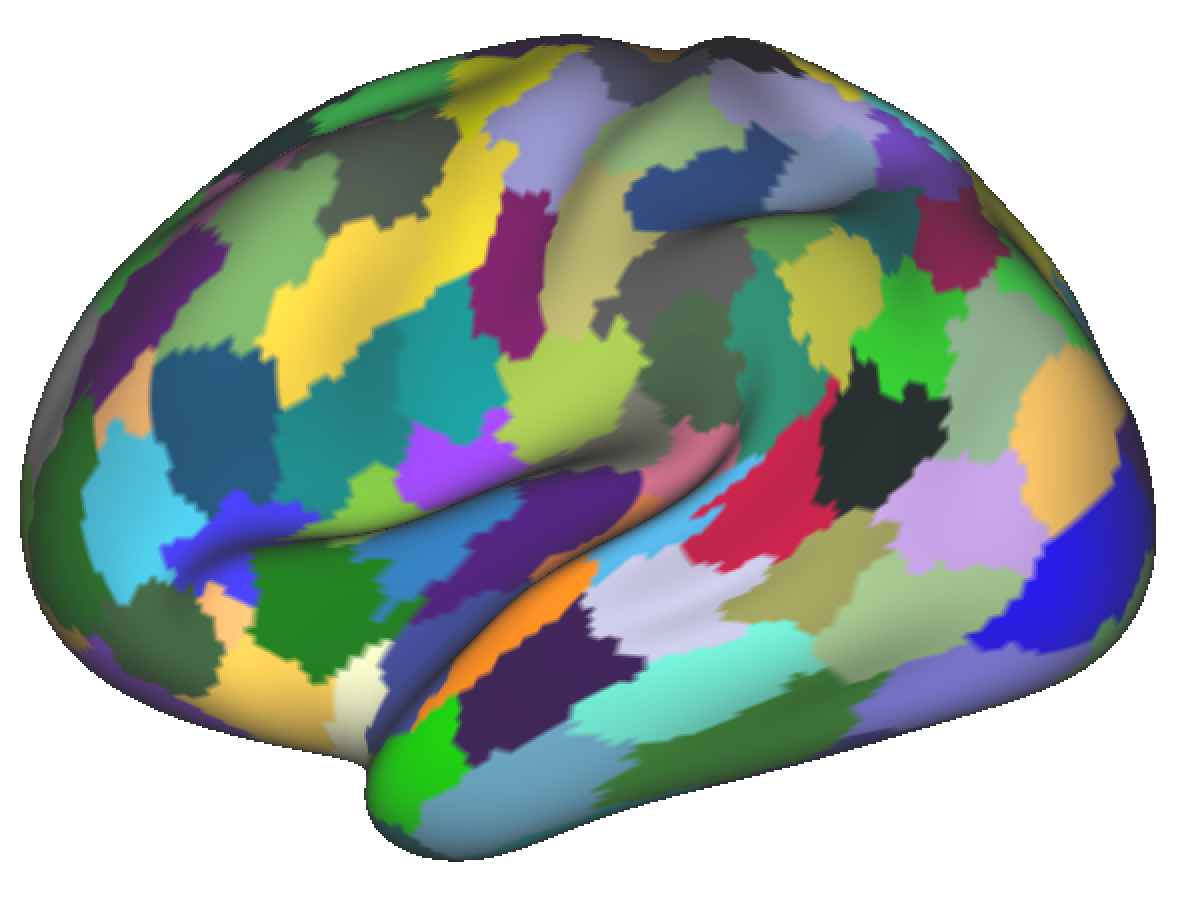} & 
\includegraphics[width=0.2\textwidth]{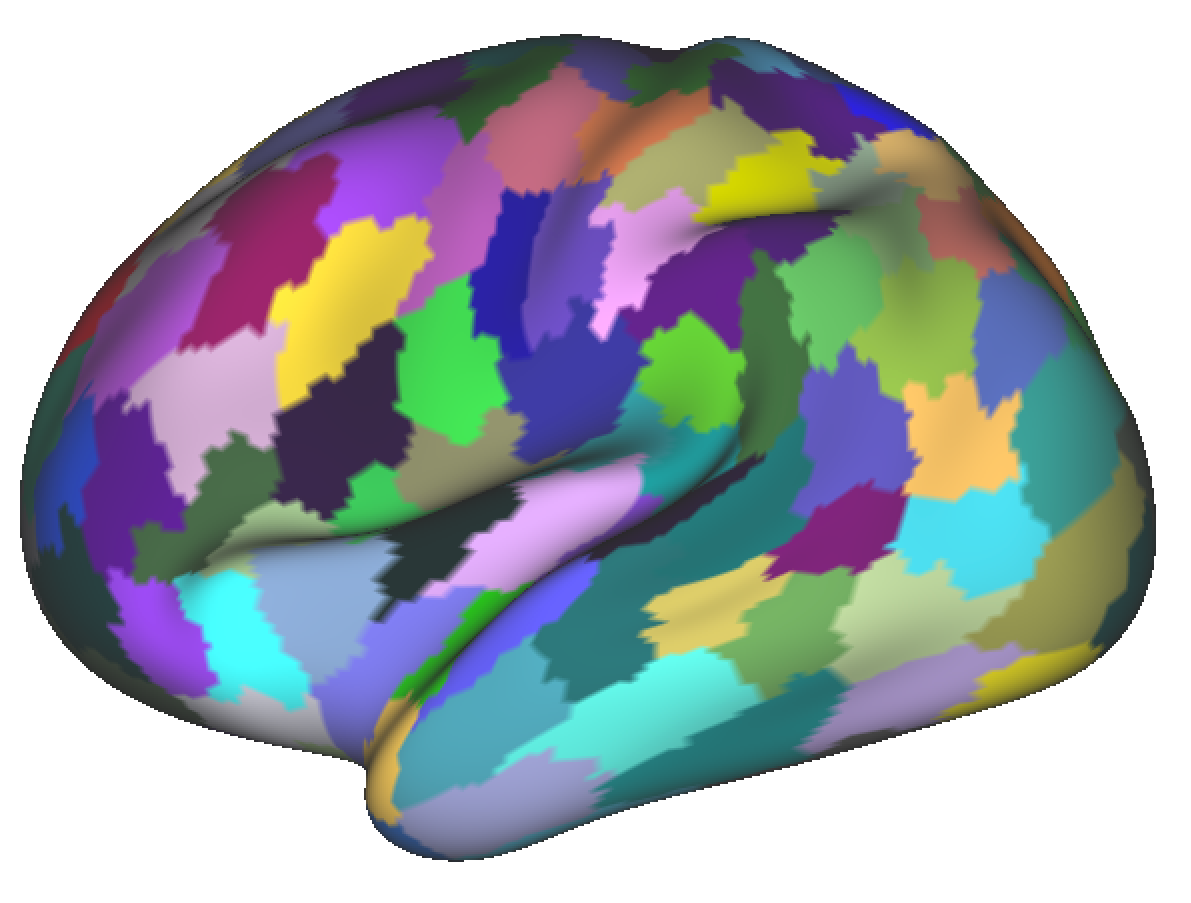} \\

\raisebox{1.5em}{\rotatebox{90}{\textit{N-Cuts}}} &
\includegraphics[width=0.2\textwidth]{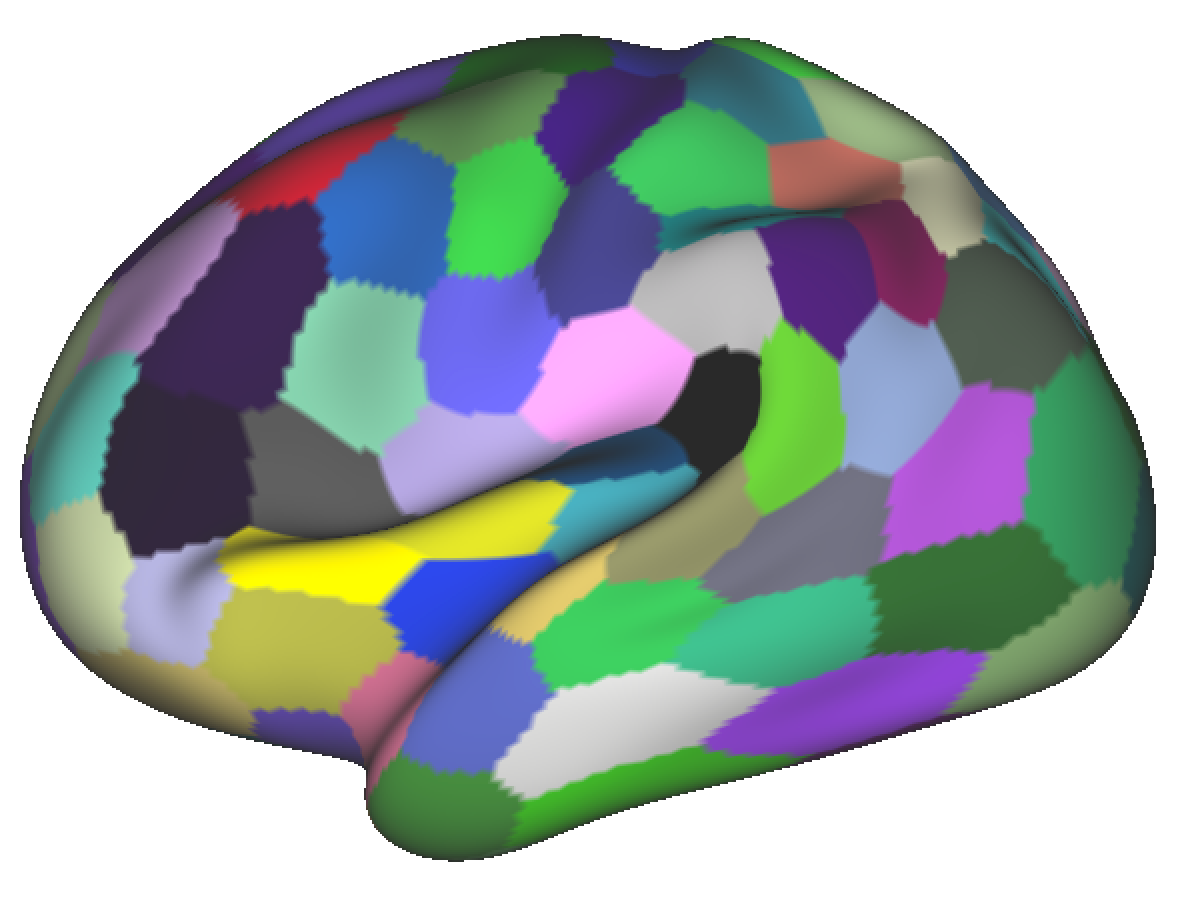} &
\includegraphics[width=0.2\textwidth]{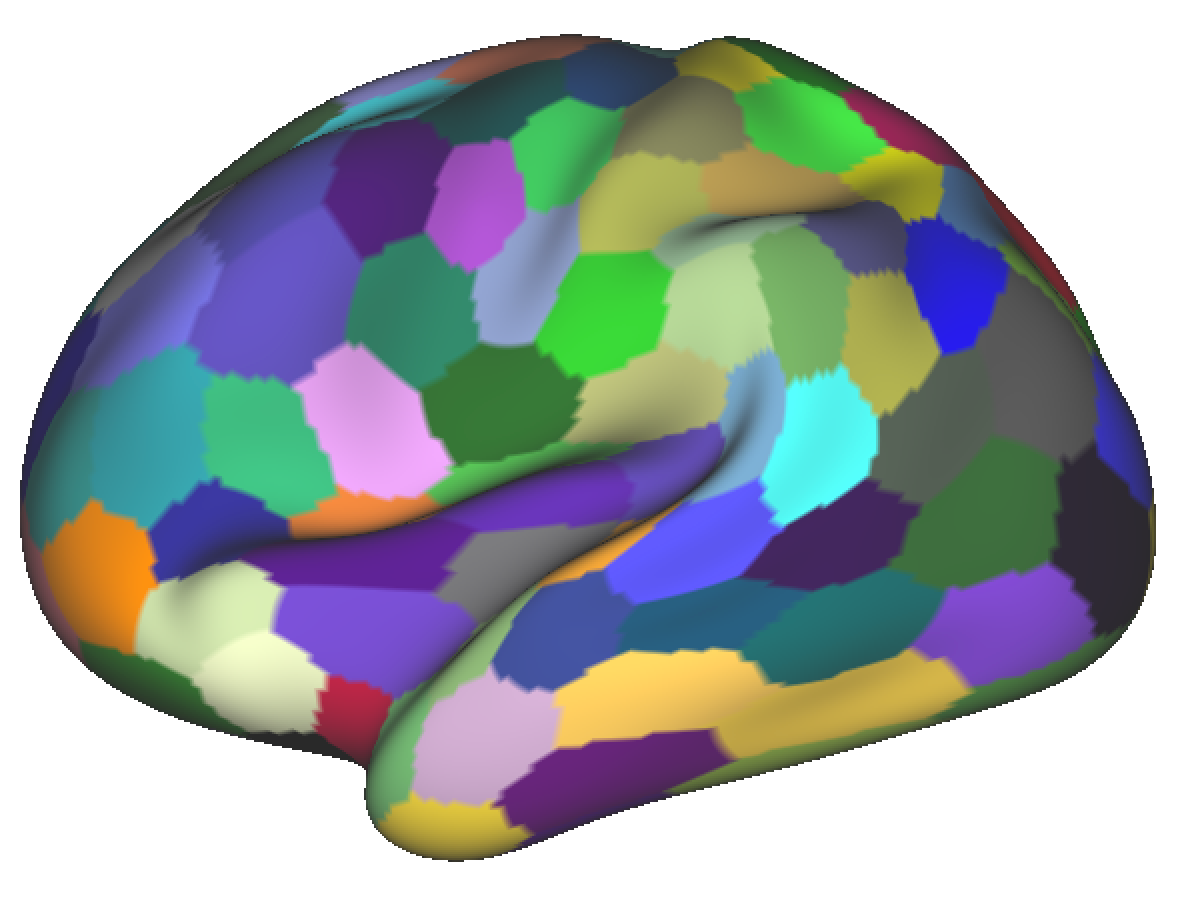} & 
\includegraphics[width=0.2\textwidth]{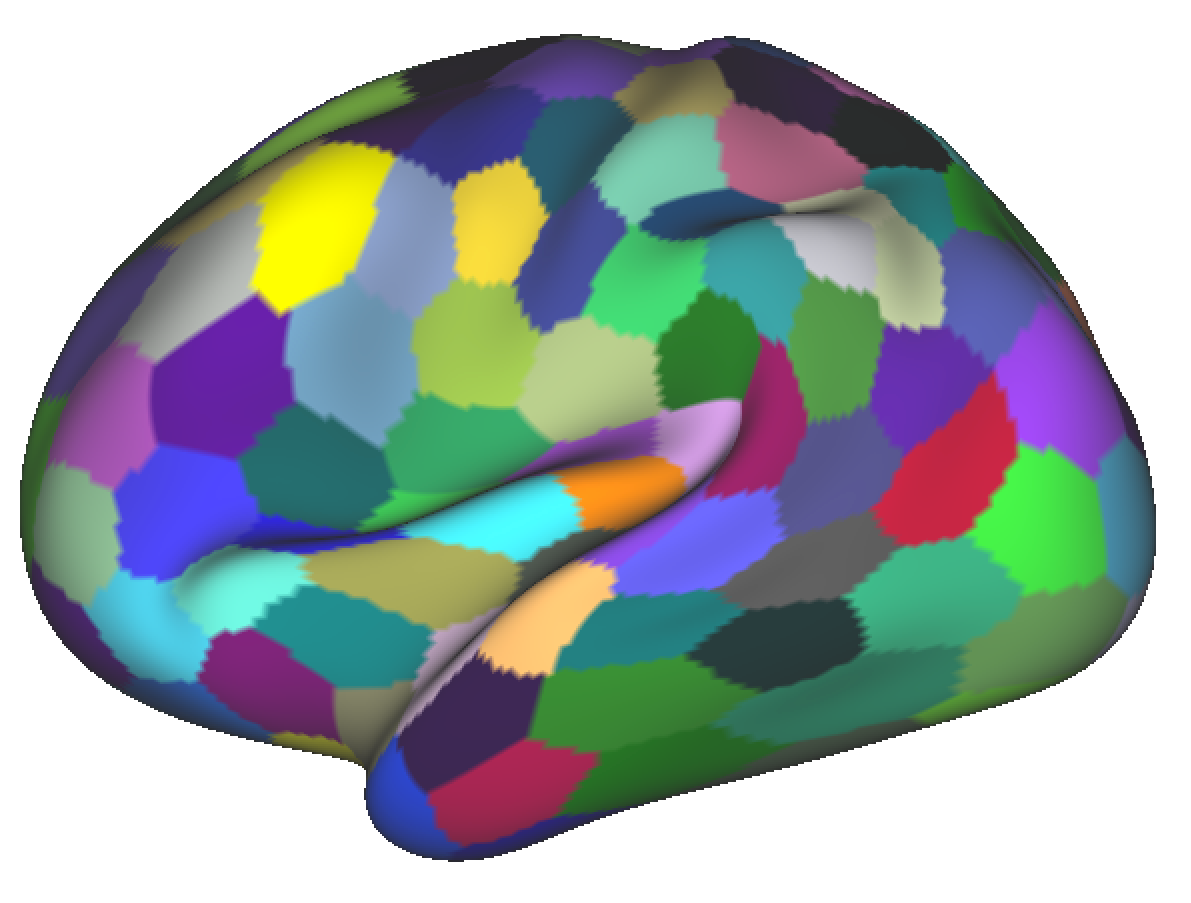} \\

\raisebox{0.9em}{\rotatebox{90}{\textit{Geometric}}} &
\includegraphics[width=0.2\textwidth]{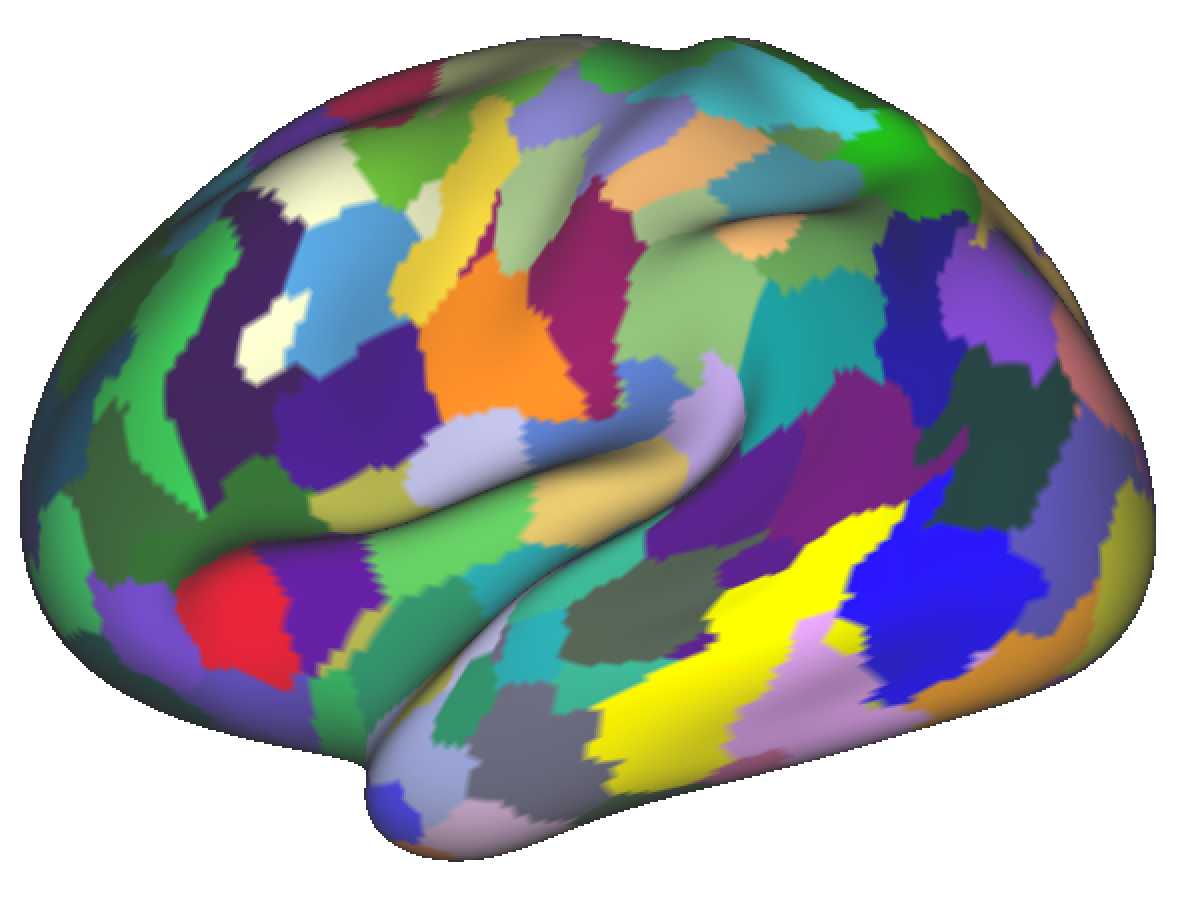} &
\includegraphics[width=0.2\textwidth]{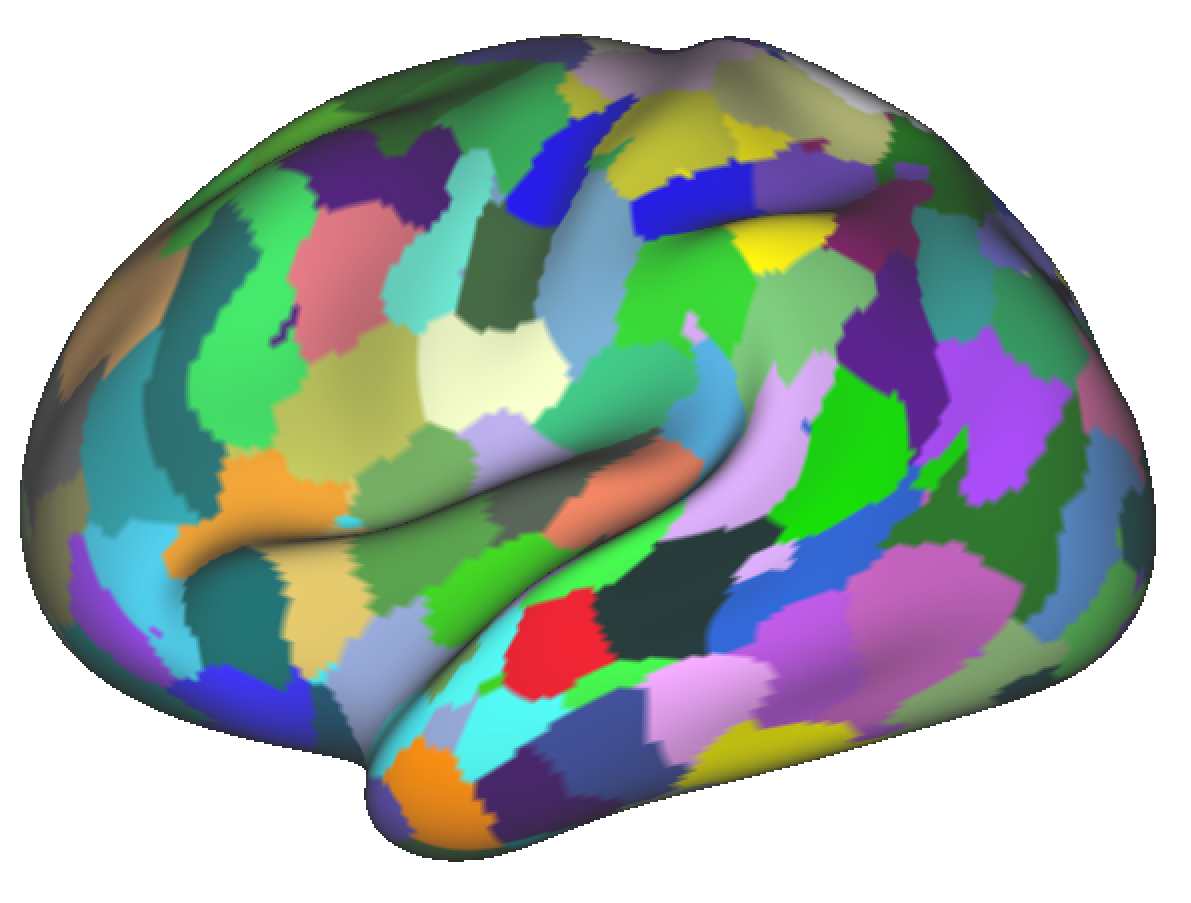} & 
\includegraphics[width=0.2\textwidth]{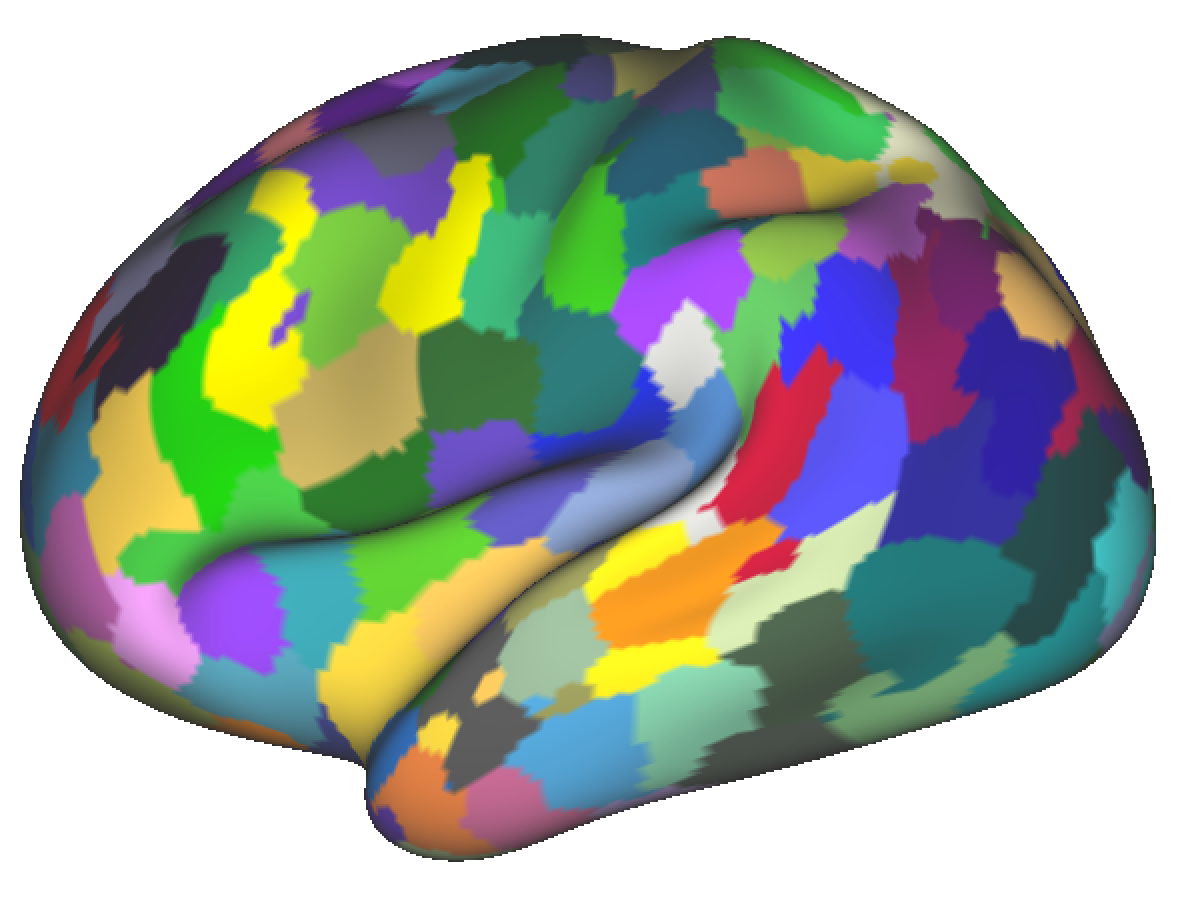} \\

\raisebox{1.2em}{\rotatebox{90}{\textit{Random}}} &
\includegraphics[width=0.2\textwidth]{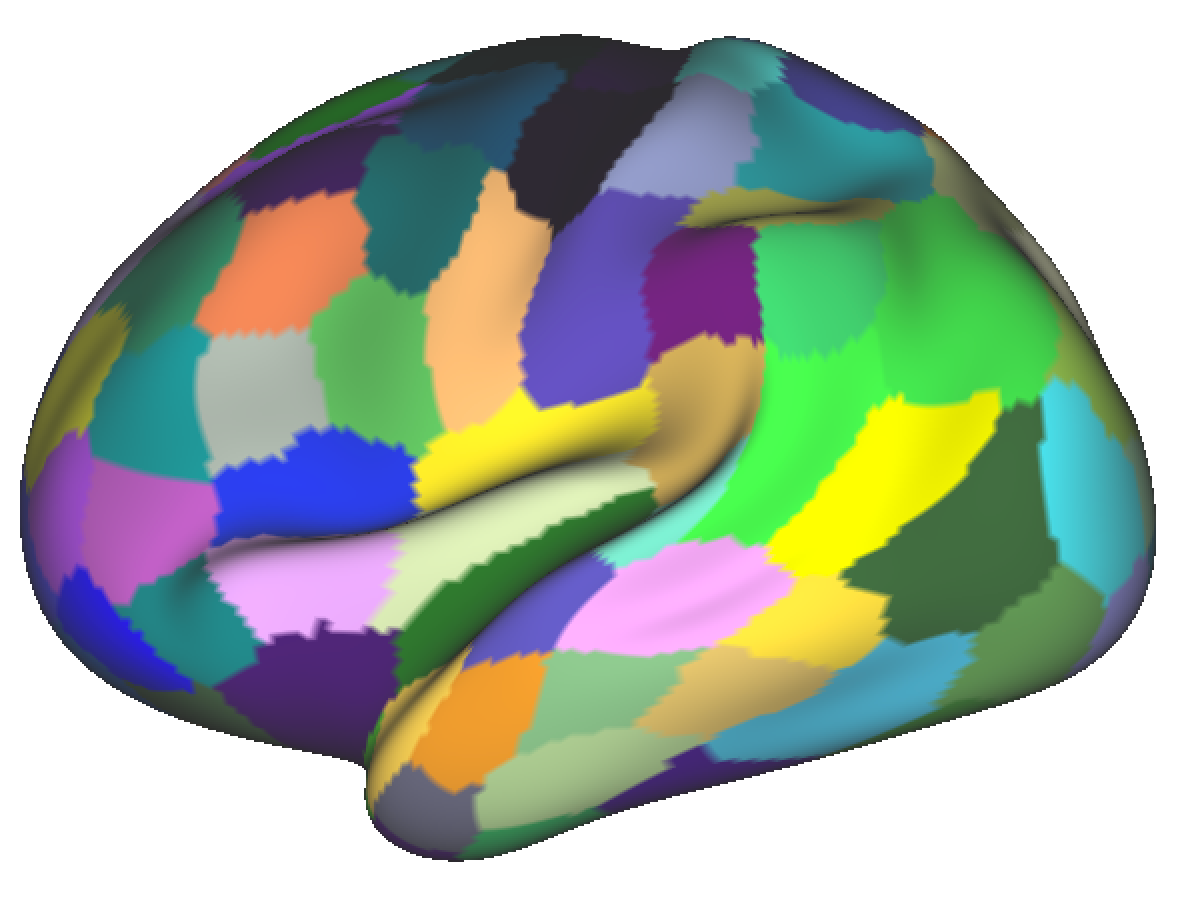} &
\includegraphics[width=0.2\textwidth]{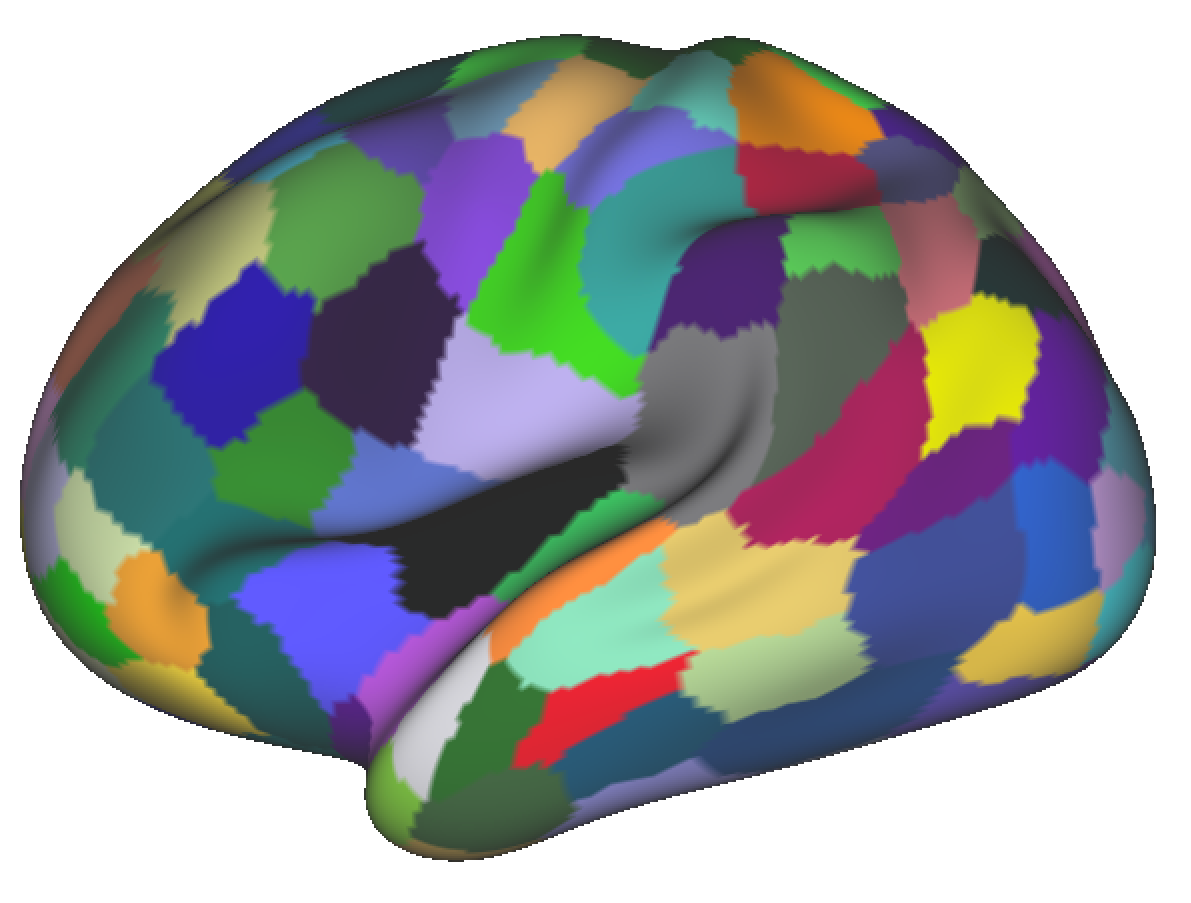} & 
\includegraphics[width=0.2\textwidth]{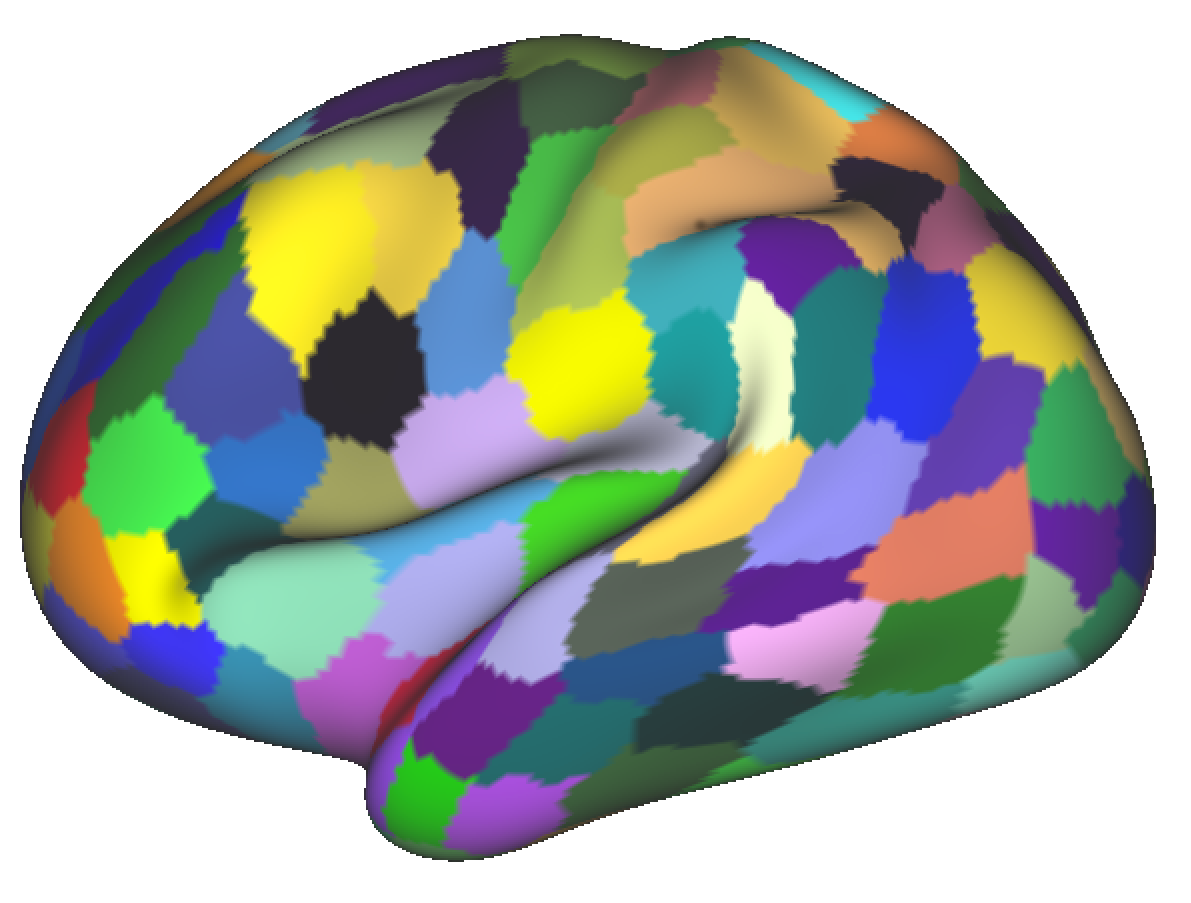} \\

\end{tabular}
\caption[Parcellations of the left lateral cortex derived from one subject.]{Parcellations of the lateral cortex of the left hemisphere derived from one subject. $K$ is determined by the number of eigenvectors used in the proposed framework.}
\label{fig:dmri-subject-visual}
\end{figure}

\subsection{Visual and Inter-Modality Assessment}
\begin{figure}[!tb]
\centering
\begin{tabular}{cccc}
\includegraphics[width=0.21\textwidth]{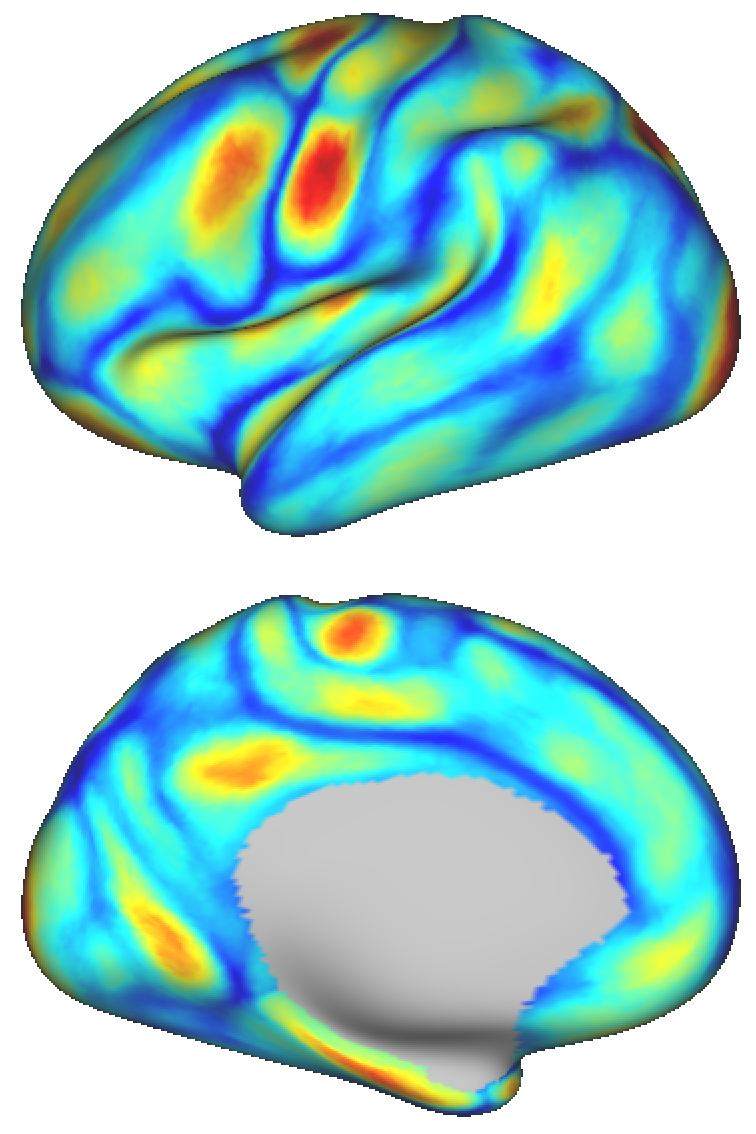}         & \includegraphics[width=0.21\textwidth]{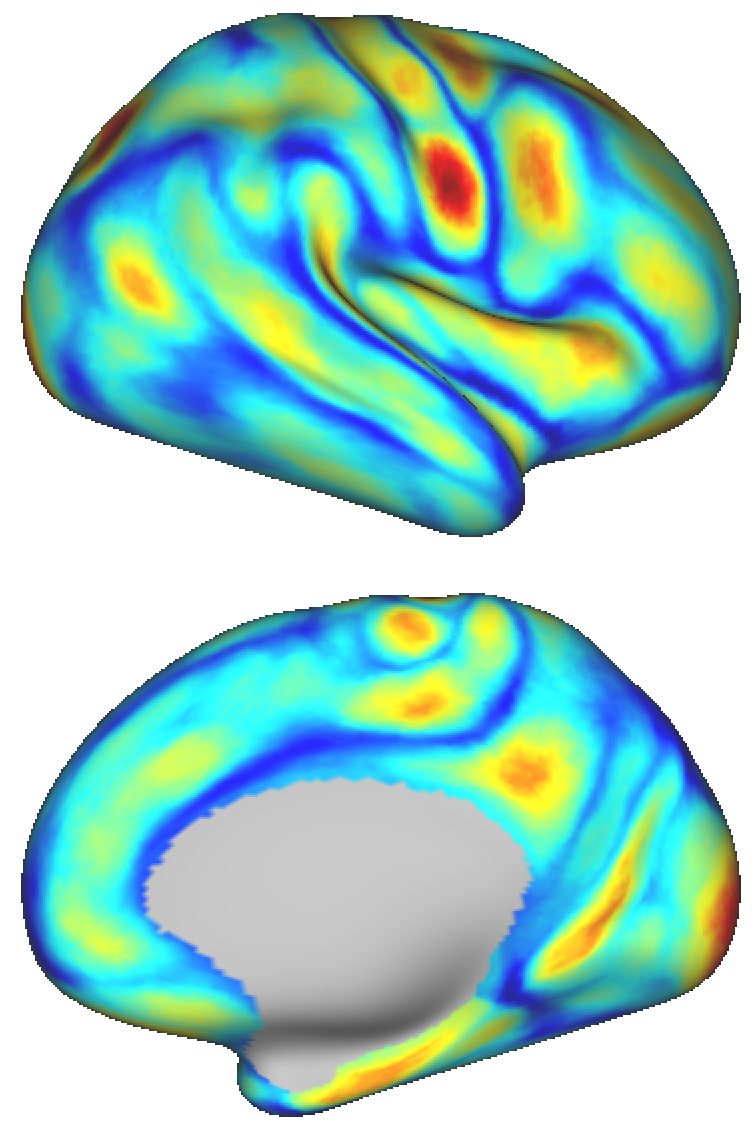}         & ~~ \includegraphics[width=0.21\textwidth]{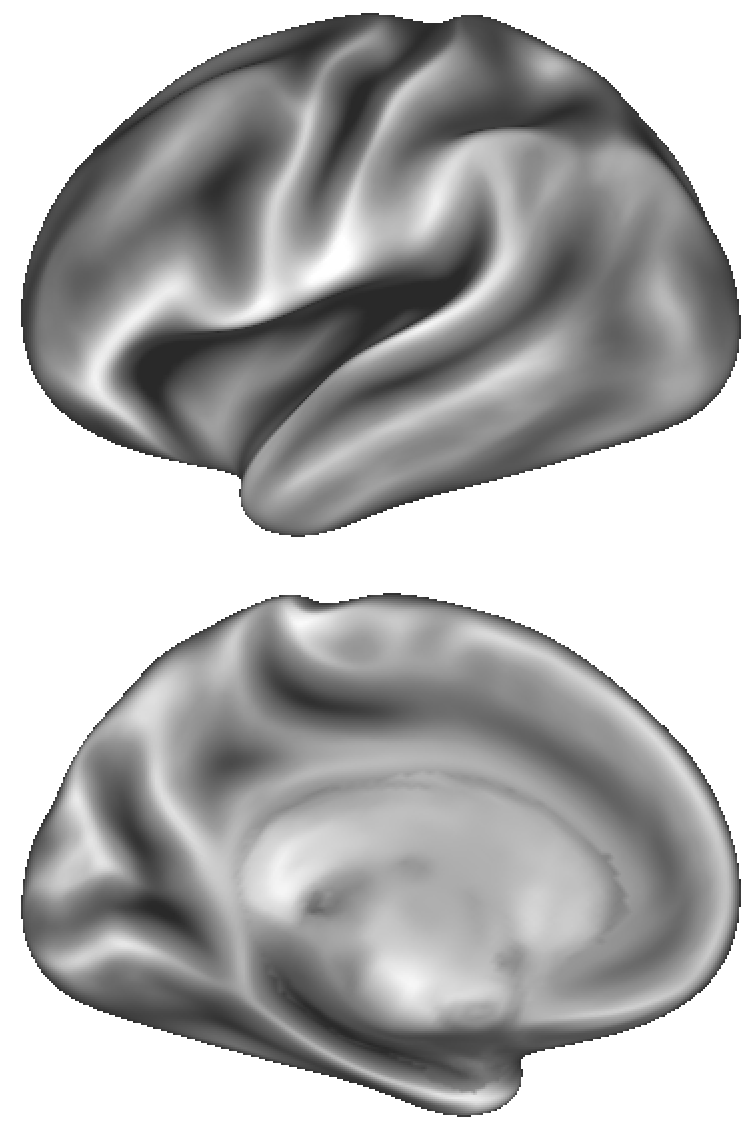}          		     & \includegraphics[width=0.21\textwidth]{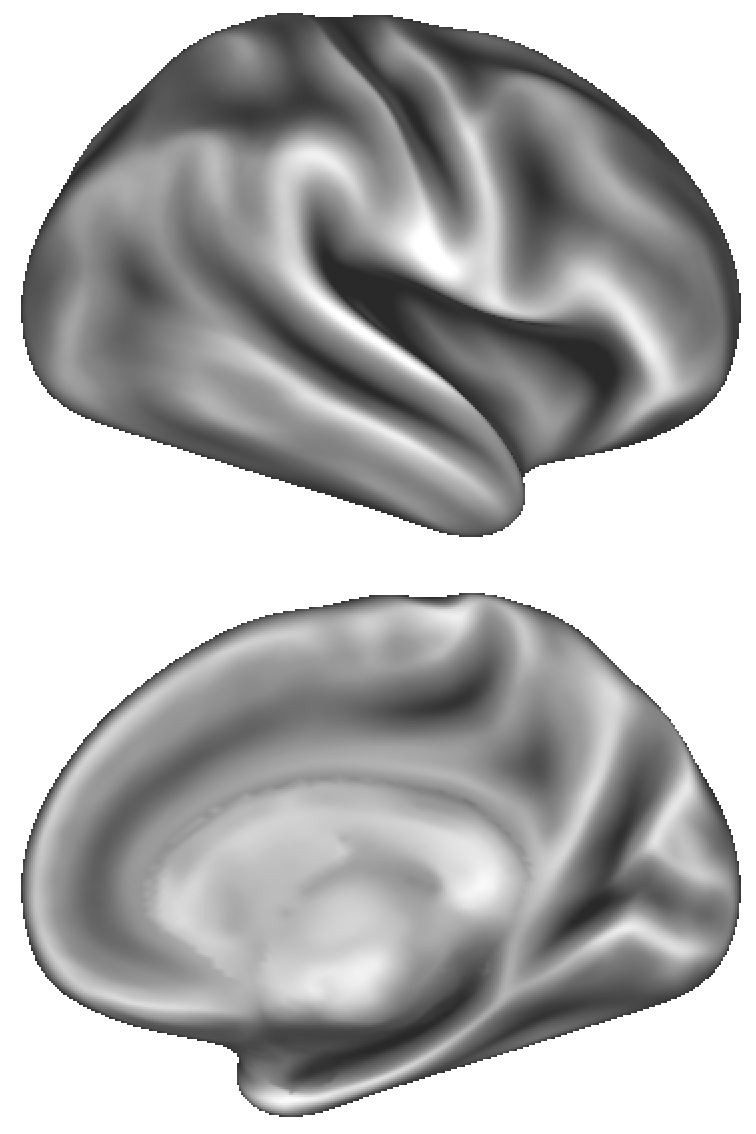}         			 \\
\multicolumn{2}{c}{\includegraphics[width=0.3\textwidth]{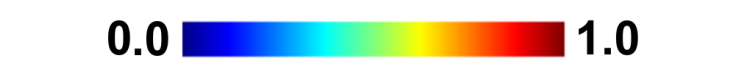}} & \multicolumn{2}{c}{\includegraphics[width=0.3\textwidth]{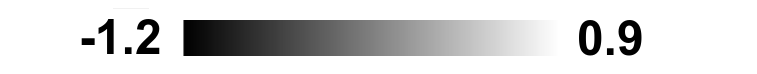}} \\
\multicolumn{2}{c}{(a)}    & \multicolumn{2}{c}{(b)}   
\end{tabular}
\caption[Inter-subject consistency and the sulcal depth maps.]{(a) Inter-subject consistency map obtained from all subjects/resolutions. Values are normalised to $[0,1]$ for better visualisation. Hotter colours indicate a higher consistency (better match) across subjects. (b) Average sulc map obtained from the complimentary sulc images provided by HCP for all subjects. The bright regions represent gyral crowns and dark regions represent buried cortex (the darker the shading the deeper the sulcus).}
\label{fig:dmri_consistency}
\end{figure}

Parcellations of the lateral cortex of the left hemisphere derived from one subject for varying resolutions are given in Fig.~\ref{fig:dmri-subject-visual} for visual inspection. Parcellation resolutions are determined by the three sets of eigenvectors ($d = 10, 15, 20$) used in the proposed framework. The consistency/variability of the proposed parcellations across subjects/resolutions is shown with a consistency map in Fig.~\ref{fig:dmri_consistency}(a). A complimentary sulc map is also provided alongside the consistency map to show the average cortical-folding organisation across subjects, in which bright regions represent gyral crowns and dark regions represent buried cortex (the darker the shading the deeper the sulcus)~\cite{VanEssen13}. This map allows to better understand the degree of gyral bias towards the proposed parcellations (see Discussion).

\begin{figure}[!t]
\centering
\begin{tabular}{cccc}
\multicolumn{2}{c}{\textit{Subject 1}} & \multicolumn{2}{c}{\textit{Subject 2}} \\
\multicolumn{2}{c}{\textit{K=260}} & \multicolumn{2}{c}{\textit{K=254}} \\
\includegraphics[width=0.2\textwidth]{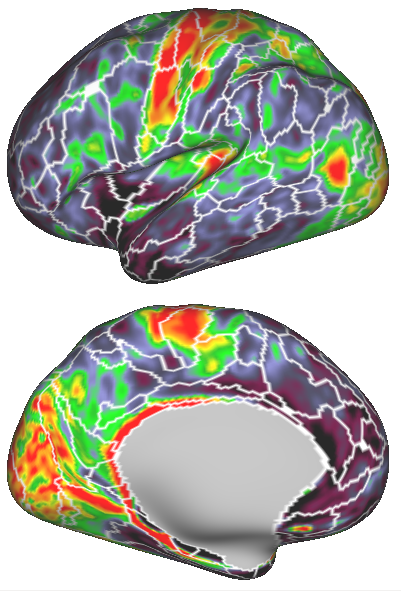}       & 
\includegraphics[width=0.2\textwidth]{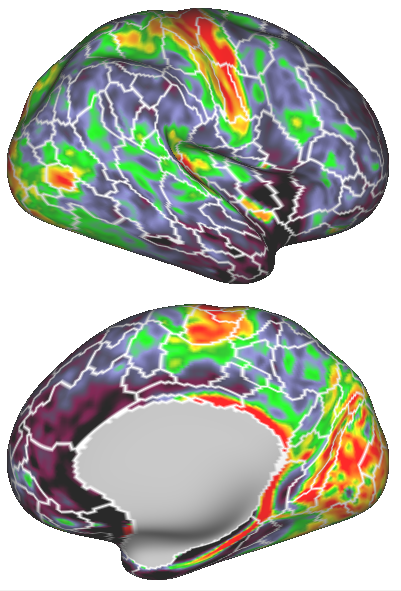}       & ~~~
\includegraphics[width=0.2\textwidth]{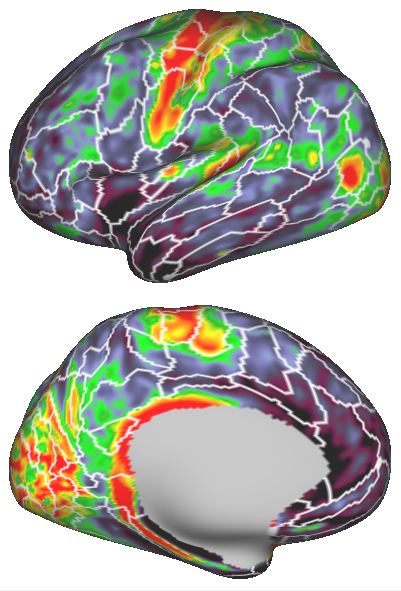}       & 
\includegraphics[width=0.2\textwidth]{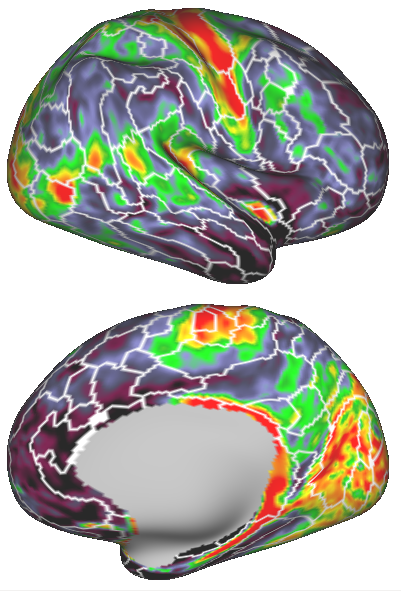}       \\
\\
\multicolumn{2}{c}{\textit{Subject 3}} & \multicolumn{2}{c}{\textit{Subject 4}} \\
\multicolumn{2}{c}{\textit{K=226}} & \multicolumn{2}{c}{\textit{K=238}}  \\
\includegraphics[width=0.2\textwidth]{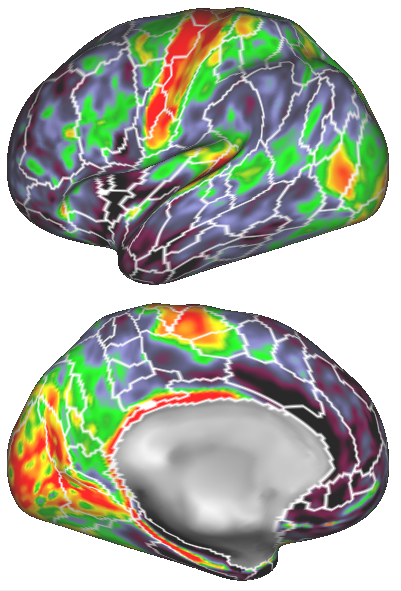}       & 
\includegraphics[width=0.2\textwidth]{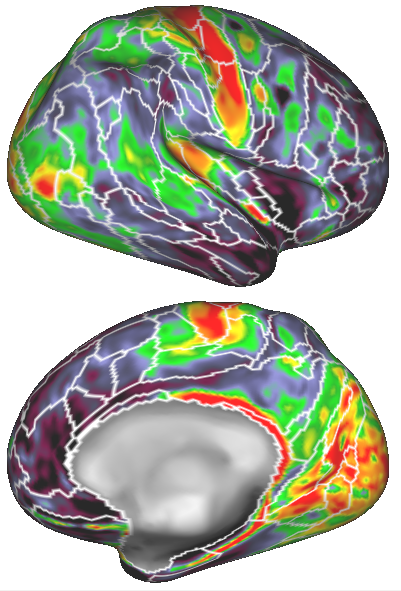}       & ~~~
\includegraphics[width=0.2\textwidth]{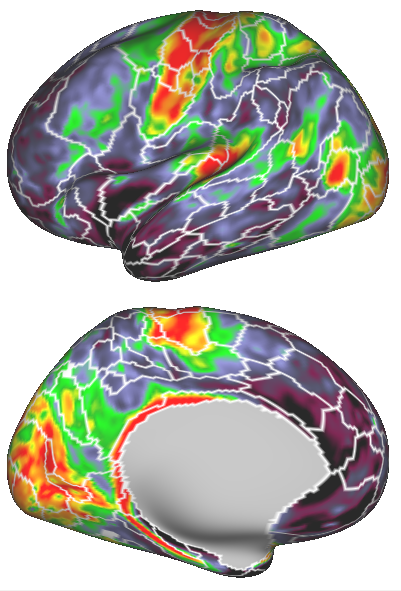}       & 
\includegraphics[width=0.2\textwidth]{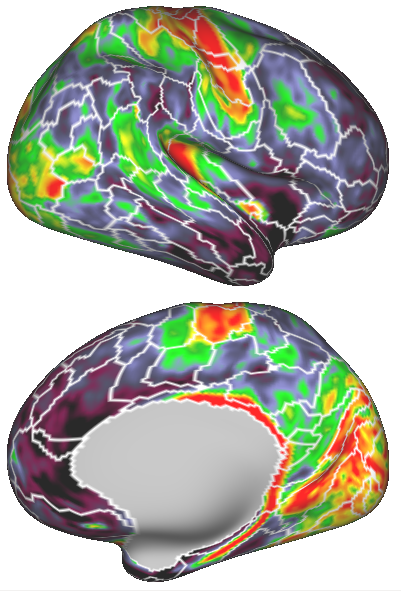}       \\

\end{tabular}
\caption[Parcellations of four different subjects compared to myelin maps.]{Parcellations of four different subjects obtained via the proposed approach are overlaid onto subject-specific myelin maps. The white borders show the parcellation boundaries delineated at the highest computed resolution (using $d = 20$ eigenvectors).}
\label{fig:dmri_visual_myel}
\end{figure}

\begin{figure}[!t]
\centering
\begin{tabular}{cccc}
\multicolumn{2}{c}{\textit{Subject 1}} & \multicolumn{2}{c}{\textit{Subject 2}} \\
\multicolumn{2}{c}{\textit{K=260}} & \multicolumn{2}{c}{\textit{K=254}} \\
\includegraphics[width=0.2\textwidth]{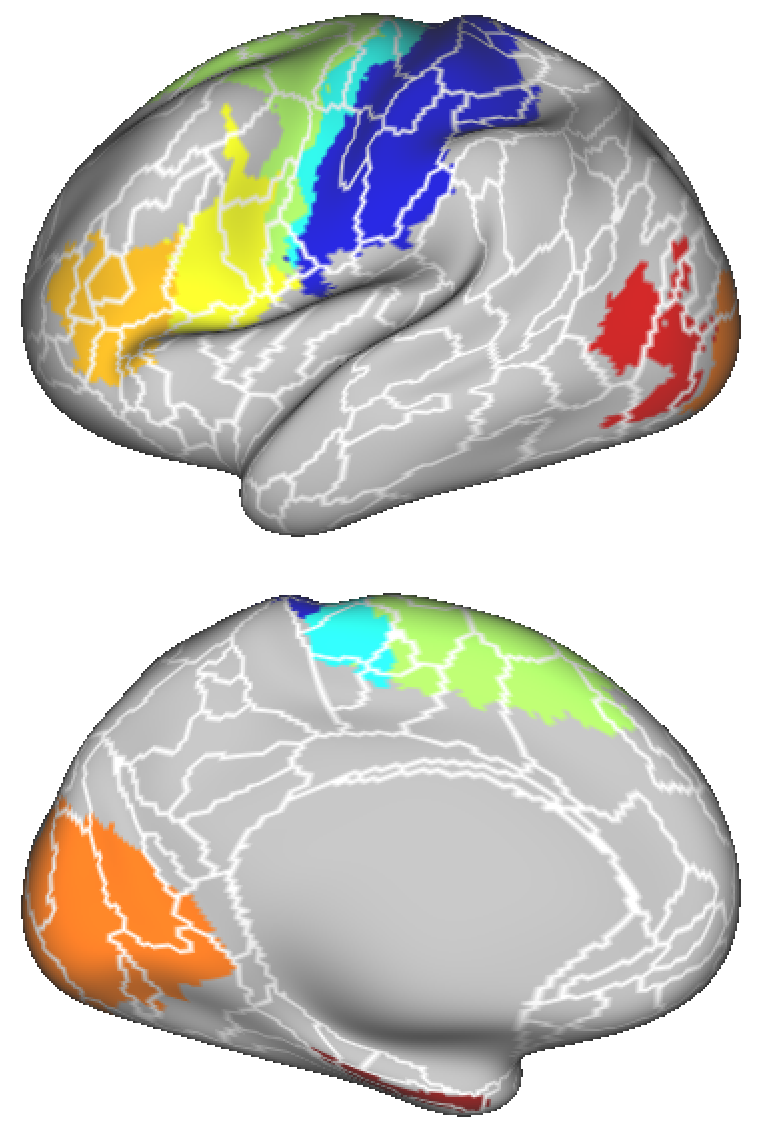}       & 
\includegraphics[width=0.2\textwidth]{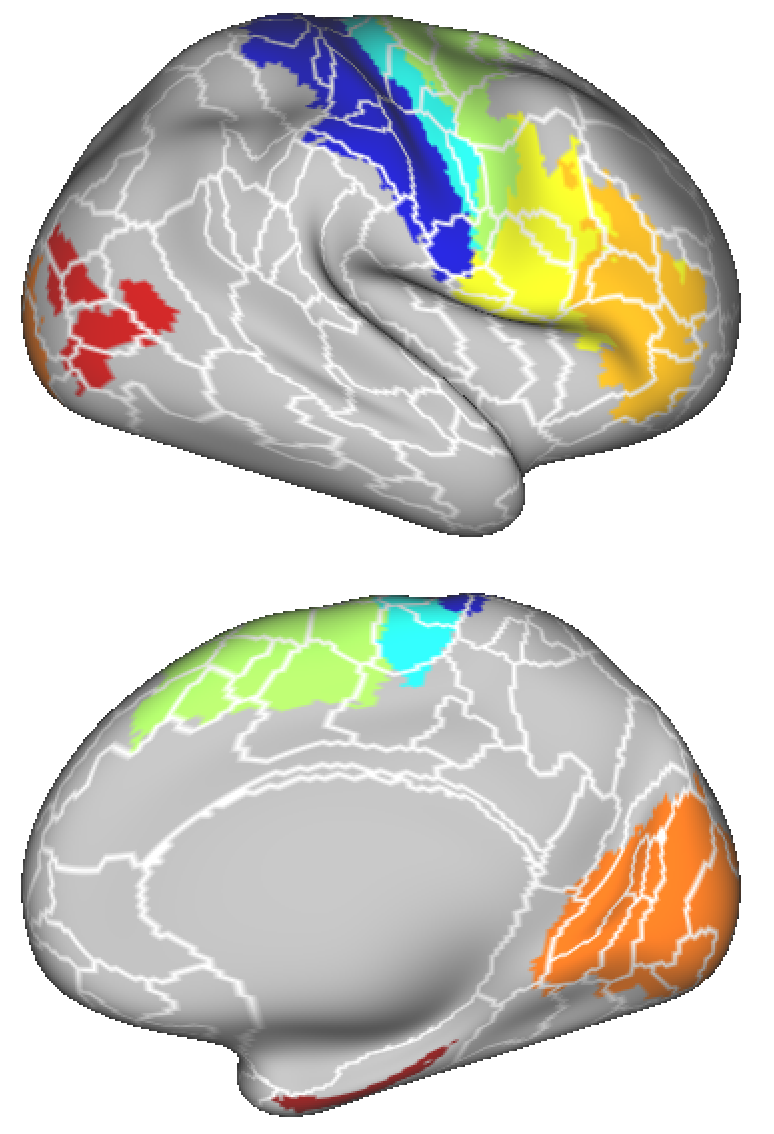}       & ~~~
\includegraphics[width=0.2\textwidth]{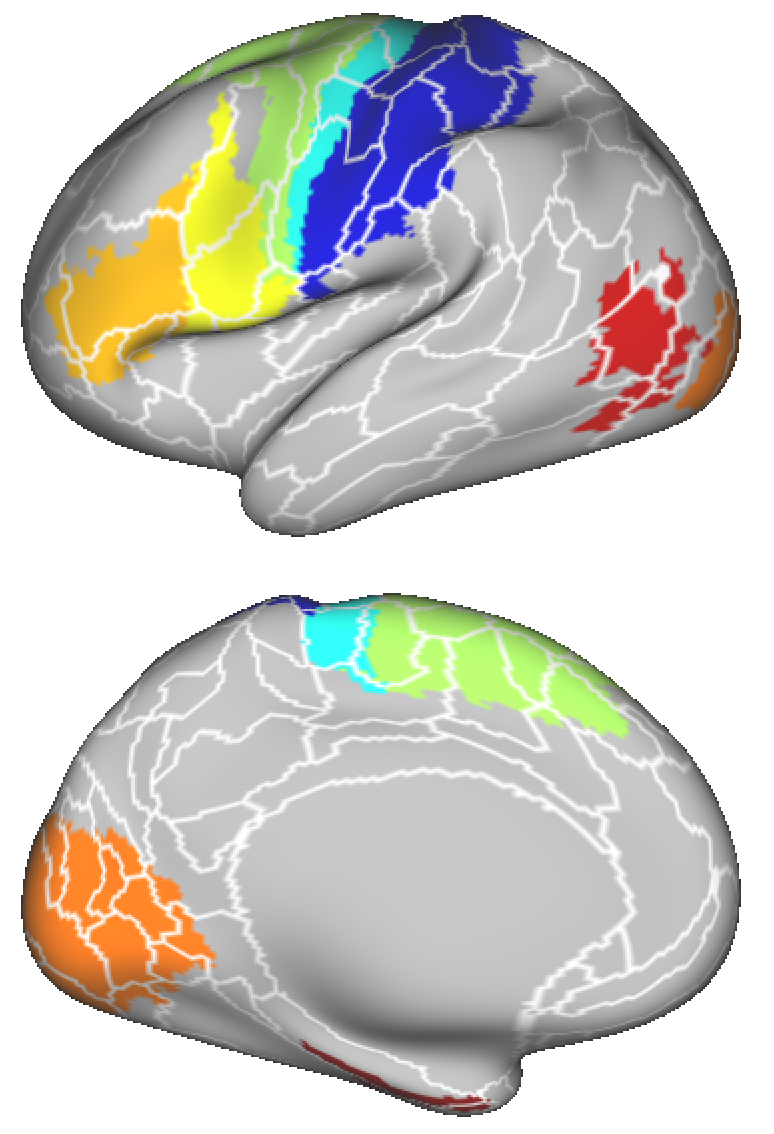}       & 
\includegraphics[width=0.2\textwidth]{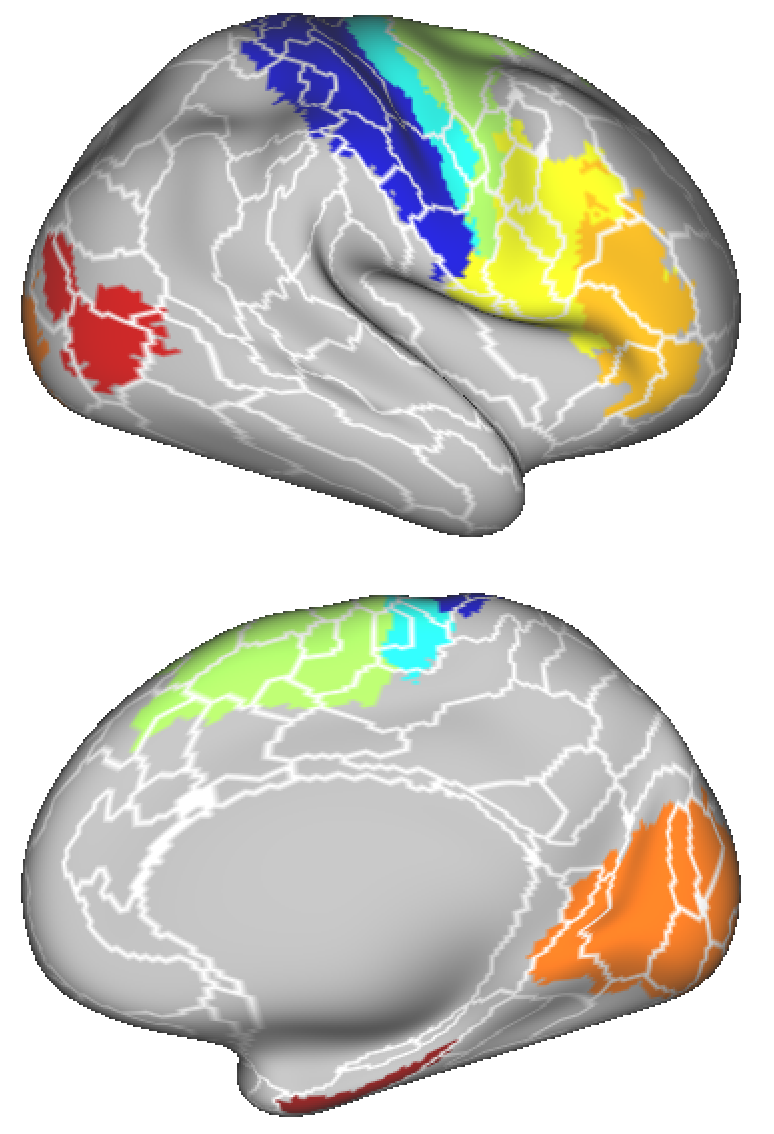}       \\
\\
\multicolumn{2}{c}{\textit{Subject 3}} & \multicolumn{2}{c}{\textit{Subject 4}} \\
\multicolumn{2}{c}{\textit{K=226}} & \multicolumn{2}{c}{\textit{K=238}} \\
\includegraphics[width=0.2\textwidth]{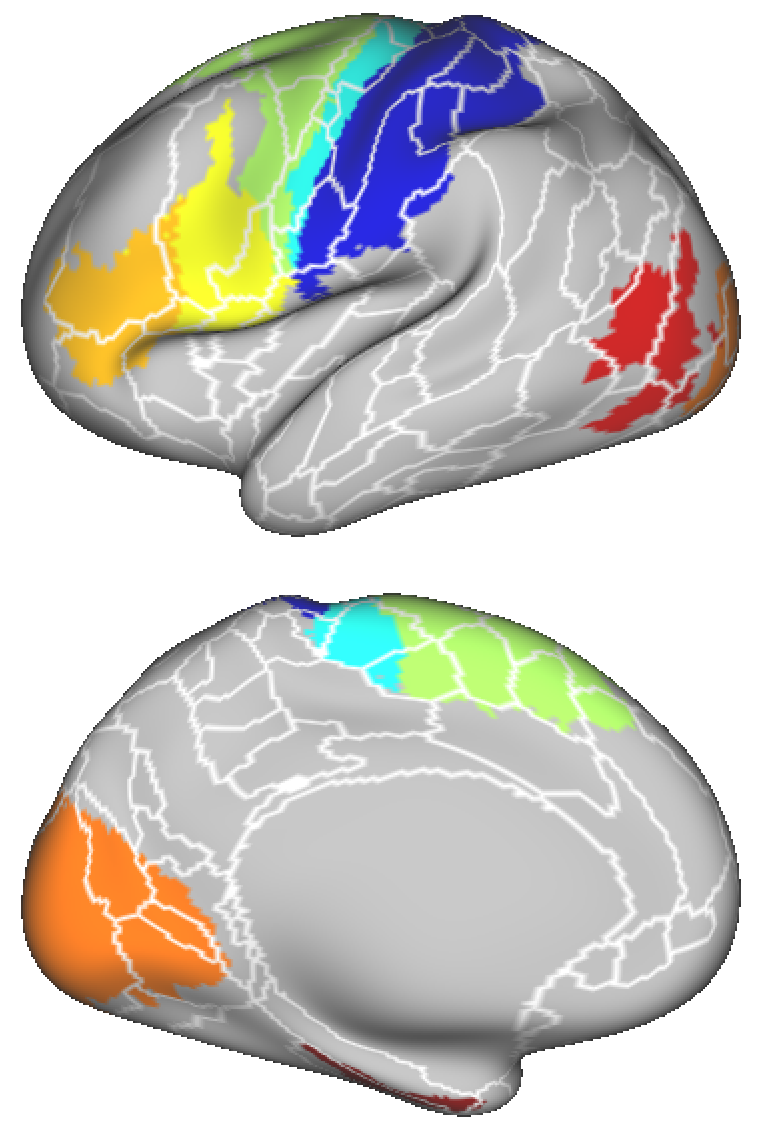}       & 
\includegraphics[width=0.2\textwidth]{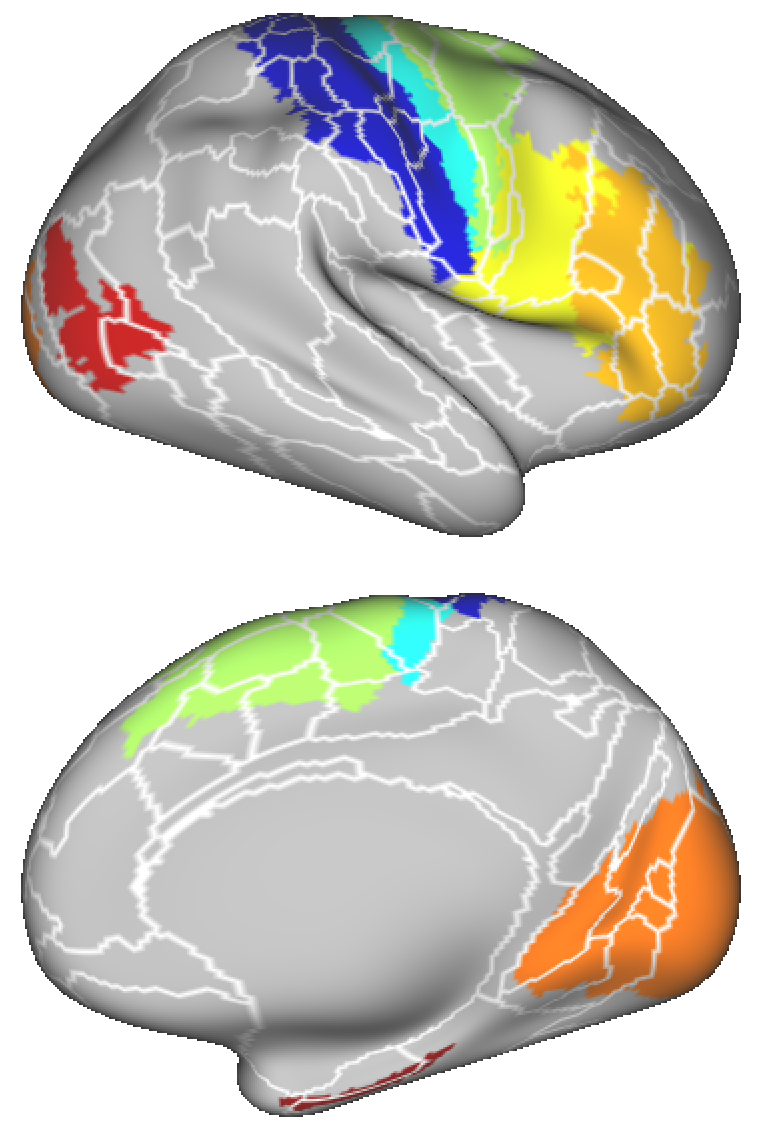}       & ~~~
\includegraphics[width=0.2\textwidth]{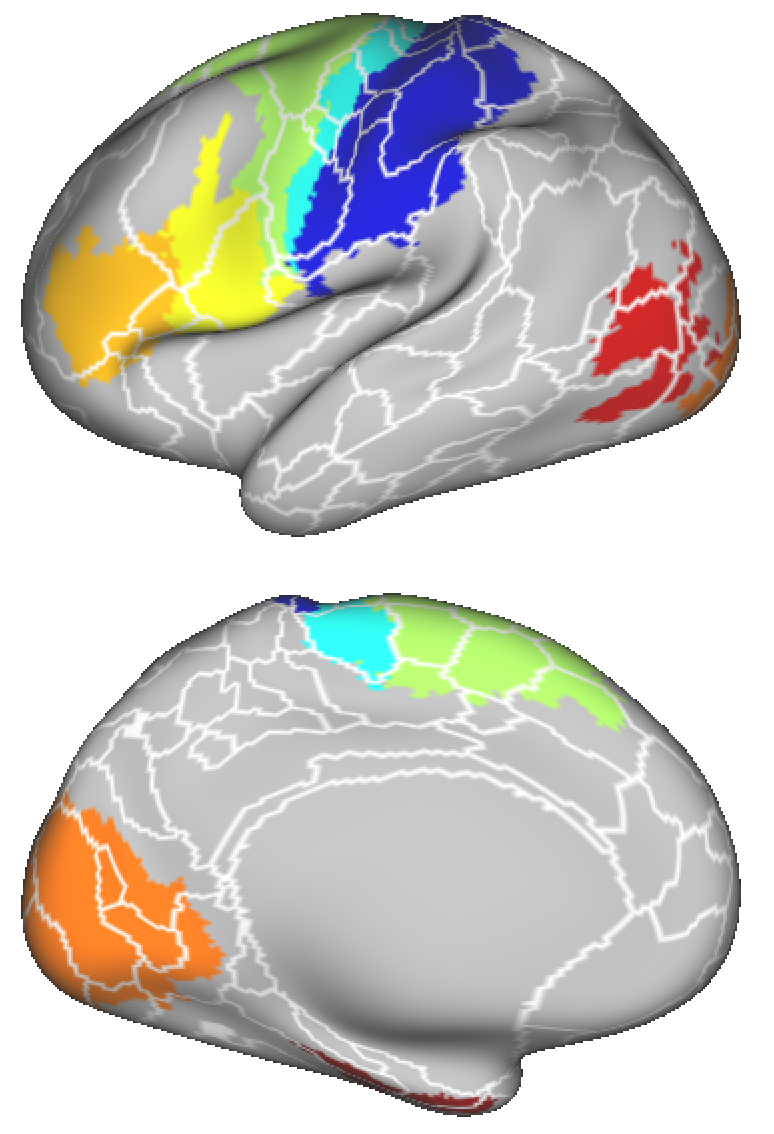}       & 
\includegraphics[width=0.2\textwidth]{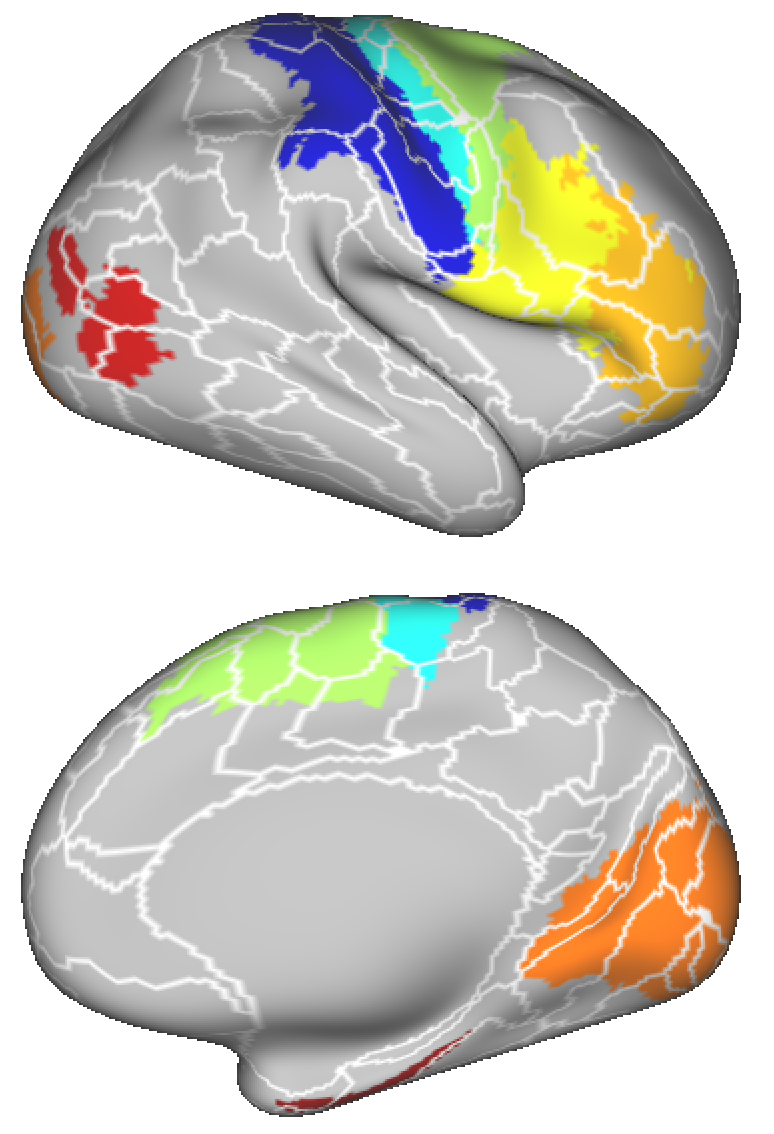}       \\
\\
\multicolumn{4}{c}{\includegraphics[width=0.4\textwidth]{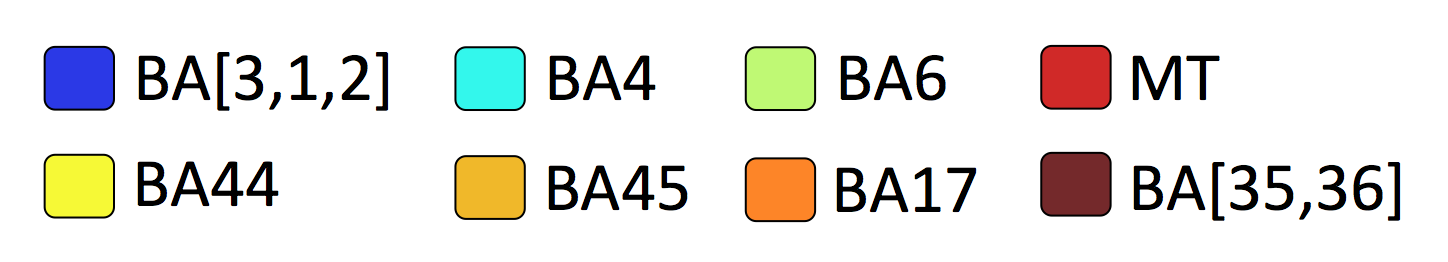}} 
\end{tabular}
\caption[Parcellations of four different subjects compared to the Brodmann atlas.]{Parcellations of four different subjects obtained via the proposed approach are overlaid onto subject-specific cytoarchitectonic regions labelled with the Brodmann atlas. The white borders show the parcellation boundaries delineated at the highest computed resolution (using $d = 20$ eigenvectors).}
\label{fig:dmri_visual_brod}
\end{figure}

\begin{figure}[!t]
\centering
\begin{tabular}{cc}
\includegraphics[width=0.4\textwidth]{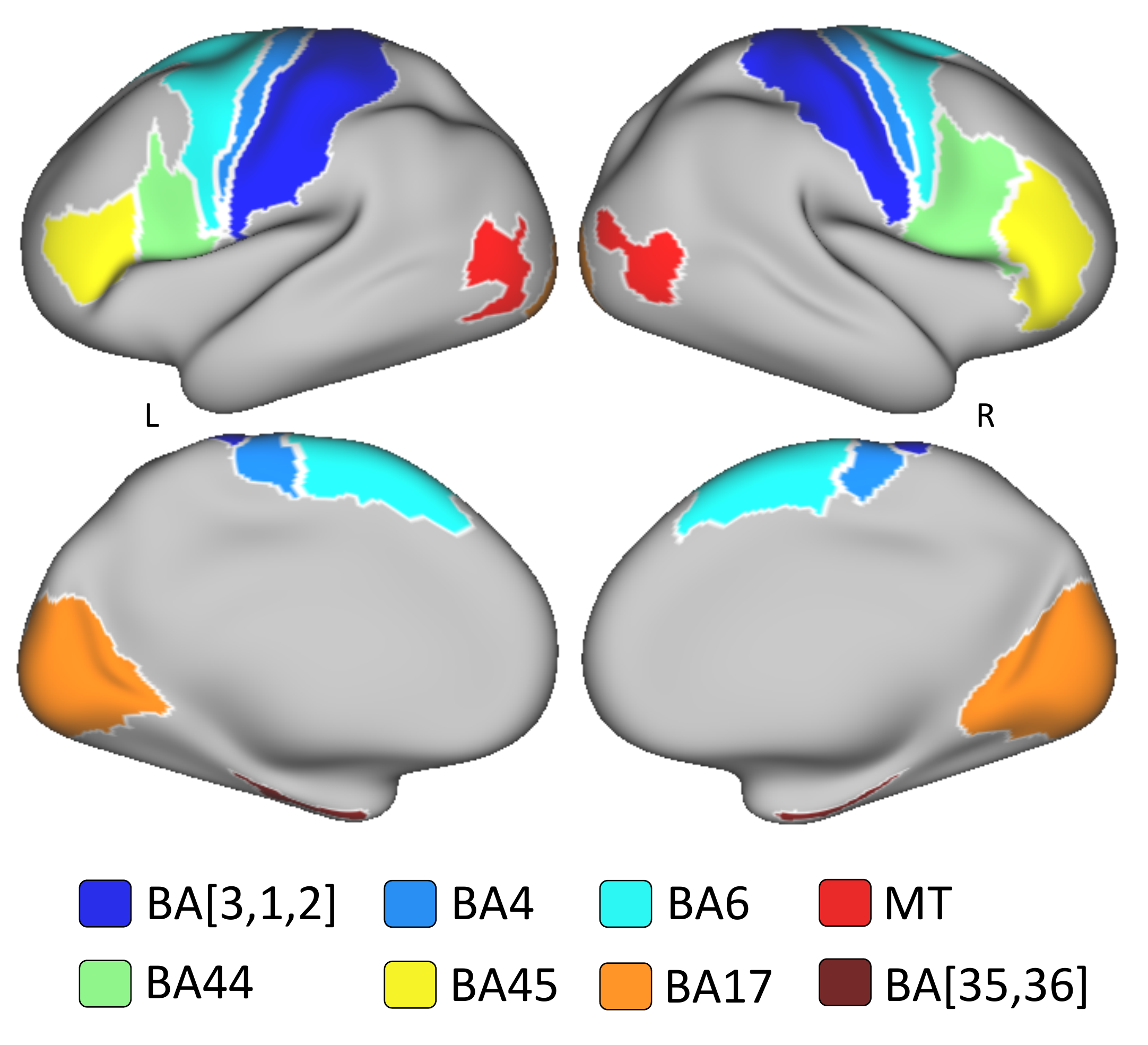}          & \raisebox{1.5em}{\includegraphics[width=0.41\textwidth]{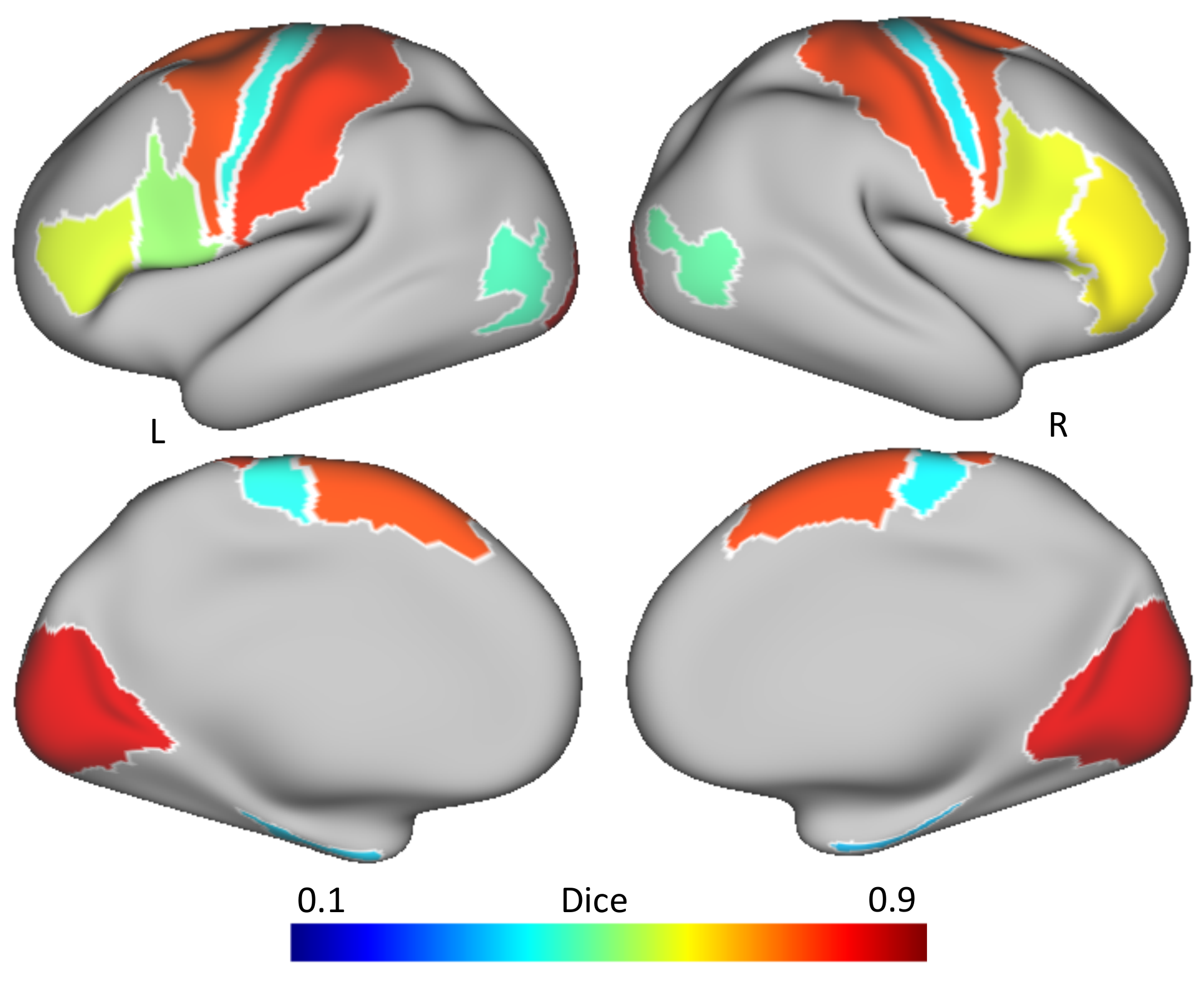}}        \\
 (a)             & (b)           \\
\multicolumn{2}{c}{\includegraphics[width=0.45\textwidth]{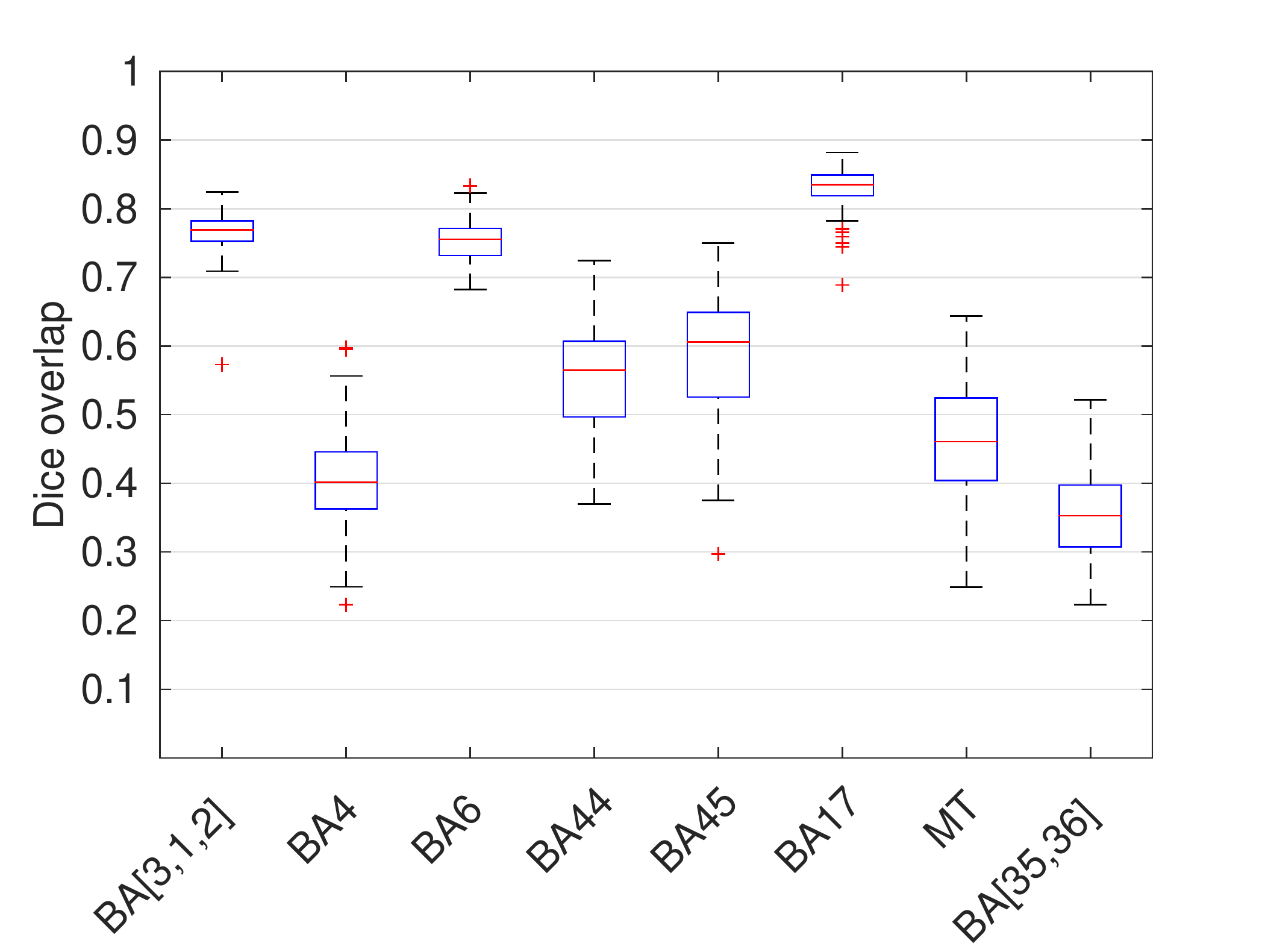}} \\
\multicolumn{2}{c}{(c)} 
\end{tabular}
\caption[Quantitative assessment of the overlap between the proposed parcellations and Brodmann areas.]{(a) Different cytoarchitectural areas as delineated from the Brodmann atlas. (b) Dice overlaps averaged across all computed resolutions (i.e. for $d=10,15,20$) and subjects are displayed on each Brodmann area. Hotter colours indicate a higher overlap. (c) Box plots show the inter-subject variability for each Brodmann area.}
\label{fig:dmri_bro}
\end{figure}

Myelin maps and Brodmann's areas projected onto the cortical surfaces of randomly selected subjects along with the proposed parcellations are given in Figs.~\ref{fig:dmri_visual_myel} and~\ref{fig:dmri_visual_brod}, respectively. The white borders in the figures show the parcellation boundaries obtained by the proposed method at the highest computed resolution (using $d = 20$ eigenvectors). The Brodmann parcellations contain labels for the primary somato-sensory cortex (BA[3,1,2]), the primary motor cortex (BA4), the pre-motor cortex (BA6), Broca's area (BA[44,45]), the visual cortex (BA17 and MT), and the perirhinal cortex (BA[35,36]) as shown in Fig.~\ref{fig:dmri_bro}(a).  

Visual assessment of the parcellation borders reveals some alignment with highly-myelinated cortical regions and several Brodmann areas, especially around the motor area and the visual cortex. Although we only present parcellations of several subjects, it appears that a similar agreement is observed across many subjects as indicated by the inter-subject consistency map in Fig.~\ref{fig:dmri_consistency}. The degree of the alignment with respect to different Brodmann areas is further shown in Fig.~\ref{fig:dmri_bro}. Average Dice scores obtained from all computed resolutions/subjects are overlaid on each Brodmann area in Fig.~\ref{fig:dmri_bro}(b) and the box plots showing the variability across subjects are presented in Fig.~\ref{fig:dmri_bro}(c). The highest overlap is observed within the primary somato-sensory cortex (BA[3,1,2]), pre-motor cortex (BA6) and primary visual cortex (BA17). 

\section{Discussion}
\label{sec:manifold-discussion}
In this chapter, we introduced a new connectivity-driven parcellation approach based on dMRI. The proposed method encapsulates the local connectivity characteristics with manifold learning and uses the low-dimensional embedding to identify locations where connectivity patterns change. Particularly, these transition locations are interpreted as an abstraction of the parcellation boundaries and consequently allow to delineate distinct parcels at different scales. In the absence of a connectivity-driven cortical atlas, we showed that the proposed parcellations provide a more accurate representation of the underlying connectivity compared to different connectivity-based, anatomical, and random parcellations with respect to well-known clustering quality measures. Our additional experiments with resting-state fMRI revealed the potential of the proposed method to reflect the functional organisation of the brain, at least more faithfully than the other tested parcellations. 

The visual and quantitative comparisons with the myelin maps and cytoarchitectonic regions showed that the proposed parcellations reflect the myelo- and cyto-architecture of the cerebral cortex to some extent, especially within the motor area. Although, this trend appears to be consistent across parcellations with respect to the consistency map [Fig.~\ref{fig:dmri_consistency}(a)], such inter-modality comparisons should always be interpreted carefully. First of all, it should be noted that connectivity-driven parcellations obtained from dMRI do not necessarily align with the cortical organisation of the brain estimated via resting-state fMRI (or other neuro-anatomical properties), as each modality provides a different source of information that does not need to agree with the other~\cite{Ng13}. In addition, cortical parcellations obtained from different data sources may be plagued by modality-specific noise and biases, e.g. physiological noise and head motion may yield spurious functional correlations, while tractography errors may bias structural connectivity~\cite{craddock2013imaging,Eickhoff15}). However, some degree of agreement with certain parts of the cerebral cortex that are known to have distinct neuro-biological features, or function, would potentially express the reliability of the proposed parcellation scheme, at least for the cortical areas with which high alignment is observed. 

In addition, it is important to consider the possible limitations/biases introduced by the data processing pipelines or the evaluation techniques used to measure the agreement between different modalities. For example, the Brodmann areas are mapped onto each subject's surface using a cortical folding driven registration algorithm. As a result, a better alignment can be achieved in the motor and visual cortex, where the folding patterns are more consistent across subjects~\cite{Parisot16a}. On a similar note, the overlap-based measure used to assess the degree of agreement between parcellations and different Brodmann areas is biased by the size and shape of the parcels. Evenly sized/shaped parcels are easier to match with their target parcels, while differences in Dice scores will be much more striking when comparing small parcels over big ones. For example, despite the fact that the parcellation boundaries usually align well with the primary motor cortex (BA4) as shown in Fig.~\ref{fig:dmri_visual_brod}, the quantitative results given in Fig.~\ref{fig:dmri_bro}(c) indicates less favourable overlapping scores for BA4 compared to the other motor-related Brodmann areas (BA[3,1,2] and BA6), which typically span a larger proportion of the motor cortex.

Another critical consideration point is the gyral bias inherent to probabilistic tractography, which is known to affect the estimated structural connectivity~\cite{VanEssen2013mapping} and to influence the parcellation boundaries, such that they naturally align well with cortical folding. The sulc map in Fig.~\ref{fig:dmri_consistency}(b) can provide a clearer intuition regarding the impact of the gyral bias on the connectivity-driven parcellations, as it shows how the cortical sheet is folded. When compared to the consistency map (Fig~\ref{fig:dmri_consistency}), it can be clearly seen that the parcellation boundaries have a strong alignment with the gyral crowns (brighter areas). In other words, parcellation boundaries follow the cortical folding patterns, as the underlying streamlines that are used to estimate structural connectivity typically terminate in gyri rather than sulci~\cite{VanEssen2013mapping}. 

In general, several other limitations should also be taken into account when interpreting parcellations derived from dMRI or comparing them with other modalities. First, dMRI provides a very indirect measurement of white matter connectivity (i.e. diffusion of water molecules in the brain). In addition, the processing methods (e.g. tractography) can suppress existing structural connections, and thus, alleviate the reliability of the connectome analysis. For example, while log transformation helps reduce tractography's bias towards short-range connections, it does not necessarily guarantee an accurate estimation of the long range connections~\cite{girard2014towards,Parisot16b}. Other limitations include the dominance of large fiber bundles, impaired detection of crossing and kissing fibers, as well as difficulty to determine the origin or termination of the tracts~\cite{VanEssen2013mapping,Ng13}. Despite all these drawbacks, dMRI constitutes a very important aspect of brain mapping, as it remains the best non-invasive way of measuring the physical (structural) connectivity within the brain. While the approximated tracts may not be entirely accurate, the similarity between the estimated connectivity patterns to characterise the anatomical organisation of the brain could still be valid~\cite{Knosche11}. Consequently, the proposed parcellations hold a great potential for brain mapping studies. 

On the other hand, when parcellations are considered from a neuro-biological point of view, using only a single modality might be rather limited in order to reveal the complex structure of the cerebral cortex, which consists of a mosaic of multiple properties nested at different levels of detail~\cite{Glasser16}. In addition, considering the limitations inherent to different modalities and processing techniques, the integration of multiple modalities to the parcellation generation task may potentially provide more accurate and robust cortical segregation of the cerebral cortex, as shown in the recently proposed multi-modal cortical parcellation~\cite{Glasser16}. A prospective future work, therefore, would be to approach the parcellation problem from a multi-modal point of view and develop a flexible parcellation framework that would allow combining information from different modalities.

While the proposed parcellations can successfully represent the structural organisation of the cerebral cortex on a single subject basis, the high variability in parcel size, shape, and number do not allow to construct spatial correspondence across subjects, which is generally a default prerequisite for group-wise connectome analysis. Nevertheless, the proposed approach can be potentially expanded for the generation of a parcellation that can reliably represent the shared characteristics within a population. For example, one can derive an average feature set from the subject-level spectral embeddings by first bringing them to a common space~\cite{Langs14,Langs15} and then applying boundary mapping. Such an approach would provide more robust parcellations, as less distinct connectivity patterns are likely to be averaged out and not propagated to the parcellation stage.

However, it should be noted that averaging (on top of feature reduction and alignment) may also lead to loss of important information regarding connectivity features. As a result, averaging, although being highly popular, may not be the most reliable solution to the parcellation problem at the group level. Next chapter investigates different ways of obtaining group-wise parcellations within the context of functional connectivity and proposes a new algorithm based on spectral decomposition for a whole-brain group-wise parcellation of the cerebral cortex, which can effectively reveal the fundamental properties of a population while preserving individual subject characteristics.

\chapter{Joint Spectral Decomposition of the Cerebral Cortex}
\label{chapter:joint}

This chapter is based on:

S. Arslan, S. Parisot, and D. Rueckert. \textit{Joint Spectral Decomposition for the Parcellation of the Cerebral Cortex Using Resting-State fMRI}. Information Processing in Medical Imaging (IPMI) Lecture Notes in Computer Science, vol. 9123, pp. 85-97, 2015.

\section*{Abstract}
\textit{This chapter describes a robust cerebral cortex parcellation method based on spectral graph theory and resting-state fMRI based connectivity that generates reliable groupwise parcellations across multiple subjects. Our method combines within- and inter-subject connectivity patterns to construct a multi-layer graph, which effectively captures the fundamental properties of the whole group as well as preserves individual subject characteristics. Spectral decomposition of this joint graph is used to obtain whole-brain parcellations of the cerebral cortex. Using rs-fMRI data collected from 100 healthy subjects, we show that our proposed algorithm computes more reproducible parcellations across different groups of subjects and at varying levels of resolution compared to two other state-of-the-art spectral methods. We also report that our group-wise parcellations are functionally more consistent, thus, can potentially be used to represent a population in network analysis.}

\section{Introduction}
The human brain is assembled into interacting regions that coordinate the nervous system. Identification of these regions is critical for a better understanding of the functional organisation of the brain and to reveal the interactions of underlying subsystems~\cite{Sporns05} that are responsible for different functions, including but not limited to, reasoning, memory, visual processing, and movement coordination. Functional connectivity studies have identified several subsystems, each of which spans across different cortical areas and associated with a different functional ability~\cite{salvador2005neurophysiological}. This has further advanced the analysis of the functional architecture of the brain by constructing graphical models of the connections within individual subsystems and their interactions with each other at different levels of detail~\cite{Yeo11,Power11}. 

In whole-brain connectivity analysis, a critical stage is the parcellation of the cerebral cortex in order to compute a set of network nodes that can effectively represent cortical regions with uniform connectivity patterns. It is of great importance for a parcellation framework to be able to cluster vertices with similar functional connectivity together, thus, the average signal can reliably represent each part of the cortical region. It is also highly critical to generate a reliable group-wise representation that reflects the common functional characteristics of the population, yet is tolerant to changes in the functional organisation at the individual subject level that may emerge due to functional and anatomical variability across subjects. In this chapter, our motivation is to compute parcellations consisting of functionally homogeneous and spatially contiguous regions which can be used as the network nodes for a whole-brain connectivity analysis. To this end, we present a robust cerebral cortex parcellation framework based on spectral graph theory and resting-state fMRI connectivity (RSFC).

Parcellating the cerebral cortex based on RSFC can potentially identify functional organisation of the brain without relying on the anatomical features or an external stimulus/cognitive process~\cite{Smith13}. RSFC-based cortical parcellation literature mainly consists of methods that subdivide the cerebral cortex into varying number of regions according to the requirements of the subsequent network analysis and topological network features across the cerebral cortex~\cite{deReus13}. Some techniques~\cite{Beckmann04,Heuvel08,Yeo11,Power11} parcellate the cortex at a very coarse level (less than hundred subregions), with the aim of identifying the functional communities, or resting-state networks, spanning across the cortex or some fractions of it. In this case, these parcellations cannot be reliably used for network node identification, as each region itself constitutes a gross, complex functional system, possibly exhibiting non-uniform functional patterns~\cite{Smith11}. 

Other methods~\cite{Craddock12,Shen13,Gordon16} typically generate a few hundred clusters without losing the ability of capturing the functional organisation of the cortex. This is generally achieved by (1) computing an average dataset from the RSFC data of each individual subject and submitting this to a parcellation algorithm or (2) first deriving individual subject parcellations and then generating a group-wise parcellation with a second-level algorithm running atop the subject-level parcellations. The group-wise parcellations generated by the former approach are not likely to reflect the individual RSFC characteristics as the inter-subject variability is usually over-smoothed during the averaging process. While the latter method allows the propagation of subject-level RSFC patterns into the group level, it still suffers from the adaptability of group representation to individual single subjects, as it is very unlikely that a group-wise parcellation obtained in such a setting would match with the subject-level parcellations. Besides, this method might result in slower computation as the parcellation should be computed for all subjects individually before generating a group-wise representation.  

In this chapter, we propose an alternative approach for obtaining group-wise parcellations: We make use of spectral graph decomposition techniques and represent the population in a multi-layer graph which effectively captures the fundamental properties of the whole population while preserving individual subject characteristics. We show that the parcellations obtained in this setting are (a) more reproducible across different groups of subjects and (b) better reflect functional and topological features shared by multiple subjects in the group compared to the parcellations obtained by the other two approaches. Using rs-fMRI data collected from 100 healthy subjects, we show that our proposed algorithm computes highly reproducible parcellations across different subsets of the same population and at varying levels of detail with an average Dice score of 0.85, achieving up to 11$\%$ better group-to-group reproducibility. We also report that our group-wise parcellations are functionally more consistent, thus, can be reliably used to represent the population in network analysis. These aspects of the proposed method differentiate it from the previous parcellation algorithms and constitute our main contributions in this chapter. Finally, our framework can be used to generate parcellations with different number of subregions, allowing users to conduct a network analysis at different levels of detail.

In the remainder of this chapter, we first summarise the methodology (Section~\ref{sec:joint-method}), starting with the dataset and following with a detailed explanation of the joint spectral decomposition method. In Section~\ref{sec:joint-exper}, we describe the experimental setup and provide details of the evaluation measures and comparison methods used to validate the proposed approach. In Section~\ref{sec:joint-results} we provide visual and quantitative results to show the effectiveness of the proposed method over other methods. Finally, in Section~\ref{sec:joint-conclus}, we conclude the chapter by discussing the impact/limitations of the proposed method and provide some insight towards future research directions.

\section{Methodology}
\label{sec:joint-method}
\subsection{Data}
We conducted our experiments on Dataset 1, which contains 100 subjects (54 female, 46 male healthy adults, ages 22-35). The details of the dataset are provided in Chapter~\ref{chapter:background} and briefly summarised here. The data for each subject was acquired in two sessions that were held on different days and divided into four runs of approximately 15 minutes each. Data was preprocessed and denoised by the HCP minimal preprocessing pipelines~\cite{Glasser13}. As an additional step, 15-minute scans of each subject were temporally concatenated, after being normalised to zero-mean and unit-variance, yielding 60-minute rs-fMRI datasets for all subjects.

\subsection{Joint Spectral Decomposition}
\begin{figure}[!bht]
\centering
\begin{tabular}{c}
\includegraphics[width=\textwidth]{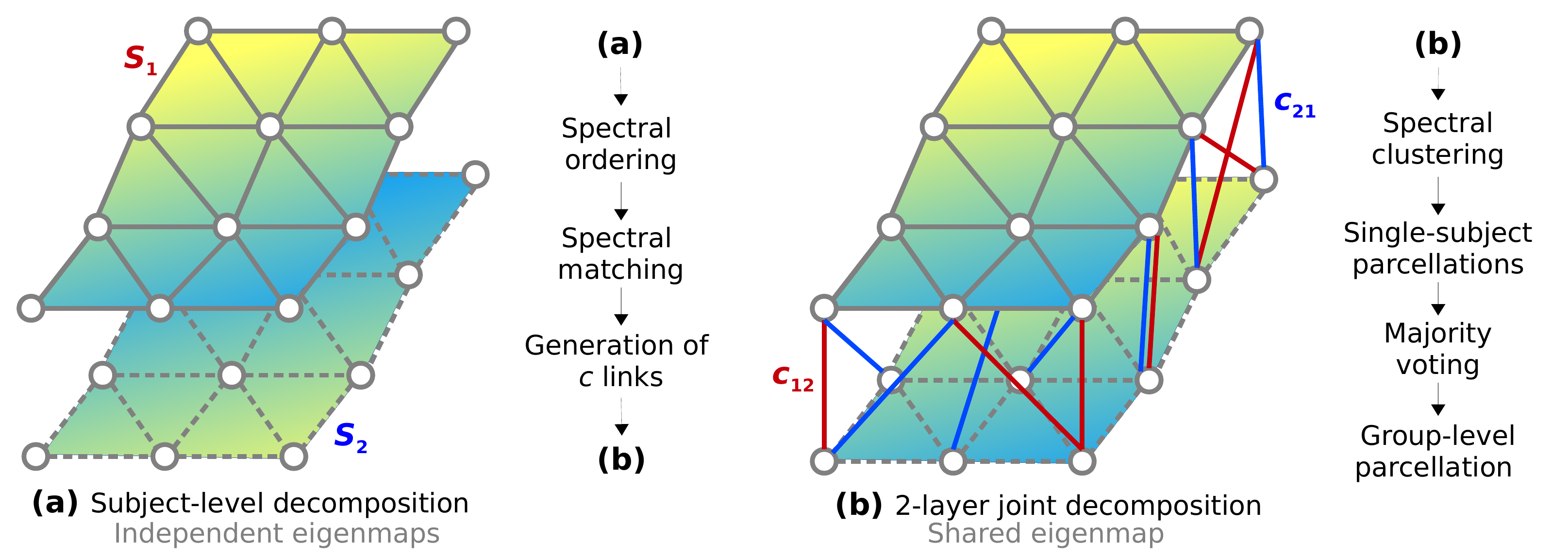}\\
\end{tabular}
\caption[Visual representation of the parcellation pipeline.]{Visual representation of the parcellation pipeline with an emphasis on (a) single-subject and (b) joint spectral decomposition, illustrated on the patches cropped from the cortical surfaces $S_1$ and $S_2$. The red and blue edges correspond to the mappings $c_{12}$ and $c_{21}$, obtained by matching the closest vertices in $S_1$ and $S_2$, respectively. }
\label{fig:spectral_matching}
\end{figure}

We propose a general framework based on spectral decomposition to identify whole-cortex parcellations that can effectively capture the connectivity across multiple subjects. At the subject level, the cerebral cortex is represented as an affinity matrix, in which the within-subject connectivity is encoded as the edge weights between the interacting cortical units. Each individual affinity matrix is transformed to the spectral domain via an eigenspace decomposition and the corresponding eigenvectors are used to identify connectional similarities between subjects. Within- and inter-subject connectivity is then combined into a multi-layer graph, which is capable of representing the fundamental properties of the underlying functional organisation of the population. Similar to the subject-level graph decomposition, this multi-layer graph can then be decomposed into its eigenvectors, creating a feature matrix in the spectral domain (e.g. a spectral embedding) that can be fed into a clustering algorithm for producing the final parcellations. A visual summary of the approach is given in Fig.~\ref{fig:spectral_matching}.

\subsubsection{Sparse Affinity Matrix} 
\label{sec:sparse_affinity}
\begin{figure*}[!bt]
\centering
\begin{tabular}{c}
\includegraphics[width=\textwidth]{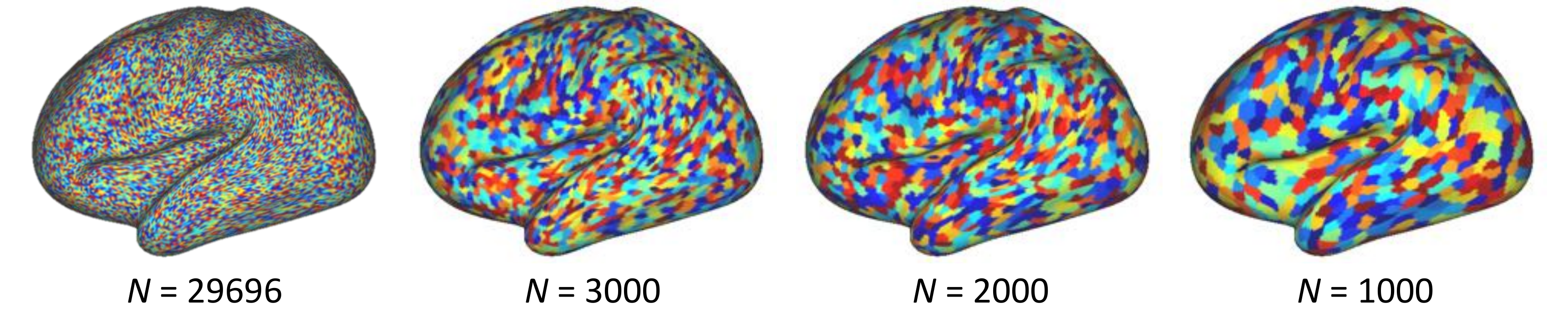} \\
(a)\\
\includegraphics[width=\textwidth]{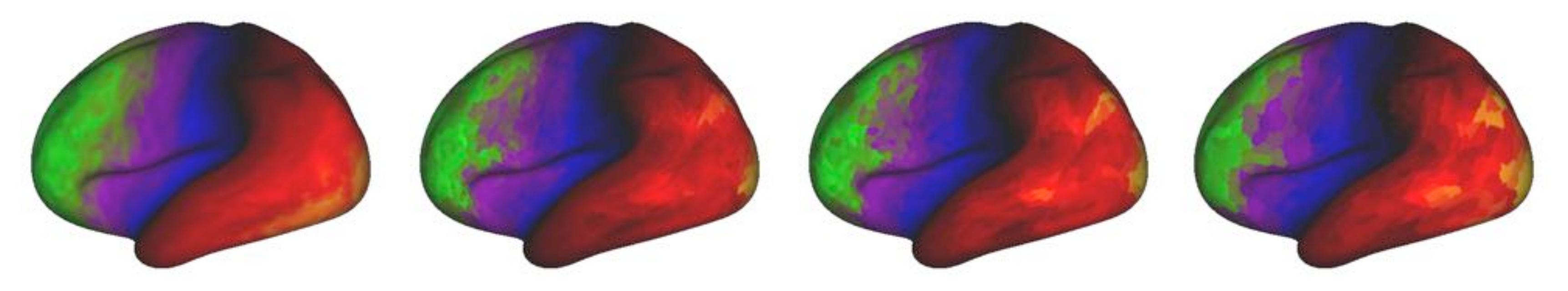} \\
(b)\\
\end{tabular}
\caption[Visualisation of spectral embedding at different spatial resolutions.]{(a) Cortical meshes visualised at different spatial resolutions. The first column shows the original mesh, in which each vertex is associated with a node in the affinity matrix $W$. It is followed by three supervertex parcellations at decreasing resolutions, denoted by $N$. (b) Eigenvectors associated with the first non-zero eigenvalues are superimposed onto the cerebral cortex. Spectral embedding are obtained by decomposing the corresponding $N \times N$ affinity matrices. }
\label{fig:sv_eig}
\end{figure*}

The cerebral cortex is represented as a smooth, triangulated mesh with no topological defects, in which each vertex is associated with a timeseries that represents the fMRI signal. Typically, an affinity matrix can be defined to capture the functional similarity across the cortex by modelling the cortical vertices and their associations as a weighted adjacency graph. However, mapping each cortical vertex in the mesh to a node in this graph can become computationally prohibitive due to high dimensionality of the data at the vertex level. For this reason, we pre-cluster cortical vertices into \textit{supervertices} with the supervertex generation algorithm introduced in Chapter~\ref{chapter:multi-level}. In order to maintain the spatial correspondence across subjects, we define the supervertices on the average Conte69 cortical mesh provided by the HCP~\cite{VanEssen12} and use the same set of supervertices for all subjects in the dataset. Supervertex parcellations obtained for different resolutions are shown in Fig.~\ref{fig:sv_eig}(a). Each supervertex is associated with the average timeseries of its constituent vertices, which further helps alleviate the impact of low SNR inherent in fMRI. For all supervertex pairs $(\mathit{v}_i,\mathit{v}_j)$ an edge in the affinity matrix $W\in{\mathbb R}^{N \times N}$ is defined as follows.

\begin{equation}
  W_{ij} =
  \begin{cases}
    r(t_i,t_j) & \text{if $i$ and $j$ are neighbours} \\ 
    0 & \text{otherwise}
  \end{cases}
\end{equation}

where $N$ denotes the number of nodes in the network (i.e. supervertices) and $r(t_i,t_j)$ is the Pearson's correlation coefficient between the timeseries $t_i$ and $t_j$ of $\mathit{v}_i$ and $\mathit{v}_j$, respectively. Edges with negative correlations are set to zero in order to ensure that the affinity matrix is positive-semidefinite (i.e. all $W_{ij} \geq 0$) for each subject. Here, the benefit of introducing a spatial constraint and defining an edge only between the adjacent vertices in the mesh is two-fold: (1) It ensures that the resulting parcellations are spatially contiguous after applying spectral clustering and (2) it yields a sparse affinity matrix, and hence, avoids computational overheat during spectral decomposition.

\subsubsection{Spectral Embedding}\label{sec:spectral_embed}
Whereas parcellating the cortical surface into supervertices provides a means of \textit{spatial} dimensionality reduction, the spectral decomposition of the affinity matrix yields a low-dimensional embedding, and hence, reduces the \textit{temporal} resolution without losing the ability to represent the functional connectivity patterns encoded in the affinity matrix. A spectral embedding is typically obtained by using the first eigenvectors obtained through the spectral decomposition of the graph Laplacian as explained in Chapter 5. Given the affinity matrix $W$, the graph Laplacian can be defined as $L=D^{-1/2}(D-W)D^{-1/2}$, where $D$ is a diagonal matrix with each entry $D_{ii}={\sum}_j W_{ij}$ representing the degree of $v_i$. $L$ is a diagonalisable matrix which can be factorised as $L = U{\Lambda}U^{-1}$, where $U = (u_0,u_1,...,u_{N-1})$ is the eigensystem, with $u_i$ representing each eigenvector and ${\Lambda}$ is a diagonal matrix that contains the eigenvalues, represented as ${\Lambda}_{ii} = {\lambda}_i$. The first eigenvectors obtained after decomposing a randomly selected subject with respect to different spatial resolutions are shown in Fig.~\ref{fig:sv_eig}(b). This figure indicates that the intrinsic connectivity at the subject level can be captured by the spectral embedding, even if the spatial resolution of the affinity matrix is immensely reduced by the use of supervertices.

Eigenvectors are powerful tools in terms of encapsulating valuable information extracted from the decomposed matrix in a lower dimensionality. In particular, sorting the eigenvalues as $0 = {\lambda}_0 \leq {\lambda}_1 \leq {\lambda}_2 \leq \cdots \leq {\lambda}_{N-1} $ and organising the corresponding eigenvectors accordingly, we can use the set of first $d$ eigenvectors after omitting the eigenvector corresponding to eigenvalue 0 to define a spectral feature matrix, i.e. $F = (u_1, u_2, \cdots, u_d)$, which is capable of representing the most important characteristics of the decomposed matrix. Thus, each node in the adjacency graph can be represented by its corresponding row in $F$, without losing any critical information, i.e. $\mathit{v}_i \mapsto (u_{1}(i), \ldots, u_{d}(i))$.

\subsubsection{Spectral Matching}\label{sec:spectral_matching}
The idea of spectral matching is computing a mapping between two embeddings by comparing their eigenvectors~\cite{Lombaert13}. Our observations on the cortical surfaces in the spectral domain revealed that eigenvectors show similar characteristics across subjects. This attribute can be exploited to obtain a joint spectral embedding that reflects functional features shared by the subjects in the group, while preserving individual subject characteristics. 

However, spectral coordinates are not directly comparable. There might exist a sign ambiguity when computing eigenvectors\footnote{Given the general eigenvector problem, if $Ax = \lambda x$, then $A(-x)=\lambda (-x)$.}, requiring sign checks of each eigenvector in the spectral embeddings. Similarly, the ordering of the eigenvectors may change~\cite{Jain06} if the eigenvectors correspond to the same eigenvalue, as the solver may not necessarily compute the eigenvectors in the same order in both embeddings~\cite{Lombaert11}. As a result, the same cortical information represented with the eigenvector $u_i$ in $F_1$ can be decoded in the eigenvector $u_j$ in $F_2$, without the necessity of being in the same order or having the same sign. Therefore, an additional correction must be carried out in order to find the corresponding eigenvectors on both cortical surfaces. We overcome this problem by aligning two spectral embeddings using Procrustes analysis~\cite{Bookstein97,Langs10,Langs15}. This technique computes a linear transformation between a reference (e.g. $F_1$) and a target matrix (e.g. $F_2$), so that the points in the target matrix best match the points in the reference matrix with respect to a goodness-of-fit criterion, e.g. the sum of squared distances between two matrices. This typically handles the sign ambiguity and ensures that both embeddings have the same ordering. 

After spectral ordering, the spectral matching problem can be solved by computing pairs of closest vertices on cortex $S_1$ and cortex $S_2$ with respect to their spectral feature matrices $F_1$ and $F_2$ as illustrated in Fig.~\ref{fig:spectral_matching}. For points (supervertices) $x$ on $S_1$ and $y$ on $S_2$, the mappings $c_{12}: x_i \mapsto y_{c_{12}(i)}$ and $c_{21}: y_i \mapsto x_{c_{21}(i)}$ for $i = 1, \cdots, N$ can be established by running a \textit{k}-nearest neighbour search with respect to $F_1$ and $F_2$. These mappings provide a set of inter-subject \textit{links} connecting cortical surfaces to each other based on the similarity of their functional connectivity captured in the low dimensional space. 

\subsubsection{Definition of a Multi-Layer Graph Across Subjects}
The use of spectral matching to find the mappings between pairs of cortical surfaces can be extended to generate a multi-layer graph for representing the whole group of subjects. A critical part in such a setting is the definition of edge weights that constitute the links between cortical surfaces. Using the correlations of rs-fMRI timeseries for this purpose is not sensible, since the mapped supervertices do not belong to the same subject. Instead, we use the correlations of the connectivity fingerprints to compute the connection strength of inter-subject links. A connectivity fingerprint is a feature vector that shows how a cortical subunit (e.g. a vertex, supervertex, or parcel) is functionally connected to the rest of the cerebral cortex~\cite{Cohen08}. It is defined by correlating the rs-fMRI timeseries associated with a supervertex with the rest of the supervertices and applying Fisher's $r$-to-$z$ transformation to the resulting correlation coefficients. Connectivity fingerprints effectively reflect the functional organisation of the cerebral cortex and intuitively, we expect the fingerprints of linked supervertices to be similar as they are mapped to each other based on their functional connectivity patterns. 

After defining inter-subject links for all subject pairs, we combine them with the individual sparse affinity matrices within a multi-layer graph $\mathbb{W} = (W^{(i,j)}~|~\forall~i,j \in [1,M])$, in which the diagonal entities ${W}^{(i,i)}=W$ corresponds to single-subject affinity matrices reflecting the \textit{within-subject} connectivity and the off-diagonal entities ${W}^{(i,j)}$ ($i \neq j$) denote the set of links $c_{ij}$ and $c_{ji}$ between cortical surfaces $S_i$ and $S_j$, representing the \textit{inter-subject} connectivity. $\mathbb{W}$ is an $M$-layer graph, with the size of $(N \times M) \times (N \times M)$ where $M$ is the number of subjects in the group and $N$ is the number of supervertices. A small patch taken from a $2$-layer graph is illustrated as an example in Fig.~\ref{fig:spectral_matching}(b). 

\subsubsection{Generation of Group-wise and Subject-Level Parcellations}
The joint spectral decomposition of $\mathbb{W}$ provides a shared spectral embedding $\mathbb{F}$, representing every subject in the group. We compute the joint-embedding by following the steps described in \textit{Spectral Embedding} (Section~\ref{sec:spectral_embed}), since $\mathbb{W}$ holds the same properties as the subject-level sparse affinity matrices, i.e. it is symmetric and positive-semidefinite. Due to the fact that each eigenvector in the joint embedding characterises the shared connectivity patterns in the population, it can be used in a spectral clustering setting in order to compute parcellations across subjects. At this stage, applying \textit{k}-means to the low-dimensional embedding can be the typical strategy for its simplicity and applicability~\cite{Arslan15a}, however, \textit{k}-means can be rather unstable due to its dependency on a random initialization. As a result, the parcellations generated by \textit{k}-means may not be reproducible on a run-to-run basis. To overcome this, we replaced \textit{k}-means with an alternative technique that discretises the spectral embedding by learning a rotation of the eigenvectors~\cite{Yu03} as suggested in~\cite{Craddock12,Thirion14}. 

The output of spectral clustering is a label vector $\mathbb{L}$ of length $N \times M$ that assigns a parcel to each supervertex on all cortical surfaces used to define $\mathbb{W}$. By dividing $\mathbb{L}$ into sequential sub-vectors of length $N$, we can obtain a parcellation for each subject and identify brain regions that are consistent across subjects. A simple majority voting across the individual subject parcellations can then be used to generate the group parcellation. 

\section{Experiments}
\label{sec:joint-exper}
\subsection{Evaluation}
We evaluate our parcellation framework with respect to three criteria: (a) reproducibility, (b) fidelity of the parcellations to the underlying connectivity data, and (c) functional consistency between subject-level and group-wise parcellations. 

Reproducibility is a widely-accepted technique for evaluating the robustness of a parcellation algorithm with respect to the extent of alignment in boundaries between different parcellations~\cite{Craddock12,Blumensath13,Shen13}. We use two well-known reproducibility measures, Dice coefficient and adjusted Rand index, to evaluate group-to-group reproducibility. Dice coefficient measures the overlap between two parcellations~\cite{Dice45}, while adjusted Rand index~\cite{Hubert85} evaluates the agreement of two parcellations without the necessity of matching parcels \textit{a priori}, and hence, can be more effective when parcellations with different numbers of clusters are compared~\cite{Milligan86}. The details of these techniques are previously given in Chapter~\ref{chapter:multi-level}. 

In order to assess how well a parcellation fits the underlying data, we make use of two cluster validity techniques, homogeneity and Silhouette analysis, both of which are previously introduced in Chapter~\ref{chapter:multi-level} and only summarised here. Homogeneity measures a parcellation's ability to group vertices with similar connectivity into same clusters and is calculated as the average similarity between every pair of vertices assigned to a parcel. A global homogeneity value is obtained by averaging the homogeneity values across all parcels~\cite{Craddock12}. Silhouette analysis measures how well vertices fit in their assigned parcels. Different from homogeneity, not only it evaluates the compactness of parcels, but also their degree of separation from each other. It is achieved by combining within-parcel dissimilarity defined as the average distance to all other vertices in the same parcel and inter-parcel dissimilarity obtained from those assigned to other parcels. A global score is obtained by averaging the Silhouette coefficients across all vertices. 

\begin{figure}[bth!]
\centering
\includegraphics[width=0.75\textwidth]{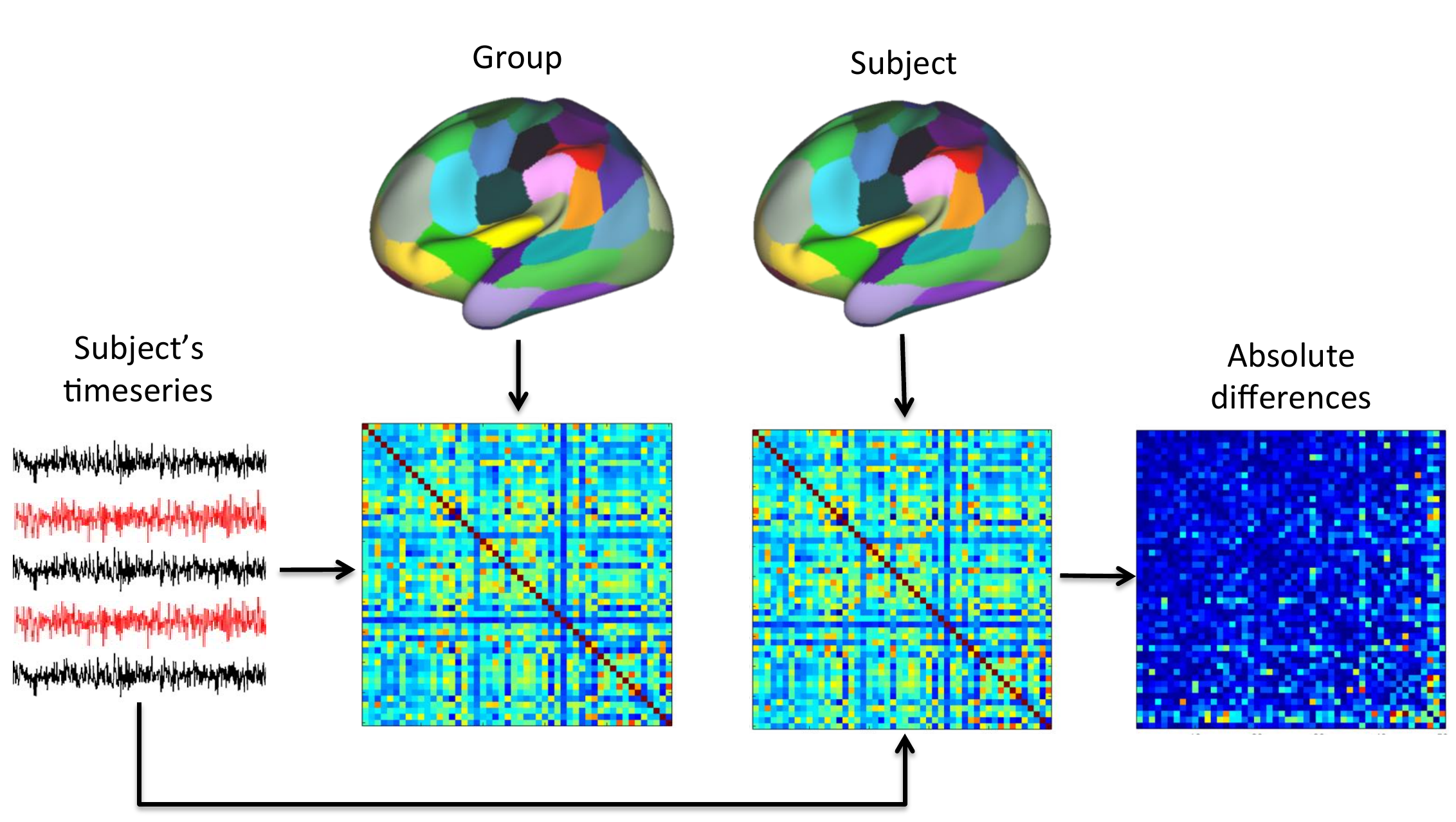}
\caption[Illustration of how to compute the sum of absolute differences.]{Illustration of how to compute the sum of absolute differences (SAD) between a group-wise and subject-level parcellation. }
\label{fig:sad-illustrate}
\end{figure}

Another critical performance measure for group-wise parcellations is their ability to represent individual subjects in terms of functional consistency. We expect the variability in connectivity networks to be consistent and minimal in order to reliably use the group-wise representation in place of each subject in the group. To this end, we evaluate functional consistency by computing the sum of absolute differences between functional connectivity networks, when we replace a subject-level parcellation with its corresponding group-wise parcellation without changing the underlying RSFC data. A connectivity network is simply defined by cross correlating the average timeseries associated with each parcel. In order to compare two networks, we first match the single- and group-level parcellations using the Dice-based similarity method and exclude the non-matching parcels from the comparison in order to allow an objective comparison of all methods. This procedure is illustrated in Fig.~\ref{fig:sad-illustrate}.

\subsection{Comparison Methods}
We compare our framework with two other parcellation strategies, both are driven by spectral clustering with normalised cuts (NCUTS)~\cite{Craddock12}. The first strategy consists in performing clustering for each subject individually and applying a second level clustering to subject-level parcellations in order to obtain group-wise parcellations (i.e. two-level approach). The second approach consists of computing an average matrix from the population and submitting this matrix to a clustering algorithm (i.e. group-average approach). We chose NCUTS as the clustering algorithm for the comparison methods, as our method also relies on NCUTS to compute parcellations from the multi-layer graph. Therefore, it provides a consistent means for assessing the performance of the proposed framework (joint spectral decomposition) with respect to two alternative strategies (two-level and group-average).

\begin{figure}[bt!]
\centering
\includegraphics[width=\textwidth]{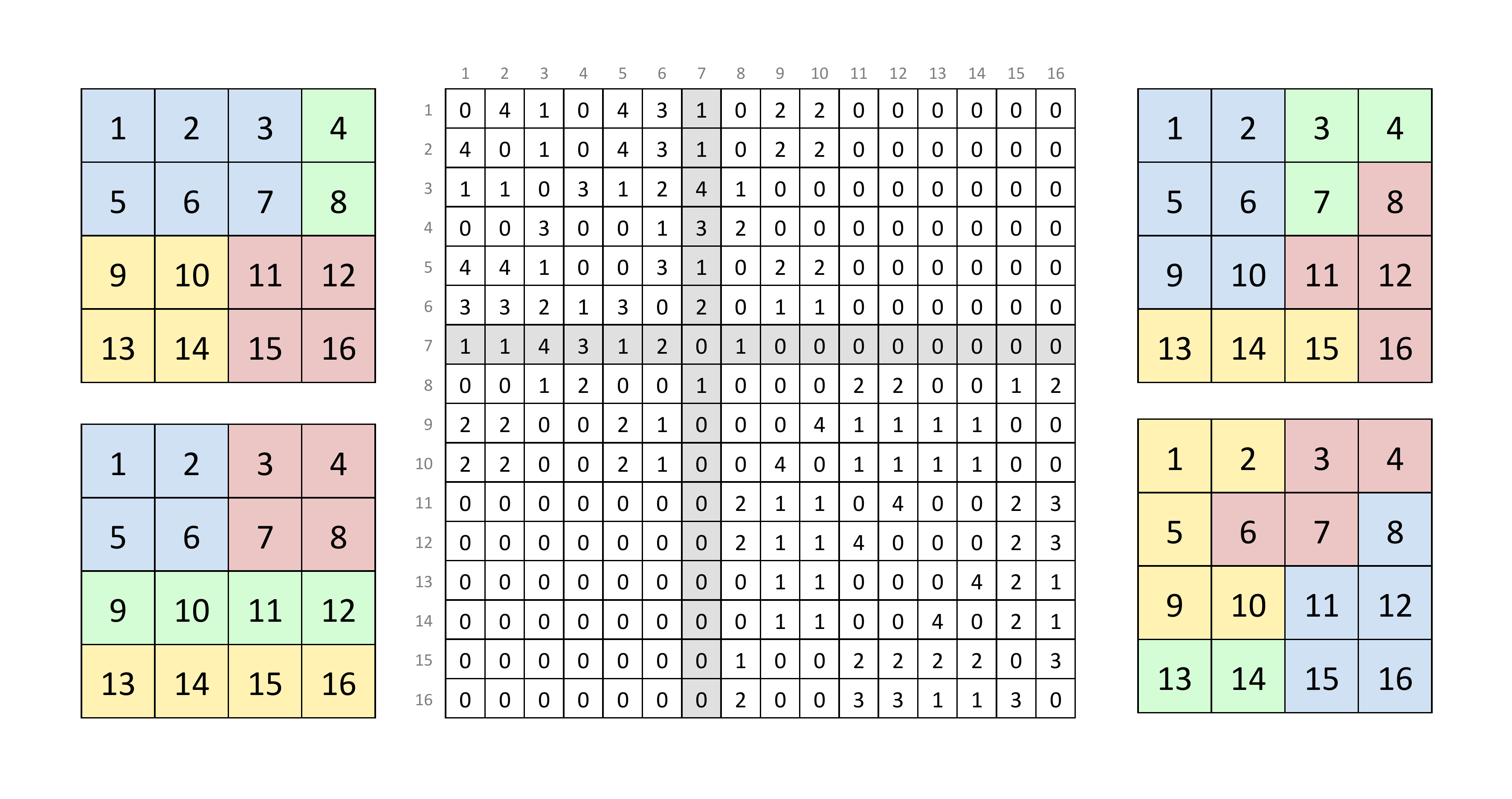}
\caption[Illustration of how to compute a symmetric adjacency matrix from four toy parcellations.]{Illustration of how to compute a symmetric $16\times16$ adjacency matrix $A$ from four toy parcellations ($4\times4$ matrices) where each colour represents a different label/parcel. For example, vertex $v_7$ (with its corresponding row and column highlighted in $A$) is assigned to the same parcel as $v_3$, $v_4$, and $v_6$, for 4, 3, and 2 times, respectively, giving $A_{7,3}=4$, $A_{7,4}=3$ and $A_{7,6}=2$. $A_{7,1}=A_{7,2}=A_{7,5}=A_{7,8}=1$, since it shares the same label with $v_1$, $v_2$, $v_5$, and $v_8$ in just one parcellation, while the rest of the entries in row 7 of the adjacency matrix are 0, since there does not exist a shared label between the other vertices and $v_7$ in any of the toy parcellations. }
\label{fig:2-level-illustration}
\end{figure}

\paragraph{Two-level approach.} This approach is similar to majority voting, in the sense that vertices assigned to the same region across subject-level parcellations are clustered together. As a result, group-wise parcellations obtained via this method can capture the shared characteristics of the population as approximated by the individual parcellations. To this end, a graphical model of the `parcel stability' is computed across all individual parcellations~\cite{Craddock12, Heuvel08}. We obtain subject-level parcellations by applying NCUTS to individual affinity matrices as defined in \textit{Sparse Affinity Matrix} (Section~\ref{sec:sparse_affinity}). We then construct an adjacency matrix (i.e. stability graph), in which an edge between vertices $v_i$ and $v_j$ is weighted by the number of times both vertices are assigned to the same parcel across all individual subject parcellations. Since the initial parcellations are spatially contiguous, only vertices sharing the same cluster membership can be connected in the adjacency matrix, and hence, the spatial integrity of the second level parcellation is also guaranteed. Following the steps described in \textit{Spectral Embedding} (Section~\ref{sec:spectral_embed}), a low dimensional representation of the stability graph is computed and submitted to NCUTS. An illustration is provided in Fig.~\ref{fig:2-level-illustration} that explains the construction of a stability graph with 4 toy parcellations. This method will be referred to as \textit{Two-Level} throughout the rest of the chapter.

\paragraph{Group-average approach.} This technique aims to capture shared patterns between individuals within a population, by computing a group average representation of connectivity. This is achieved by averaging the individual affinity matrices (after applying Fisher's $r$-to-$z$ transformation) and then submitting the average matrix to NCUTS. This method will be referred to as \textit{Group-Avr} throughout the rest of the chapter.

\subsection{Experimental Setup}
In our experiments, we subdivide the dataset into two equally-sized, mutually exclusive groups (Subgroup 1 and Subgroup 2) by random selection and compute group-level parcellations for each subgroup (of 50 subjects) by running the algorithms separately on the left and right hemispheres. For each method, we generate parcellations containing $K = 100$ to $500$ regions (i.e. 50 to 250 per hemisphere, in increments of 50). This process was repeated for 100 times, each time forming new subgroups of 50 subjects and generating two new group-wise parcellations for the same population. We assess reproducibility by comparing parcellations obtained from Subgroup 1 and Subgroup 2 in each repetition. For the homogeneity and Silhouette analysis, we only evaluate the parcellations obtained from Subgroup 1, but with respect to the data from Subgroup 2 so as to reduce the possible bias towards parcellation generation methods.  

\subsection{Parameter Selection}
Other than the number of parcels $K$, the proposed algorithm has three external model parameters: (1) the number of eigenvectors retained in the spectral feature matrix for spectral matching, $d$, (2) the number of supervertices used to pre-cluster the cortical vertices, $N$, and (3) the number of links that connect each supervertex to other cortical surfaces, $k$. 

The eigenvectors corresponding to the smallest eigenvalues represent the most important characteristics of the input dataset. However, there is still no consensus on how to select the optimal number of eigenvectors in the low-dimensional embedding. We retained the first 50 eigenvectors for the spectral matching process, since our experiments showed that further increasing $d$ did not change the mappings between subjects, thus had no impact on the final parcellations. 

The number of supervertices ($N$), and the number of links that interconnect subjects ($k$), have a more critical role in the proposed framework as they might directly affect the computed parcellations. Decreasing the spatial dimensionality through an initial clustering stage to some extent is primarily beneficial to alleviate the impact of noise at the vertex level. A very coarse initial parcellation, however, would lead to over-smoothing of the subject-level connectivity, and may consequently generate a bias towards capturing features regarding the spatial geometry of the multi-layer graph, rather than the underlying function. Similarly, $k$ determines the number of links that map one subject to another, which are consequently used to define the inter-subject connectivity in the multi-layer graph. Therefore, $k$ can be considered as a weighting parameter that controls the influence of the inter-subject connectivity over the intra-subject connectivity. The more number of links are constructed between subjects, the more the multi-layer graph is dominated by inter-subject variability, which may consequently lead to less reproducible group-wise parcellations. In our experiments, we selected these parameters as $N = 2000$ and $k = 10$. We further analyse the impact of the selection of these parameters to parcellation performance in the following section.

\section{Results}
\label{sec:joint-results}
\begin{figure}[t!]
\centering
\begin{tabular}{cc}

\includegraphics[height=5cm]{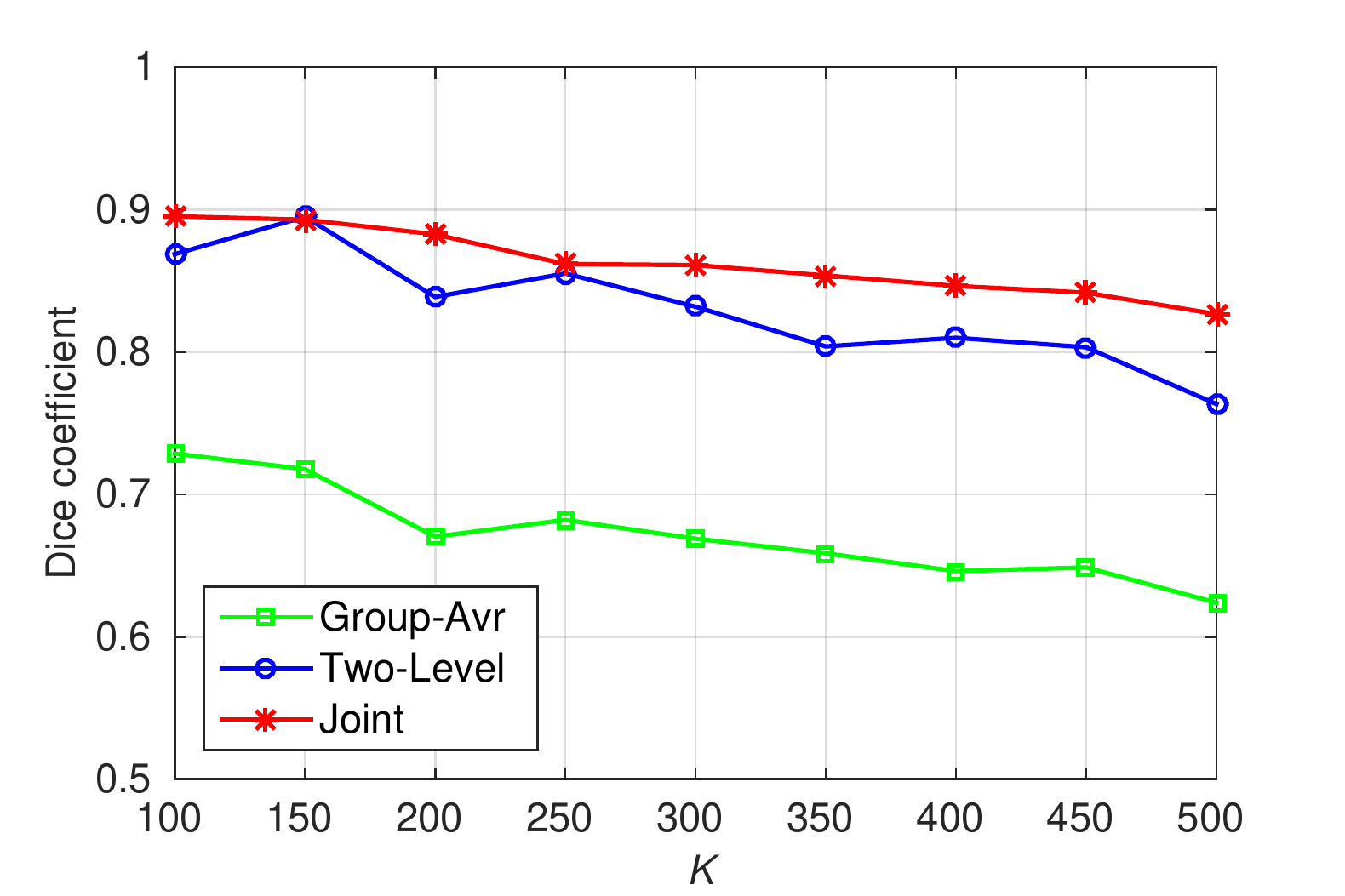} \kern-2.0em & 
\includegraphics[height=5cm]{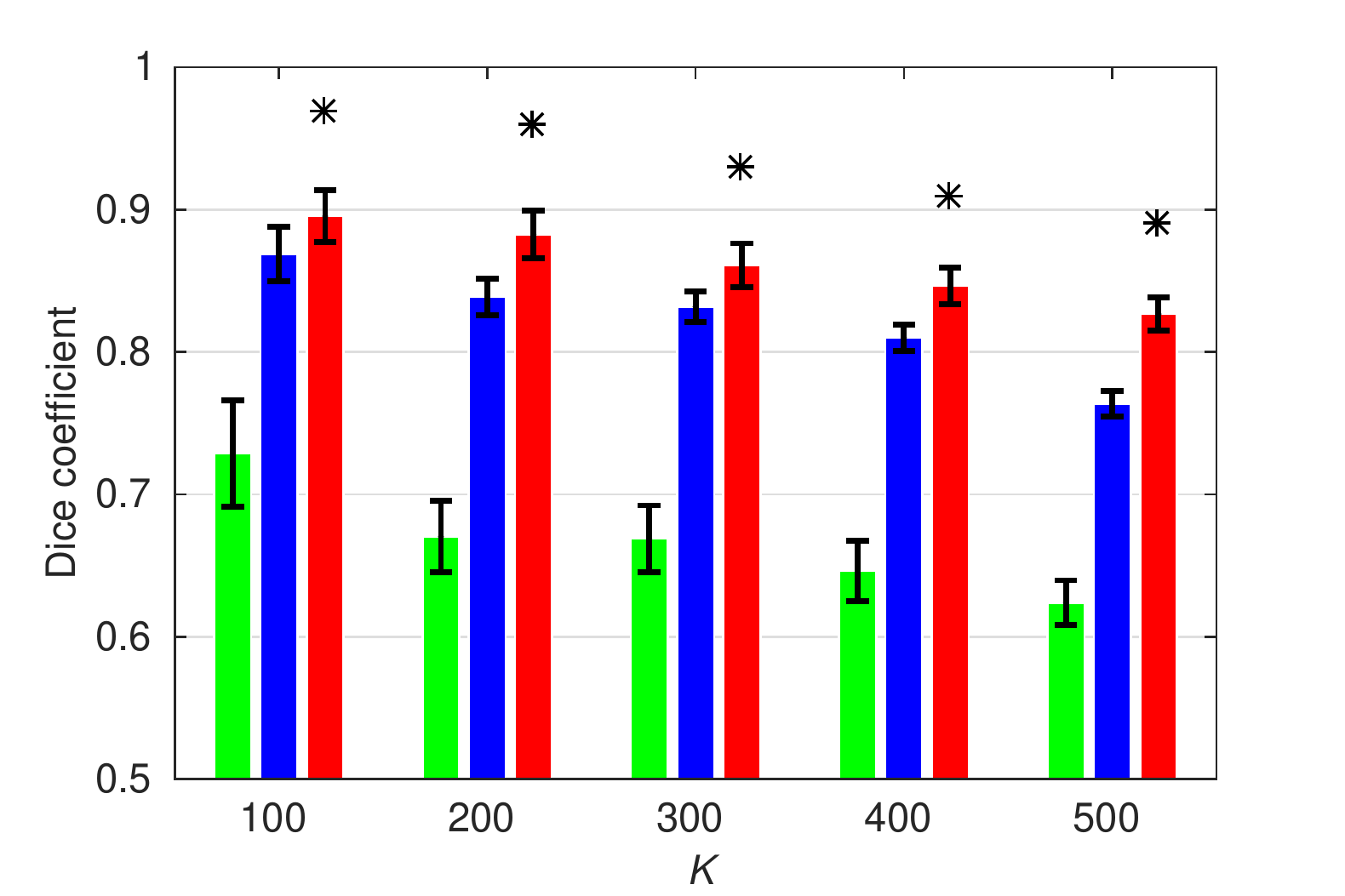} \\
\includegraphics[height=5cm]{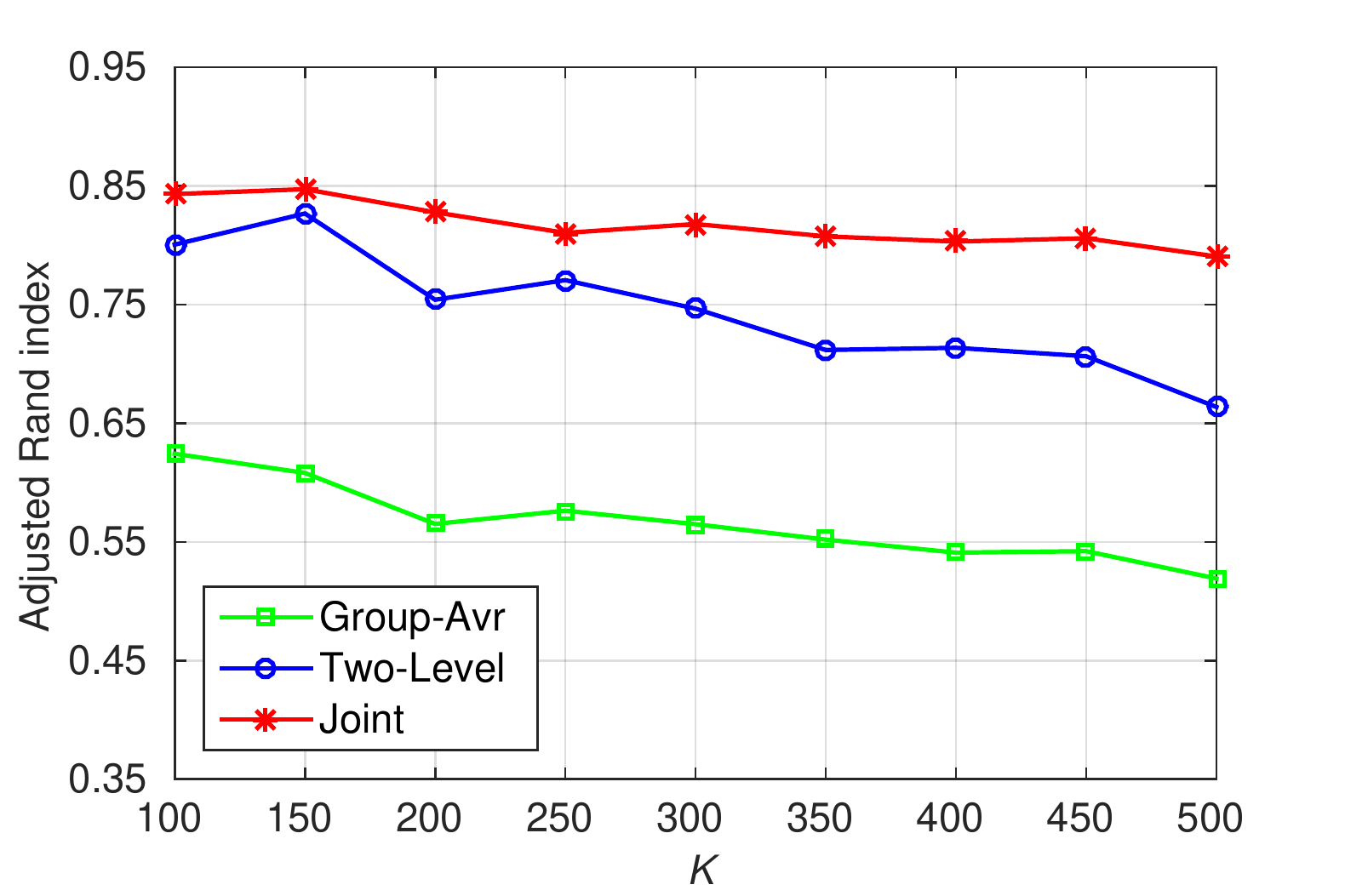} \kern-2.0em & 
\includegraphics[height=5cm]{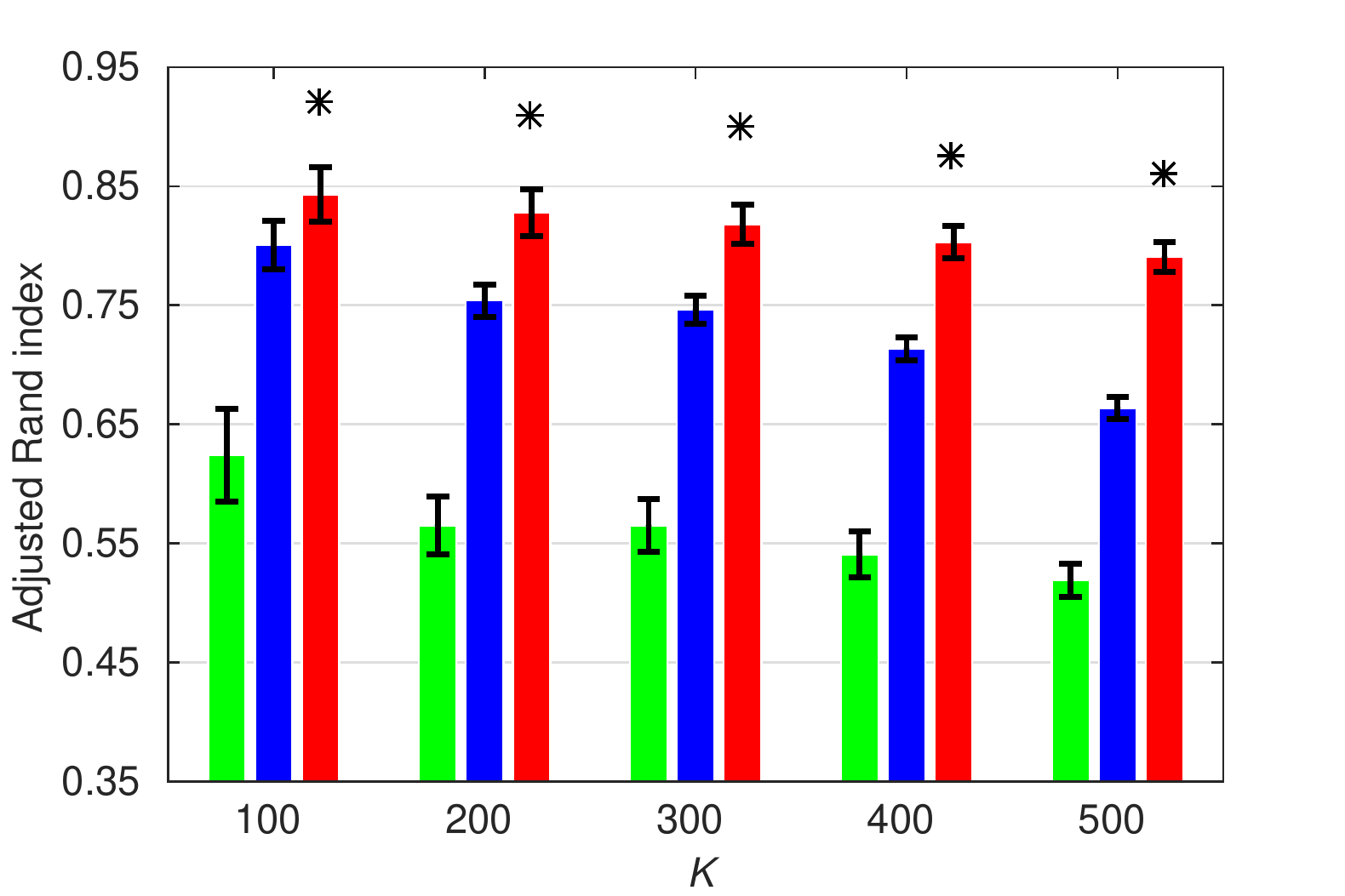}

\end{tabular}
\caption[Group-to-group reproducibility results for different number of parcels.]{Group-to-group reproducibility results for different number of parcels, measured by Dice similarity (\textit{top}) and adjusted Rand index (\textit{bottom}). Reproducibility is computed between two equally-sized, mutually exclusive groups of 50 subjects, randomly selected from the dataset. This is repeated for 100 times, each time forming two subgroups and measuring reproducibility between the parcellations generated for each subgroup. Whiskers in the bar graphs indicate the variability across repetitions as measured by standard deviation. Stars (*) show statistical significance between the winner and the runner-up based on Wilcoxon signed rank test with $p < 0.01$. }
\label{fig:repro}
\end{figure}

We present the group-wise reproducibility results obtained by the proposed algorithm as well as the comparison methods in Fig.~\ref{fig:repro}. Both Dice coefficients and adjusted Rand indices indicate that our joint spectral decomposition approach is able to obtain more reproducible parcellations at each level of resolution, with at least an average Dice score of 0.82. \textit{Group-Avr} provides the least favourable results compared to the other two approaches. This might indicate that methods that propagate information from subject-level parcellations in order to generate group-wise representations yield more reproducible results, as they already provide a means of spatial smoothing. There is a general decreasing trend in all methods with the increasing parcellation resolution, which can be attributed to the fact that, as $K$ gets larger, the functional variability across subjects gets more prominent, thus, reducing the similarity between the common characteristics within different groups and leading to less reproducible group parcellations. 

\newcommand*\rot{\rotatebox{90}}
\newcommand{\resultH}{2.5cm}

\begin{figure*}[!ht]
\centering
\begin{tabular}{cccc}
\rot{\rlap{~~~$K$ = 100}} &
\includegraphics[height=\resultH]{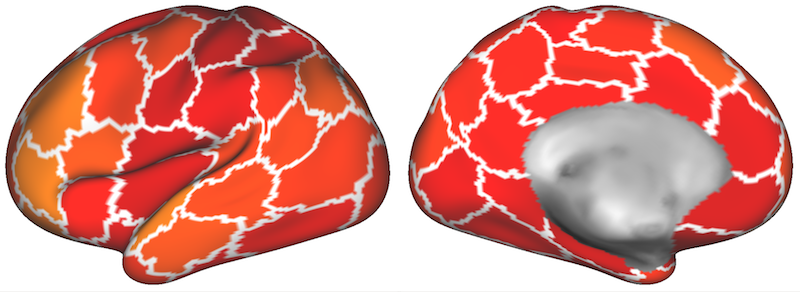} &
\includegraphics[height=\resultH]{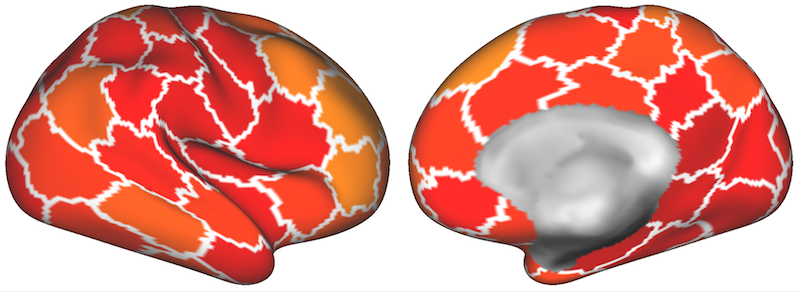} & 
~ \vspace{0.5cm} \\
\rot{\rlap{~~~$K$ = 200}} &
\includegraphics[height=\resultH]{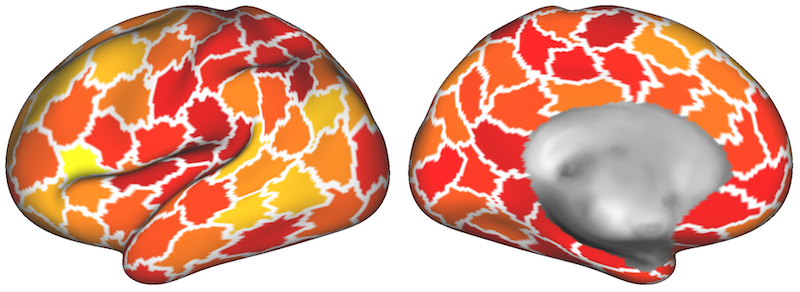} &
\includegraphics[height=\resultH]{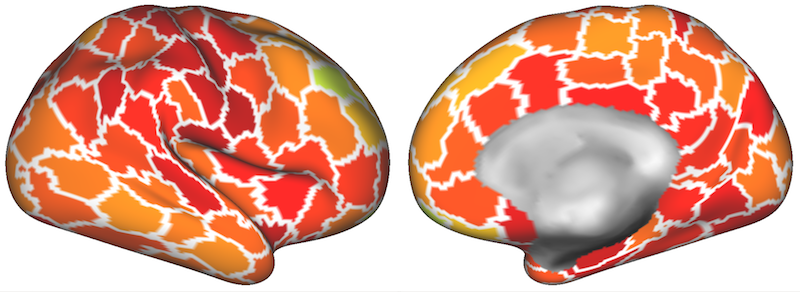} & 
~ \vspace{0.5cm} \\
\rot{\rlap{~~~$K$ = 300}} &
\includegraphics[height=\resultH]{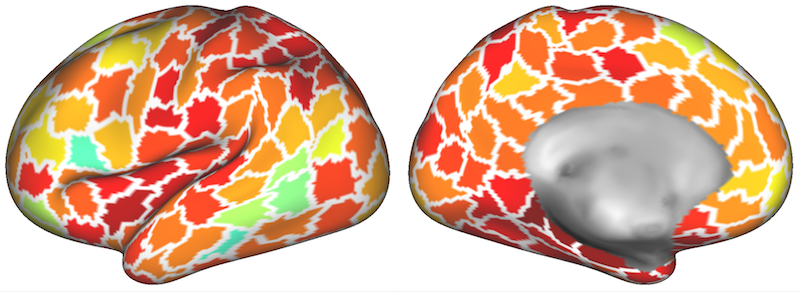} &
\includegraphics[height=\resultH]{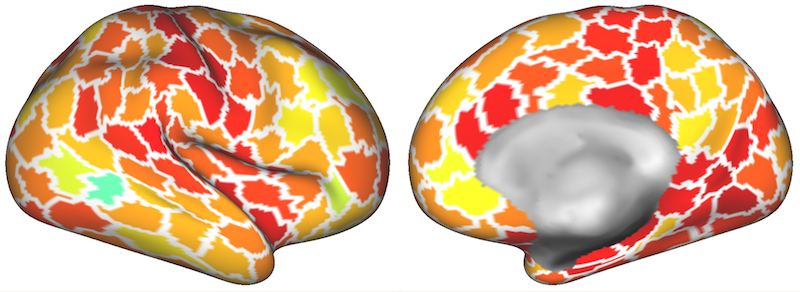} & 
\includegraphics[height=\resultH]{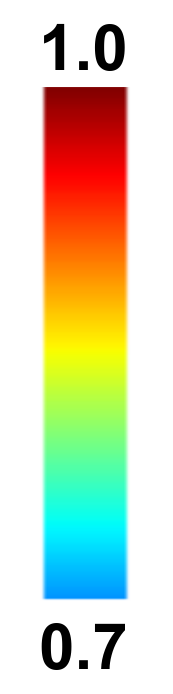} 
\vspace{0.5cm} \\
\rot{\rlap{~~~$K$ = 400}} &
\includegraphics[height=\resultH]{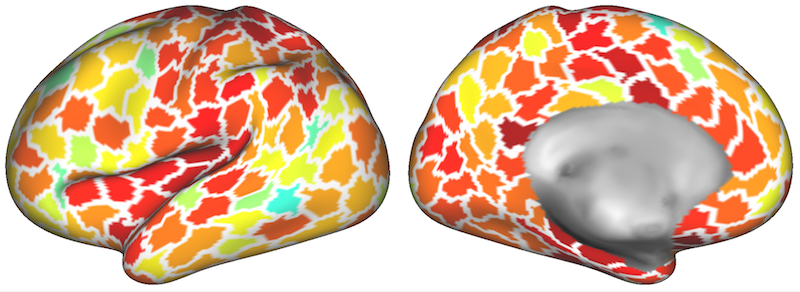} &
\includegraphics[height=\resultH]{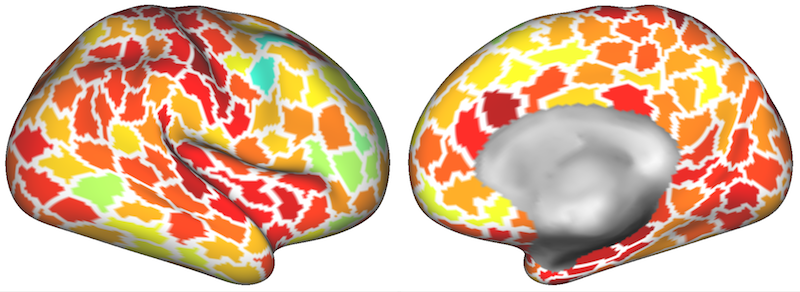} & 
~ \vspace{0.5cm} \\
\rot{\rlap{~~~$K$ = 500}} &
\includegraphics[height=\resultH]{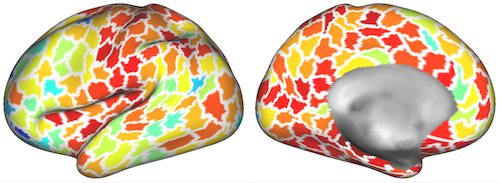} &
\includegraphics[height=\resultH]{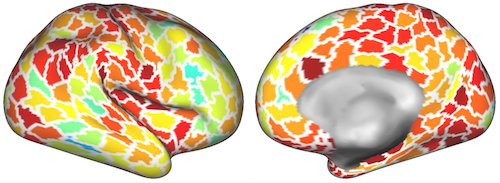} & 
~ \\
\end{tabular}
\caption[Group-wise whole-brain parcellations obtained by the joint spectral decomposition method.]{Group-wise whole-brain parcellations obtained by the joint spectral decomposition method for $K$ = 100, 200, 300, 400, and 500 parcels. The colouring indicates the average reproducibility score (Dice coefficients) of each parcel across subject-level parcellations.}
\label{fig:visual_results}
\end{figure*}

In Fig.~\ref{fig:visual_results}, we present the group-wise parcellations obtained by our approach from one group of 50 randomly selected subjects with different number of parcels and visualize the reproducibility of each group-level parcel across individual subject parcellations. Cross-subject reproducibility is measured by Dice similarity. In order to compute how reproducible a region is across different subjects, we follow the following procedure: For each group-level parcel we find its match in a subject-level parcellation and record their Dice score. We repeat the same process for all 50 subjects and average the Dice scores to obtain the reproducibility measure for that group-level parcel. Hot colours indicate a higher reproducibility across single-subject parcellations. Due to functional and structural variability between different individuals and the varying levels of SNR in the rs-fMRI data, it is not possible to obtain high Dice scores for each part of the cerebral cortex~\cite{Gordon16}. In general, the similarity between the group-wise parcellation and the individual subject parcellations drops with increasing resolution, as the impact of inter-subject variability gets stronger at higher level of granularity. Nevertheless, our approach is robust enough to achieve an average Dice score of at least 0.7 for each group-wise parcel. 

\begin{figure}[th!]
\centering
\begin{tabular}{cc}

\includegraphics[height=5cm]{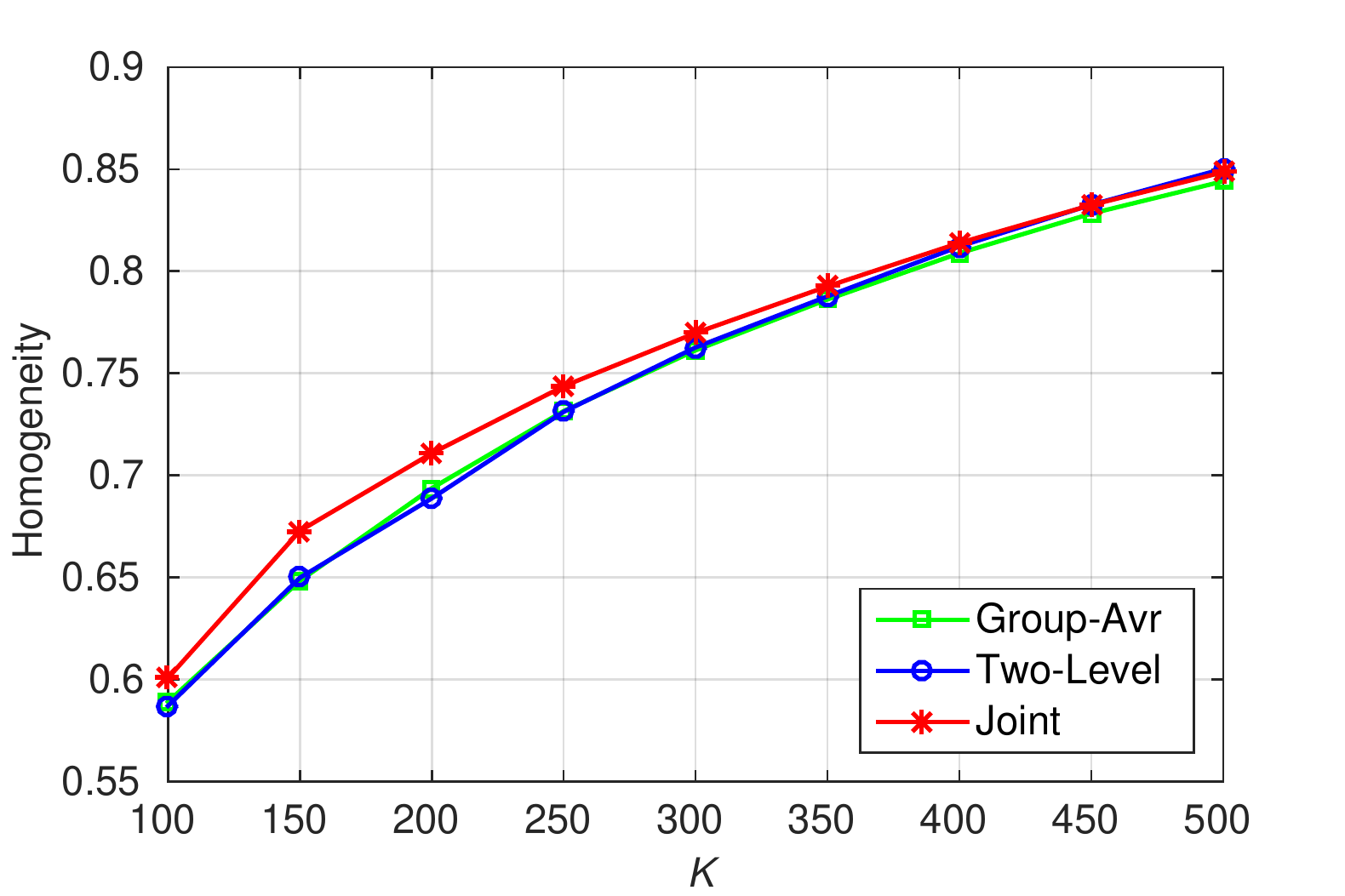} \kern-2.0em & 
\includegraphics[height=5cm]{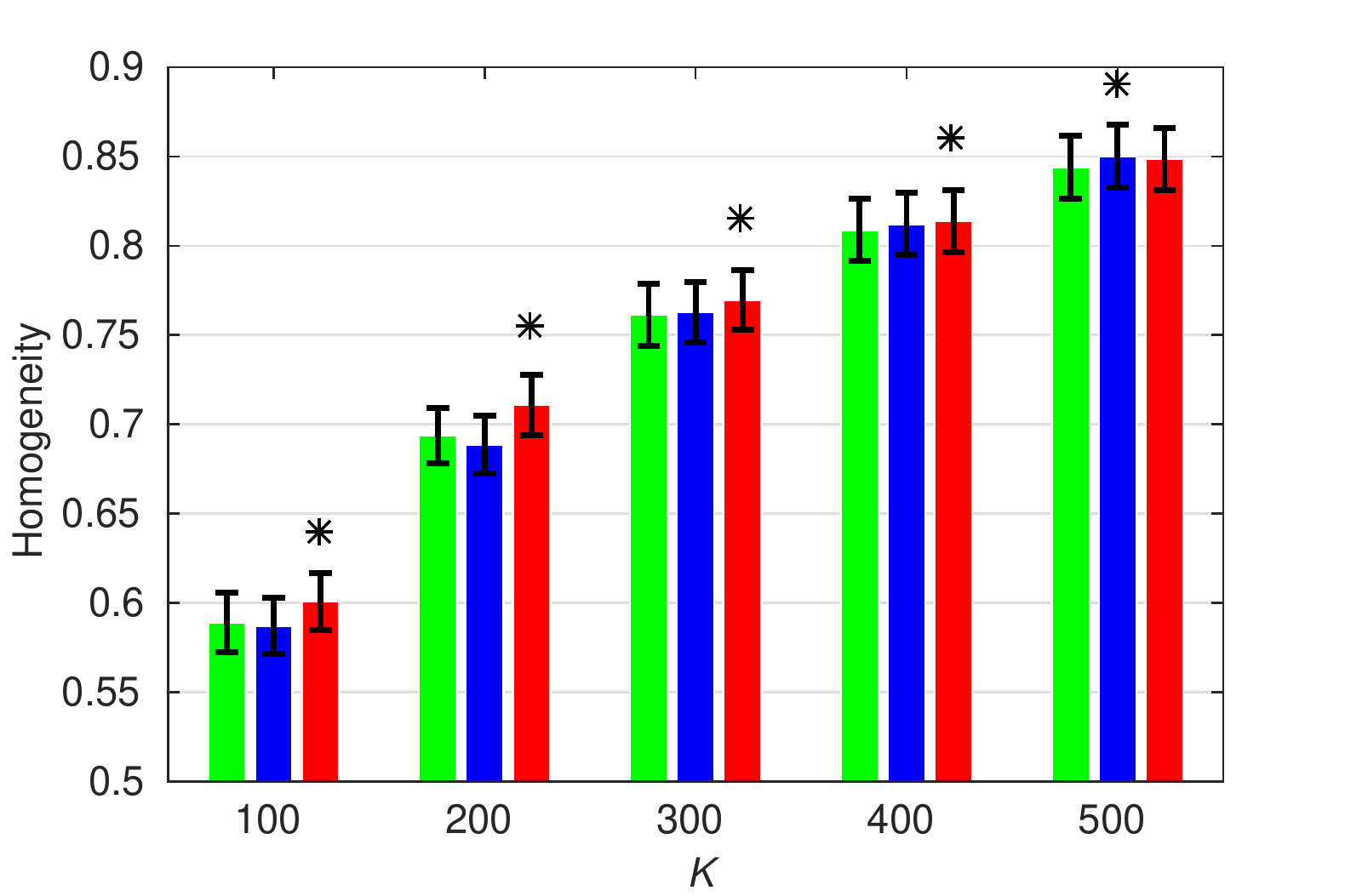} \\
\includegraphics[height=5cm]{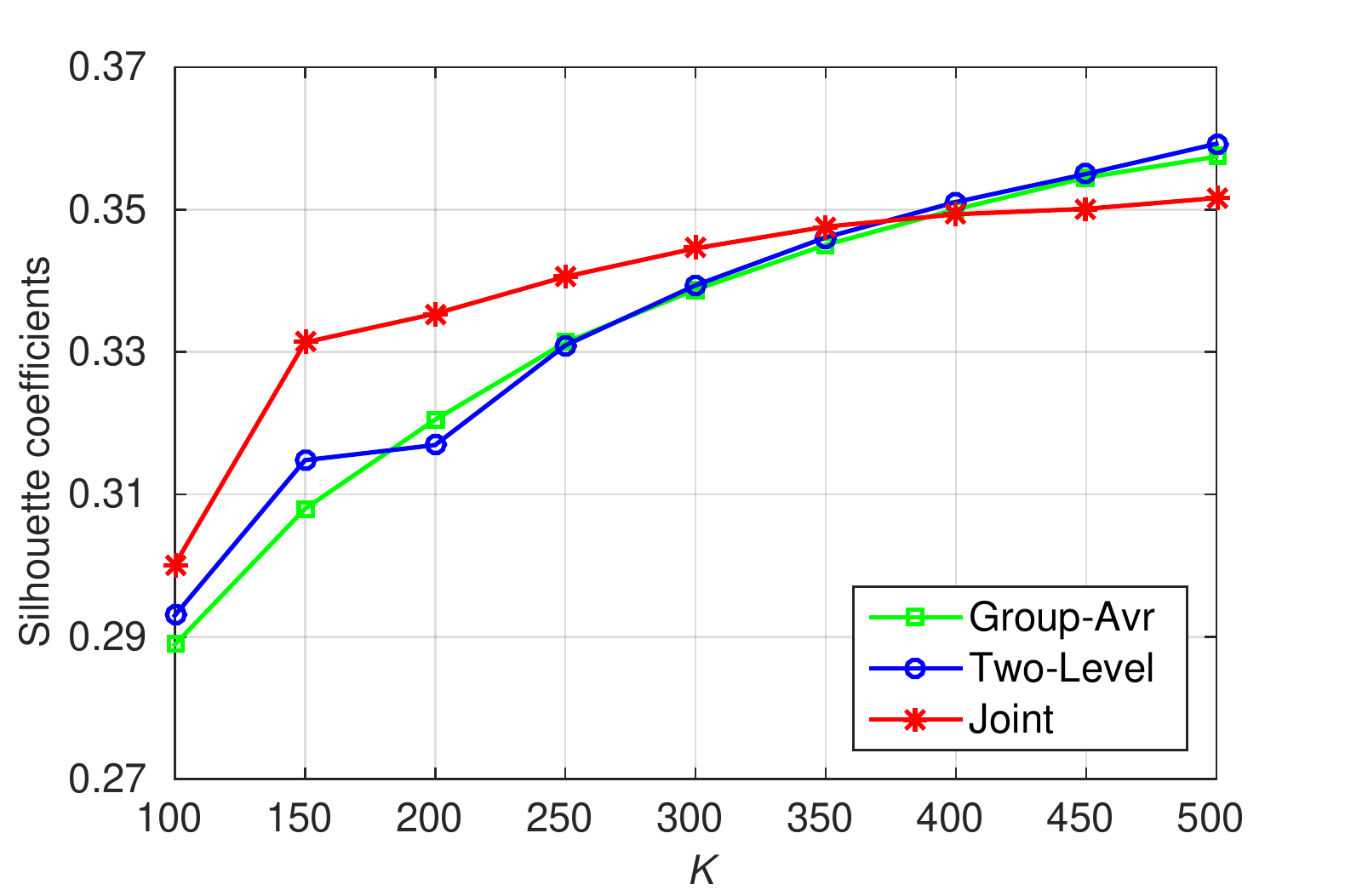} \kern-2.0em & 
\includegraphics[height=5cm]{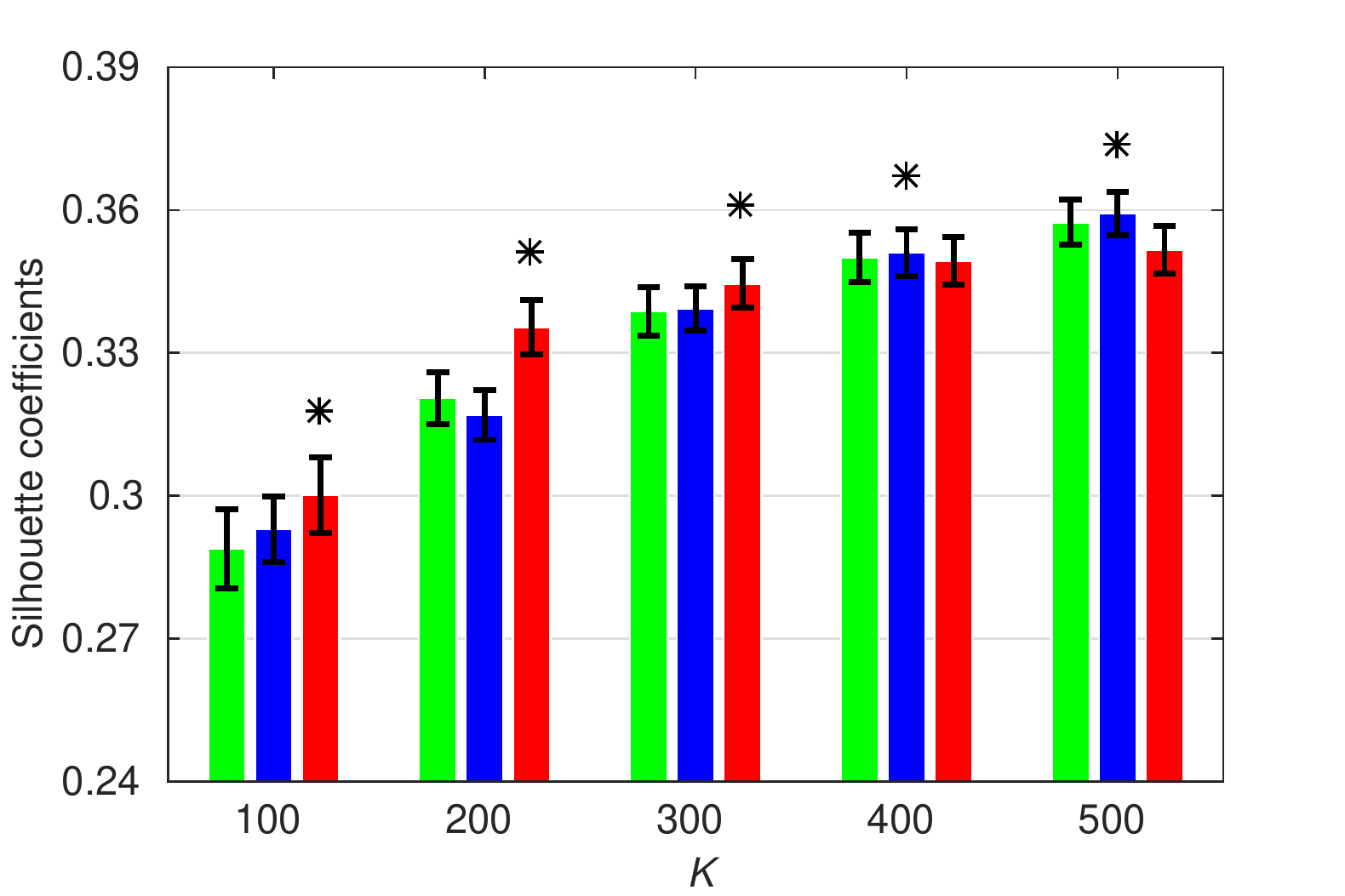}

\end{tabular}
\caption[Homogeneity and Silhouette analysis results for different number of parcels.]{Homogeneity (\textit{top}) and Silhouette analysis (\textit{bottom}) results for different number of parcels. Both measures are computed for 100 different group-wise parcellations, each of which is obtained from a set of 50 randomly selected subjects. Whiskers in the bar graphs indicate the variability across runs as measured by standard deviation. Stars (*) show statistical significance between the winner and the runner-up based on Wilcoxon signed rank test with $p < 0.01$. }
\label{fig:homo_silh}
\end{figure}

Goodness-of-fit of the parcellation approaches in terms of homogeneity and Silhouette coefficients is reported in Fig.~\ref{fig:homo_silh}. Although, homogeneity results show a relatively similar performance for all of the approaches, the proposed method consistently generate the most homogeneous parcellations for almost all resolutions, apart from the highest level (i.e. $K=500$) where \textit{Two-Level} appears to perform slightly better. Similar trends are observed with Silhouette coefficients, in which the proposed approach shows a more favourable performance for most of the resolutions, particularly for $K=100$ to $300$, but is outperformed by the other approaches at the highest level of granularity. As a general trend in both measures, all methods show a performance increasing with the number of parcels computed. This might be linked to the fact that both measurements are dependent on the size of the parcels (e.g. smaller parcels yield better results)~\cite{Gordon16,Craddock12}. 

\begin{figure}[!t]
\centering
\includegraphics[height=5cm]{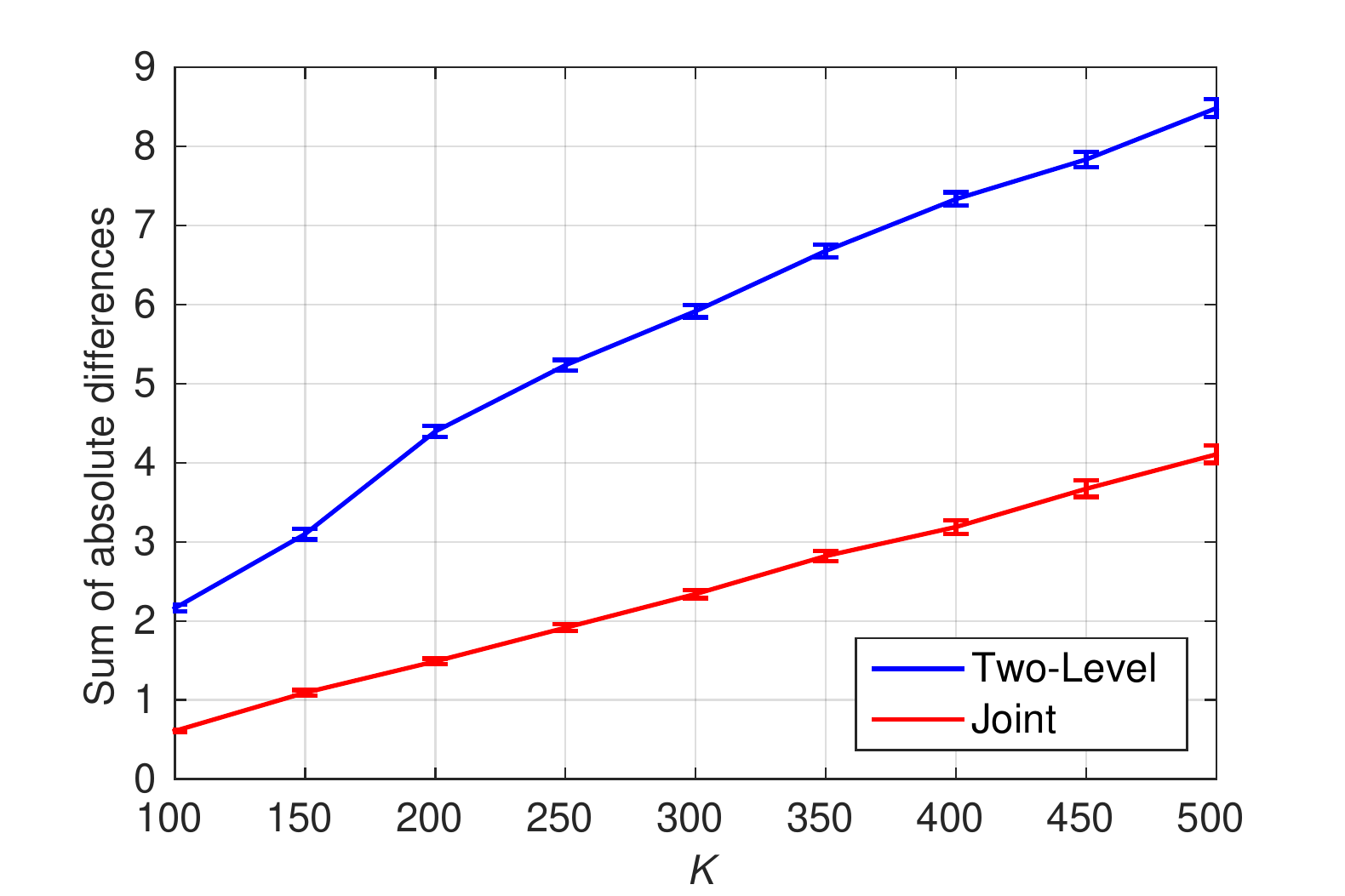}
\caption[Functional consistency results for different number of parcels.]{Functional consistency results of the proposed method (\textit{Joint}) and \textit{Two-Level} for different number of parcels. The sum of absolute differences (SADs) are computed between the connectivity networks obtained by a group-wise parcellation and its corresponding subject-level parcellations, and repeated for 100 different group parcellations, each of which is obtained from a set of 50 randomly selected subjects. The results shown here are the average SADs across all runs, normalised by the number of parcels at each level of granularity. Whiskers in the bar graphs indicate the variability across runs as measured by standard deviation. Lower SADs indicate better performance. }
\label{fig:consistency}
\end{figure}

In Fig.~\ref{fig:consistency}, we present the average sum of absolute differences (SADs) between the functional connectivity networks obtained by the individual parcellations and their group-wise representations. We excluded \textit{Group-Avr} from this experiment, since it does not provide individual subject parcellations as part of its pipelines. The results show a clear tendency in favour of the proposed method at all resolutions, indicating that the group-wise parcellations obtained by \textit{Joint} are functionally more consistent, and thus, can potentially represent the population with a higher reliability in network analysis. As expected, SADs show a steadily-rising trend for both approaches, due to increasing variability between the group- and subject-level parcellations with respect to network dimensionality, which directly depends on the parcellation resolution. 

\subsection{Parameter Analysis}
We analyse the impact of the external model parameters on the parcellation performance. We run our algorithm with different values for each parameter, while fixing the value of the other parameter and observe the changes in four different performance measures, namely, group-to-group reproducibility, homogeneity, Silhouette coefficients, and sum of absolute differences, over a spectrum of parcellation resolutions. This analysis is conducted based on the experimental setting described above, but repeated for a fewer number of runs (i.e. 10) and only on the left hemisphere. We consider any combination of the following parameter values $k=\{1, 10, 50\}$ and $N=\{1000, 2000, 3000\}$ per hemisphere. We present the results obtained for each parameter in Fig.~\ref{fig:param-anal}.

\newcommand{\paramH}{5.1cm}
\begin{figure}[th!]
\centering
\begin{tabular}{cc}

\includegraphics[height=\paramH]{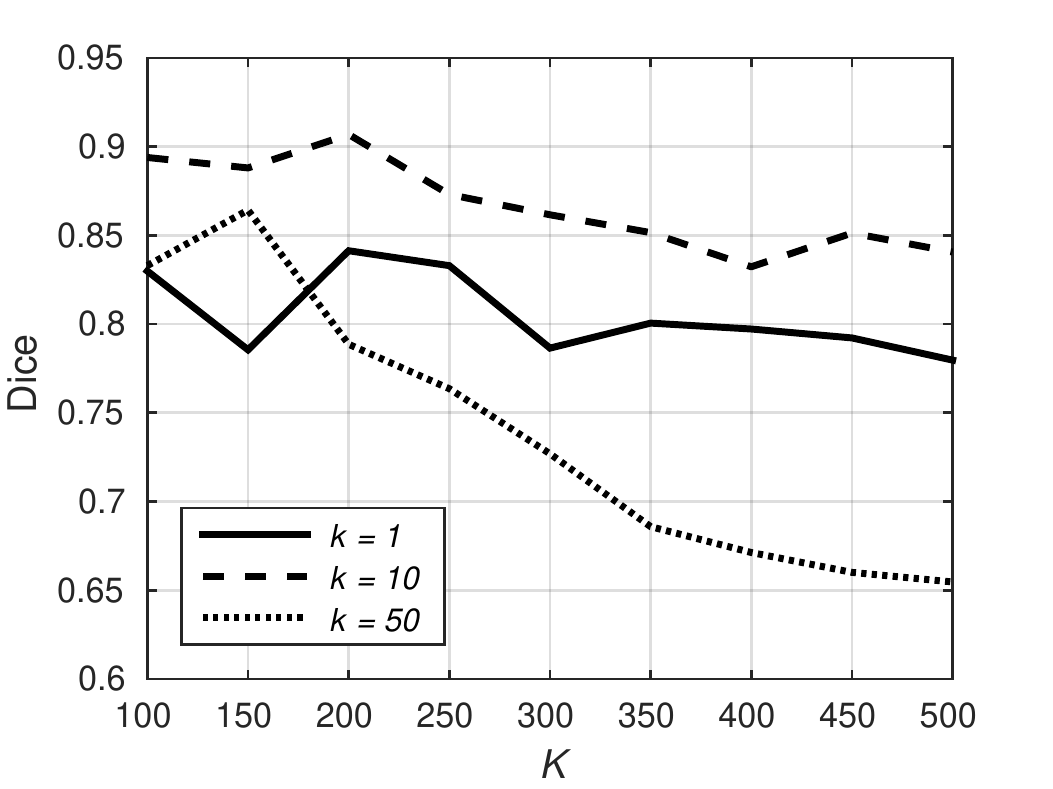} \kern-1.0em & 
\includegraphics[height=\paramH]{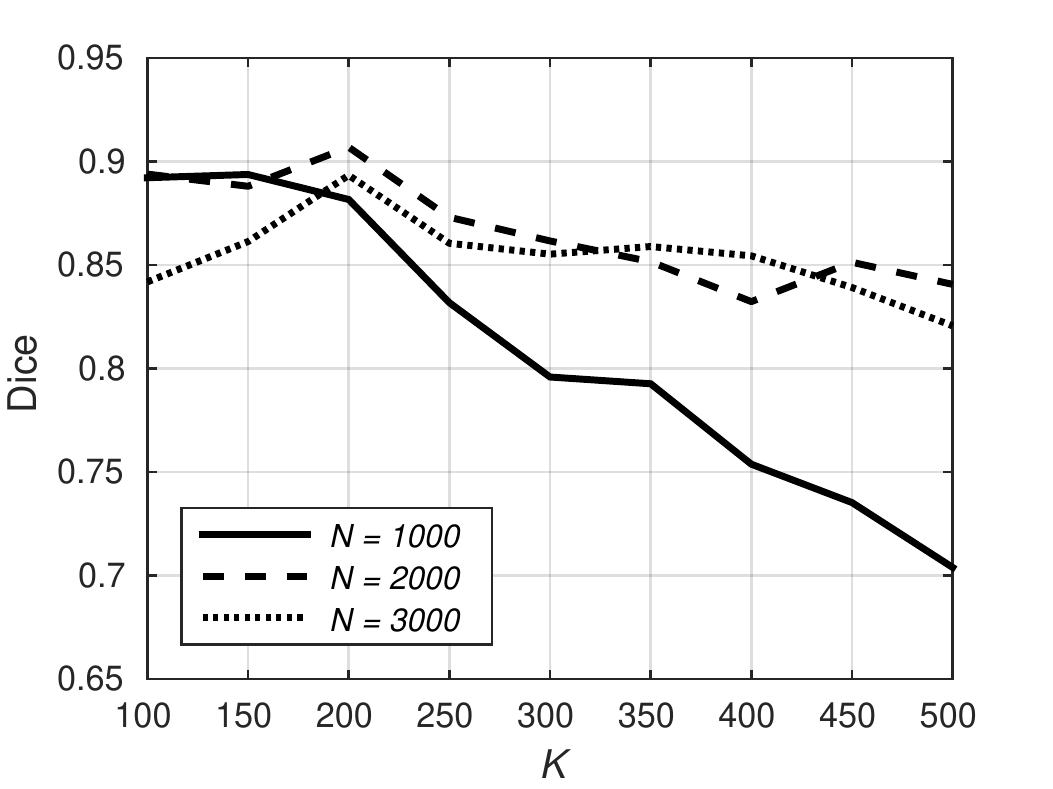} \\
\includegraphics[height=\paramH]{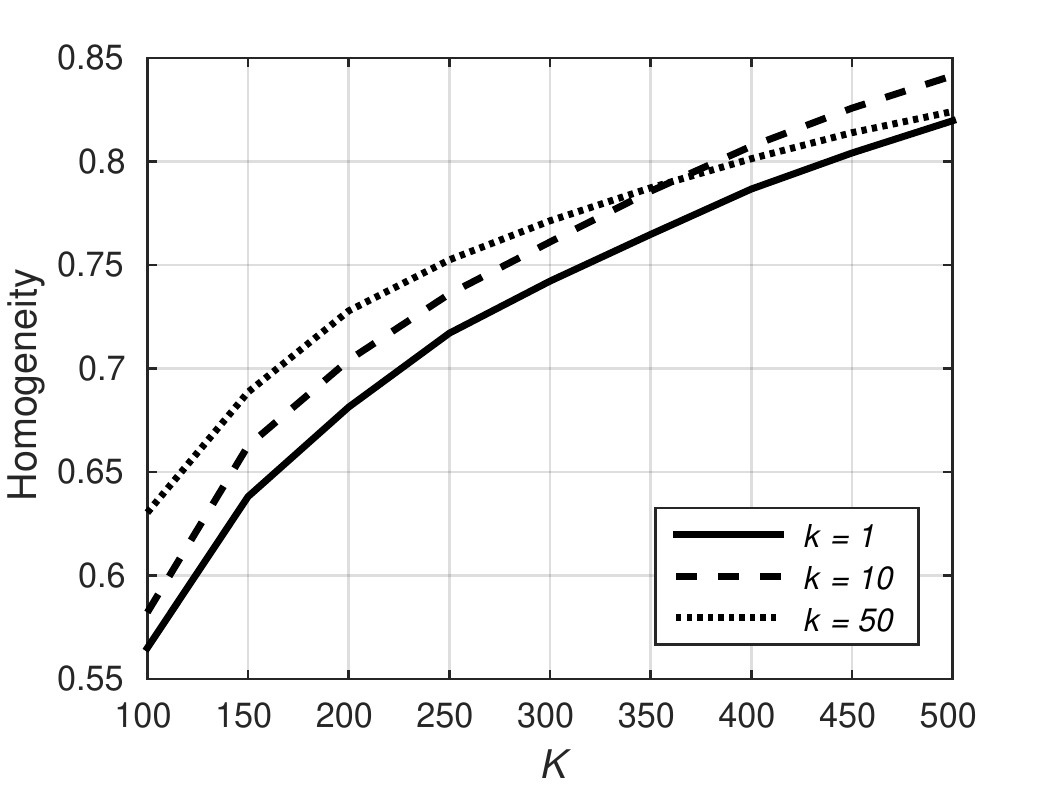} \kern-1.0em & 
\includegraphics[height=\paramH]{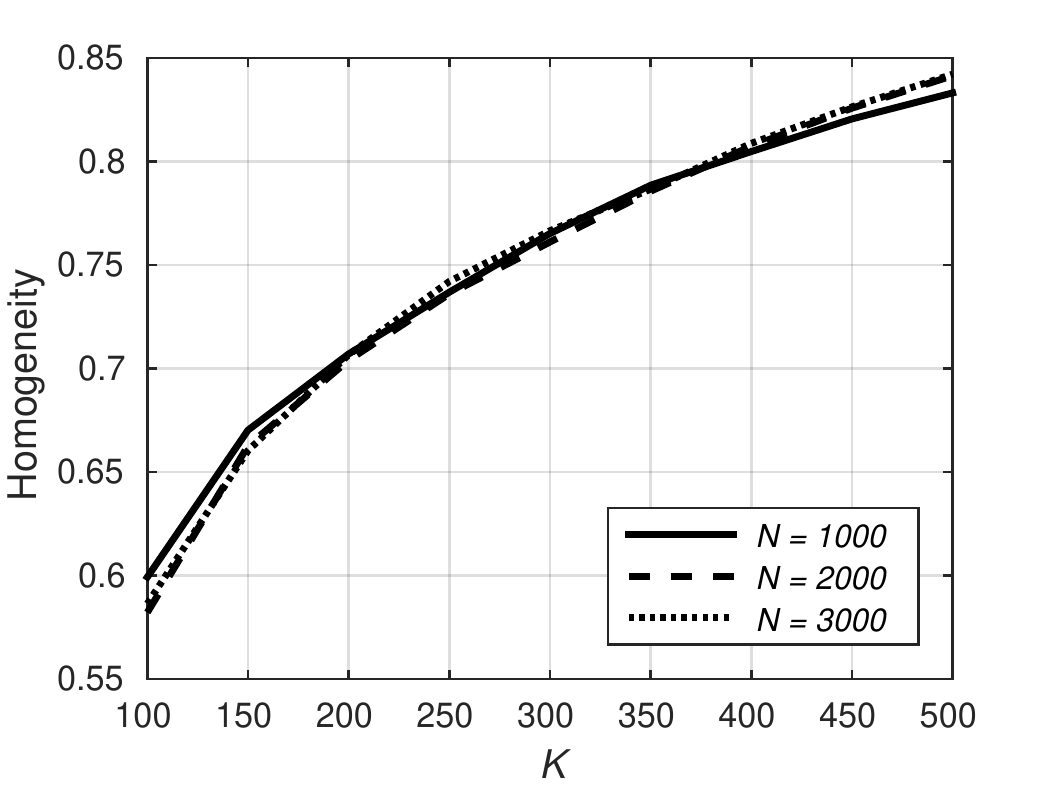} \\
\includegraphics[height=\paramH]{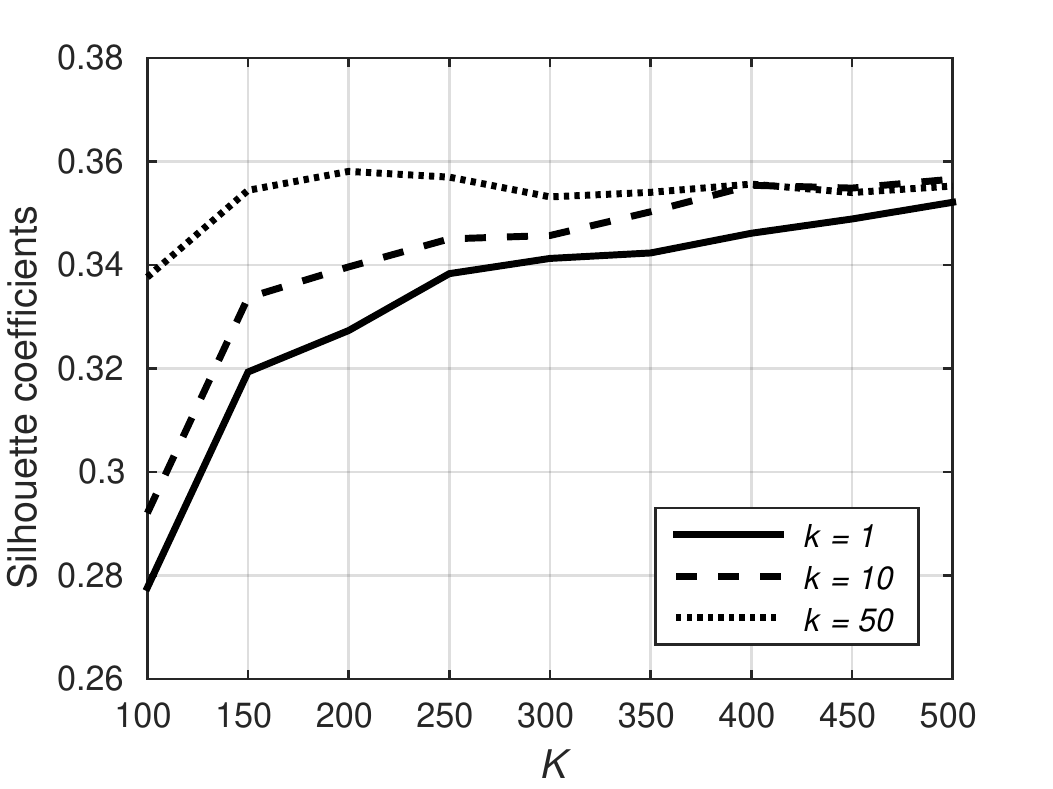} \kern-1.0em & 
\includegraphics[height=\paramH]{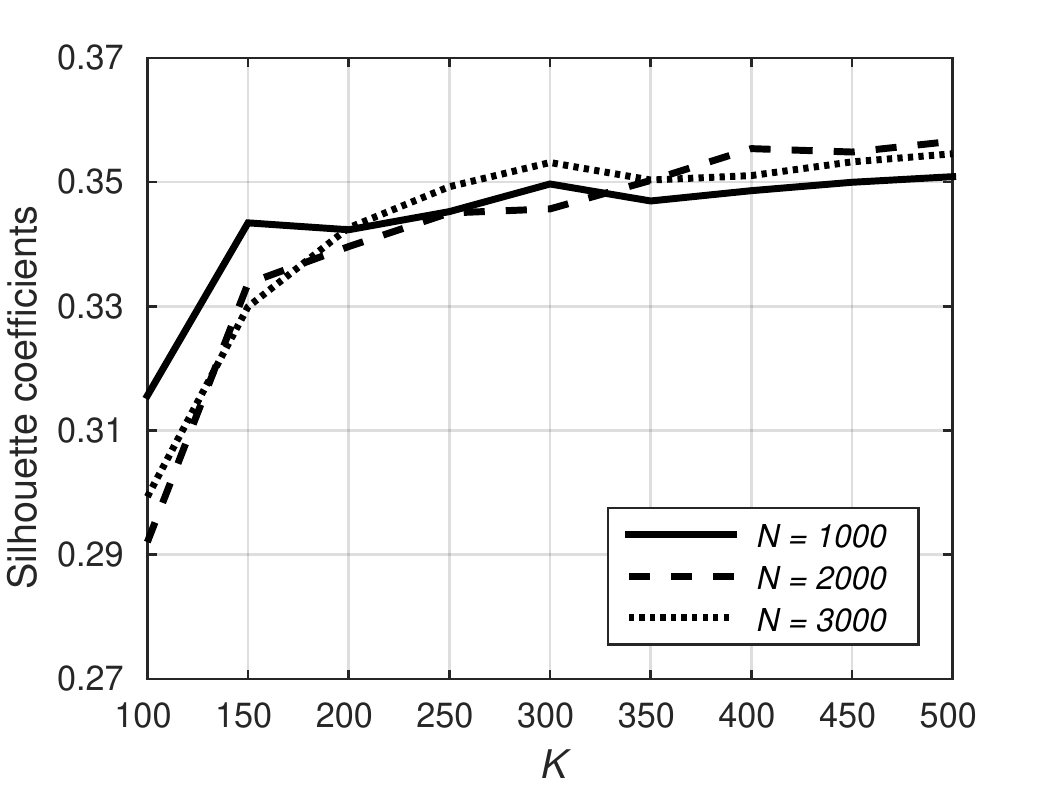} \\
\includegraphics[height=\paramH]{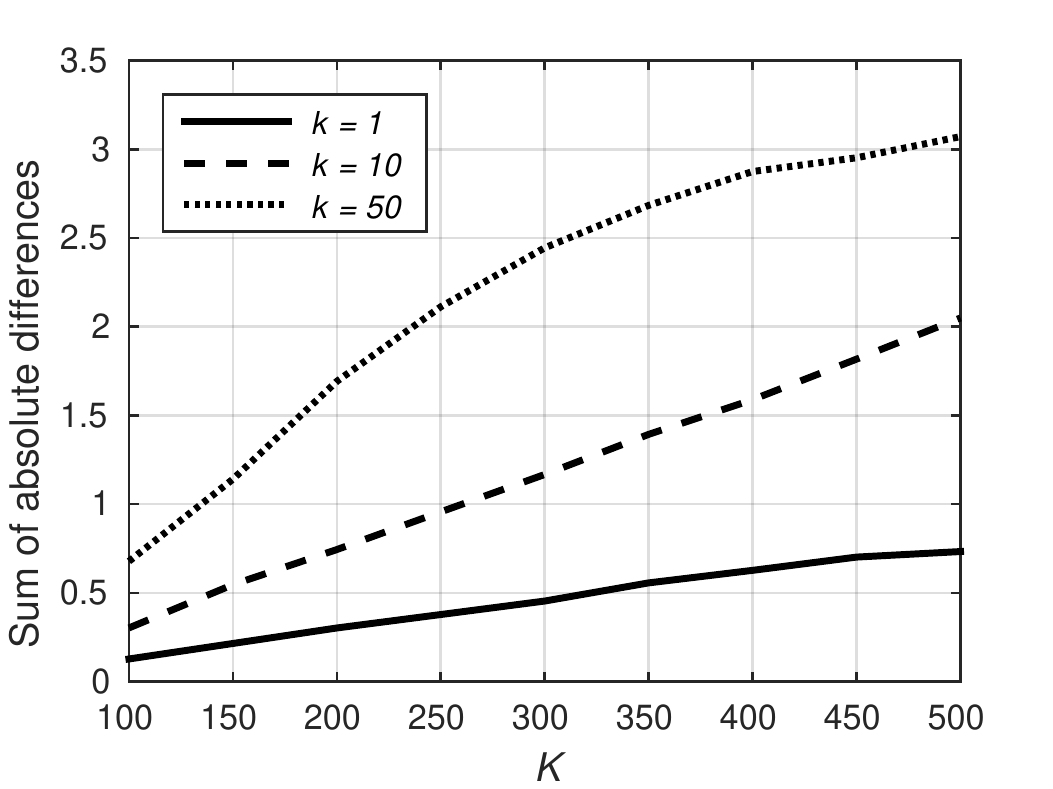} \kern-1.0em & 
\includegraphics[height=\paramH]{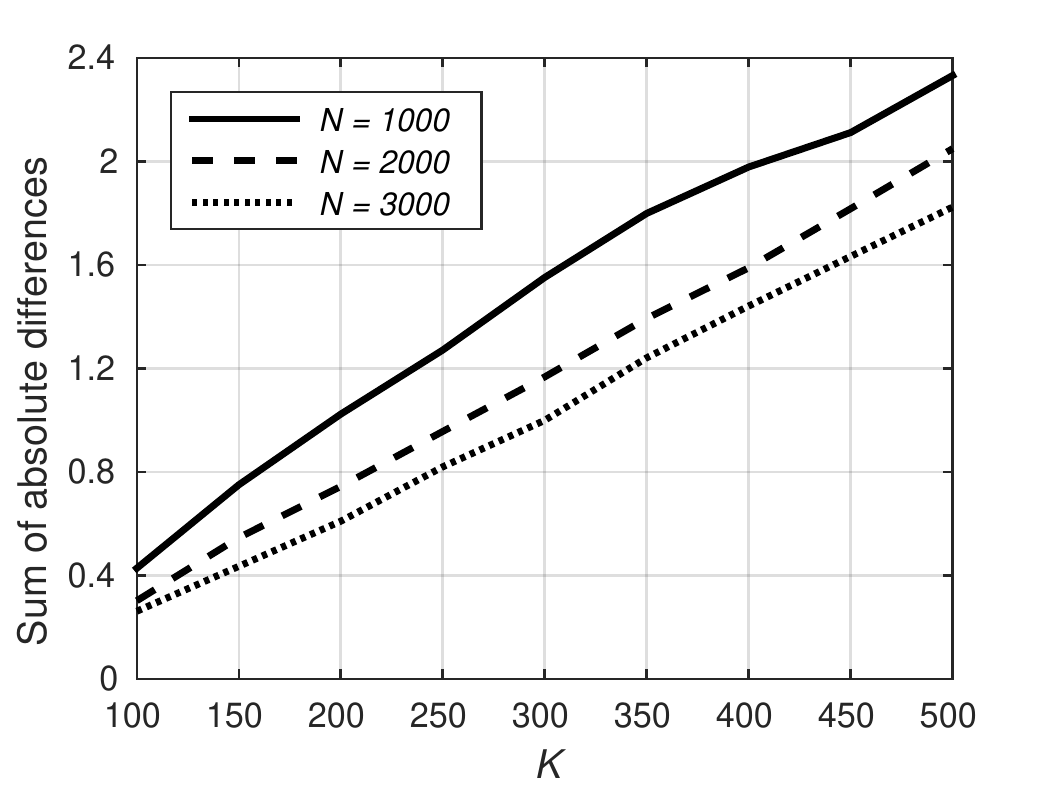} \\

\end{tabular}
\caption[Evolution of different evaluation measurements with respect to the number of (a) inter-subject links and (b) supervertices.]{Evolution of Dice-based group-to-group reproducibility, homogeneity, Silhouette coefficients, and sum of absolute differences as a function of (\textit{left}) the number of inter-subject links $k$ and (\textit{right}) the number of supervertices $N$, with respect to varying number of parcels $K$.}
\label{fig:param-anal}
\end{figure}

The first parameter is the number of inter-subject links $k$, which is used to define the mappings in spectral matching and stands for as a weight parameter between the intra- and inter-subject connectivity represented in the multi-layer graph. Increasing the value of this parameter leads to an inclination towards inter-subject connectivity in the graphical model, yielding less reproducible group-wise parcellations and higher sum of absolute differences between subject and group parcellations. However, lowering $k$ does not always reduce the group-to-group reproducibility and having around 10 links per supervertex appears to produce more robust parcellations. Considering the fidelity of the parcellations to the underlying data, increasing $k$ helps obtain more faithful representations of the functional connectivity within the population as indicated by the homogeneity values and Silhouette coefficients. 

The second parameter is the number of supervertices, $N$, used for spatial dimensionality reduction of intra-subject connectivity matrices to alleviate the impact of noise at the vertex level. Using too coarse parcellations lowers reproducibility, which might be attributed to the fact that functional connectivity at the subject level is over-smoothed, and hence, the parcellation framework relies more on the inter-subject connectivity for the generation of group-wise parcellations. A similar trend can also be observed in the SAD results, where the dissimilarity between subject and group-wise networks increases more rapidly with decreasing supervertex resolutions. The homogeneity and Silhouette analysis results are less significantly affected by the number of supervertices, which might also indicate that the fidelity of the model is more dependent on the number of links between the subjects. 

It is worth noting that, computational performance also remains as a critical consideration point regarding the selection of $k$ and $N$. Together with the number of subjects in the population, both parameters have a direct impact on the size of the multi-layer graph, and hence, can dramatically increase the computational space/time requirements for joint spectral decomposition. In Table~\ref{tab:param-anal}, we present the computational performance measures of the decomposition process for different combinations of $k$ and $N$. Table~\ref{tab:param-anal}(a) shows the amount of memory allocated for each multi-layer graph based on a different parameter setup, while Table~\ref{tab:param-anal}(b) presents the computation time during the execution of their spectral decomposition. Each multi-layer graph is defined as a sparse matrix from a set of 50 subjects and the spectral decomposition is performed using the \textit{eigs} function in Matlab (ver R2015a), on a 32-Core CPU server with 256GB RAM, operated by 64bit Ubuntu. 

\begin{table}[hb!]
\centering
\caption[Computational performance of the proposed method.]{(a) Size of memory and (b) computation time for the spectral decomposition of multi-layer graphs obtained with varying $k$ and $N$ (mb - mega bytes, gb - gigabytes, min - minutes).}
\begin{tabular}{cc}
\includegraphics[width=0.48\textwidth]{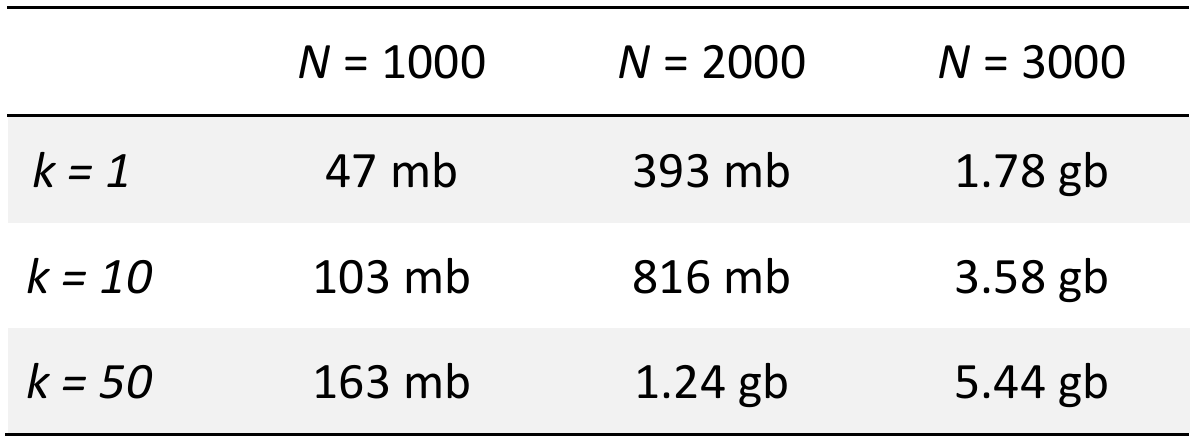}  & 
\includegraphics[width=0.48\textwidth]{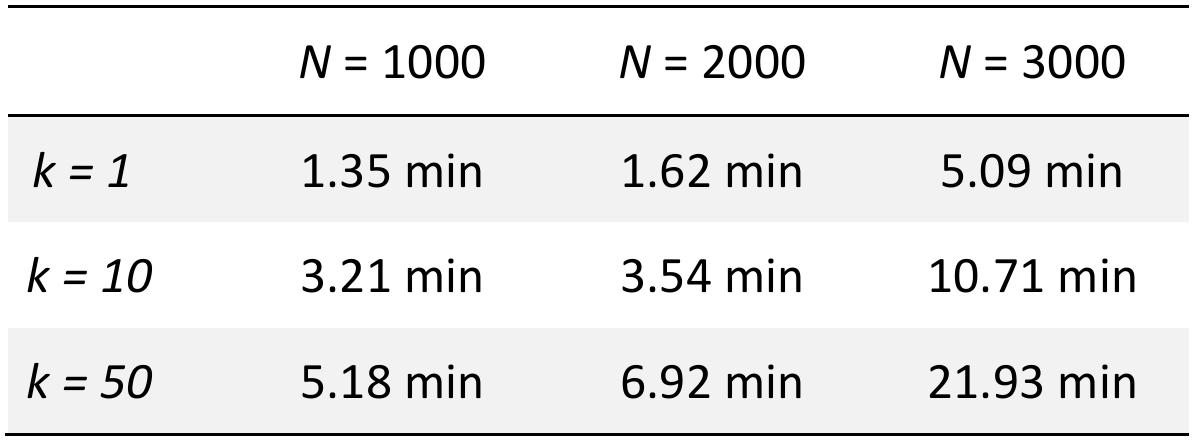} \\
(a) & (b) \\
\end{tabular}
\label{tab:param-anal}
\end{table}



\section{Discussion}
\label{sec:joint-conclus}
In this chapter, we presented a spectral graph decomposition approach to parcellate the cerebral cortex with respect to functional connectivity estimated from resting-state fMRI. Our experiments demonstrated that the proposed algorithm can produce robust parcellations with higher reproducibility and can better reflect functional and topological features shared by multiple subjects compared to other parcellation methods. The performance of our framework can be attributed to the proposed multi-layer graphical model, which combines functional connectivity captured at the subject level with the general functional tendency of the population.

Group-wise parcellations obtained by our approach can be used to define network nodes in connectome analysis, as they represent functionally uniform regions across the cerebral cortex. In general, such parcellations can be useful for comparing different groups of healthy subjects, for example, in order to identify how connectivity changes through ageing or from one gender to the other. This could further enable generation of a functional connectivity atlas, assuming that a very large cohort is used to minimise the inter-subject variability and noise-related artefacts. In addition, analysing connectivity patterns in a group of subjects with a brain disorder could help derive biomarkers in order to better understand disease-related differences in the brain connectivity. 

Spectral clustering can effectively capture the shared connectional features across the cerebral cortex, however due to hard spatial constraints imposed to the connectivity data, parcellations can be biased by the geometry of the cortical mesh~\cite{Thirion14}. Since the normalised-cuts optimisation simultaneously maximises the inter-cluster dissimilarity and within-cluster similarity, parcellations obtained via spectral clustering tend to consist of uniformly sized regions. This constitutes a well-known limitation of spectral techniques, as parcels of uniform size cannot effectively adapt to cortical regions that are naturally irregular in shape and size~\cite{Eickhoff15}. One solution to overcome this could be to perform parcellation at a higher resolution, i.e. over-parcellate the cortex, and look for consistently identified boundaries across different scales. However, it should be noted that, spectral clustering algorithms do not inherently provide parcellations that are hierarchically consistent, hence, parcellations obtained at different resolutions do not necessarily share similar boundaries. 

We developed our approach from functional connectivity captured in the form of spontaneous resting-state fMRI correlations, however, the proposed joint decomposition framework can easily be generalized to operate on other types of connectivity data, such as structural connectivity estimated from dMRI. Although, the relationship between structural and functional connectivity is an active research area~\cite{Honey09,Heuvel09,Ng13}, a comparison of parcellations derived from different sources of connectivity is under-explored, and can provide an interesting future work towards locating cortical areas that are consistently assigned to the same regions across different parcellations/resolutions. In addition to comparing dMRI and rs-fMRI driven parcellations, combining them could also be an interesting future direction towards identifying the cortical organisation of the brain more accurately than by using just a single modality.
 
In the absence of a gold standard parcellation, we evaluated the accuracy of parcellations with respect to widely-accepted clustering validity techniques, such as reproducibility, parcel homogeneity, Silhouette analysis, and sum of absolute differences. While these measures are particularly important to show the robustness of the parcellations and their ability to fit the connectivity in the population, they do not directly provide a means of assessing the reliability of parcellations from a network analysis point of view. We also limited our experiments only to cover the spectral parcellation techniques, so as to demonstrate the advantage of the proposed joint decomposition framework over the two other alternative approaches. The following chapter provides a more comprehensive evaluation of the state of the art by incorporating other connectivity-driven, anatomical, and random brain parcellations from the literature. We also expand the scope of evaluation by including simple network analysis tasks to our experiments and further assess the agreement of the parcellations with complementary information, such as task fMRI and the myelo-/cyto-architecture of the cerebral cortex. 

\chapter{A Systematic Comparison of Parcellation Methods}
\label{chapter:survey}
This chapter is based on:

S. Arslan, S. I. Ktena, A. Makropoulos, E. C. Robinson, D. Rueckert, and S. Parisot, \textit{Human Brain Mapping: A Systematic Comparison of Parcellation Methods for the Human Cerebral Cortex}, NeuroImage, 2017. (\textit{In Press})


\section*{Abstract}
\textit{This chapter provides a systematic comparison of connectivity-driven, anatomical, and random parcellation methods proposed in the thriving field of brain parcellation. Using functional MRI data and a plethora of quantitative evaluation techniques investigated in the literature, we evaluate 24 group-wise parcellation methods at different resolutions. We assess the accuracy of parcellations from four different aspects: (1) reproducibility across different groups, (2) fidelity to the underlying connectivity data, (3) agreement with fMRI task activation, myelin maps, and cytoarchitectural areas, and (4) network analysis. This extensive evaluation of different parcellations generated at the group level highlights the strengths and shortcomings of the various methods and aims to provide a guideline for the choice of parcellation technique and resolution according to the task at hand. The results obtained in this study suggest that there is no optimal method able to address all the challenges faced in this endeavour simultaneously.
} 

\section{Introduction}
In this chapter, we propose a systematic comparison of existing parcellation methods using publicly available resources and evaluation measures that are widely used in the literature through a structured experimental pipeline. We focus on rs-fMRI, as the majority of data-driven parcellation methods we are investigating have been developed and tested using this modality. We aim to provide some insight into the reliability of parcellations in terms of reflecting the underlying mechanisms of cognitive function, as well as, revealing their potential impact on network analysis. 

There have been several recent studies that provide an overview of the current state of the parcellation literature~\cite{Eickhoff15,Thirion14,deReus13}. In particular, Eickhoff et al.~\cite{Eickhoff15} review many design choices that may emerge in the process of parcellating the brain, including but not limited to definition of ROIs, choice of clustering method, selection of the number of clusters, and the evaluation methods, as well as discuss possible limitations and pitfalls of connectivity-driven approaches. Thirion et al.~\cite{Thirion14} specifically focus on the selection of a clustering algorithm for brain parcellation and provide a detailed analysis of three widely-used clustering algorithms with respect to their appropriateness for parcellation. de Reus and van den Heuvel~\cite{deReus13} approach the problem from a different perspective and address the impact of parcellation model/resolution on network analysis. However, to the best of our knowledge, this is the first large-scale systematic comparison of the state-of-the-art parcellation methods that encompasses all these different aspects in a unified experimental setting.  

The main contributions of our study are the following: (1) We evaluate 24 group-level methods using publicly available datasets provided by the Human Connectome Project~\cite{VanEssen13}. (2) Our experiments consist of quantitative assessments of parcellations at both subject and group levels and for different resolutions. (3) We evaluate parcellations not only from a data clustering point of view but also with regards to network analysis and multi-modal consistency. Our evaluation includes reproducibility (e.g. Dice coefficient and adjusted Rand index), cluster validity analysis (e.g. Silhouette coefficient and parcel homogeneity) and multi-modal comparisons with task fMRI activation, myelin and cytoarchitectural maps. In addition, we devise simple network-based tasks (such as gender classification and individual subject prediction) to evaluate the potential impact of parcellations on capturing representative features at different scales. 

The remainder of this paper is organised as follows: Section~\ref{sec:materials_methods} summarises the procedures pursued during the generation and evaluation of parcellations. Experimental results are presented in Section~\ref{sec:results}. In Section~\ref{sec:discussion}, we discuss the reliability and applicability of parcellations for network analysis and summarise the impact/limitations of this study. 

\section{Materials and Methods}
\label{sec:materials_methods}
\begin{figure}[t!]
\centering
\includegraphics[width=0.8\textwidth]{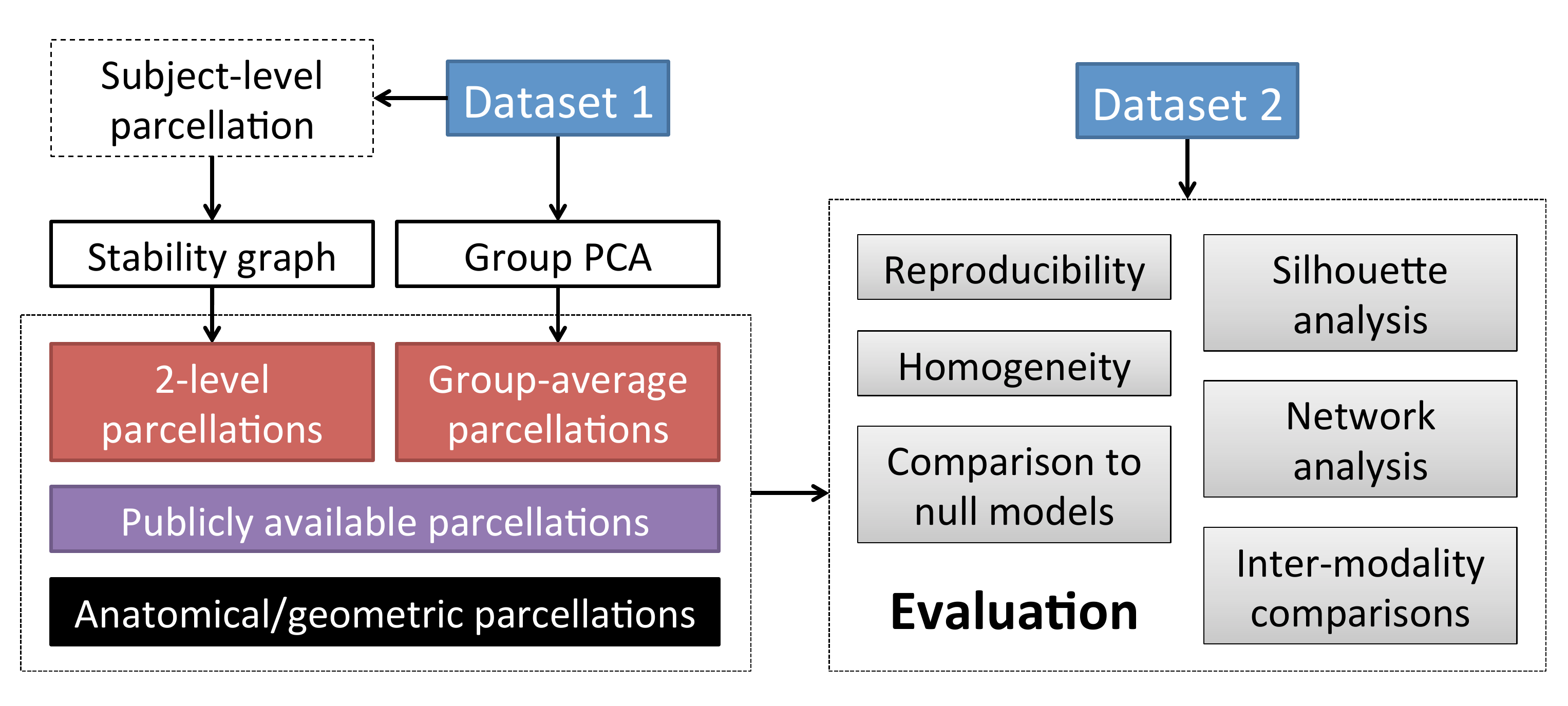}
\caption{Visual outline of parcellation generation and evaluation steps.}
\label{fig:generation-analysis-pipelines}
\end{figure}

A summary of the processing pipelines is given in Fig.~\ref{fig:generation-analysis-pipelines}. A brief description of the parcellation methods is provided in Tables~\ref{tab:group-level-methods-computed} and~\ref{tab:group-level-methods-provided}, respectively.

\subsection{Data}
This study is carried out using data from the publicly available Human Connectome Project (HCP) database~\cite{VanEssen13}. All connectivity-driven parcellations are derived from Dataset 1, which consists of 100 subjects (54 female, 46 male adults, aged 22-35). For evaluation purposes, we use Dataset 2 comprising randomly selected 50 male and 50 female adults of age 22-35. The evaluation is performed on Dataset 2 so as to reduce the possible bias towards parcellations computed from Dataset 1 with respect to the provided ones. The details of both datasets are previously given in Chapter~\ref{chapter:background} and briefly summarised here.

We use rs-fMRI as our primary data modality for the generation and evaluation of parcellations. This is because most  methods selected for this study were developed for rs-fMRI driven parcellation, and rs-fMRI allows test-retest measurements across acquisitions, subjects, and groups. The rs-fMRI scans for each subject were conducted in two sessions, consisting of a total of four runs of approximately 15 minutes each. All subjects were preprocessed by the HCP structural and functional minimal preprocessing pipelines~\cite{Glasser13}. The output of these pipelines for each subject is a standard set of cortical vertices with inter-subject correspondence. Following these preprocessing steps, each timeseries was temporally normalised to zero-mean and unit-variance.			

\subsection{Parcellation Methods}
In order to provide a comprehensive evaluation of the state of the art on surface-based brain parcellation at the group level, we gathered 24 parcellation methods from the literature. The methods included in this study satisfy at least one of the following criteria: 

\begin{enumerate}
\item An implementation is publicly available.  
\item Pre-computed parcellations are publicly available. Both surface-based and volumetric parcellations are considered. 
\item The method can easily be re-implemented.
\end{enumerate}

All the computed and publicly available methods considered in this study are respectively presented in Tables~\ref{tab:group-level-methods-computed} and~\ref{tab:group-level-methods-provided}, along with their associated names, and briefly explained below.
 
\begin{table}[t!]
\centering
\includegraphics[width=\textwidth]{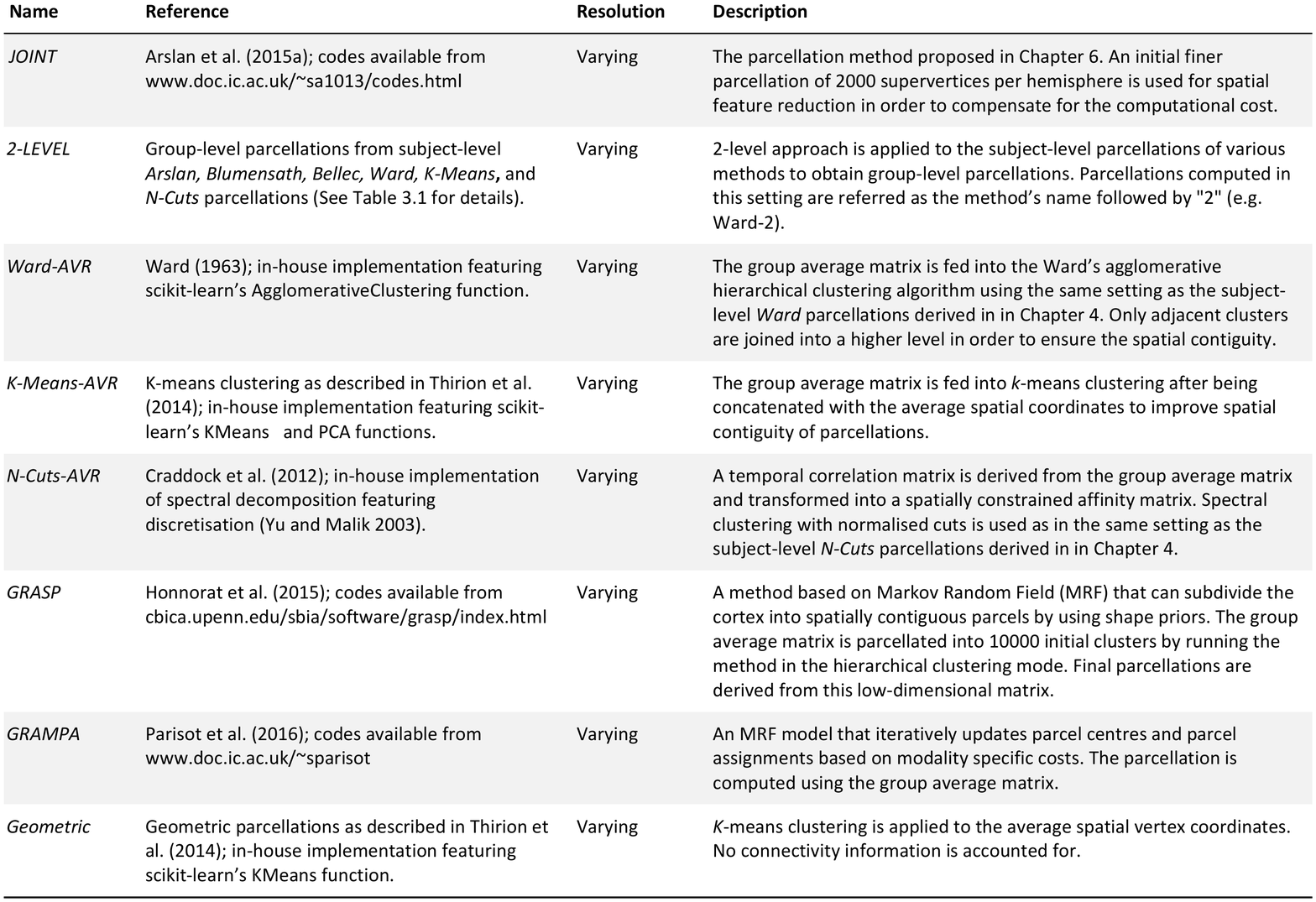}
\caption{Computed parcellation methods.}
\label{tab:group-level-methods-computed}
\end{table}

\subsubsection{Computed Parcellations}
The two more popular ways of computing a data-driven group-wise parcellation~\cite{Craddock12} are (1) performing parcellation for each subject individually and applying a second level clustering algorithm to subject-level parcellations (i.e. \textit{2-level approach}), and (2) computing a representative feature matrix from the population, for instance by concatenating BOLD timeseries across subjects, and submitting this combined matrix to a parcellation method (i.e. \textit{group-average approach}).

The \textit{2-level} approach is explained in the previous chapter in more detail and summarised here. It is based on the idea of identifying shared features of the population as approximated by the individual parcellations. For this purpose, an adjacency matrix is constructed from subject-level parcellations, in which an edge between vertices $v_i$ and $v_j$ is weighted by the number of times both vertices are assigned to the same parcel across all individual subject parcellations. This matrix is then subdivided into different number of regions using spectral clustering with normalised cuts~\cite{Heuvel08,Craddock12}. We use this approach to generate a group-level parcellation from the individual subject parcellations obtained by (1) the proposed hierarchical clustering method introduced in Chapter 4, which will be denoted as \textit{Arslan} in the remainder of this chapter, and (2) other connectivity-driven subject-level methods such as \textit{K-Means}, \textit{Ward}, \textit{N-Cuts}, \textit{Blumensath}, and \textit{Bellec}, all of which are previously explained in Chapter 4 and summarised in Table 4.1. The only difference is in the implementation of \textit{K-Means}, in which we combined the PCA components with spatial vertex coordinates to improve spatial contiguity of parcellations.

The \textit{group-average approach} aims to capture shared patterns between individuals within a population by computing a group average representation of connectivity. This may be achieved by concatenating the timeseries of each subject and applying PCA for dimensionality reduction before parcellation~\cite{Thirion14,Smith14}. However, using the full-concatenated timeseries with traditional PCA quickly becomes computationally prohibitive when the population's size increases. To overcome this, we follow the methodology employed by the HCP for the generation of group average matrices.  We use FSL's incremental group PCA~\cite{Smith14}, a technique developed for computing `pseudo timeseries' that can (to good approximation) estimate  the real PCA output applied to the original combined dataset, while relying on a limited amount of memory. We apply this technique to generate group-level pseudo timeseries from both Dataset 1 and Dataset 2. Group-level parcellations are computed from each of these datasets using our in-house implementations of clustering techniques (\textit{K-Means}, \textit{Ward} and \textit{N-Cuts}) as well as connectivity-driven methods for which implementations are available~\cite{Honnorat15,Parisot16b}.

Alternative to these two most commonly-used approaches, we also incorporate parcellations obtained by the joint spectral decomposition technique which is introduced in Chapter 6 and will be denoted as \textit{JOINT} hereafter. The only difference in the parameter setting is the number of subjects used to construct the multi-layer graph (50 vs. 100 subjects). 

Finally to obtain a baseline for the connectivity-driven methods, `geometric parcellations' are derived using \textit{k}-means clustering of the average spatial coordinates of all cortical vertices as described in~\cite{Thirion14} and Chapter 4.

\begin{table}[t!]
\centering
\includegraphics[width=\textwidth]{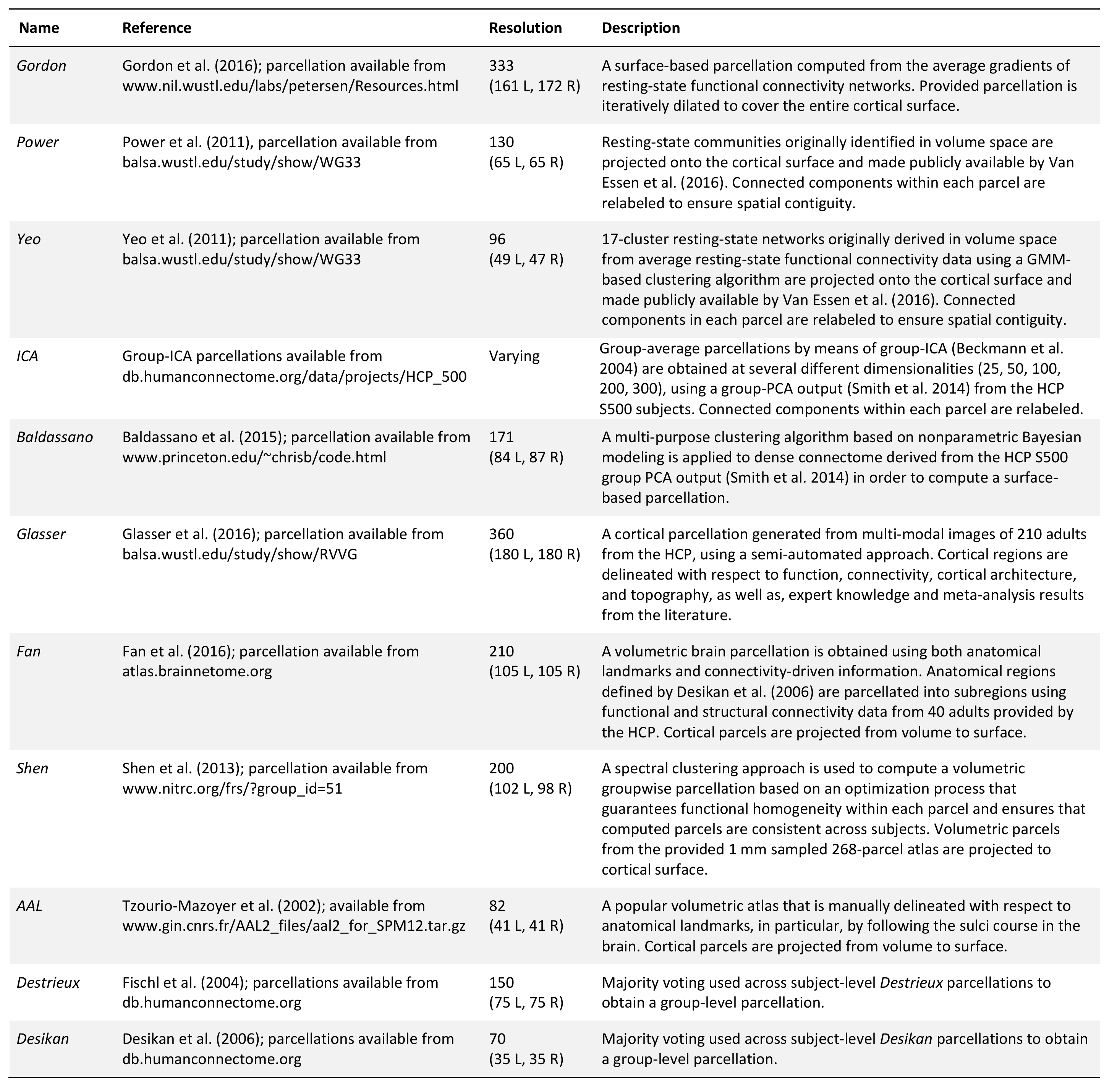}
\caption{Pre-computed, publicly available parcellation methods.}
\label{tab:group-level-methods-provided}
\end{table}

\subsubsection{Publicly available parcellations} Several pre-computed, publicly available group-level parcellations are included in this study~\cite{Gordon16,Yeo11,Power11,Baldassano15,Fan16,Shen13,Smith14,Glasser16}. Details on the method and the resolution of the parcellations are provided in Table~\ref{tab:group-level-methods-provided}. In particular, it should be noted that the parcellations provided by~\cite{Baldassano15} (\textit{Baldassano}) and the \textit{ICA} parcellations~\cite{Beckmann04,Smith14} are computed from a much larger HCP cohort (group average of 500 subjects) which can comprise our evaluation dataset. This may introduce a bias in the evaluation of both methods. 

The methods proposed by~\cite{Yeo11} (\textit{Yeo}) and~\cite{Power11} (\textit{Power}), as well as the \textit{ICA} parcellations~\cite{Beckmann04,Smith14} were originally developed for identifying communities or resting-state networks, hence do not naturally provide spatially contiguous parcellations. Since this can affect the evaluation measures, we overcome this by relabelling connected components within each parcel. We then remove very small parcels and slightly dilate the remaining ones to adjust for vertices lost. \textit{K-Means} (both 2-level, and group-average versions) and another connectivity-driven approach, \textit{GRAMPA}, can also provide spatially disjoint parcels. In our experiments, we do not apply any post-processing to the parcellations derived by these methods, as we aim to obtain roughly the same number of regions for all computed parcellations for the sake of consistency. Nonetheless, we perform additional experiments to analyse the impact of relabelling connected components for these methods and discuss how their performance changes compared to the original parcellations.

The multi-modal parcellation of the human cerebral cortex~\cite{Glasser16} (\textit{Glasser}) is computed through expert manual annotation of imaging data from several modalities, including function, connectivity and cortical architecture. 

We also incorporate anatomical atlases in our study, including the volumetric Automated Anatomical Labelling (\textit{AAL}) brain atlas~\cite{TzourioMazoyer02} and two surface-based parcellations provided by the HCP~\cite{Fischl04,Desikan06}, namely \textit{Destrieux} and \textit{Desikan}. We obtain a group-wise representation of the \textit{Destrieux} and \textit{Desikan} parcellations using majority voting across their subject-level versions. 

Several parcellations are only available in volume space~\cite{TzourioMazoyer02,Shen13,Fan16}. We use volume-to-surface and surface-to-surface sampling techniques to project volumetric parcels onto the HCP average cortical atlas (Conte69)~\cite{VanEssen12}. \textit{AAL}~\cite{TzourioMazoyer02} and the volumetric parcellation by~\cite{Shen13} (\textit{Shen}) are projected onto the cortical surface generated from the Colin27 brain~\cite{Holmes98} using FreeSurfer~\cite{Fischl12}, which is then registered to the Conte69 standard space using multimodal surface matching~\cite{Robinson14}. Our last volumetric parcellation, \textit{Fan}~\cite{Fan16}, is provided in the HCP volumetric space, and is therefore directly projected onto the HCP's standard surface. Finally, all volumetric parcellations are post-processed and each parcel is slightly dilated to fill holes that may have emerged during projection. Unfortunately, volume-to-surface resampling is not a straightforward process, and hence, it is impossible to retain all volume-based parcels after projection. However, we ensure that the parcellation boundaries and relative positions of parcels to each other remain as faithful to the original atlas as possible. 

\subsection{Parcellation Evaluation Techniques}
We gather here the most commonly used evaluation techniques from the literature to evaluate parcellations with respect to varying resolutions. These techniques can be separated into four categories with regards to the parcellation aspects they assess: (1) reproducibility, (2) clustering validity measures, e.g. homogeneity and Silhouette analysis, (3) multi-modal comparisons with cytoarchitecture, task fMRI activation, and myelination, and (4) network analysis. A summary of the evaluation techniques is given in Table~\ref{tab:evaluation-techniques}.

\begin{table}[t!]
\centering
\includegraphics[width=\textwidth]{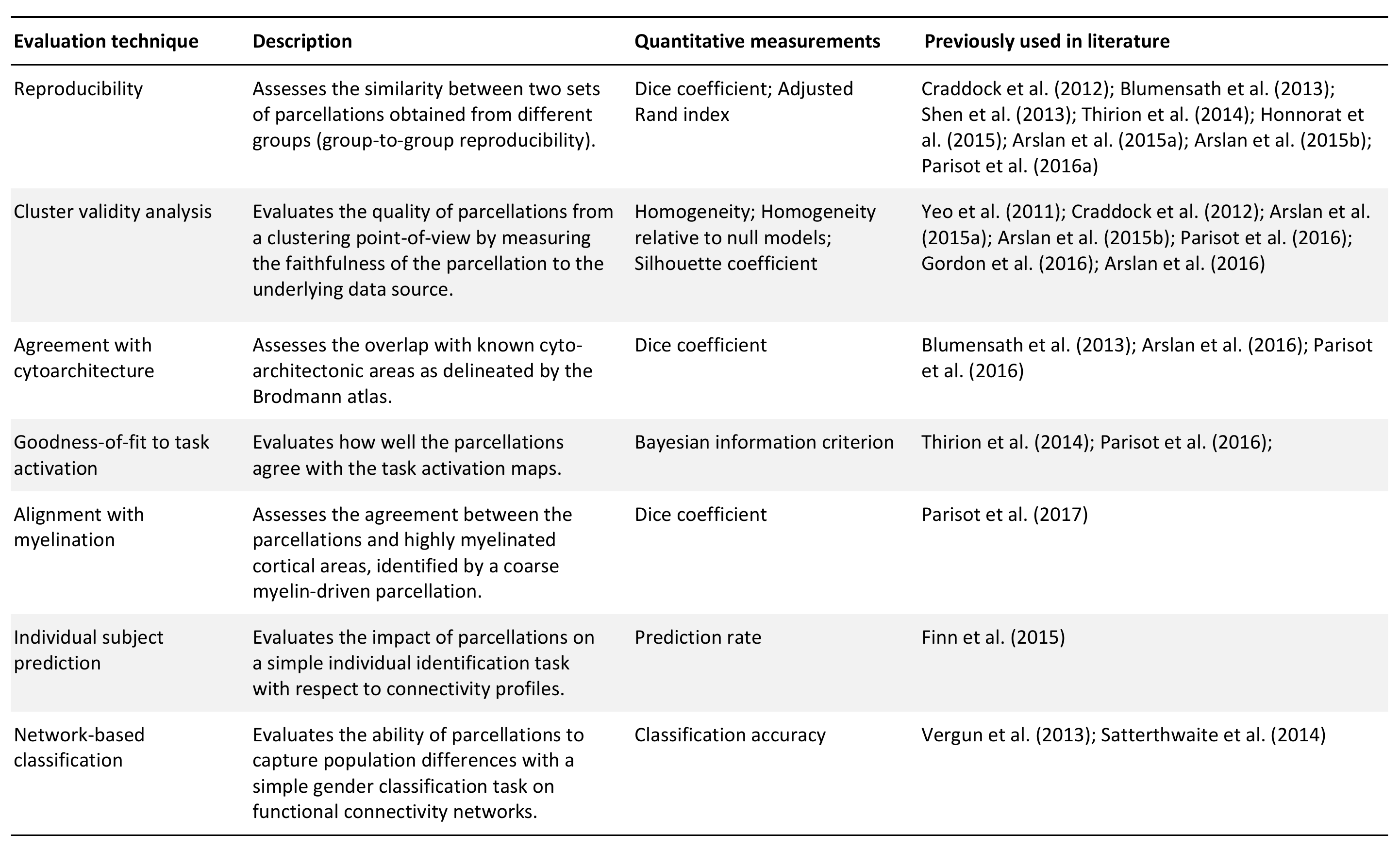}
\caption{Parcellation evaluation techniques.}
\label{tab:evaluation-techniques}
\end{table}

\subsubsection{Reproducibility}
\label{sec:repro}
We use reproducibility techniques to measure the degree of similarity between two parcellations obtained from different groups. A robust parcellation method should yield very similar parcellations for different groups of subjects, assuming that the groups comprise a relatively large number of subjects with similar characteristics (e.g. healthy adults from same age range). To this end, we compare the parcellations obtained from Dataset 1 with the ones derived from Dataset 2 (i.e. group-to-group reproducibility). Unfortunately, we are limited to performing the reproducibility analysis only for the computed parcellations, as only one parcellation/atlas is publicly available from each external source. 

For measuring reproducibility on a group-to-group basis we adapt the two widely-used measures, Dice coefficient~\cite{Dice45} and adjusted Rand index (ARI)~\cite{Hubert85}, both of which have been previously used for reproducibility analysis in Chapters 4 and 6. Dice coefficient relies on a parcel-to-parcel comparison by measuring the overlap between pre-matched regions while ARI provides a more direct evaluation of parcellation configurations and can be more reliable when parcellations with different resolutions are compared~\cite{Milligan86}. Similarly to the subject-level analysis in Chapter 4, we also use the modified version of Dice coefficient~\cite{Blumensath13}, that runs a merging stage to account for the over-parcellated regions from one parcellation to the next. All measures provide reproducibility scores within a range of $[0,1]$, where 1 implies a perfect match (identical parcellations) and 0 implies that the parcellations do not agree on any of the labels. Please refer to Chapter 4 for a more detailed explanation of each technique.

\subsubsection{Cluster Validity Measures}
\label{sec:validity}
This second category of validation measures aims to evaluate parcellations from a clustering point of view. Among many tools targeted at evaluating clustering solutions, we focus on two measures that are extensively being used for brain parcellations, namely homogeneity and Silhouette coefficients. In addition, we adopt the evaluation technique proposed in~\cite{Gordon16} that compares parcellations to a set of `null models' obtained by randomly relabelling the parcellation without altering the relative parcel locations with respect to each other. 

\paragraph{\textit{Homogeneity}:} This technique measures the degree of similarity between vertices that are assigned to the same parcel and is extensively used in the literature for the evaluation of parcellations ~\cite{Craddock12,Shen13,Gordon16,Arslan15b,Parisot16a,Honnorat15}. Parcellating the cerebral cortex into highly homogeneous regions might be particularly important for network analysis, as the network nodes are typically represented by the average signal (e.g. BOLD timeseries) within each parcel~\cite{Shen13,Gordon16}. In this study, we utilise the same homogeneity technique presented in Chapter 4 for the subject-level parcellations. That is, we first compute the similarity between every pair of vertices (represented by their connectivity fingerprints) within a parcel using Pearson's correlation coefficient. The average of pair-wise correlations are then used to measure the homogeneity of each parcel. A global homogeneity value is obtained by averaging the homogeneity scores across all parcels. Different from the subject-level analysis in Chapter 4, we use a weighted arithmetic mean, where each parcel's contribution to the global homogeneity is proportional to its size. This is to avoid possible unfairness that can emerge from the parcel size distributions of the tested parcellations. Smaller parcels tend to have a higher homogeneity than large ones, such that, a parcellation mostly composed of many small parcels and a few large regions may likely perform better than one with similarly sized parcels. 

We obtain homogeneity results from the average connectivity fingerprints of all subjects, as well as on a per subject basis by using each subject's connectivity fingerprints and then averaging across subjects.

\paragraph{\textit{Comparison to null models}:} While computing homogeneity by means of a weighted mean reduces the bias towards small parcels, homogeneity values remain dependent on the resolution of parcellations so that fair comparison between different resolutions is not possible. An alternative is proposed in~\cite{Gordon16} which consists of comparing a parcellation with the so-called `null models' of the same resolution.

\begin{figure}[t!]
\centering
\includegraphics[width=\textwidth]{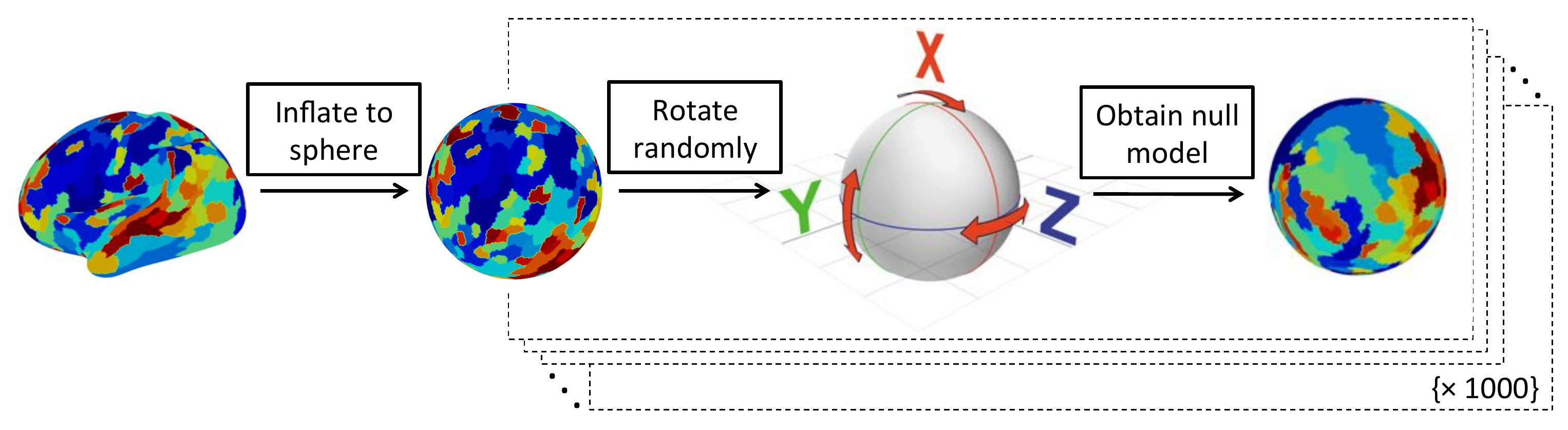}
\caption{Illustration of the generation of null models.}
\label{fig:null-rotation}
\end{figure}

In order to obtain such null models, we perform the procedure illustrated in Fig.~\ref{fig:null-rotation}. For each hemisphere, we project the parcellation onto a standard spherical surface provided by the HCP and randomly rotate each point in this sphere around the x, y, and z axes. This process moves each parcel to a new location on the cortical surface without altering their relative positions. We then measure the homogeneity of the rotated parcellation and repeat the same process for 1000 different null models. Parcels that move to the medial wall, where no connectivity information is available, are discarded from computations. The advantage of this approach is that it reduces the observed biases with respect to parcel shape and size, as the parcellations are compared to their rotated versions, which have the same resolutions and similar parcel shapes.

In order to quantitatively evaluate parcellations with respect to their null models, we (1) count the number of rotated parcellations with lower homogeneity scores than the original parcellation and (2) compute the difference between the homogeneity of the original parcellation and the mean homogeneity score of null models, scaled by their standard deviation (i.e. $z$ scores relative to null models)~\cite{Gordon16}. 
 
\paragraph{\textit{Silhouette analysis}:} Another popular clustering validity tool that can be used to quantify parcellation reliability is the Silhouette coefficient (SC)~\cite{Rousseeuw87}. It is typically used as an indicator of how well vertices fit in their assigned cluster~\cite{Yeo11,Craddock12}. This is achieved by measuring the similarity of a vertex to the other vertices within the same cluster, with respect to its dissimilarity to the out-of-cluster vertices. As a result, SC not only evaluates the compactness (homogeneity) of a parcellation, but also its ability to separate regions with distinct features. For our group-level Silhouette analysis, we rely on the same SC definition as given in Chapter 4. A global score is obtained for each parcellation by averaging the Silhouette coefficients across all vertices. SC can range within $[-1, +1]$, in which a value close to 1 indicates that the vertex is clustered with a high degree of confidence.

Similarly to the homogeneity analysis, Silhouette coefficients are obtained from the average connectivity fingerprints of all subjects. However, we also perform the same analysis on a per subject basis by using each subject's connectivity fingerprints and then averaging across subjects. 

\subsubsection{Comparisons with Other Modalities}
The previously proposed measurements assess the accuracy of parcellations from a clustering point of view. However, when defining regions of interest for neuro-anatomical purposes, the consistency of these areas with well-defined neuro-biological features also constitutes a critical aspect of evaluation. To this end, we expand our comparisons to those with other modalities. We test the parcellation quality by evaluating their agreement with task activation maps and myelination patterns, and their overlap with well-known cortical regions delineated from cytoarchitectonic features.

It should be noted that \textit{Glasser} is the only parcellation that is simultaneously driven by task fMRI, myelination, and neuro-anatomist input, and as a result, will develop an inevitable positive bias towards these modalities. Therefore, the performance of this approach with respect to the inter-modality comparisons should be interpreted by taking this into account.

\paragraph{\textit{Goodness-of-fit to task activation}:} Bayesian information criterion (BIC) is proposed in~\cite{Thirion14} as a means of quantifying the agreement of parcellations with task activation. Each vertex is associated with a task activation map (or the concatenated task activation maps of all subjects as in our case, since we consider group-wise parcellations). The BIC criterion measures the goodness-of-fit of a probabilistic model of the concatenated task activation maps by penalising the negative log likelihood by the complexity of the model, determined by the number of parcels.

\begin{figure}[bh!]
\centering
\begin{tabular}{cc}
\includegraphics[width=0.45\textwidth]{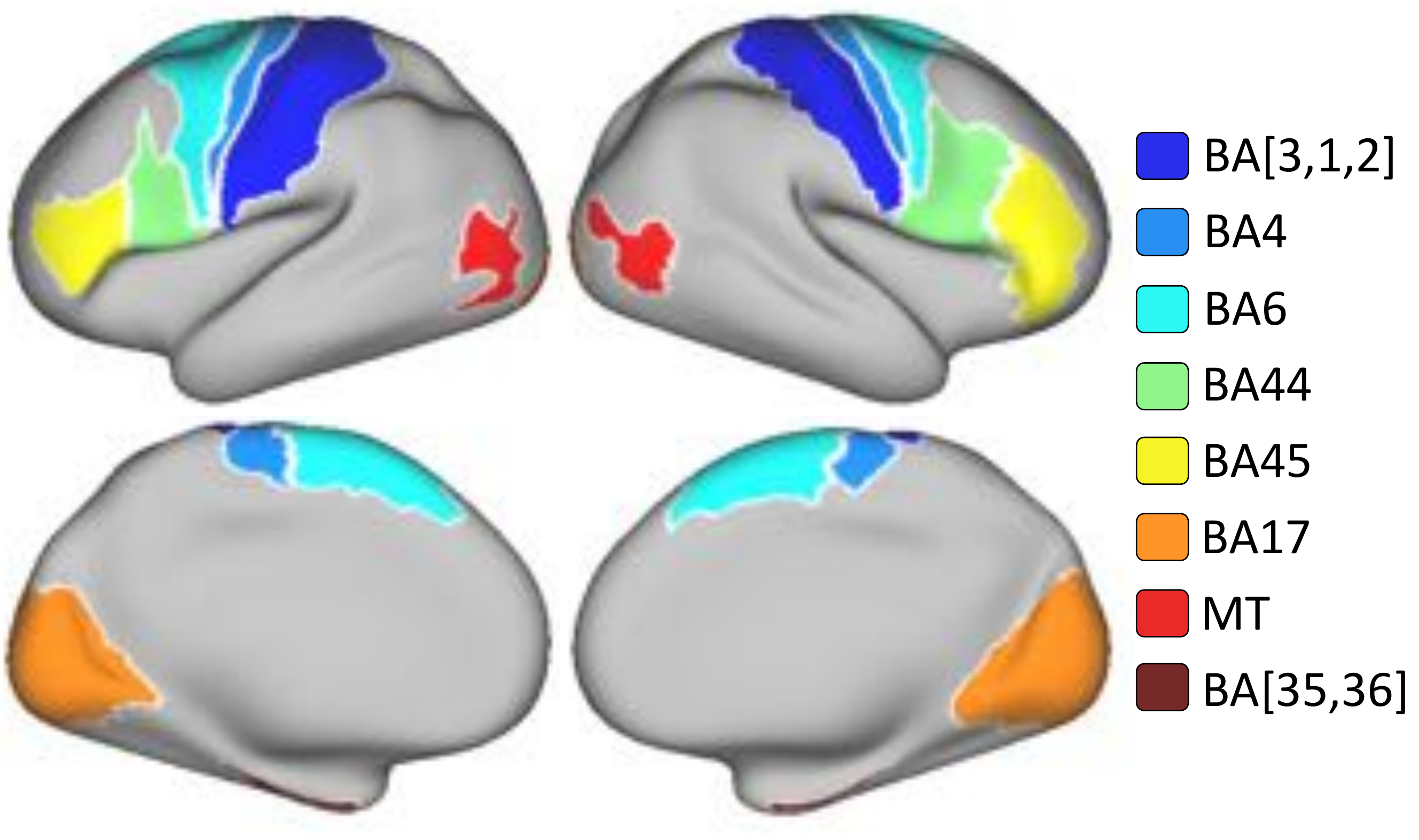} &
\includegraphics[width=0.4\textwidth]{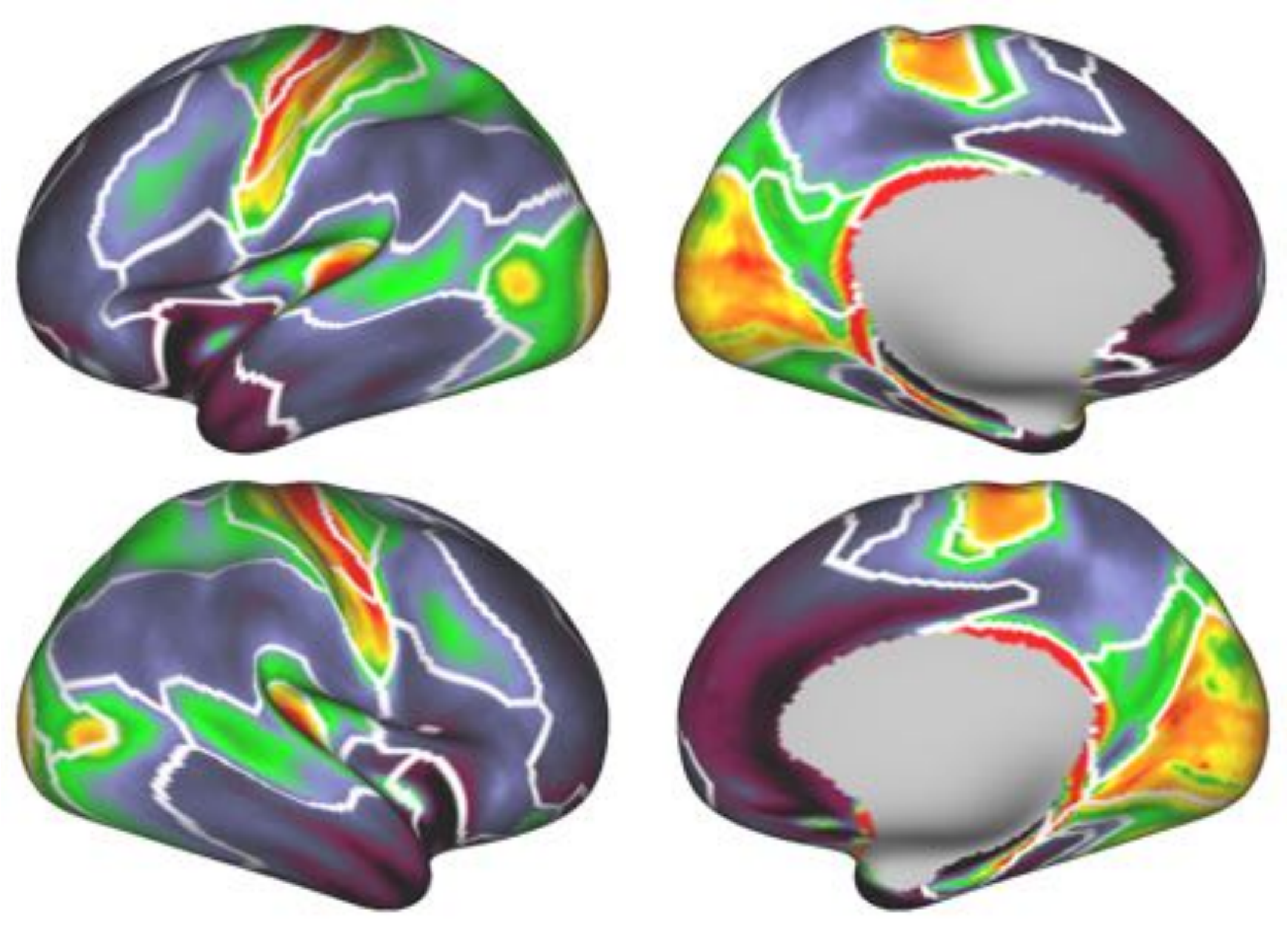} \\
(a) & (b)
\end{tabular}
\caption[Cyto- and myelo-architecture of the cerebral cortex.]{Cyto- and myelo-architecture of the cerebral cortex as defined respectively by (a) Brodmann's areas and (b) a coarse-resolution myelin-driven parcellation.}
\label{fig:bro}
\end{figure}

\paragraph{\textit{Overlap with cytoarchitectonic areas}:} We measure the agreement of our parcellations with the Brodmann areas~\cite{Brodmann09}. Although functional connectivity obtained from BOLD timeseries does not necessarily reflect the cytoarchitecture of the cerebral cortex~\cite{Wig14}, agreement (to a certain extent) with some known cytoarchitectonic areas could indicate a parcellation's ability to reflect the underlying cortical segregation~\cite{Gordon16}. Our standpoint to include comparisons with the cytoarchitecture is to show the extent of such agreement with at least certain areas, such as the motor and visual cortex, for which several parcellation techniques report a noticeable alignment~\cite{Blumensath13,Wig14,Gordon16}. To this end, we use the Brodmann atlas provided by the HCP, which contain labels for the primary somatosensory cortex (BA 3, 1, and 2), the primary motor cortex (BA 4), the premotor cortex (BA 6), Broca's area (BA 44, 45), the visual cortex (BA 17 and MT), and the perirhinal cortex (BA 35, 36) as shown in Fig.~\ref{fig:bro}(a).  

Quantitative comparisons are performed with the joined Dice coefficient approach explained in Chapter 5. In brief, overlapping parcels are matched with the Brodmann areas if their overlap ratio is $ \geq 0.5$. We then compute the Dice coefficient between the matching pairs. It is important to note that several parcels can be matched to the same area and therefore merged into a larger parcel. 

\paragraph{\textit{Agreement with structured myelination patterns}:} Strong similarities have been observed between myelin maps and resting-state fMRI gradients~\cite{Glasser11}. We should therefore expect the boundaries of rs-fMRI driven parcellations to align with myelination patterns. To evaluate this, we compute a coarse-resolution myelin-driven parcellation (25 parcels) using the method described in~\cite{Parisot16b} from the average myelin map of all subjects. This method simply regroups vertices with similar myelin values and, as shown in Fig.~\ref{fig:bro}(b), effectively delineates the major changes in myelination across the cortex. 

We compare the parcellations obtained by different methods to these coarse parcellations using the joined Dice coefficient approach described above. To this end, we only consider the highly myelinated cortical areas, i.e., cortical areas with a mean myelin value below a certain threshold are discarded. 

\subsubsection{Network Analysis}
Parcellations can significantly reduce the dimensionality of the dense human connectome without eliminating valuable information about the interactions between different brain regions and the mechanisms through which these interconnections give rise to complex cognitive processes. It has been common practice in recent neuroscience studies to explore several neurological~\cite{tijms2013single,fornito2015connectomics} and neuro-developmental disorders~\cite{jafri2008method, liu2008disrupted,dennis2011altered, fornito2012schizophrenia} from a network perspective. These disorders have often been linked to a disruption or abnormal integration of spatially distributed brain regions that would normally be part of a single large-scale network, leading to their characterisation as disconnection syndromes~\cite{catani2005rises}. Apart from the clinical value of network analysis, efforts to explore potential correlations between connectivity patterns and certain phenotypes like fluid intelligence~\cite{smith2016linking}, or to predict an individual's biological age~\cite{qiu2015manifold, robinson2008multivariate, pandit2014whole} have been made. Therefore, a parcellation method can also be evaluated in terms of its ability to capture the inter-individual variability and to reveal patterns that explain observed cognitive performance. 

Once the parcellation has been generated, a network representation can easily be obtained by mapping each network node to a parcel. The edge weights in functional networks usually represent the statistical dependency between the brain regions underlying the connected nodes. In our analysis of functional networks, we use temporal correlation of the representative timeseries as an estimate of the connection strength between two brain parcels. 

We explore different ways in which the underlying parcellation can affect network analysis through (1) a network-based classification task and (2) individual subject identification from connectivity profiles.

\paragraph{\textit{Network-based classification}:} Several studies suggest differences in both structural and functional connectivity between genders~\cite{gong2011brain}. More specifically, in terms of functional connectivity derived from rs-fMRI data, which is the focus of the current survey, significant differences in the topological organisation of functional networks have been found between males and females~\cite{tian2011hemisphere}. For this reason we choose a gender classification task to evaluate the impact of the parcellation on network-based classification tasks. We employ linear Support Vector Machines (SVMs)~\cite{burges1998tutorial}, a well-established classifier from the machine learning literature and a 10-fold cross-validation procedure to get an estimate of each method's performance. Previous studies~\cite{robinson2008multivariate, vergun2013characterizing, satterthwaite2014linked} have used SVM as a machine learning classifier, which is designed to make predictions based on high-dimensional data, to investigate sex differences in functional connectivity.

Given a set of \textit{p}-dimensional feature vectors, SVM aims to identify a $(p-1)$-dimensional hyperplane that represents the largest separation or margin between the feature vectors of the two classes. The hyperplane is chosen in a way that the distance from the nearest data points of each class is maximised. The weights assigned to the normalised features to obtain a low-dimensional representation of the feature vectors can, additionally, be used to describe how heavily weighted the connectivity feature is within the multivariate model~\cite{satterthwaite2014linked}.

Since node correspondences are ensured with group-wise parcellations, an embedding of each subject's connectivity matrix can be employed to get a general vector representation~\cite{varoquaux2013learning}, rendering the use of the aforementioned classifier straightforward. This approach is often referred to as `bag of edges'~\cite{craddock2013imaging} and has been widely used when the underlying parcellation is the same among all subjects. Connectivity networks are generated using the same set of nodes for all subjects in Dataset 2. The nodes correspond to non-overlapping regions specified by the provided or data-driven parcellations obtained from Dataset 1. In order to explore the performance of different parcellation methods in capturing population differences, we show the results of a gender prediction task with $r$-to-$z$ transformed full correlation networks. Before the classification step, dimensionality reduction through PCA is performed~\cite{pereira2009machine} and the components explaining 100\% of the variance~\cite{robinson2010identifying} in the training data are preserved for both training and testing.

\paragraph{\textit{Individual subject identification}:} Most of the human neuroimaging studies typically focus on identifying patterns of brain connectivity that are shared across individuals. However, it has been recognised that brain structure~\cite{Burgel06} and function~\cite{Mueller13} may exhibit a great amount of individual variability. Despite many shared characteristics discovered across individuals, evidence suggests that the human connectome possesses connectional traits that are unique to each subject~\cite{Mueller13,Barch13}. A recent study by Finn et al.~\cite{Finn15} has shown that connectivity networks obtained from rs-fMRI can be used as `fingerprints' to identify individuals with a high accuracy. Results suggest that the intrinsic connectivity captured during rest (or task) is (1) robust and reliable across different acquisitions, and (2) distinct enough to successfully distinguish one individual from another. Finn et al.~\cite{Finn15} explored several factors that may affect the identification accuracy, including the impact of the parcellation used to obtain the connectivity networks, however, this has been covered rather limitedly. Here, we adapt these experiments to the focus of the current study and explore the extent to which the underlying parcellation and its resolution affect a similar subject identification task.

\begin{figure}[!t]  
\centering
\includegraphics[width=0.55\textwidth]{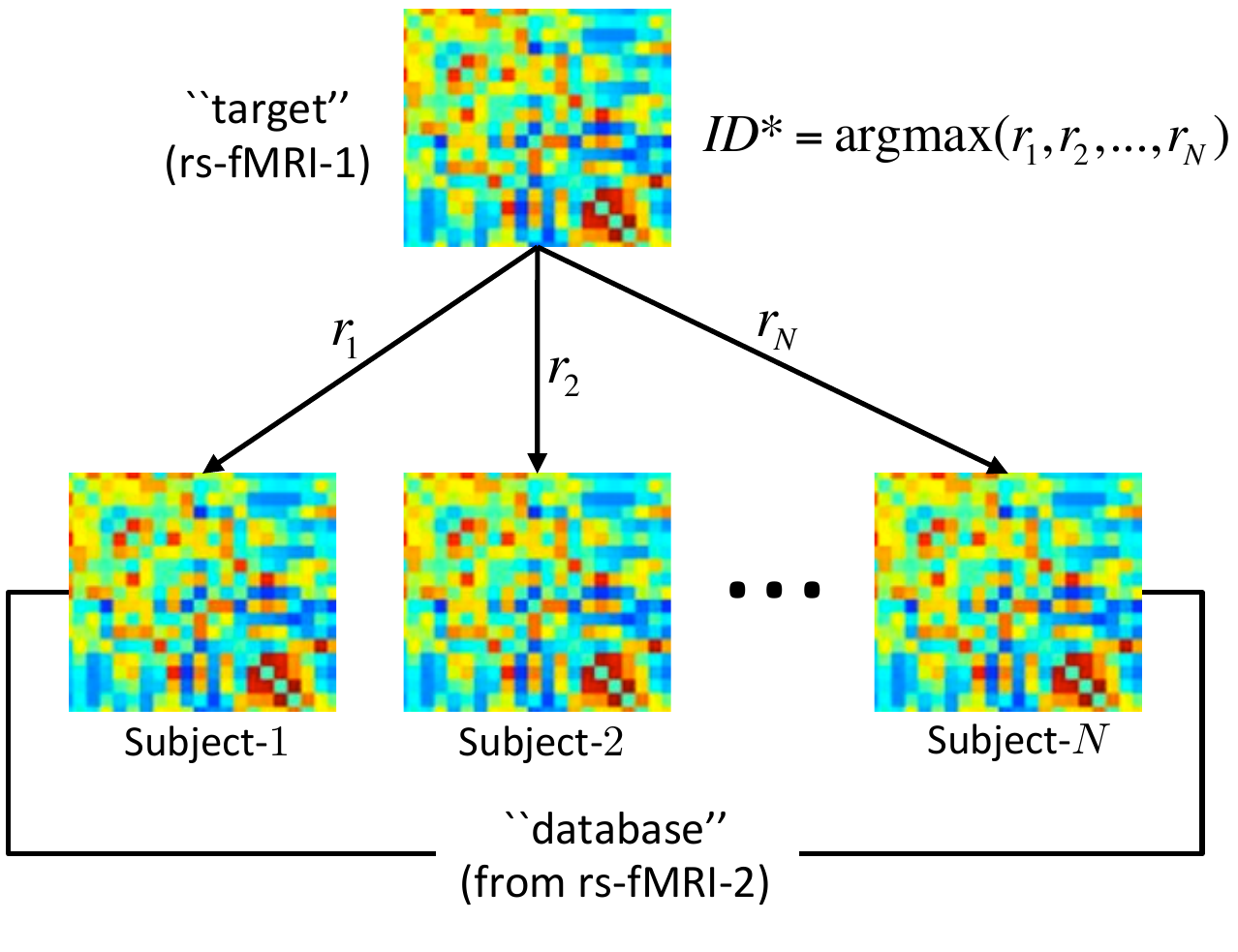} 
\caption[Individual subject identification procedure.]{Individual subject identification procedure. Given a `target' matrix, we compute the correlations between this matrix and all the `database' connectivity matrices. The predicted identity $ID*$ corresponds to the subject with the highest correlation coefficient ($r$). }
\label{fig:prediction_outline}
\end{figure}

For each subject, two $r$-to-$z$ connectivity networks are generated from two different acquisitions (rs-fMRI-1 and rs-fMRI-2) using one of the group-level parcellations and  Pearson's correlation to estimate the connection strength of the edges between each pair of nodes. rs-fMRI-1 networks are denoted as the `target set', whereas networks obtained from the rs-fMRI-2 timeseries are referred to as the `database'. The identification is performed on a per subject basis as illustrated in Fig.~\ref{fig:prediction_outline}. One matrix is selected from the target set and compared with all the other matrices in the database to find the matrix with the maximum similarity. If the target and the predicted matrix belong to the same subject, that iteration is assigned a score of 1, and 0 otherwise (i.e. actual and predicted matrices correspond to different subjects). The similarity is measured after embedding all edges in a one-dimensional vector~\cite{varoquaux2013learning}. At the end of this process, the prediction (success) rate is calculated as the ratio of correctly identified subjects to the population size. The identification task is repeated for all group-level parcellations and available resolutions to assess their impact on the success rate.

\section{Results}
\label{sec:results}
\subsection{Experimental Setup}
All computed parcellations are generated from Dataset 1, in which each subject is represented by two 30-min rs-fMRI scans (rs-fMRI 1 and rs-fMRI 2) that were obtained by concatenating the timeseries of two 15-min scans acquired on the same day~\cite{Glasser13}. The 2-level parcellations are generated from the individual subject parcellations obtained from the rs-fMRI 1 set. This set is also submitted to MIGP to obtain the group-PCA matrix, which is subsequently used to compute parcellations using the group-average approach. rs-fMRI 2 set is only used for the individual subject identification task and are not involved in the generation or evaluation of parcellations otherwise.

Dataset 2 is primarily used to evaluate the group-wise parcellations (publicly available and computed ones from Dataset 1). A second set of group-level parcellations is also generated using Dataset 2 in order to assess reproducibility across different groups. It is worth noting that, this second set is solely used to assess group-to-group reproducibility and excluded from any other stage of the analysis pipelines.

Most of the pre-computed/publicly available parcellations comprise a fixed number of regions, while the methods for which an implementation is available can be explored at different resolutions, allowing us to assess the sensitivity of quantitative measures with respect to the number of parcels. For these methods, we generate parcellations containing between 50 and 500 regions (i.e. 25 to 250 per hemisphere), in increments of 50. 

Finally, results are reported using the following naming scheme: parcellations obtained using the 2-level approach will be referred to via their associated method name followed by `2' (e.g. \textit{Ward-2}), whereas parcellations derived from the group-average approach will be accompanied by the `AVR' suffix (e.g. \textit{Ward-AVR}). 
 
\subsection{Reproducibility Results}
\begin{figure}[!tb]  
\centering
\begin{tabular}{ll}
\includegraphics[width=0.5\textwidth]{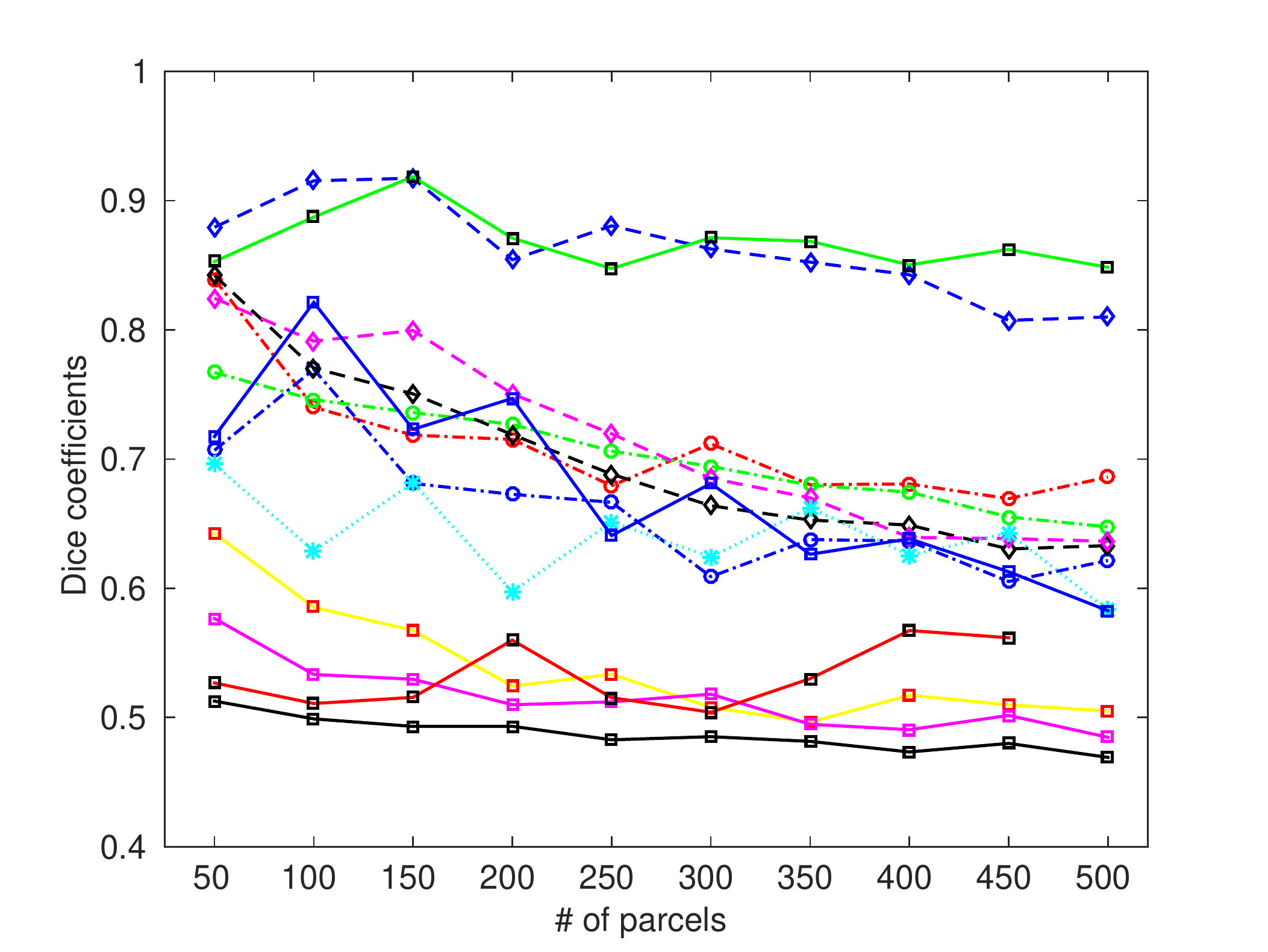} \kern-2.5em & 
\includegraphics[width=0.5\textwidth]{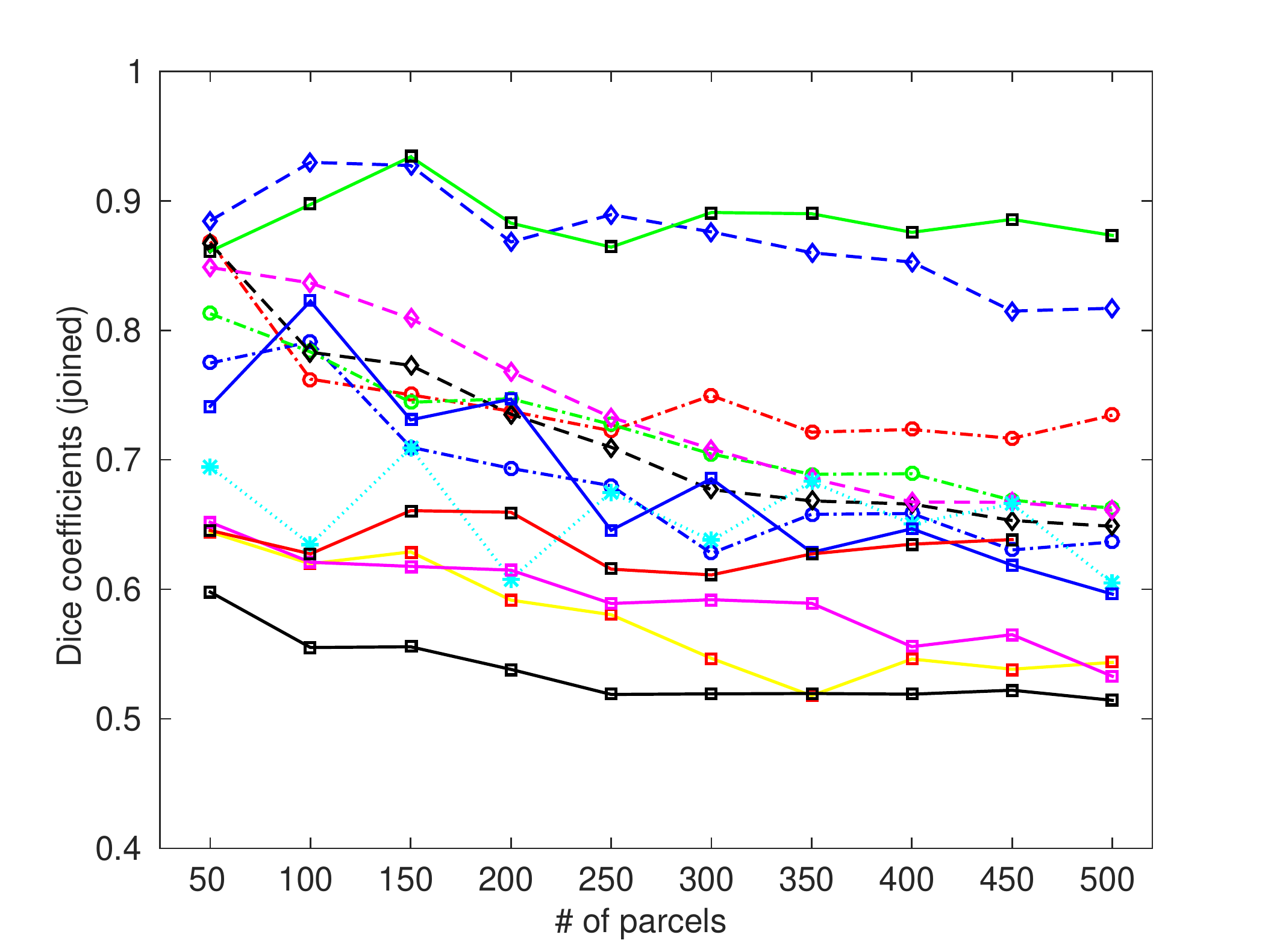} \\

\includegraphics[width=0.5\textwidth]{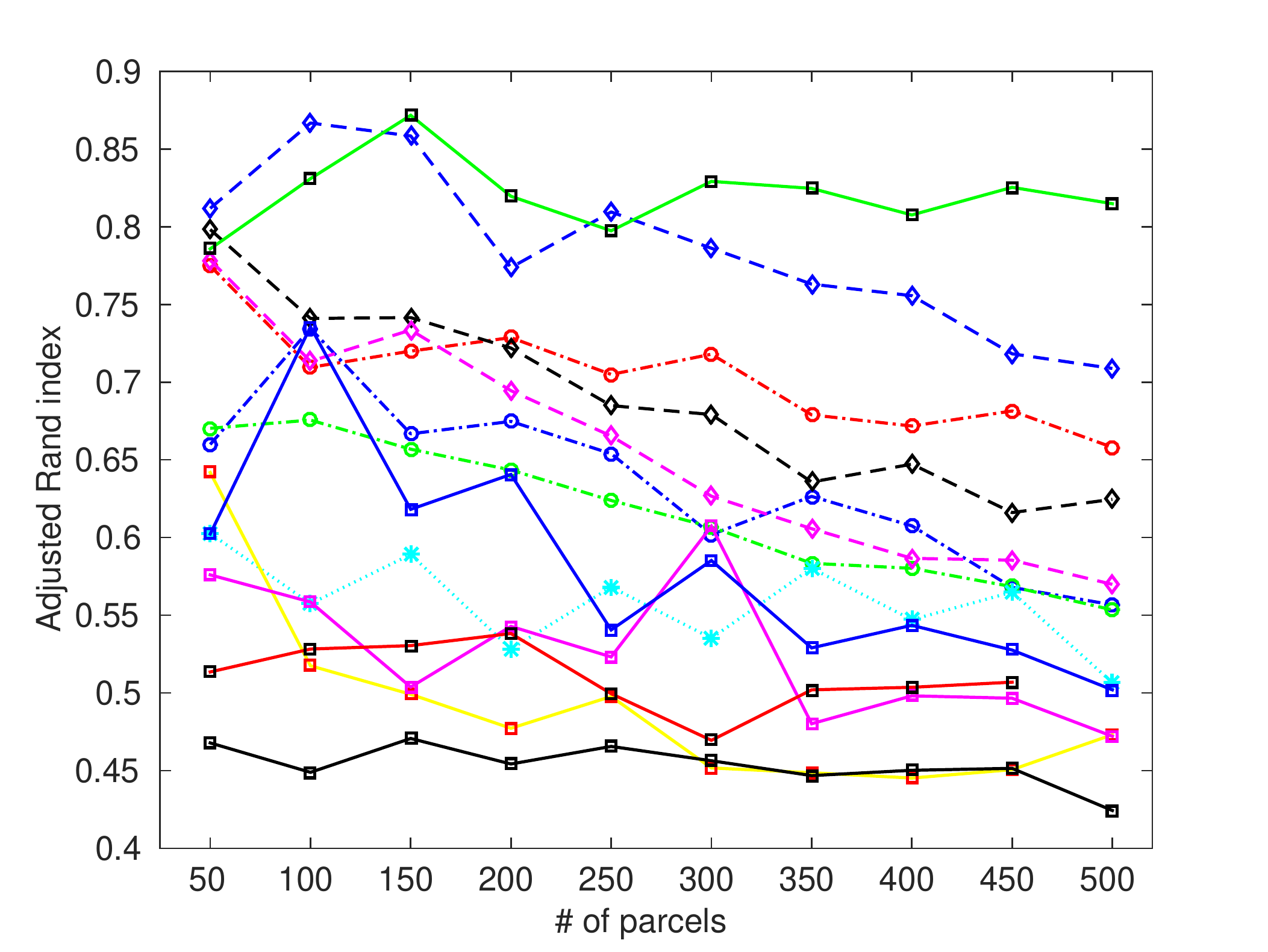} \kern-2.5em &  \kern 1em
\raisebox{0.13\height}{\includegraphics[width=0.17\textwidth]{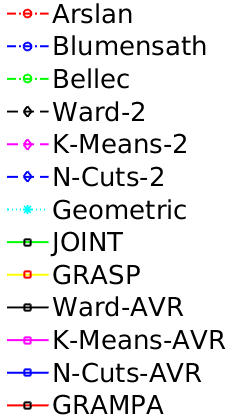}} \\

\end{tabular} 
\caption[Group-level reproducibility results.]{Group-level reproducibility results. Reproducibility values for each method are obtained using Dice coefficient (top, left), Dice coefficient after joining over-parcellated regions (top, right), and adjusted Rand index (bottom). }
\label{fig:group-reproducibility}
\end{figure}

The reproducibility values (Fig.~\ref{fig:group-reproducibility}) are only reported for methods that allow the derivation of multiple parcellations. As expected, the spectral techniques have the best reproducibility, with \textit{N-Cuts} and \textit{JOINT} leading the others. In general, more favourable results are achieved by 2-level parcellations. This may be attributed to the fact that these parcellations are obtained from a set of individual parcellations that already provide a means of spatial smoothing. Furthermore, the parcellations are computed using normalised cuts, a technique known to increase the reproducibility of parcellations~\cite{Craddock12,Blumensath13}. Among the parcellations derived from the average matrix, \textit{Ward-AVR} shows the least favourable performance. MRF-based methods (i.e. \textit{GRASP} and \textit{GRAMPA}) and \textit{K-Means-AVR} also have a relatively poor performance. While joining over-parcellated regions generally increases reproducibility for the group-average approaches, it has a lesser impact on the 2-level parcellations as most of them only show a marginal improvement.

\subsection{Cluster Validity Results}
\begin{figure}[!t]  
\centering
\begin{tabular}{cc}
\includegraphics[width=0.5\textwidth]{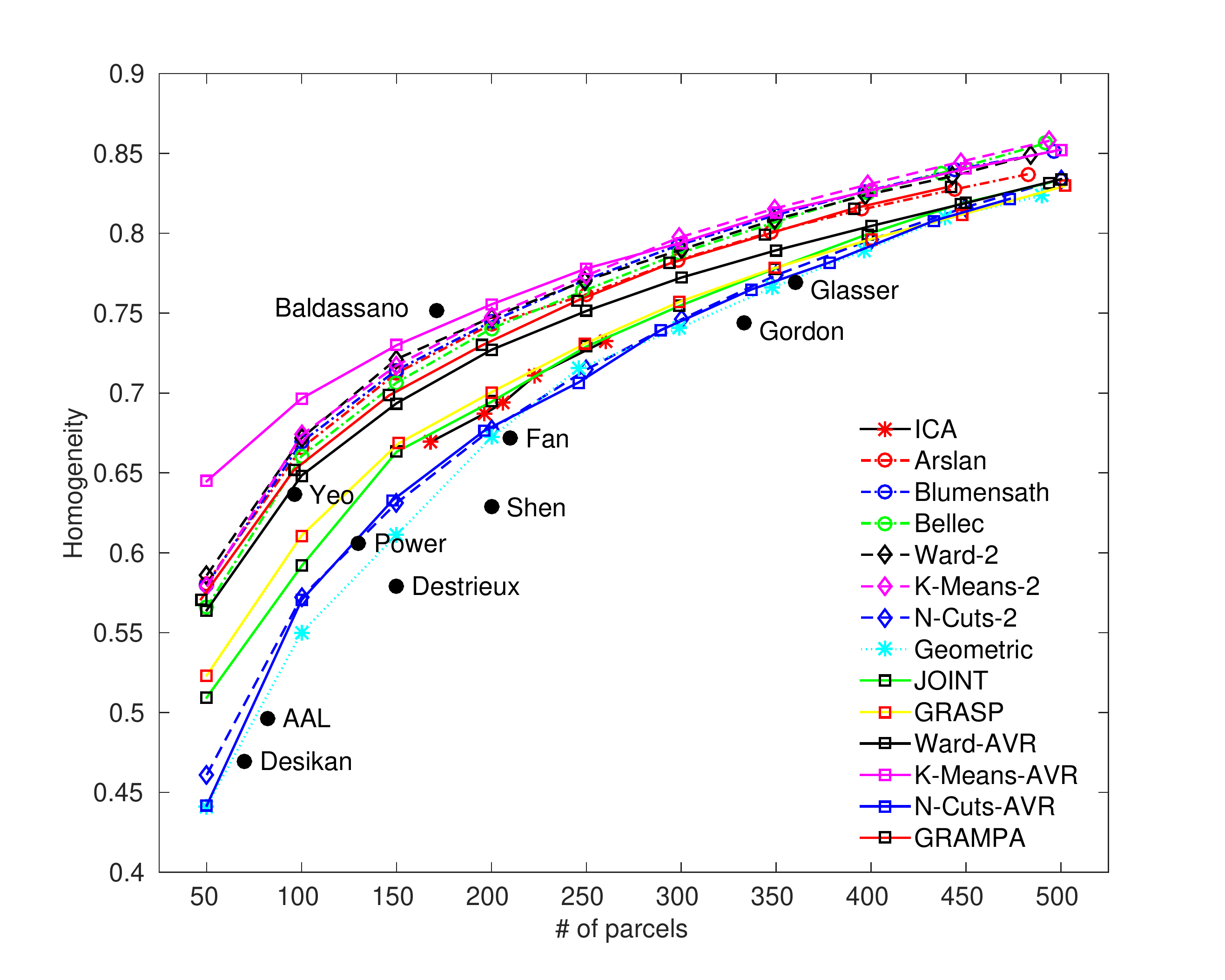} \kern-2.5em &
\includegraphics[width=0.5\textwidth]{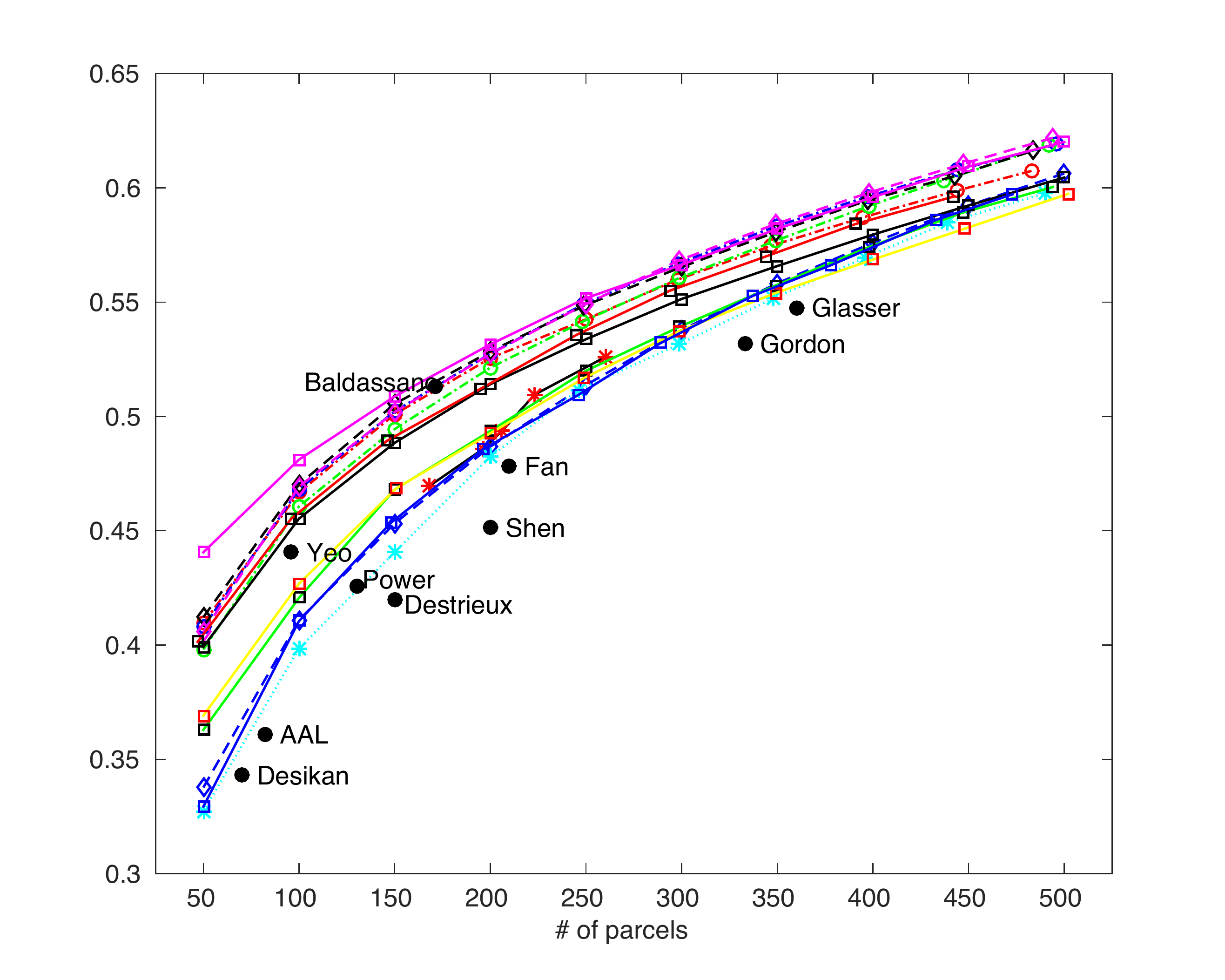} \\
(a) & (b) \\
\end{tabular}
\caption[Group-level homogeneity results.]{Group-level homogeneity results obtained (a) from the average connectivity fingerprints of all subjects and (b) on a per subject basis by using each subject's connectivity fingerprints and then averaging across subjects. Whereas lines show the homogeneity values for all computed resolutions, black dots correspond to the homogeneity scores obtained from the publicly available parcellations with fixed resolutions.}
\label{fig:group-homogeneity}
\end{figure}

\begin{figure}[!t]  
\centering
\begin{tabular}{c}
\kern-2.0em  \includegraphics[width=1.10\textwidth]{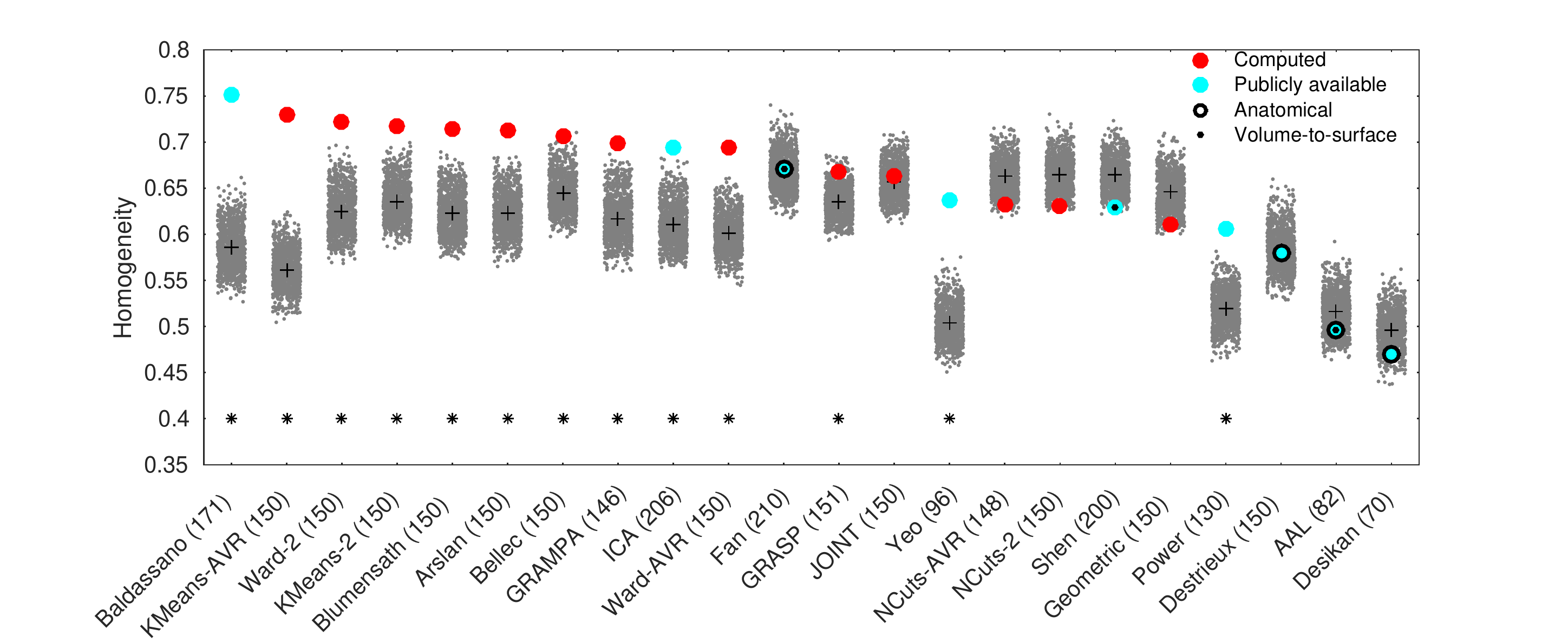} \\
\includegraphics[width=0.82\textwidth]{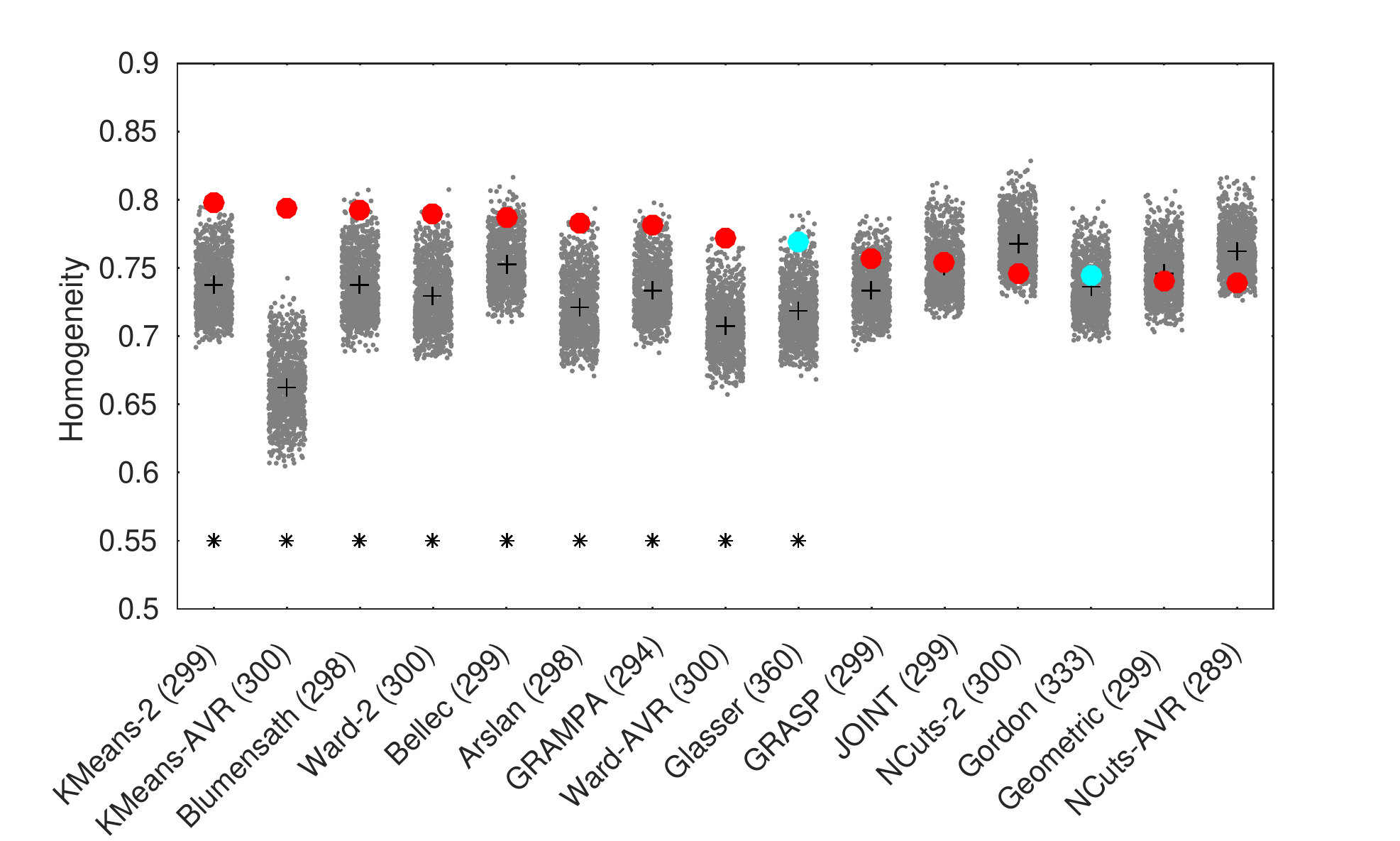}
\end{tabular}
\caption[Homogeneity of the parcellations and their respective 1000 null models.]{Homogeneity of each parcellation (red and purple dots) and their respective 1000 null models (gray dots). Null models yield different homogeneity scores due to variation across parcel size and location. Cross ($+$) indicates the average homogeneity obtained by each set of null parcellations. Asterix ($*$) indicates that the computed homogeneity is higher than at least 950 of its null parcellations (i.e. $p<0.05$). \textit{Top:} Results of publicly available parcellations with relatively low resolutions (comprising around or fewer than $200$ regions) and the computed parcellations with 150 parcels. \textit{Bottom:} Results of publicly available parcellations with higher resolutions (e.g. comprising around or greater than $300$ regions), with the computed parcellations having a fixed resolution of 300 parcels. The exact number of parcels for each method is indicated aside the method name in parentheses.}
\label{fig:group-null}
\end{figure}

\begin{figure}[!t]  
\centering
\includegraphics[height=0.5\textwidth]{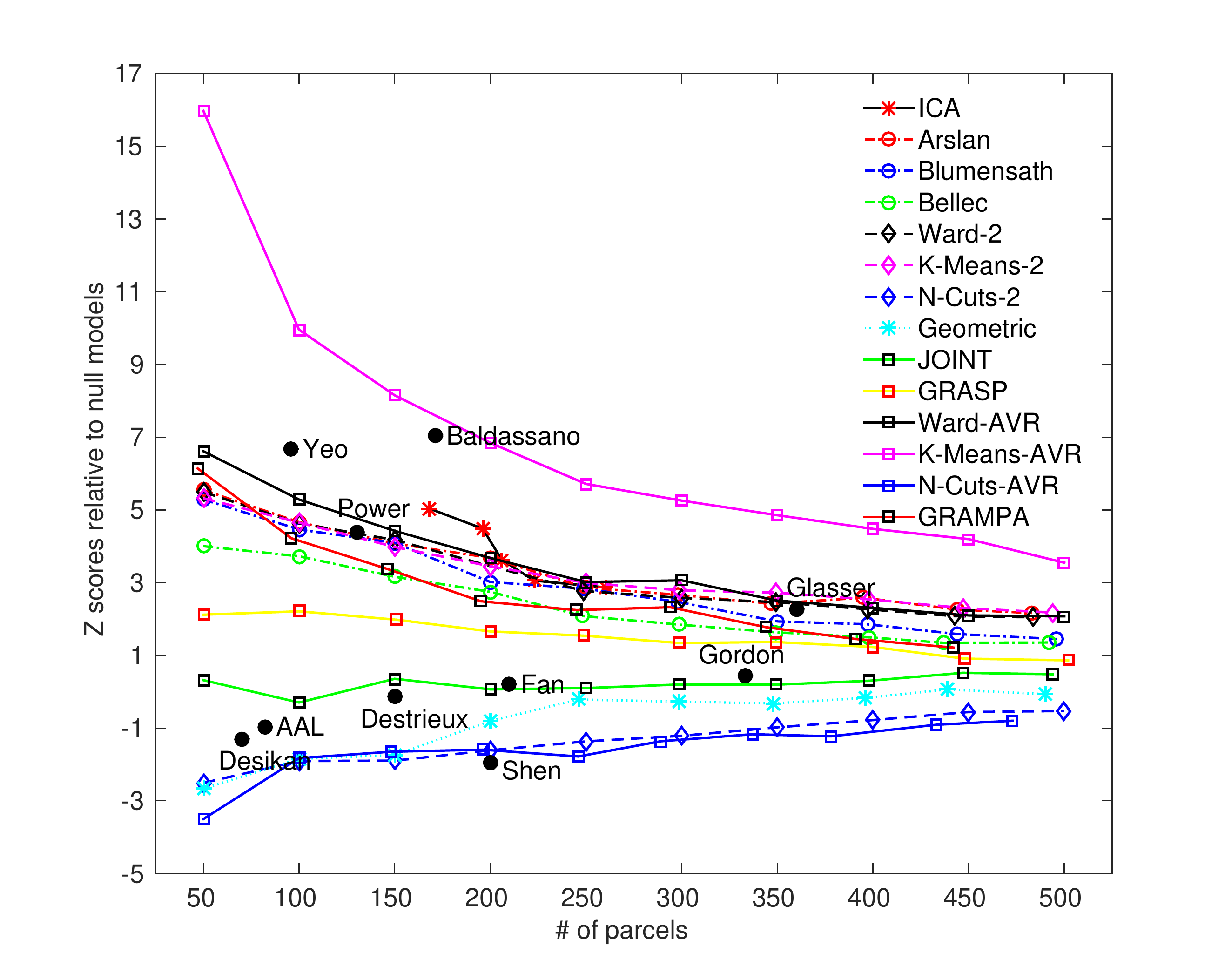} 
\caption[Difference between the actual homogeneity and the homogeneity distribution of null models.]{Difference between the actual homogeneity and the homogeneity distribution of null models. Lines show the $z$ scores relative to null models for all computed resolutions, while black dots correspond to the $z$ scores obtained from the publicly available parcellations with fixed resolutions. }
\label{fig:group-null-diff}
\end{figure}

Clustering validity results in terms of parcellation homogeneity are summarised in three figures. First of all, homogeneity values obtained by each method/resolution are given in Fig.~\ref{fig:group-homogeneity}. The homogeneity of each method for a set of selected resolutions together with the homogeneity of their respective null parcellations are presented in Fig.~\ref{fig:group-null}. The difference between the homogeneity of the computed parcellations and the homogeneity distribution of null models measured as $z$ scores is shown in Fig.~\ref{fig:group-null-diff}. 

Homogeneity results in Fig.~\ref{fig:group-homogeneity} show a relatively poor performance for most of the provided parcellations. The methods that generate the most reproducible parcellations (e.g. spectral methods \textit{JOINT}, \textit{N-Cuts-2}, and \textit{N-Cuts-AVR}) as well as \textit{Geometric} also obtain poor homogeneity values. In general, other connectivity-driven computed parcellations tend to generate highly homogeneous parcellations with the group-average and 2-level methods obtaining very similar results. Among them, \textit{K-Means-AVR} especially excels at lower resolutions, but is outperformed by \textit{Baldassano}, one of the publicly available parcellations based on functional connectivity when similar resolutions are considered. It should be noted, though, that \textit{Baldassano} is obtained from a larger HCP cohort (500 subjects) which may contain our evaluation set and positively bias homogeneity results.

As shown in Fig.~\ref{fig:group-null} and~\ref{fig:group-null-diff}, we observe similar performance trends for most of the computed parcellations when compared to null models. Anatomical parcellations (\textit{AAL}, \textit{Destrieux}, and \textit{Desikan}), and some of the provided parcellations (\textit{Fan}, \textit{Gordon}, and \textit{Shen}), regardless of their respective resolutions perform similar to or worse than their null models. Among the publicly available parcellations, \textit{Baldassano} is on par with \textit{K-Means-AVR}, while \textit{Yeo}, \textit{Power}, and \textit{ICA} also yield good results. 

\begin{figure}[!t]  
\centering
\begin{tabular}{cc}
\includegraphics[width=0.5\textwidth]{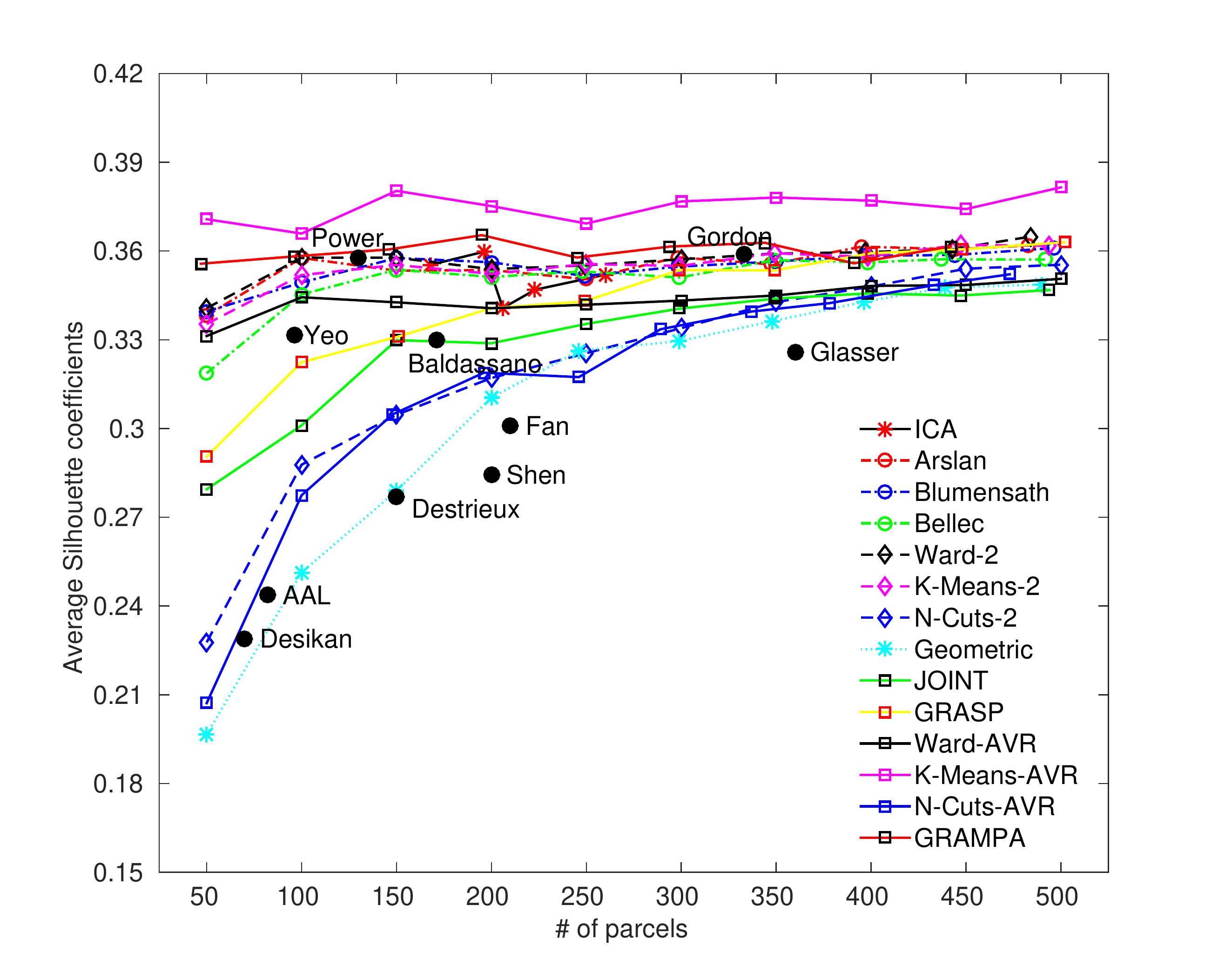} \kern-2.5em &
\includegraphics[width=0.5\textwidth]{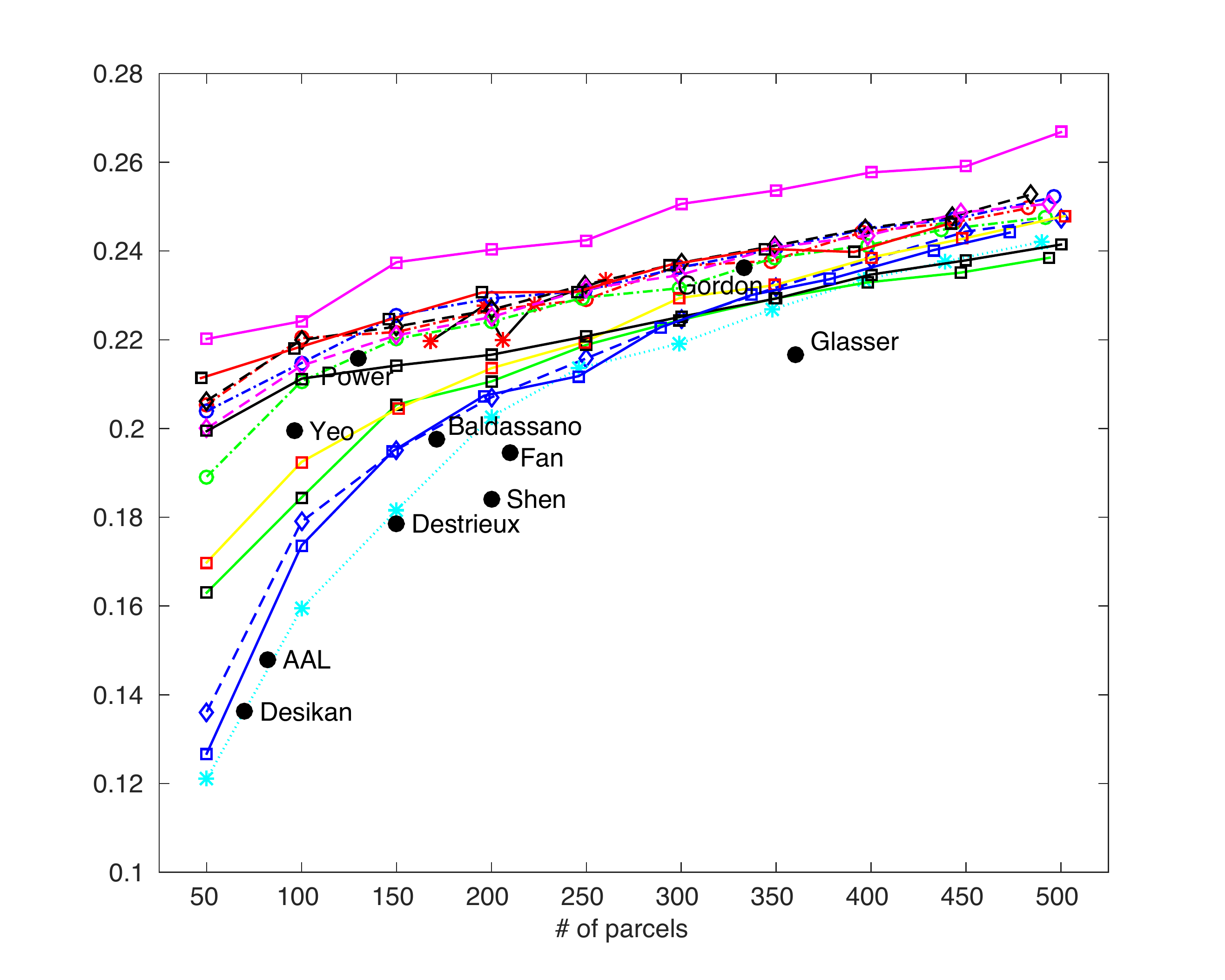} \\
(a) & (b) \\
\end{tabular}
\caption[Group-level Silhouette analysis results.]{Group-level Silhouette analysis results obtained (a) from the average connectivity fingerprints of all subjects and (b) on a per subject basis by using each subject's connectivity fingerprints and then averaging across subjects. Lines show the Silhouette coefficients (SC) for all computed resolutions, while black dots correspond to the SC obtained from the publicly available parcellations with fixed resolutions.}
\label{fig:group-silhouettes}
\end{figure}

Group-level Silhouette coefficients (Fig.~\ref{fig:group-silhouettes}) mostly follow the tendency observed in homogeneity. \textit{K-Means-AVR} outperforms the other approaches at all resolutions. It is followed by another group-average technique, \textit{GRAMPA}, which shows a good performance at low levels of granularity. All 2-level approaches, apart from \textit{N-Cuts-2}, perform equivalently well and produce more distinct parcels than most of the group-average methods. In contrast to the homogeneity results, \textit{Gordon} and \textit{Power} are the top-performing provided parcellations. Interestingly, despite producing homogeneous parcellations, \textit{Baldassano}, \textit{Yeo}, and \textit{ICA} show an average performance in terms of Silhouette coefficients. This shows that generating homogeneous parcellations does not necessarily guarantee a good separation between parcels. Overall, spectral techniques perform poorly but still surpass the anatomical and geometric parcellations. 

\begin{figure}[!ht]  
\centering
\begin{tabular}{cc}
\includegraphics[width=0.5\textwidth]{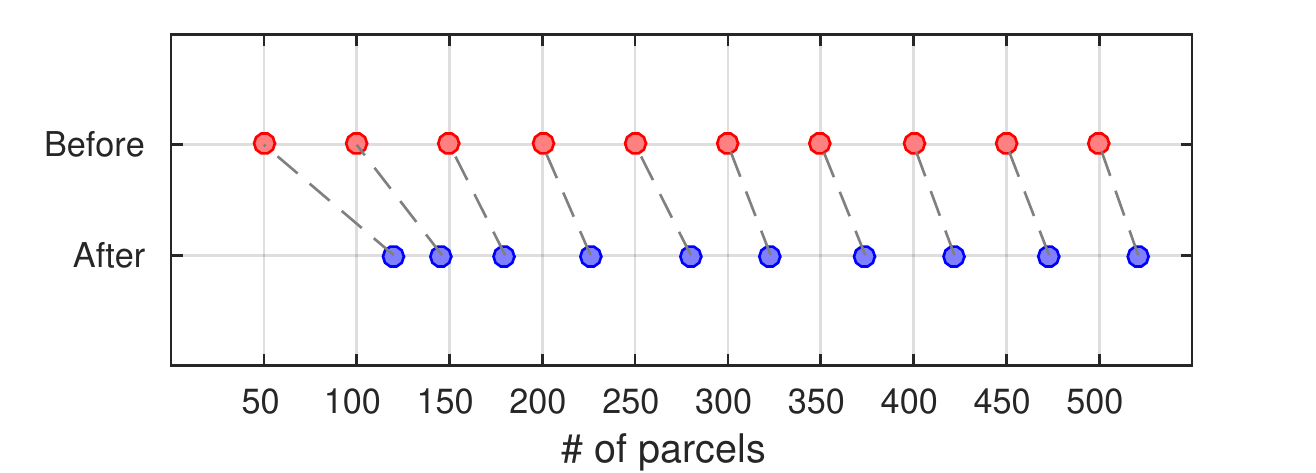} & \kern-2em 
\includegraphics[width=0.5\textwidth]{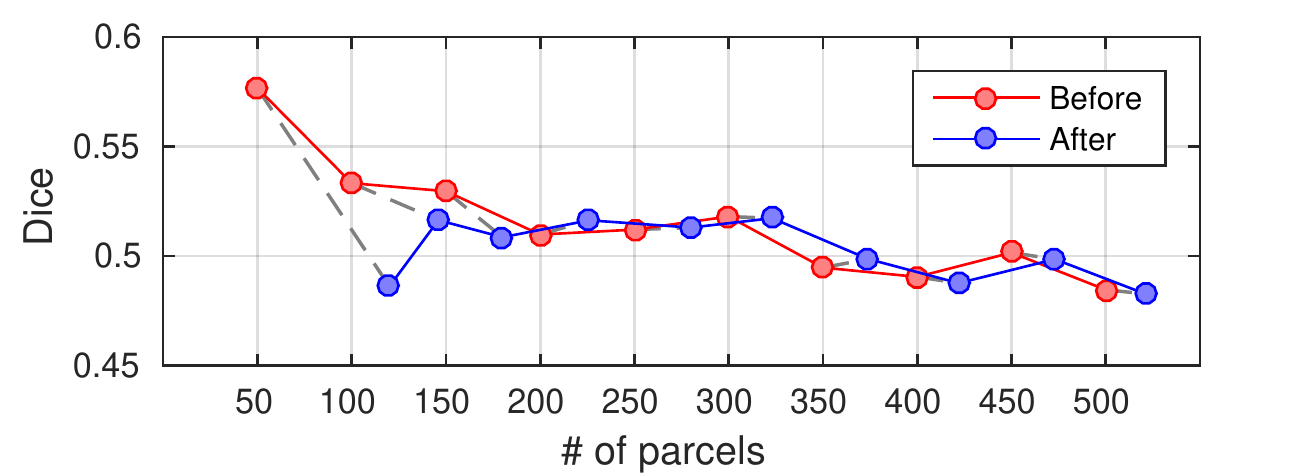} \\
(a) & \kern-2em (b) \\
\includegraphics[width=0.5\textwidth]{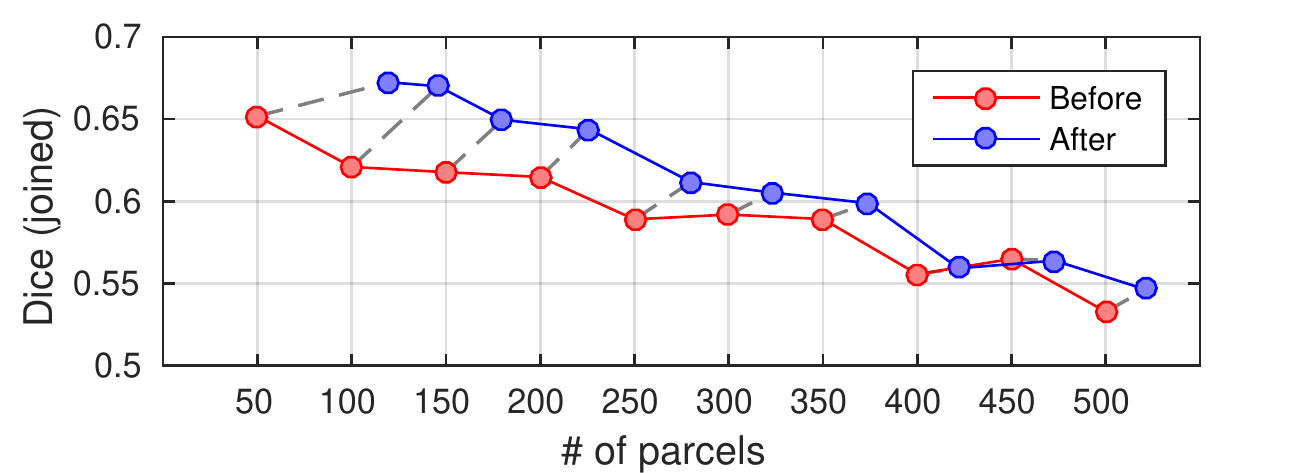} & \kern-2em 
\includegraphics[width=0.5\textwidth]{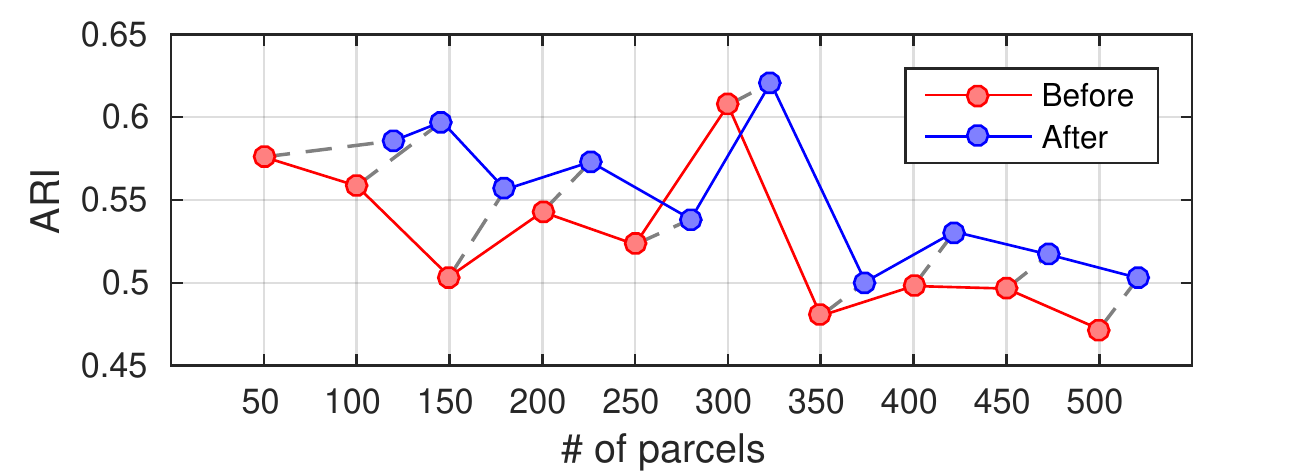} \\
(c) & \kern-2em (d) \\
\includegraphics[width=0.5\textwidth]{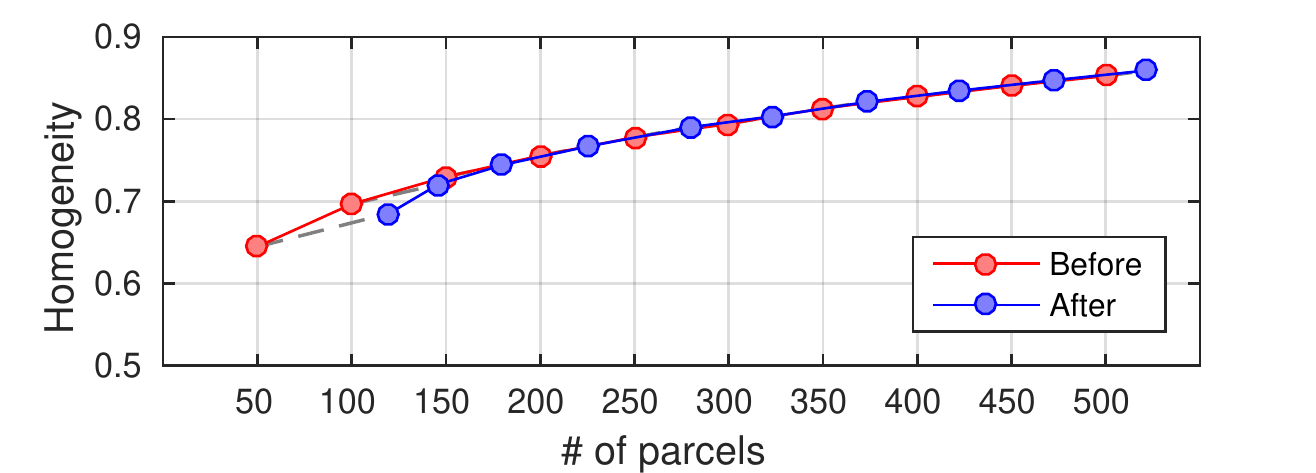} & \kern-2em 
\includegraphics[width=0.5\textwidth]{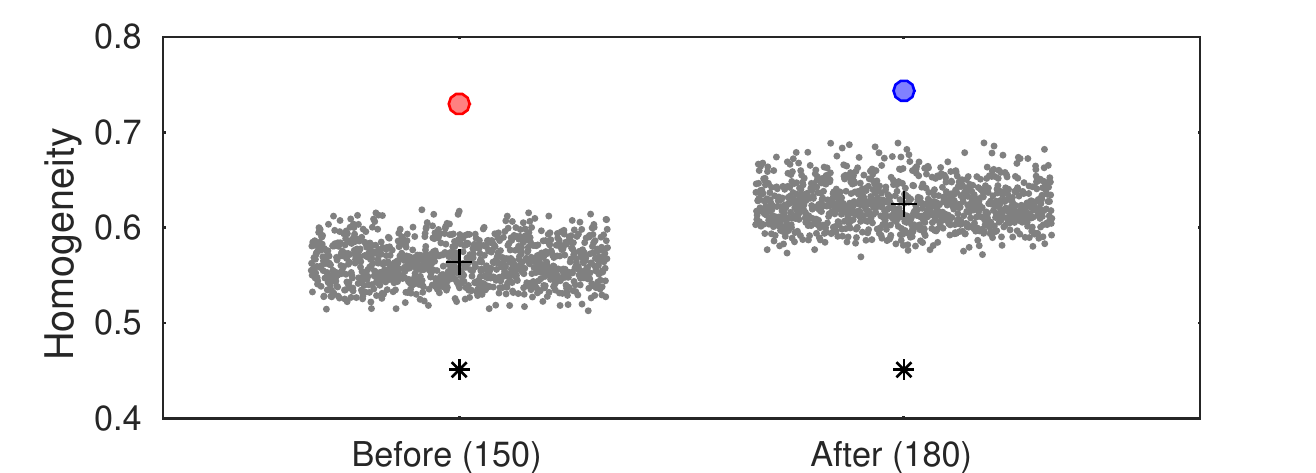} \\
(e) & \kern-2em (f) \\
\includegraphics[width=0.5\textwidth]{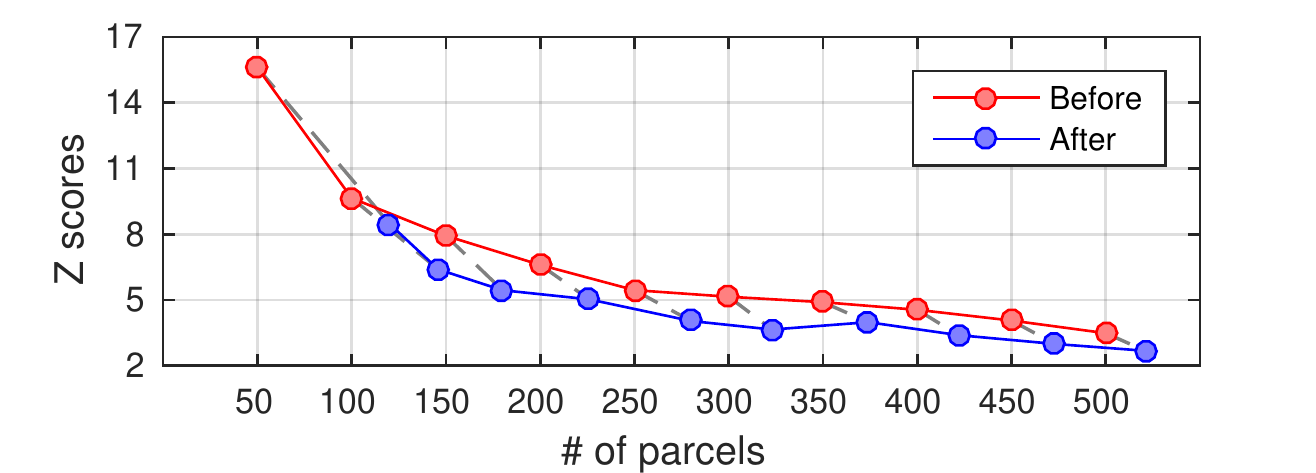} & \kern-2em 
\includegraphics[width=0.5\textwidth]{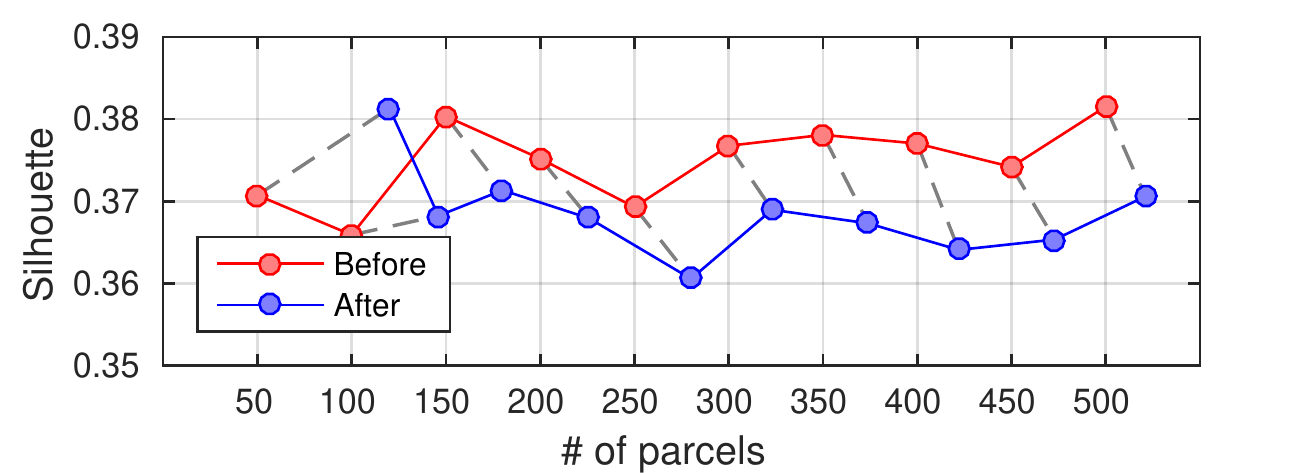} \\
(g) & \kern-2em (h) \\
\end{tabular}
\caption[Quantitative evaluation measures obtained from \textit{K-Means-AVR} after relabelling spatially disjoint regions.]{Quantitative evaluation measures obtained from \textit{K-Means-AVR} parcellations, \textit{before} and \textit{after} disjoint parcels are split into spatially contiguous regions. Points representing the original and relabelled parcellations (shown in red and blue, respectively) are matched with dashed lines for ease of comparison. The blue points correspond to the number of parcels acquired at each resolution after splitting, and therefore, are plotted further to the right with respect to the red points, which align with the resolutions of the original parcellations (50 to 500, in increments of 50) along the x axis. (a) The number of parcels before and after the split process. (b-d) Group-to-group reproducibility obtained via Dice similarity, joined Dice similarity, and adjusted Rand index. (e-h) Clustering accuracy measured via parcel homogeneity, comparison to null models (only for one resolution), $z$ scores relative to null models, and Silhouette analysis.}
\label{fig:kmeans_avr_extra}
\end{figure}

\paragraph{Impact of relabelling connected components in disjoint parcellations.} Considering all computed parcellation methods, two k-means variants, \textit{K-Means-AVR},  and \textit{K-Means-2} as well as \textit{GRAMPA} can generate spatially disjoint parcellations. In particular, \textit{K-Means-AVR} yields many discontinuous parcels, which significantly increases the total number of parcels after the relabelling process as shown in~\ref{fig:kmeans_avr_extra}. This unsurprisingly yields more homogeneous regions, as homogeneity depends on the resolution and likely to increase when the cortex is parcellated into more subregions (i.e. homogeneous regions still stay homogeneous when subdivided). On the other hand, as we alter the clustering configuration unnaturally by forcing parcels to split, fidelity to the underlying data is negatively affected, yielding lower Silhouette coefficients and $z$ scores relative to null models. However, this change in the spatial structure of the parcellations in general appears to yield a positive impact on reproducibility, as indicated by the joined Dice coefficients and adjusted Rand indices. The other two methods, \textit{GRAMPA} and \textit{K-Means-2}, only produce few parcels that are discontinuous, thus relabelling does not lead to a significant change in their evaluation measures. 

\subsection{Multi-Modal Comparisons}
\begin{figure}[!t]  
\centering
\includegraphics[height=0.5\textwidth]{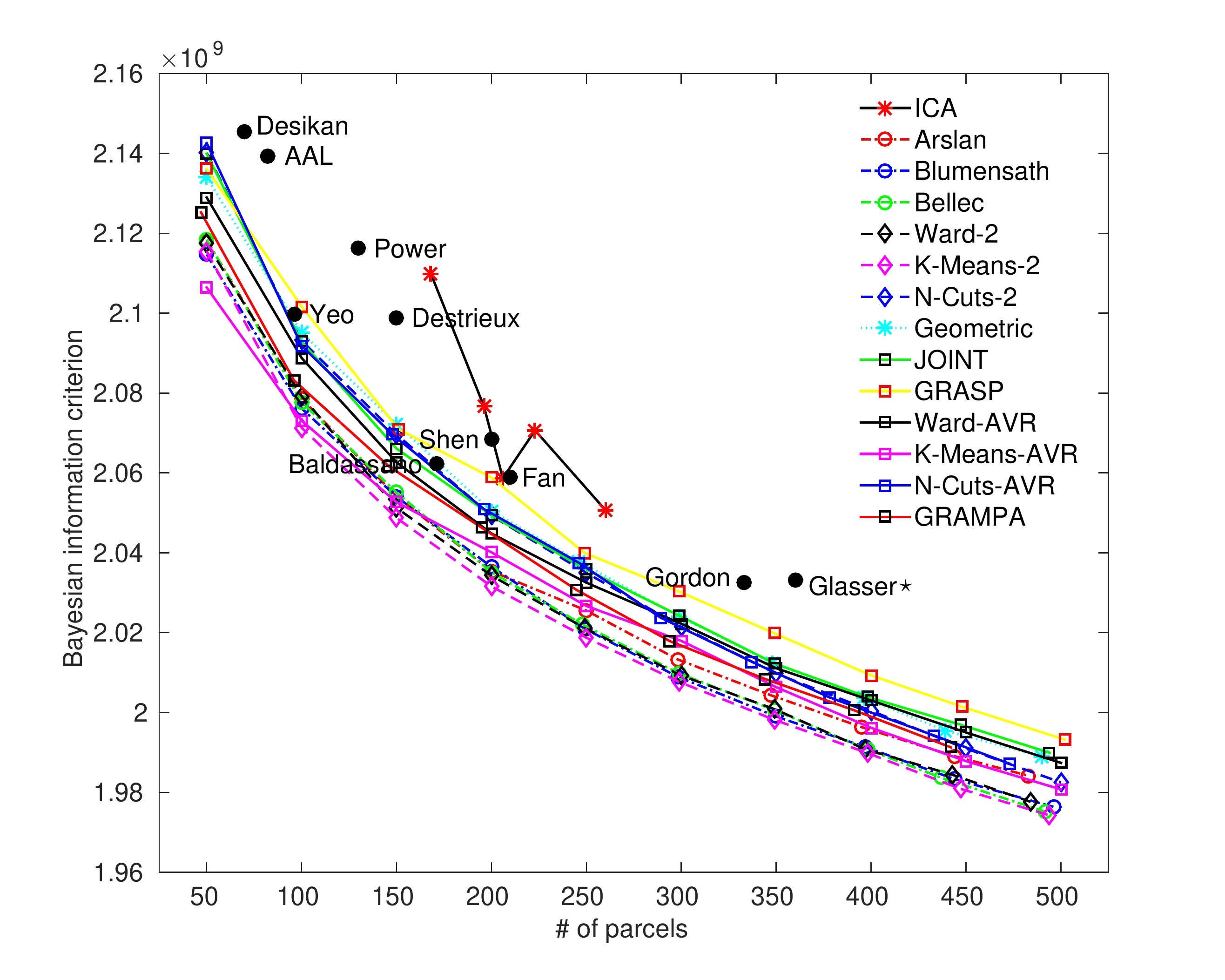} 
\caption[Agreement of group-level parcellations with task activation.]{Group-level Bayesian information criterion (BIC) results for measuring the agreement with task activation. Lines show the BIC values for all computed resolutions, while black dots correspond to the BIC obtained from the publicly available parcellations with fixed resolutions. Lower BIC indicates higher agreement with the task activation. $\star$: It should be noted that \textit{Glasser} is derived from group average task activation maps, which can influence this evaluation.}
\label{fig:group-bic}
\end{figure}

The agreement with concatenated single-subject task activation maps is reported in Fig.~\ref{fig:group-bic}. In general, all provided parcellations yield relatively poor BIC values compared to the computed parcellations with similar resolutions. The 2-level approaches tend to yield better results than their group-average (AVR) counterparts, in particular at higher resolutions, with \textit{K-Means-2} showing the best performance for most resolutions. This could be linked to the fact that the parcellations are derived from the subject level, where the individual task activation is also estimated from. The only provided methods that show a competitive performance are \textit{Yeo} and \textit{Baldassano}, while \textit{GRASP} yields the worst results amongst the computed parcellations. \textit{Glasser}  has a poor performance despite being driven by task average data. This can be attributed to the fact that it is generated from a different dataset which does not necessarily reflect the single subject task data in our evaluation set. Furthermore, this parcellation is obtained using a different registration method (based on cortical folding, myelination and functional connectivity) than our evaluation set, which can also negatively influence the results. 

\begin{figure}[!t]  
\centering
\begin{tabular}{c}
\includegraphics[height=0.7\textwidth]{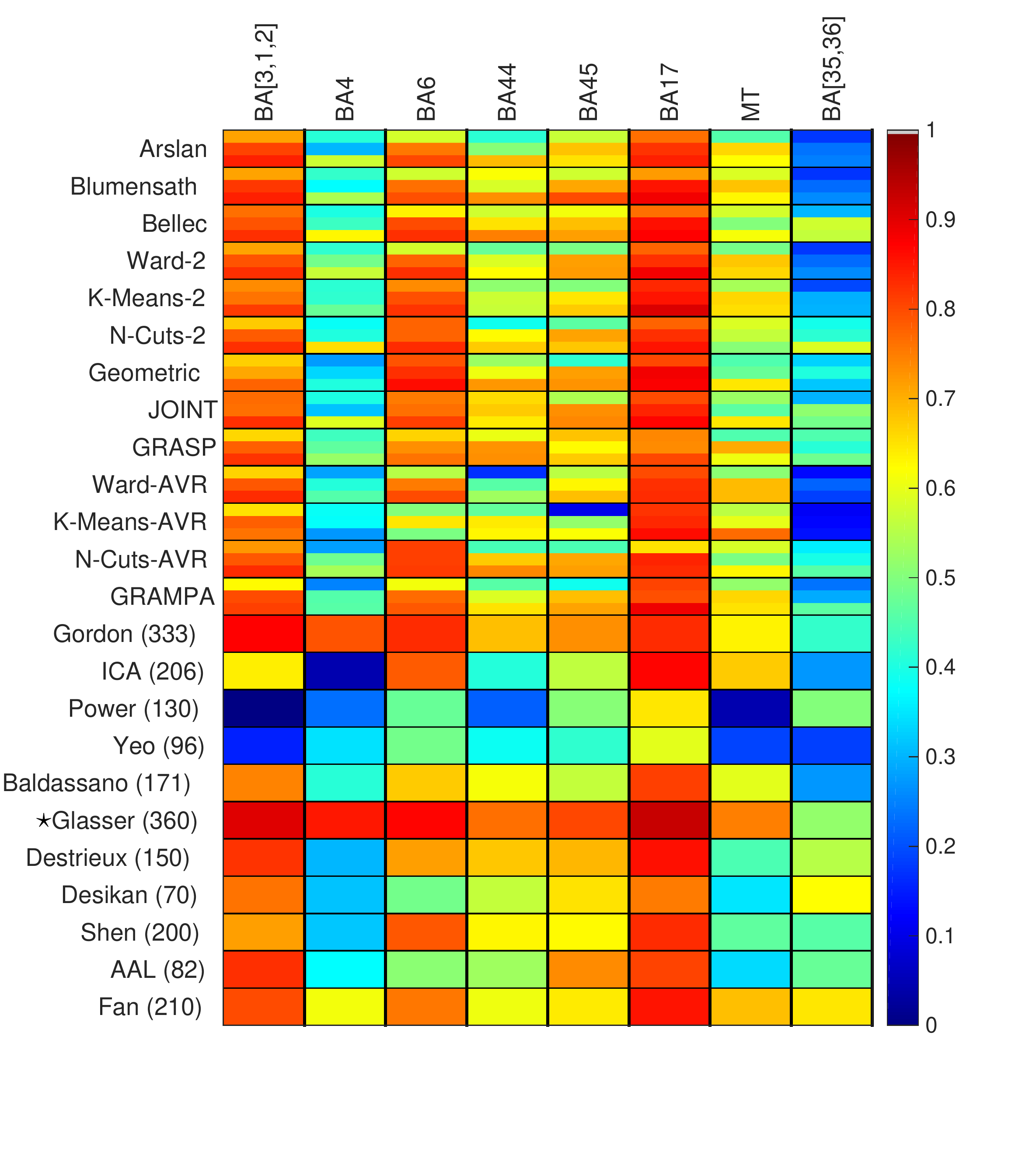} \\
\includegraphics[height=0.45\textwidth]{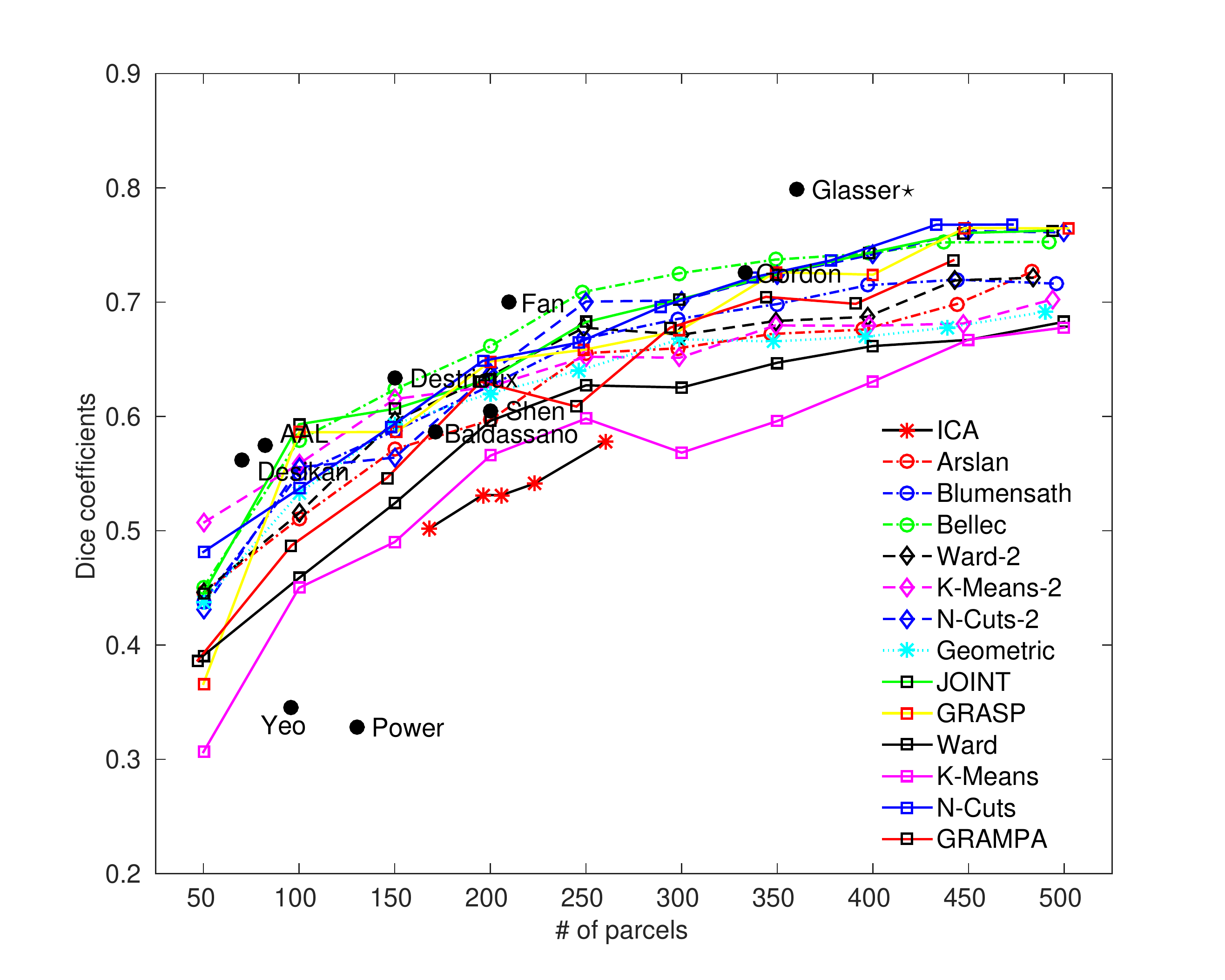} 
\end{tabular} 
\caption[Agreement of group-level parcellations with the Brodmann atlas.]{\textit{Top}: Agreement of all group-level parcellations with several Brodmann areas. For the computed parcellations (top 13 rows), each cell shows Dice coefficients for $K$ = 100, 200, and 300 regions, respectively from top to bottom. For the other parcellations, resolutions are indicated aside their names in parentheses. \textit{Bottom}: Average Dice coefficients for each method/resolution.$\star$: It should be noted that \textit{Glasser} uses expert knowledge and priors from the neuro-anatomical literature for the delineation of parcellation borders, which can influence this evaluation.}
\label{fig:group-ba}
\end{figure}

The overlap between the groupwise parcellations and the Brodmann areas (BA) for all resolutions is given in Fig.~\ref{fig:group-ba}. Many methods show a high degree of agreement with the primary somato-sensory cortex (BA[3,1,2]), pre-motor cortex (BA6), and primary visual cortex (BA17). Relatively low measures are obtained for the rest of the Brodmann areas, especially for the perinatal cortex (BA[35,36]). Overall, \textit{Glasser} shows the best performance and yields the highest overlap for most areas. Similarly, other provided parcellations \textit{Fan} and \textit{Gordon}, as well as the anatomical parcellations show a relatively high performance. \textit{Yeo}, \textit{Power}, and \textit{ICA} yield the lowest overlap measures, and in contrast to the general tendency in the group, do not align well with BA[3,1,2]. Interestingly, \textit{K-Means-AVR} produces the poorest results amongst the computed parcellations for all resolutions. This can be linked to the fact that \textit{K-Means-AVR} parcels are not necessarily spatially contiguous and may be spread across the cortex. In particular, the 2-level approaches perform better than the group-average ones, with \textit{Bellec} leading them at almost all levels of granularity.

\begin{figure}[!t]  
\centering
\includegraphics[height=0.5\textwidth]{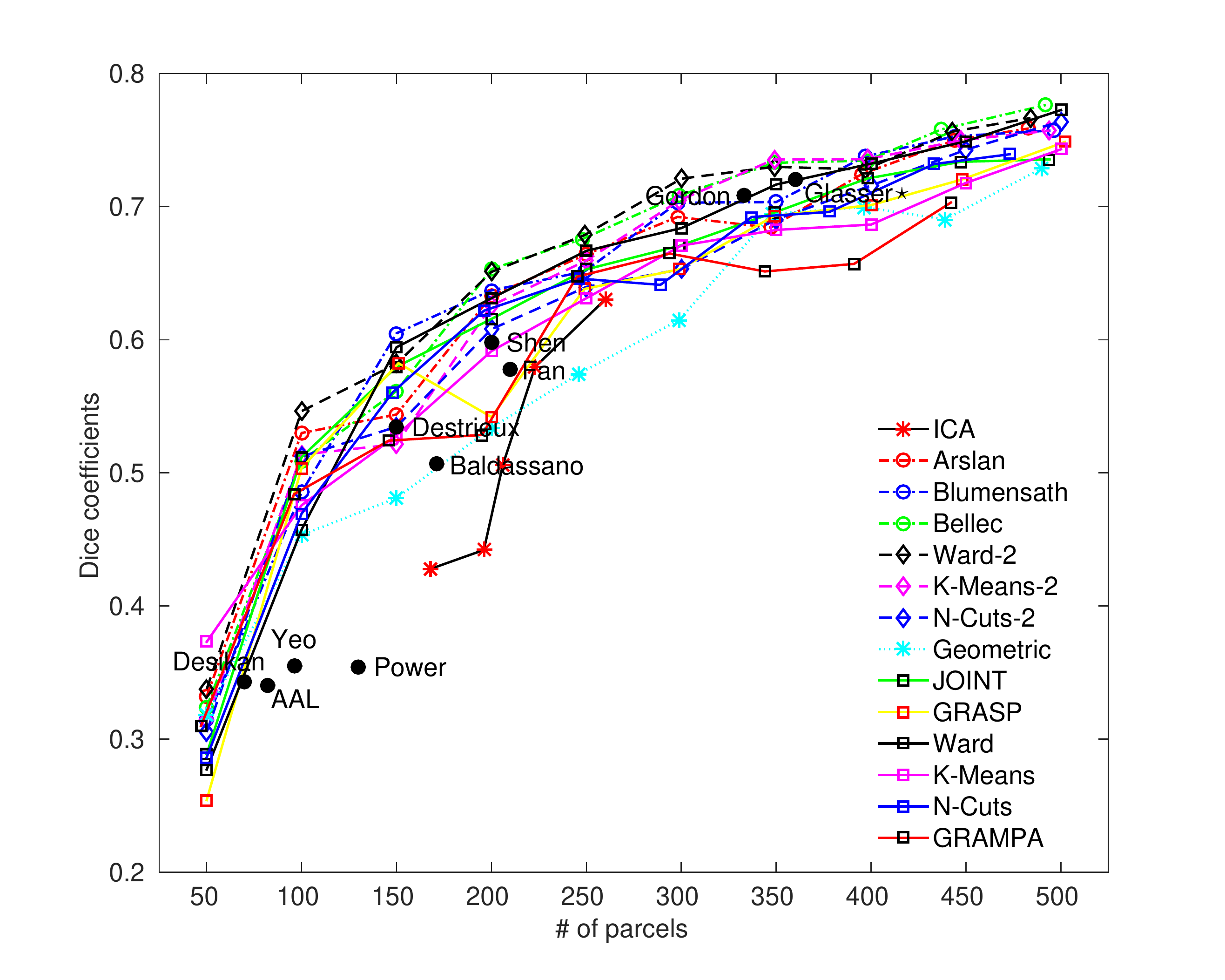} 
\caption[Agreement of group-level parcellations with highly myelinated cortical areas.]{Dice-based overlap measures of all group-level parcellations with highly myelinated cortical areas, derived from a coarse parcellation of the average myelination map. $\star$: It should be noted that \textit{Glasser} is derived from myelin maps and is therefore expected to have a good performance here.}
\label{fig:group-myelin}
\end{figure}

Average overlap scores with myelin based parcellations are given in Fig.~\ref{fig:group-myelin}. In general, the 2-level approaches show similar performance and outperform the group-average methods for most resolutions. \textit{Bellec}, \textit{Ward-2} and \textit{K-Means-2} have the highest agreement among the computed parcellations, while \textit{GRAMPA} and \textit{Geometric} yield relatively poor measures.\textit{Glasser} and \textit{Gordon} show the best performance amongst provided parcellations and outperform most of the other approaches when similar resolutions are considered. This is to be expected for \textit{Glasser} since it is derived from myelin maps. Other provided parcellations generally yield relatively low measures.

\subsection{Network Analysis}

\begin{figure}[!th]  
\centering
\begin{tabular}{ll}
\kern-1em \includegraphics[height=0.6\textwidth]{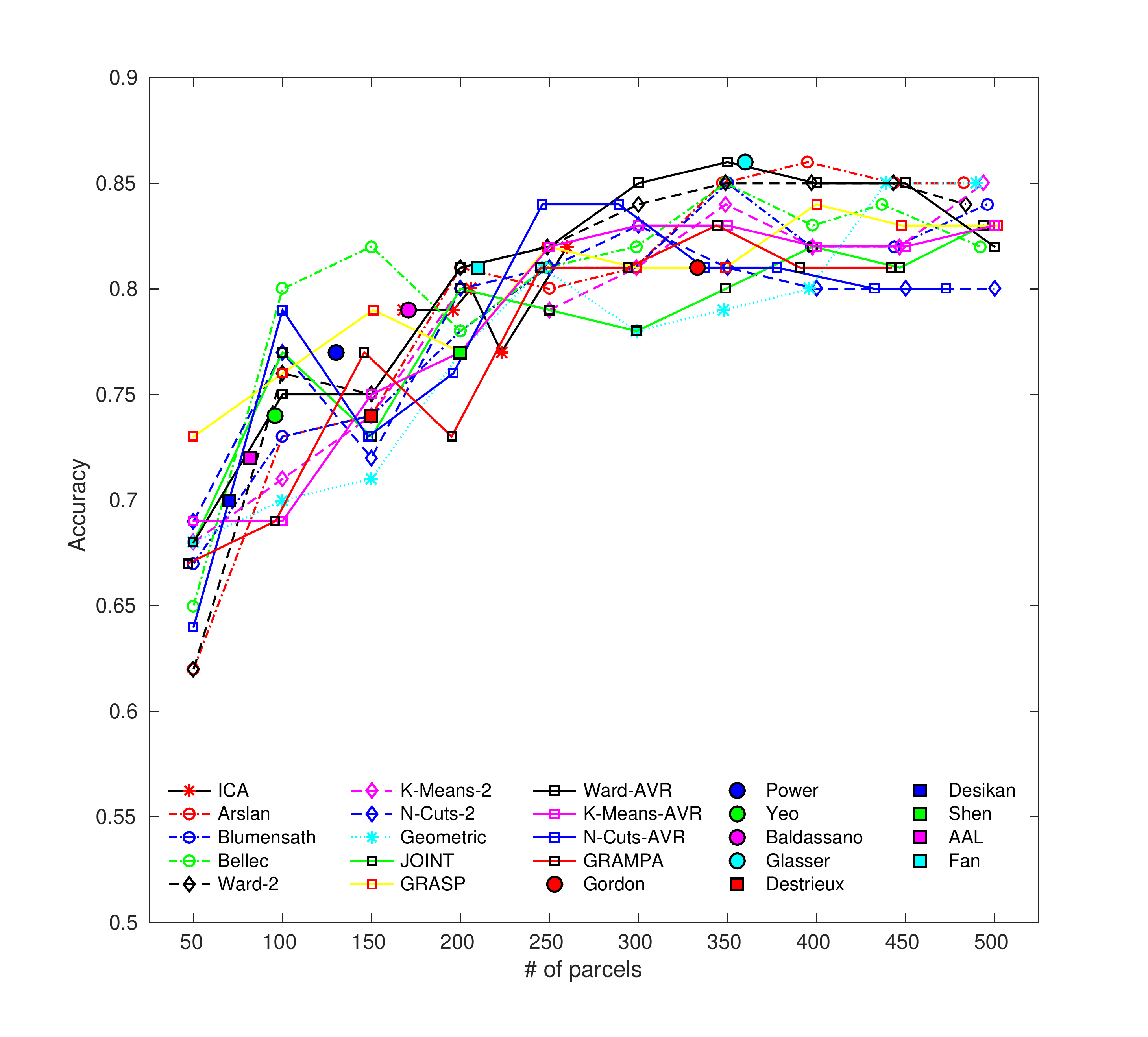} \kern-3.0em &
 \includegraphics[height=0.6\textwidth]{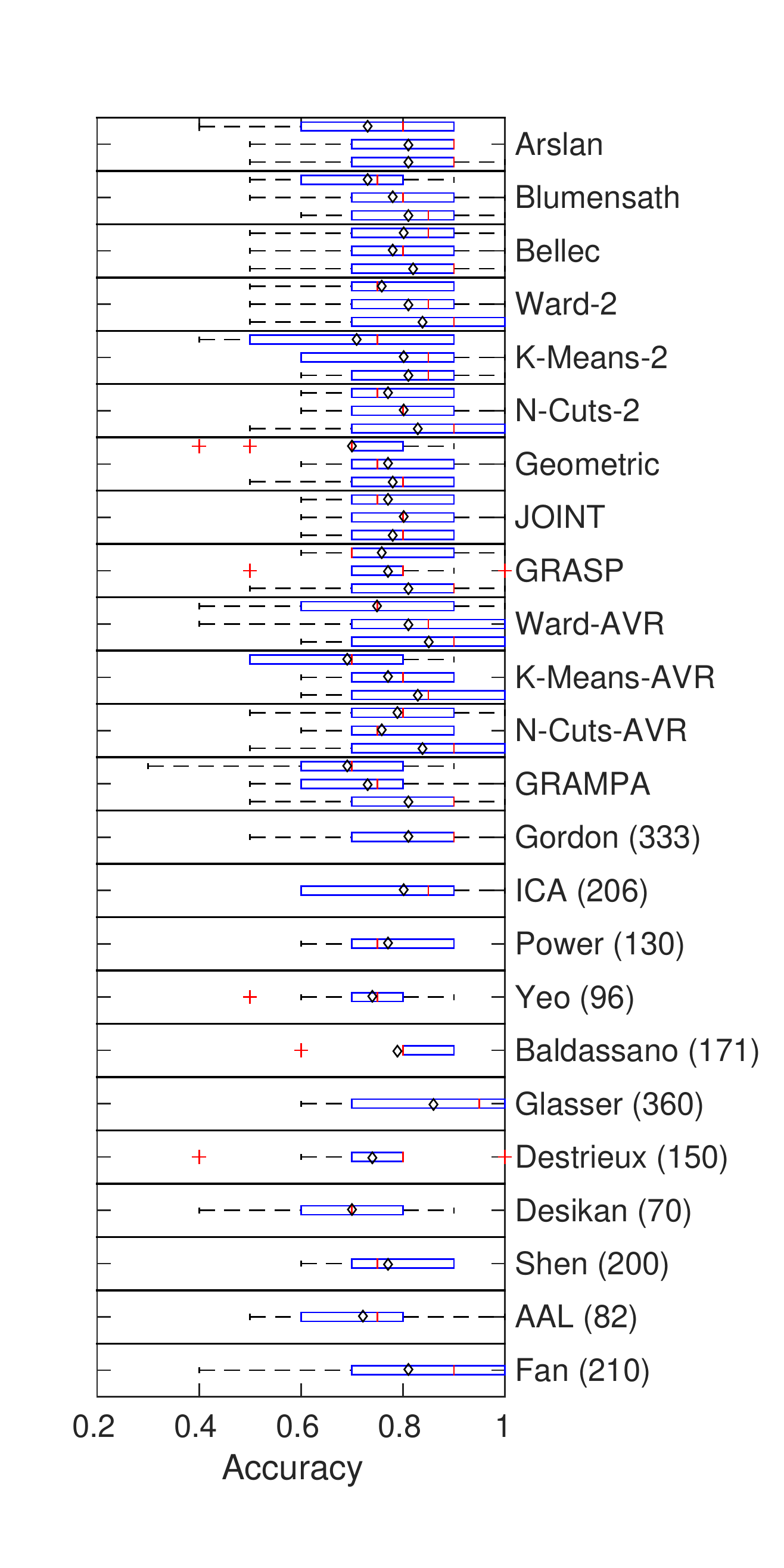}
\end{tabular}
\caption[Gender classification results.]{Gender classification results. \textit{Left:} Average Gender classification accuracy on 100 subjects with SVM. \textit{Right:} Variation across results is shown with respect to 10-fold cross-validation.}
\label{fig:network-svm}
\end{figure}

The results obtained for gender classification are illustrated in Figure~\ref{fig:network-svm}. Although there is no single winner across all different resolutions, anatomical parcellations are generally outperformed by several data-driven methods with similar number of parcels. Overall, results obtained with SVM are not very consistent across resolutions, since there is no obvious upward/downward trend with increasing resolution. In fact, most methods demonstrate a similar average performance, being able to classify males and females with above 60\% accuracy for granularities below 150 parcels and above 70\% for higher resolutions. More specifically, \textit{Geometric} tends to perform poorly compared to the rest of the methods, both at lower and higher resolutions. The highest SVM classification accuracy (86\%) is achieved with \textit{Ward-AVR} and \textit{Glasser} at the scale of 350 and 360 parcels, respectively. Moreover, we can observe that increasing the resolution of the parcellation in data-driven approaches beyond a certain value (350 parcels) does not necessarily provide additional information about population differences. However, lower resolutions lead to lower classification scores, perhaps due to the fact that functional information valuable for the discrimination between the two classes fades by averaging the signal in larger parcels. Interestingly, \textit{N-Cuts-AVR}, \textit{Bellec} and \textit{Arslan} perform quite well for several resolutions, while \textit{GRASP} yields the top accuracy among all methods for 50 parcels across the cortex. It is also worth mentioning that the parcellations provided by \textit{Yeo}, \textit{Shen} and \textit{Gordon} have below average performance, while \textit{Fan} and \textit{Glasser} have good performance compared to parcellations with similar resolutions.

\begin{figure}[!bht]  
\centering
\includegraphics[height=0.5\textwidth]{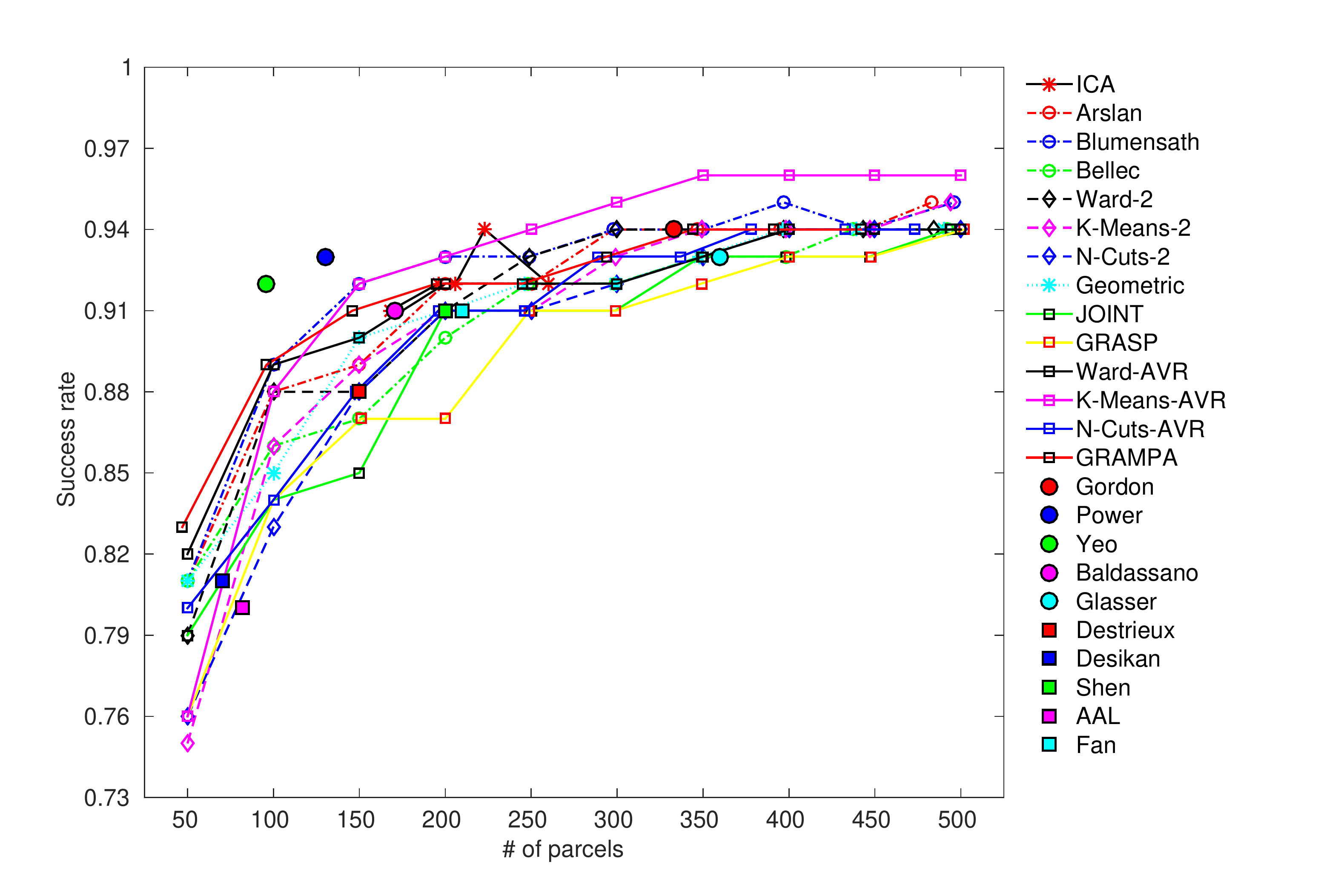}
\caption{Individual subject prediction results.}
\label{fig:network-predict}
\end{figure}

Individual subject identification results presented in Fig.~\ref{fig:network-predict} show a clearer trend than gender classification: connectivity networks derived from relatively high-resolution parcellations yield a higher success rate. In particular, \textit{K-Means-AVR} appear to obtain the highest accuracy at most of the granularity levels, whereas for lower resolutions, \textit{Yeo}, \textit{Power} and \textit{ICA} outperform the other approaches. The rest of the methods have similar performances with \textit{GRASP}, anatomical and spectral parcellations generally having the lowest success rate for most resolutions. The scale of the parcellation appears to have a much stronger impact on the identification accuracy than the parcellation scheme itself. This is likely due to the fact that the variability across individuals cannot be as effectively captured by low-resolution parcellations.

\section{Discussion}
\label{sec:discussion}
In this chapter, we proposed a large-scale comparison of existing group-level parcellation methods using state-of-the-art evaluation measures and publicly available data provided by HCP. The generation and evaluation of the parcellations is based on resting-state functional connectivity, which is thought to express the network behaviour underlying high level cognitive processes. We considered several criteria simultaneously to evaluate the quality of the parcellations, such as reproducibility, parcel homogeneity, and Silhouette analysis. These measurements assessed the performance from a cluster quality point of view. The neuro-biological interpretation of the obtained parcels is also investigated by comparing parcel boundaries with well-defined neuro-biological properties, such as cytoarchitecture and myelination, as well as task activation. In addition, we devised two simple network analysis task, i.e. gender classification and subject identification, in order to measure the impact of the underlying parcellation on network analysis. 

Our experiments show that there is no clear trend in favour of a specific method - or type of method - regarding all evaluation metrics considered. For instance, \textit{k}-means clustering appears to be largely leading in terms of clustering quality. It, however, shows a poor performance regarding reproducibility and agreement with other modalities. In addition, while cortical delineation intrinsically requires a relatively large number of parcels, this does not appear to be a requirement for effective network analysis. This may suggest that different types of parcellations are to be investigated depending on the task at hand (e.g. one should use different methods when considering network analysis or cortical delineation). 

We observe that connectivity-driven parcellations have a much better agreement with the underlying rs-fMRI connectivity compared to anatomical and random parcellations as expected. The benefit of using connectivity to parcellate the brain is not as clear regarding the delineation of cortical areas (agreement with other modalities and established brain delineations) and subsequent network analysis. In particular, anatomical parcellations appear to yield equivalent or better results with respect to cytoarchitecture. A general conclusion regarding network analysis would be to use any parcellation available, or learn it from the data, as it seems to have a limited impact. However, while this may be true for simple analysis of healthy subjects, it would have to be investigated further in the context of largely different brains (such as large age gaps or diseases).  

\subsubsection{Connectivity estimated from resting state fMRI and its impact on parcellations}
Resting state fMRI is the most commonly used state-of-the-art technique to map whole-brain functional connectivity, with its high spatial resolution favouring its application over alternative electro-physiological recordings, like EEG and MEG. Its effectiveness to map the function of the brain has been consistently shown across a wide range of studies~\cite{Damoiseaux06,salvador2005neurophysiological,Heuvel08,Power11}. However, the true biological interpretation of the BOLD signals is still unknown~\cite{Eickhoff15}, and its low temporal resolution (commonly at the order of seconds) is a limiting factor for the observation of high-frequency patterns. The structure of functional connectivity can also be a shortcoming in itself. Several sources of noise can influence BOLD signals, including head movement, cardiac and respiratory pulsations. On top of this, one known issue with functional connectivity is the possible indirect connections mediated by third-party regions~\cite{Smith11,Eickhoff15}. As a result, it can be difficult to separate the actual signal from the noise, (or the direct connections from the indirect) and to assess the reliability of functional connectivity derived from fMRI and their interpretation. In particular, the noise in the data and the different hypotheses and processing decisions taken for a given parcellation technique could explain why different connectivity driven parcellation methods perform better or worse with respect to certain evaluation measures. For example many techniques alter the structure of the connectivity network by  applying thresholding to exclude negative and weak correlations from the connectivity matrices with the aim of obtaining more robust parcellations~\cite{Heuvel08,Power11,Craddock12,Arslan15a}.

\subsubsection{Evaluation of parcellations from a clustering point of view} 
When parcellations are evaluated, both implicit constraints inherent to the method and explicit constraints imposed to the data should be taken into consideration, as they yield inevitable biases towards the computed parcellations~\cite{Blumensath13}. It is, therefore, highly critical to evaluate clustering accuracy from different perspectives. 

\textit{K}-means, hierarchical clustering, and spectral clustering (as well as their variants) are frequently used to obtain connectivity-driven parcellations, ultimately serving the task of brain mapping~\cite{Eickhoff15}. Their impact on the parcellation configuration as well as their limitations and advantages over each other have been extensively reviewed in~\cite{Thirion14,Eickhoff15} and further discussed in Chapter 3. In general, our results align with the previous literature (as well as with those derived from subject-level parcellations in Chapter 4) regarding the performance of these clustering algorithms. For example, \textit{k}-means generally provides the best performing groupings of the data, but suffers from low reproducibility due to the fact that it does not inherently rely on hard spatial constraints. On the contrary, spectral techniques are usually dominated by spatial constraints and consequently yield highly reproducible parcellations, but at the cost of inaccurate alignment with the brain's underlying functional organisation. Hierarchical clustering offers the advantage of generating reproducible parcellations to a certain extent, while still capturing the functional connectivity with high fidelity. 

Several parcellations computed on a different dataset yield relatively good cluster quality results. One can infer from this observation that similar characteristics shared by healthy adults can be robustly detected across different datasets as long as the analysis is performed on a large cohort (for example \textit{ICA} and \textit{Baldassano} are originally obtained from a group of 500 subjects where this number increases to 1000 for \textit{Yeo}). It should be also noted that, \textit{ICA} and \textit{Baldassano} can also comprise some subjects from our test dataset as they are computed from a larger HCP cohort. This may constitute an important factor promoting a more favourable performance for these two methods compared to the others. 

Predictably, anatomical parcellations yield the lowest performance in terms of clustering quality. However, they allow a more intuitive neuro-biological interpretation which can make network analysis more insightful. On top of that, our network-based experiments show that a better quality clustering does not necessarily benefit network analysis. One limitation is their relatively low resolution which is typically addressed by partitioning each parcel into subunits without altering the anatomically delineated boundaries. This can be achieved randomly~\cite{Hagmann08,Honey09} or using functional connectivity~\cite{Patel08, Fan16}. This approach is adapted by \textit{Fan}, but appears to provide a limited improvement compared to anatomical parcellations.  

\subsubsection{Agreement of parcellations with other neuro-biological properties of the cortex}
The anatomical parcellations based on cortical folding, i.e. \textit{Desikan} and \textit{Destrieux}, as well as the anatomo-functional atlas based on the Desikan parcels (i.e. \textit{Fan}) interestingly show a high degree of agreement with the cytoarchitecture of the cerebral cortex. Although these results may reflect a better alignment between anatomy and cytoarchitectural atlases than with rs-fMRI, this might also be linked to registration errors as the Brodmann maps are registered to each individual subject based on cortical folding. While we can expect a good overlap in the motor and visual cortex, where the folding patterns are more consistent across subjects, stronger misalignments could occur in other regions. 

Similar observations can be made for connectivity-driven parcellations, in which case a higher degree of alignment is found within the motor and visual cortex. Despite the fact that functional connectivity obtained from BOLD timeseries is not necessarily expected to reflect the cytoarchitecture of the cerebral cortex, these results agree with several rs-fMRI based studies that report similar findings regarding these regions~\cite{Blumensath13,Wig14,Gordon16}. On the other hand, a more consistent agreement can be expected between the connectivity-driven parcellations and highly myelinated areas, as the gradients in rs-fMRI-driven connectivity have been observed to align well with the myelination patterns~\cite{Glasser11}.

One should also take into consideration the reliability of the evaluation techniques used to compare the different modalities. For example, overlap-based measures, such as the Dice coefficient, are biased by the size of the parcels.  Evenly sized/shaped parcels are easier to match with their target parcels, while differences in Dice scores will be much more striking when comparing small parcels over big ones. This bias can lead to more favourable results for some of the parcellations, such as \textit{Geometric}, \textit{N-Cuts}, and \textit{Random}, all of which comprise more uniformly shaped/sized parcels than the rest of the approaches. Although such quantitative measures can provide a means of comparing different methods, the quality of a parcellation with respect to cytoarchitecture or myelin content should also be visually assessed before drawing any conclusion. 

Similarly, the Bayesian information criterion has a bias towards more complex models, i.e. parcellations with higher resolution are always favoured~\cite{Thirion14}. It should be also noted that there may exist redundant and contradictory information in the different tasks/contrasts which could bias the results. On top of that, the SNR in the task activation maps is low, therefore, it is likely that the results might be compromised by noise. Finally, our experiments have compared group-level parcellations to single subject level task activation maps. While the objective is to evaluate whether these group parcellations provide a good representation of the population, one could also consider comparing to group average task activation maps. This would alleviate single subject noise and could yield better results, in particular for provided parcellations. For example, the \textit{Glasser} parcellation is expected to have a much better performance with respect to group level task maps on which it is derived. 

Additionally, this multi-modal parcellation (\textit{Glasser}) can give a clearer intuition on the behaviour of inter-modality comparisons. This method does not only rely on resting-state functional connectivity, but also embodies information from task activation, myelin content, and the cortical architecture. It yields very good overlap with the Brodmann areas and myelin content, especially on some parts of the cortex (e.g. motor cortex, highly-myelinated areas), indicating that the overlap measures used for multi-modal comparisons do provide accurate information.

\subsubsection{Impact of parcellations on network analysis}
When network analysis is concerned, our experiments based on two different tasks, gender classification and individual subject prediction, yield different results. The former does not favour any particular method to subdivide the brain into regions that would better reflect population differences. On the other hand, when individual subject prediction task is taken into consideration, we observe a dominance of connectivity-driven approaches, with a rather robust trend of increasing performance with respect to parcellation resolution. As the difference between the two sets of results can be linked to the specific network analysis tasks at hand, their ability to assess the parcellation quality should be interpreted separately.

Although classification analysis has previously been applied in studies of functional connectivity to predict demographic measures including gender~\cite{satterthwaite2014linked, robinson2008multivariate} and age~\cite{vergun2013characterizing}, our experiments suggest that the classification score alone is not a valuable tool for the evaluation of parcellation quality. Instead, the number of features selected (edges in the connectivity matrix) to achieve the same classification performance might be a better means of evaluation. Provided that a larger number of subjects is available, a good parcellation should give a sparse selection of features and a more interpretable result. The results do not favour any parcellation, either anatomy, cytoarchitecture, or data driven, to better reflect population differences. In fact, anatomical parcellations appear to perform as well as data-driven approaches when the similar resolutions are considered. This could be attributed to the specific task at hand, since anatomical and, more specifically, cerebral volume differences have been reported between males and females that significantly influence the volume of white and gray matter~\cite{leonard2008size}. Therefore, volume/anatomy-related differences and sex-related differences are hard to disentangle under the current experimental setting, despite the fact that all subjects have been registered to the same anatomical space. 

When individual subject identification is concerned, connectivity-driven approaches appear to show a consistently more favourable performance compared to the other parcellation schemes. However, not particularly the parcellation scheme itself, but its resolution boosts the performance of the prediction task, as any parcellation with more than 200 regions can accurately predict at least 90 out 100 subjects. This may be linked to the fact that, parcellations with fewer regions may not be able to capture relevant features to distinguish one subject from another. Similar findings are also reported in~\cite{Finn15}, where higher accuracy is obtained by using a more fine-grained parcellation. It is also worth noting that, more robust trends are observed with the prediction task than gender classification. While one can infer that the classification task is simply more sensitive than the prediction task, this could also be attributed to the fact that capturing characteristic features on a single-subject basis is much easier than identifying population differences from a whole group of subjects.

\subsubsection{Limitations}
\begin{figure}[!th]  
\centering
\includegraphics[width=\textwidth]{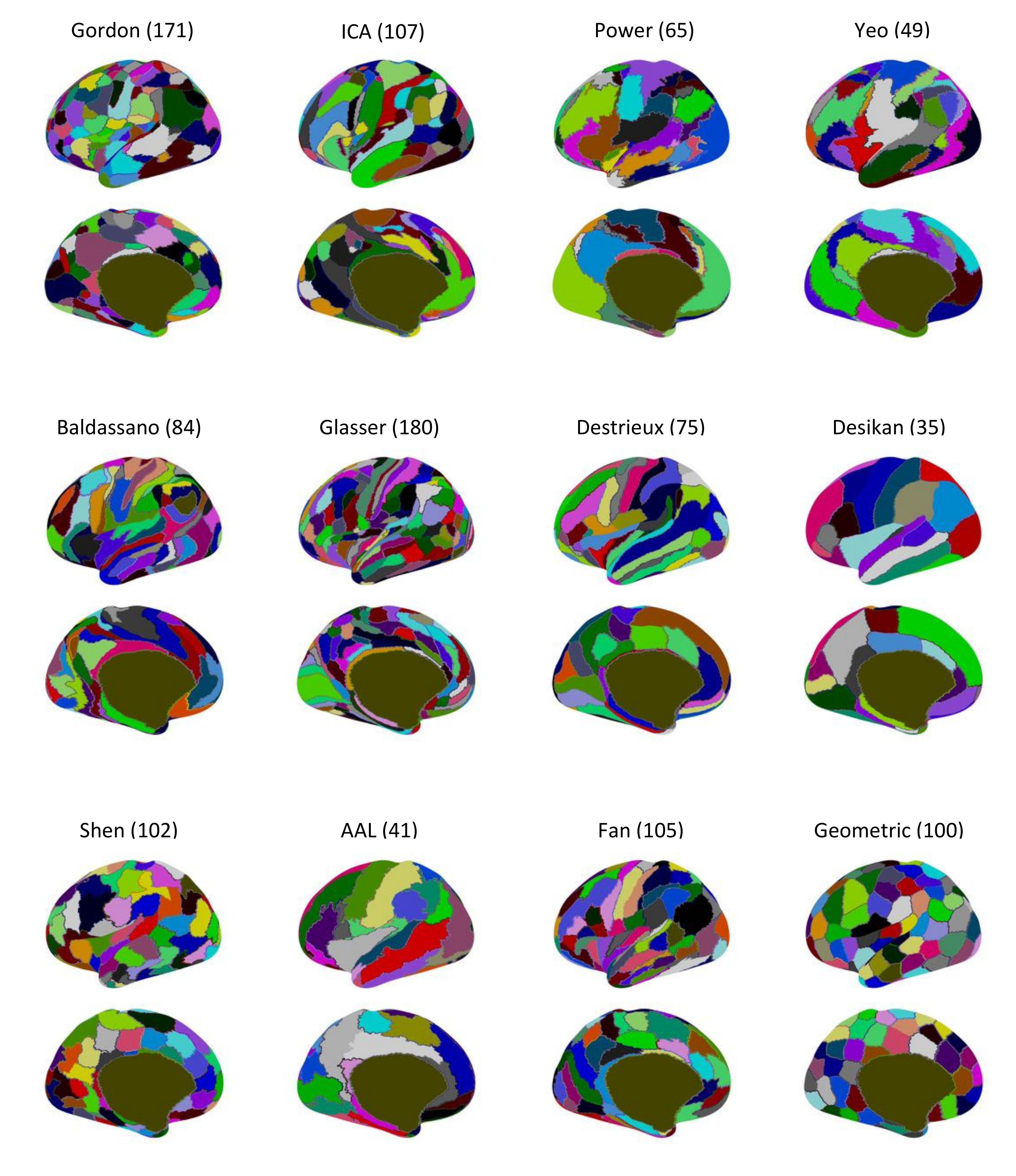} 
\caption[The first half of the parcellations considered in this study as projected on the left hemisphere.]{The first half of the parcellations considered in this study as projected on the left hemisphere. The number of parcels is shown along with the name of the method above each parcellation. Computed parcellations are only shown for a fixed resolution of 100 regions.}
\label{fig:all-parcels-1}
\end{figure}

\begin{figure}[!th]  
\centering
\includegraphics[width=\textwidth]{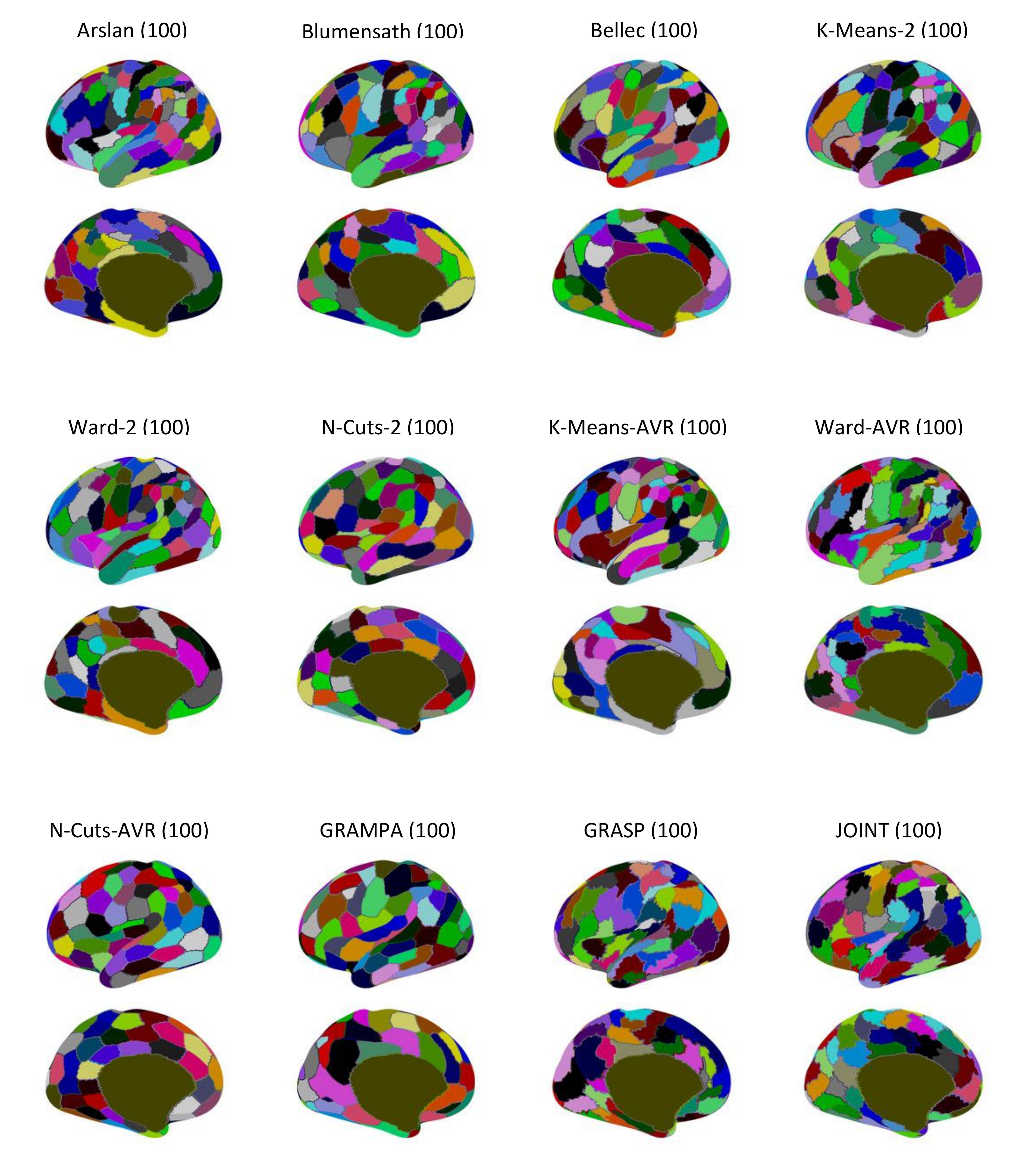} 
\caption[The second half of the parcellations considered in this study as projected on the left hemisphere.]{The second half of the parcellations considered in this study as projected on the left hemisphere. The number of parcels is shown along with the name of the method above each parcellation. Computed parcellations are only shown for a fixed resolution of 100 regions.}
\label{fig:all-parcels-2}
\end{figure}

In this paper, we considered both surface-based and volumetric parcellations. Whilst efforts are made to be fair to all methods, several important methodological choices have been made, which may have an impact on the evaluation and possibly promote some parcellations over the others. In particular, parcellations are not the products of the same processing pipeline. Most of the publicly available parcellations have been generated under different assumptions, from different sets of subjects with varying cohort size and after being subject to a series of processing steps. Additional processing was applied to certain methods to make  parcellations comparable on a more standard basis. Parcellations that do not naturally provide spatially contiguous cortical areas (e.g. \textit{Yeo}, \textit{Power}, \textit{ICA}) were relabelled while methods that do not cover the entire cortical surface (e.g. \textit{Gordon}) were dilated. Similarly, we used the group-average \textit{Glasser} parcellation in our experiments, despite the fact that this method also provides individual parcellations tailored to each subject. In particular, the performance of parcellations sampled from a volumetric space should be interpreted carefully due to the complicated transformation steps. Nevertheless, we believe these parcellations are an essential aspect of our evaluation. Please see Figs.~\ref{fig:all-parcels-1} and~\ref{fig:all-parcels-2}, for figures showing group-wise parcellations used in this study, respectively. All the parcellations and evaluation code are made publicly available via the web page: \url{http://biomedia.doc.ic.ac.uk/brain-parcellation-survey}, in case one may need access to these parcellations for their own analysis on a different dataset.

\chapter{Conclusions}
\label{chapter:conclusion}

This thesis presents a number of different connectivity-driven parcellation methods to subdivide the human cerebral cortex into non-overlapping, spatially contiguous, and homogeneous regions that can be used for studying the functional and structural organisation of the human brain. Our technical contributions offer new solutions to different aspects of the parcellation problem. As features for the proposed methods, we rely on the connectivity information obtained from rs-fMRI and dMRI, the most-widely used neuroimaging techniques for mapping the human brain. We propose multi-scale parcellations both for identifying the organisation of individual brains and representing shared patterns of connectivity within a relatively large cohort of healthy subjects. 

In Chapter 4, we introduce a new parcellation method for segregating the cerebral cortex on a single subject basis. The proposed methodology relies on two different clustering approaches combined into a two-layer framework, in which the first layer stands as a spatial feature reduction stage and provides a relatively high-level representation of the cortex. Running atop this initial parcellation, the second layer computes a spectrum of parcellations, nested in a hierarchical fashion, yielding subject-level parcellations at different levels of granularity. We show that the proposed parcellations subdivide the cerebral cortex into regions with distinct patterns of connectivity and with higher reproducibility and homogeneity compared to another two-layer approach. We further discuss the advantages and limitations of our method over several other data-driven clustering techniques, and show that the functional organisation of the individual human brain can be more accurately modelled by connectivity-driven parcellations with respect to random or anatomical ones. 

In Chapter 5, we again address the subject-level parcellation problem, but from the perspective of a different modality, i.e. structural connectivity acquired via dMRI. We reformulate the parcellation problem as a feature reduction problem and utilise a non-linear manifold learning technique to identify connectivity characteristics of an individual subject that may not be directly visible in high dimensional space. Assuming that sharp changes in connectivity patterns residing in this low-dimensional embedding correspond to parcellation boundaries, we use edge detection techniques to subdivide the cerebral cortex into spatially contiguous and non-overlapping regions. Our experiments show that the proposed method is capable of providing distinct parcels with uniform connectivity profiles at varying resolutions with respect to several other parcellation techniques. We further show that our parcellations can potentially represent both the structural and functional organisation of the brain and hence provide a more reliable set of network nodes for connectome analysis.

In Chapter 6, we propose an alternative methodology to group-wise parcellation approaches currently used in the literature. We suggest combining within- and between-subject connectivity patterns into a multi-layer graphical model in order to better capture the fundamental connectional characteristics of the whole population, while still preserving connectivity profiles unique to each subject. Using spectral decomposition techniques, we partition this joint graph into cortical parcellations at varying levels of granularity. Experimenting with resting-state functional connectivity, we show that the proposed framework is able to delineate more reproducible parcellations across different groups of subjects and can more accurately model the underlying data with respect to two popular ways of obtaining group-wise parcellations. We also report that individual subjects can be more reliably represented by our group-wise parcellations compared to the other approaches.

In Chapter 7, we provide a large-scale systematic comparison of connectivity-driven, anatomical, and random parcellation methods proposed for brain mapping purposes. We rely on rs-fMRI and widely-used quantitative evaluation techniques to evaluate 24 group-wise parcellation methods at different resolutions. We assess the parcellation accuracy from different aspects, including reproducibility, fidelity to the underlying data, similarity with task activation and myelo-/cyto-architecture of the cortex, as well as performance in network analysis. Taking the experimental results into consideration, we extensively review the strengths and limitations of the various methods over each other, with the aim of providing a guideline for choosing the most suitable parcellation for future connectome studies. We conclude that there is not a single `best' parcellation that can simultaneously address all the challenges and yield high performance measures for all the tested cases. 

\section{Future Work}
In this thesis, we developed both individual and group-wise parcellation methods -as well as evaluated many more- with the aim of providing some insight into different levels of representing the brain's functional and structural organisation. 

Analysing individual parcellations in collaboration with group-wise models can potentially enhance our capacity in understanding inter-subject variability. While we explored the variations between individuals with respect to subject-level parcellations, it could be highly interesting to locate cortical regions that are most consistent and/or least similar across subjects by comparing individual parcellations to a group representation. Similarly, a subject-to-group analysis from a network theoretic point of view could provide a reliable means to reveal the cortical areas with similar/different connectivity patterns. In this context, understanding the source of variability across subjects constitutes an additional challenge. While differences in connectivity can be linked to a possible impairment of cognitive abilities or brain functioning, these could also be associated with genetic variations, between-subject topological differences, or simply caused by imaging/processing artefacts or low SNR in the data~\cite{Langs14,Dubois16,Gordon16individual}. Given the proposed methods at multiple levels, studying this variability could be an interesting challenge to tackle as a follow-up research project. 

The results presented as part of this thesis may indicate parcellation techniques and/or resolutions that are more appropriate for the problem under investigation. For example, one can prefer a more homogeneous parcellation to define network nodes in connectome analysis, as they are more likely to represent regions with uniform patterns of connectivity and reduce information loss. Similarly, a more reproducible parcellation can be selected for comparing different groups of subjects, for instance, to identify how connectivity changes longitudinally. While our experiments only covered healthy subjects, one can use a data-driven parcellation as reference for analysing connectivity patterns in a cohort with a specific brain disorder (after learning a parcellation for this new population), assuming the organisation of the brain is not too altered, for example, due to existence of a tumour. This could help derive biomarkers in order to better understand disease-related differences in brain connectivity, and thus, constitutes one of our primary future directions. 

Another point worth noting is the fact that, not only the parcellation scheme itself, but its resolution might have an impact on network analysis, depending on the task at hand~\cite{arslan2017human}. A recent study suggests that increasing the parcellation resolution yields more reliable biomarkers for studying brain disorders~\cite{Abraham16}. Similarly, using more ROIs for network analysis is reported to improve the performance of age prediction tasks~\cite{Liem16}. This might be linked to the fact that parcellations with fewer ROIs may not be able to capture structural patterns of interest from the underlying data due to their resolutions. In this case, connectivity-driven parcellations provide a greater flexibility to study the impact of resolution on connectome analysis, as they allow to subdivide the cortex at different resolutions, as opposed to pre-computed parcellations/atlases with fixed resolutions. As a matter of fact, parcellation resolution should also be taken under consideration in future work that aims to identify biomarkers for studying brain disorders.


The parcellations we presented in this thesis can be used to model the functional/structural organisation of the brain and/or derive distinct features for network analysis. However, additional information might be required to enhance the information provided by rs-fMRI or dMRI alone. Recent evidence suggests that a single modality is too limited to reveal the complex structure of the cerebral cortex, which consists of a mosaic of multiple properties nested at different levels of detail~\cite{parisot2017flexible,Glasser16,Eickhoff15}. From a neuro-biological point of view, the integration of other modalities to the parcellation generation task may provide more accurate and robust cortical segregation of the cerebral cortex, as shown in the recently proposed multi-modal cortical parcellation~\cite{Glasser16}. A prospective future work therefore would be to use a similar technique and expand the current evaluation pipeline towards parcellations obtained from different modalities and their combinations. Last but not least, features obtained from multiple modalities may yield more robust biomarkers for the prediction of neuro-cognitive disorders. In this context, a recent study has shown that multimodal integration of anatomical features (such as cortical thickness) and functional connectivity improves brain-based age prediction~\cite{Liem16}.
 
Deep learning has recently gained a lot attention due to its notable performance in classification and pattern analysis tasks~\cite{plis2014deep}. These tasks are also critical for neuroimaging and connectomics research, and thus, making deep models attractive to the researchers working in the related fields. For example, deep learning is actively being used for exploring neurological disorders and shown to be an effective tool for identifying brain-based biomarkers (see~\cite{Vieira17} for a review), studying disease-related differences in populations~\cite{Parisot17}, and functional network modelling~\cite{Ktena17}. These works are mainly made possible by the advancements in \textit{geometric deep learning}, which allows to generalise structured deep models to non-Euclidean domains, such as graphs~\cite{bronstein2017geometric}. Since brain connectivity can be represented as a graph of interacting units (possibly derived from a parcellation), one of our future works is to use convolutional neural networks (CNNs) through geometric deep learning in order to explore within- and inter-subject connectivity. On the other hand, while CNNs that are supervised by anatomical parcellations are already being used for brain segmentation problems~\cite{Lee11,de2015deep}, learning a connectivity-based parcellation directly from the data is still not possible due to lack of a ground truth parcellation. Yet, unsupervised deep learning models, such as auto-encoders can be used to learn a latent representation of the underlying data, which in turn, may be used for parcellation purposes. For example such a non-linear representation can potentially be considered to substitute the Laplacian eigenmaps used in Chapter 5. Auto-encoders in conjunction with CNNs can be used to reconstruct the input brain signals (e.g. timeseries) in an unsupervised setting and feature maps obtained throughout such a deep model may reveal the intrinsic connectivity patterns and potentially be used to derive parcellations. 

\addcontentsline{toc}{chapter}{List of Publications}
\chapter*{List of Publications}

\begin{itemize}

\item \textbf{S. Arslan}, S. I. Ktena, A. Makropoulos, E. C. Robinson, D. Rueckert, S. Parisot, \textit{Human Brain Mapping: A Systematic Comparison of Parcellation Methods for the Human Cerebral Cortex}, NeuroImage, 2017, \textit{In Press}.

\item S. I. Ktena, \textbf{S. Arslan}, S. Parisot, D. Rueckert, \textit{Exploring Heritability of Functional Brain Networks with Inexact Graph Matching}, International Symposium on Biomedical Imaging (ISBI), 2017.

\item S. Parisot, B. Glocker, S. I. Ktena, \textbf{S. Arslan}, M. D. Schirmer, D.  Rueckert, \textit{A Flexible Graphical Model for Multi-modal Parcellation of the Cortex}, NeuroImage, 2017, NeuroImage, vol. 162, pp. 226-248, 2017. 

\item \textbf{S. Arslan}, S. Parisot, D. Rueckert, \textit{Boundary Mapping through Manifold Learning for Connectivity-Based Cortical Parcellation}. International Conference on Medical Image Computing and Computer Assisted Intervention (MICCAI), vol. 9900 of LNCS. Springer, pp. 115-122, 2016. -Received a travel award-

\item S. Parisot, \textbf{S Arslan}, J. Passerat-Palmbach, W. M. Wells, D. Rueckert, \textit{Group-wise Parcellation of the Cortex through Multi-scale Spectral Clustering}, NeuroImage, vol. 136, pp. 68-83, 2016. -Featured on issue cover-

\item \textbf{S. Arslan} and D. Rueckert, \textit{Multi-level Parcellation of the Cerebral Cortex Using Resting-State fMRI}, International Conference on Medical Image Computing and Computer Assisted Intervention (MICCAI), vol. 9351 of LNCS. Springer, pp. 47-54, 2015.

\item \textbf{S. Arslan}, S. Parisot, D. Rueckert, \textit{Joint Spectral Decomposition for the Parcellation of the Cerebral Cortex using Resting-State fMRI}, Information Processing in Medical Imaging (IPMI) Lecture Notes in Computer Science, vol. 9123, pp. 85-97, 2015. -Oral presentation-

\item S. Parisot, \textbf{S. Arslan}, J. Passerat-Palmbach, W. M. Wells III, D. Rueckert, \textit{Tractography-Driven Groupwise Multi-Scale Parcellation of the Cortex}, Information Processing in Medical Imaging (IPMI) Lecture Notes in Computer Science, vol. 9123, pp. 600-612, 2015.

\item \textbf{S. Arslan}, S. Parisot, D. Rueckert, \textit{Comparing Connectivity-based Groupwise Parcellations Generated from Resting-state fMRI and DTI Data: Preliminary results}, Symposium on Big Data Initiatives for Connectomics Research, London, 2015.

\item S. I. Ktena, \textbf{S. Arslan}, D. Rueckert, \textit{Gender Classification and Manifold Learning on Functional Brain Networks}," Symposium on Big Data Initiatives for Connectomics Research, London, 2015.

\item \textbf{S. Arslan}, S. Parisot, D. Rueckert, \textit{Supervertex Clustering of the Cerebral Cortex Using Resting-state fMRI}," Organization for Human Brain Mapping (OHBM), Honolulu, 2015.

\item \textbf{S. Arslan}, S. Parisot, D. Rueckert, \textit{How to Represent Subregions in a Parcellated Brain for fMRI Analysis?}," Organization for Human Brain Mapping (OHBM), Honolulu, 2015.

\end{itemize}


\bibliographystyle{abbrv}
\bibliography{bibliography/bibliography}

\end{document}